\newcommand{\Sp}{{\it Spitzer}}

\newcommand{\HST}{{\it HST}}

\newcommand{\gsim}{\gtrsim}

\documentclass{aa}

\usepackage{graphicx}
\usepackage{txfonts}
\begin{document}

\title{Clustering of red and blue galaxies around high-redshift 3C radio sources as seen by the Hubble Space Telescope}
\titlerunning{Clustering of galaxies around high-redshift 3C radio sources}

\author{Zohreh Ghaffari
  \inst{1}\fnmsep\thanks{Corresponding author:{ghaffari@astro.ruhr-uni-bochum.de}}
  \and
  Martin Haas\inst{1}
  \and
  Marco Chiaberge\inst{2,3}
  \and
  S. P. Willner\inst{4}
  \and
  Rolf Chini\inst{1,5}
  \and 
  Hendrik Hildebrandt\inst{1}
  \and\\
  Roberto de Propris\inst{6}
  \and
  Michael J. West\inst{7}
}

\institute{Astronomisches Institut, Ruhr--Universit\"at Bochum,
  Universit\"atsstra{\ss}e 150, 44801 Bochum, Germany
  \and
  AURA for the European Space Agency, Space Telescope Science Institute, 
  3700 San Martin Drive, Baltimore, MD 21218, USA
  \and
  Center for Astrophysical Sciences, Johns Hopkins University,
  3400 N. Charles Street, Baltimore, MD 21218, USA
  \and
  Harvard-Smithsonian Center for Astrophysics, 60 Garden St.,
  Cambridge, MA 02138, USA
  \and
  Instituto de Astronom\'{i}a, Universidad Cat\'{o}lica del
  Norte, Avenida Angamos 0610, Casilla
  1280 Antofagasta, Chile
  \and
  Finnish Centre for Astronomy with the European Southern Observatory,
  University of Turku, V\"ais\"al\"antie 20, 21500 Piikki\"o, Finland
  \and
  Lowell Observatory, 1400 West Mars Hill Road,
  Flagstaff, AZ 86001, USA
} 

\date{Received 12 September 2020; accepted 24 June 2021}


\abstract
    {
      To properly understand the evolution of high-redshift galaxy clusters, 
      both passive and star-forming galaxies have to be considered. 
      Here we study the clustering environment of 21  
      radio galaxies and quasars at $1<z<2.5$ from the third Cambridge catalog (3C). We use 
      optical and near-infrared Hubble Space Telescope images with a 2$\arcmin$ field-of-view, 
      where the filters encompass the rest-frame 4000\,\AA\ break. 
      Passive red and star-forming blue galaxies were separated in the color--magnitude diagram
      using a redshift-dependent cut derived from galaxy evolution models.
      We find that about 16 of 21 radio sources inhabit 
      a galaxy overdensity on scales of 250\,kpc ($30\arcsec$) projected radius. 
      The sample shows a diversity of red and blue overdensities and also sometimes a deficiency of blue galaxies in the center.
      The following tentative evolutionary trends are seen: extended proto-clusters with only weak overdensities at $z > 1.6$, 
      red overdensities at $1.2<z<1.6$, and 
      red overdensities with an increased deficit of central blue galaxies
      at $z<1.2$.
      Only a few 3C sources show a blue overdensity tracing active star-formation in the cluster centers; 
      this rarity 
      could indicate that
      the powerful quasar activity may quench star-formation in the vicinity of most radio sources.
      The derived number of central luminous red galaxies and the radial density profiles are comparable to those 
      found in local clusters, indicating that some 3C clusters are already mass-rich and compact. 
    }


    {}
    

    \maketitle
    %

    \section{Introduction} \label{sec:introduction}
    Detecting galaxy clusters at high-redshift ($z>1$) has proven to be a difficult task. Surveys based on X-ray imaging \citep{Fassbender2011a, Mehrtens2012, Mantz2018} or the Sunyaev-Zel'dovich effect \citep{Carlstrom11} readily detect clusters at $z < 1$, but only a few dozen have been found beyond this redshift (e.g., \citealt{Bleem2015,Hennig2017,Barrena2018,Khullar19,Aguado2019, Barrena20,Huang2020}). This may imply either a low space density of high-$z$ clusters or a lack of substantial intra-cluster gas that can be detected by the above methods.

    Studies using density enhancements and the red sequence (RS) method \citep{Gladders2000} again identify numerous clusters at $z < 1$, but only a small number above this redshift \citep{Eisenhardt08,Muzzin09,Wilson09,Papovich2010,Hildebrandt11}. This suggests that such clusters may indeed be rare at these redshifts and beyond or they might not be recognized because of their low contrast relative to the surrounding field or a lack of prominent red sequences.
    An alternative to blind searches for galaxy overdensities is to look near regions where one presumes that dense environments already exist, using tracers such as radio galaxies (RGs) and active galactic nuclei (AGN). These imply the presence of 
    supermassive black holes hosted by large galaxies and, therefore, regions where the average density of the universe 
 is likely to be high (e.g., \citealt{Miley08}).

 Numerous teams have found galaxy overdensities (ODs) around distant
 AGN and RGs (e.g., \citealt{Best00,Best03,Pentericci2000,Venemans07,Haas09,Falder10,Falder11,Galametz12,Wylezalek13,Hatch14,Kotyla16,Ghaffari17}). A similar strategy, using submillimeter galaxies as distant
mass beacons, has found proto-clusters up to $z \sim 5$ \citep{Steidel1998,Capak2011,Cai2017,Martinache2018,Miller2018, Miller2020}. The studies using RGs as mass beacons revealed that
galaxy ODs are found around about 50\% of the RGs. Some RGs appear to be
surrounded by fairly evolved clusters while others are not.
However, local AGNs 
do not appear to be strongly associated with clusters and groups (e.g., \citealt{Wethers21}). 
The lack of overdensities around some radio galaxies and AGNs might indicate lower mass groups or regions where the collapse of structure is not well advanced.

    To properly understand the evolution of high-$z$ galaxy clusters, both passive and star-forming galaxies have to be considered, and this should be done ideally with a complete cluster or proto-cluster sample. 
One suitable high-$z$ radio sample is the third Cambridge catalog (3C), selected at 178~MHz. 
There are 
two versions: a smaller one complete down to 10~Jy \citep{Laing83} and a larger one, 
used here, which is complete to about 7.5~Jy and contains 64 radio galaxies and quasars at $1 < z < 2.5$ \citep{Spinrad85}.
This latter sample has been used for two other recent studies.
Using optical and near-infrared images from a Hubble Space Telescope (\HST) snapshot survey of 22 sources from the 3C sample, 
\citet{Kotyla16} (K16) detected overdensities around half of 
these objects.
\citet{Ghaffari17} (G17) combined \Sp-IRAC imaging at 3.6 and 4.5 $\mu$m of the complete 3C sample with optical data from the PanStarrs1 survey and again found that about 50\% of radio sources in the high-$z$ 3C sample lie within overdensities, with colors typical of early-type galaxies, and none were found beyond $z=1.5$. Unlike previous studies, G17 
did not 
presume
a population of evolved RS galaxies.
The G17 IRAC cluster study revealed that the ODs should be measurable
with a 2$\arcmin$ field-of-view  such as provided by the \HST\ images
\citep{Hilbert16}. This led us to perform a new 
analysis of the environments of 3C sources at high redshifts 
based on the radial density profiles within the \HST\
images and separating red and blue galaxies. In comparison to the K16
study, we consider fainter galaxies and 
use different color and magnitude selection criteria.

    This paper is organized as follows.
    Section~\ref{sec:data}
    describes 
    the observations and data.
    Section~\ref{sec:results} presents our 
    main results, which include 
    color--magnitude diagrams of the galaxies, surface density maps and radial profiles around the 3C sources,
    statistical analysis of the detected over- and under-densities, 
    a classification scheme for 
    color-based cluster morphologies, 
    and a comparison with local clusters.
    Section~\ref{sec:discussion} discusses the diversity of red and blue galaxy clustering, and addresses evolutionary trends.
    Section~\ref{sec:summary} summarizes the results.

    Throughout this work we adopt a standard $\Lambda$CDM cosmology with $H_0 = 70$\,km\,s$^{\rm -1}$\,Mpc$^{\rm -1}$,
    $\Omega_{\Lambda} = 0.73$, and $\Omega_{m} = 0.27$ \citep{Spergel07}.
    With this cosmology, a projected angular distance of 30$\arcsec$
    corresponds within 3 percent to 250\,kpc over the entire redshift range $1 < z < 2.5$. 
    All magnitudes are AB, where zero mag corresponds to 3631\,Jy.

    \begin{table}
      \renewcommand{\thetable}{\arabic{table}}
      \caption{The sample.
      }
      \label{tab_sample}
      \footnotesize
      \begin{tabular}{lcccccc}
        \hline
        Name & RA (J2000) & Dec (J2000)& Redshift& Type$^1$\\
        \hline
        3C\,068.1 &  02 32  28.942 & $+$34 23  46.810  & 1.238 & Q \\
        3C\,186   &  07 44  17.539 & $+$37 53  17.387  & 1.067 & Q \\
        3C\,208.0 &  08 53  08.601 & $+$13 52  54.790  & 1.110 & Q \\
        3C\,210   &  08 58  10.047 & $+$27 50  52.955  & 1.169 &   G \\
        3C\,220.2 &  09 30  33.557 & $+$36 01  24.431  & 1.158 & Q \\
        3C\,230   &  09 51  58.894 & $-$00 01  27.206  & 1.487 &   G \\
        3C\,255   &  11 19  25.278 & $-$03 02  50.554  & 1.355 & Q \\
        3C\,257   &  11 23  09.474 & $+$05 30  17.986  & 2.474 &   G \\
        3C\,268.4 &  12 09  13.661 & $+$43 39  20.732  & 1.398 & Q \\
        3C\,270.1 &  12 20  33.951 & $+$33 43  11.503  & 1.532 & Q \\
        3C\,287   &  13 30  37.708 & $+$25 09  10.987  & 1.055 & Q \\
        3C\,297   &  14 17  24.098 & $-$04 00  48.834  & 1.406 &   G \\
        3C\,298   &  14 19  08.190 & $+$06 28  34.806  & 1.437 & Q \\
        3C\,300.1 &  14 28  31.274 & $-$01 24  07.546  & 1.159 &   G \\
        3C\,305.1 &  14 47  09.393 & $+$76 56  20.778  & 1.132 &   G \\
        3C\,322   &  15 35  01.332 & $+$55 36  53.039  & 1.681 &   G \\
        3C\,324   &  15 49  48.793 & $+$21 25  37.326  & 1.206 &   G \\
        3C\,326.1 &  15 56  10.168 & $+$20 04  21.079  & 1.825 &   G \\
        3C\,356   &  17 24  19.023 & $+$50 57  40.824  & 1.079 &   G \\
        3C\,432   &  21 22  46.277 & $+$17 04  37.718  & 1.785 & Q \\
        3C\,454.1 &  22 50  32.914 & $+$71 29  18.312  & 1.841 &   G \\

        \hline
      \end{tabular}\hspace*{-5.0cm}
      ~\\
      $~^1$ Type denotes quasar (Q) or radio galaxy (G), depending on
      whether or not broad emission lines have been identified in their spectra.
    \end{table}
    \section{Images, source catalogs, and photometry} \label{sec:data}



    We use the \HST\ sample of 22 radio sources from the 3C catalog
    originally 
    imaged by \citet{Hilbert16} 
    using the F606W and F140W filters. 
    This sample was randomly selected from the full 3C catalog based on visibility in the telescope scheduling.
    We excluded 3C418 
    because of its low galactic latitude and 
    severe stellar contamination. 
    The sample is shown in Table 1.
    K16 extracted a catalog of galaxies from the F140W images using Source Extractor \citep{Bertin96} and then derived colors from the F606W images within apertures measured from the segmentation image in F140W. All source photometry was placed on the AB system and corrected for Galactic extinction \citep{Schlafly11}. Stars were removed using Source Extractor's stellarity parameter.

    Our study uses a deeper source catalog, created with the Source Extractor
tool in the same manner as described by K16. The catalog consists of two
samples, one detected in both filters (6371 sources) and one in IR only
(2484 sources). The full catalog contains about 450 galaxies per 3C field.

    \begin{figure}
      \vspace{-5mm}
      \includegraphics[width=0.98\linewidth]{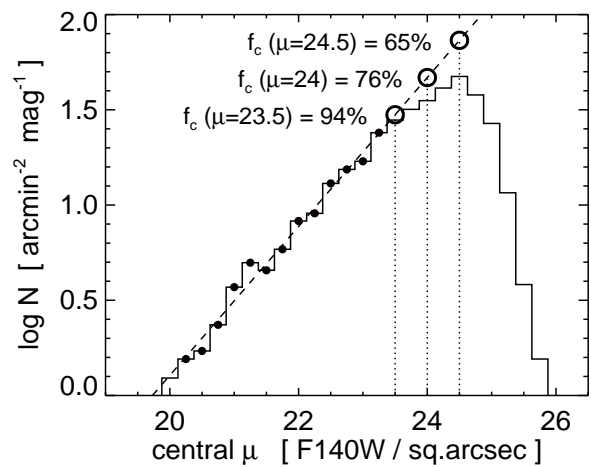}
      \caption{ 
          Log($N$) versus central surface brightness $\mu$ for F140W. 
          The dashed line is a linear fit for $20 < \mu < 23.25$ (small dots).  
        At $\mu \ge 23.5$, the histogram declines below the dashed line due to incompleteness.
        The completeness fractions are calculated as $f_c = 10^{\rm data} / 10^{\rm fit}$
        for three values of $\mu$ as labeled and marked with the dotted vertical lines and open circles at the intersection with the dashed line.}
      \label{fig:compl_1}
    \end{figure}

    \begin{figure} 
      \vspace{-5mm}
      \includegraphics[width=0.98\linewidth]{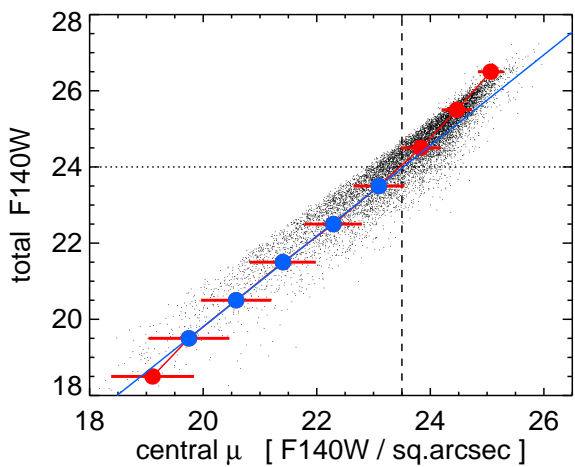}
      \caption{Total F140W magnitude vs. $\mu$. Each galaxy is plotted by a tiny black dot, altogether forming the ``galaxy cloud''.
        The vertical dashed line marks the surface brightness completeness limit at $\mu_{cl} \sim 23.5$. 
        Because the galaxy cloud does not exhibit a sharp boundary at this completeness limit, we proceed as follows: 
        The thick red and blue dots mark the average and standard deviation of $\mu$ in the total F140W bins (e.g., $18 < F140W < 19$ plotted at $F140W=18.5$).
        The solid blue line is a linear fit through the thick blue dots, which are brighter than $\mu_{cl}$.
        The intersection of the solid blue line with $\mu_{cl}$  yields a total F140W value $\sim$24 (horizontal dotted line).
        At F140W $\sim$ 24, a large fraction of the galaxy cloud lies right of $\mu_{cl}$ and 
        therefore suffers from surface brightness incompleteness.
        The total F140W completeness limit (100\%) is likely about 1 mag brighter at 23 mag, 
        where the galaxy cloud is brighter than 
        $\mu_{cl}$.}
      \label{fig:compl_2}
    \end{figure}

    \begin{figure}
      \vspace{-5mm}
      \includegraphics[width=\linewidth]{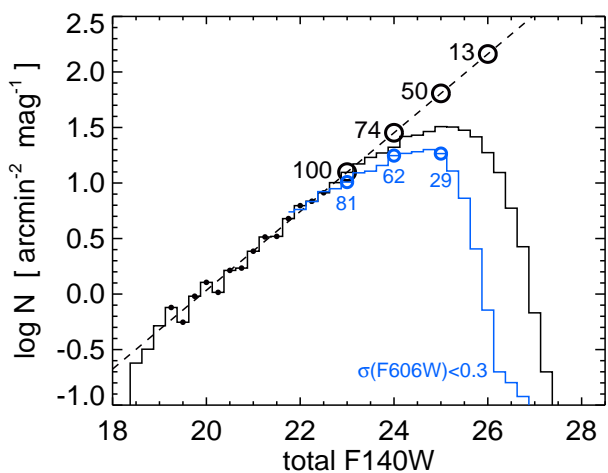}
      \caption{ 
          Log($N$) versus total F140W  magnitude
        for all galaxies detected in F140W (black histogram). 
        To fit the dashed line, only data points brighter than 22.5 mag were used (small dots).
        The black numbers indicate the completeness fractions $f_c = 10^{\rm data} / 10^{\rm fit}$ in percent.
        The completeness limit $f_c \sim 100$\% lies at 23 mag.
        To estimate the completeness for objects with colors, 
        the blue histogram shows galaxies detected in both filters.
      }
      \label{fig:compl_3}
    \end{figure}



    To estimate detection and completeness limits in our dataset, we followed the procedure described by \citet{Andreon2000}. 
  In contrast to unresolved stars, galaxies are extended and at a given total magnitude 
  the surface brightness decreases with increasing source size; 
  therefore the source detection actually depends on the detectable surface brightness.
    For each galaxy we measured the total magnitude from Source Extractor and a central surface brightness within a 5 pixel aperture 
    (equivalent to 0$\farcs$2 in F606W and 0$\farcs$64 in F140W). 
    Figure~\ref{fig:compl_1} shows
    the number N of detected galaxies as a function of central surface brightness $\mu$.
    We fit a straight line to the log(N) vs $\mu$ plot at high surface brightness; 
    deviations from this relation at about 23.5 mag arcsec$^{-2}$ in F140W 
    indicate the onset of incompleteness
    in the initial detection of objects. 
    As Fig.~\ref{fig:compl_2} shows, the surface brightness incompleteness means that we start to miss objects at about F140W $\sim$ 24 mag. 
    The true 100\% completeness limit is likely to be about one mag brighter, where the entire galaxy cloud is brighter than the central surface-brightness limit.
    Fig.~\ref{fig:compl_3} shows the equivalent to Fig.~\ref{fig:compl_1} for galaxies detected in F140W and F606W.
    Table~\ref{tab_completeness} summarizes our findings. As a final test, we introduced artificial galaxies 
    in our images using the IRAF
    {\it artdata} program (a mixture of half spheroids and half disks) and repeated the detection procedures. 
    The resulting incompleteness fractions are consistent with those shown in Table~\ref{tab_completeness}.
    For comparison, 
    an $L^*$ galaxy has F140W $\sim$ 22 mag at $z \sim 1-1.5$ (e.g., \citealt{Gabasch04}).


    \begin{figure*}
      \hspace{-5.0mm}\includegraphics[width=0.380\linewidth,clip=true]{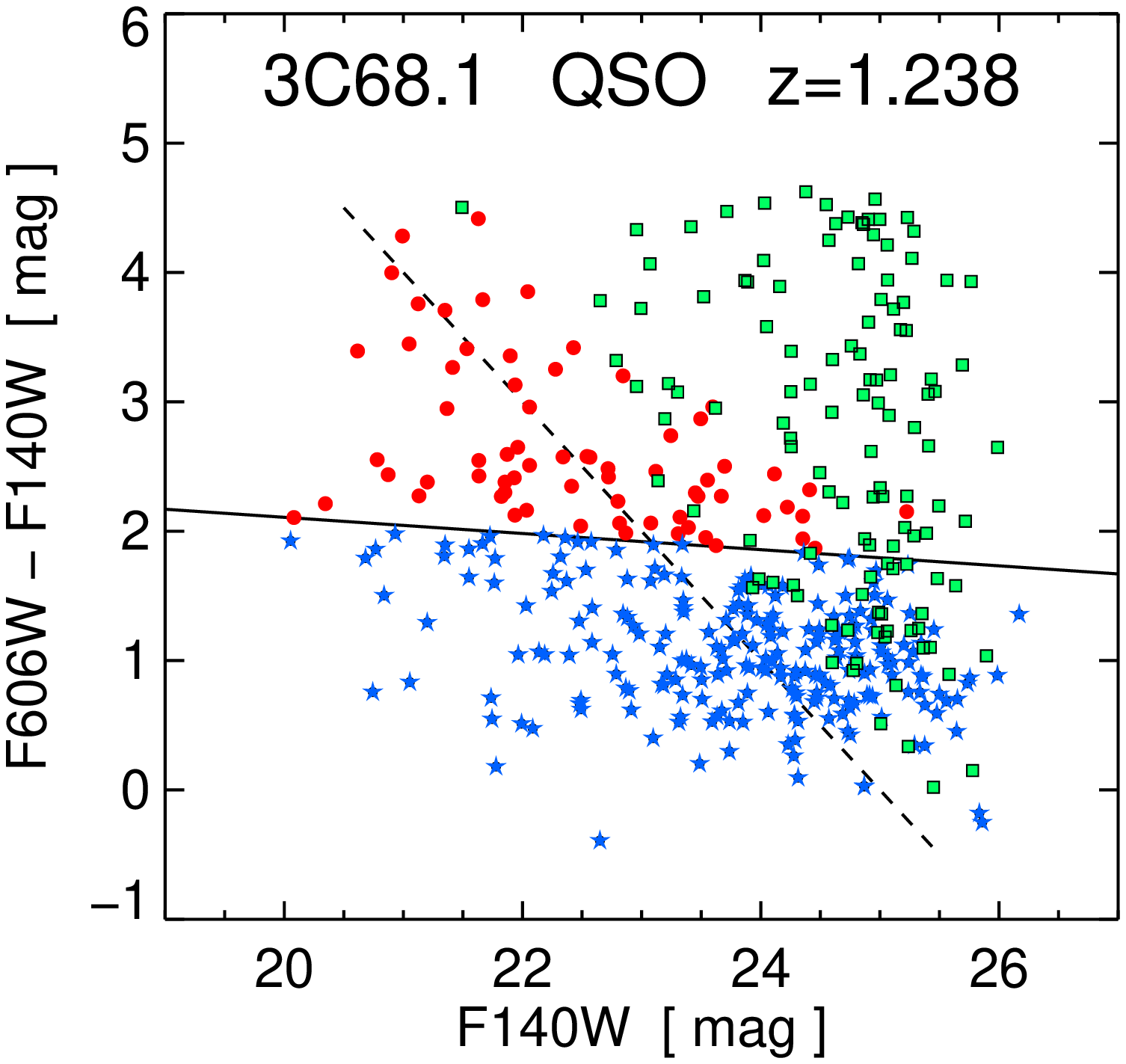}
      \hspace{-0.0mm}\includegraphics[width=0.295\linewidth,clip=true]{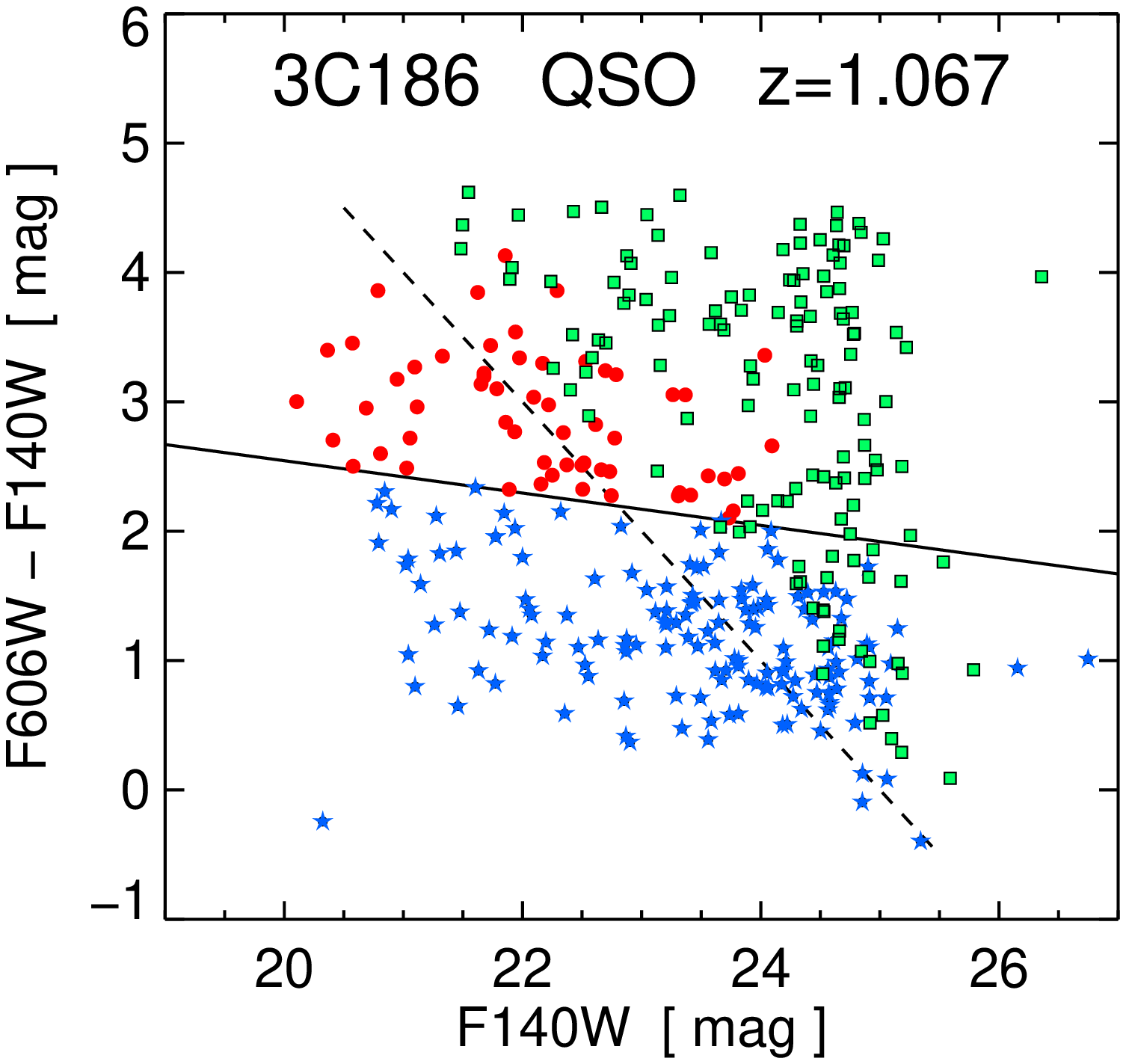}
      \hspace{-0.0mm}\includegraphics[width=0.295\linewidth,clip=true]{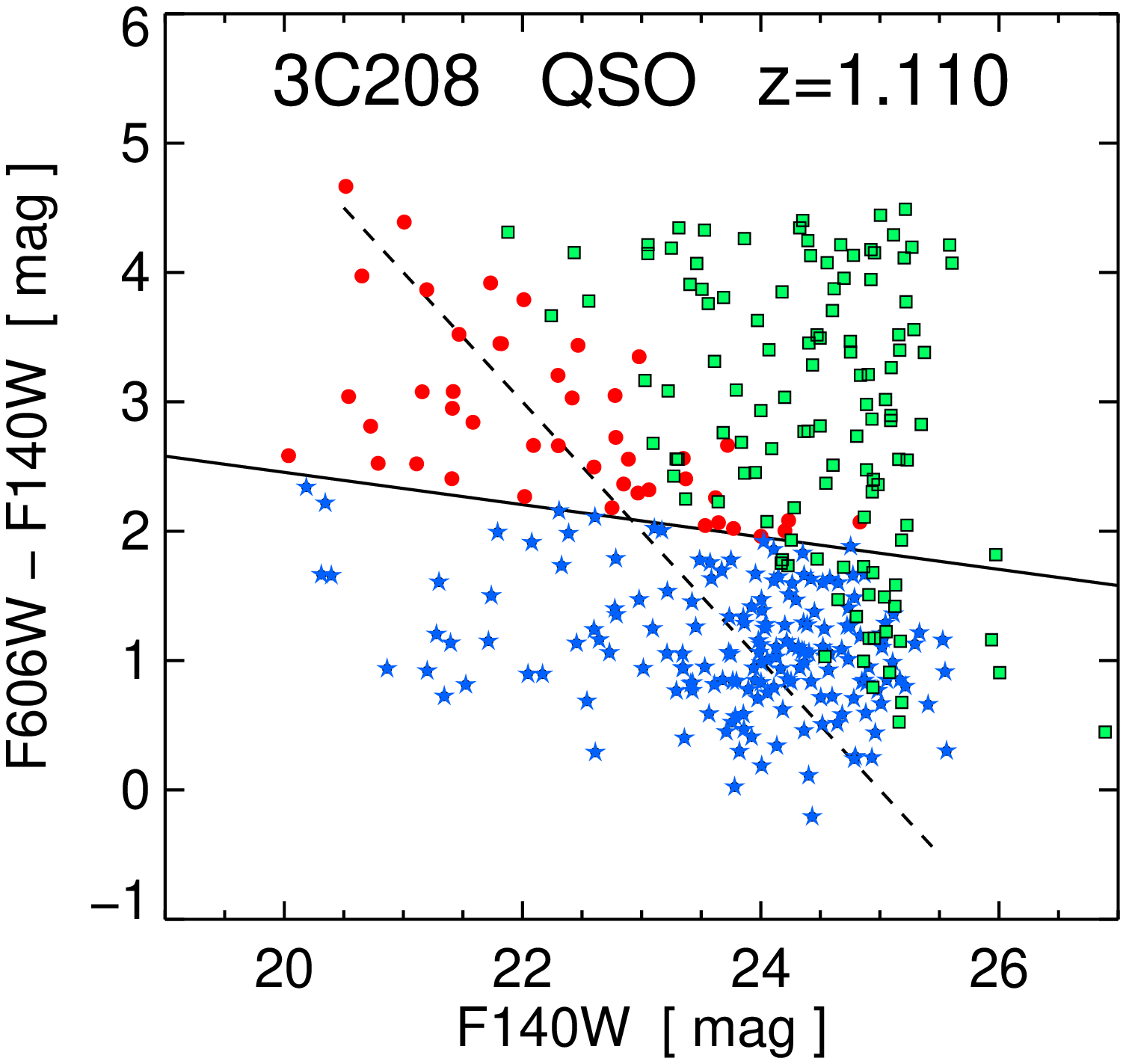}

      \hspace{-5.0mm}\includegraphics[width=0.380\linewidth,clip=true]{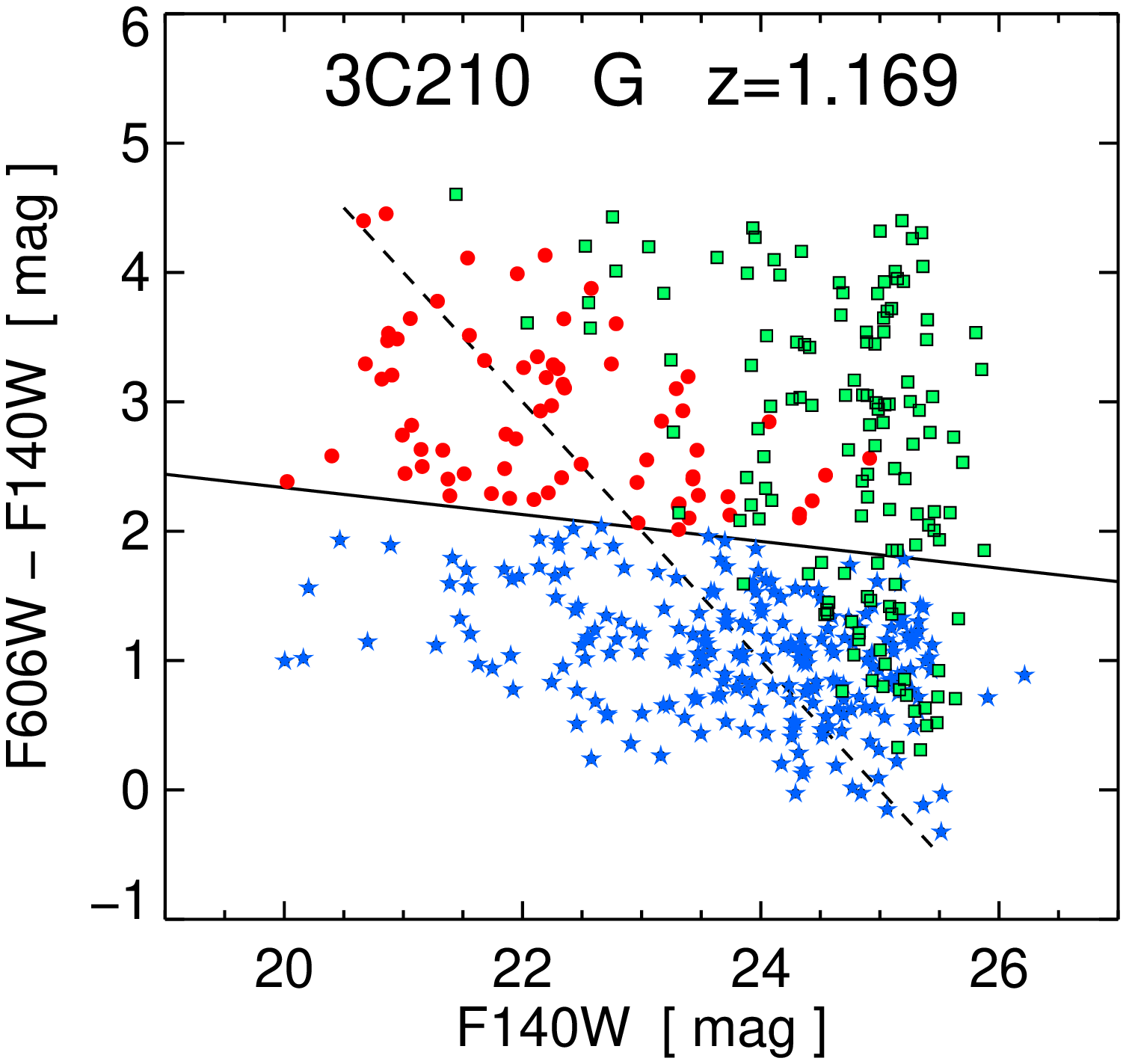}
      \hspace{-0.0mm}\includegraphics[width=0.295\linewidth,clip=true]{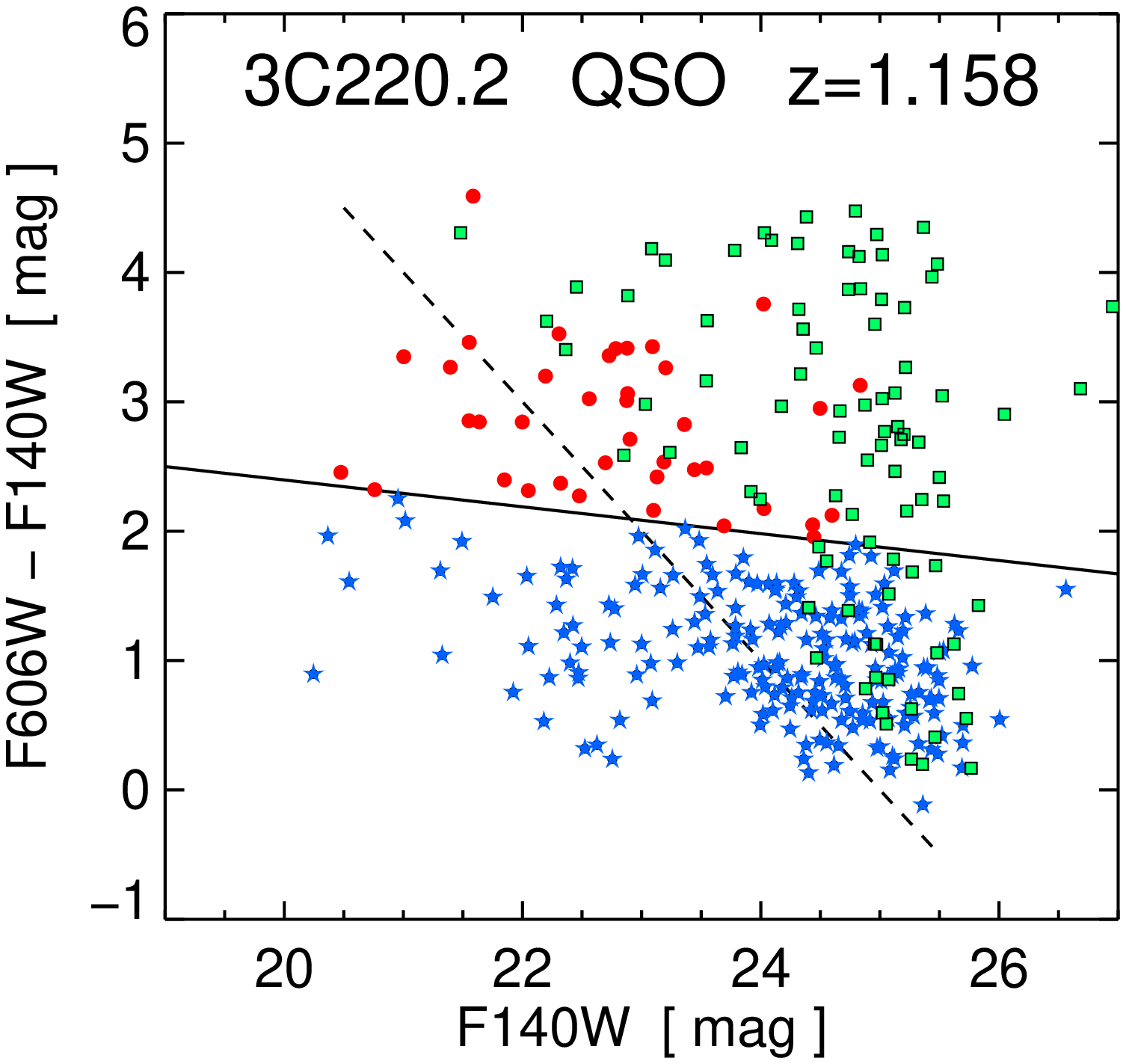}
      \hspace{-0.0mm}\includegraphics[width=0.295\linewidth,clip=true]{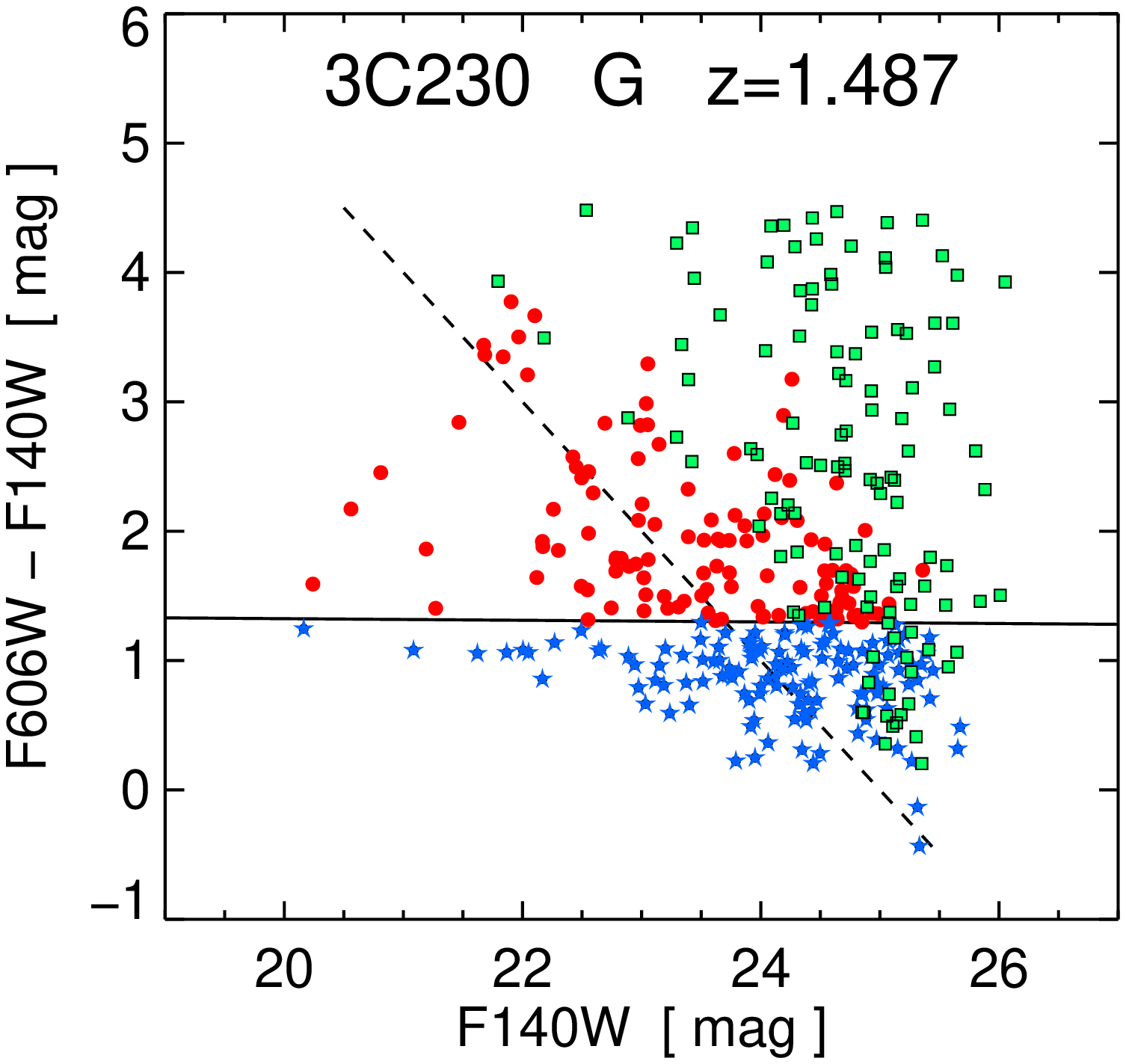}

      \hspace{-5.0mm}\includegraphics[width=0.380\linewidth,clip=true]{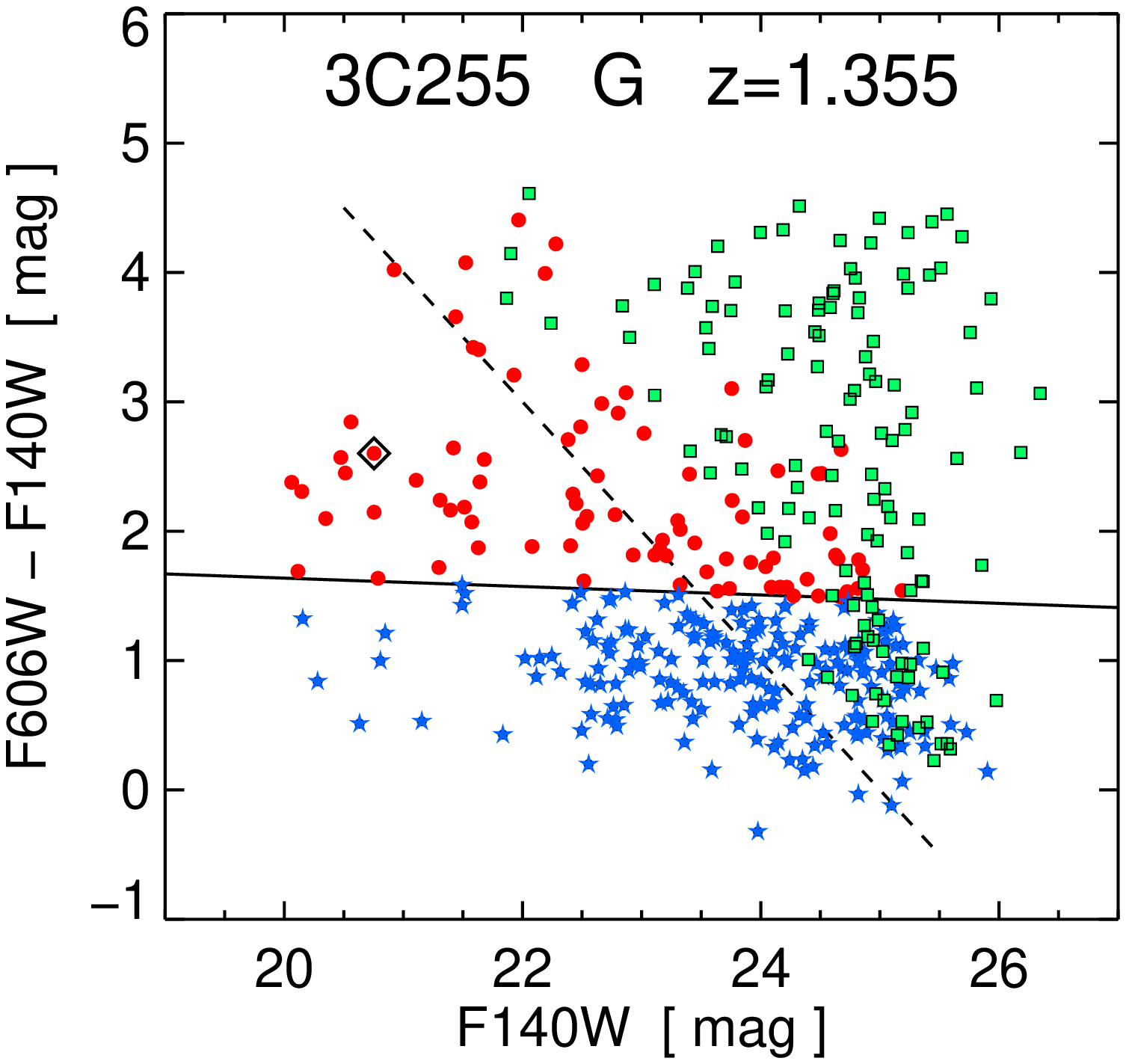}
      \hspace{-0.0mm}\includegraphics[width=0.295\linewidth,clip=true]{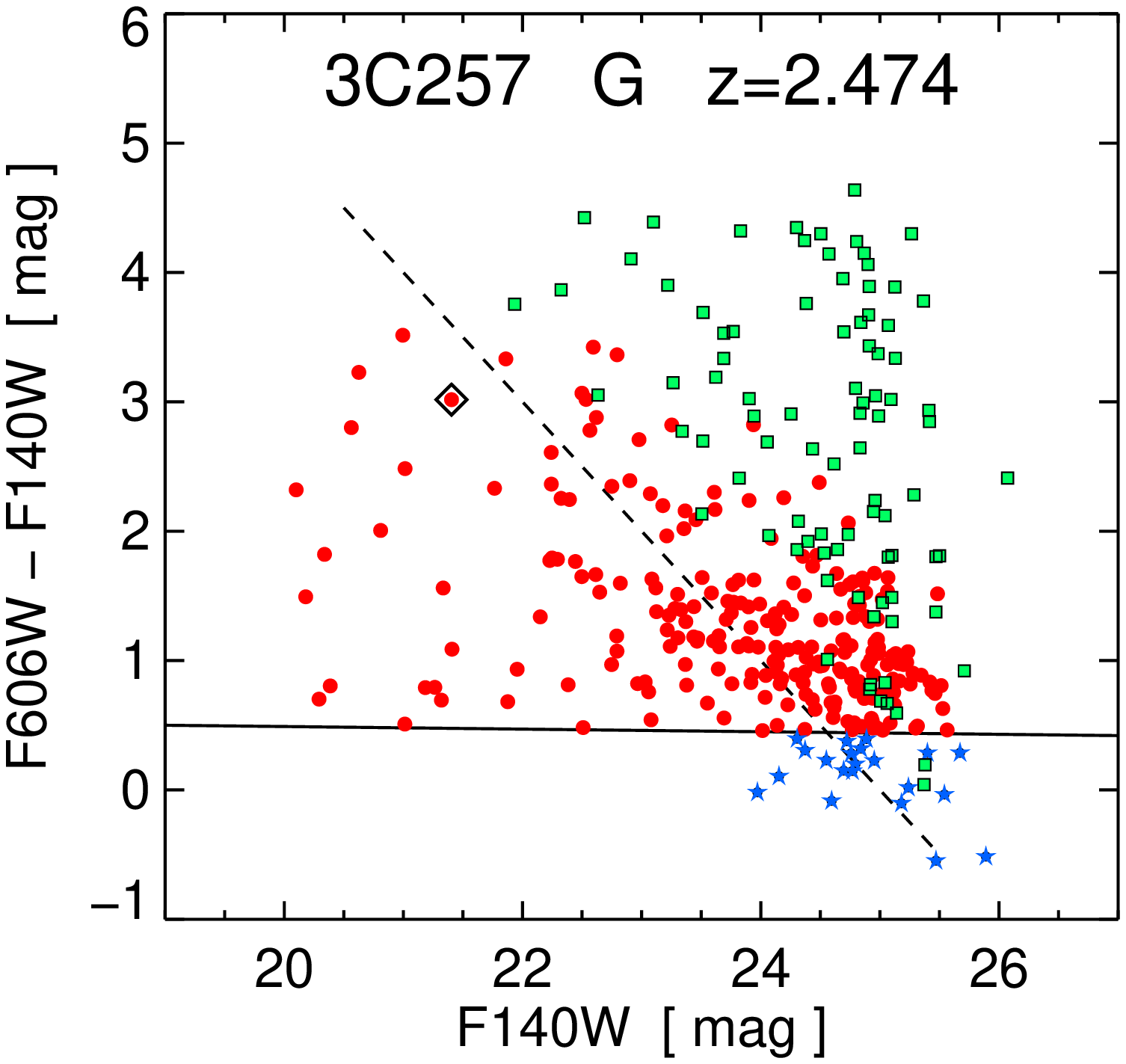}
      \hspace{-0.0mm}\includegraphics[width=0.295\linewidth,clip=true]{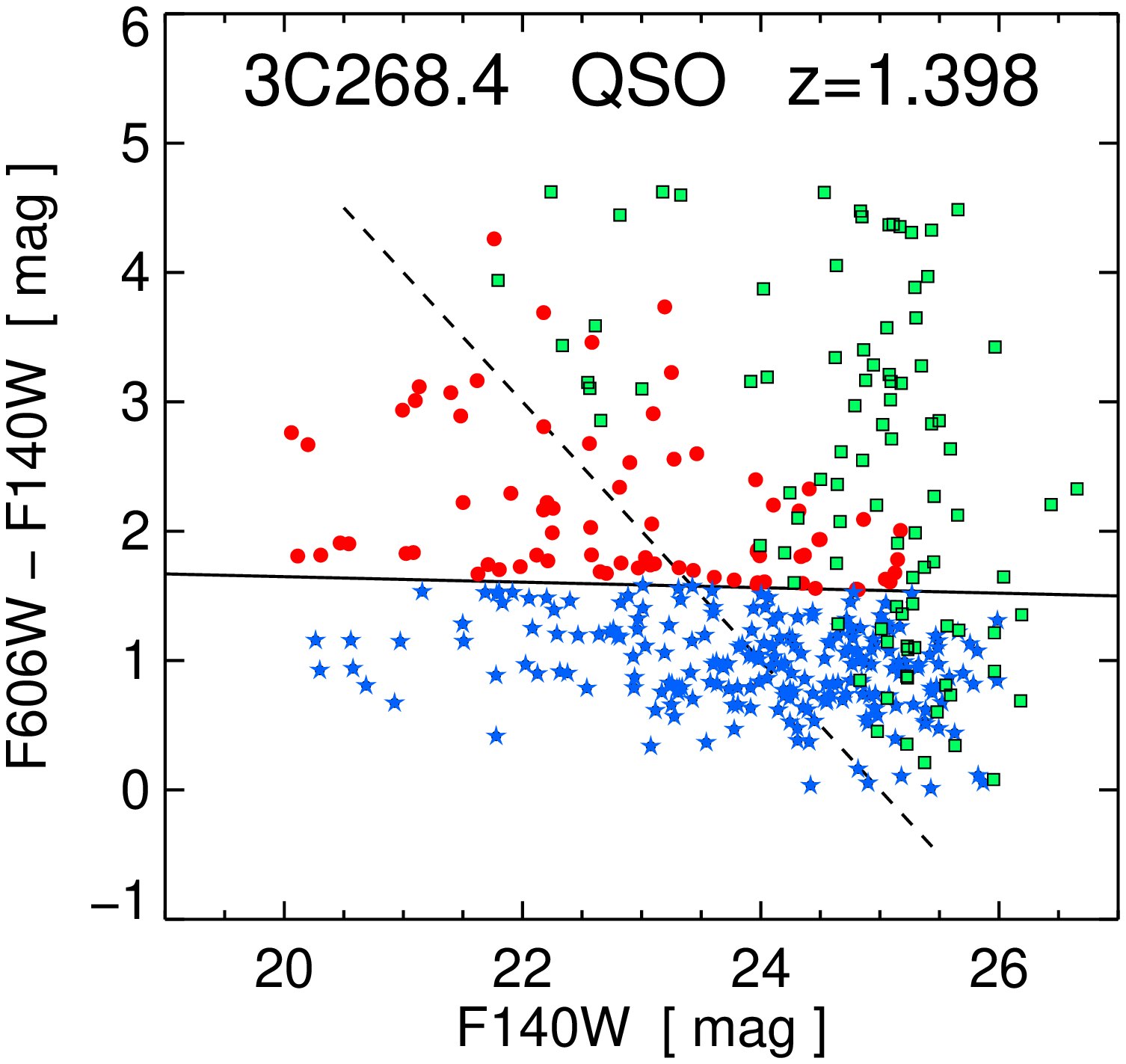}

      \hspace{-5.0mm}\includegraphics[width=0.380\linewidth,clip=true]{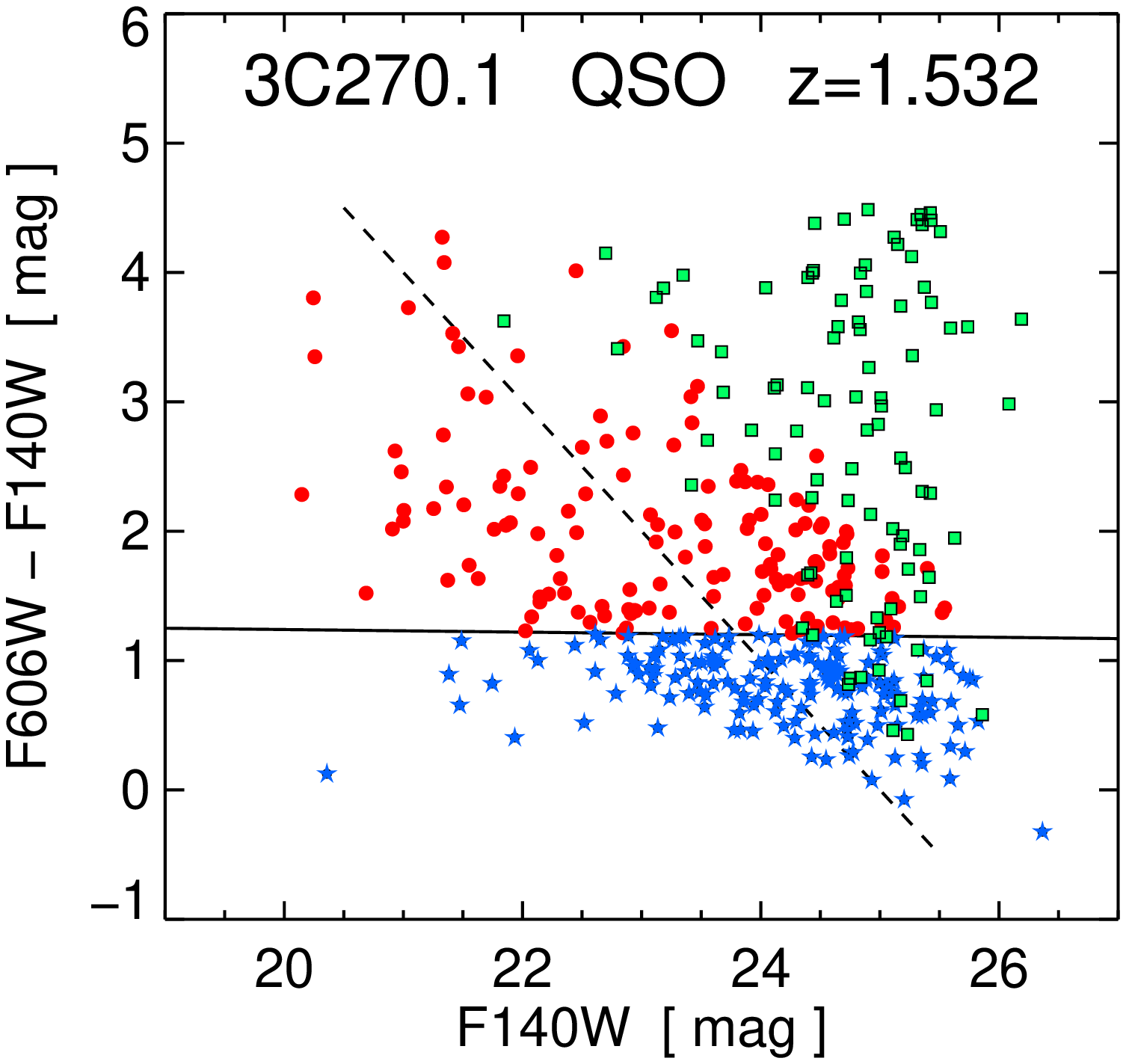}
      \hspace{-0.0mm}\includegraphics[width=0.295\linewidth,clip=true]{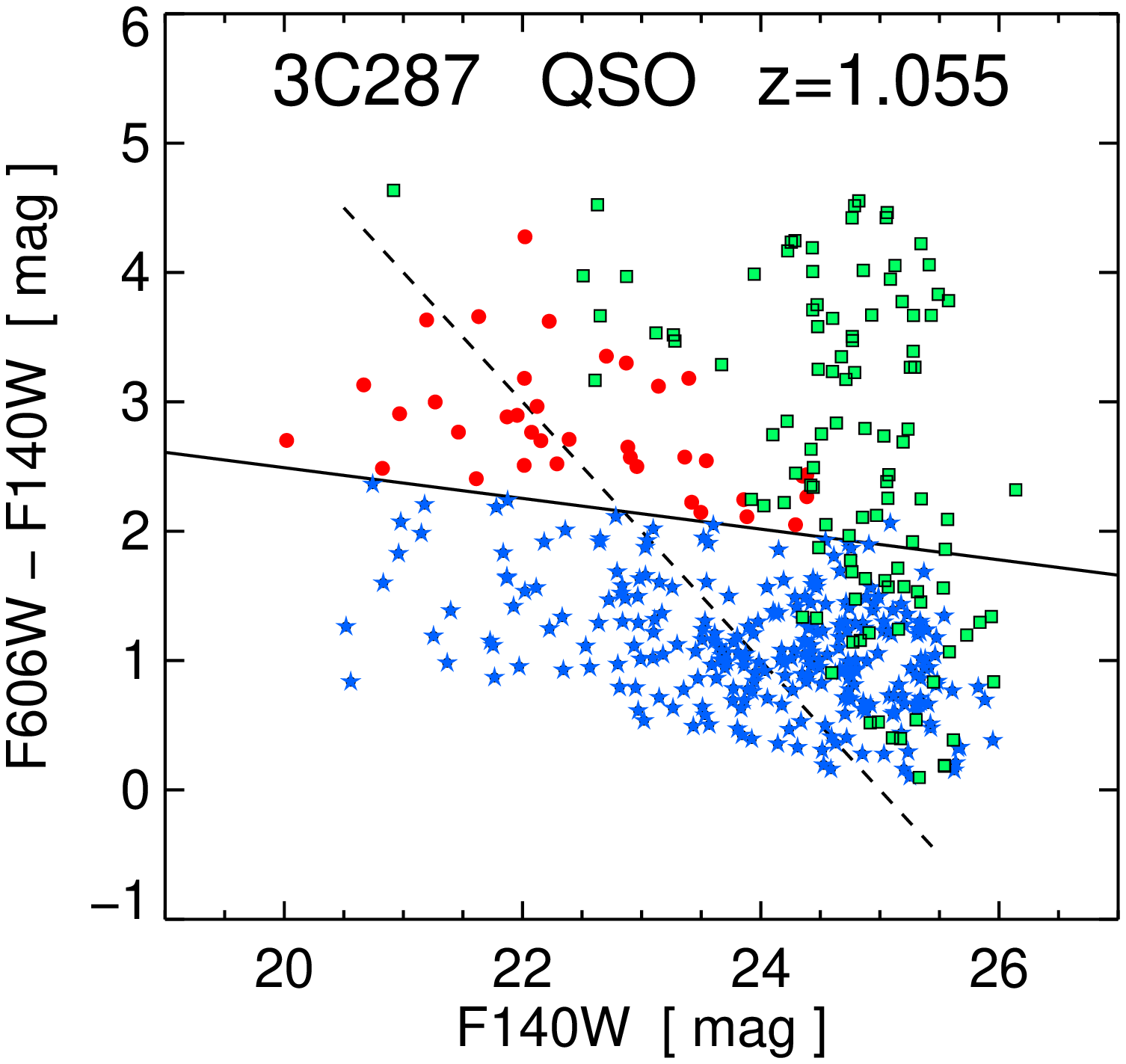}
      \hspace{-0.0mm}\includegraphics[width=0.295\linewidth,clip=true]{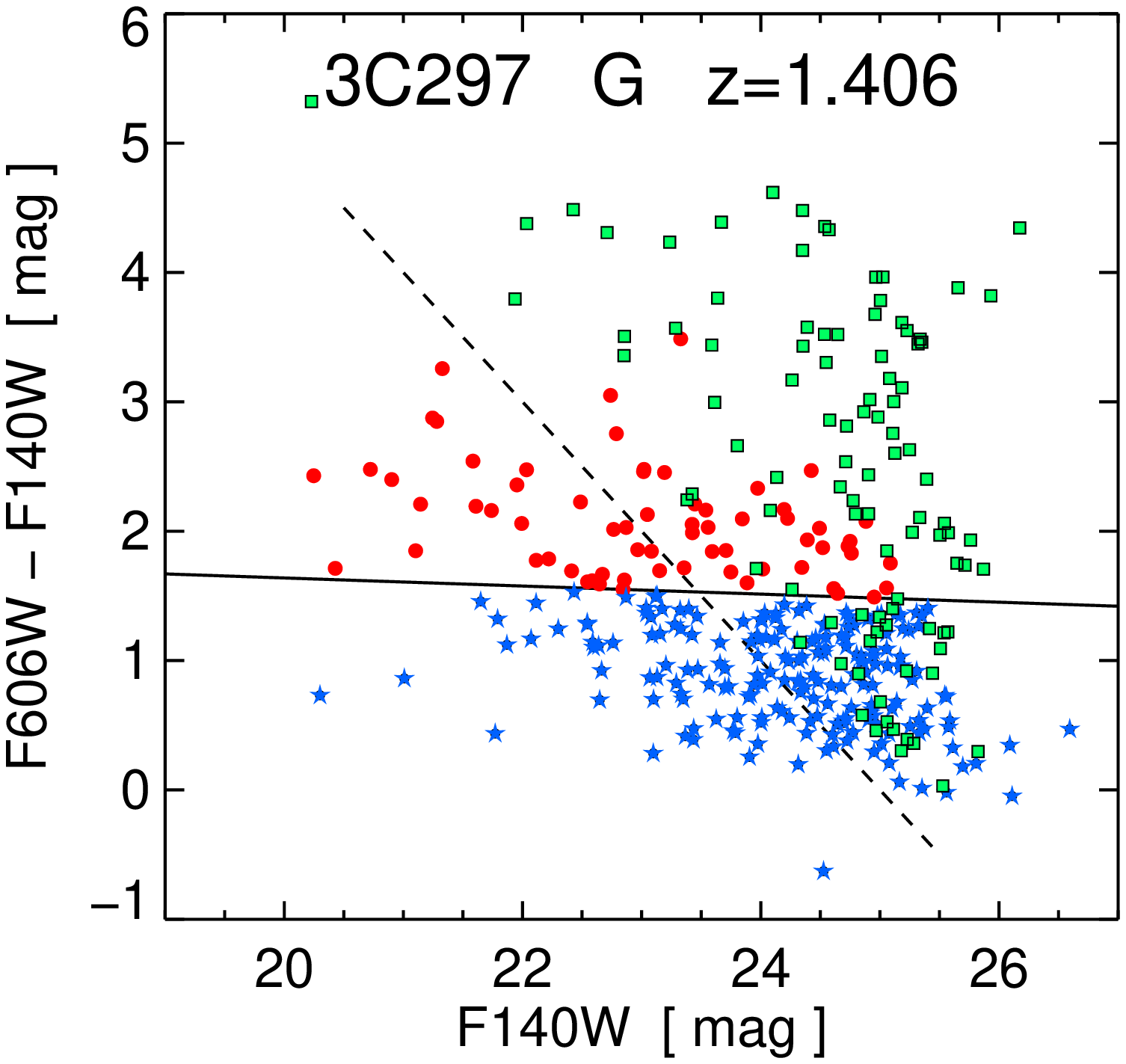}

      \caption{Color--magnitude diagrams (CMDs) for the 3C fields. 
        The labels list the 3C name, type (galaxy or quasar), and redshift.
        The 3C source is shown  
        as a black diamond 
        if it fits into the plot range (3C\,255, 3C\,257, 3C\,322, 3C\,326.1).
        Red dots and blue stars denote sources in the two-filter catalog, separated by the redshift-dependent color cut (solid black line).
        The F140W error bars are typically smaller than the size of the symbols but slightly exceed them at the faint end ($1\sigma <0.3$\,mag).
        The color error bars may be larger, reaching $1\sigma = 0.45$\,mag, 
        but 
        are not plotted here to avoid confusion.
        The green symbols mark the sources of the single-filter sample, henceforth denoted the $IR$-$only$ sample, 
        most of which are fainter than 24\,mag.
        Although their colors are unknown, to illustrate how numerous these sources are they have been randomly assigned colors to fill the upper right 
        quadrant of the CMD.
        The dashed line marks the 75\% color completeness limit, 
        derived from the 
        3$\sigma$ detected two-filter sample at F606W = 25 mag and F140W = 23 mag 
        (Table~\ref{tab_completeness}) and assuming a slope of $-$1.
      }
      \label{fig:cmd}
    \end{figure*}

    \addtocounter{figure}{-1}

    \begin{figure*}
      \hspace{-5.0mm}\includegraphics[width=0.380\linewidth,clip=true]{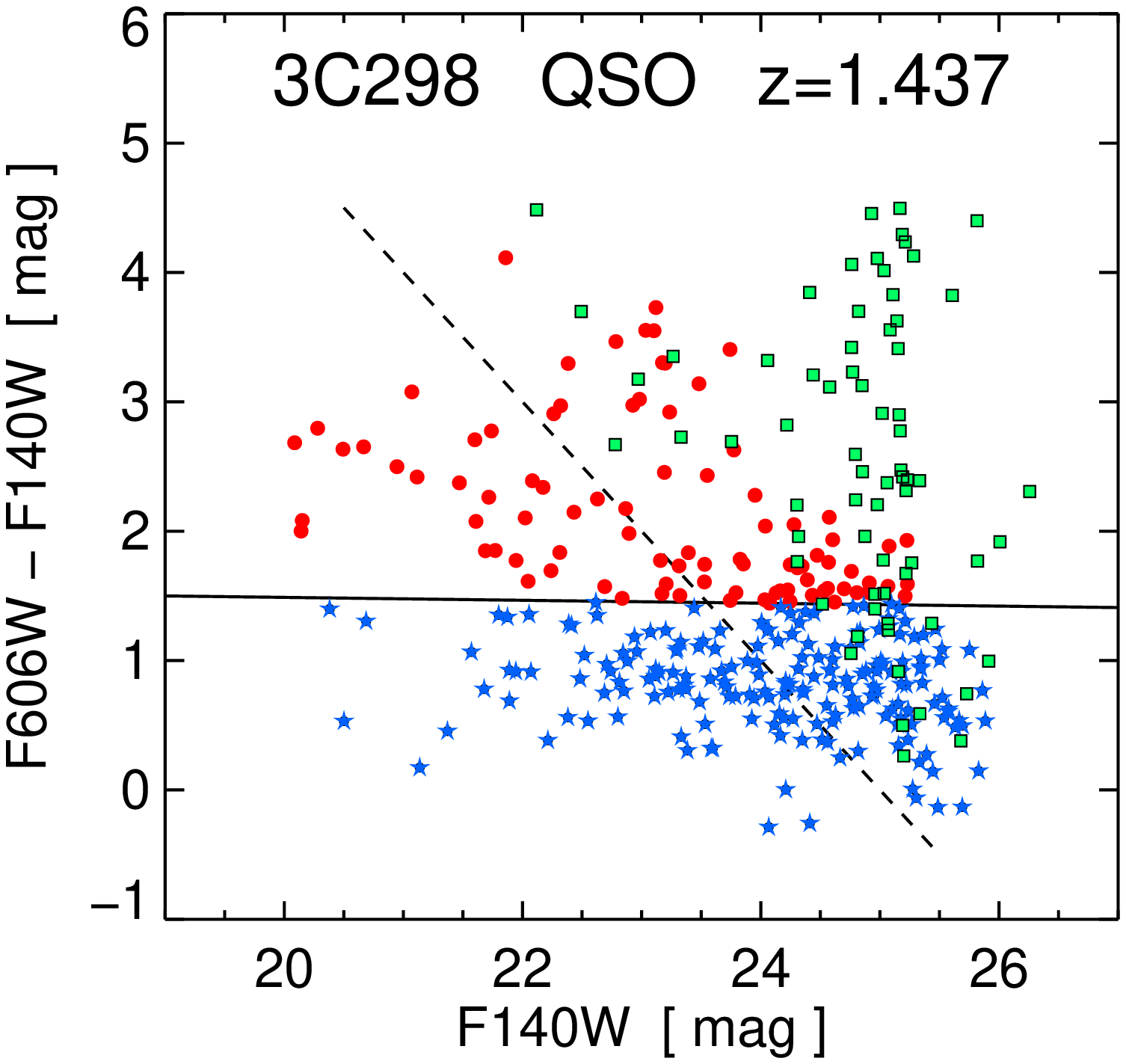}
      \hspace{-0.0mm}\includegraphics[width=0.295\linewidth,clip=true]{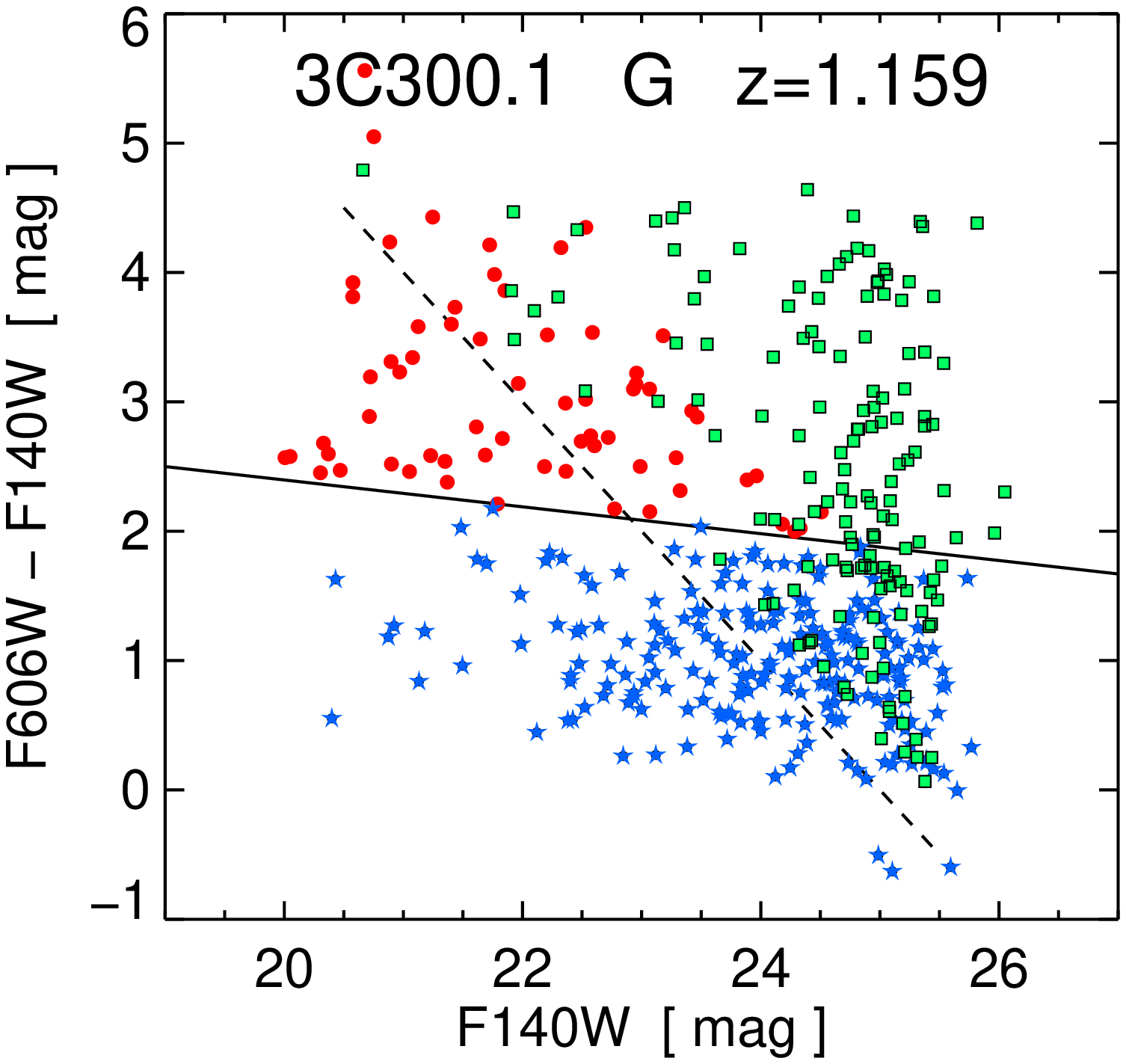}
      \hspace{-0.0mm}\includegraphics[width=0.295\linewidth,clip=true]{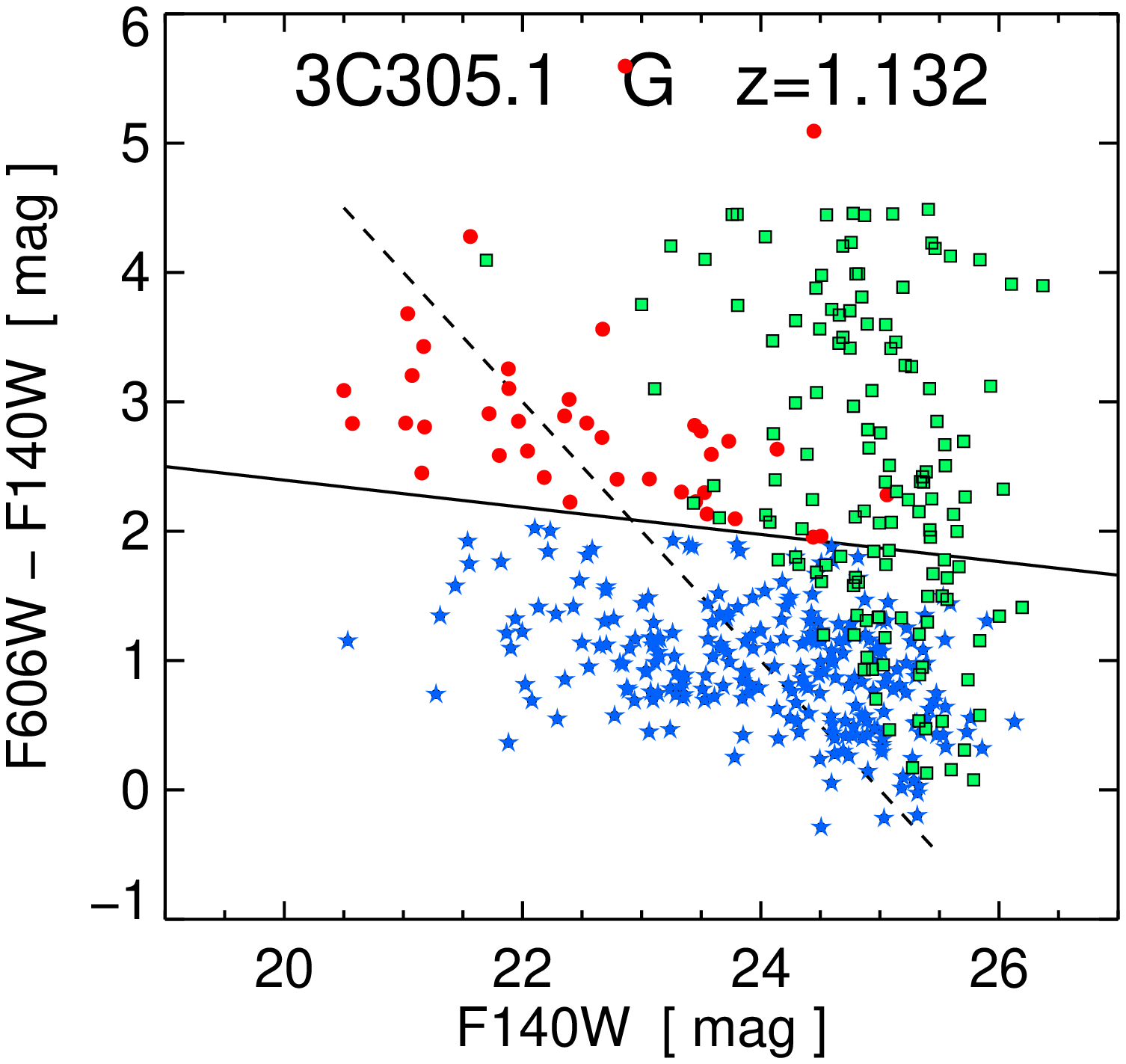}

      \hspace{-5.0mm}\includegraphics[width=0.380\linewidth,clip=true]{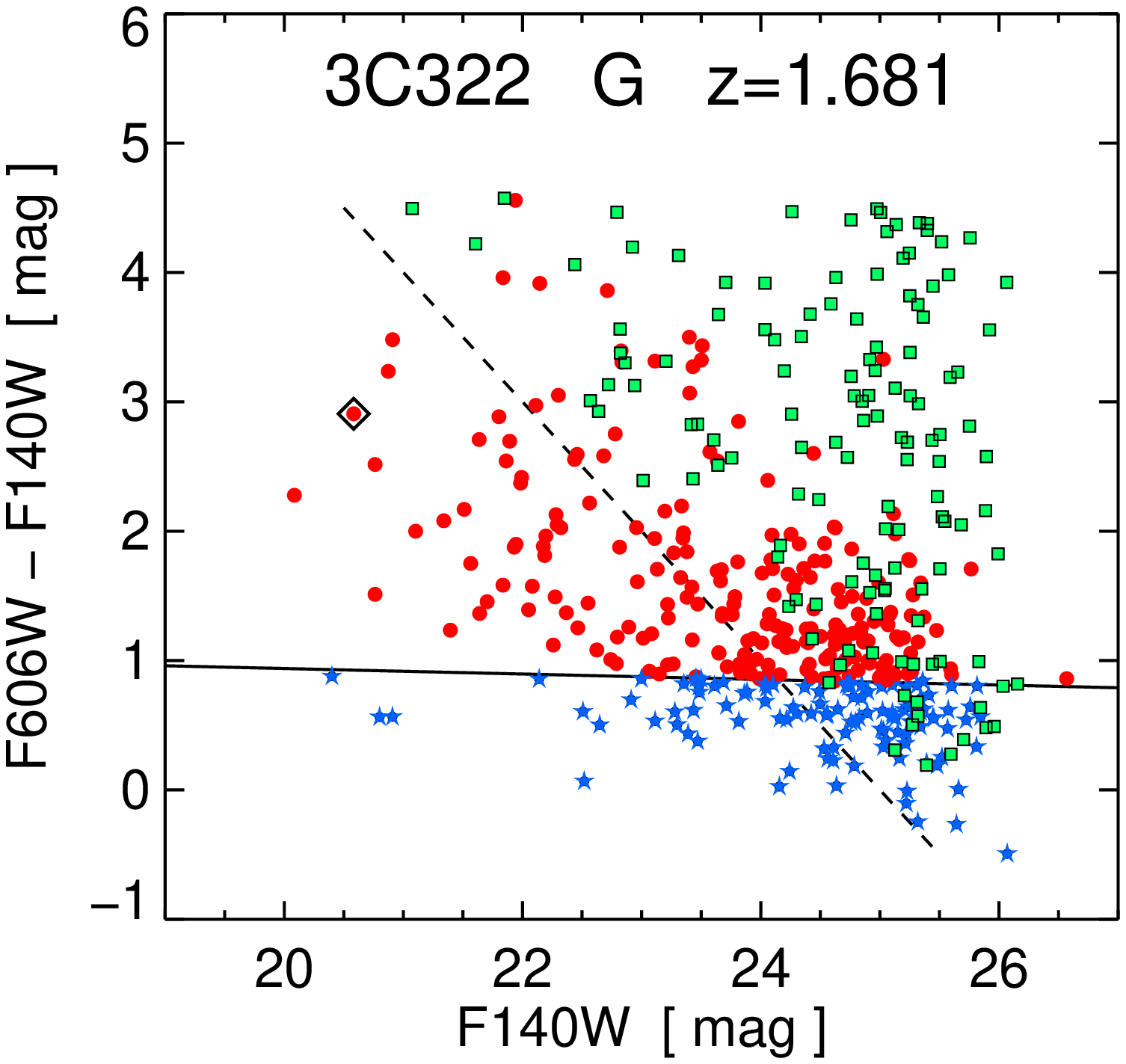}
      \hspace{-0.0mm}\includegraphics[width=0.295\linewidth,clip=true]{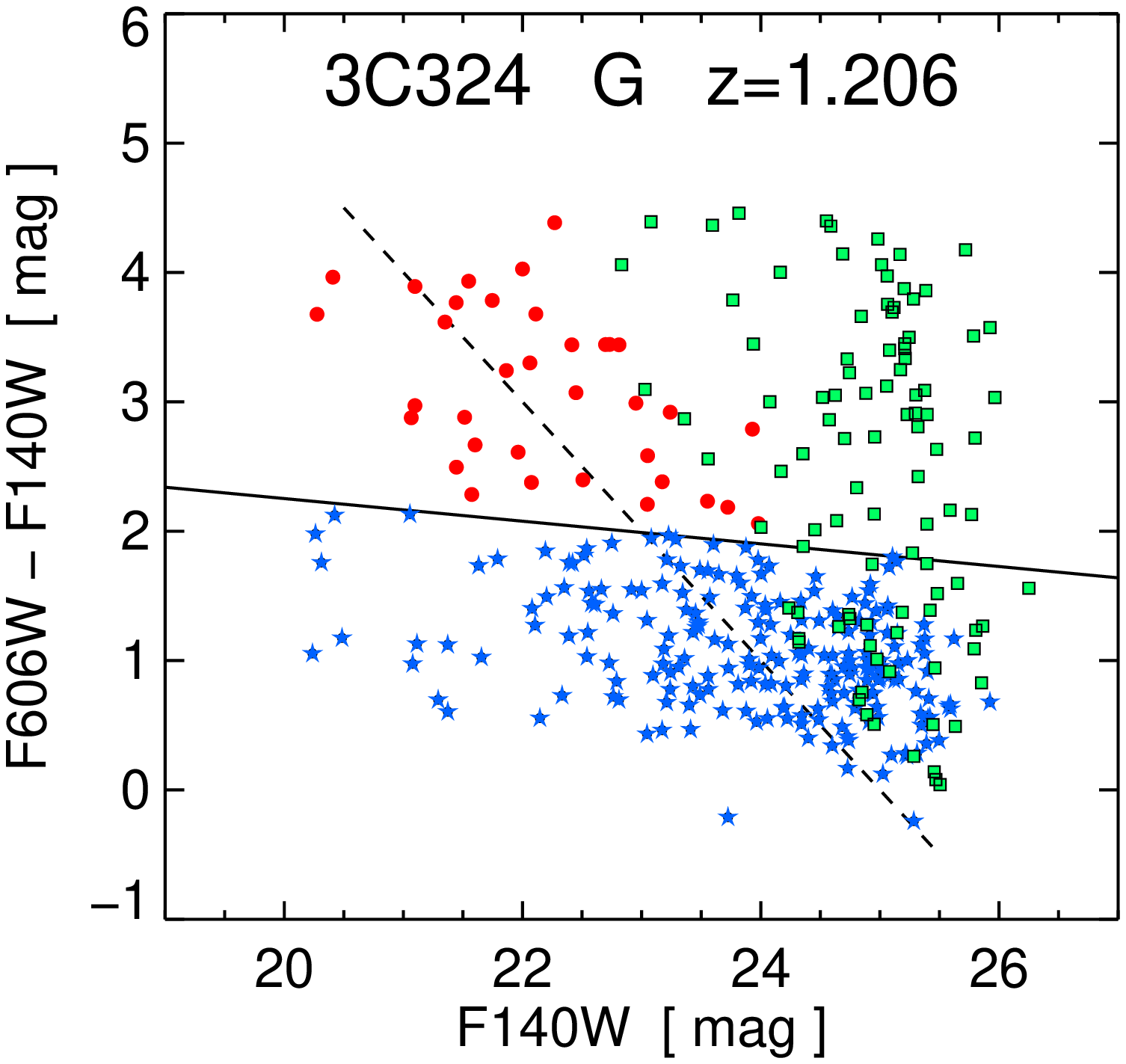}
      \hspace{-0.0mm}\includegraphics[width=0.295\linewidth,clip=true]{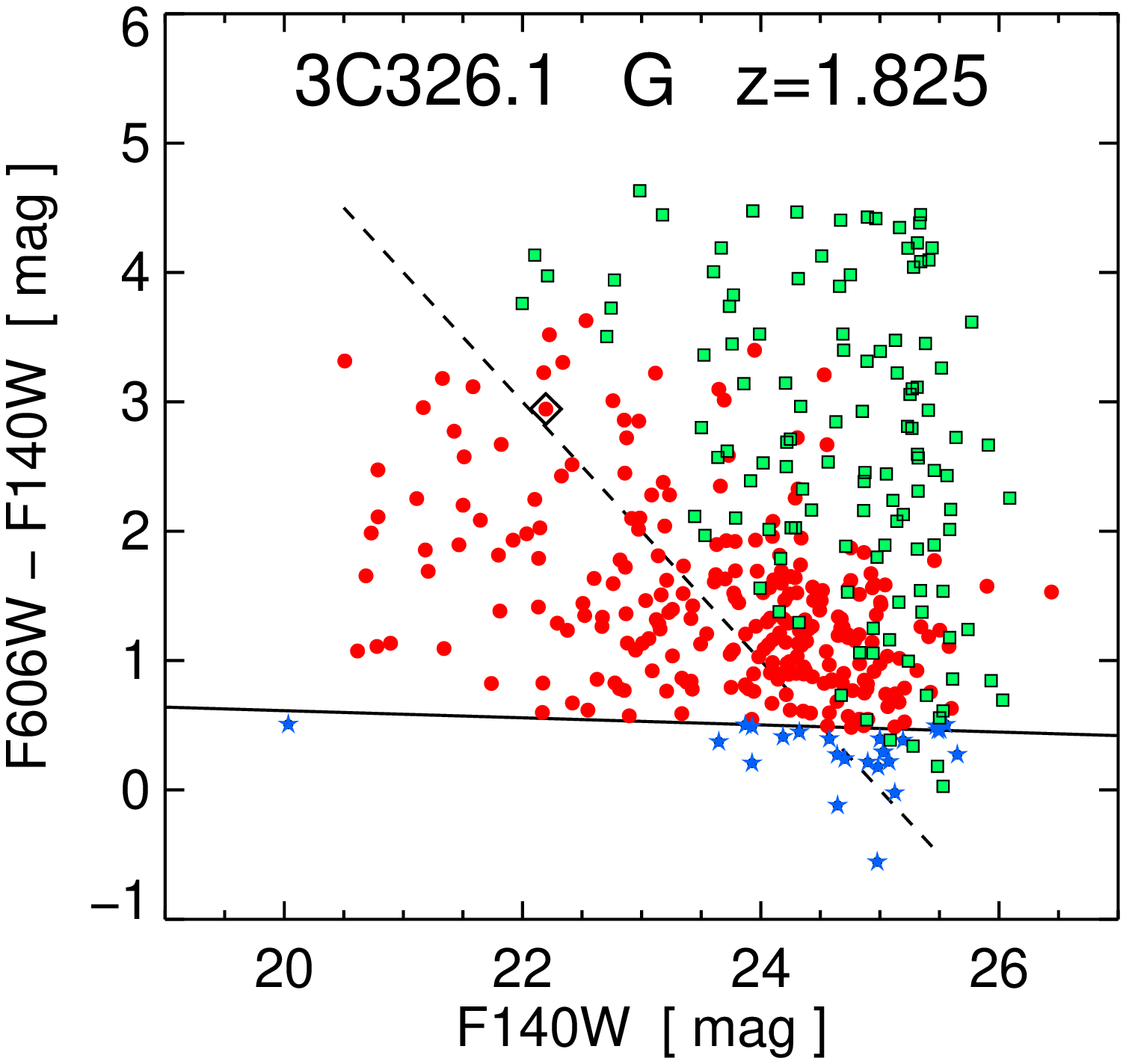}

      \hspace{-5.0mm}\includegraphics[width=0.380\linewidth,clip=true]{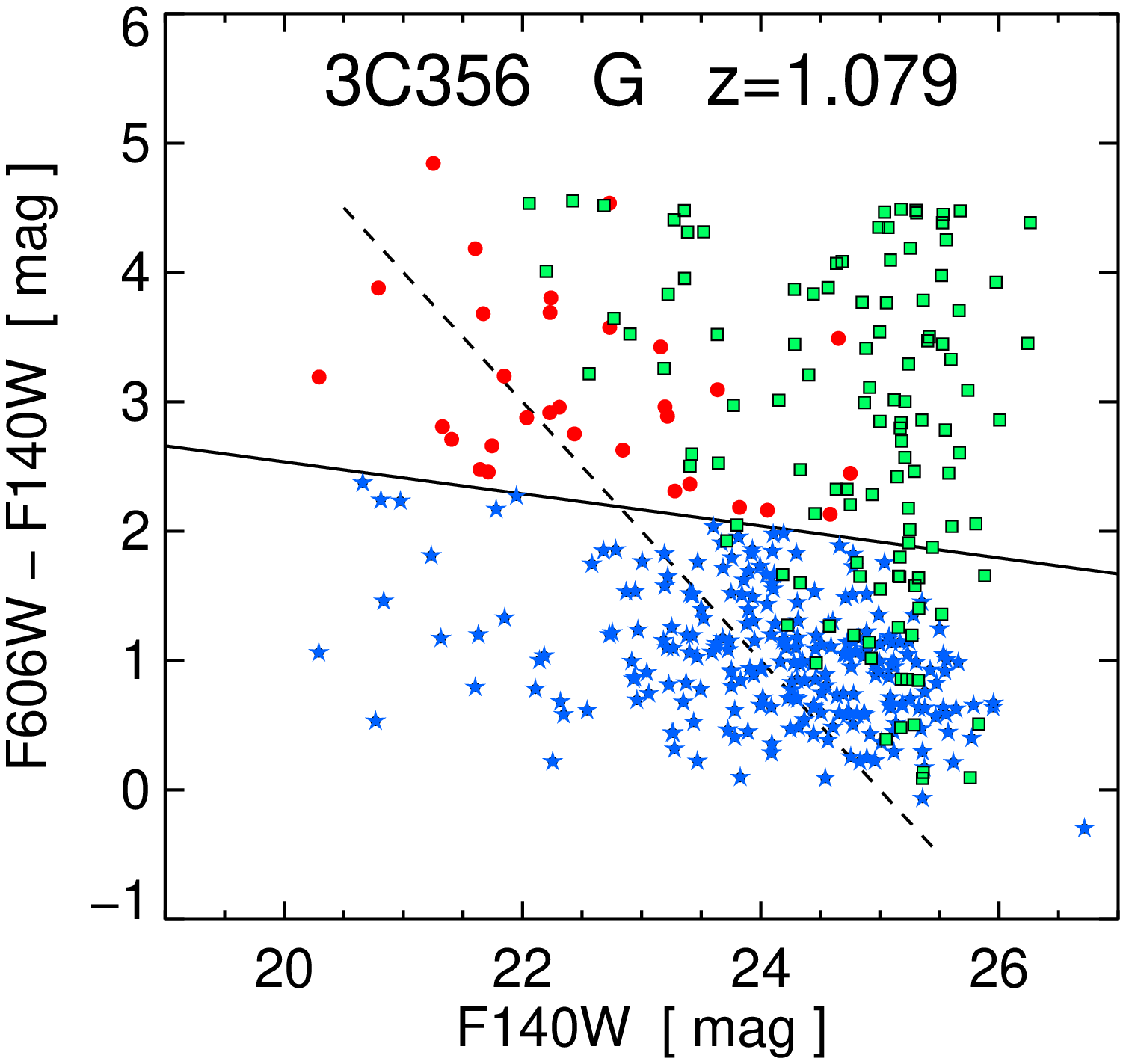}
      \hspace{-0.0mm}\includegraphics[width=0.295\linewidth,clip=true]{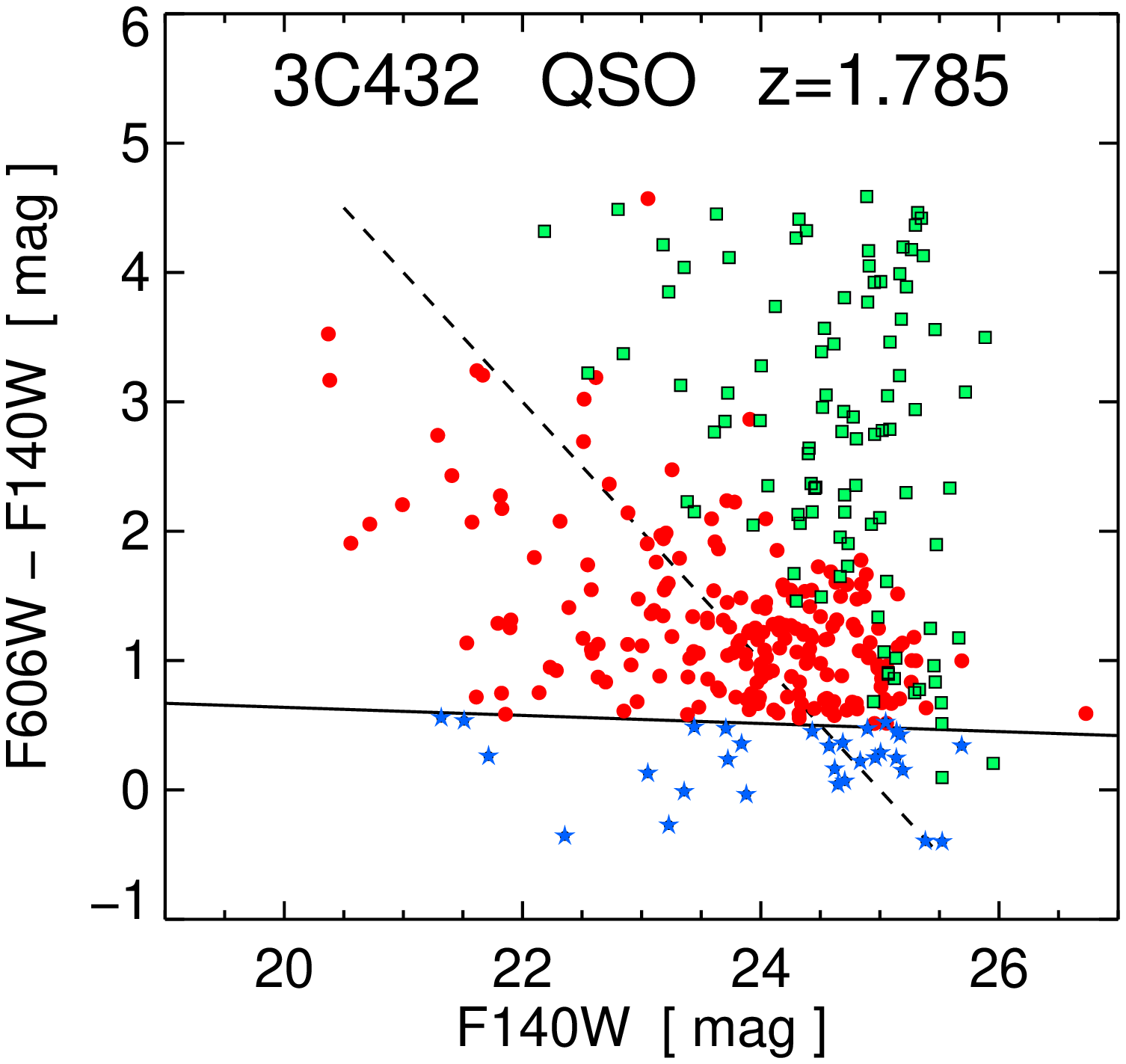}
      \hspace{-0.0mm}\includegraphics[width=0.295\linewidth,clip=true]{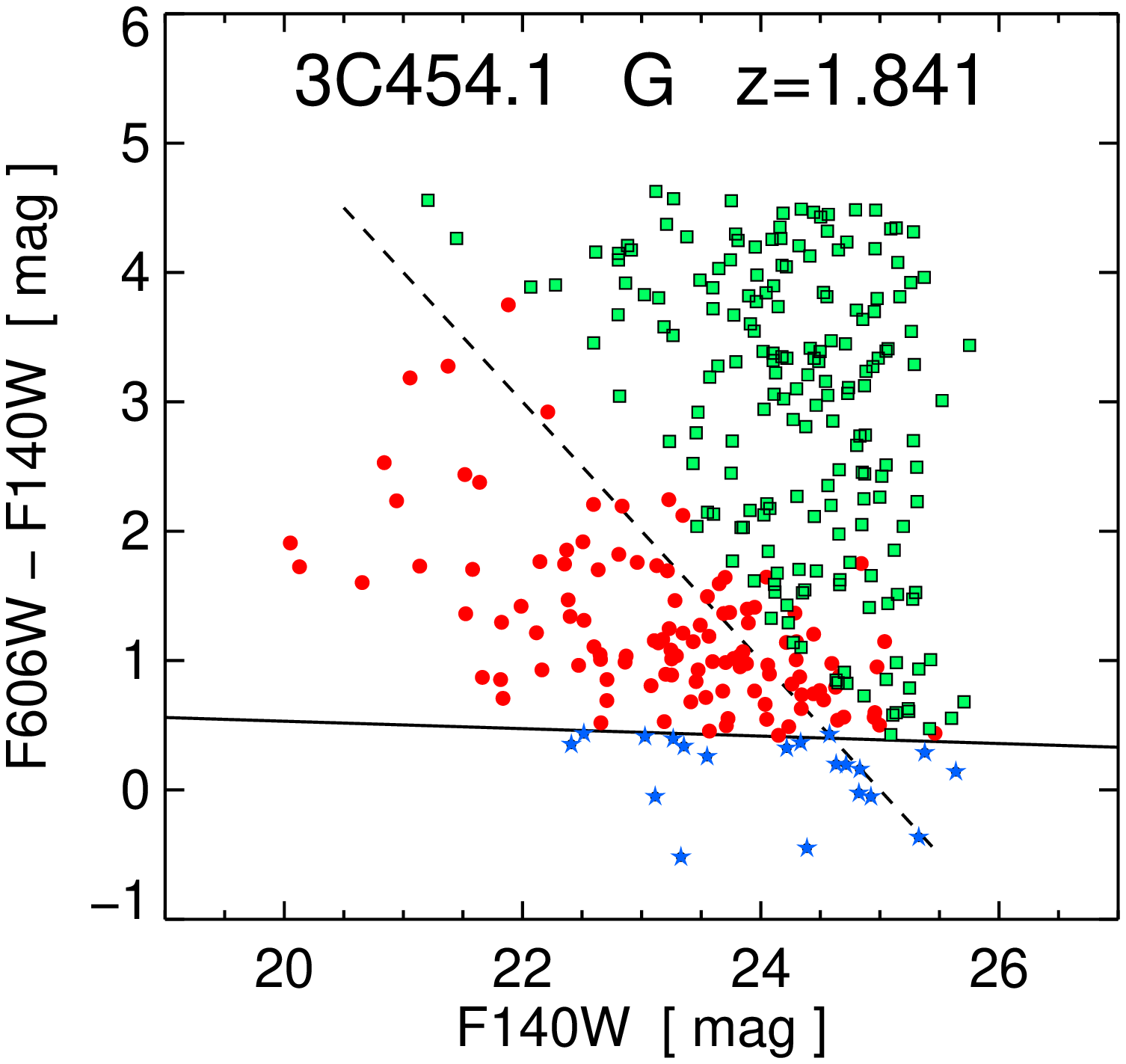}
      \caption{continued.
      }
    \end{figure*}

    \begin{table}
      \renewcommand{\thetable}{\arabic{table}}
      \caption{Completeness fractions in percent.
      }
      \label{tab_completeness}
      \setlength\tabcolsep{5.0pt}
      \footnotesize
      \begin{tabular}{l|l|rrrrrr}
        \hline
        Row & Filter / mag$~^*$           & 22 & 23 & 24 & 25 & 26 & 27 \\
        \hline

        1 &  F606W 3-$\sigma$ det.    & 100  & 100   & 100  &  99 &  69 &  35 \\
        2 &  F606W + F140W $~^1$      & 100  & 100   &  97  &  73 &  30 &  1  \\
        \hline
        \hline
        3 &  F140W 3-$\sigma$ det.    & 100  & 100   &  74  &  50 &  13 &  -  \\
        4 &  F140W + F606W $~^1$      & 100  &  81   &  62  &  29 &  -  &  -  \\
        5 &  F140W + F606W $~^2$      & 100  &  65   &  45  &  20 &  -  &  -  \\
        \hline
        6 &  F140W + F606W blue $~^3$ & 100  & 100   &  78  &  36 &   3 &  -  \\

        \hline 
      \end{tabular}\hspace*{-5.0cm}
      ~\\
      $~^*$ Magnitude in F606W (row 1--2) and F140W (row 3--6)\\
      $~^1$ Sources detected with 3-$\sigma$ in both filters\\
      $~^2$ Sources detected with 3-$\sigma$ in F140W and with 10-$\sigma$ in F606W, that is $\sigma$(F606W)\,$<$\,0.1 mag.\\
      $~^3$ Sources detected with 3-$\sigma$ in both filters and $ {\rm F606W} - {\rm F140W} < 1$
    \end{table}

    \section{Analysis and results} \label{sec:results}

    We split our sample into four 
    subsets: $all$ (all galaxies, including single band detections), $IR$-$only$ (galaxies detected in F140W but not in F606W), $red$ and $blue$ where these latter were selected from color-magnitude diagrams as described below.

    \subsection{Color--magnitude diagrams} \label{sec:cmds}
    Figure 4 shows color-magnitude diagrams for all 21 3C fields. To separate samples of red and blue galaxies, we adopted the RS as identified by K16, based on an evolutionary model from the GALEV tool \citep{Kotulla09}. Given the relatively large color errors and intrinsic scatter of the RS and uncertainties in the adopted metallicities and formation redshifts, we adopted a color of 1 mag below the RS for galaxies to be defined blue (with the exception of 3C257 where we use a color difference of 0.5 mag as there would otherwise be no ``blue'' galaxies in this cluster).


    Sources detected only in F140W, must be redder than the color selection line shown in Fig.~\ref{fig:cmd}. We stacked the F606W images of these objects (based on the positions in F140W) and derived their median color. This is about 0.5 mag redder than objects detected in both filters.


    \begin{figure*}

      \hspace{-0mm}\includegraphics[width=0.245\textwidth, clip=true]{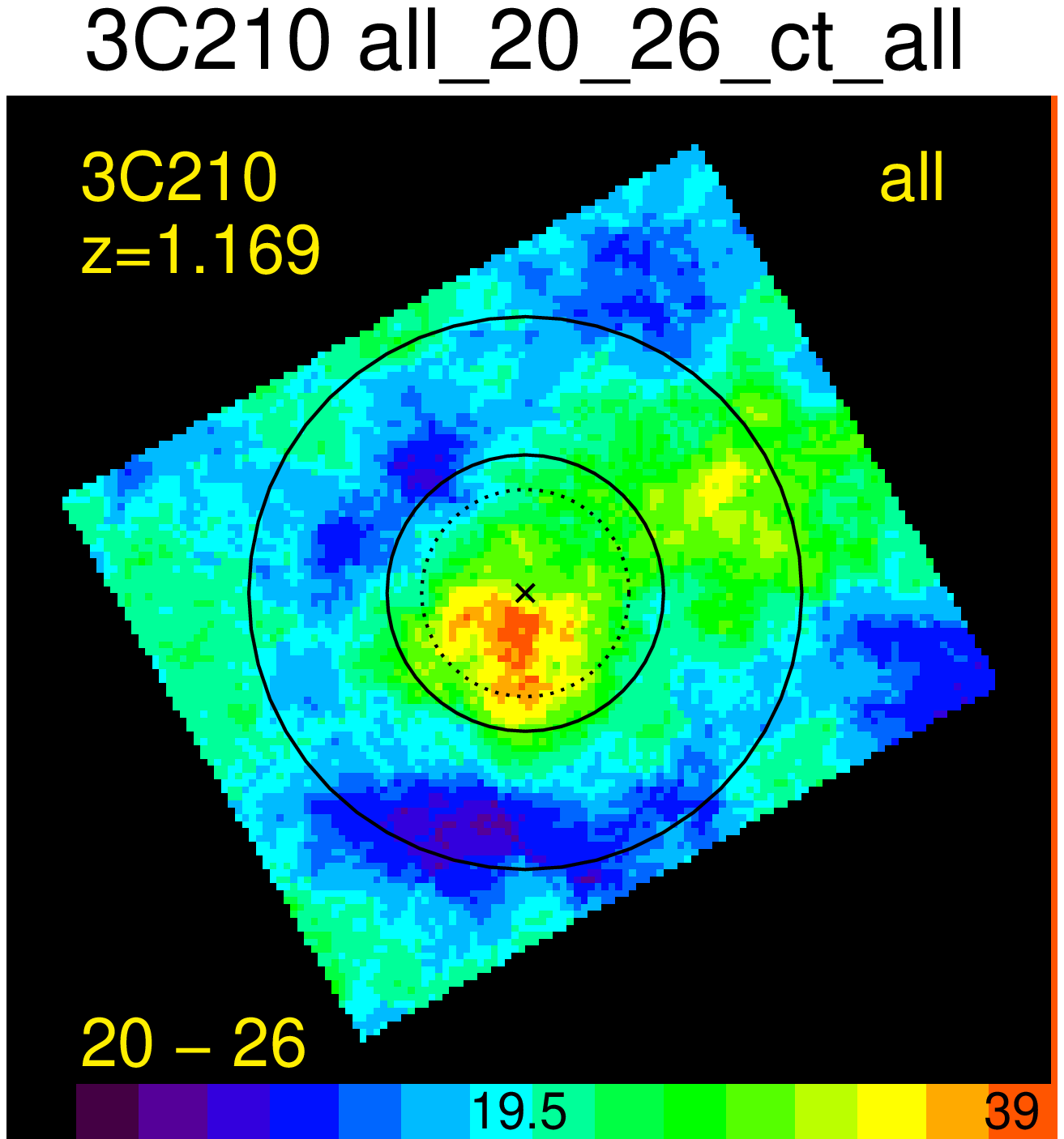}           
      \includegraphics[width=0.245\textwidth, clip=true]{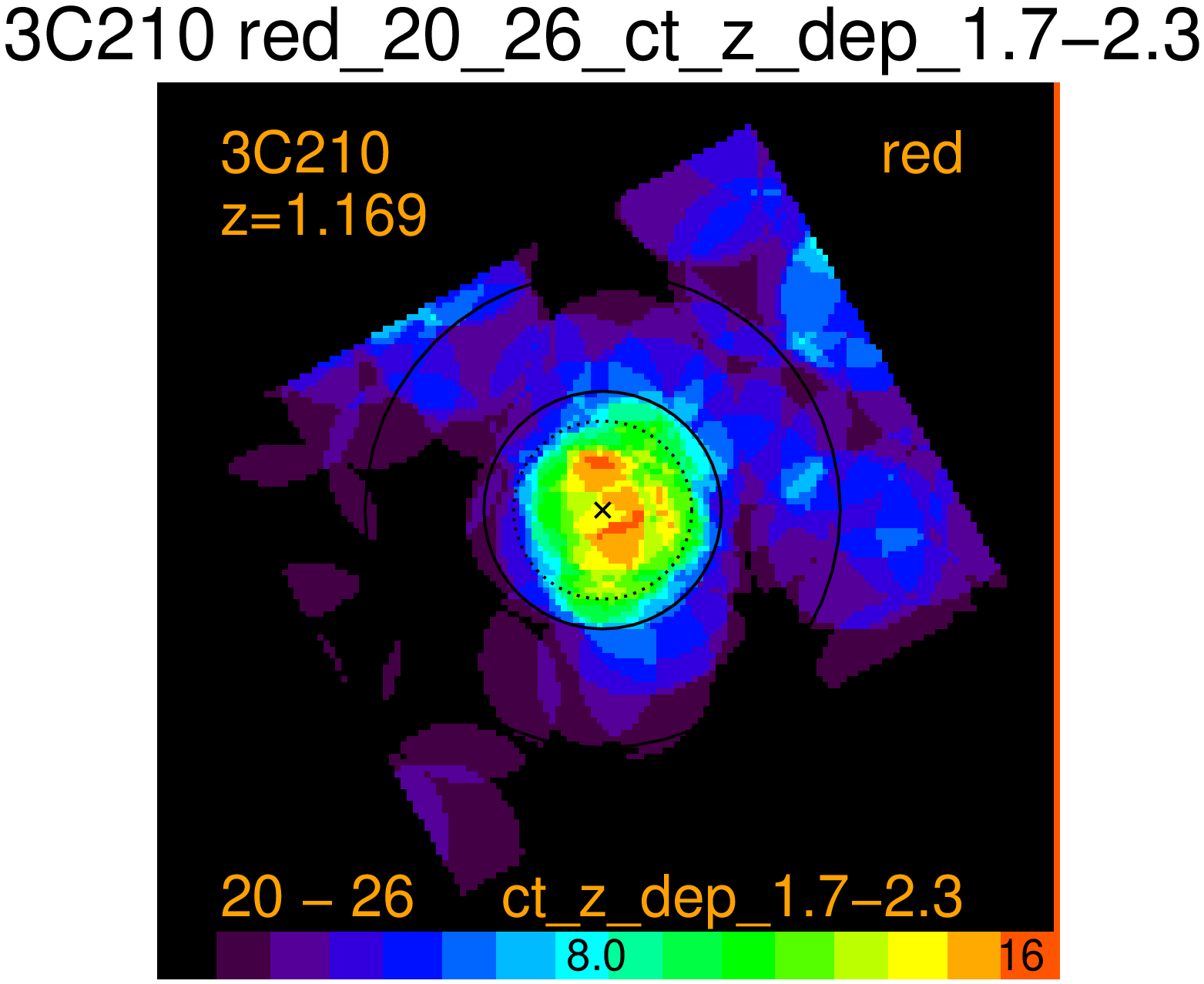} 
      \includegraphics[width=0.245\textwidth, clip=true]{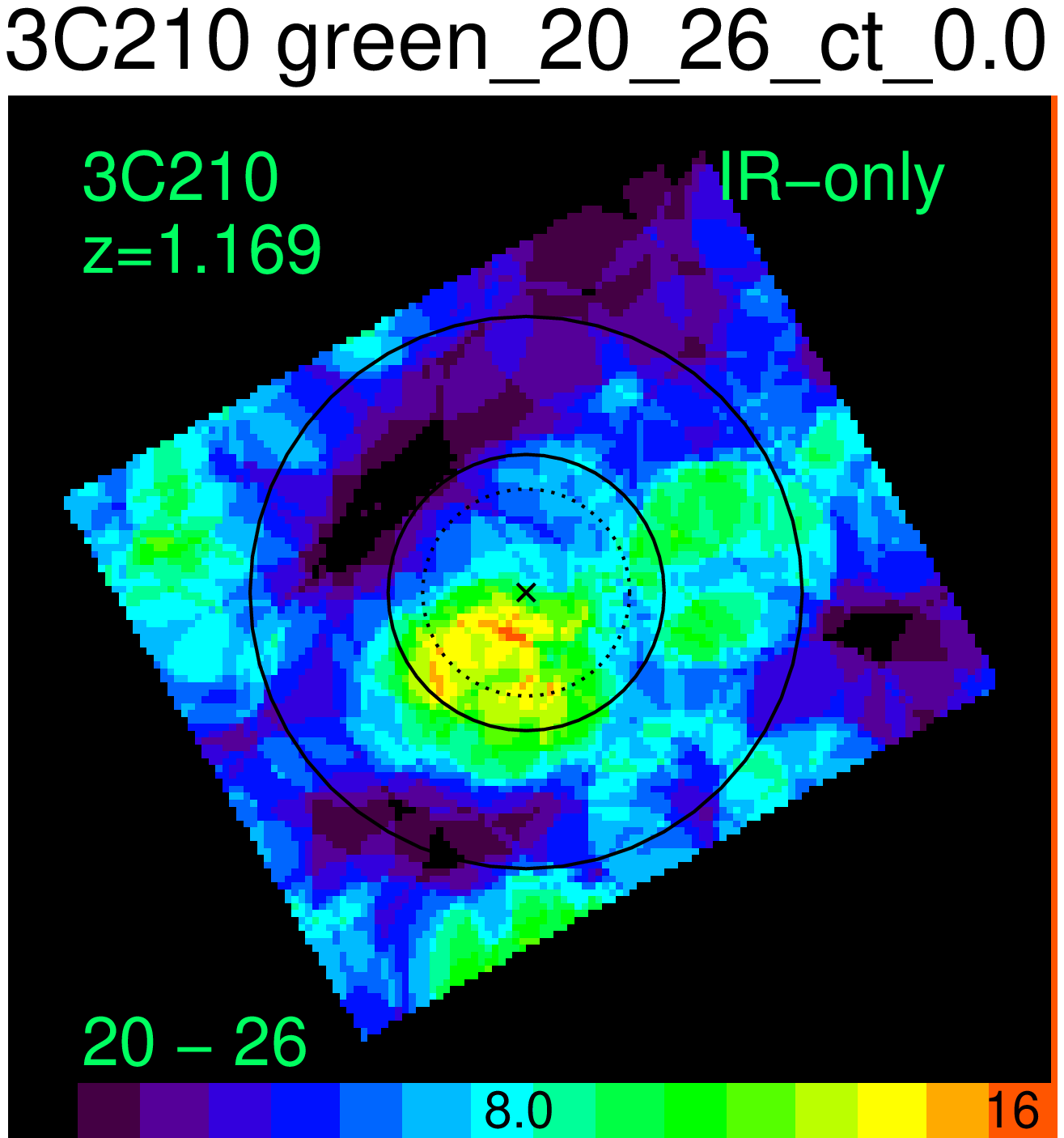}         
      \includegraphics[width=0.245\textwidth, clip=true]{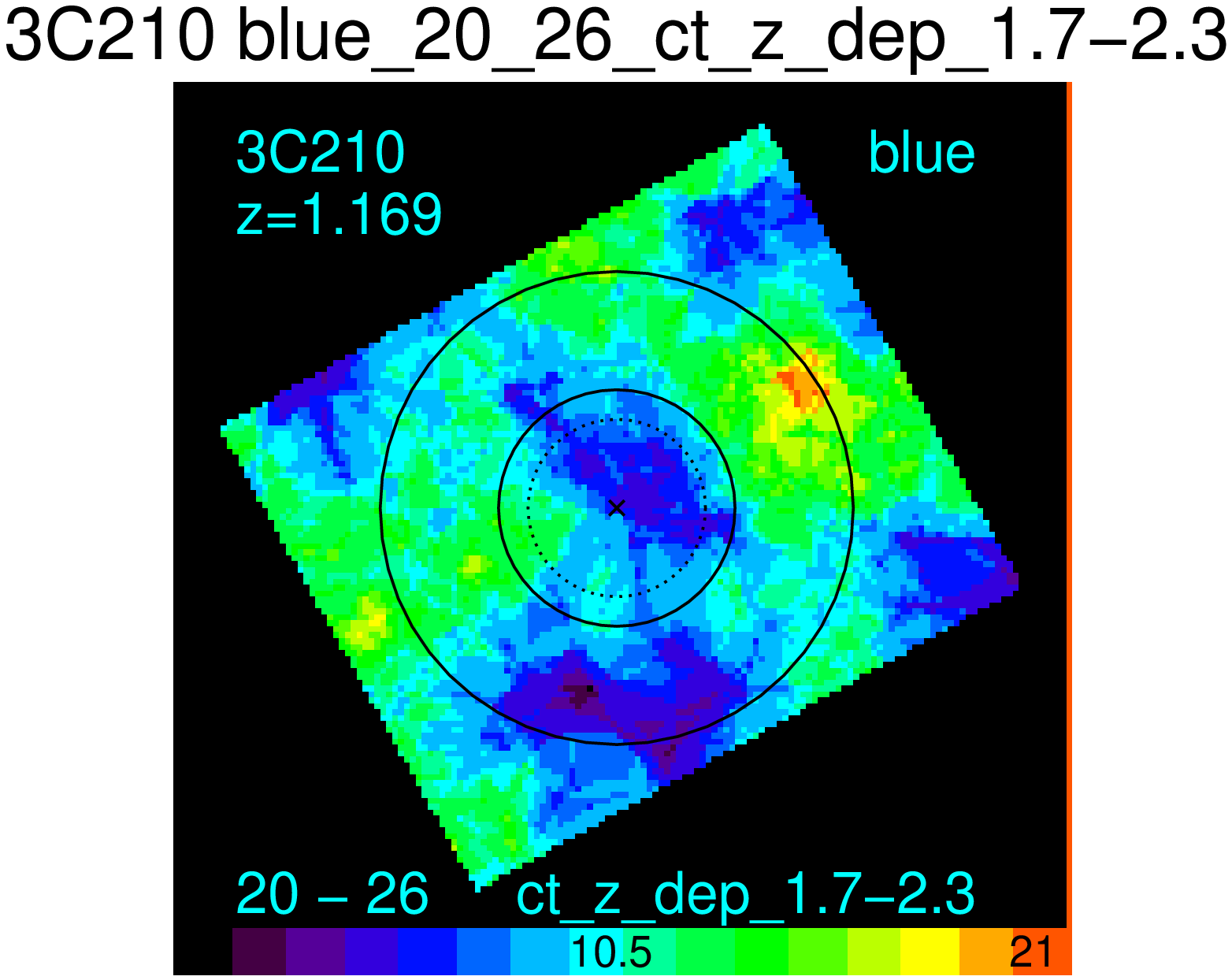}

      \hspace{-0mm}\includegraphics[width=0.245\textwidth, clip=true]{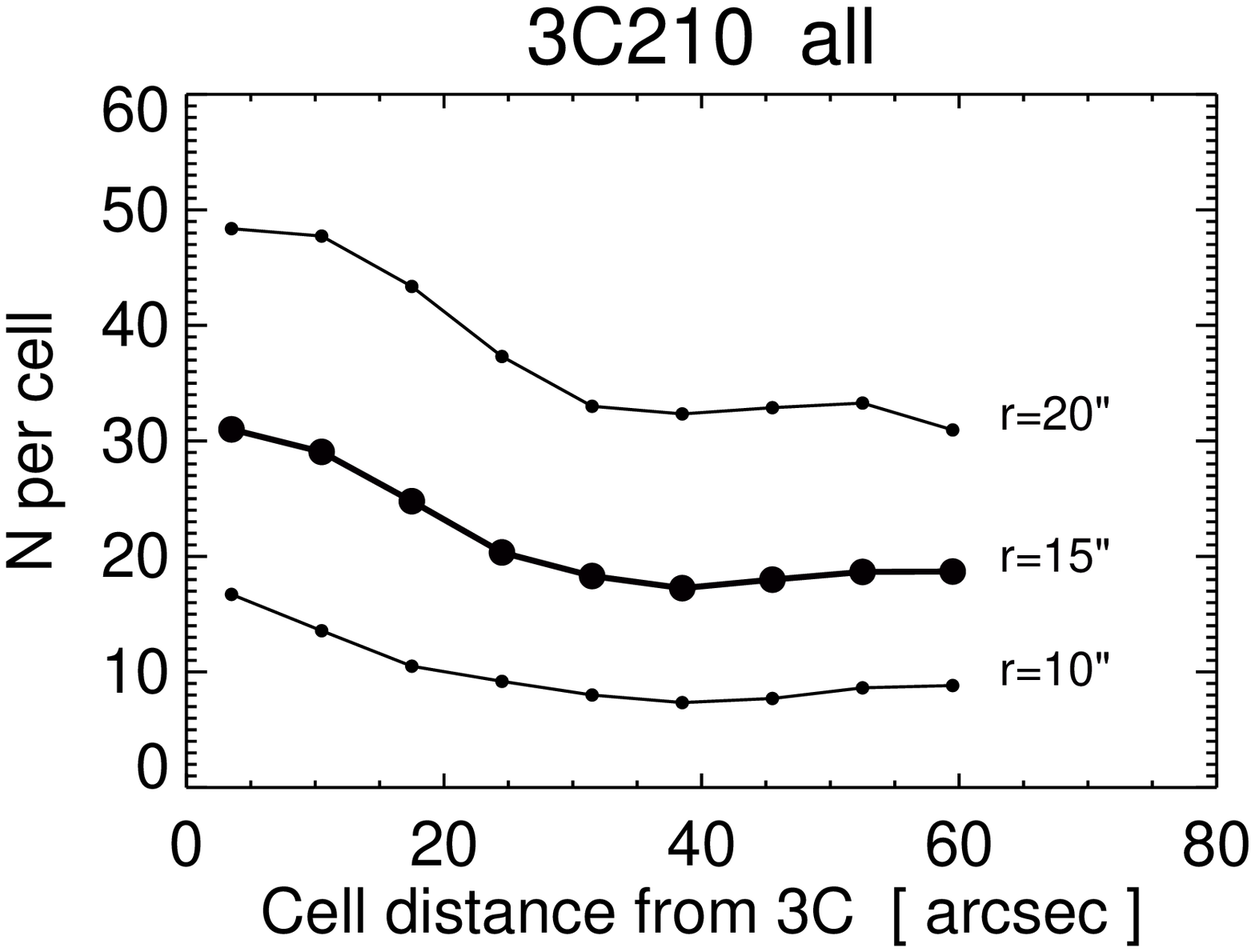}
      \includegraphics[width=0.245\textwidth, clip=true]{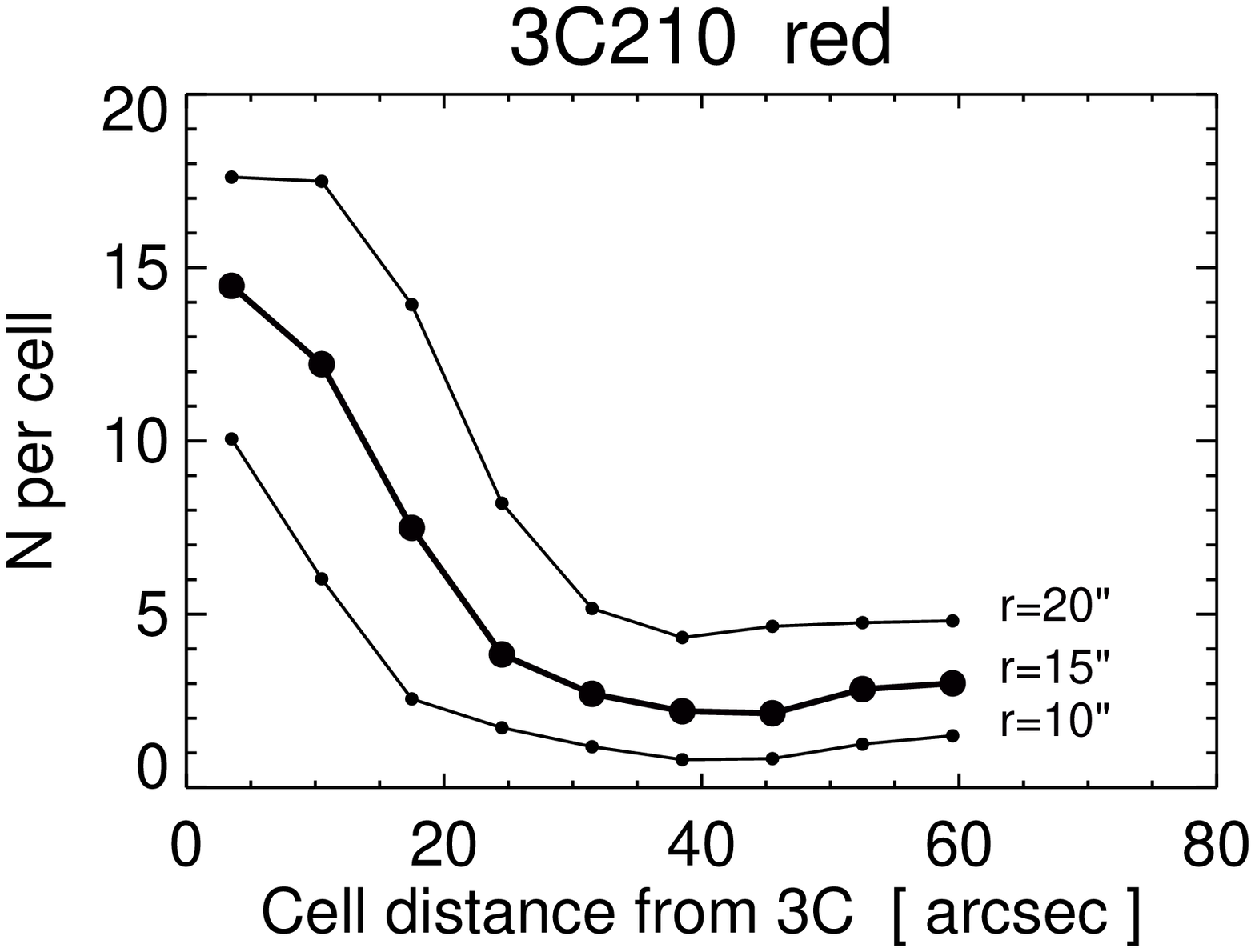}
      \includegraphics[width=0.245\textwidth, clip=true]{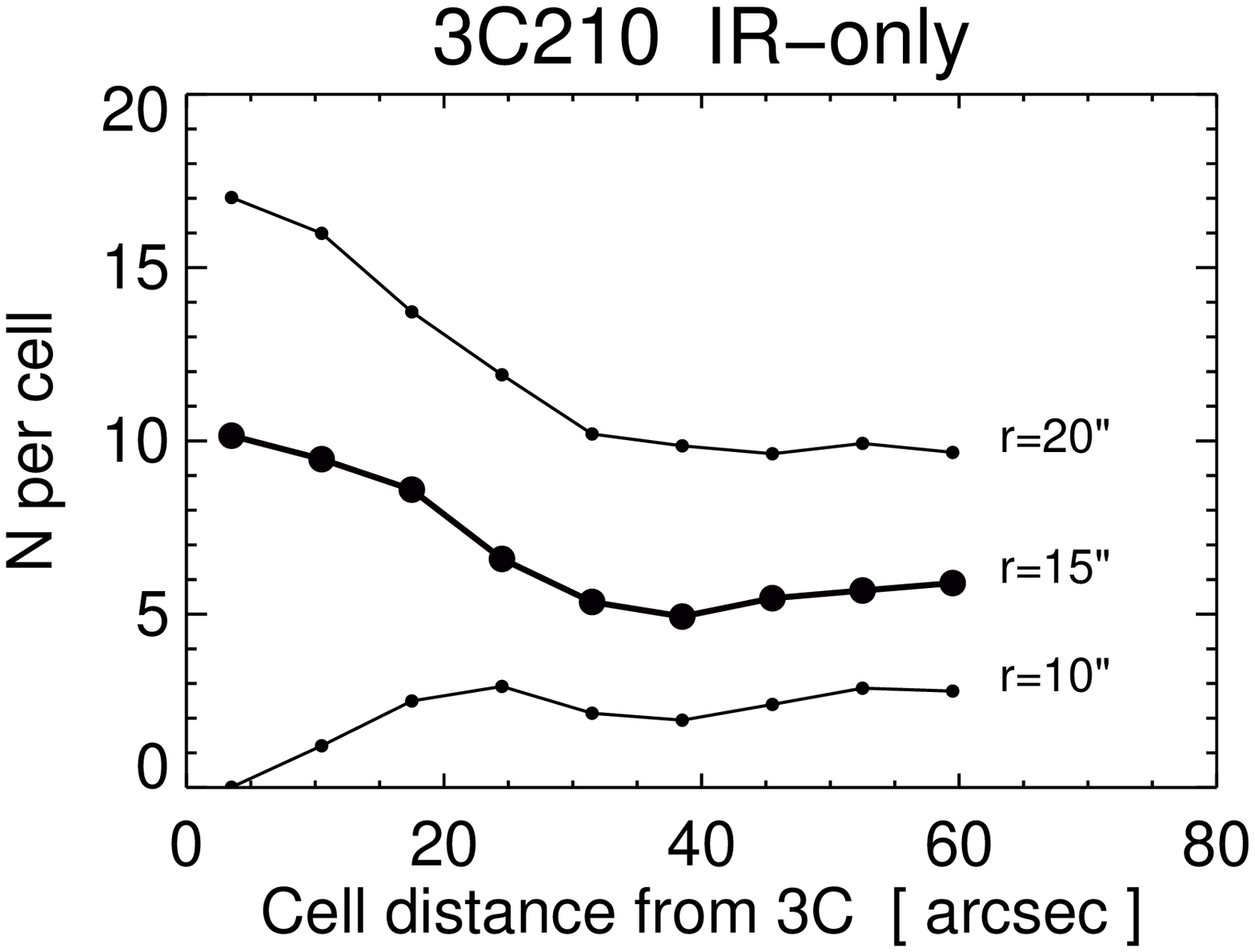}
      \includegraphics[width=0.245\textwidth, clip=true]{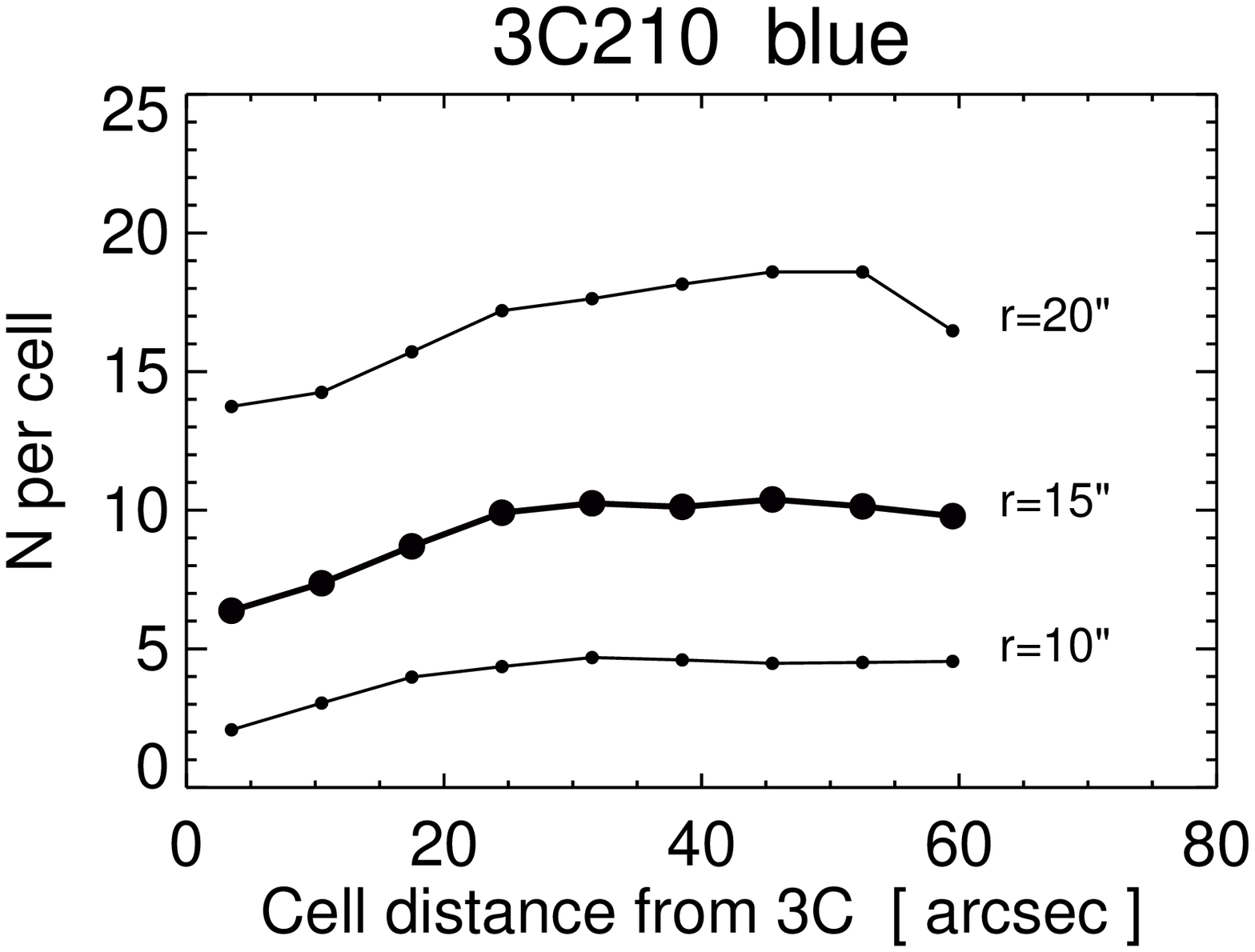}

      \caption{Surface density maps and radial density profiles of the 3C\,210 field. 
        The top row 
        shows the surface density maps
        for four selection criteria. These are from left to right:
        $all$ galaxies, $red$, $IR$-$only$ and $blue$ galaxies;
        for $red$ and $blue$ galaxies we used the redshift-dependent color threshold as labeled in the bottom right corner.
        The magnitude range is $20 < \rm{F140W} < 26$ as labeled in the bottom left corner.
        The map size is 2.5$\arcmin$, north is up, east to the left. 
        The black cross marks the position of the 3C source.
        The 3C source  is used only as a signpost for a density enhancement and is therefore excluded from the 
        density maps and all further density calculations.
        The dotted circle indicates the cell size ($r_{\rm c}=15\arcsec$),
        and the two solid circles mark a radius of 20$\arcsec$ and 40$\arcsec$ around the 3C source.
        The color bar at the bottom of the maps gives the linear range used for the map:
        from $N = 0$ (black, left end) to the maximal surface density per cell as labeled at the right end of the bar (red); 
        light blue gives the numbers as labeled in the middle of the color bar.
        %
        The bottom row 
        shows the mean radial surface density per cell 
        as a function of distance from the 3C source.
        %
        The radial surface density is plotted for three different cell sizes.
        The thick line with large dots marks the curve for cell radius $r_{\rm c} = 15\arcsec$, which was used for the maps shown.
        The two other curves are for $r_{\rm c} = 10\arcsec$ and $r = 20\arcsec$, without area normalization, to avoid confusion of the three curves.
        %
      }
      \label{fig:sd_maps_3C210}
    \end{figure*}



    \subsection{Surface density: maps and radial profiles} \label{sec:maps}

    To produce surface density maps, we counted galaxies in circular cells of radius $15\arcsec$ centered on a 150 x 150 grid with $1\arcsec$ spacing. Cells falling partially outside of the \HST\ images were masked. We also tried circular cells of radius $10\arcsec$ and $20\arcsec$ to test the influence of cell size.

    Fig.~\ref{fig:sd_maps_3C210} shows surface-density maps and radial density profiles for the 3C\,210 field. 
    It is typical for fields with a clear central OD (COD).\footnote{We distinguish between OD in general, anywhere in the map,  
      and central OD, which is OD around the 3C source. A negative COD means under-density (UD, CUD)}
    Visual inspection of the maps leaves no doubt that 3C\,210 lies in the center of an overdensity of $red$ galaxies
    with a radial extent of 15$\arcsec$--30$\arcsec$. 
    Beyond $r=30\arcsec$, essentially no $red$ galaxies are found.
    In contrast, the $blue$ galaxies avoid the central $r = 20\arcsec$ area and are found in a few clumps at $r > 20\arcsec$.
    The $IR$-$only$ galaxies show a clear density enhancement with a peak slightly shifted to the south of 3C\,210 
    but still within $30\arcsec$ around 3C\,210.
    The $IR$-$only$ population is about 24--26\,mag and may include both $blue$ and $red$ galaxies (Fig.~\ref{fig:cmd}), though
    the stacking results suggest mainly faint $red$ galaxies.
    The OD near 3C\,210 shows up even when using $all$ galaxies without any color constraints.
    Maps with cell radius $r_{\rm c} = 10\arcsec$ and $r_{\rm c} = 20\arcsec$ are of sharper and shallower contrasts respectively 
    but are similar to the $r_{\rm c} = 15\arcsec$ maps used for further analysis.
    The surface density maps and radial density profiles of the sample for $20 < F140W < 26$ are shown 
    in Figs.~\ref{fig:sd_maps_1} to \ref{fig:sd_maps_7}.
    The radial density profiles are listed in Table~\ref{tab_rsf}.
    


    \begin{table}

      \renewcommand{\thetable}{\arabic{table}}
      \caption{Summary of central over- and under-densities of the 3C clusters, sorted by redshift.
        The horizontal lines divide the sample at $z=1.2$ and $z=1.6$.
        The columns 3--6
        mark  a COD with ``$+$'' and ``$++$'' if detected with S/N larger than 3 and 5, respectively, with entries from Table~\ref{tab_od}. 
        CUDs are marked with ``$-$'' and ``$--$''.
        Column 7 
        indicates if a red-sequence was found by K16.
        Column 8 assigns the color-dependent morphology classification described in Sect.~\ref{sec:discussion_classification}.        
        A ``?'' notes uncertain cases due to a small number of galaxies 
        for the COD/CUD calculation.
      }
      \label{tab_classification}
      \setlength\tabcolsep{4.0pt}
      \footnotesize
      \begin{tabular}{lc|cccc|c|c}
        \hline
                  (1)       & (2)  & (3)  & (4) & (5)  & (6)  & (7)  &  (8)  \\
        Name &   $z$      & all  & red      & IR-only  & blue    & K16  & Class                         \\
        \hline                                                                            
        3C\,287   & 1.055 &      &          &          &   $--$  &      &       V                       \\
        3C\,186   & 1.067 &      &   $+$    &   $+$    &   $--$  & RS   &       I                       \\
        3C\,356   & 1.079 &      &   $++$   &   $+$    &   $--$  &      &       I                       \\
        3C\,208.0 & 1.110 & $++$ &   $++$   &   $+$    &   $+$   &      &     III                       \\
        3C\,305.1 & 1.132 &      &   $+$,?  &          &   $-$   &      &       I?                      \\                                                        
        3C\,220.2 & 1.158 &      &   $++$   &          &         &      &      II                       \\                                                        
        3C\,300.1 & 1.159 & $++$ &   $++$   &   $+$    &         & RS   &      II                       \\
        3C\,210   & 1.169 & $++$ &   $++$   &   $++$   &   $-$   & RS   &       I                       \\
        \hline                                                                                
        3C\,324   & 1.206 & $++$ &   $++$   &   $++$   &   $++$  &      &     III                       \\
        3C\,068.1 & 1.238 & $+$  &          &   $++$   &         & RS   &      II                       \\
        3C\,255   & 1.355 & $+$  &   $++$   &          &         & RS   &      II                       \\
        3C\,268.4 & 1.398 & $++$ &   $++$   &          &         &      &      II                       \\
        3C\,297   & 1.406 & $+$  &          &   $++$   &         &      &      II                       \\
        3C\,298   & 1.437 &      &          &          &         &      &      VI                       \\
        3C\,230   & 1.487 & $++$ &   $++$   &   $++$   &         & RS   &      II                       \\
        3C\,270.1 & 1.532 &      &   $++$   &          &         &      &      II                       \\                                                        
        \hline                                                        
        3C\,322   & 1.681 &      &          &   $+$,?  &         &      &      II? $\rightarrow$ VI     \\
        3C\,432   & 1.785 &      &   $-$    &          &  $++$,? &      &      IV? $\rightarrow$ VI?    \\                                                        
        3C\,326.1 & 1.825 & $++$ &   $+$    &   $++$   &  $-$,?  &      &       I?                      \\
        3C\,454.1 & 1.841 & $+$  &          &          &         &      &      VI                       \\
        3C\,257   & 2.474 & $+$,?&   $+$,?  &   $+$,?  &         & RS   &      II?                      \\

        \hline
      \end{tabular}\hspace*{-5.0cm}
    \end{table}

   
    Most surface-density maps reveal several density enhancements which,
    if not caused by fore- or background sources, indicate a clumpy structure of the forming galaxy clusters.
    The radial surface-density profiles vary among the 3C fields.
    Both the maps and the profiles depend on the parameters like magnitude and color 
    and slightly on the cell size.
    Most overdensities 
    have radii $r<30\arcsec$, and we used the density difference between the area
    within this radius and that at $30\arcsec < r < 60\arcsec$ to parameterize the degree of central overdensity (COD).
    Table~\ref{tab_classification} summarizes which CODs are significant.

    In most cases the 3C source lies within $15\arcsec$ of the central density enhancement.
    However, there are  few exceptions, for example 3C\,297 and 3C\,300.1, 
    which both have a 
    $red$ OD shifted toward the south.
    One could think of shifting the adopted cluster center position, but that position is not unique.
    3C\,300.1, for instance, exhibits an $IR$-$only$  COD shifted in a different direction, 
    so that the density enhancements of $red$ and $IR$-$only$ do not coincide. 
    Therefore, we kept the 3C position as the cluster center. 


    \subsection{Significance and frequency of overdensities}\label{sec:frequency}

    \begin{figure*}
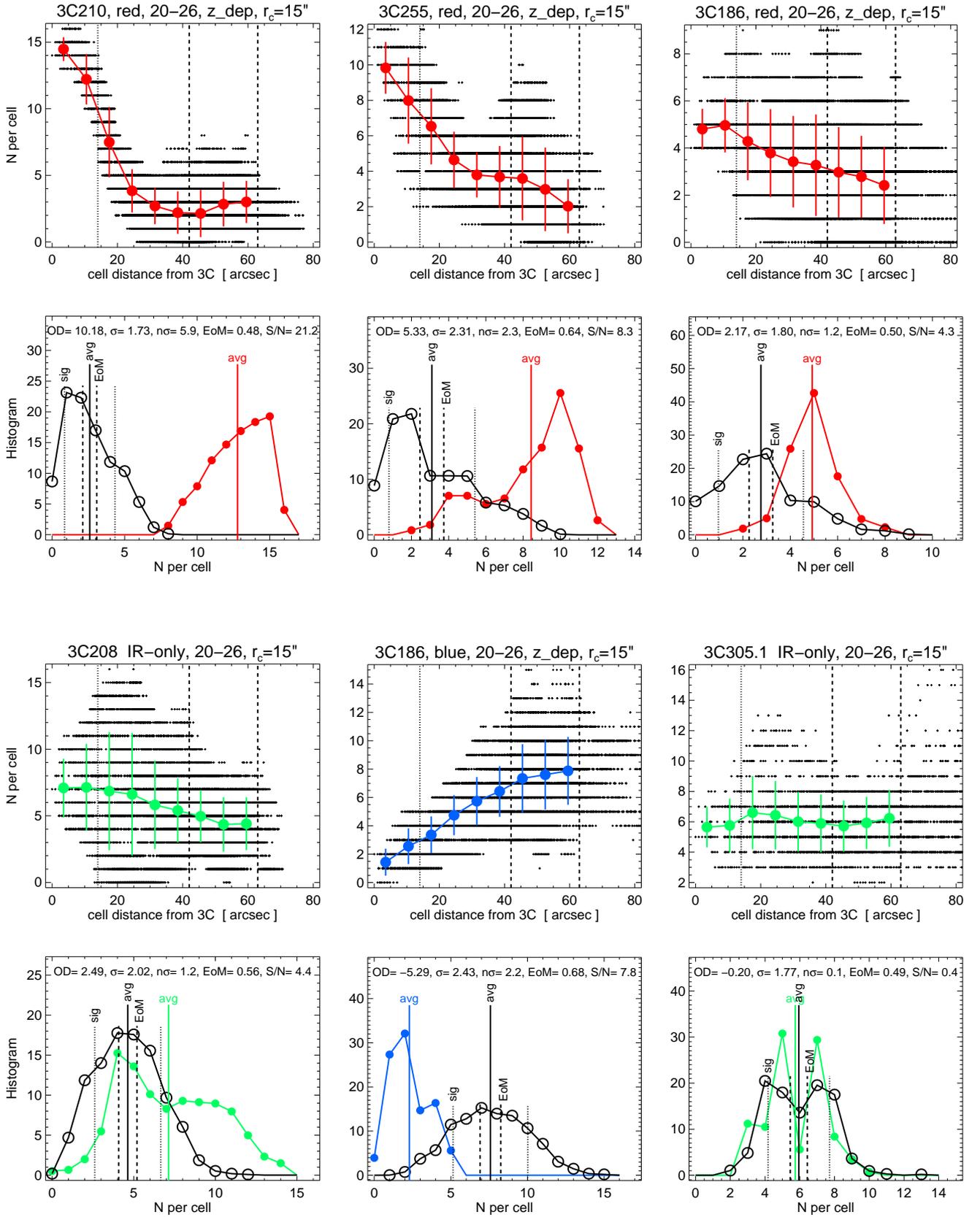

      \includegraphics[width=6.6cm, clip=true]{3C210_plot_cell_sigma_red_20_26_z_dep_grid_15_1.ps_page_1}
      \includegraphics[width=5.7cm, clip=true]{3C255_plot_cell_sigma_red_20_26_z_dep_grid_15_1.ps_page_1}
      \includegraphics[width=5.6cm, clip=true]{3C186_plot_cell_sigma_red_20_26_z_dep_grid_15_1.ps_page_1}

      \includegraphics[width=6.6cm, clip=true]{3C210_plot_cell_sigma_red_20_26_z_dep_grid_15_1.ps_page_2}
      \includegraphics[width=5.7cm, clip=true]{3C255_plot_cell_sigma_red_20_26_z_dep_grid_15_1.ps_page_2}
      \includegraphics[width=5.7cm, clip=true]{3C186_plot_cell_sigma_red_20_26_z_dep_grid_15_1.ps_page_2}

      \vspace{10mm}

      \includegraphics[width=6.6cm, clip=true]{3C208_plot_cell_sigma_green_20_26_ct_0.0_grid_15_1.ps_page_1}
      \includegraphics[width=5.7cm, clip=true]{3C186_plot_cell_sigma_blue_20_26_z_dep_grid_15_1.ps_page_1}
      \includegraphics[width=5.7cm, clip=true]{3C305.1_plot_cell_sigma_green_20_26_ct_0.0_grid_15_1.ps_page_1}

      \includegraphics[width=6.6cm, clip=true]{3C208_plot_cell_sigma_green_20_26_ct_0.0_grid_15_1.ps_page_2}
      \includegraphics[width=5.7cm, clip=true]{3C186_plot_cell_sigma_blue_20_26_z_dep_grid_15_1.ps_page_2}
      \includegraphics[width=5.7cm, clip=true]{3C305.1_plot_cell_sigma_green_20_26_ct_0.0_grid_15_1.ps_page_2}

      \caption{Radial surface density profiles and histograms from the cell counts in Sect.~\ref{sec:maps} for six examples.
        Top panel: the small "+" symbols mark individual cells, 
        while the colored lines and 
        filled symbols
        show the mean radial 
        profile and standard deviation.
        The vertical black dotted and dashed lines mark the radial limits used 
        to define the central region 
        and periphery.
        Bottom: histogram of the cell counts (made from the "+" in the top row),
        colored with 
        filled circles for the center and 
        black open circles for the periphery.
        The vertical lines 
        indicate the average (avg),
        and for the periphery also 
        the standard deviation (1$\sigma$) around the avg and
        its uncertainty 
        (= error of the mean, $ \rm EoM = \sigma / \sqrt{13}$, 
        assuming N$_{\rm ic}=13$ independent cells in the periphery).
        These statistical values 
        are written at the top of the histogram (S/N = OD / EoM).
      }
      \label{fig:od_histograms}
    \end{figure*}

    \begin{figure*}

      \includegraphics[width=9.5cm, clip=true]{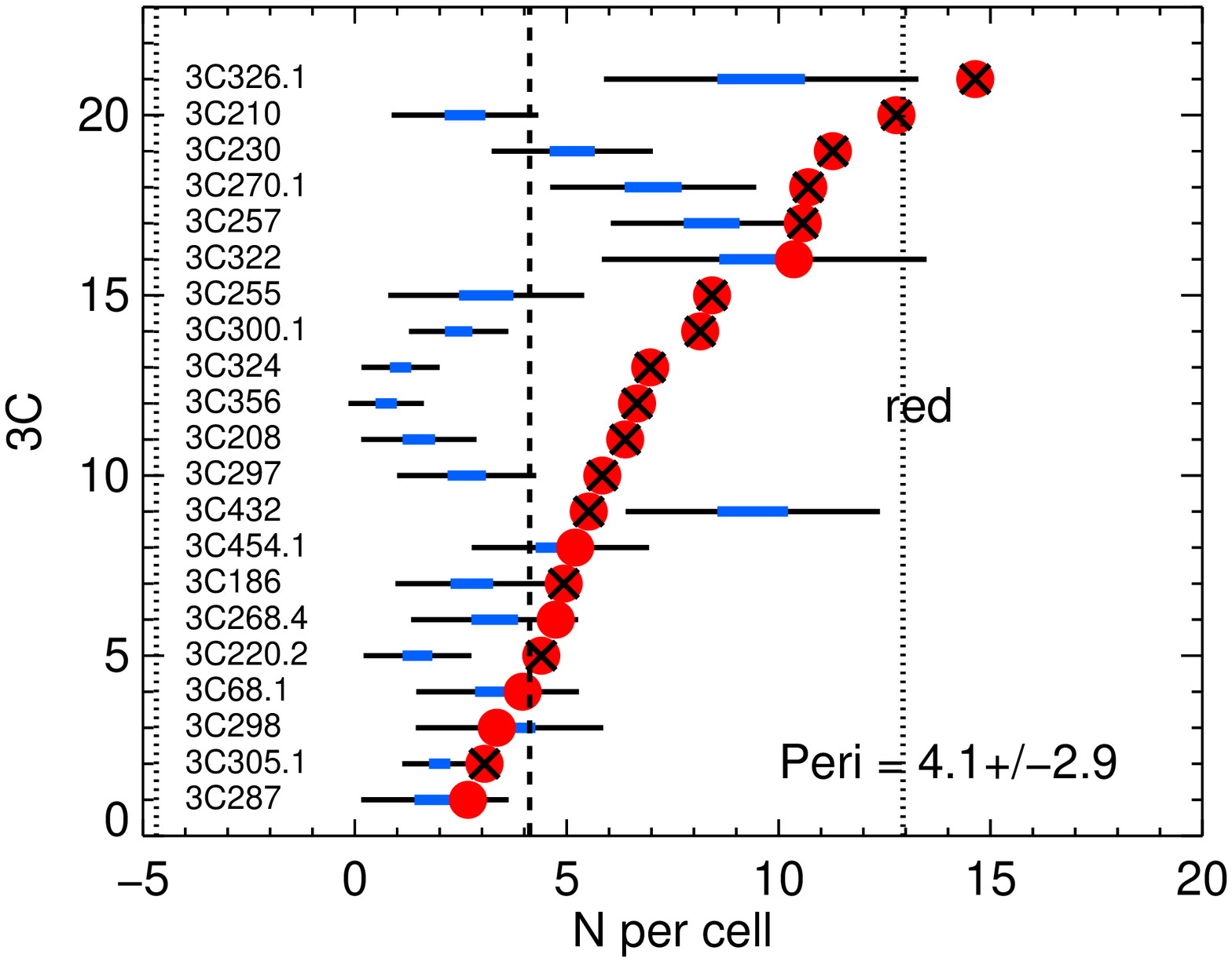}
      \includegraphics[width=8.3cm, clip=true]{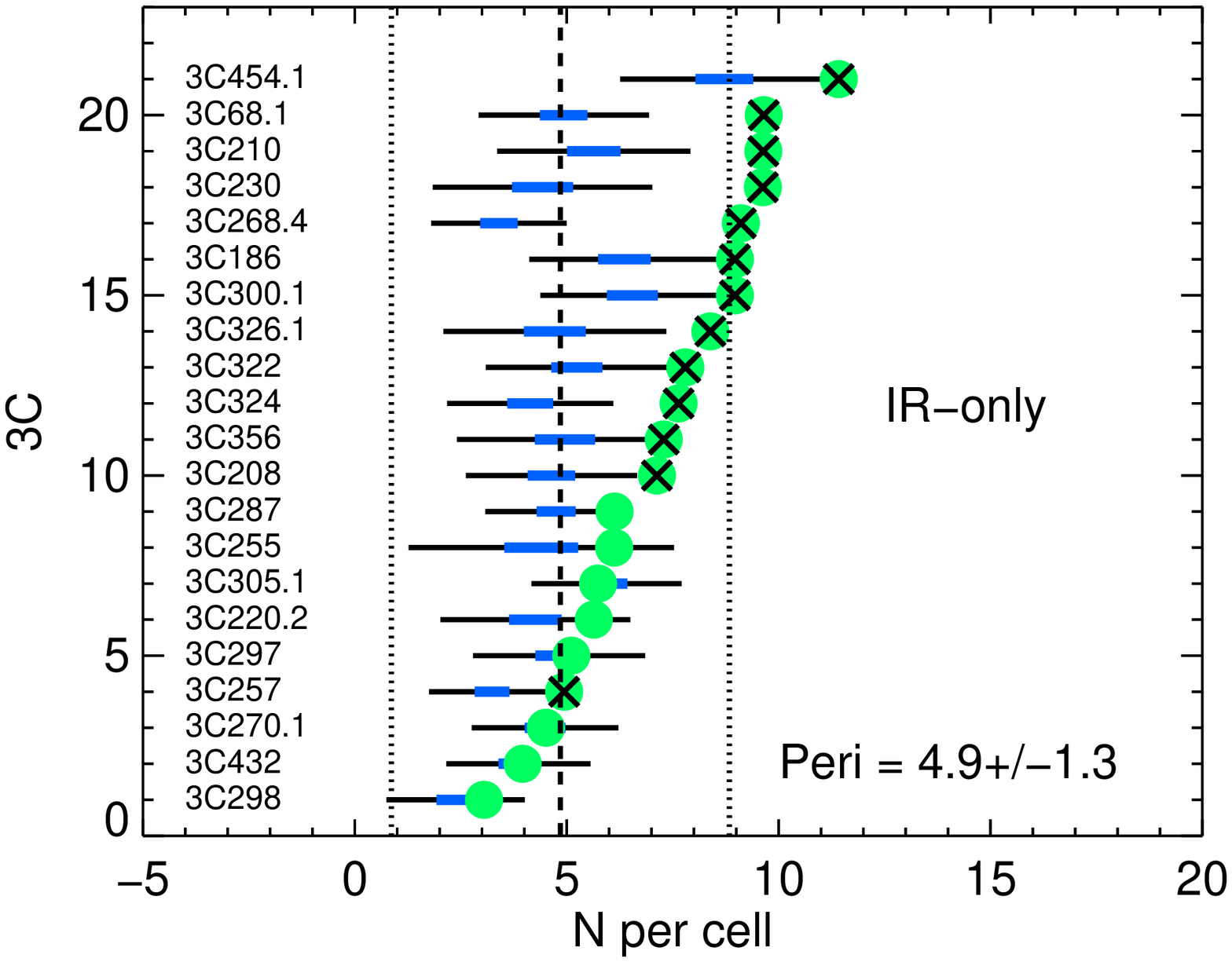}

      \includegraphics[width=9.5cm, clip=true]{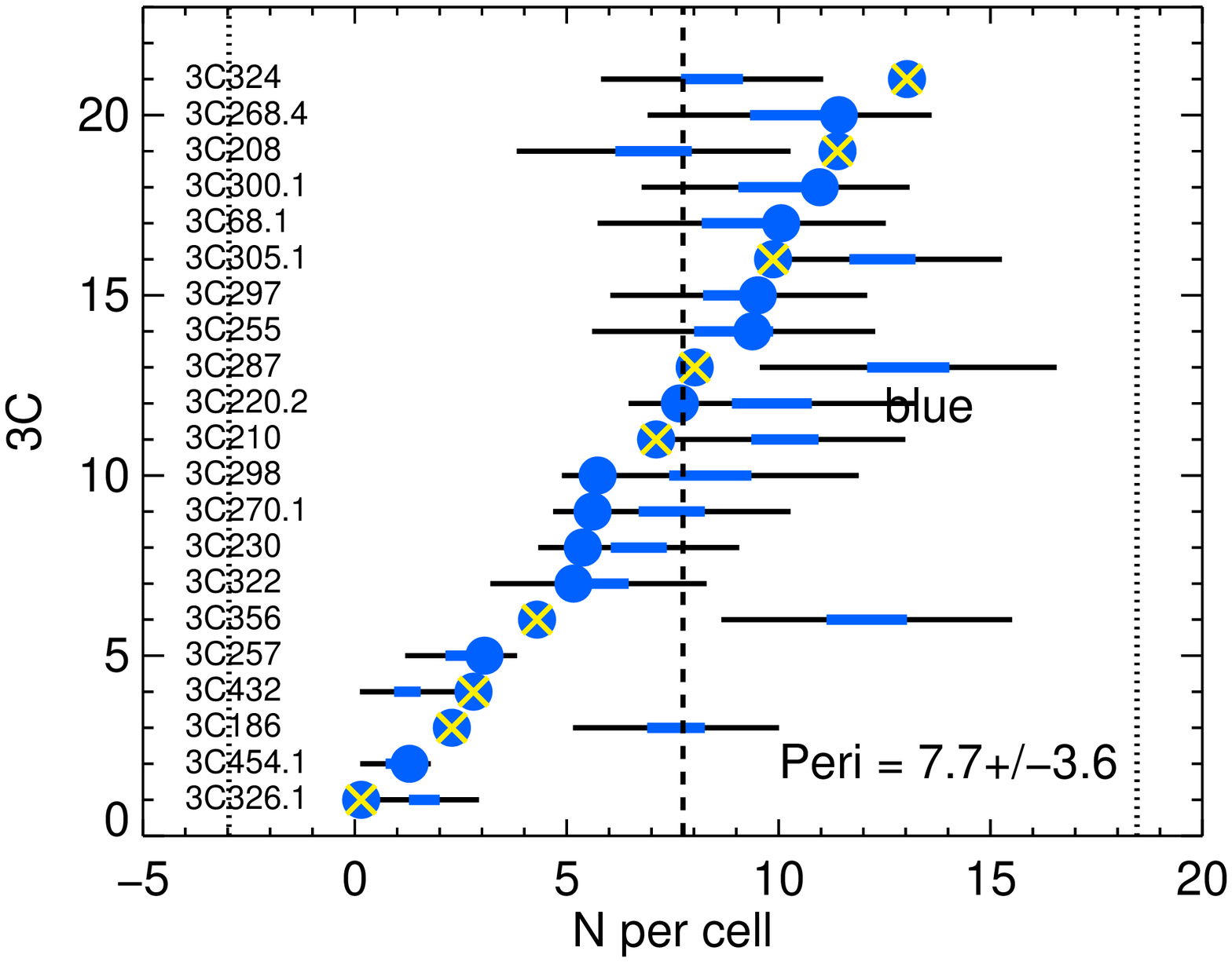}
      \includegraphics[width=8.3cm, clip=true]{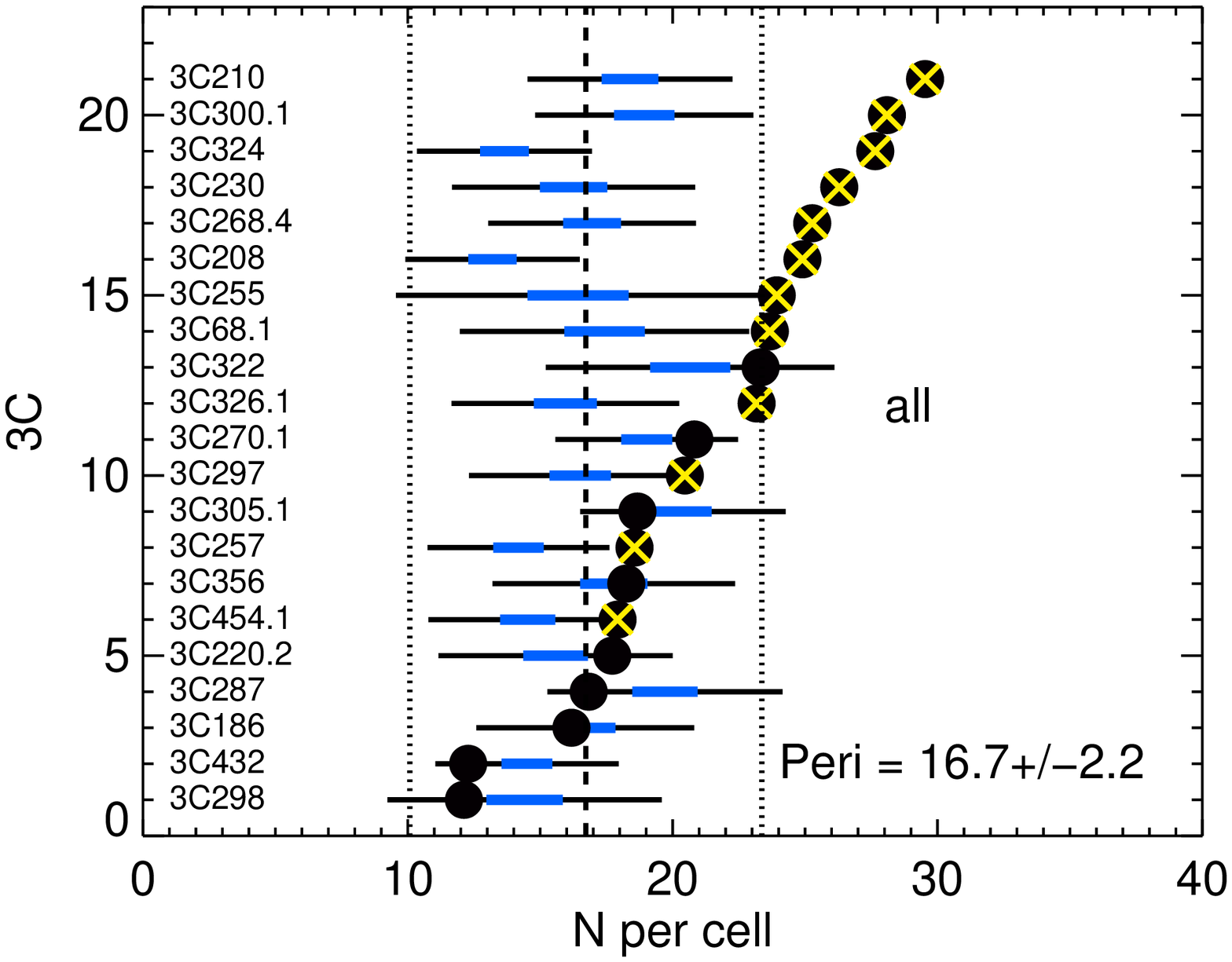}

      \caption{Frequency of central overdensities for the $red$, $IR$-$only$, $blue$, and $all$ samples 
        for the brightness range $20 < \rm{F140W} < 26$,
        $r_{\rm c} = 15\arcsec$. 
        In each panel, the 3C sources are sorted by the central surface density 
        (vertical axis).
        For each 3C source, the 
        horizontal axis shows the 
        surface density of the center (SDC, 
        filled circle) and of the periphery (SDP, horizontal bar with standard deviation as black bar 
        and EoM as thicker blue bar) using values from Table~\ref{tab_od}. 
        Crosses within the symbols indicate whether
        the COD is significant according to approach 1, 
        as described in the text.
        The vertical lines mark the average of the periphery calculated over all 3C sources (long-dashed line) and the 3$\sigma$ range (dotted
        line); 
        the corresponding values are 
        shown 
        in the lower right corner.
      }
      \label{fig:sd_statistics}
    \end{figure*}

    We have calculated the central overdensities as $ \rm COD = \rm SDC - \rm SDP$. 
    The surface density of the center (SDC) and the periphery (SDP) 
    were calculated as the average 
    of the first two and last four data points of the radial density profiles, respectively.
    The cells entering the SDC calculation reach to $r=14 + 15 = 29\arcsec$, and 
    the cells entering the SDP calculation reach  $r=42 - 15 = 27\arcsec$, but the overlap 
    is negligible.    
    Throughout this paper, ``periphery'' means the outer region surrounding the center as measured on the \HST\ images; 
    it does not imply that the area is beyond the cluster.
 
    To estimate the significance of the COD, a simple approach would be to require 
    that the COD is larger than three times the standard deviation of the periphery $\sigma_{\rm SDP}$. 
    Unfortunately, this approach fails in most cases, because the subclustering in the periphery increases $\sigma_{\rm SDP}$.
    The situation would become even worse, when using external comparison fields, because of the cosmic field-to-field variance. 
    However, for our purpose it is relevant
    whether there is a local density enhancement centered on the 3C source, 
    irrespective of the presence of other subclustering in the periphery.
    Therefore, 
    we consider the overdensity as significant, if it is above three times the error of the mean (EoM) of SDP. \footnote{
      EoM = $\sigma_{\rm SDP} / \sqrt{N}$. SDP has been calculated from $N$ cells. 
      They overlap and thus are not independent.
      Therefore, we approximated the number of independent cells ($N_{\rm ic}$) 
      from the areas of periphery ($A_{\rm p}$) and cell ($A_{\rm cell}$):
      $N_{\rm ic} = A_{\rm p} / A_{\rm c}$.
      $A_{\rm p}$ is the area of the \HST\ image (reduced by 15$\arcsec$ for cell radius and 5$\arcsec$ for border) 
      minus the central area of $r\sim30\arcsec$. 
      $N_{\rm ic} \approx 13$
      for
      $A_{\rm p} = 103 \times 116 - \pi \cdot 30^2$ square arcsec
      and
      $r_{\rm c} = 15\arcsec$.
    }
    Fig.~\ref{fig:od_histograms} illustrates that this approach is reasonable and well justified.
    Table~\ref{tab_od} lists the surface overdensity for all four catalogs.
    Fig.~\ref{fig:sd_statistics} graphically summarizes the frequency of overdensities;
    it also shows that the alternative use of the average of the
    periphery calculated over all 3C sources (vertical dashed and dotted lines) 
    fails in many cases to reveal overdensities
    because of the large 
    field-to-field variations of the periphery, 
    and therefore for each 3C field its individual periphery is used to quantify the central overdensity relative to the periphery.

\begin{figure*}

    \hspace{-2.5mm}\includegraphics[height=11.65cm, clip=true]{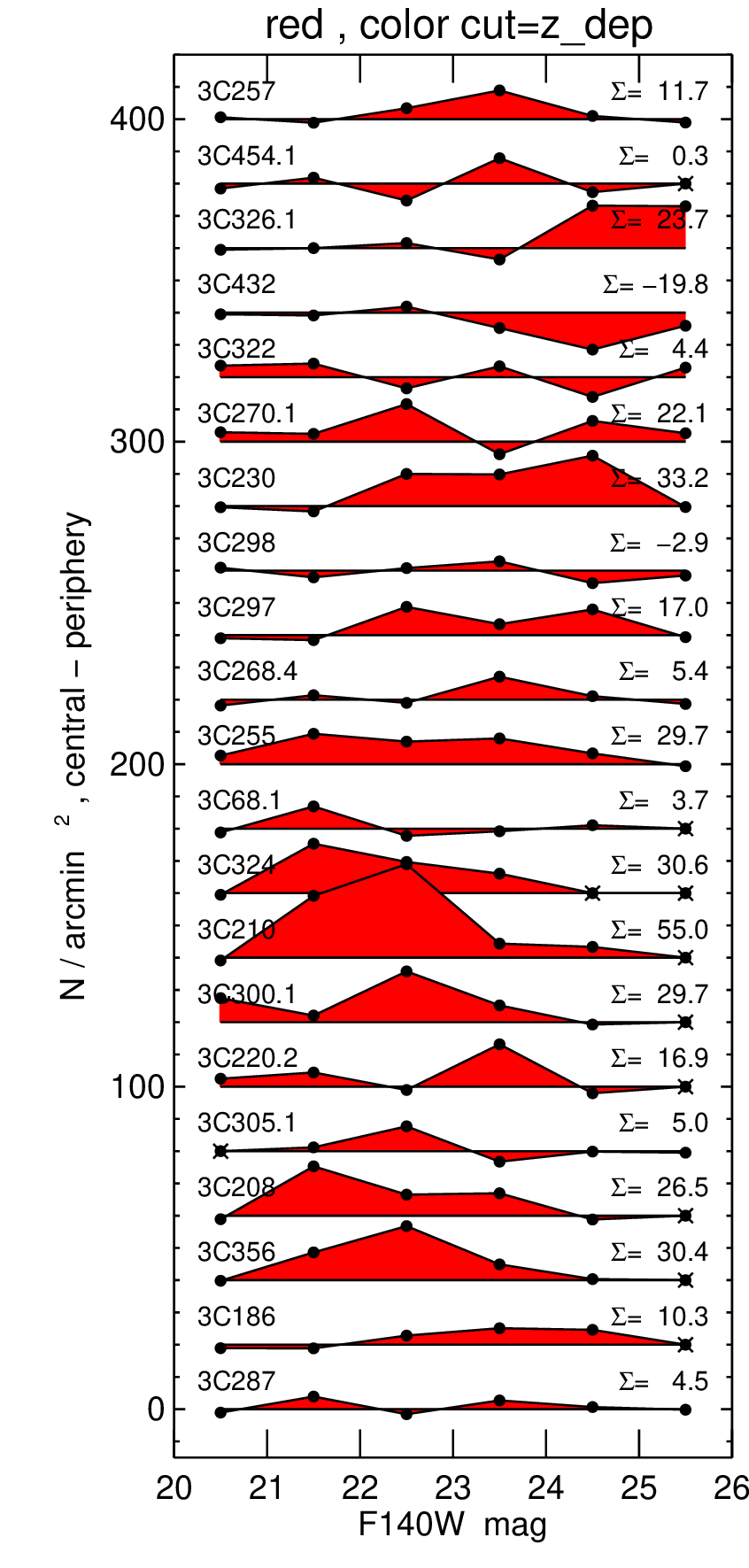}
    \hspace{-0.5mm}\includegraphics[height=11.65cm, clip=true]{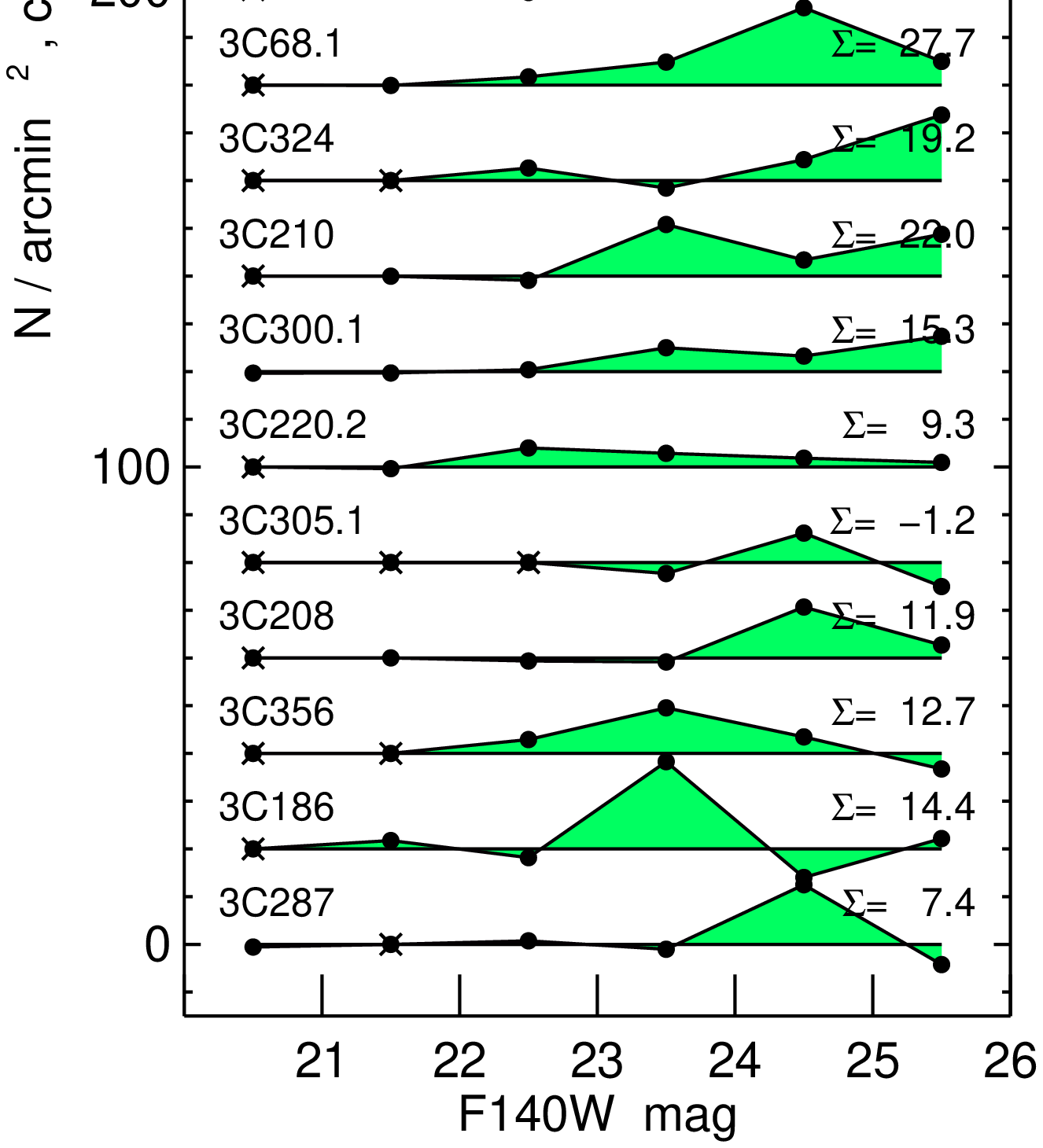}
    \hspace{-0.5mm}\includegraphics[height=11.65cm, clip=true]{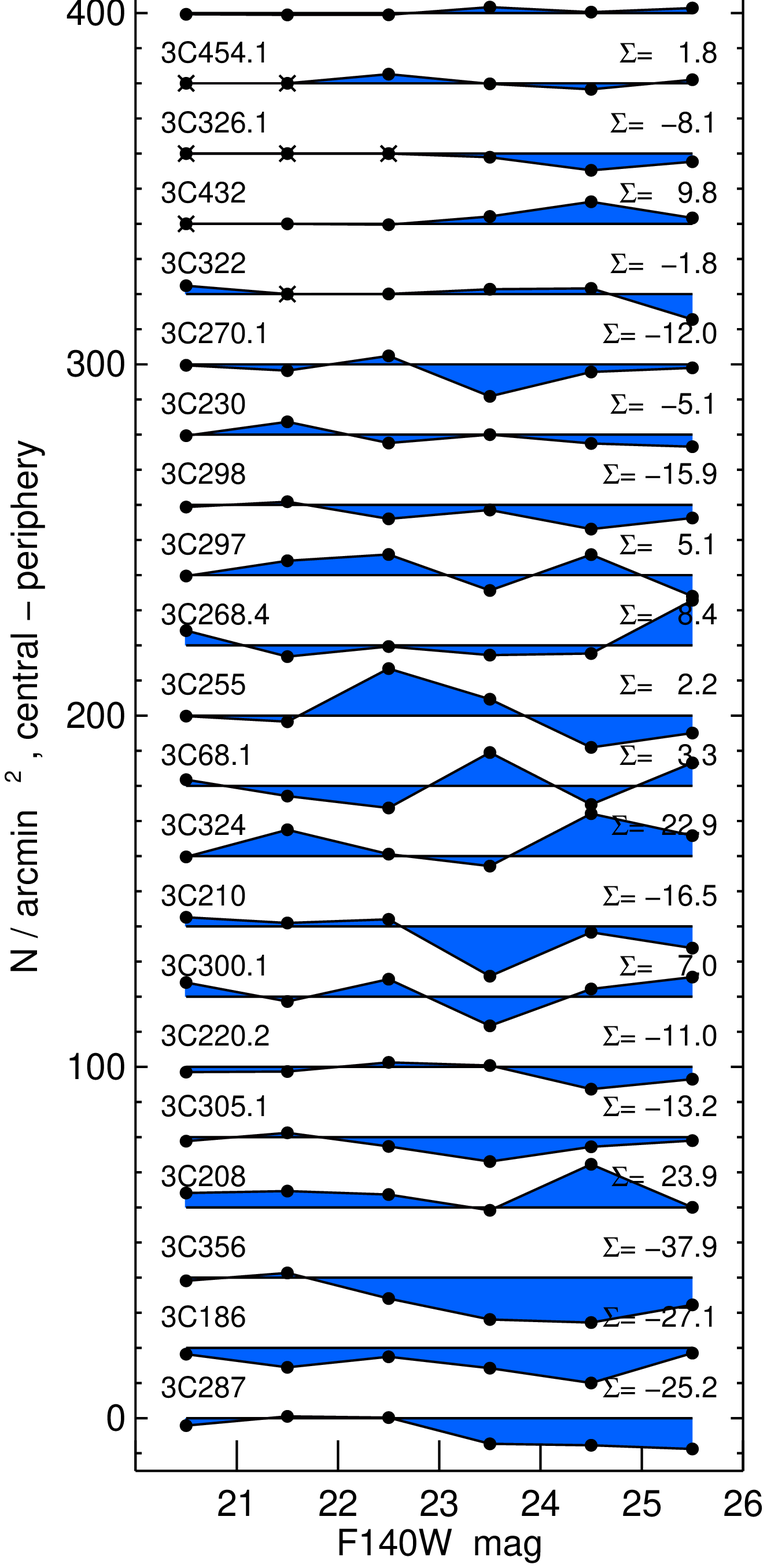}
    \hspace{-0.5mm}\includegraphics[height=11.65cm, clip=true]{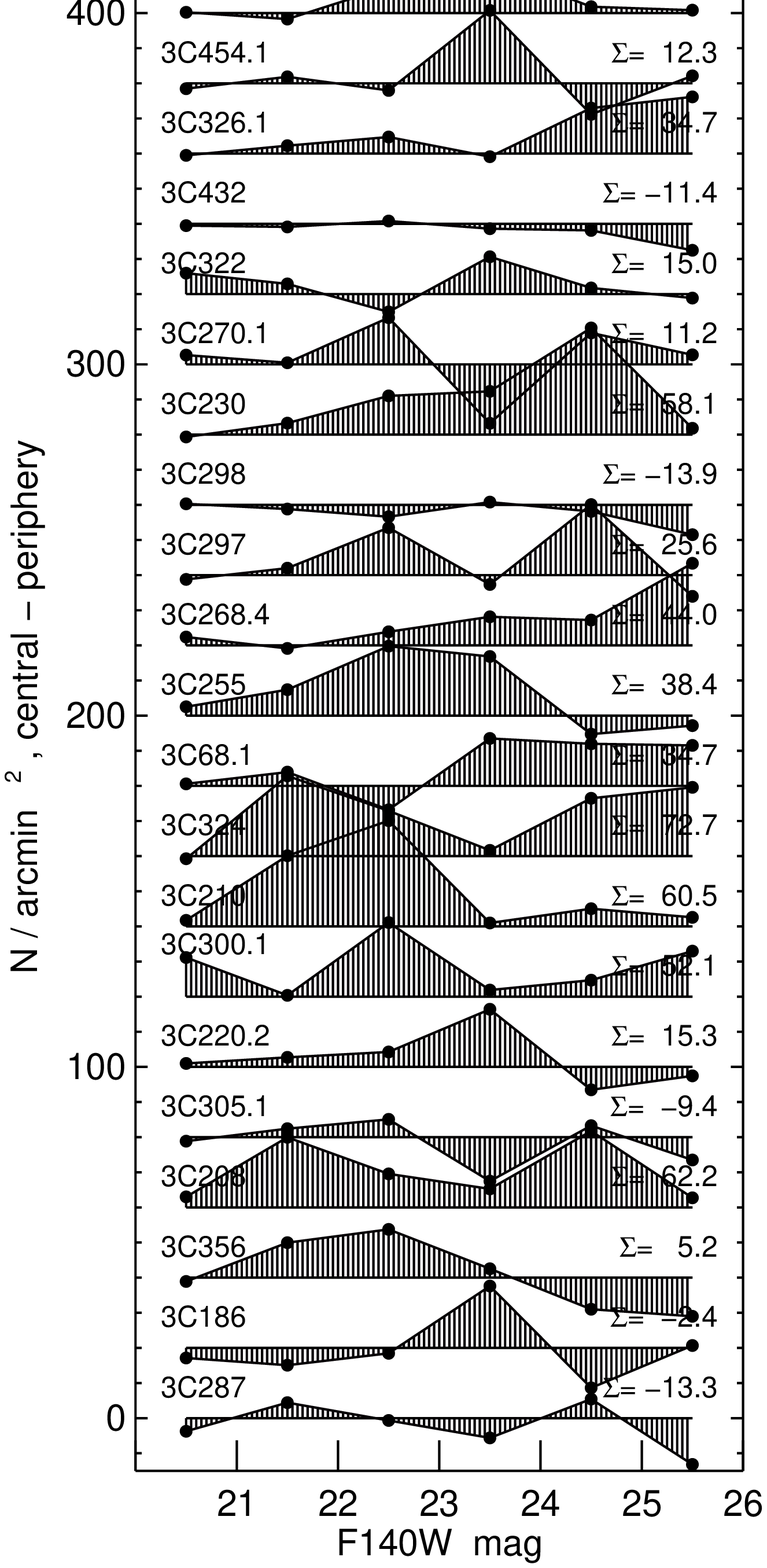}
    
    \caption{Central overdensities (COD) versus brightness 
      for the $red$, $IR$-$only$, $blue$ and $all$ subsamples.
      Each row plots 
      the COD around the labeled 3C 
      source,
      which are sorted by redshift.
      The rows are subsequently shifted vertically by 20 / arcmin$^{\rm 2}$.
      The total COD integrated over the brightness
      range 
      is labeled as $\Sigma$.
      The 3C sources themselves have been excluded from the counts.
      The brightness bins have a width of 1\,mag,
       which results in small numbers of objects per bin.
      A COD of $N = 5$ per square arcmin translates to 1 galaxy per cell of $r = 15\arcsec$.
    }
    \label{fig:od_versus_mag_all}
\end{figure*}

    Some 3C sources exhibit a significant COD without any doubt, as confirmed by inspection of the maps 
    (e.g., Fig.~\ref{fig:sd_maps_3C210}).
    Others show no COD.
    If we use a different range of luminosities, however, 
    the radial profiles of some of these sources indicate a weak COD (Fig.~\ref{fig:od_versus_mag_all}). 
    Examples are 3C\,68.1, 3C\,257, 3C\,270.1, 3C\,322 3C\,326.1, 3C\,454.1.
    In addition to the CODs derived using the standard parameters, Table~\ref{tab_od} 
    lists these fine-tuned cases in the column ``comments''. 

    About 16 
    of the 21 observed 3C sources show a significant COD;
    that means they lie in a clustering of $red$ or $IR$-$only$ galaxies within a radius of 250\,kpc centered on the 3C source. 
    In about 5 of those fields,
    the $red$ and $IR$-$only$ COD is accompanied by a $blue$ CUD (negative $blue$ COD),
    which means that the $blue$ galaxies avoid the center and hence are overdense at
    radius 250--500\,kpc.

    Most 
    of the detected 
    CUDs are at  $z < 1.5$ (Table~\ref{tab_classification}).
    When the $blue$ CUDs are accompanied by a $red$ COD, 
    they are $\sim$2\,mag fainter than the $red$ CODs (e.g.,  3C\,356, 3C\,210).
    This is
    consistent with the expectation that the passive (giant) red cluster galaxies are brighter
    than the star-forming $blue$ galaxies (even at restframe optical wavelengths).

    The $blue$ CODs at  magnitude 20--23 
    indicate the presence of luminous star-forming galaxies in the clusters (e.g., 3C\,208, 3C\,324). 
    Sources with neither $blue$ COD nor CUD are 3C\,305.1 and 3C\,220.2. 
    At $z > 1.5$,  the detection of $blue$ CODs or CUDs is rare; this could be due to 
    the redshift-dependent blue color cut which leaves few $blue$ galaxies (Fig.~\ref{fig:cmd}).

    The surface density maps often show several over- and under-densities in the same field 
    (Figs.~\ref{fig:sd_maps_1} to \ref{fig:sd_maps_7}). 
    Without spectroscopic redshift measurements, we cannot exclude the possibility 
    that some galaxy ODs are at a different redshift, such as a group projected on the line of sight. 
    However, the probability of projected galaxy groups unrelated to a single 3C source is small:



Firstly, 
      the central ODs are fairly well centered on the 3C source (within 15$\arcsec$ radius).
      They are often accompanied by additional ODs in the surroundings.
      The $red$ surface density maps show between 1 and 4 ODs per map, 
      on average $n1 = 2$ ODs.
      The area of the surface density maps is $a_{\rm map} =  2.5$ square arcmin. 
      The probability $P1$ that at least one of the OD centers falls inside $r=15\arcsec$ around the 3C source is 
      
      $$ P1 = n1 * a_{15} / a_{\rm map} = 0.15 = 15\%,$$ 
      where $a_{15}$ is the area inside $r=15\arcsec$. 
      A $red$ COD is found in 14--15 of the 21 fields. 
      The 
      joint
      probability of finding 14 or more chance projections is
      $P < 6 \cdot 10^{-6}$.
      Therefore, chance projections 
      cannot account for the frequency of $red$ CODs.

      Secondly, some $red$ CODs are accompanied by $blue$ central underdensities with $r>30\arcsec$. 
      A pure projection scenario requires 
      that
      a $red$ OD falls close to the 3C source ($P1$) 
      and 
      additionally
      that all ($n2$) blue clumps  
      lie 
      outside $r=30\arcsec$ 
      region
      around the 3C source ($P2$).
      The $blue$ surface density maps show between 1 and 6 ODs per map, 
      on average $n2 = 4.5$ ODs.
      The probability of $n2$ out of $n2$ $blue$ ODs centered in the periphery 
      at $r>30\arcsec$ is: 
      
      $$P2 = [(a_{\rm map} - a_{30}) / a_{\rm map}]^{n2} = 0.185 = 18.5\%.$$
      
      The combined probability 
      of finding both 
       a $red$ COD and a $blue$ CUD is then 
      $P_{comb} = P1 * P2 = 0.15 * 0.185 = 0.028 = 2.8\%$.  
      This estimate is for a single 3C field with $red$ COD and $blue$ CUD.
      Our sample contains 3 out of 21 such fields. 
      The probability of finding that many (or more) chance projections is 
      $P  = 0.022 = 2.2\%$.
      Again, chance projections 
      make at most a minor contribution to the joint CODs/CUDs.

    In addition,     
    the cumulative CODs/CUDs completely disappear 
    when the sky position of the 
    centroid 
    is randomly altered (Sect.~\ref{sec:cod}).
    This rules out a pure projection origin.




\subsection{Brightness dependence of the overdensities} \label{sec:od_vs_mag}

Fig.~\ref{fig:od_versus_mag_all} shows the  CODs versus brightness of the four
galaxy types $all$, $red$, $IR$-$only$ and $blue$.
The figure reveals COD brightness trends which depend on the galaxy colors:

\begin{itemize}

\item  [1)]
  $Red$ galaxies: at $z < 1.3$, most 3C sources show
  a striking COD of  
  $red$ galaxies in the $20-24$ magnitude range.
  At higher redshift (upper half of the panel) the CODs weaken and shift to 21--25\,mag.
  The shift is consistent with the increase of the distance modulus $m - M$ from $z=1$ to $z=1.5$.
  This is reminiscent of a wave running through the parameter space, 
  whereby the amplitude decreases with increasing redshift (probably due to incompleteness at faint magnitudes).
  A few 3C sources lack $red$ CODs in all brightness bins.

\item  [2)]
  $IR$-$only$ galaxies: nearly all 3C sources show an 
  $IR$-$only$ COD in at least one magnitude bin,
  mostly at  magnitude 23--26.
  The $IR$-$only$ sample contains faint, presumably red galaxies,
  beyond the completeness limit of the $red$ sample.
  Therefore the $IR$-$only$ CODs 
  are likely to be a continuation 
  of the $red$ CODs toward faint magnitudes.
  3C sources lacking a $red$ COD but showing a prominent $IR$-$only$ 
  COD are 3C\,186 ($z = 1.1$), 3C\,68.1 ($z = 1.2$),
  3C\,298 ($z = 1.5$), and 3C\,454.1 ($z = 1.8$). 
  These CODs are also supported by the surface density maps; 
  3C\,186 and 3C\,454.1 are located between two density enhancements.

\item  [3)]
  $Blue$ galaxies: they appear as CODs and UDs (underdensities = negative CODs).
  The UDs indicate a lack of $blue$ galaxies in the center compared to the periphery.
  Most UDs are at  $z < 1.5$. 
  When the $blue$ UDs are accompanied by a $red$ COD, 
  they are $\sim$2\,mag fainter than the $red$ CODs (e.g.,  3C\,356, 3C\,210).
  This is
  consistent with the expectation that the passive (giant) red cluster galaxies are brighter
  than the star-forming $blue$ galaxies (even at rest-frame visible wavelengths).
  On the other hand, the $blue$ CODs at  magnitude 20--23, 
  indicate the presence of luminous $blue$ starforming galaxies (e.g., 3C\,208, 3C\,324). 
  At $z > 1.5$,  the detection of $blue$ CODs or UDs is rare; this could be due to 
  the redshift-dependent blue color cut which leaves few $blue$ galaxies (Fig.~\ref{fig:cmd}).

\item [4)]
  $All$ galaxies: along the entire brightness range, most 3C sources show a clear COD in at least two magnitude bins.
  For some 3C sources the CODs extend over three to four magnitudes, 
  allowing for an estimate of the luminosity function (LF) 
  of those sources contributing to the CODs. 

\end{itemize}
To 
summarize, the overdensities show trends with brightness and redshift, albeit with a large spread.
At $z<1.5$, 
prominent $red$ CODs are found for galaxies in the 20--23 mag range 
and these continue to fainter magnitudes (23--26 mag) as $IR$-$only$ CODs; 
this may be explained by the incompleteness of our two-filter detections.
For $blue$ galaxies, 
while 
some CODs are found, the majority of 3C sources exhibit 
$blue$ underdensities, 
mostly at faint magnitudes.
For the five 3C sources at $z>1.5$ (starting with 3C\,322), any overdensities are small,
and we suggest that this is at least partly  due to incompleteness of our shallow imaging. 


    \subsection{Cumulative overdensities}\label{sec:cod}

\begin{figure}

      \includegraphics[width=0.235\textwidth, clip=true]{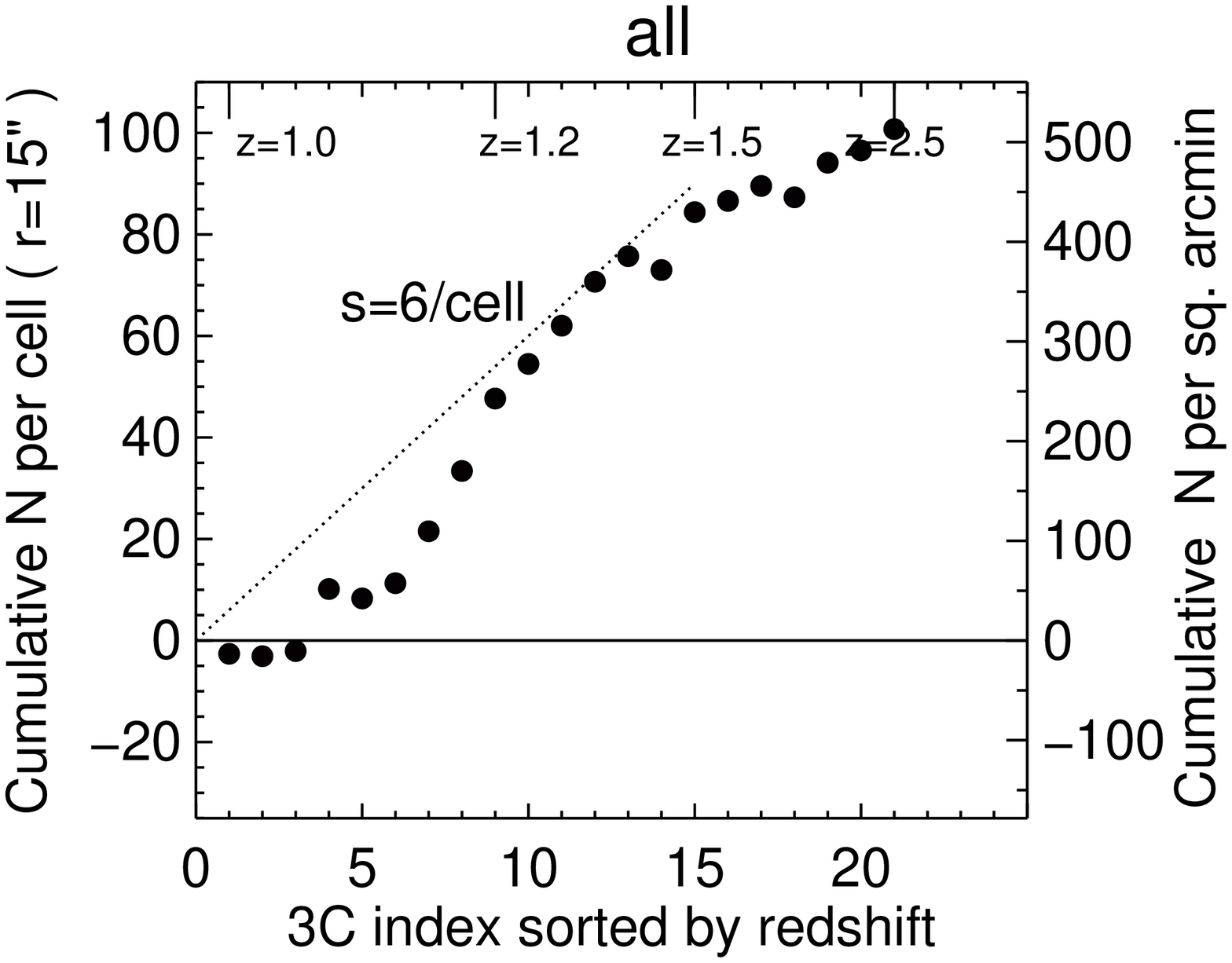}
      \hspace{1mm}\includegraphics[width=0.235\textwidth, clip=true]{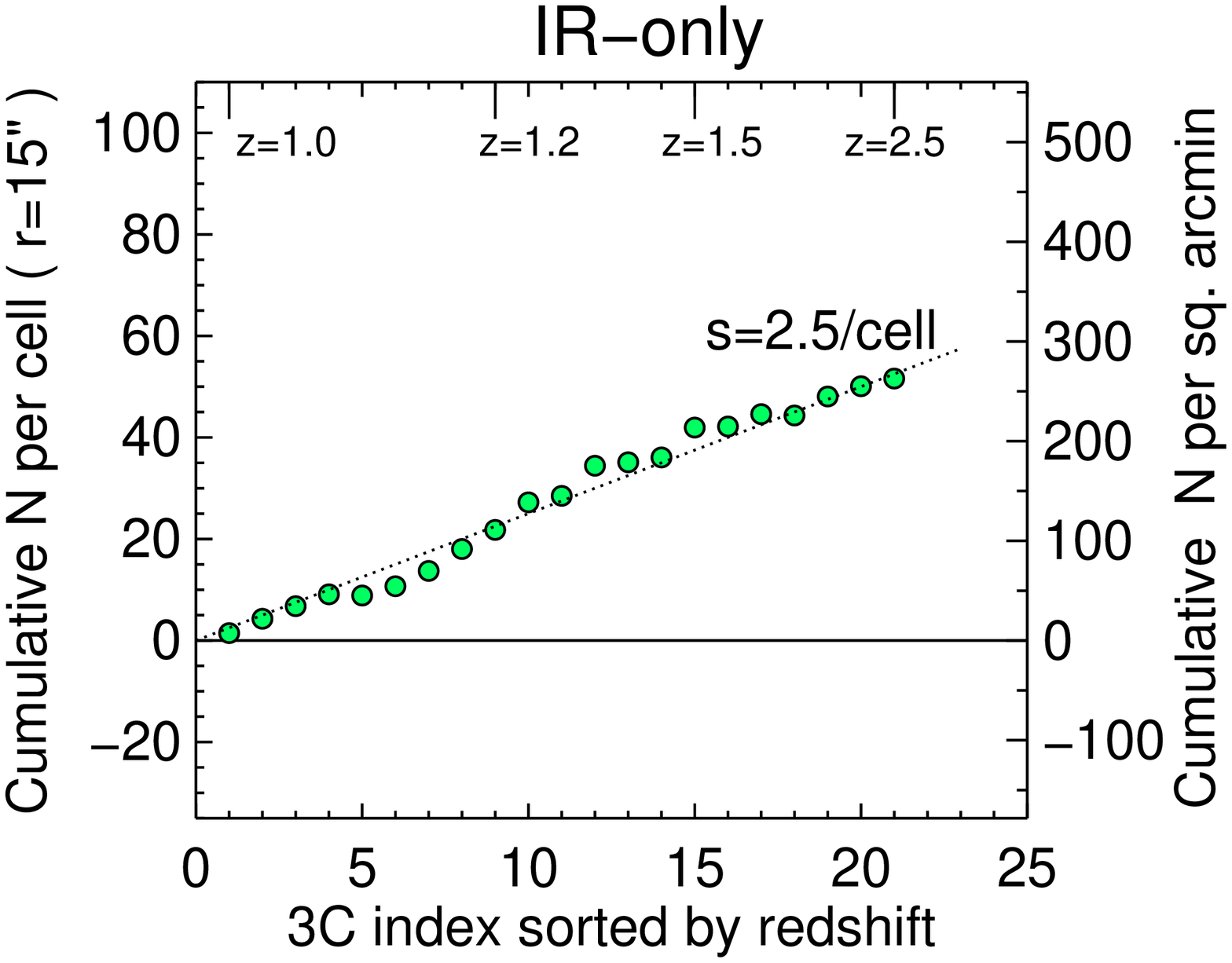}

      \includegraphics[width=0.235\textwidth, clip=true]{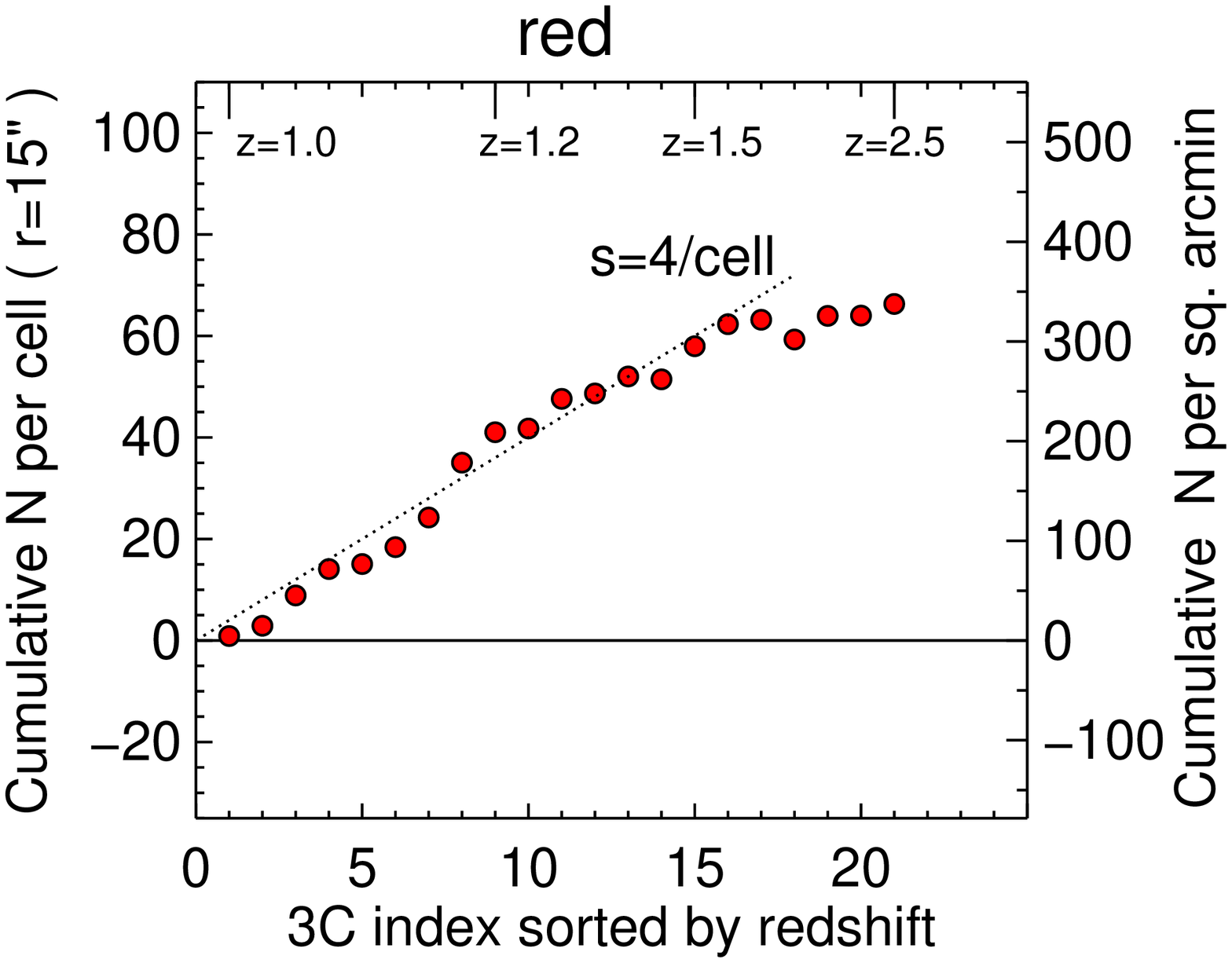}
      \hspace{1mm}\includegraphics[width=0.235\textwidth, clip=true]{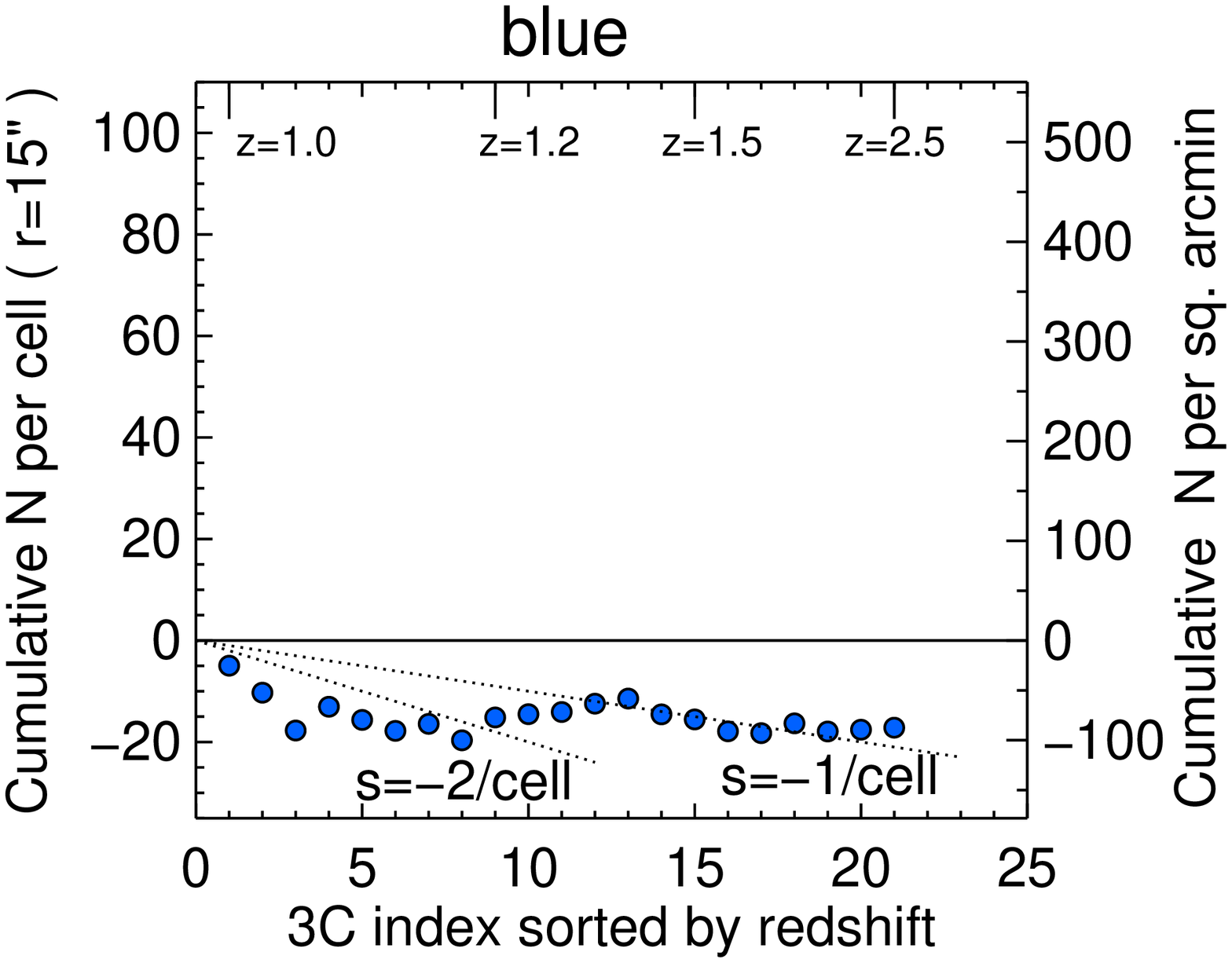}

      \caption{Cumulative central overdensities of the 
      $all$, $red$, $IR$-$only$ and $blue$
      samples 
        for the brightness range $20 < \rm{F140W} < 26$ and
        cell radius $r_{\rm c} = 15\arcsec$.
        The 3C sources are 
        not included in 
        the counts.
        The 
        horizontal axis displays  
        the index of each 3C source, sorted by redshift, as labeled at the top of each panel.
        To guide the eye, the diagonal dotted lines mark slopes of an average central overdensity of
        $s$ galaxies per cell per 3C field, 
        where the value of $s$ is shown in each panel.
      }
      \label{fig:cod_from_maps}
    \end{figure}

    Fig.~\ref{fig:cod_from_maps} shows the 
    cumulative distribution of overdensities (CCOD) as a function of redshift \footnote{
      We sort the 3C sources 
      by redshift and then 
      successively add the measured individual CODs:
      CCOD(i) $=$ CCOD(i-1) + COD(i), where COD(i) is the COD of the $i^{ th}$ 3C source.}. 
    The CCOD slope gives the average COD strength.
    Most overdensities consist of red galaxies 
    with a relative deficit of blue galaxies close to the 3C source, 
    indicating the presence of a relatively mature cluster population in the inner few hundred kpc.


    \begin{figure}

      \includegraphics[width=0.237\textwidth, clip=true]{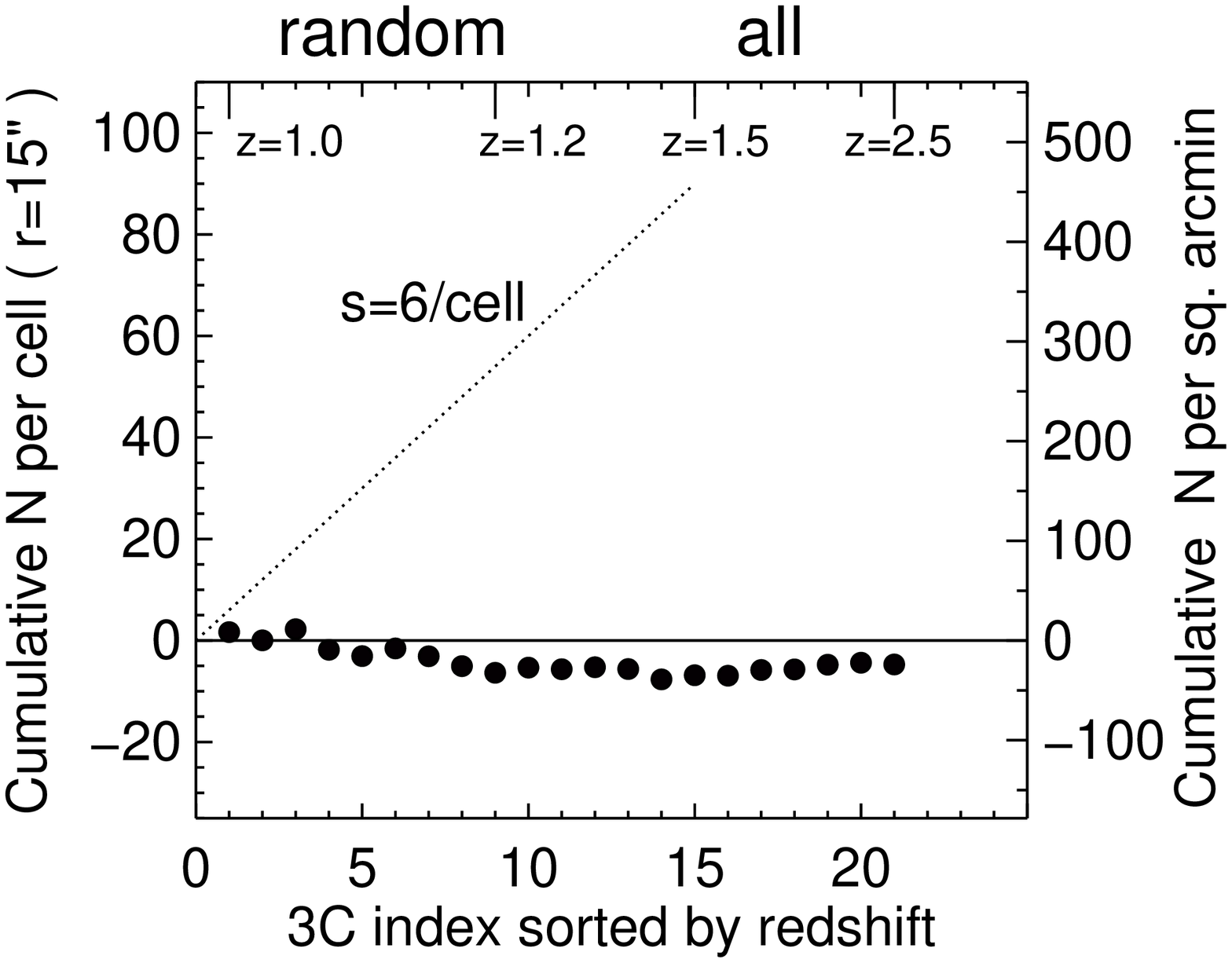}
      \hspace{1mm}\includegraphics[width=0.237\textwidth, clip=true]{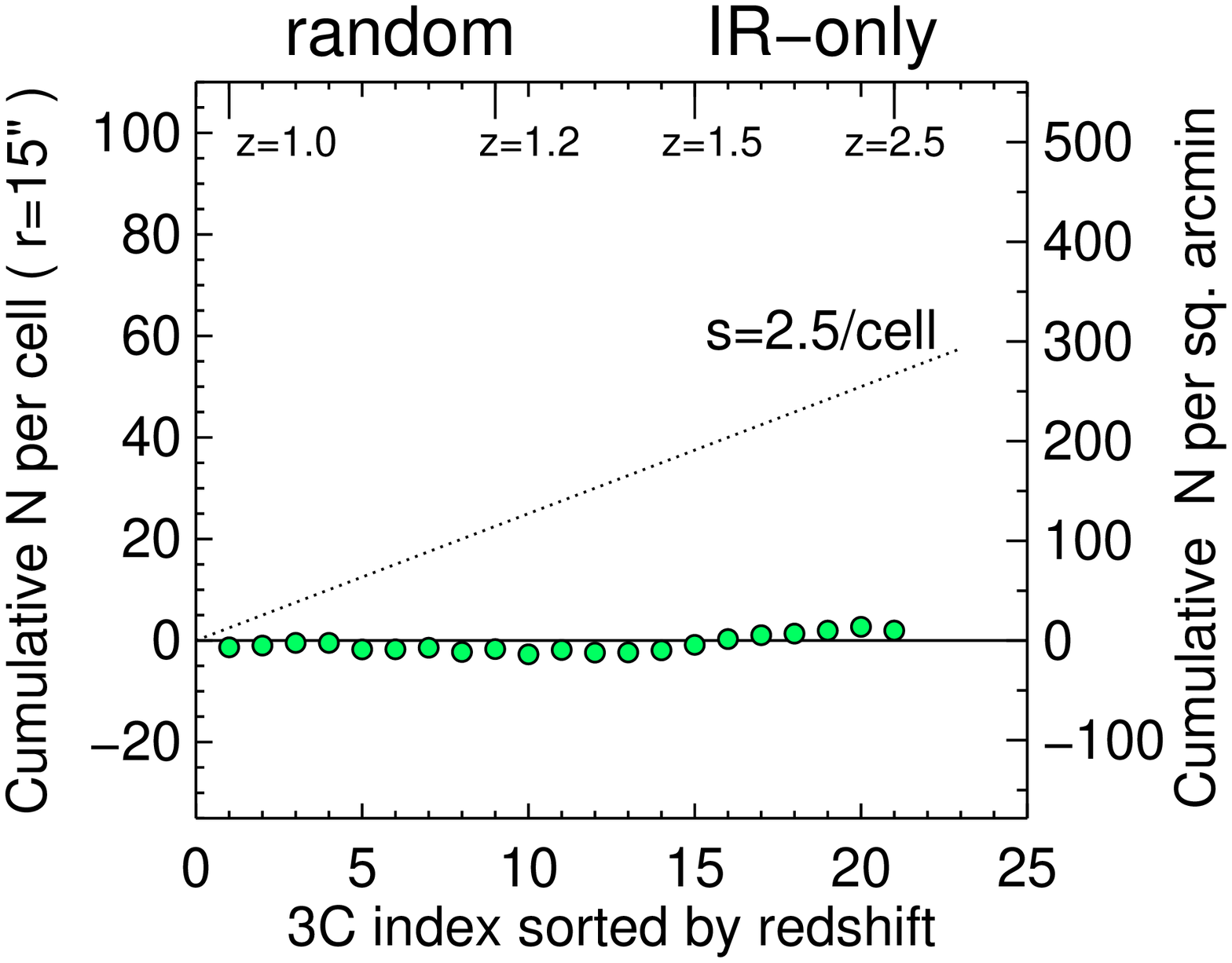}

      \includegraphics[width=0.237\textwidth, clip=true]{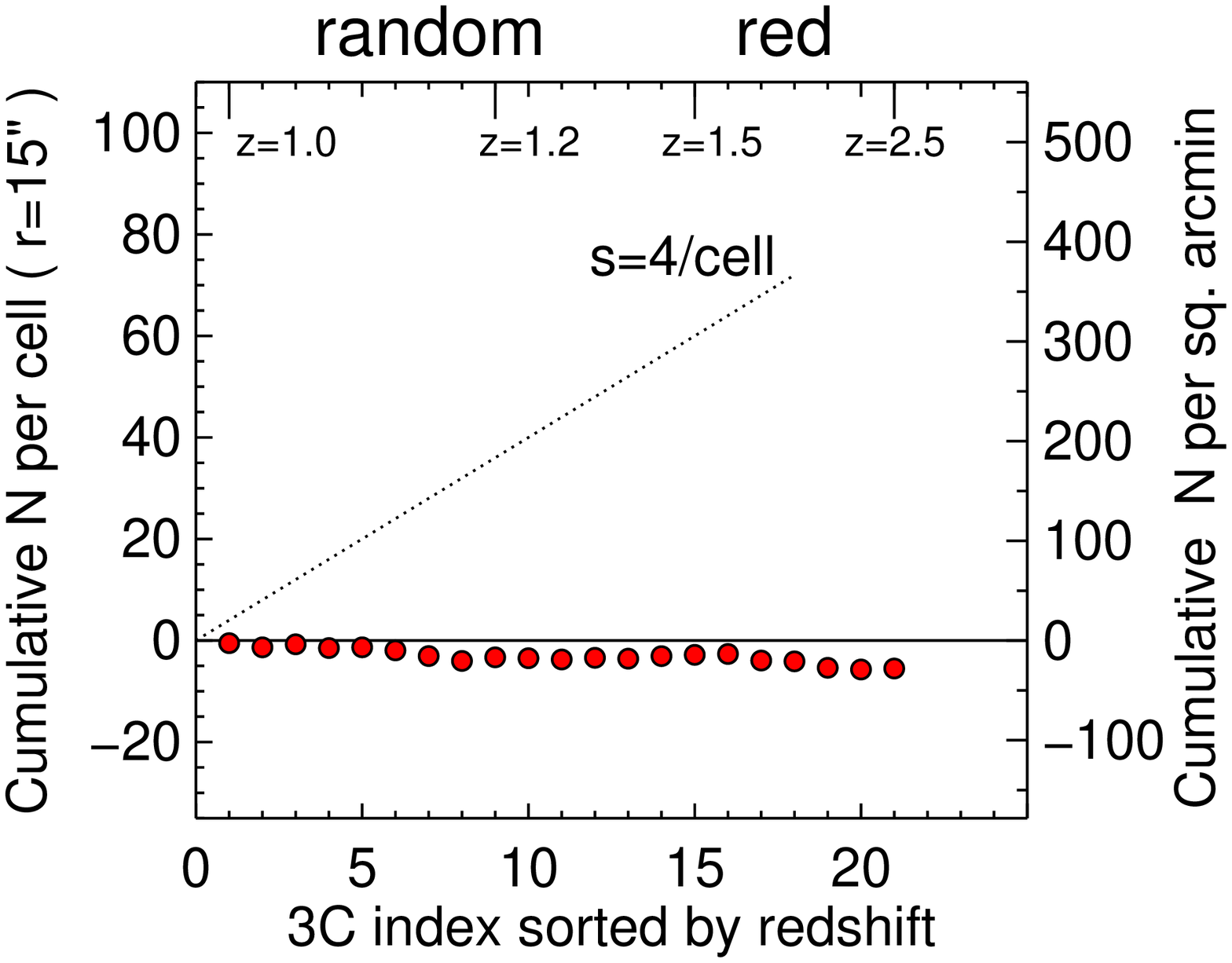}
      \hspace{1mm}\includegraphics[width=0.237\textwidth, clip=true]{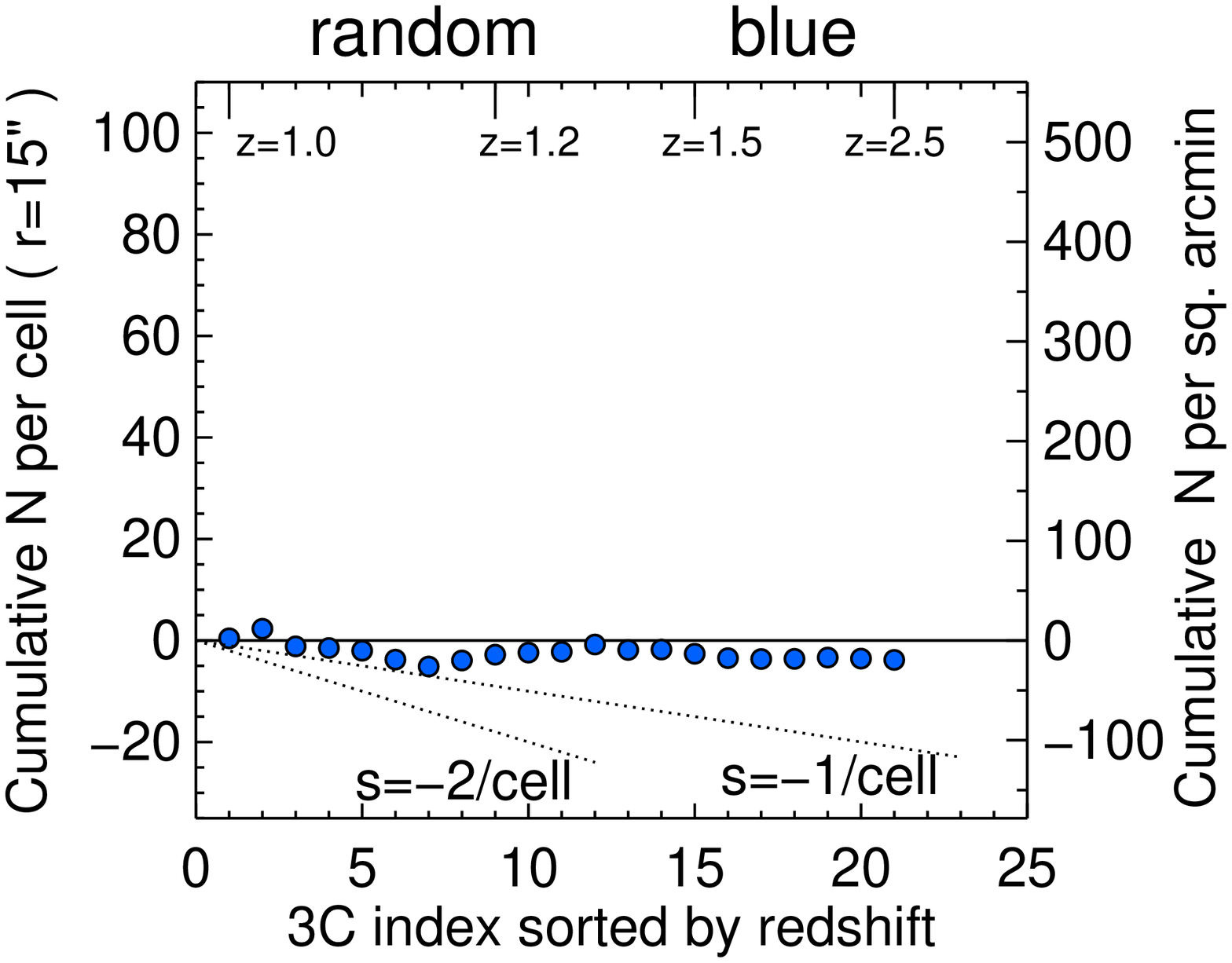}

      \caption{Result of calculating the central overdensities and their cumulative diagrams
        for randomized positions of the 3C sources.
        The average counts per cell were determined from 10 
        random realizations for all four subsamples, as indicated.
        We note the disappearance of the cumulative overdensity in comparison to Fig.~\ref{fig:cod_from_maps}.
        The dotted lines mark slopes as in Fig.~\ref{fig:cod_from_maps}.
      }
      \label{fig:cod_from_maps_random}
    \end{figure}


\begin{figure*} 
  \vspace{5mm}

  \includegraphics[height=3.46cm,clip=true]{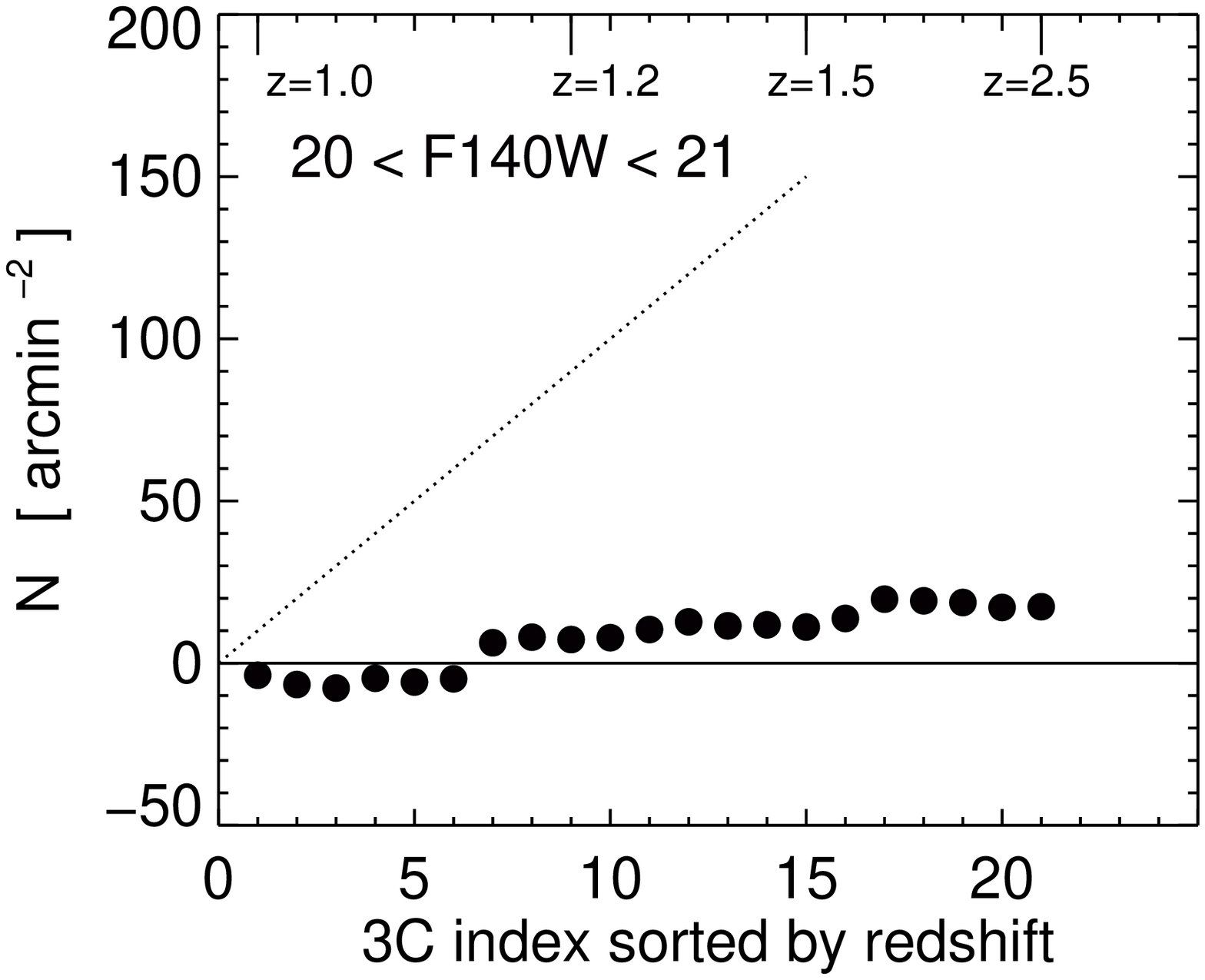}
  \includegraphics[height=3.46cm,clip=true]{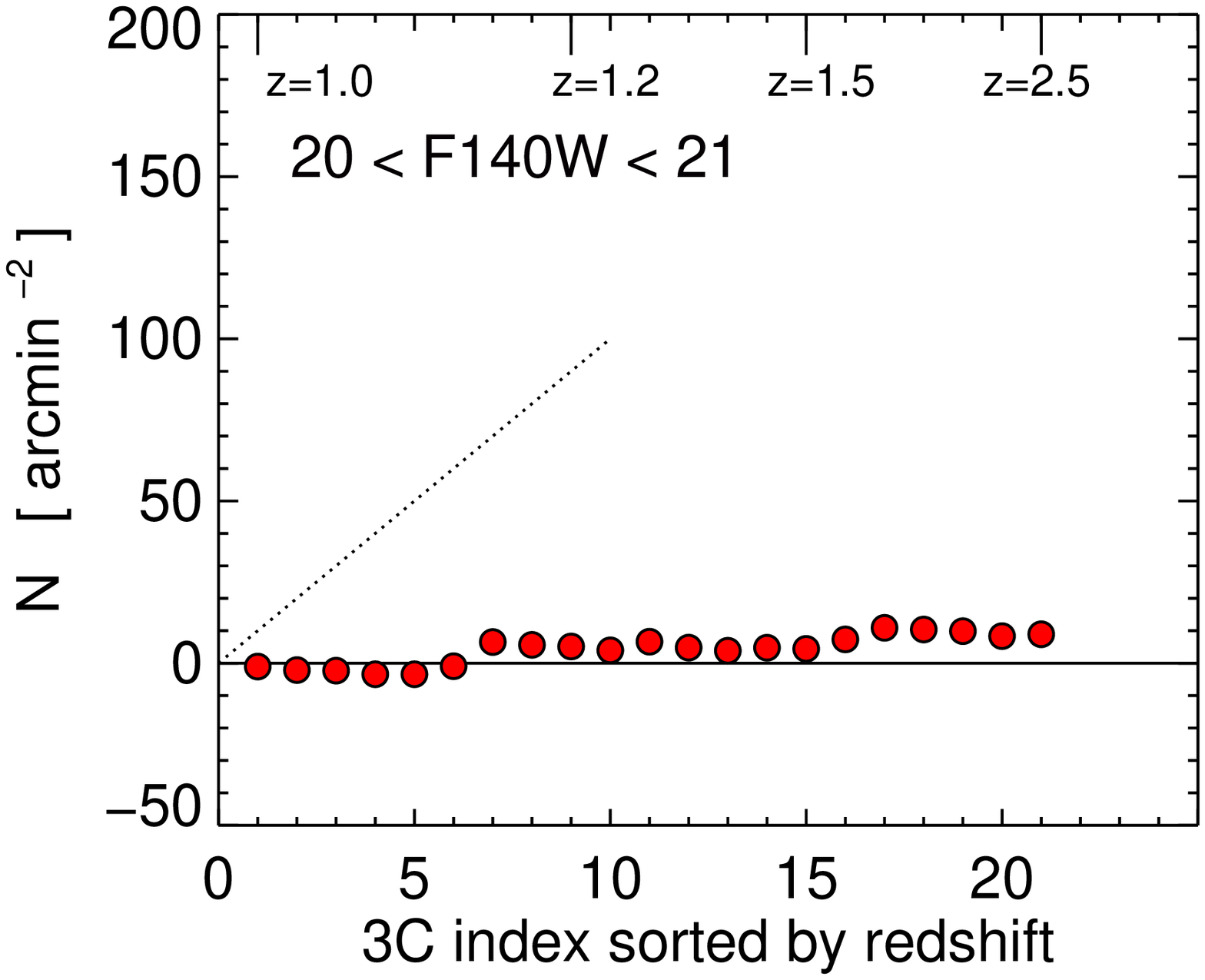}
  \includegraphics[height=3.46cm,clip=true]{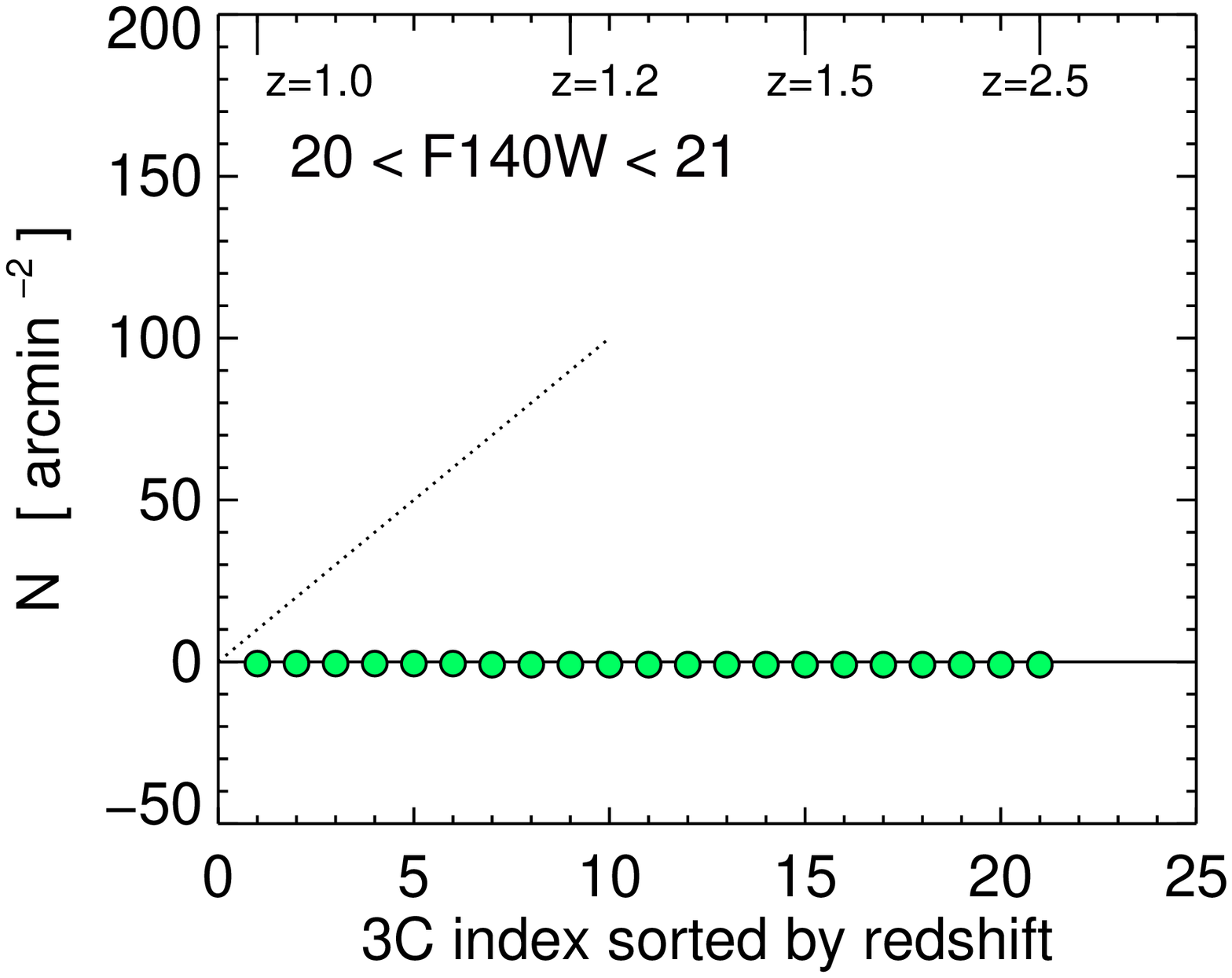}
  \includegraphics[height=3.46cm,clip=true]{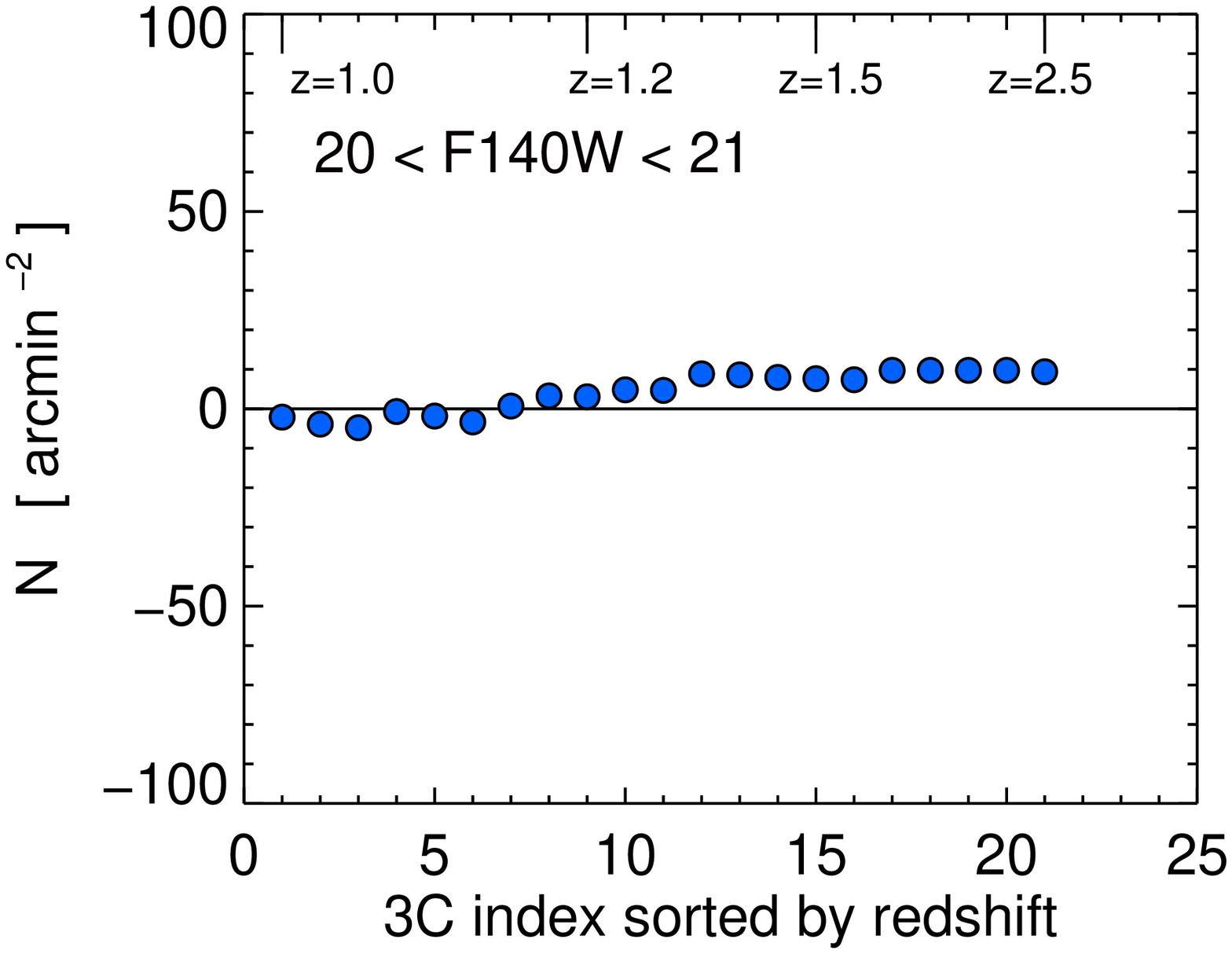}

  \includegraphics[height=3.46cm,clip=true]{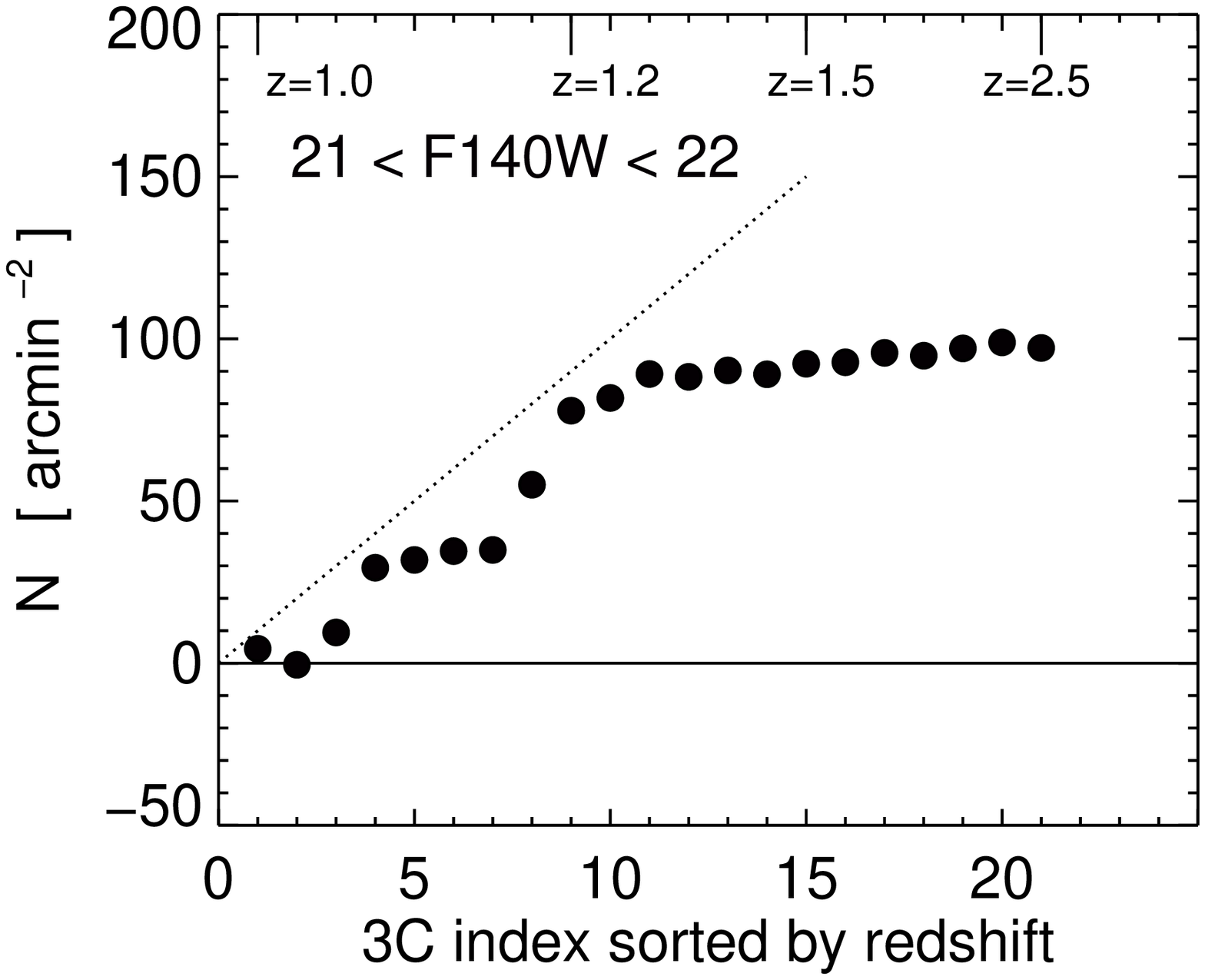}
  \includegraphics[height=3.46cm,clip=true]{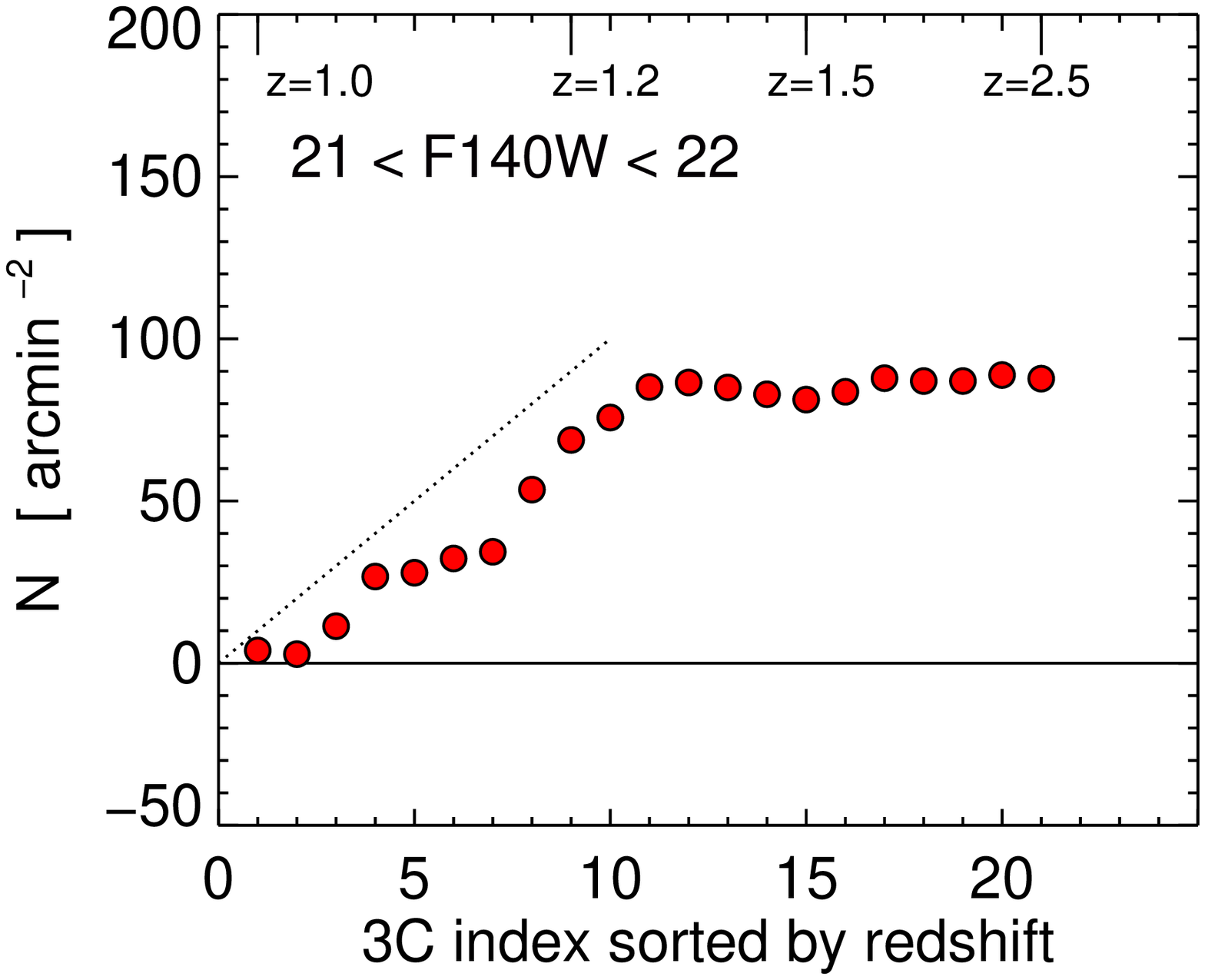}
  \includegraphics[height=3.46cm,clip=true]{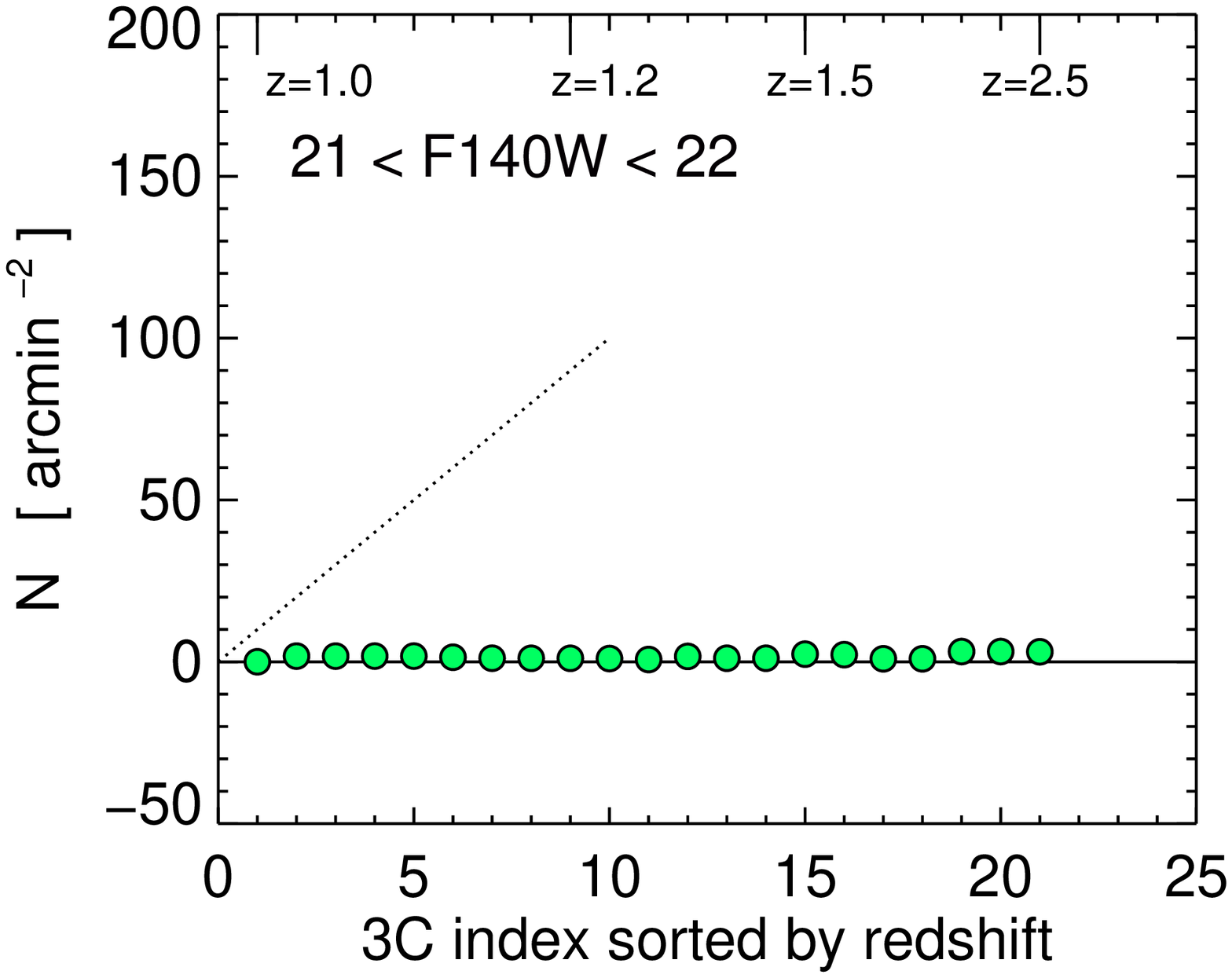}
  \includegraphics[height=3.46cm,clip=true]{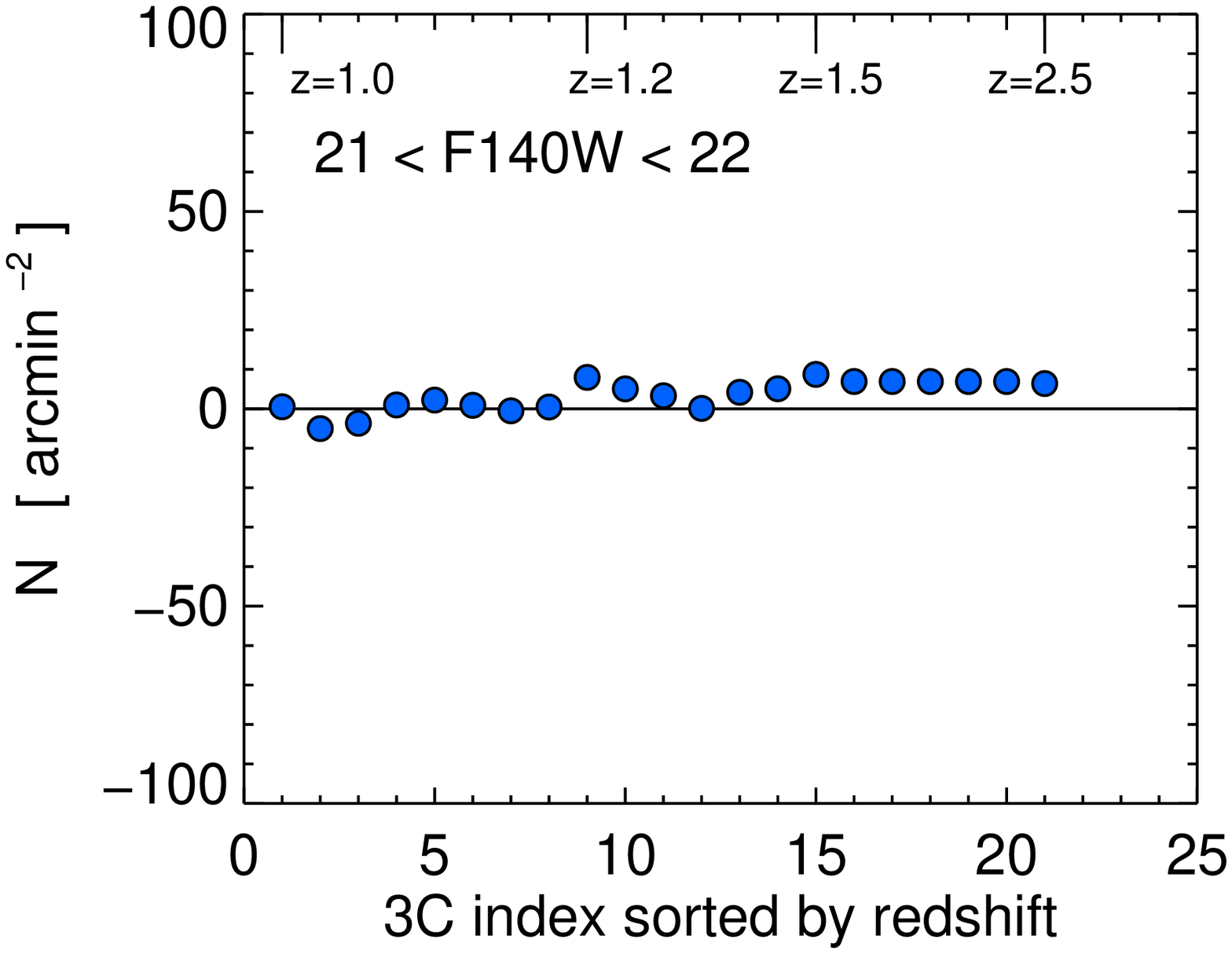}

  \includegraphics[height=3.46cm,clip=true]{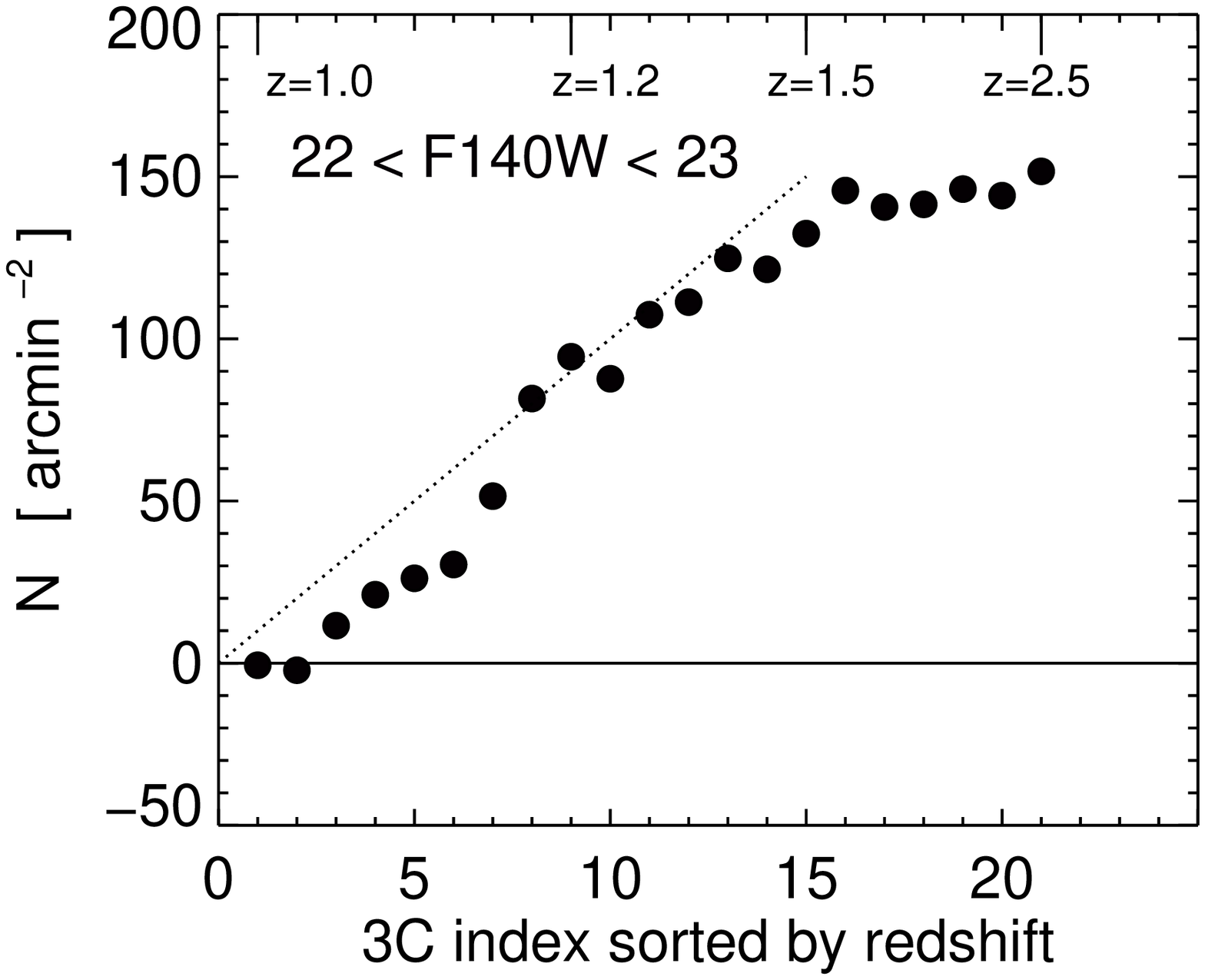}
  \includegraphics[height=3.46cm,clip=true]{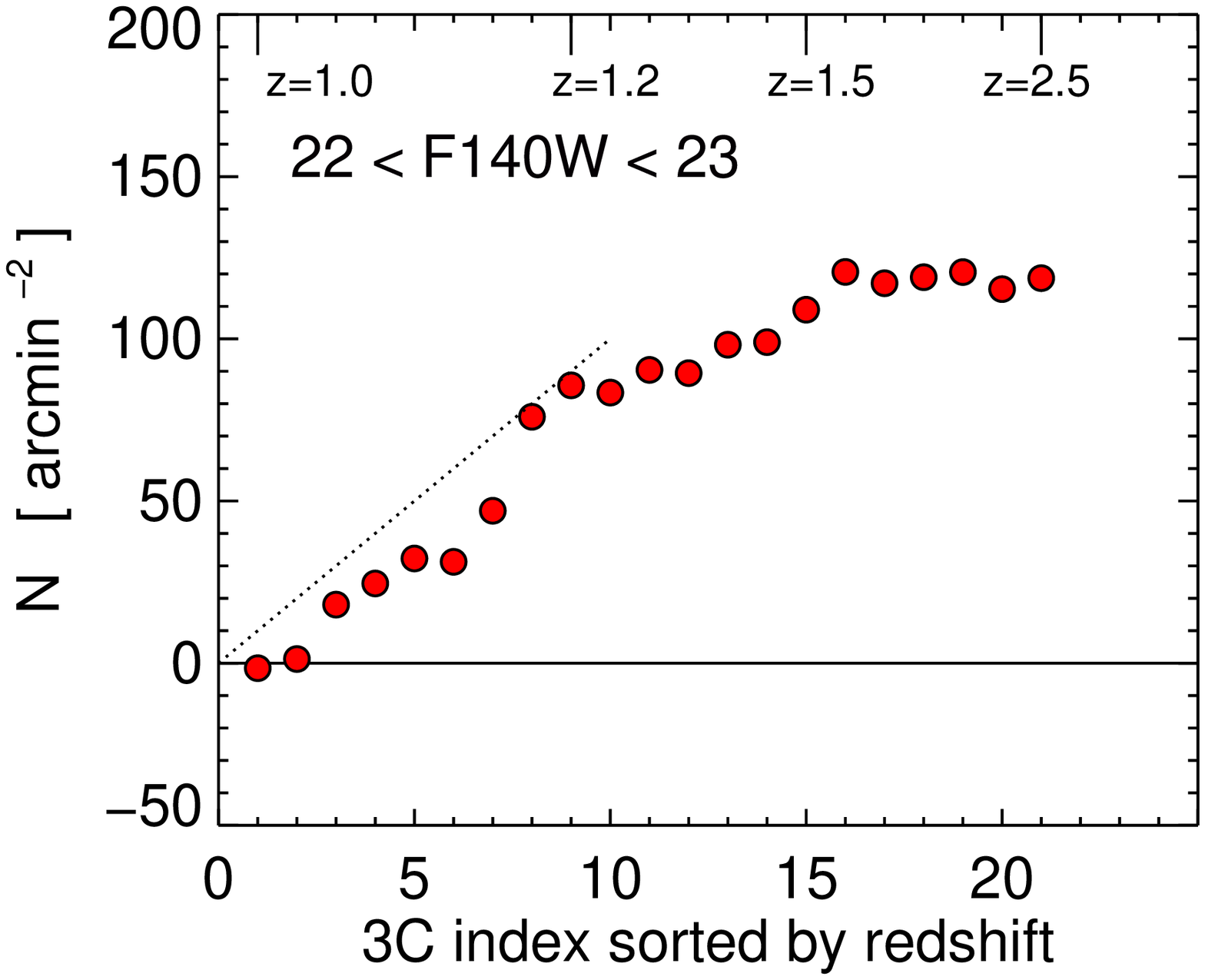}
  \includegraphics[height=3.46cm,clip=true]{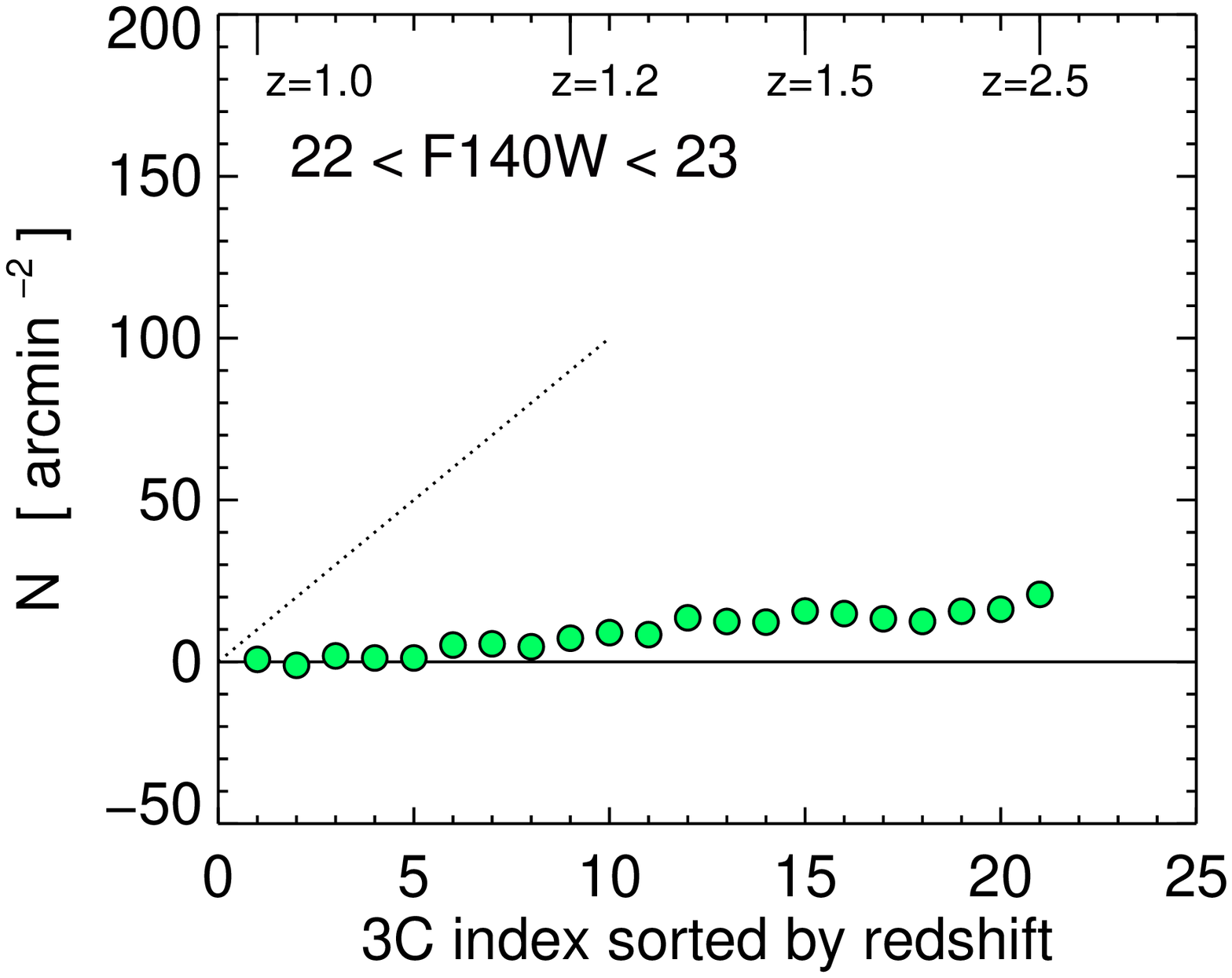}
  \includegraphics[height=3.46cm,clip=true]{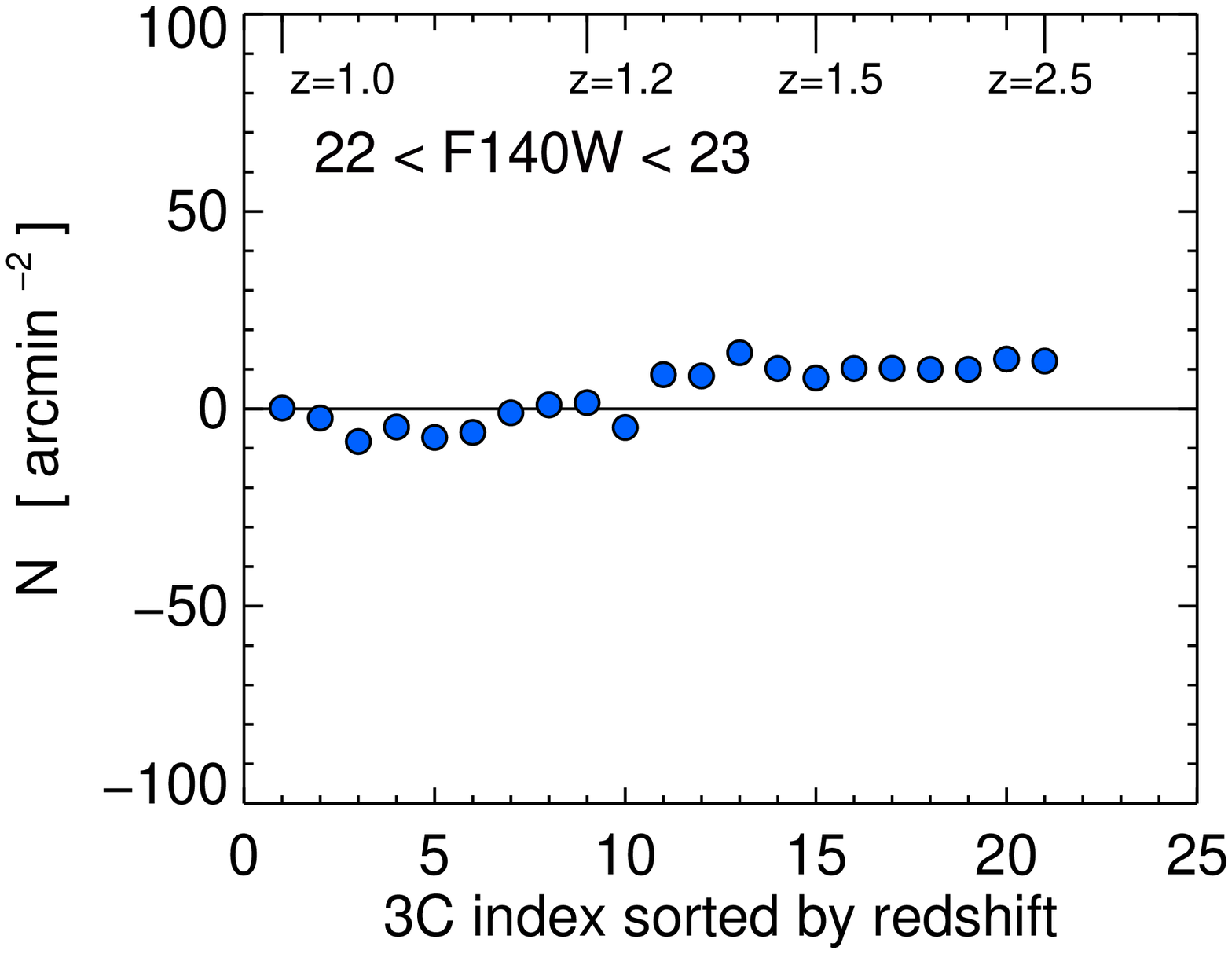}

  \includegraphics[height=3.46cm,clip=true]{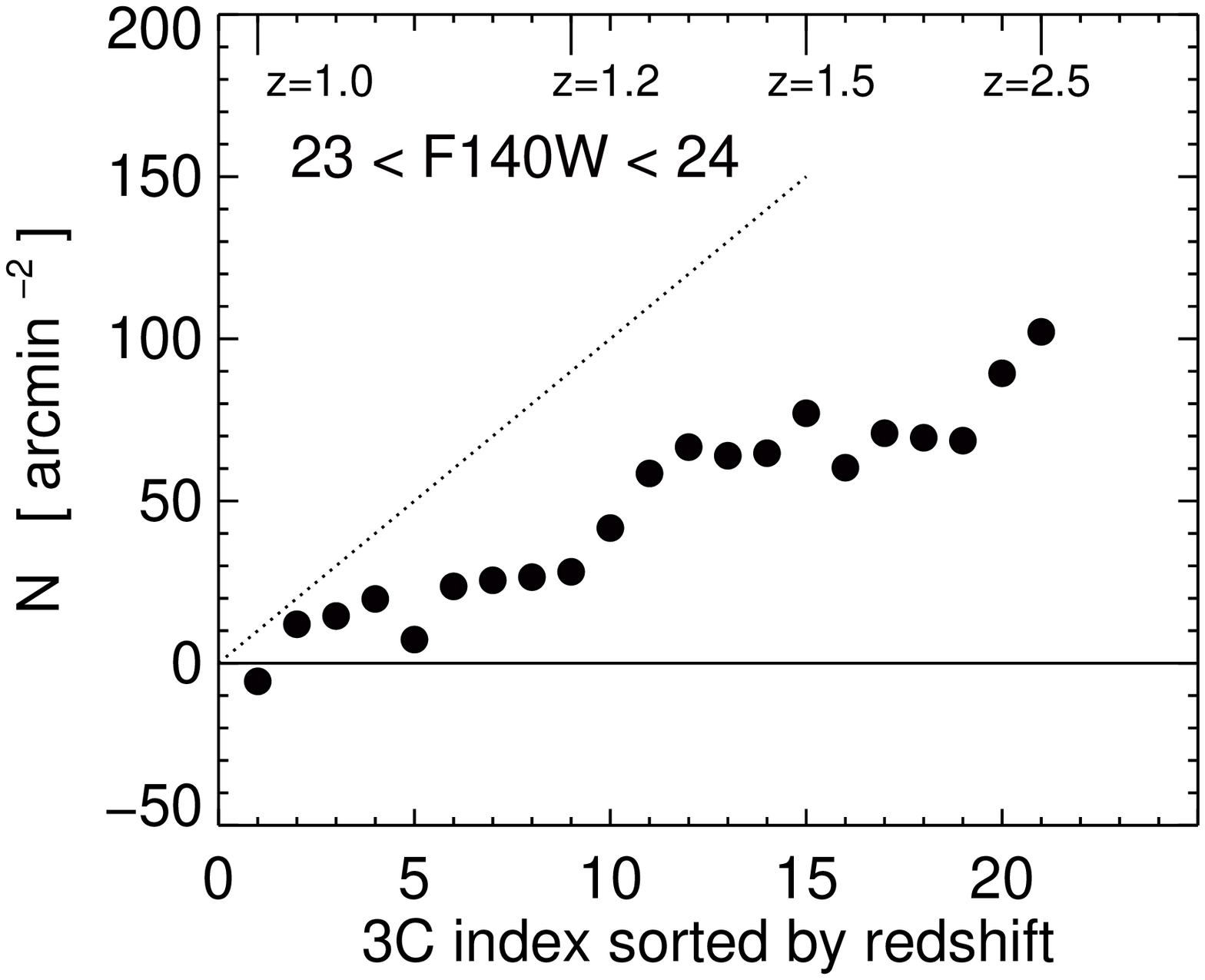}
  \includegraphics[height=3.46cm,clip=true]{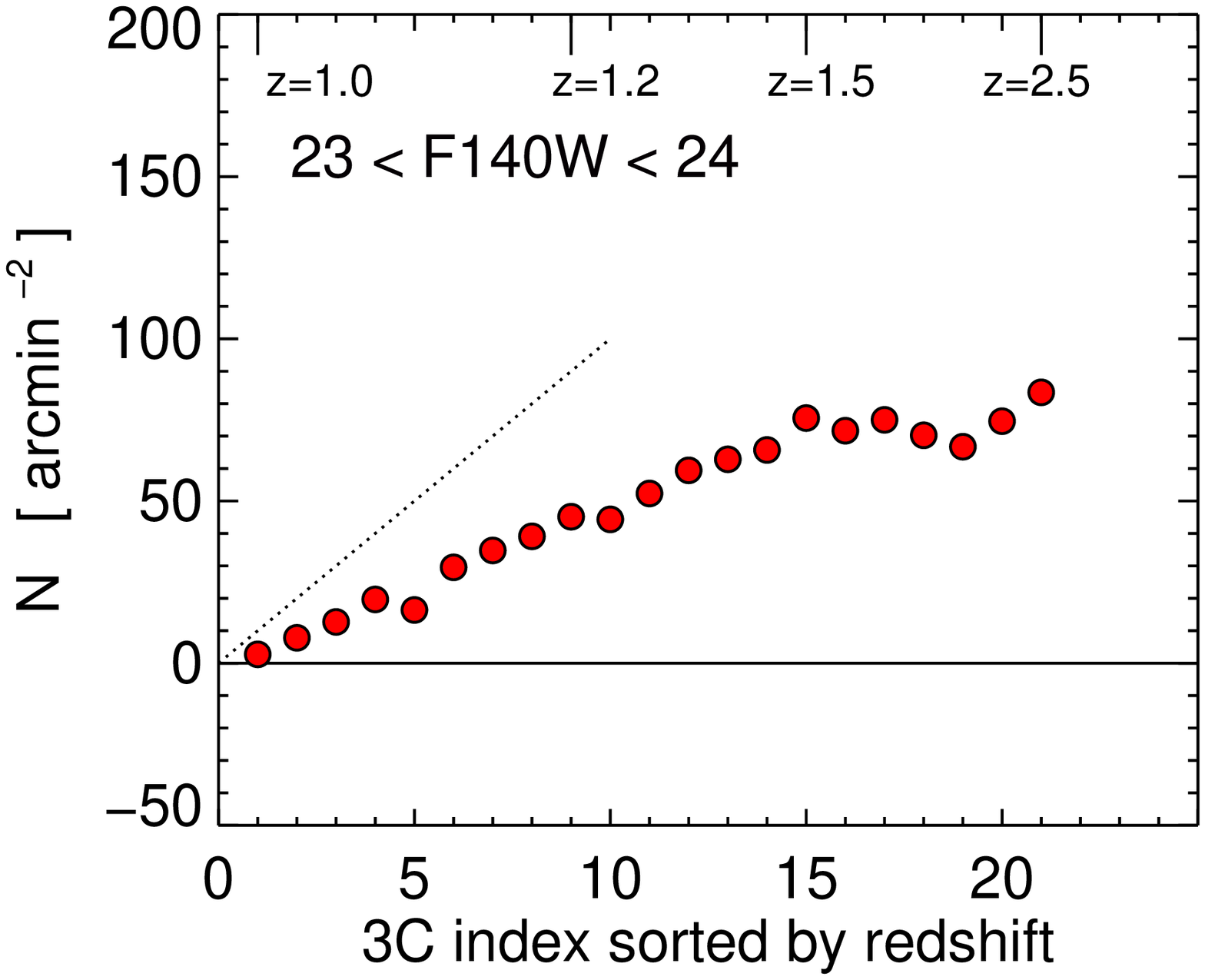}
  \includegraphics[height=3.46cm,clip=true]{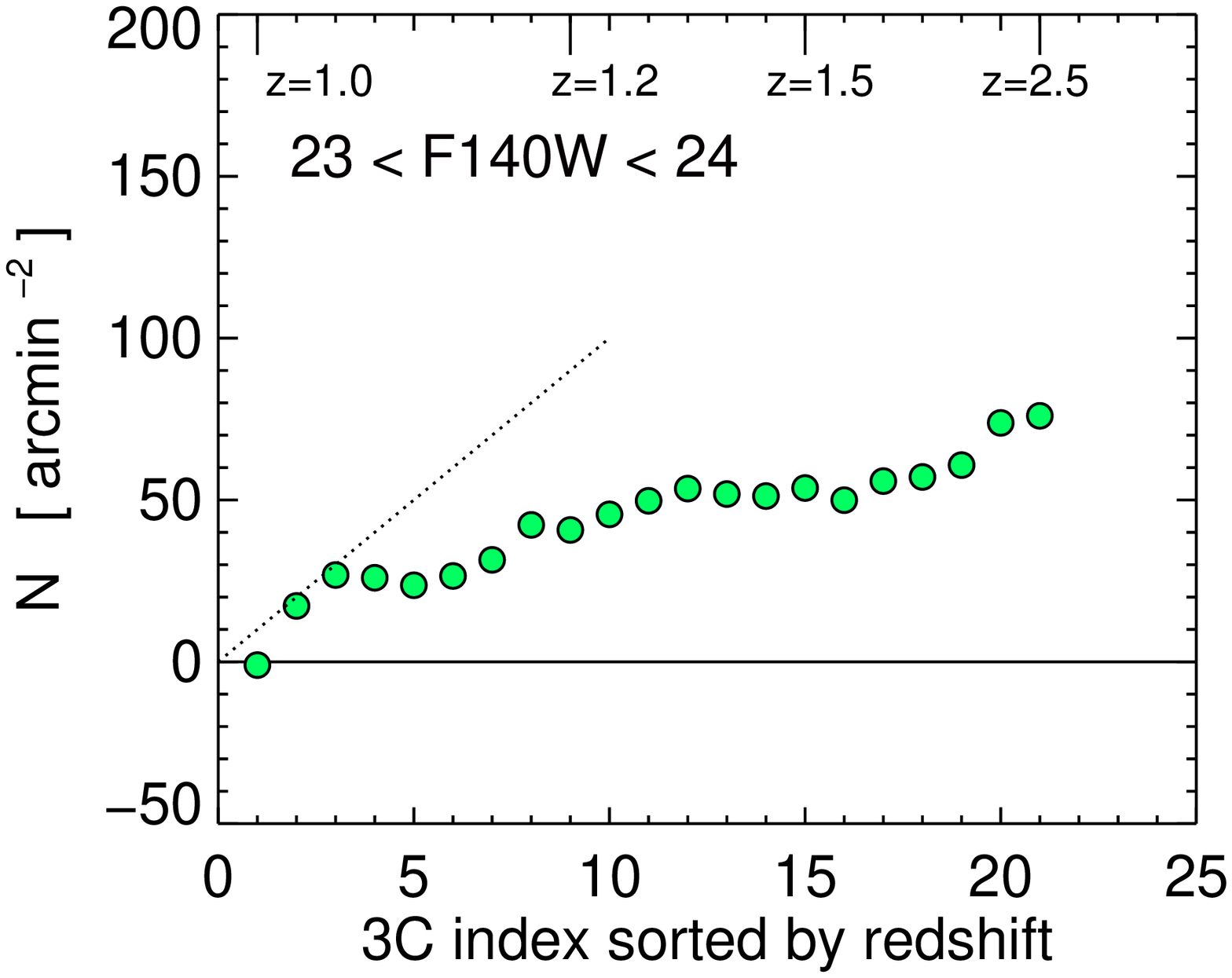}
  \includegraphics[height=3.46cm,clip=true]{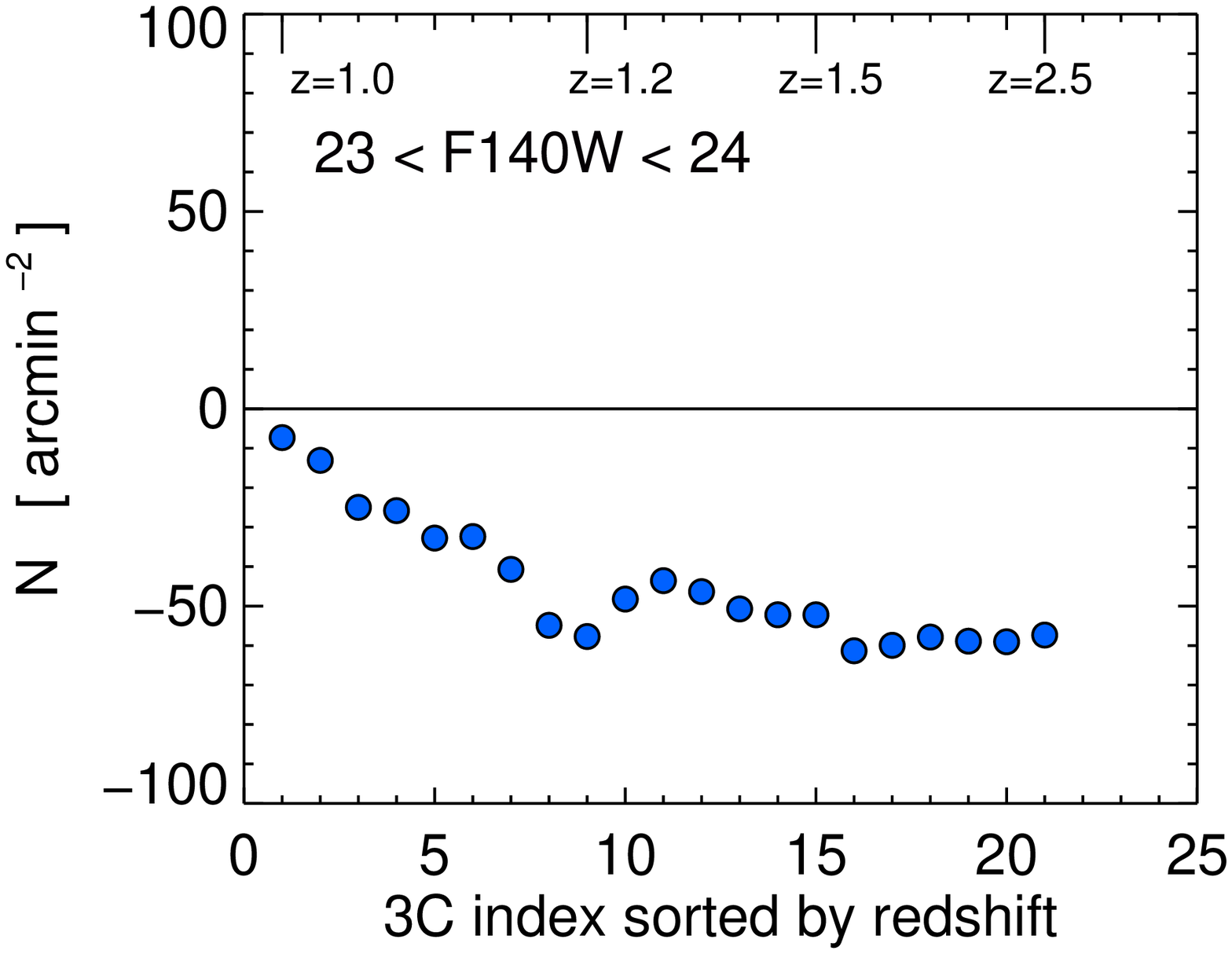}

  \includegraphics[height=3.46cm,clip=true]{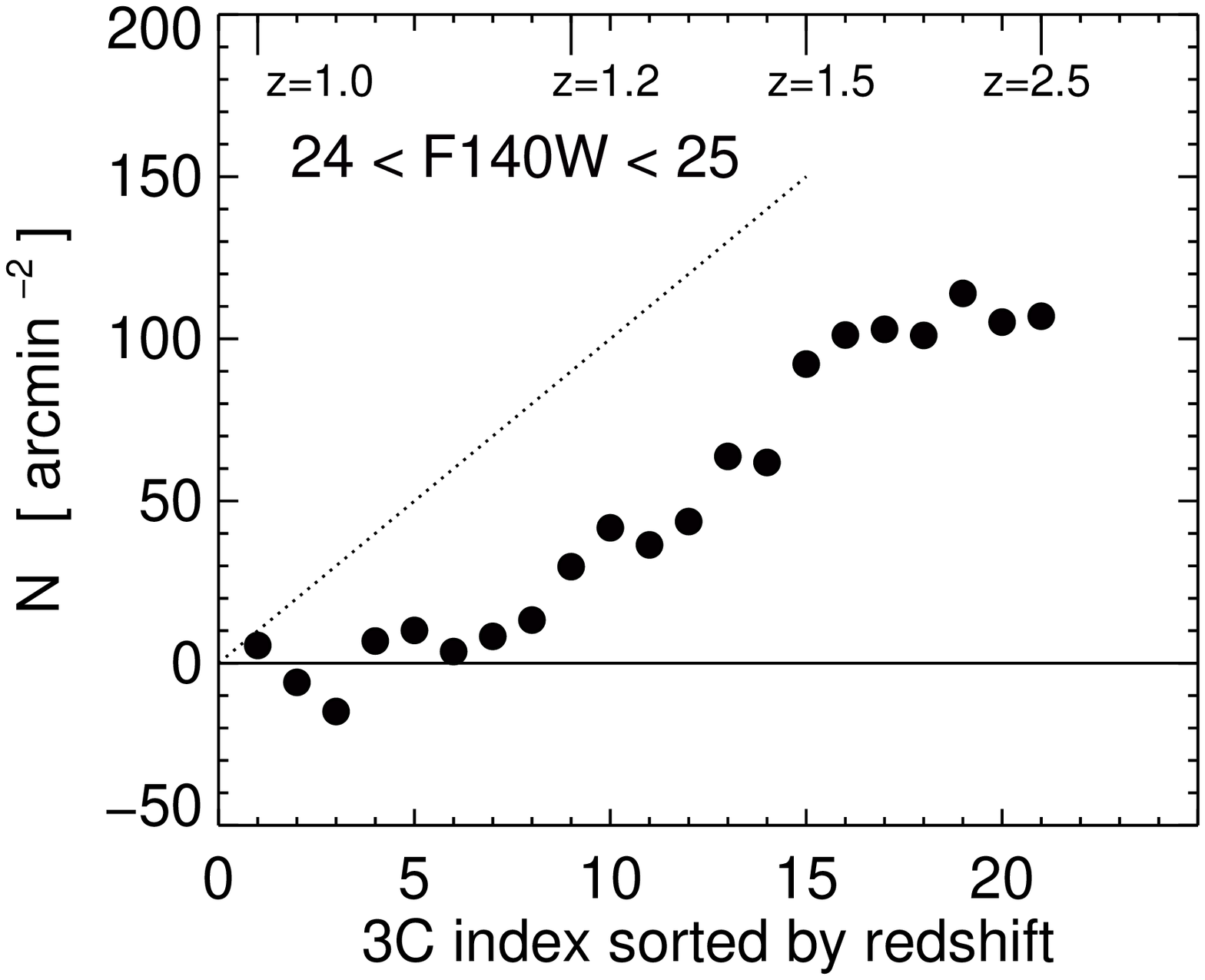}
  \includegraphics[height=3.46cm,clip=true]{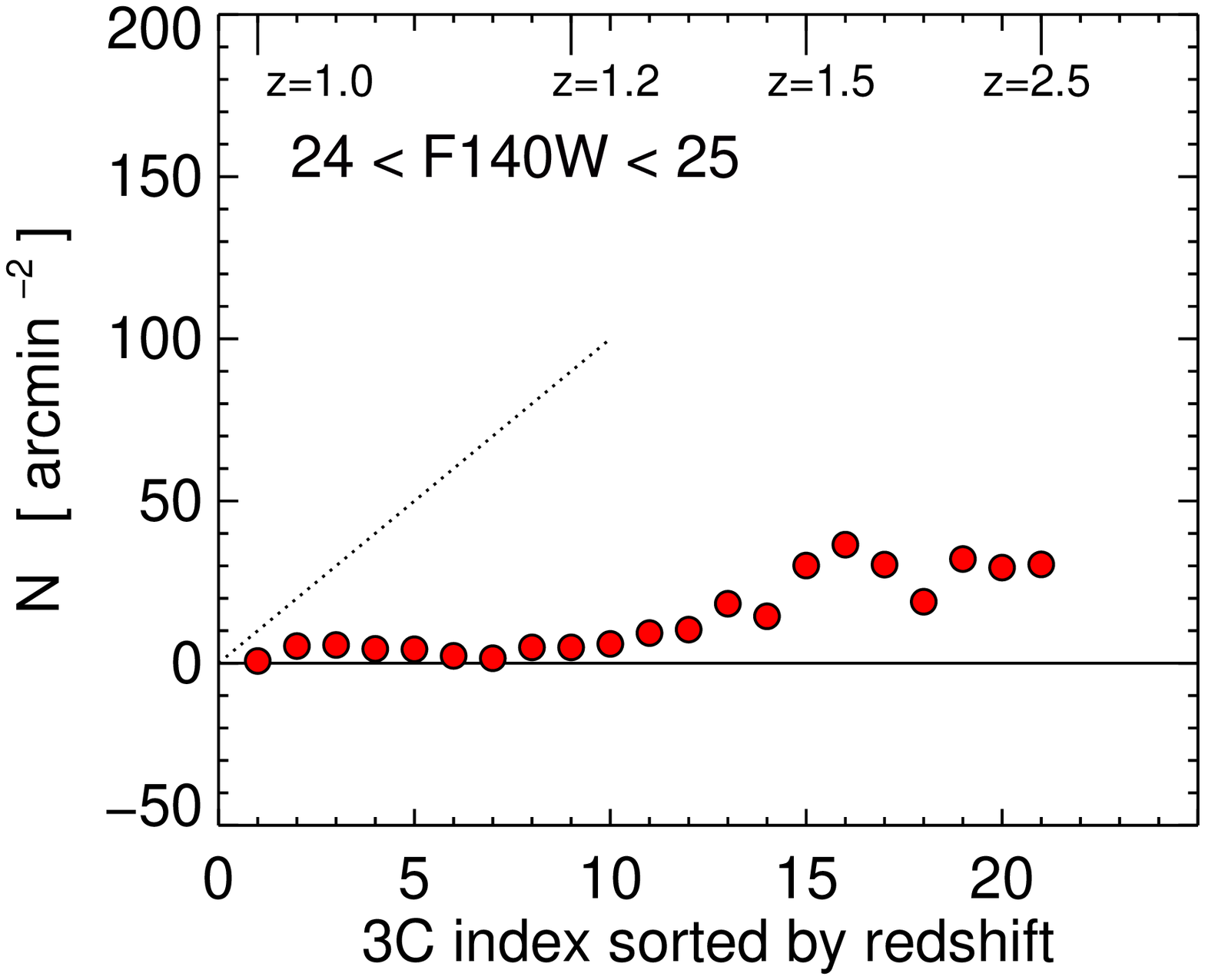}
  \includegraphics[height=3.46cm,clip=true]{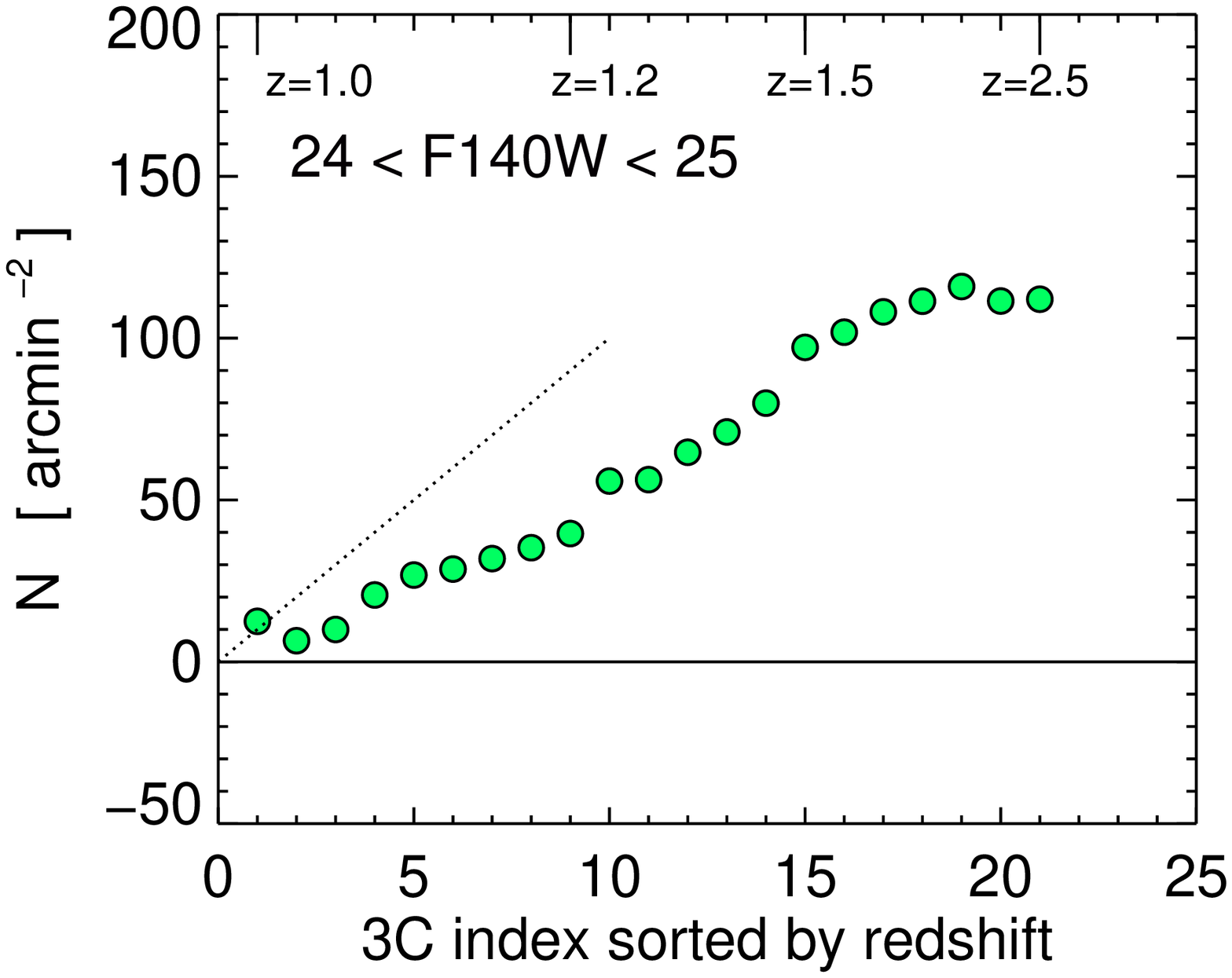}
  \includegraphics[height=3.46cm,clip=true]{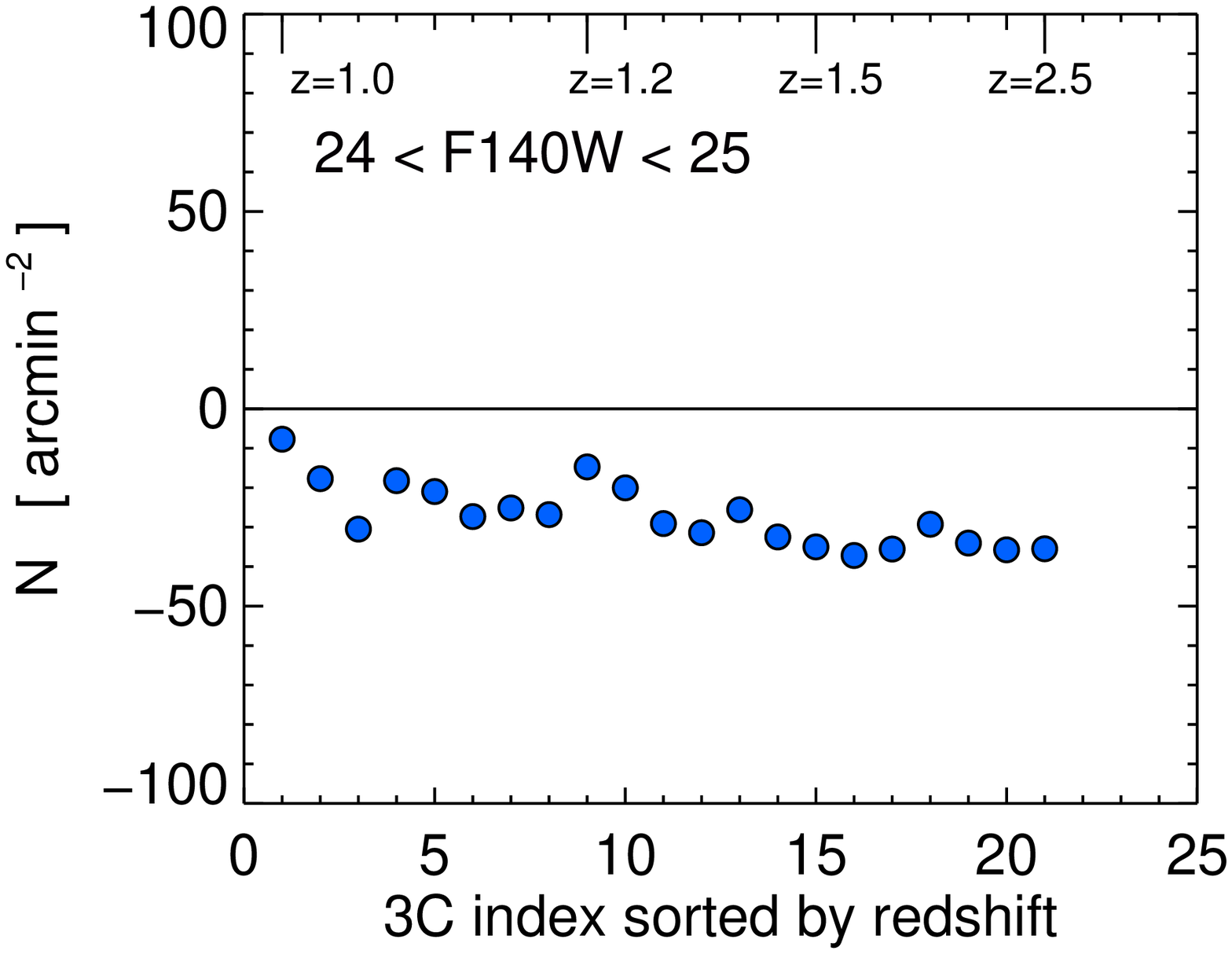}

  \includegraphics[height=4.15cm,clip=true]{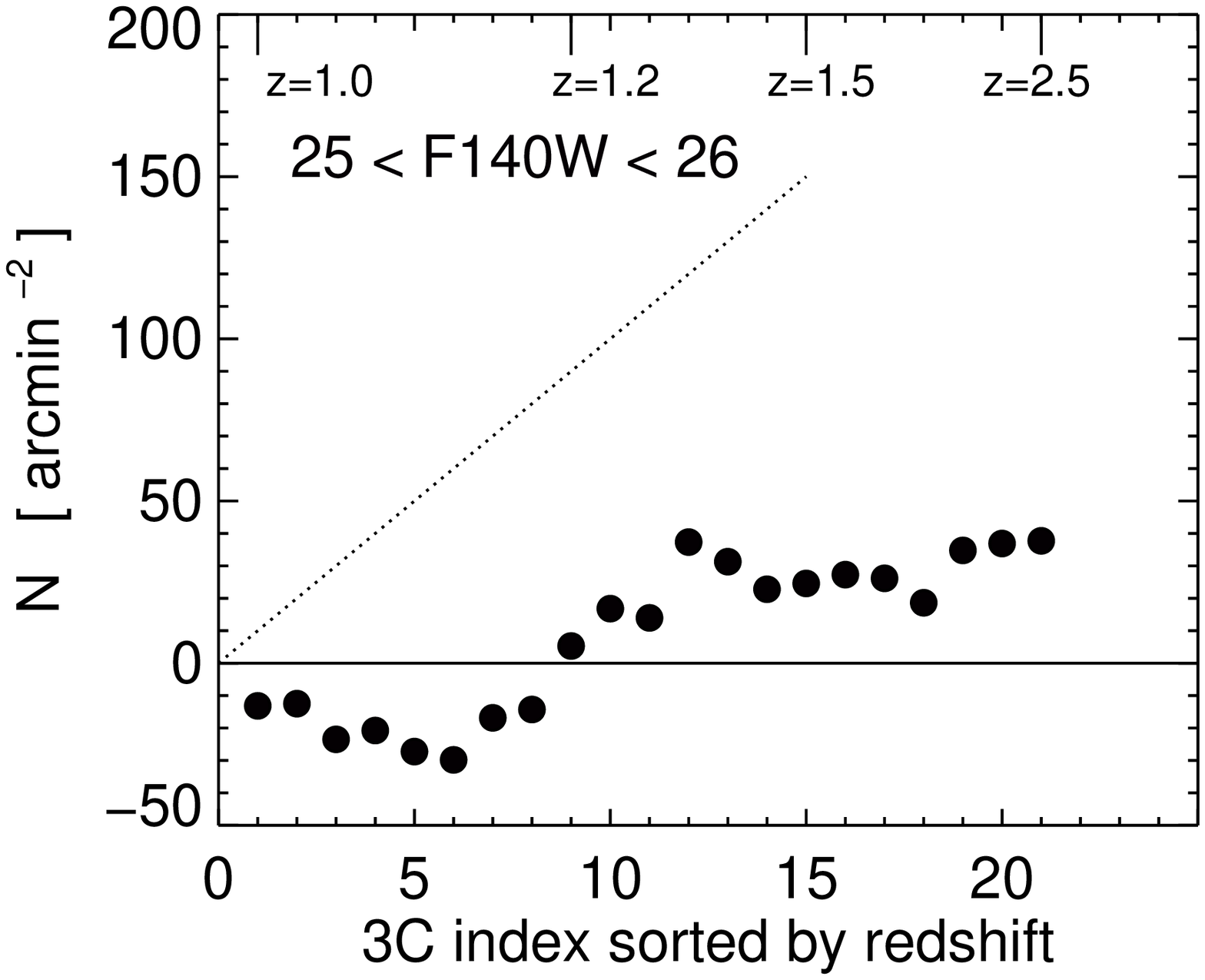}
  \includegraphics[height=4.15cm,clip=true]{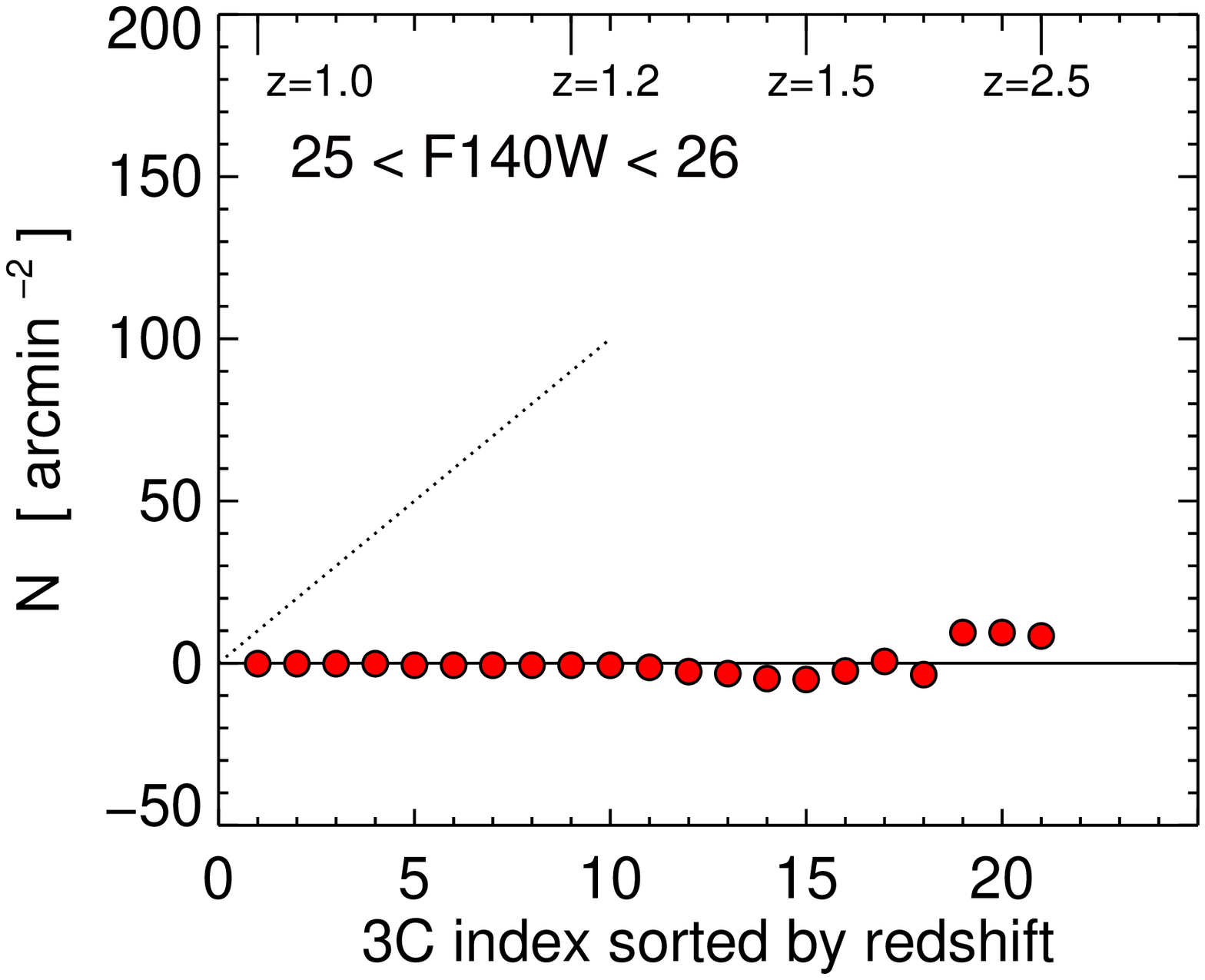}
  \includegraphics[height=4.15cm,clip=true]{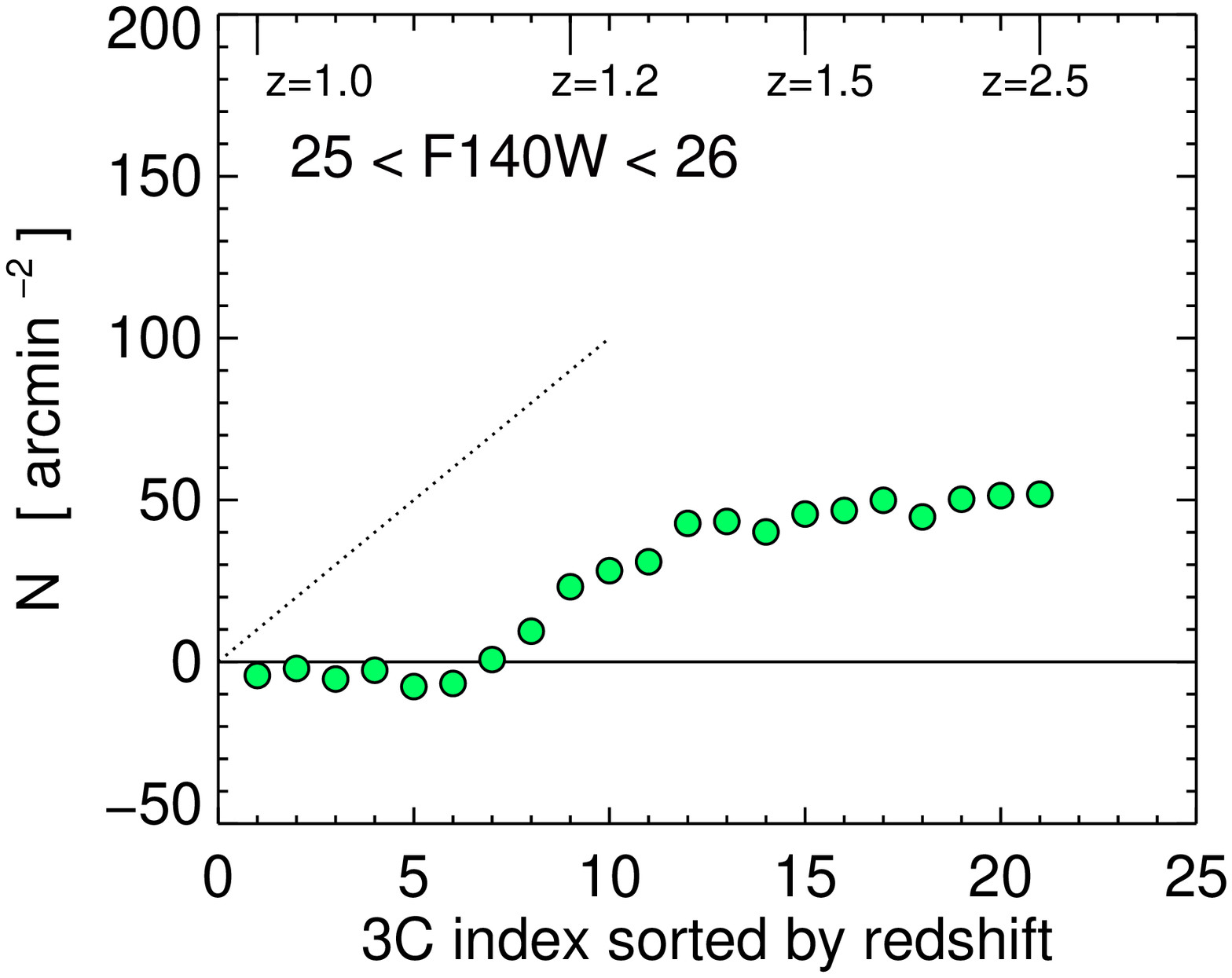}
  \includegraphics[height=4.15cm,clip=true]{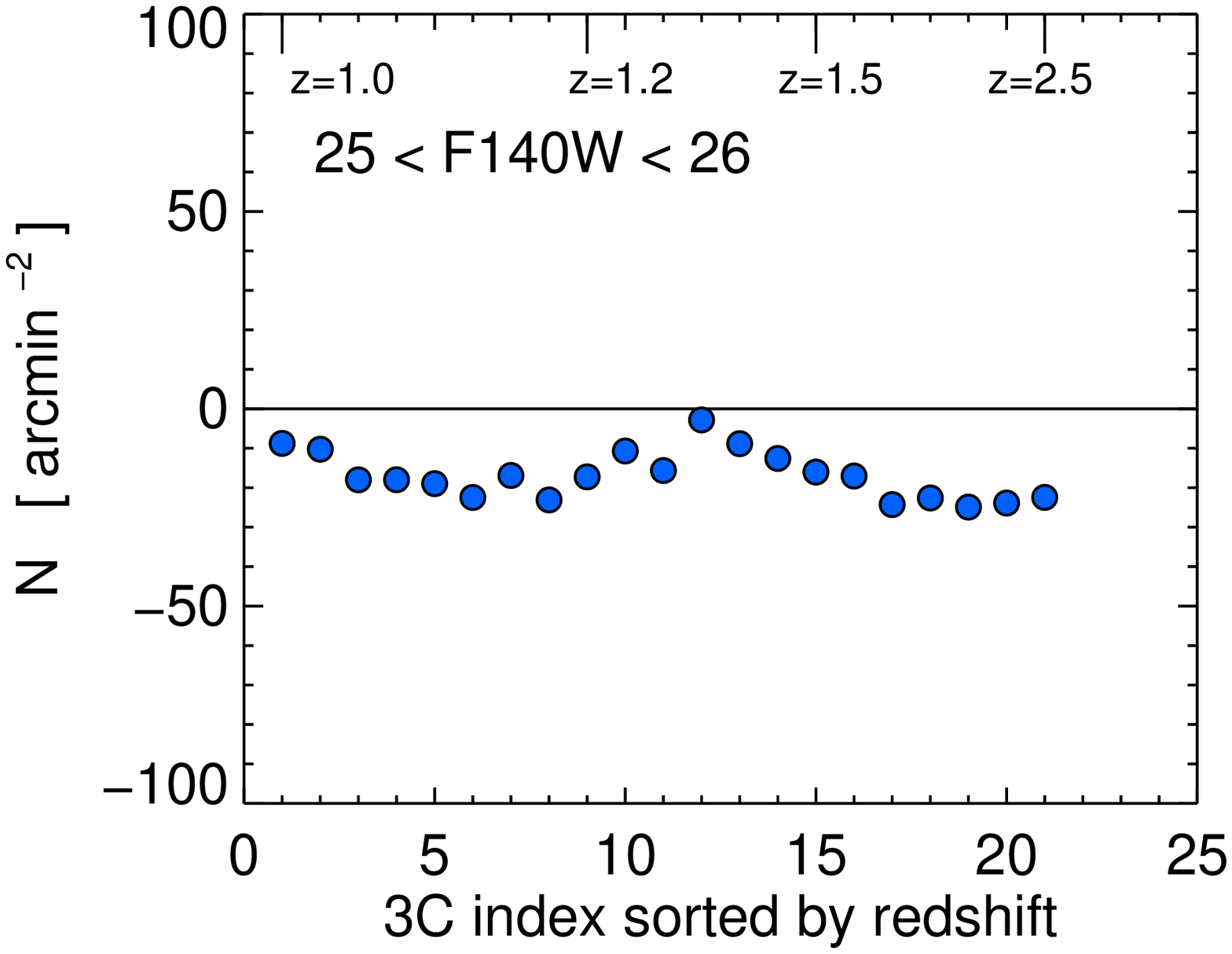}

  \caption{Cumulative central overdensities for different brightness and the samples $all$, $red$, $IR$-$only$, and $blue$ (from left to right).
    The 3C sources are excluded from the counts.
    The horizontal axis is the index of the 3C source, sorted by redshift, as labeled at the top of each panel.
    The diagonal dotted line marks a slope of a cumulative central overdensity of
    $s = 10$ galaxies per square arcmin per 3C field ($\sim$2 galaxies per cell of 15$\arcsec$ radius, for comparison with Fig.~\ref{fig:cod_from_maps}).
    For the $blue$ sample, the vertical axis is shifted and zoomed compared to the other three samples.
  }
  \label{fig:cod_versus_brightness}

  \vspace{5mm}
  
\end{figure*}

    We estimated the statistical significance of the CCOD
    in two ways.
    First, we calculated the cumulative surface densities separately for the central and peripheral 
    regions. A two-sided KS test yields a probability $<$$10^{\rm -4}$ 
    that two distributions drawn from the same population would be so different. 
    (An analogous example is shown in Fig.~10 of G17.)
    Secondly, we performed a randomized experiment.
    Instead of the true 3C position, 
    a random position was chosen as the new ``center'' position and 
    the radial surface density analysis was performed using 
    this location. 
    In all cases this experiment led to a CCOD with constant slope near zero and 
    a noisy curve around this slope;
    the noise may be due to the fact that 
    the random position 
    is sometimes 
    close to the true 3C position.
    We repeated this experiment 10 times and averaged the resulting CCODs.
    Fig.~\ref{fig:cod_from_maps_random} shows that the
    cumulative overdensity disappears in comparison with the corresponding
    panels in Fig.~\ref{fig:cod_from_maps}.
    The marginal negative slope for $all$ and $red$ suggests that for the randomized central positions,
    the 3C source lies, on average, in the ``new'' periphery, 
    slightly raising the ``new'' SDP and lowering the random COD.
    The randomized results are similar for other brightness bins.
    The two methods confirm that the CODs and CCODs seen in Fig.~\ref{fig:cod_from_maps} 
    are 
    associated with 
    the 3C sources.

    Fig.~\ref{fig:cod_versus_brightness} shows the CCOD for different brightness bins and 
    the samples $all$, $red$, $IR$-$only$, and $blue$; 
    the CCODs were calculated using the COD values shown in Fig.~\ref{fig:od_versus_mag_all}.
    The slopes of the brightness-dependent CCODs corroborate the trends seen in Fig.~\ref{fig:od_versus_mag_all}.
    The $red$ overdensities are essentially made up of bright galaxies 
    (21-23\,mag) with CCOD slopes ($s \sim 10$ galaxies per square arcmin per 3C field) 
    turning-over at $z = 1.2$ and $z = 1.5$, respectively.
    For fainter $red$ galaxies (23--24\,mag) the slope flattens ($s \sim 5$). 
    The flattening by a factor 
    of about two 
    is stronger than expected from the $\sim$70\% completeness 
    (row 4 of Table~\ref{tab_completeness}).
    The incompleteness of faint $red$ galaxies is further compensated for by the $IR$-$only$ sample (23--25\,mag).
    The $IR$-$only$ CCODs show only little turn-over at $z > 1.5$ 
    and their slopes ($s \sim 5$) match the bright red slopes ($s \sim 10$) flattened by the completeness fractions  
    ($\sim$50\%, row 3 of Table~\ref{tab_completeness}). 
    The $blue$ CCODs reveal that the observed deficiency of central blue galaxies is strongest 
    at 23--24\,mag and $z < 1.2$ ($s \sim -5$). 
    It is still discernable at fainter magnitudes and larger redshift but these blue galaxies suffer 
    from incompleteness ($\sim$50\%, row 6 of Table~\ref{tab_completeness}).

    To summarize, the brightness-dependent CCODs yield a consistent average picture of 
    bright and faint red overdensities across 3--4 magnitudes as well as blue underdensities.  
    Tentatively correcting the overdensity for incompleteness at faint magnitudes 
    suggests a well-established faint red galaxy population 
    in some but not in all 3C fields (a counter example is 3C\,305.1).



    \subsection{Classification scheme of 3C clusters}  \label{sec:discussion_classification}

    The clustering of red and blue galaxies around the 3C sources shows a diversity 
    of over- and  under-densities. 
    To bring some order to this jumble, sorting is required as in many fields of astronomy, 
    and for that we used simple mathematical combinatorics.
    We started 
    by
    considering all (theoretically) possible combinations, here called classes:  
    red COD,
    red COD and blue COD,
    red COD and blue CUD,
    blue COD,
    blue COD and red CUD,
    and so on. 
    We assigned each 3C source to a class, 
    combining the color-dependent COD and CUD results for the $red$ (including $IR$-$only$) 
    and $blue$ samples into a joint scheme. 
    This way the environments 
    are sorted by galaxy color and cluster morphology 
    into six classes listed in Table~\ref{tab_classification}.
    We arranged the six classes in an intuitive way,
    guided by 
    various scenarios for cluster evolution.
    In the description of the classes, we 
    assume that the red and blue galaxies are passive and star-forming, respectively.
    The classes are:
 
    \begin{itemize}

    \item [I)] $red$ COD + $blue$ CUD:
      clusters with
      a $red$ central overdensity surrounded by a peripheral $blue$ density enhancement.
      These may be the most evolved clusters in the sample, 
      hosting passive galaxies in the center and star-forming galaxies 
      further out.
      Our sample contains 
      three certain members of this class, all at $z<1.2$, 
      with two additional less certain 
      members.

    \item [II)] $red$ COD only:
      clusters with a $red$ central overdensity but neither an over- nor under-density of $blue$ galaxies.
      Compared to class I, 
      they 
      do not exhibit a 
      central underdensity of blue galaxies relative to the periphery. 
      Compared to class III, the $blue$ central overdensity has declined to the level of the periphery.
      Our sample contains 
      eight certain members of this class, all at $z<1.5$, 
      plus two less certain members at $z>1.6$.

    \item [III)] $red$ COD and $blue$ COD:
      clusters with co-spatial $red$ and $blue$ central overdensities,
      with star-forming galaxies (still) in the center.
      The 
      star formation 
      activity may be due to a younger cluster age or to the merger of clusters.
      Our sample contains 
      two certain members of this class, both at $z<1.3$.

    \item [IV)] $blue$ COD without $red$ COD:
      the center hosts a group of star-forming galaxies but not yet 
      on the RS; 
      if 
      these red galaxies do exist, their distribution is extended, indicating a young cluster or proto-cluster. 
      Our sample contains only one member of this class, 3C\,432 at $z=1.785$: 
      its $blue$ COD is based on a small excess of only 2--3 galaxies. 
      Therefore the classification is uncertain and could change to class VI.

    \item [V)] $blue$ CUD only:
      clusters with no $red$ COD but
      surrounded by peripheral $blue$ density enhancements.
      The lack of a $red$ COD 
      suggests
      that they are evolutionarily young.
      The powerful radio galaxy may have 
      cannibalized
      nearby galaxies, 
      but the periphery hosts a reservoir of mostly star-forming galaxies for 
      future
      cluster (or group) assembly.
      Our sample contains only two members of this class, both at $z<1.2$.

    \item [VI)] 
    no COD or CUD at all:
      any 
      clustering of galaxies is extended on the $r = 500$ kpc scale, 
      consistent with the expectation for extended proto-clusters.
      Our sample contains 3--6 members of this class, with 2--5 of them at $z>1.6$.
    \end{itemize}
    
    This classification scheme covers all COD/CUD combinations found in our sample. 
    One may think of additional COD/CUD combinations, but they were not seen.
    Actually, a reduced classification scheme 
    consisting 
    of only two classes
    would facilitate astrophysical interpretation;
    two such classes would be quiescent/active clusters lacking/containing star-forming galaxies in their cores. 
    To 
    be clear: we 
    do not infer the existence of six cluster 
    classes from our small sample of only 21 radio sources; rather the classes are derived 
    from mathematical combinatorics of possible red/blue COD/CUD combinations,
    and from this we find certain class preferences. 
    This scheme may allow us --- using more complete samples in the future --- to corroborate evolutionary trends. 

    \begin{figure}
      \hspace{-8mm}\includegraphics[width=10cm]{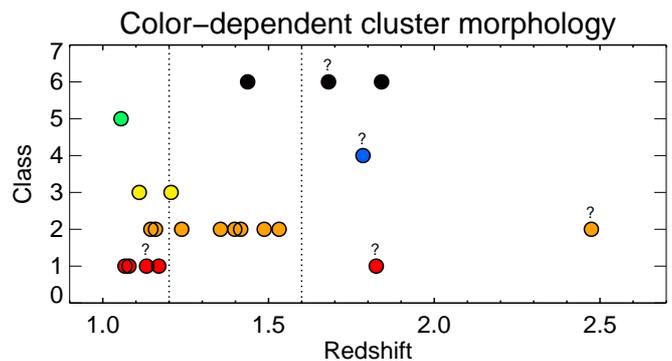}
      
      \caption
          {Classification of 3C clusters versus redshift. 
            The 
            vertical axis gives the 
            six classes (I, ..., VI)
            described in Sect.~\ref{sec:discussion_classification}.
            Values are taken from Table~\ref{tab_classification}.
            A ``?'' marks uncertain cases.
            The vertical dotted lines divide the sample at $z=1.2$ and $z=1.6$.
          }
          \label{fig:class}
    \end{figure}
    

    Except as noted, the classification at $z<1.6$ should be reliable
    because the CODs/CUDs are well discerned on the surface density maps and the radial density profiles.
    The CODs/CUDs are based on sufficient 
    galaxies in the $red$ and $blue$ samples 
    ($N \gsim 10$ with the exception of 3C\,305.1). 

    For $z>1.6$, the number of 
    detected 
    galaxies is small.
    In particular the $red$ sample 
    suffers from severe incompleteness 
    which can not be adequately compensated
    for
    by inclusion of 
    the $IR$-$only$ sample. 
    This makes recognizing CODs/CUDs and the resulting classification uncertain.

    \subsection{Comparison with local clusters}\label{sec:coma}


    \begin{figure*}
      \hspace{-3.0mm}\includegraphics[width=10.0cm,clip=true]{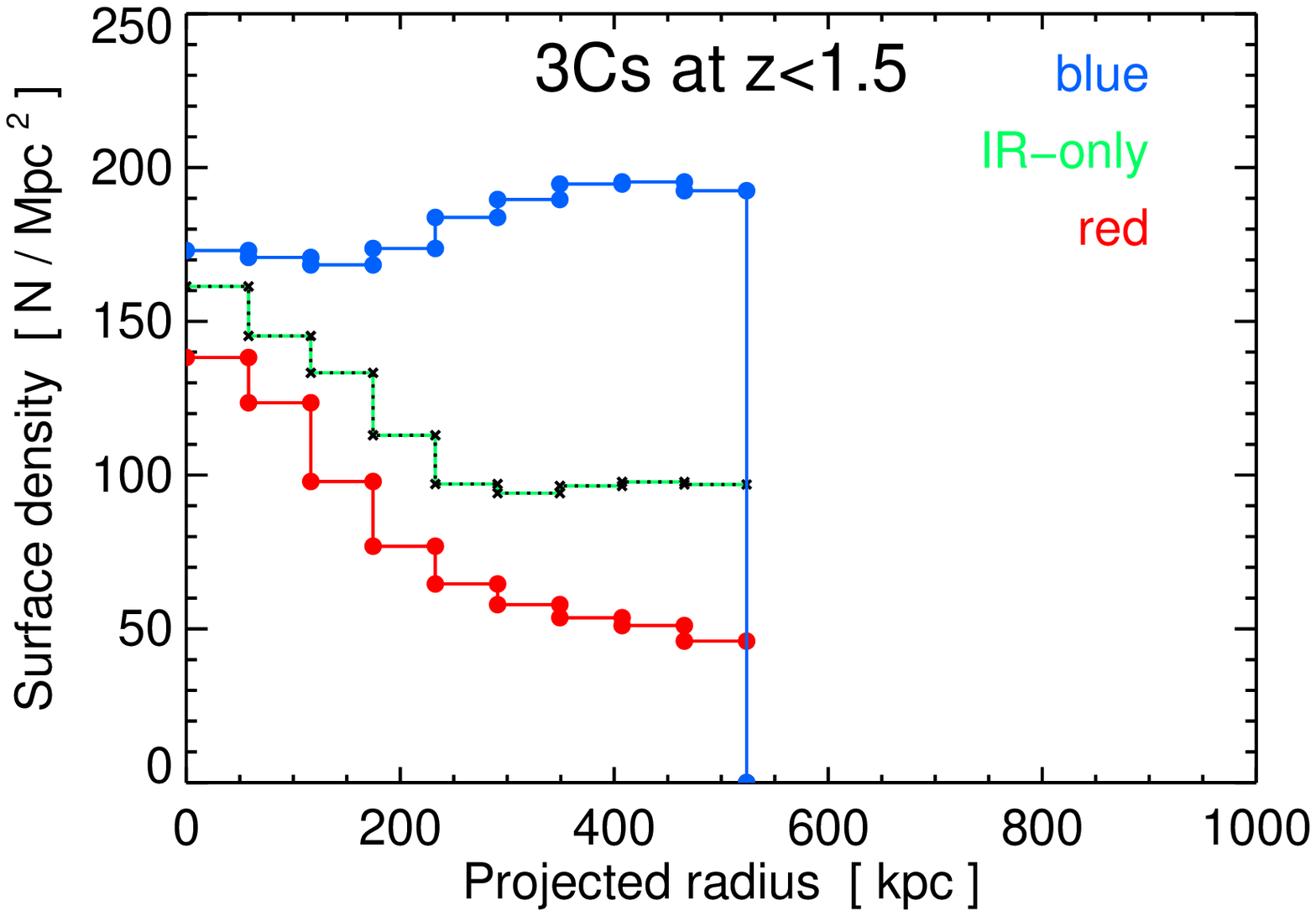}
      \hspace{-0.0mm}\includegraphics[width=8.67cm,clip=true]{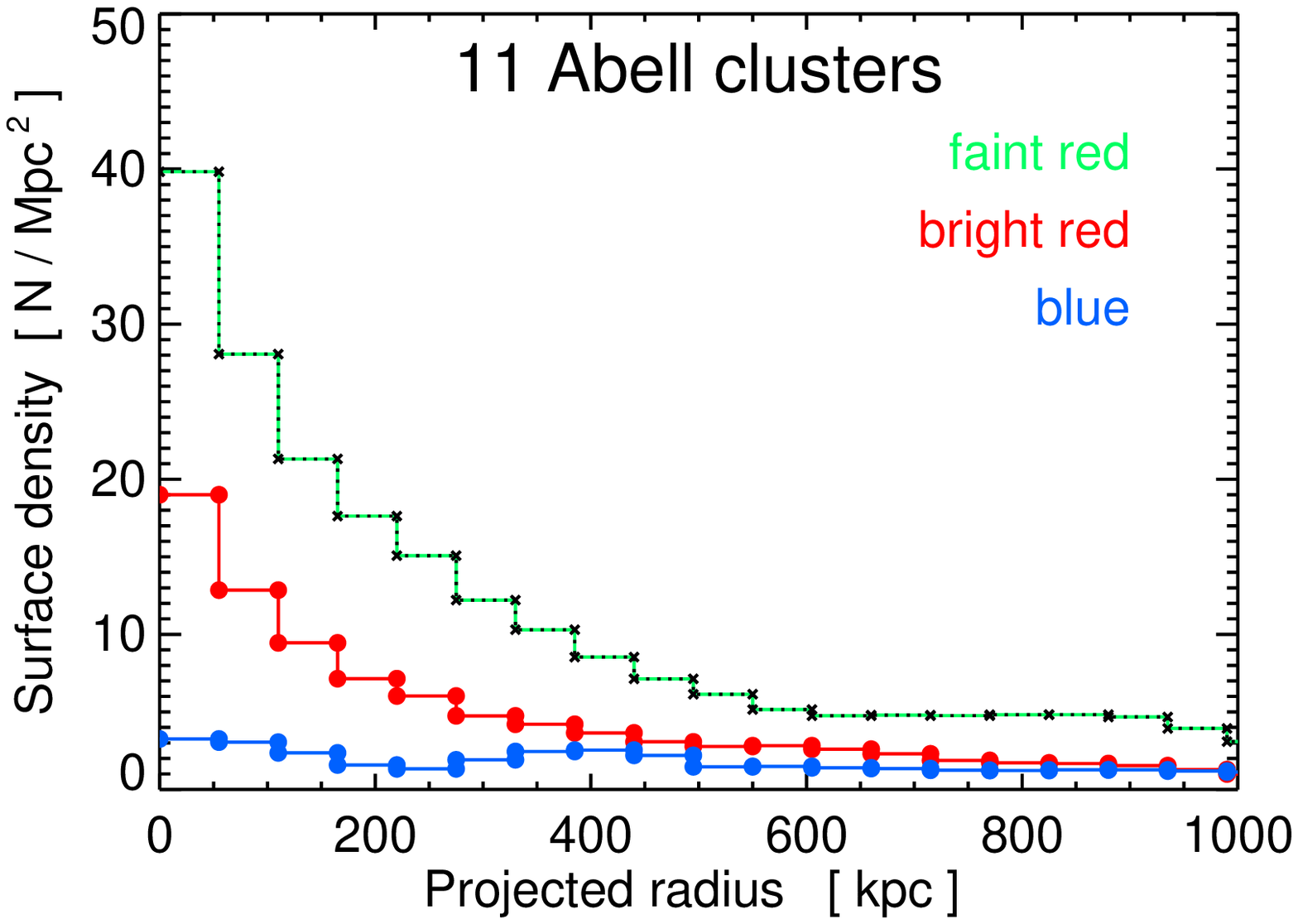}

      \hspace{-3.0mm}\includegraphics[width=10.0cm,clip=true]{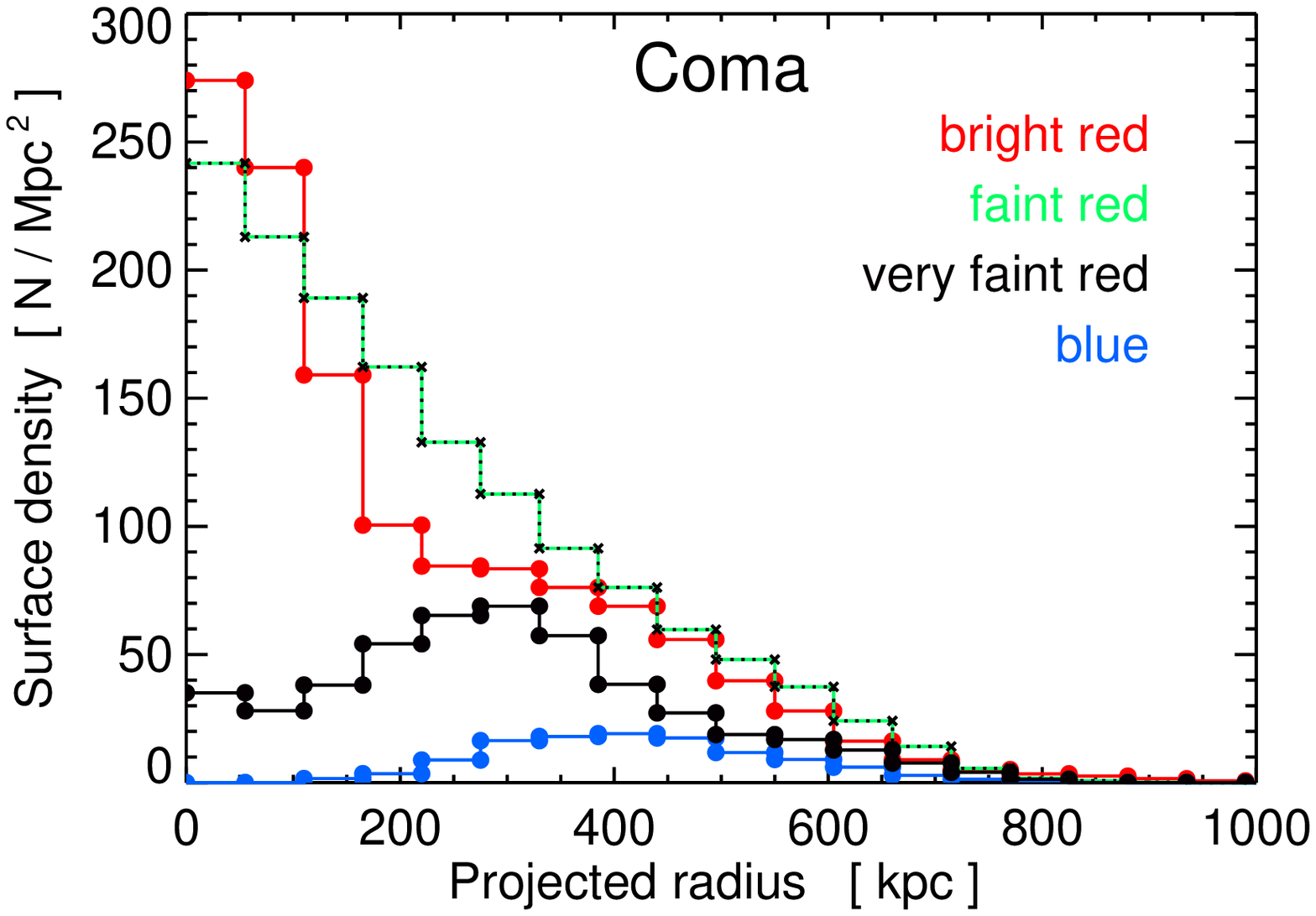}
      \hspace{-2.0mm}\includegraphics[width=8.9cm,clip=true]{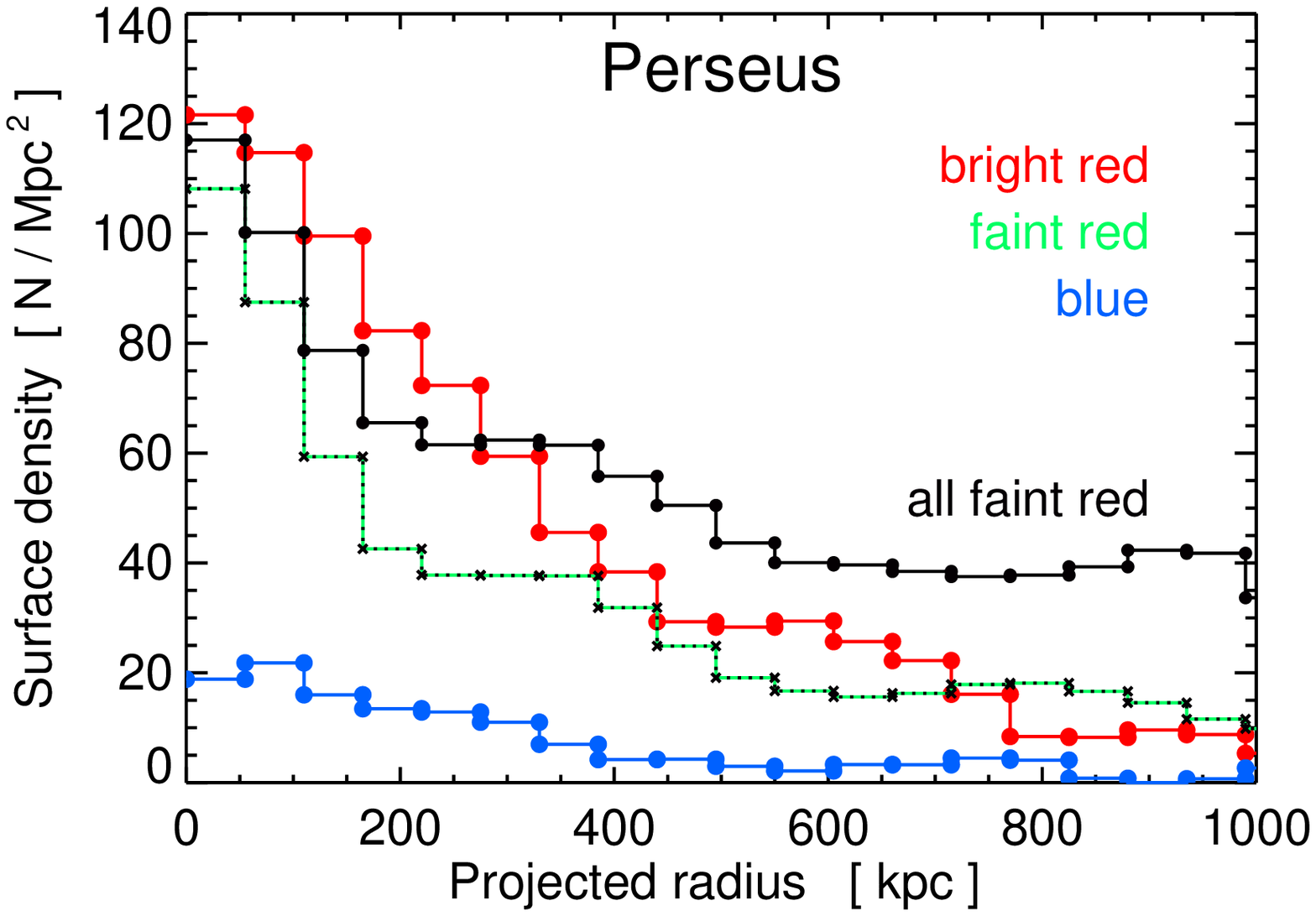}

      \caption{
        Radial surface density profiles of 3C sources and local clusters for different types of galaxies.
        The 3C profiles are averaged for $red$/$IR$-$only$/$blue$ galaxies, respectively, 
        using the 15 sources at $1 < z < 1.5$
        with brightness range $21 < \rm{F140W} < 26$ {\it (top left)}. 
        For the local clusters, only spectroscopic members are used, split into $bright$-$red$, $faint$-$red$, and $blue$. 
        For the 11 Abell clusters, the averaged profiles are shown {\it (top right)}. 
        The profiles for the Coma and Perseus clusters {\it (bottom)} show, in addition,
        spectroscopically confirmed very faint red sources of Coma and $all$ $faint$-$red$ candidate cluster members of Perseus; 
        the Perseus samples $all$ $bright$-$red$ and $all$ $blue$ are not shown
        because these are spectroscopically 90\% complete, so that their profiles remain essentially 
        the same as the profiles of the spectroscopic samples shown. The standard deviations of each radial bin are large and not shown to avoid confusion. 
      }
      \label{fig:coma}
    \end{figure*}


    \begin{table*}
      \renewcommand{\thetable}{\arabic{table}}
      \caption{The local comparison clusters. 
      }
      \label{table_local_clusters}
      \footnotesize
      \begin{tabular}{lrrrrrrrc}
        \hline
        Name & RA (J2000) & Dec (J2000)& N gal$~^5$  & c$z$ [km/s] & $m-M$ & [kpc/$\arcsec$] & richness$~^1$ & B--M Type$~^2$ \\
        \hline

        Abell\,0954   &   153.43667  &$-$0.10861 &    103  & 13623   &  37.90 &   1.684 & 0 &   II        \\
        Abell\,0957   &   153.41792  &$-$0.91444 &     44  & 28622   &  36.23 &   0.846 & 1 &   I-II        \\
        Abell\,1139   &   164.51780  &   1.49880 &    128  & 11876   &  36.03 &   0.778 & 0 &   III        \\
        Abell\,1189   &   167.76700  &   1.12830 &     42  & 28824   &  37.97 &   1.731 & 0 &   III        \\
        Abell\,1236   &   170.68708  &   0.46222 &     40  & 30533   &  38.09 &   1.821 & 0 &   II        \\
        Abell\,1238   &   170.74160  &   1.09210 &     89  & 22160   &  37.37 &   1.360 & 0 &   III        \\
        Abell\,1364   &   175.91500  &$-$1.76090 &     36  & 31859   &  38.67 &   1.882 & 1 &   III	           \\   
        Abell\,1620   &   192.44220  &$-$1.58890 &    119  & 25513   &  37.63 &   1.503 & 0 &   III        \\
        Abell\,1663   &   195.71125  &$-$2.50611 &     72  & 24827   &  37.68 &   1.539 & 1 &   II        \\
        Abell\,1692   &   198.06690  &$-$0.93180 &     90  & 25235   &  37.68 &   1.536 & 0 &   II-III        \\
        Abell\,1750   &   202.70792  &$-$1.87278 &    133  & 25647   &  37.71 &   1.552 & 0 &   II-III        \\
        \hline  
        Coma$~^3$                   &  194.89879  &  27.959389 & 310     & 6925    &  35.10 &   0.500 & 2 &   I-II        \\
        Perseus$~^4$                 &  49.950667   &  41.511696 & 373     & 5366    &  34.50 &   0.388 & 2 &   II-III      \\
        \hline
      \end{tabular}\hspace*{-5.0cm}
      ~\\
      $~^1$ Abell richness class \\
      $~^2$ Bautz-Morgan morphological type \\ 
      $~^3$ Adopted cluster center  position  between NGC\,4874 and NGC\,4889 \\ 
      $~^4$ Adopted cluster center  position  of NGC\,1275 \\ 
      $~^5$ Number of spectroscopic members down to 5 mag (SDSS $r$ band) below the third-brightest galaxy.\\
      Note: Abell cluster positions are X-ray centers from NED. Using average galaxy positions instead changes results by $<$10\%. 
    \end{table*}



    %
    %

    The number of galaxies in a typical overdensity is $<$10 per cell of 15$\arcsec$ radius (Figs.~\ref{fig:sd_maps_1}--7).
    These overdensities could be galaxy groups or clusters.
    To check whether the 3C sources are located in 
    massive clusters rather than smaller groups, 
    we compared the richness and spatial concentration of the 3C companions with those of 
    well-studied
    local clusters. 
    This does not mean that a 3C cluster will evolve into one of the chosen clusters. 
    The local sample comprises 11 Abell clusters from the 2dF cluster survey \citep{DePropris2003},  
    the Coma cluster \citep{Michard08}, and the Perseus cluster \citep{Meusinger2020}.\footnote{Coma and Perseus are Abell 
      clusters too (Abell 1656 and Abell 426), but here they are not part of the sample of 11 Abell clusters.}
    Most of the Abell clusters are relatively poor, 
    with Abell richness class 0--1, 
    while the Coma and Perseus clusters are 
    more massive systems 
    of richness class 2.
    The Abell and Coma cluster data cover a radial extent of 1\,Mpc, and Perseus is twice of that. 
    The 11 Abell clusters were drawn from the 60 clusters of the 2dF cluster survey, a  
    random collection of clusters from different available catalogs \citep{DePropris2003} and 
    selected here to have SDSS photometry available.
    The properties of 
    these
    comparison clusters are listed in Table~\ref{table_local_clusters}.
    All clusters contain spectroscopically confirmed 
    member galaxies with a reported spectroscopic completeness of 
    about 80\% (at the bright end about 95\%, 
    dropping to 50\% at the faint, red end). 

    At $z \sim 0$, the $u$ and $r$ bands cover a rest-frame part of the spectrum similar 
    to the $F606W$ and $F140W$ filters at $1<z<1.5$.
    For the galaxies 
    in Perseus, \citet{Meusinger2020} cataloged $u$ and $r$ band photometry from SDSS.
    For Coma and the Abell clusters, 
    we obtained $u$ and $r$ from the SDSS DR12 catalog.
    All photometry was corrected for Galactic foreground extinction \citep{Schlafly11}.

    For the local clusters, we selected three galaxy types: $bright$-$red$, $faint$-$red$, and $blue$ to be compared with the $red$, $IR$-$only$, and $blue$ types of the 3C sources.
    In $u-r$ vs. $r$ CMDs, the RS 
    is easily determined by eye, and we applied a color cut $u-r=0.4$ below the RS to separate $blue$ 
    and $red$ galaxies. 
    For the 3C sources we had applied a softer color cut $ \rm F606W -  \rm F140W = 1$ below the predicted RS, 
    in order to account for the large color uncertainty of both the RS and the faint galaxies in the distant Universe.  
    To separate $bright$-$red$ and $faint$-$red$, we applied a magnitude cut 2.5\,mag fainter than the third brightest galaxy in $r$.
    This cut was motivated by the transition between $red$ and $IR$-$only$ CCODs 
    at $23< \rm F140W<24$, about 2.5\,mag fainter than the bright end of the CCODs at $ \rm F140W = 21$ 
    (Fig.~\ref{fig:cod_versus_brightness}). Corresponding to the 5 mag brightness range of the galaxies in F140W, 
    for the local clusters we applied a faint end cut of 5\,mag fainter than the bright end.
    For each cluster, we performed the overdensity analysis exactly as done for the 3C sources 
    using  a cell radius of 125\,kpc (corresponding to 15$\arcsec$ for the 3C sources). 
    We removed the cD galaxies NGC~1275 (Perseus), NGC~4874, and NGC~4889 (Coma), 
    as we had done for the 3C sources by removing the 3C itself.

    Fig.~\ref{fig:coma} compares the radial density profiles of 
    fifteen 3C fields at $z < 1.5$ and the local clusters.
    The smoothness of the radial profiles 
    shows 
    that the trends are real and significant.
    Comparison of the 3C sources with the local clusters 
    reveals: 
    \begin{itemize}

    \item[1)]
      For all four clusters/samples, the $bright$-$red$ profile declines with increasing radius but with different 
      amplitudes.
      The central peak of the 3C sources is about $N = 90$ (per sq. Mpc) above the periphery level at $r = 400$\,kpc. 
      This is a factor 5 larger than the $N = 18$ peak level of the richness 0--1 Abell clusters. 
      The rich clusters Coma and Perseus have a factor 2--3 higher central peak than the 3C sources.

    \item[2)]
      The profile of the 3C $IR$-$only$ sources has a peak $N = 60$ above the periphery level,
      roughly consistent with the $N = 40$ peak of the 
      faint-red profile of the Abell clusters. 
      For the Abell clusters, the ratio of faint to bright red galaxies appears constant about 0.5, independent of the radius.
      As for the bright red galaxies, the rich clusters Coma and Perseus have a 2--3 times higher central peak than the 3C sources. 

      In contrast to the Abell clusters, 
      for Coma the ratio of faint-to-bright red galaxies changes with radius, showing a maximum around $r=250$\,kpc.
      This excess of 
      off-center faint-red galaxies in Coma continues to 1 mag fainter galaxies.
      The radial profile of the very-faint $18 < r < 19$ red sources shows that
      these sources prefer the annulus around 300 kpc and do not peak in the center. 
      This result remains also for Coma's morphologically identified member galaxies without spectra \citep{Michard08}.

      For Perseus, faint red galaxies are less numerous than 
      bright ones.
      This could be due to incompleteness of the faint red population in the spectroscopic sample as noted by \citet{Meusinger2020}.
      Both the 3C's sources and the spectroscopically confirmed faint red sources of the local clusters 
      suffer from incompleteness (50\% for the 3C sources and 10--50\% for the local spectroscopic galaxies). 
      To check this, we
      also plotted 
      the radial profile of all faint red candidate member galaxies.
      At the central peak  ($r<100$ kpc) this profile roughly agrees with the spectroscopic profile 
      but then lies a factor 1.5--2 above the spectroscopic profile. 
      Inside $r< 500$\,kpc, it is reminiscent of the IR-only profile of the 3C sources.

    \item[3)]
      The profile of the 3C blue sources exhibits a maximum around $r = 400$ kpc with a decline toward the center
      and a small enhancement at $r<100$ kpc. 
      The small central enhancement is due to the contribution of the minority of 3C sources with blue COD.
      The Abell clusters show a similar profile with a dip around $r=250$ kpc.
      The blue profiles of the individual Abell clusters show a large diversity, three/eight with/without blue galaxies in the center.
      Coma lacks any central blue galaxies (see also Fig.~4 of \citealt{Andreon1996}).
      On the other hand, Perseus shows a clear presence of blue galaxies in the center.
      All local cluster centers show a scarcity of blue galaxies compared to red galaxies.

    \end{itemize}
    To summarize, for the 3C sources 
    the radial density profiles of the three galaxy types are quantitatively 
    comparable to those of local clusters. This provides evidence 
    that at least some of
    the 3C sources are located in clusters and not just in galaxy groups.
    Other 3C sources may be surrounded by galaxy groups.


    \section{Discussion and conclusions}\label{sec:discussion}

    The clustering of red and blue galaxies shows a diversity 
    among the 3C sources.
    The majority are associated with a spatial concentration of red galaxies,
    most of them with blue galaxies in the periphery, but
    some 3C sources show also blue galaxies in the center. 
    A few 3C sources lack any COD within 250~kpc.
    3C clusters or proto-clusters may be more extended than 250~kpc radius, and therefore the CODs/CUDs 
    we measured mark only the peak of the ``cluster iceberg'', and our conclusions will refer to this peak. 

    \subsection{Central overdensity of red and $IR$-$only$ galaxies} \label{sec:discussion_red}
    The $red$ and $IR$-$only$ CODs suggest the presence of a 
    RS.
    Its bright end has already been shown for seven 3C sources 
    in our sample by K16 using a combination of morphology and colors. 
    The surface density map and overdensity analysis 
    revealed a COD of bright $red$ galaxies in 11/16 (68\%) of the 3C sources at $z<1.5$. 
    This refers to the 
    roughly $L^*$ galaxies of the RS in the central 250 kpc radius.
    The detection of $\sim$2 mag fainter RS galaxies is 
    affected by incompleteness of the data,
    a tentative completeness correction suggests their presence in some but not all 3C fields.    
  


    The more massive and luminous galaxies are expected to evolve faster than their lower mass 
    counterparts (e.g., \citealt{Bauer2005, Conselice2007}).  %
    Consequently,
    the RS may have assembled at the bright end and 
    grown successively toward the faint end.
    This growth may take place in the center or in the periphery.
    In our 3C data, the faint RS members have to be searched for in the $IR$-$only$ data. 
    Among 3C sources with a clear $red$ COD, some show co-spatial clustering of $red$ and $IR$-$only$ galaxies, 
    so that both samples show their peak within $20\arcsec$ around the 3C source
    (e.g., 3C\,210, 3C\,230).
    Some other examples show a strong $red$ COD but a broad spatial distribution of 
    the $IR$-$only$ galaxies (3C\,324, 3C\,356).



    The observed CODs shift from $red$ to $IR$-$only$ CODs at $z > 1.5$,
    consistent with the increasing distance modulus.
    A tentative correction of the $IR$-$only$ CCOD slopes for (about 50\%) incompleteness
    suggests the presence of central 
    overdensities of red galaxies also at $z > 1.5$, comparable to 
    those 
    at $z<1.5$  
    (Fig.~\ref{fig:cod_versus_brightness}).
    However, visual inspection of the surface density maps (also in narrow magnitude ranges) 
    reveals that any overdensities at $z > 1.5$ 
    are 
    far more extended 
    than
    those seen, for instance, for 3C\,210 at $z<1.5$.
    This leads us to conclude that $red$ and $IR$-$only$ overdensities at $z > 1.5$
    are less evolved compared to those at $z<1.5$.

    \subsection{Central lack of blue galaxies, but also blue overdensities} \label{sec:discussion_blue}

    Visual inspection of the surface density maps and radial density profiles 
    reveals
    a central underdensity (CUD) of blue galaxies for 5/16 of the 3C sources 
    at $z < 1.5$ (Table~\ref{tab_classification}).
    Because the central density should not be negative, 
    we conclude that the periphery is overdense in $blue$ galaxies 
    and that these avoid 
    the innermost 250\,kpc region 
    around the 3C source.
    This is most prominent for 3C\,186, 3C\,210, 3C\,287, and 3C\,356 (all at $z<1.2$),
    but it is also evident for 
    3C\,305.1.
    Three--fourths 
    of the $blue$ CUDs are accompanied by a $red$ or $IR$-$only$ COD: 
    3C\,186, 3C\,210, (3C\,305.1 marginal), and 3C\,356. 
    These properties imply a spatial segregation of $red$ and $blue$ galaxies in these high-$z$ 3C clusters.
    One of the $blue$ CUDs appears without corresponding $red$ or $IR$-$only$ COD: 
    3C\,287, but it lies between two $IR$-$only$ ODs. 
    These findings from the inspection of the surface density maps and radial density profiles are 
    consistent with the numbers on the radial density profiles in Table~\ref{tab_od}.
    The results 
    for
    $blue$ CUDs and CODs 
    remain even using
    a sharper color cut, 0.5 mag below the predicted RS.
    
    On the other hand, two 3C sources, 3C\,208 and 3C\,324 both at $z \lesssim 1.2$, show a significant $blue$ central overdensity.
    Their $blue$ COD is accompanied by a pronounced $red$ COD as well. 
    These two clusters contain both passive and star-forming galaxies in their center.

    Our 3C sample contains only one source with a $blue$ COD but no $red$ COD; 
    in fact, 
    it is even more extreme 
    because it has 
    a $red$ CUD. 
    If the $blue$ COD and $red$ CUD of 3C\,432 at $z=1.8$ are both real, then
    it indicates a clustering of blue galaxies in the center accompanied by a deficit of central red sources. 
    It is unlikely that the powerful radio AGN has 
    cannibalized 
    the red but not the blue galaxies in its vicinity.
    Therefore, 3C\,432 might be located in a compact group of star-forming galaxies, which 
    is part of an extended  proto-cluster 
    with 4--6 other subclusters of $red$ and $IR$-$only$ galaxies (Fig.~\ref{fig:sd_maps_7}). 


    \subsection{Evolutionary aspects} \label{sec:discussion_evolution}

    We have sorted the diversity of 
    red/blue CODs/CUDs into a classification scheme for the color-dependent 
    cluster morphology
    (Sect.~\ref{sec:discussion_classification}, Table~\ref{tab_classification}).
    This classification scheme was formally guided by simple mathematical combinatorics, 
    but it may reveal possible evolutionary trends of the 3C clusters.
    Figure~\ref{fig:class} plots the cluster class versus redshift.
    The most frequent classes in each redshift group are: 
    \begin{itemize}

    \item[$\bullet$]
      class I at $z<1.2$, frequency 4/8, only 2 have class II

    \item[$\bullet$]
      class II  at $1.2<z<1.6$, frequency 6/8, none has class I

    \item[$\bullet$]
      class VI  at $z>1.6$, frequency 3/5 but uncertain. 

    \end{itemize}
    This tentatively suggests two evolutionary trends: 
    \begin{itemize}

    \item[1)]
      The clusters at $z<1.6$ evolve, on average, from class II to class I. 
      The evolution refers to the spatial segregation of central red and peripheral blue galaxies,
      which 
       becomes more pronounced at lower redshifts.\footnote{
        In order to account for the galaxy ages, the color cuts become bluer with increasing redshift (Fig.~\ref{fig:cmd}).
        Varying the color cuts shows that 
        the absence of $blue$ CUDs at $1.2<z<1.6$ compared to $z<1.2$ is not an observational bias. 
      }

    \item[2)] For $z>1.6$,
      we see mostly extended proto-clusters. They 
      are likely to 
      evolve to the more concentrated clusters at $z<1.6$.
      This evolutionary trend is consistent with
      expectations for the proto-cluster/cluster evolution around $z \sim 1.6$.
      However, the classifications are uncertain, 
      as mentioned 
      in 
      Sect.~\ref{sec:discussion_classification}. 
    \end{itemize}
    
    The current sample of 21 clusters is small, and the diversity of cluster types is large. 
    Therefore it is encouraging that consistent evolutionary trends could be derived.
    The full 3C sample of 64 high-$z$ sources may  
    corroborate these tentative results.
    The classification scheme is simple and 
    can serve as a useful guide 
    for future studies.

    The high-$z$ 3C cluster sample has been selected by using massive radio galaxies as markers for cosmic mass concentrations.
    How far the classification scheme and the evolutionary trends are applicable to clusters in general needs to be investigated.
    For instance, powerful AGN activity may quench star-formation in the vicinity of the radio source.
    On the other hand, \citet{Galametz2009} looked for the frequency of AGN in galaxy clusters at $0<z<1.5$ selected by X-ray, mid-IR, and radio criteria. 
    Regardless of selection method, those $z > 0.5$ clusters show an overdensity of AGNs at $r < 500$~kpc, 
    and this AGN excess increases with redshift. 
    This suggests that any bias on the cluster properties from using radio galaxies as cluster signposts may be small.

    To follow the evolution of the 3C environment to lower redshift ($z<1$), 
    homogeneous selection of blue and red galaxies is required 
    for a proper comparison with the high-$z$ results.
    Regarding X-ray observations, most 3Cs at $z<1$ show extended X-ray emission from hot intra-cluster gas, 
    but among the high-$z$ 3C sources only a few 
    have been observed deep enough to allow 
    detection of extended cluster emission  
    \citep{Wilkes13,Massaro15,Stuardi18,Jimenez2020}.

    $Herschel$ far-infrared observations of 
    the entire high-$z$ 3C sample have revealed (ultra-)luminous star-forming activity 
    in 22/64 ($\sim$30\%) of the 3C sources \citep{Podigachoski15}.
    In contrast, the host-normalized SFR of the bulk of the 
    3C sources at $z<1$ is low \citep{Westhues2016}.
    ULIRG activity suggests major mergers of gas-rich galaxies, and indeed 
    92\% of the high-$z$ 3C sample are associated with recent or ongoing merger events \citep{Chiaberge15}. 
    Such a merger requires at least one gas-rich (and hence star-forming) galaxy close to the radio source 
    or infalling toward it from the cluster periphery.
    If there were many gas-rich 
    companion 
    galaxies in the vicinity of the radio source, 
    they should show up as a $blue$ COD, but this is not observed in the majority of our sample.
    An explanation could be that the number of gas-rich merger-food galaxies is small.
    The cannibalization of a single gas-rich 
    galaxy would be 
    sufficient for 
    a major merger which subsequently quenches star-formation in its vicinity.

        The richnesses and radial density profiles of the fifteen 3C clusters at $1<z<1.5$ are, on average, 
    comparable to those of poor local clusters (Fig.~\ref{fig:coma}). 
    About 
    one-third 
    of the 3C sources show $red$ CODs which are a factor 2--3 larger than average,
    indicating that they are already massive  
    systems (Abell richness class 2 or larger) at these redshifts. 
    If their masses grow 
    further, then their local successors 
    are likely to 
    be very rich clusters 
    (Abell class 3 or larger), which are rare. 


    \section{Summary and outlook} \label{sec:summary}


We have explored the environment of 21 high redshift 3C sources at $1<z<2.5$ using \HST\ images in the filters F606W and F140W.
These filters encompass the rest frame 4000\,\AA\ break. 
We applied redshift-dependent color cuts to separate passive red and star-forming blue galaxies.
The images have a 2$\arcmin$ field-of-view  (about 1 Mpc at the 3C's redshift), yielding about 450 galaxies per 3C environment.

To summarize our methodology and results:
\begin{enumerate}

\item
 The completeness of single-filter-detected galaxies 
 is
 around 100\% at F140W\,=\,23 and 50\% at 25 mag.
The
completeness is lower
when F606W detection is also considered.
To reduce the incompleteness of faint red galaxies, we include galaxies detected in the F140W filter only.
Based on a stacking analysis, the IR-only detected galaxies are about 0.5 mag redder than 
those detected in two filters.

\item
We analysed four galaxy samples: $all$, $red$, $blue$ and $IR$-$only$.
For each sample, surface density maps and radial density profiles show  
a  diversity 
in
the clustering of red and blue galaxies around the 3C sources.

\item
At $z<1.5$, about 80\% of the  
fields 
reveal an overdensity of red galaxies 
within 
a projected radius $r \sim 250$\,kpc around the 3C source. 
The overdensity is defined 
relative to 
the mean galaxy density of the periphery (250\,kpc $<r<$ 500\,kpc).

 \item
   The probability that fore- and back-ground galaxy groups seen in projection along the line of sight 
   could account for the central overdensity within an individual 3C field is $<$15\%.
 Experiments in which the 3C sources were assigned random locations within each image confirm that 
  the likelihood of chance projections is negligible
  for the sample as a whole.
  Therefore most of the overdensities are associated with the 3C sources.

\item
About three 3C sources show an overdensity of blue galaxies, indicating star-forming activity close to the 3C source.

\item
For five 3C sources we found a spatial segregation of 
red and blue galaxies,
with the former populating preferentially the central regions while the latter reside in 
outer regions of the 
clusters.


\item
The strength and significance of the observed overdensities 
decrease at 
$z>1.5$.
This trend may be due to the lower sensitivity and completeness of our data
at these redshifts.
Nevertheless, some sources at $z<1.5$ lack overdensities, indicating that any clustering has not yet formed the
spatial concentration found for most other 3C sources.
We suggest that these are proto-clusters in an early assembly phase. 

 \item
 At $z<1.5$, 
 the overdensities of bright red galaxies could mark the luminous end of the red-sequence.
%
 Despite incompleteness of faint $red$ and $IR-only$ galaxies, 
 our data indicate that the red-sequence continues in some but not all 3C sources 
 to about 3 mag fainter than the bright end; 
 in some 3C sources the faint red-sequence galaxies show less spatial concentration than the brighter ones.


\item
The diversity of red and blue clustering led us to present a classification scheme of  
color-dependent cluster morphology.
We find a preference of class I and II at $z<1.6$ and class VI at $z>1.6$, indicating possible 
different evolutionary states. 

       
\item
The derived number of central luminous red galaxies and the radial density profiles are comparable to those 
found in local clusters, indicating that some 3C clusters are already mass-rich and compact.

\end{enumerate}
With only two filters, the different clustering of red and blue galaxies is 
efficiently detected, hinting at possible evolutionary trends. 
These findings are based on a representative 
subset 
of the complete 3C sample of 64 high-$z$ sources.
The results make future studies with larger fields and larger samples promising, in particular
to improve the statistics of the classification scheme, 
to corroborate the evolutionary trends, and to investigate possible subclustering.
While the density maps and the 
randomized 
experiments 
strongly suggest that the density enhancements are associated with the 3C sources,
spectroscopic observations will verify cluster membership of the candidates and enable a kinematic study.

            \begin{acknowledgements}
              This work is based on observations taken by the HST-3C collaboration, program GO~13023, with the NASA/ESA HST,
              which is operated by the Association of Universities for Research in Astronomy, Inc.,
              under NASA contract NAS~5-26555.
              The 3C environment research at Ruhr-University Bochum
              is supported by funds from Deutsche Forschungsgemeinschaft (DFG grant HA3555/13-1).
            \end{acknowledgements}

            %
            \bibliographystyle{aa} 
            \bibliography{ghaffari_3c_hst.bib} 
            %

            \begin{appendix}


              \section{Surface density: maps and radial profiles of the entire sample}\label{appendix_a}

\begin{figure*}
                \hspace{-0mm}\includegraphics[width=0.245\linewidth, clip=true]{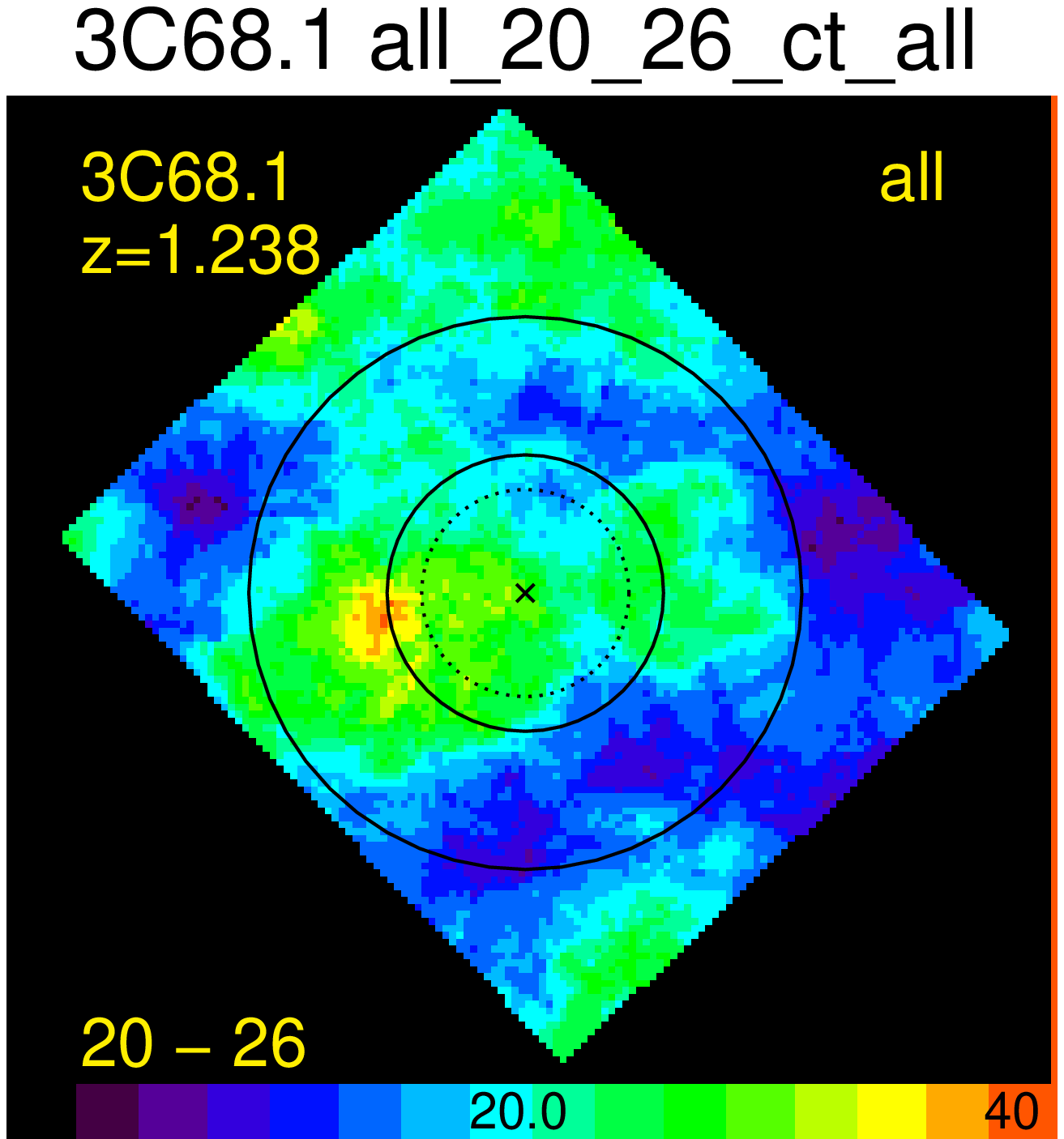}                
                \includegraphics[width=0.245\linewidth, clip=true]{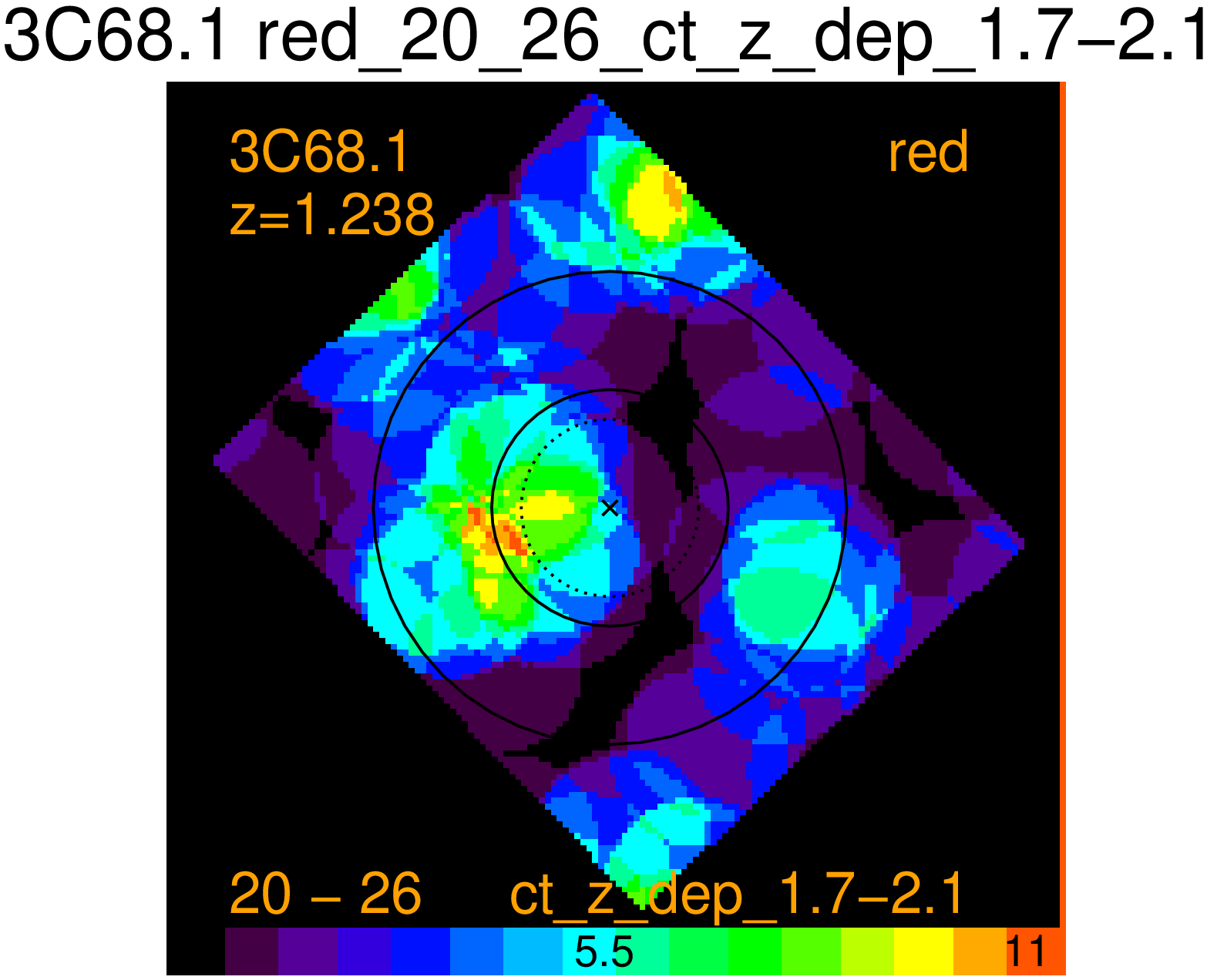}      
                \includegraphics[width=0.245\linewidth, clip=true]{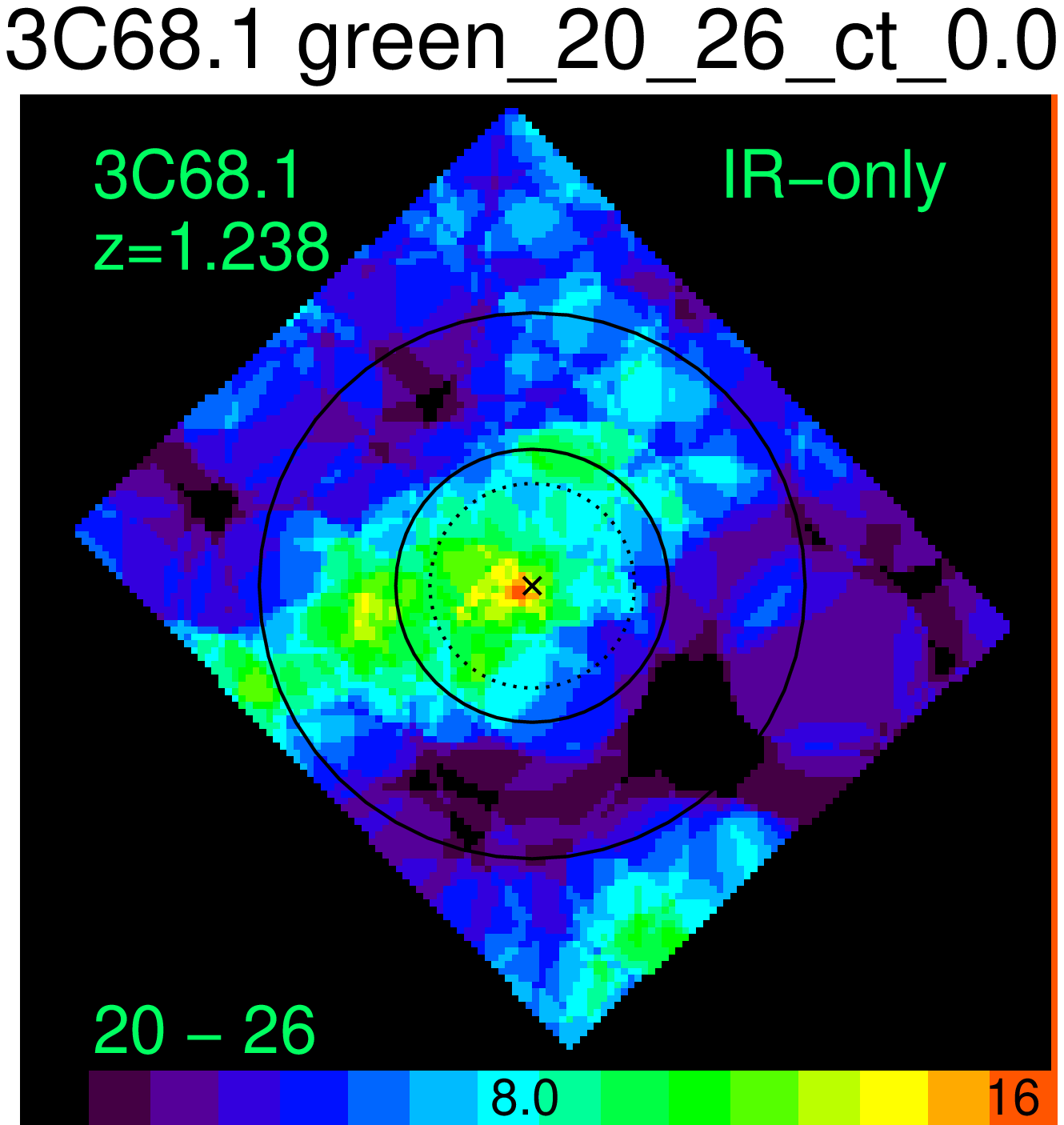}              
                \includegraphics[width=0.245\linewidth, clip=true]{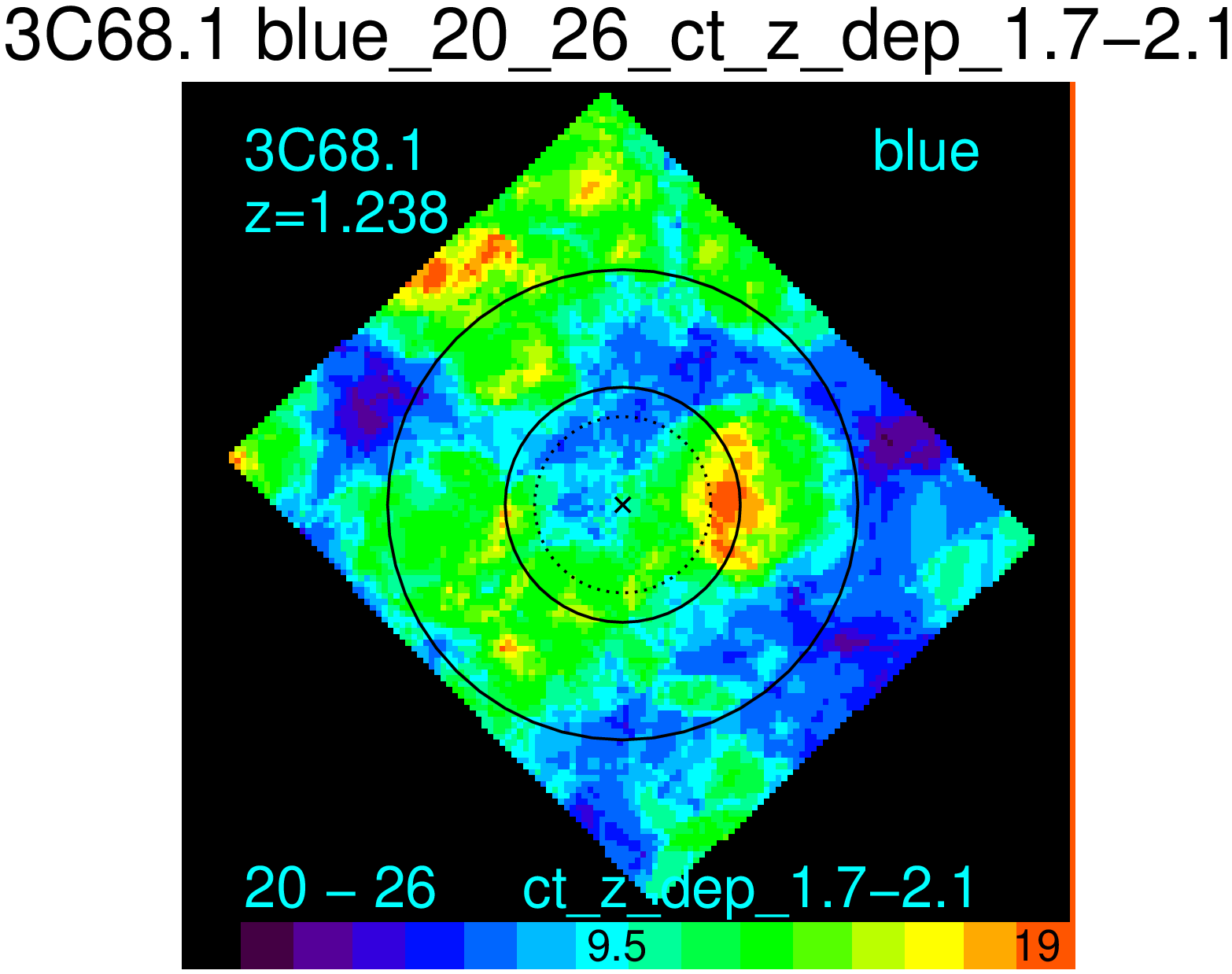}     
                
                \hspace{-0mm}\includegraphics[width=0.245\textwidth, clip=true]{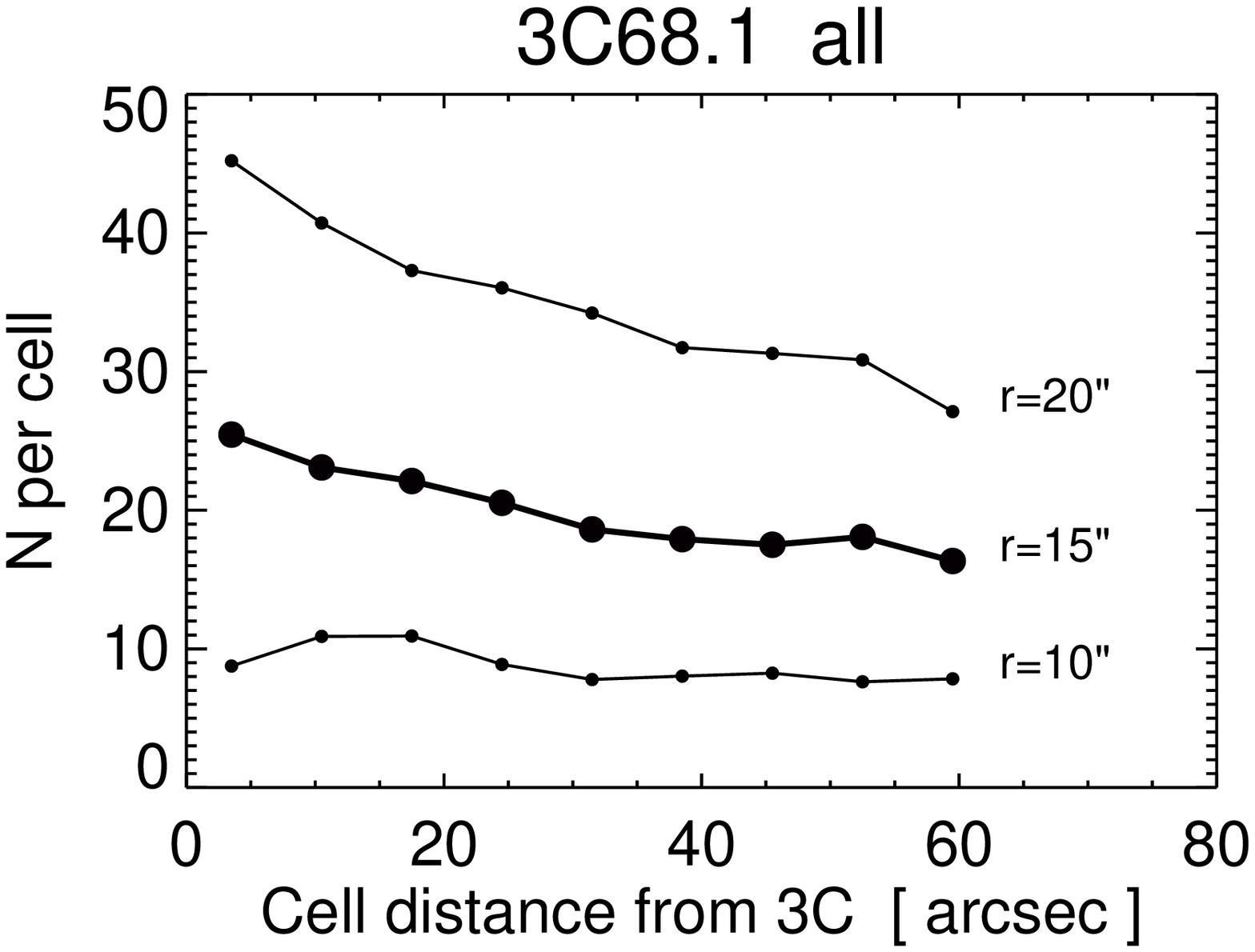}                  
                \includegraphics[width=0.245\textwidth, clip=true]{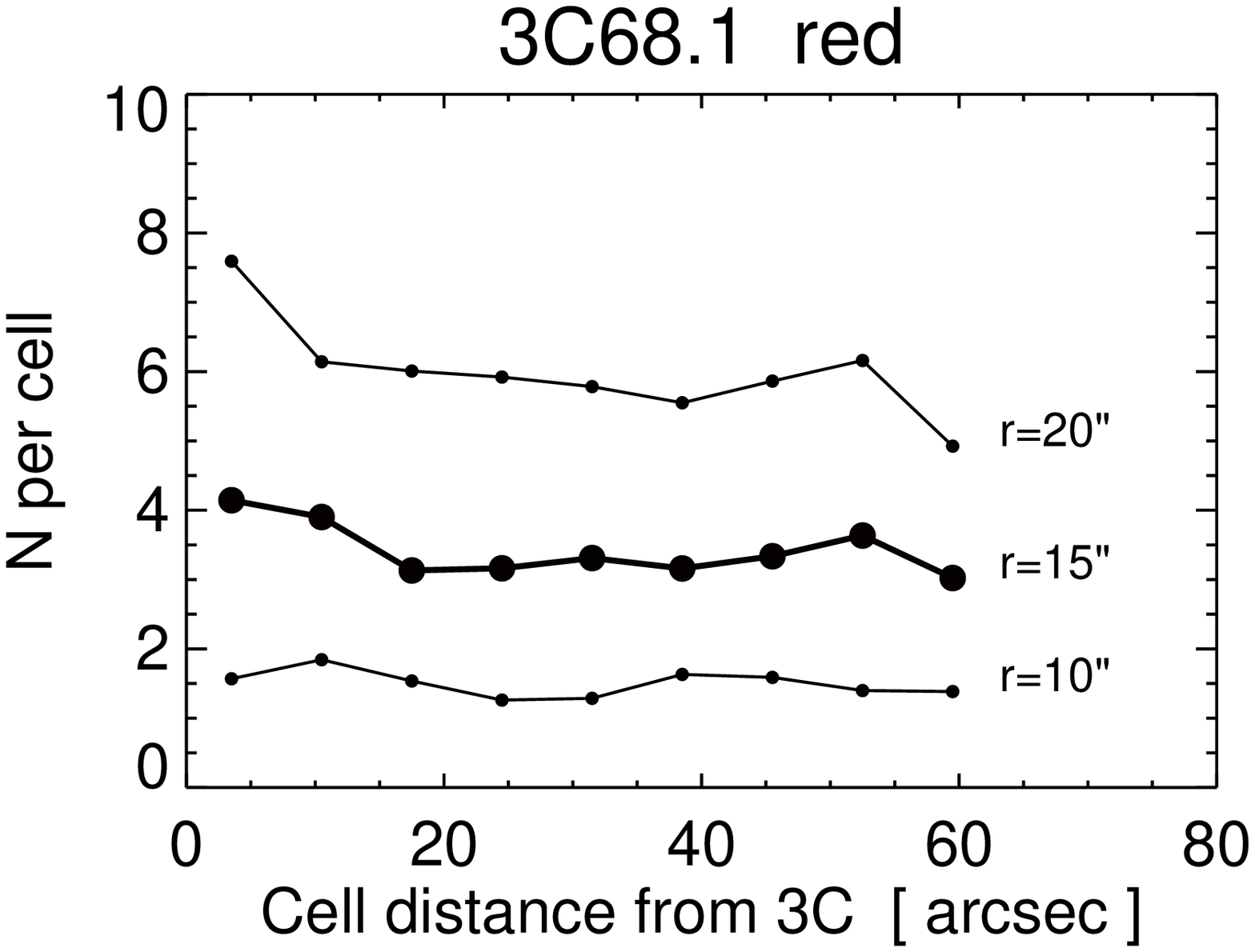}                   
                \includegraphics[width=0.245\textwidth, clip=true]{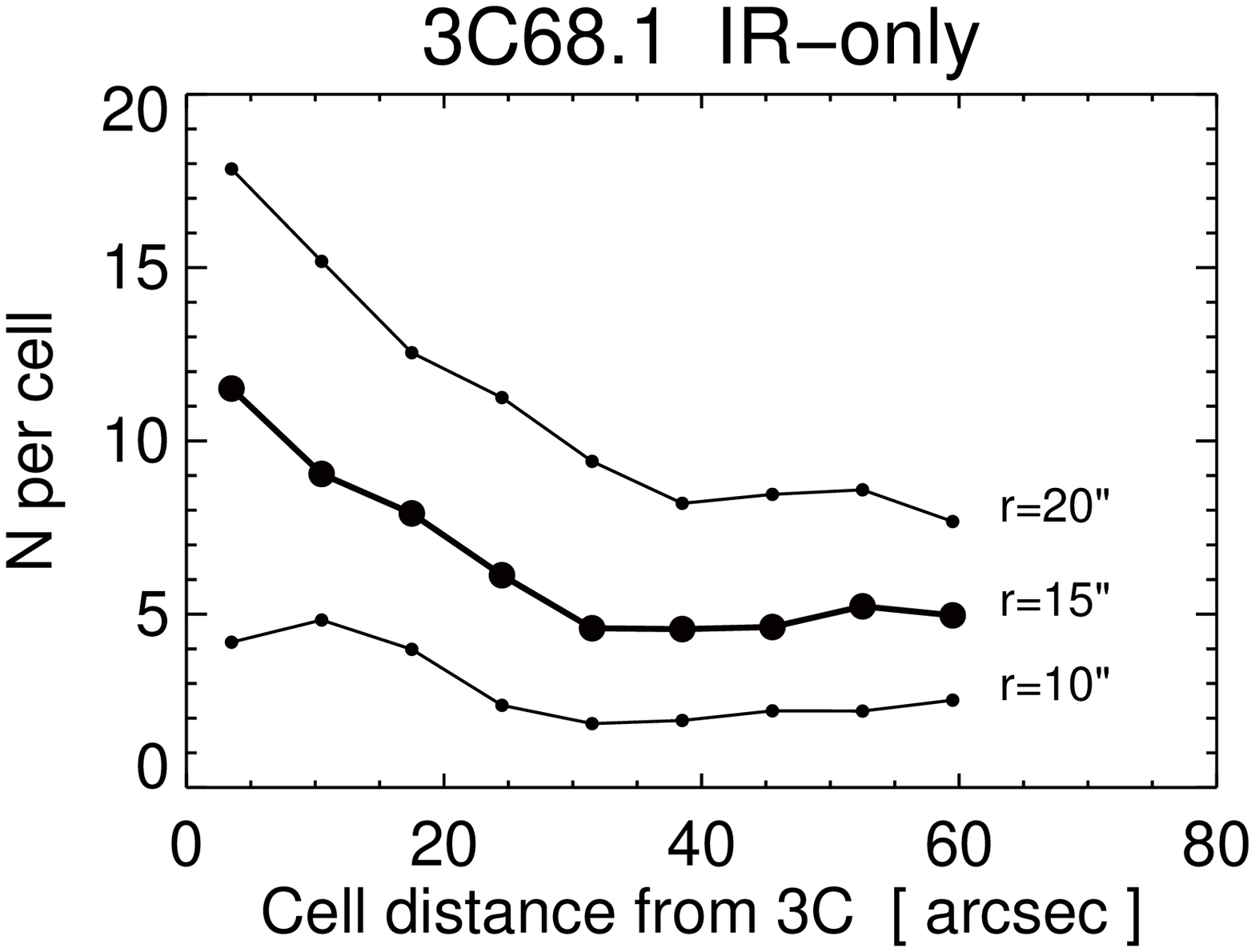}                
                \includegraphics[width=0.245\textwidth, clip=true]{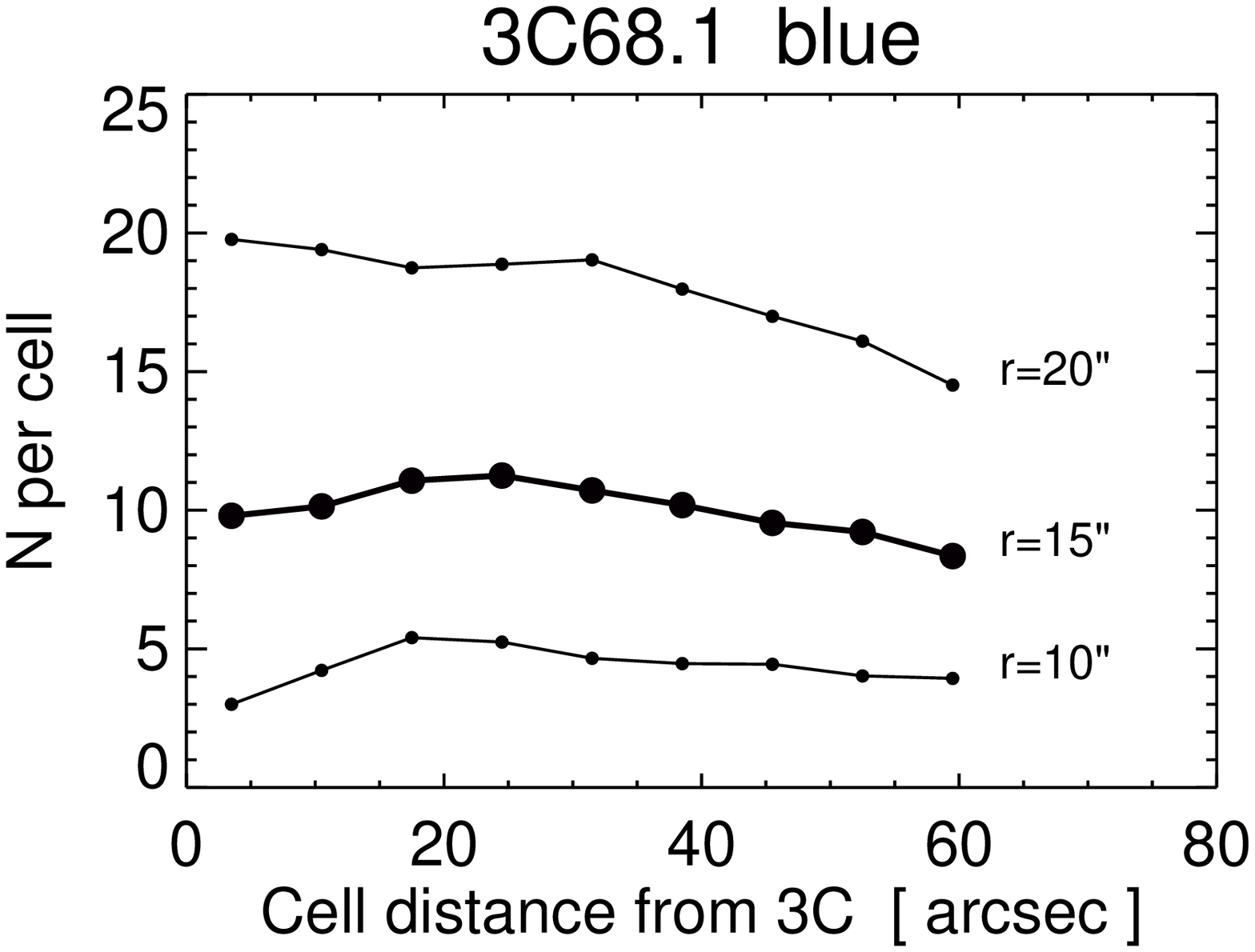}

                \hspace{-0mm}\includegraphics[width=0.245\textwidth, clip=true]{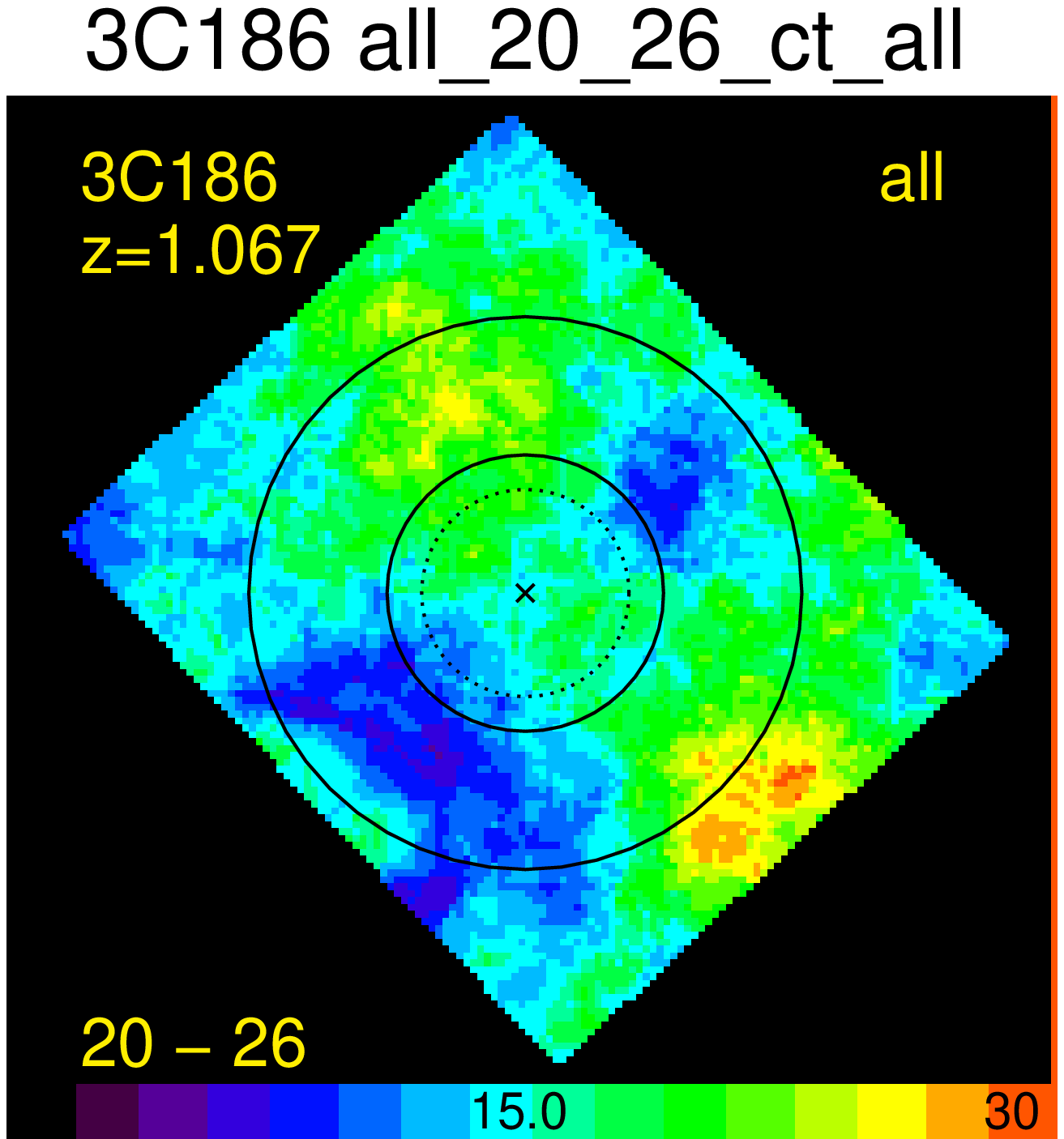}                 
                \includegraphics[width=0.245\textwidth, clip=true]{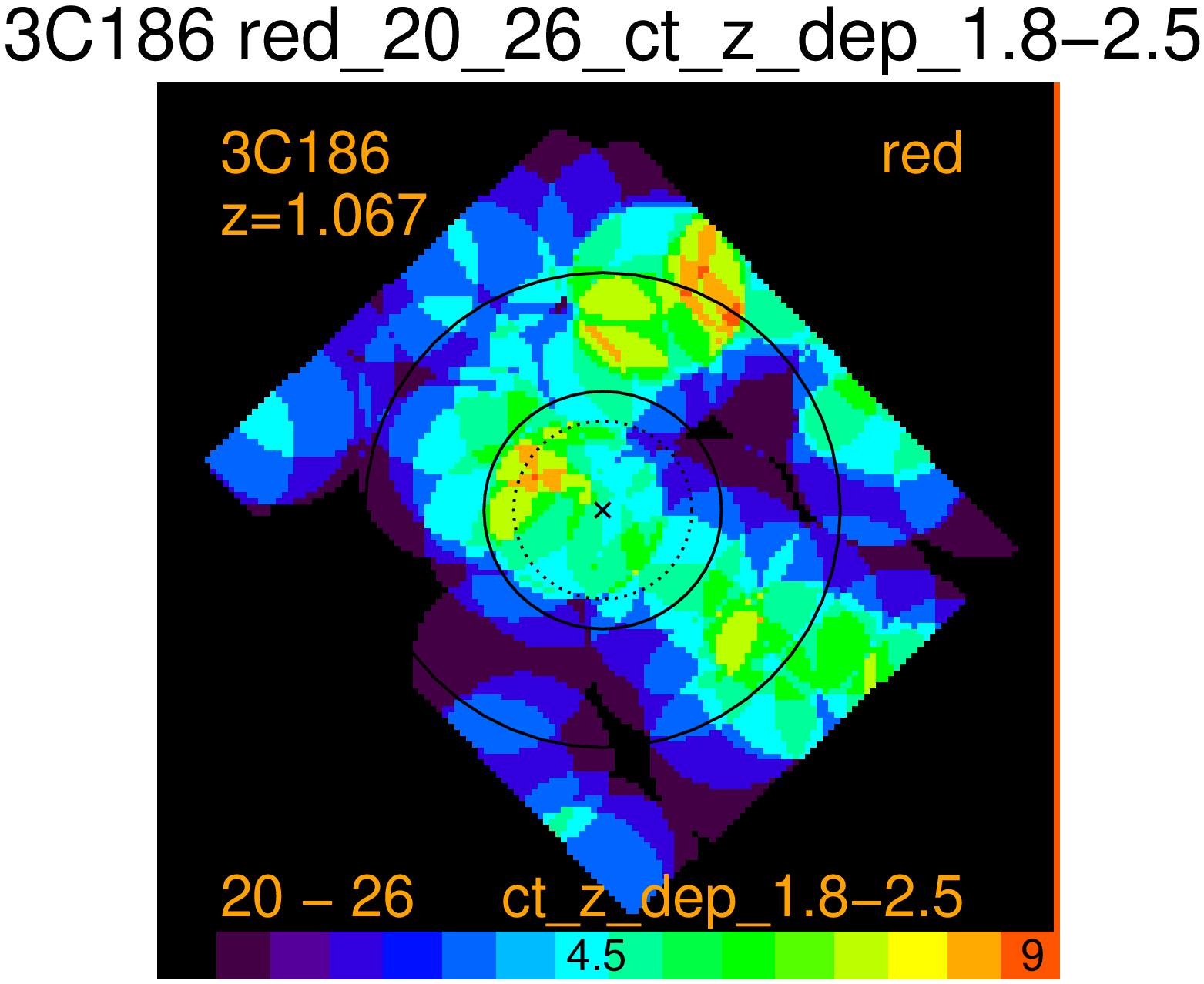}       
                \includegraphics[width=0.245\textwidth, clip=true]{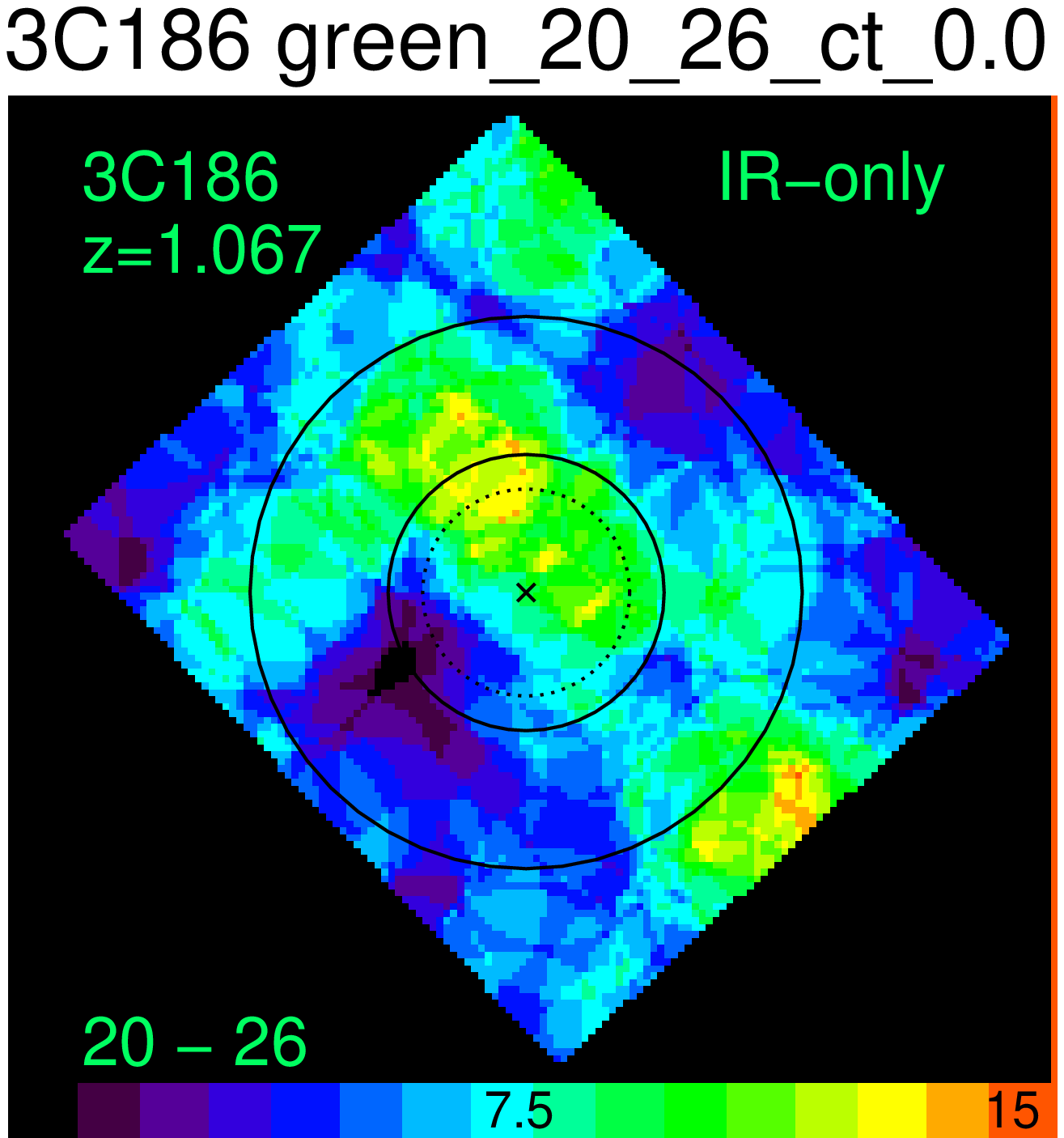}               
                \includegraphics[width=0.245\textwidth, clip=true]{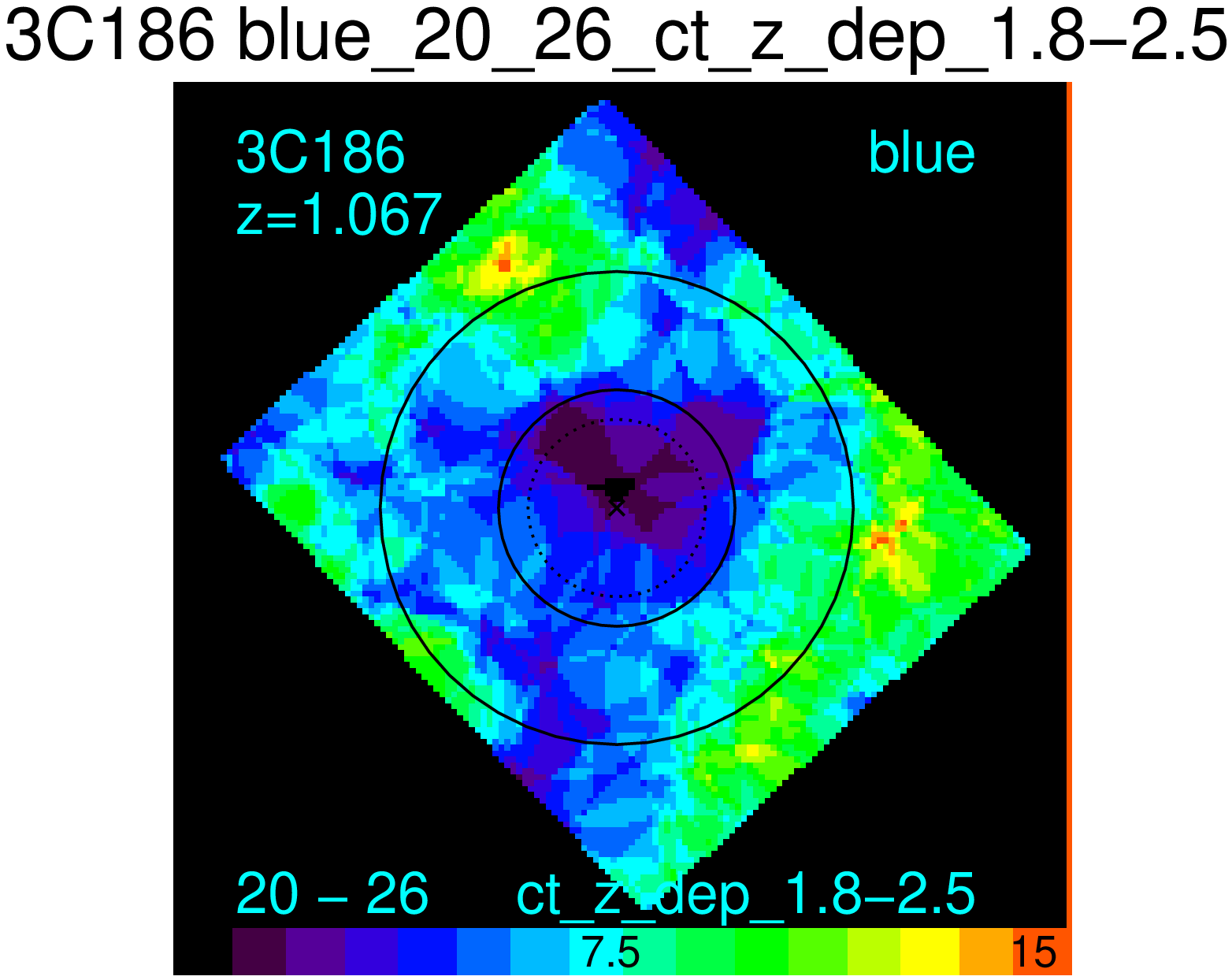}      
                
                \hspace{-0mm}\includegraphics[width=0.245\textwidth, clip=true]{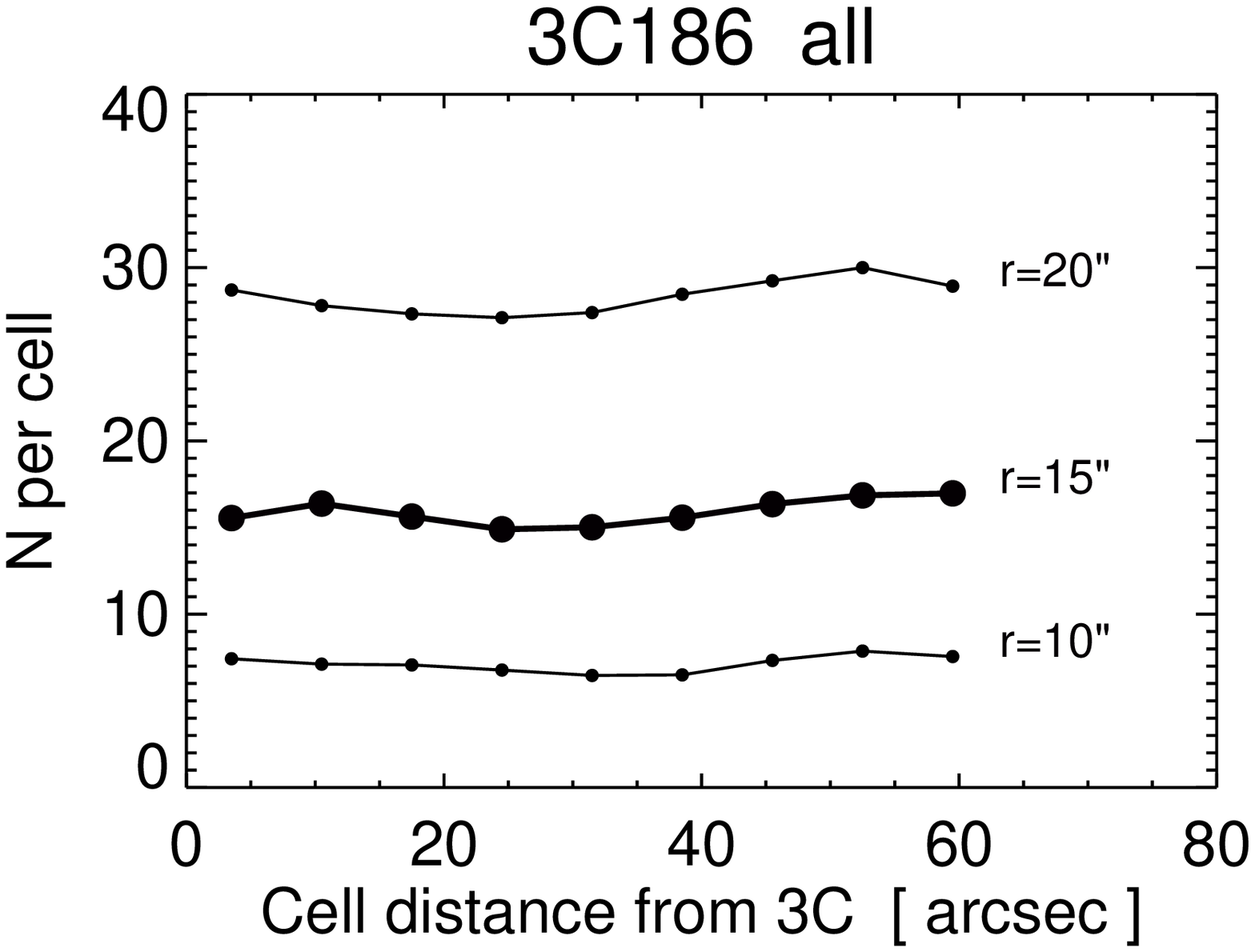}                   
                \includegraphics[width=0.245\textwidth, clip=true]{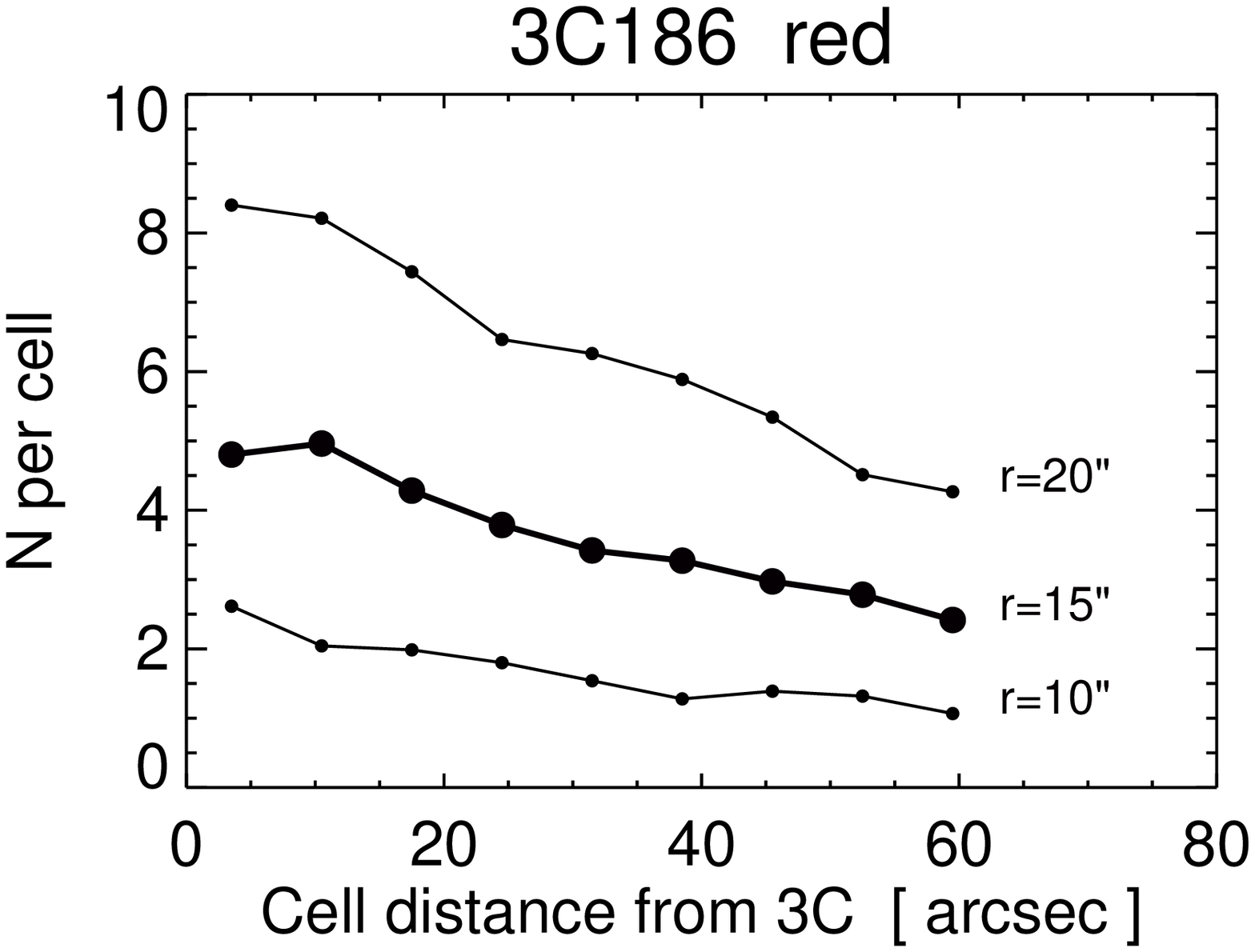}                    
                \includegraphics[width=0.245\textwidth, clip=true]{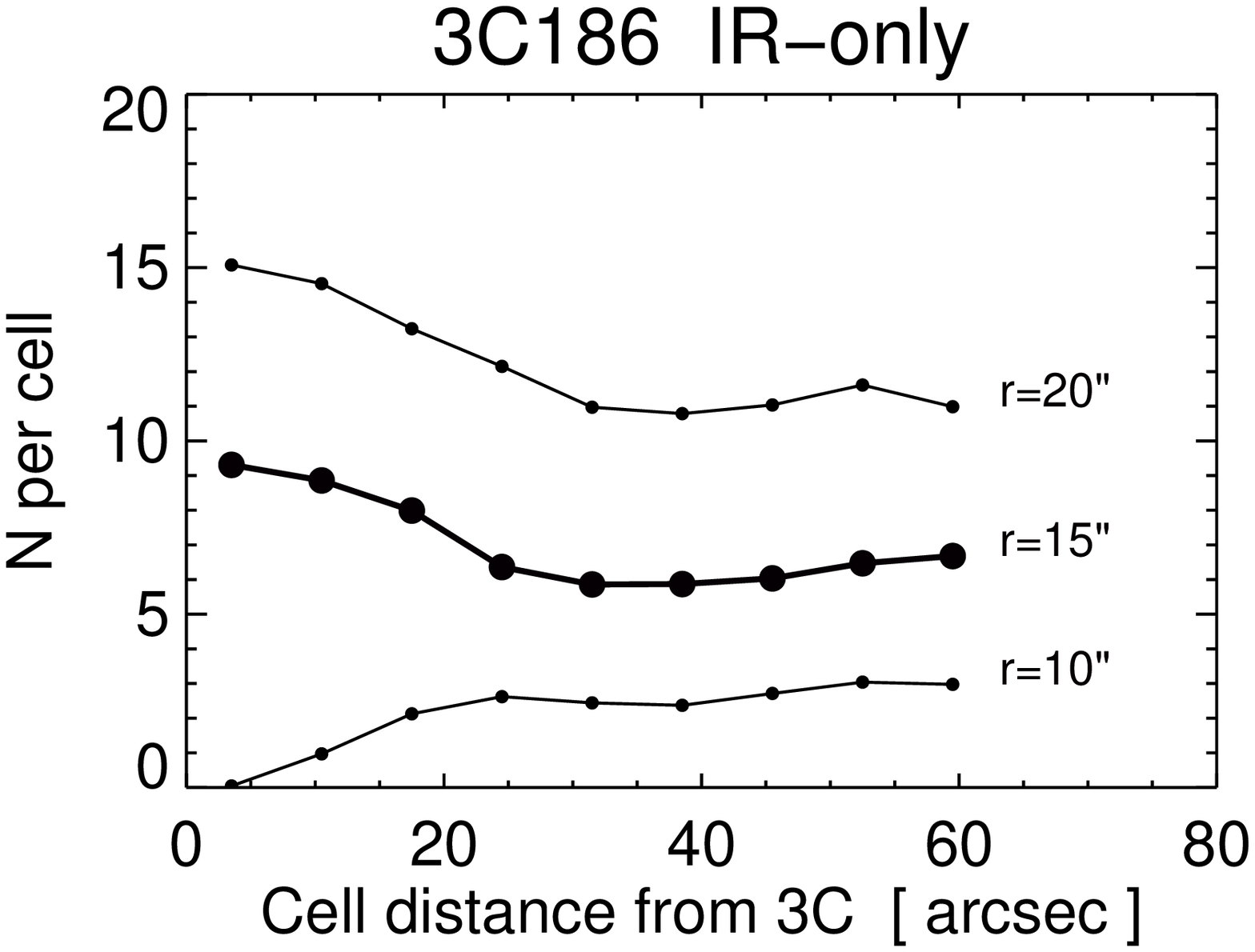}                 
                \includegraphics[width=0.245\textwidth, clip=true]{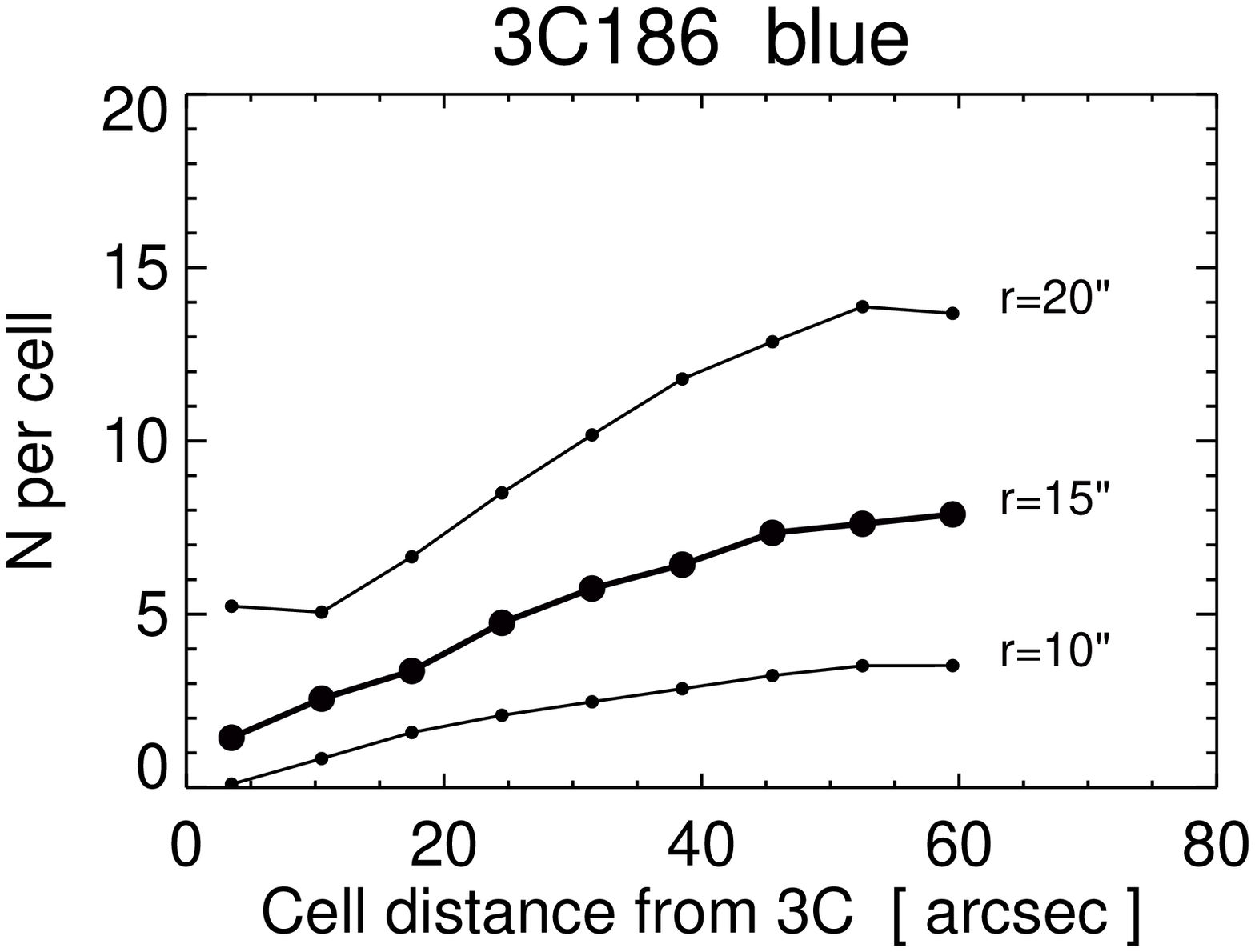}

                \hspace{-0mm}\includegraphics[width=0.245\textwidth, clip=true]{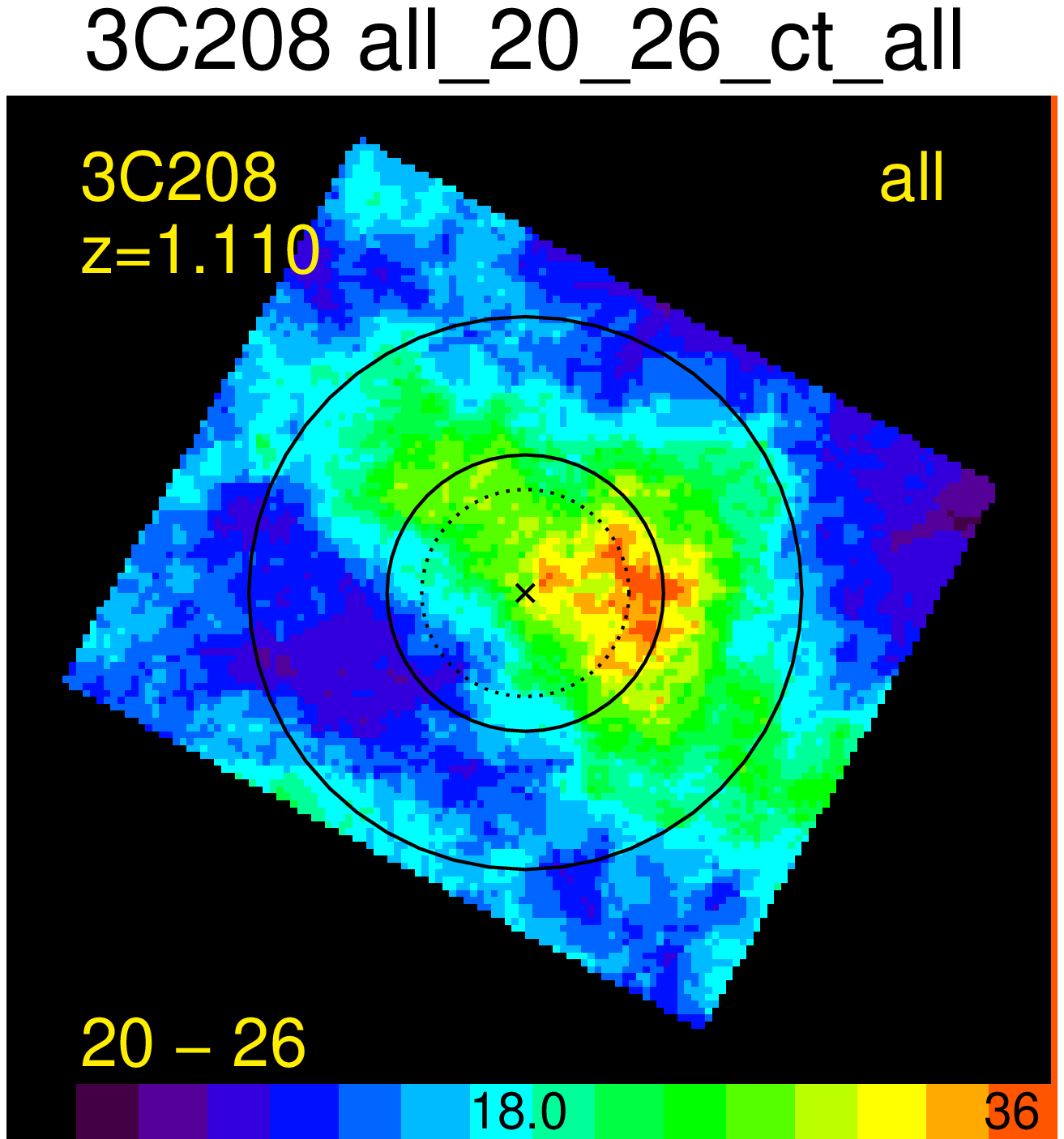}                 
                \includegraphics[width=0.245\textwidth, clip=true]{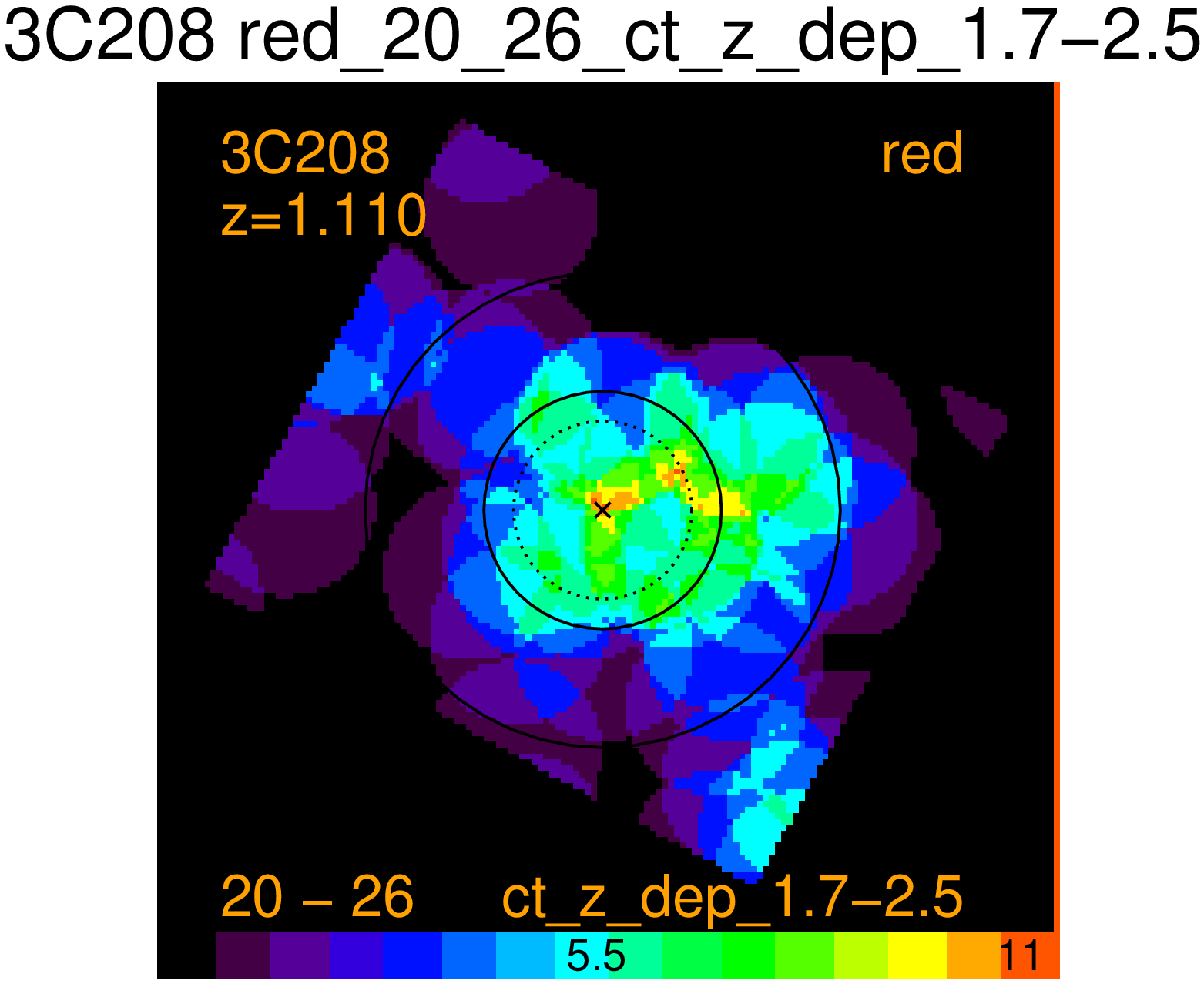}       
                \includegraphics[width=0.245\textwidth, clip=true]{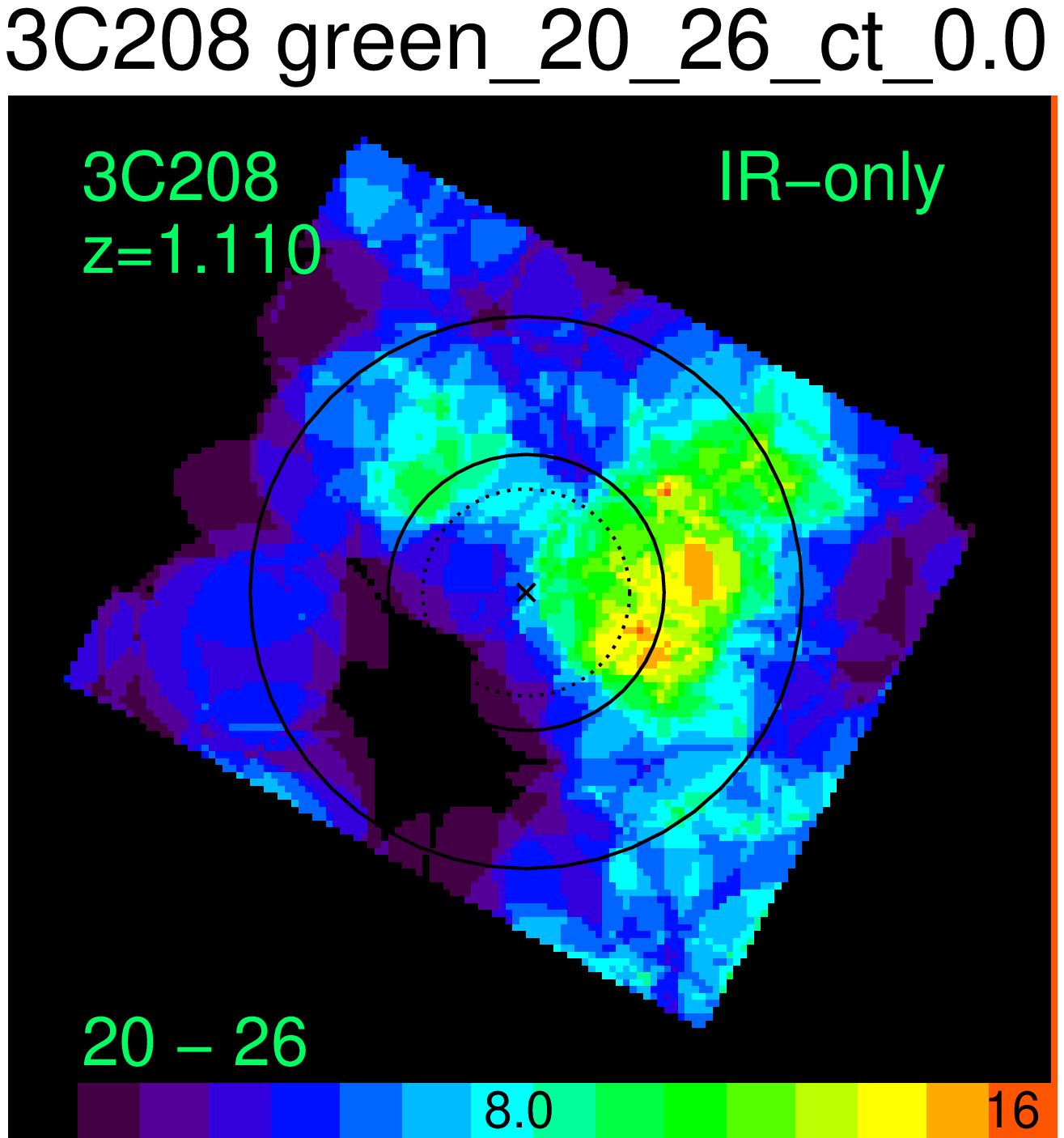}               
                \includegraphics[width=0.245\textwidth, clip=true]{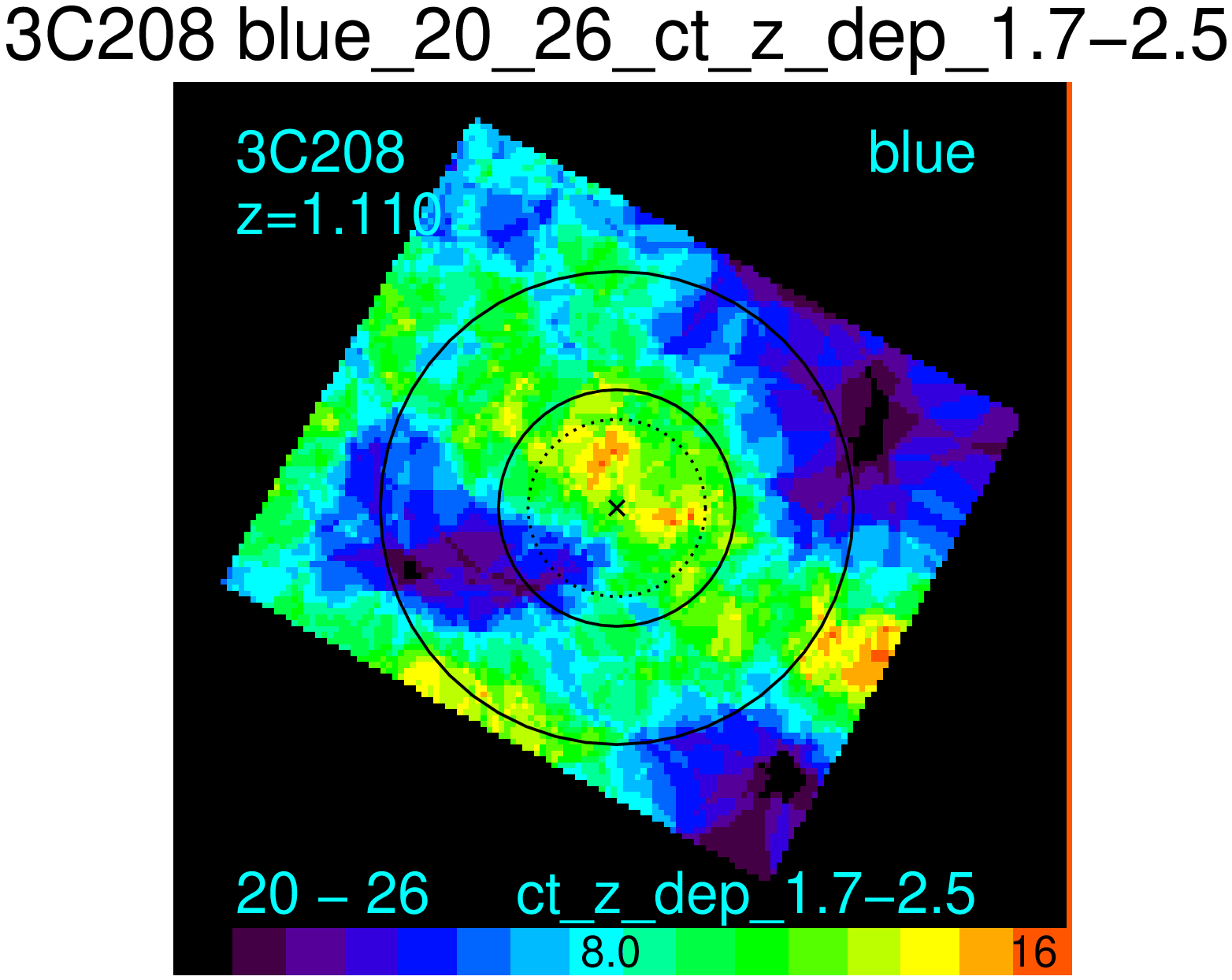}      
                
                \hspace{-0mm}\includegraphics[width=0.245\textwidth, clip=true]{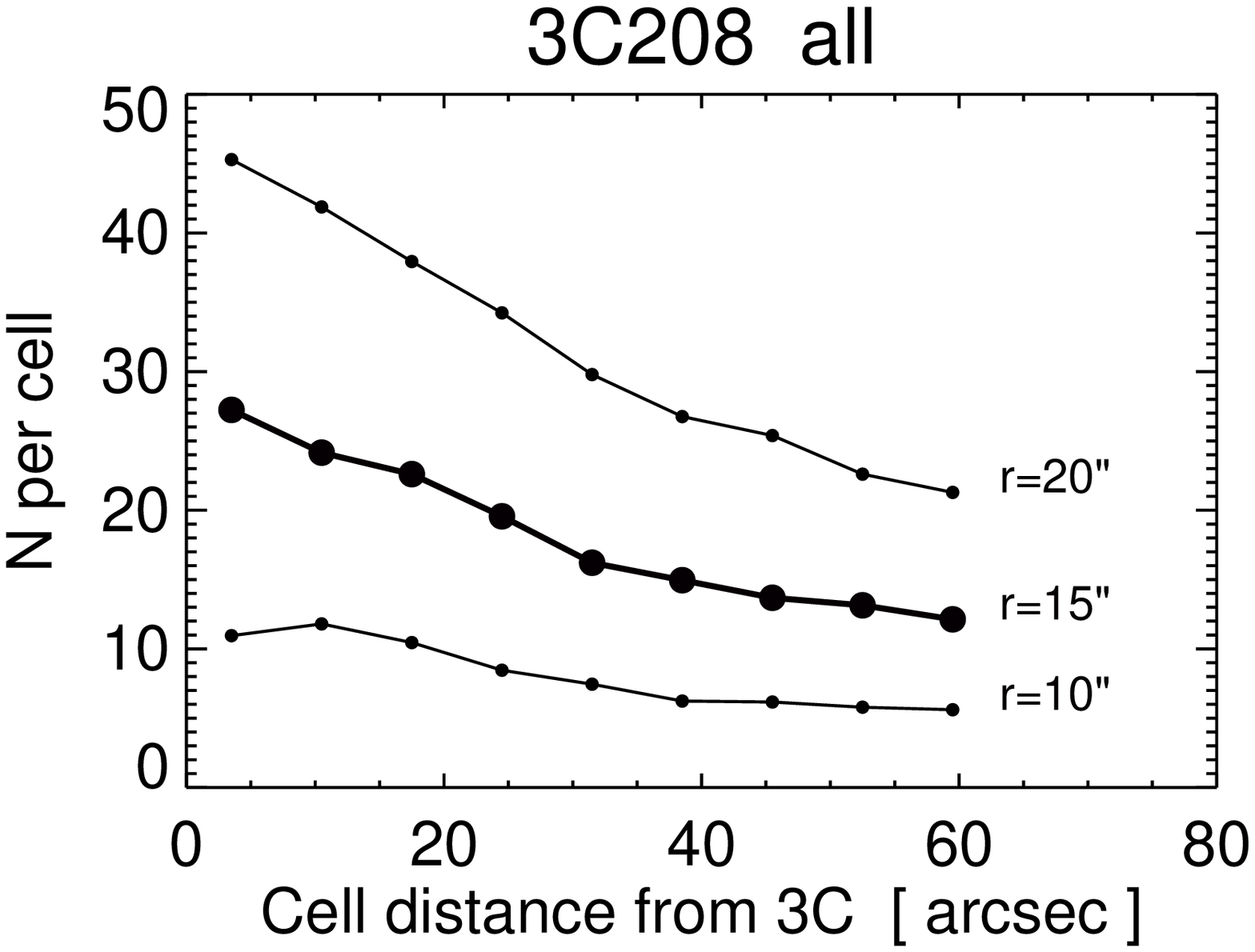}                   
                \includegraphics[width=0.245\textwidth, clip=true]{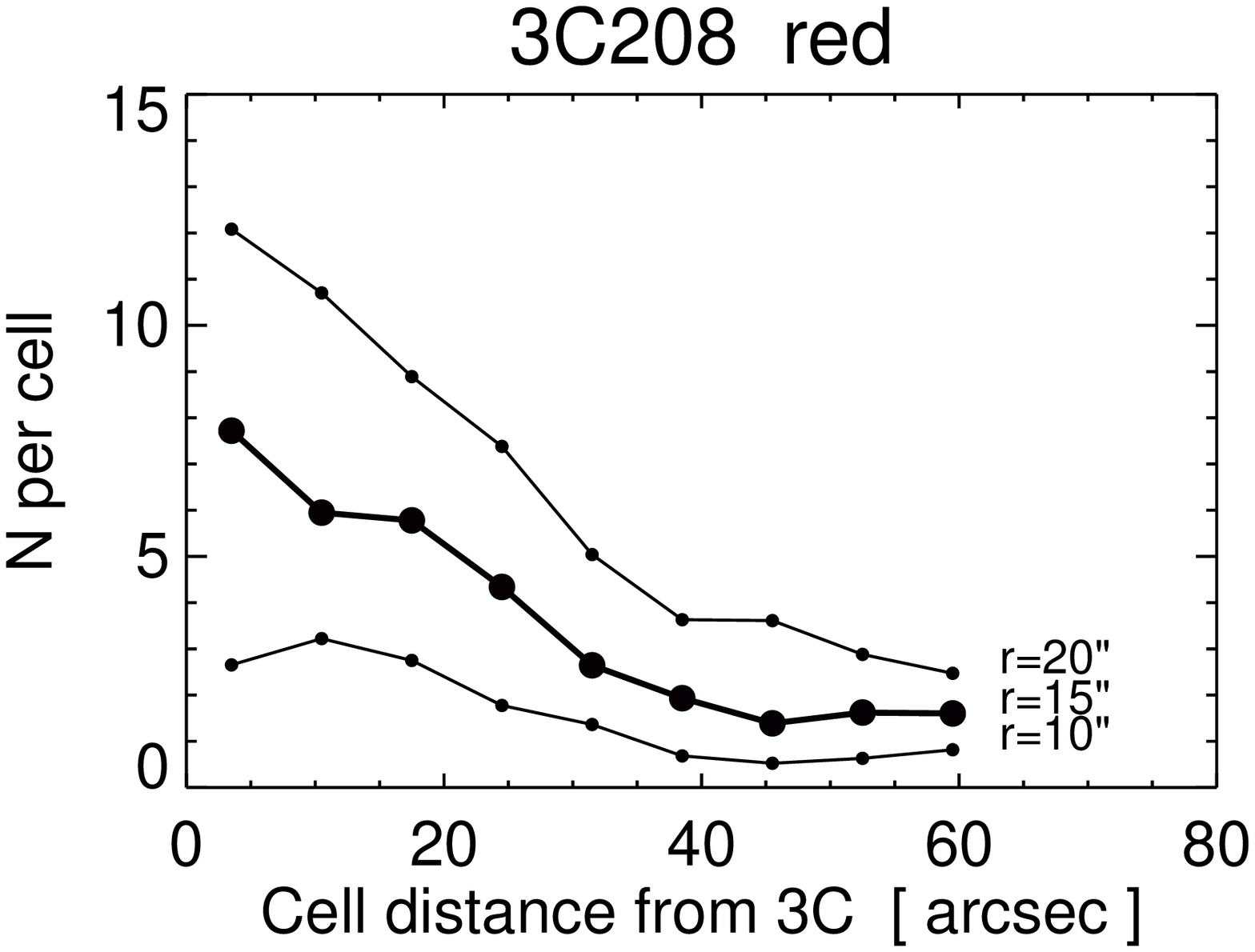}                    
                \includegraphics[width=0.245\textwidth, clip=true]{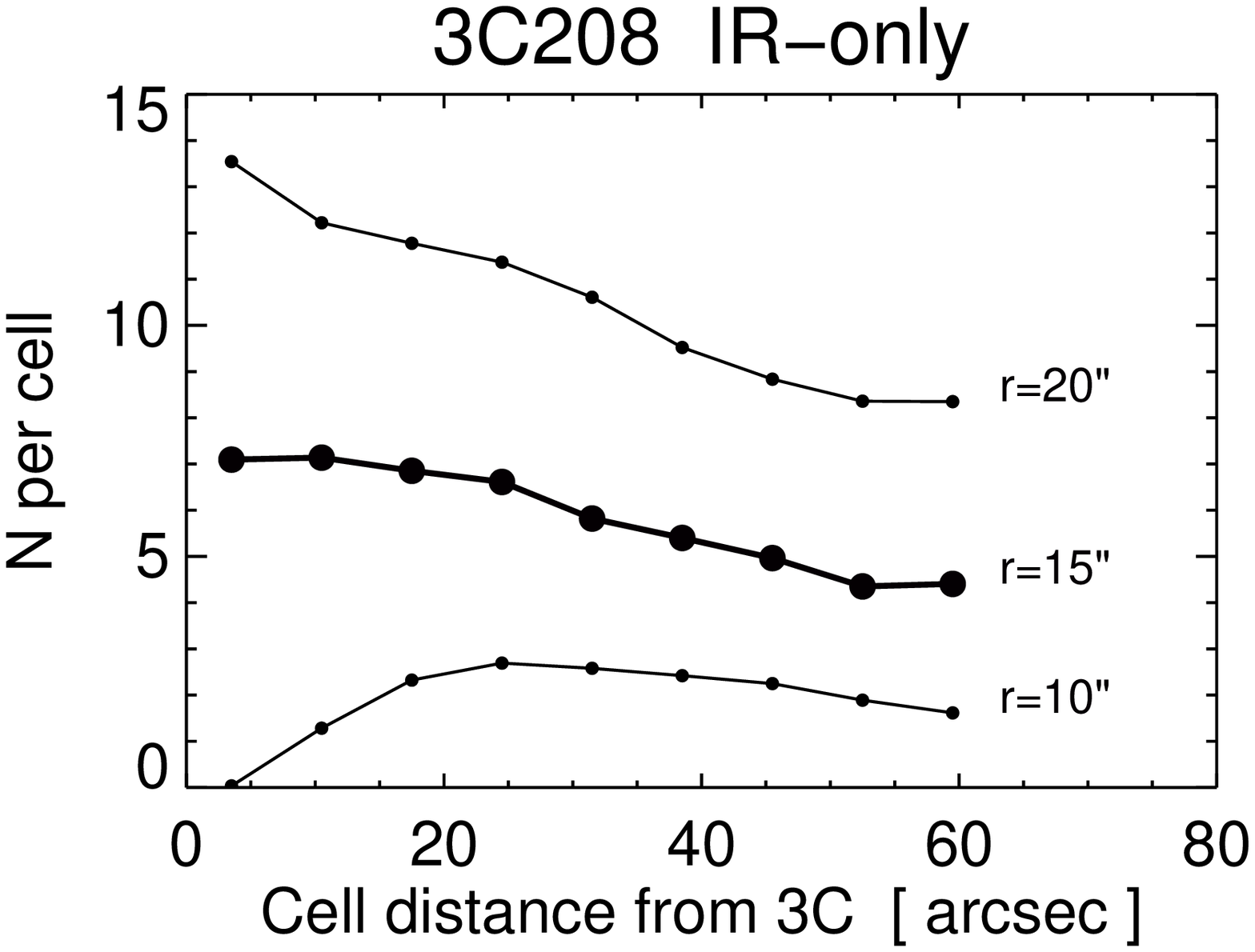}                 
                \includegraphics[width=0.245\textwidth, clip=true]{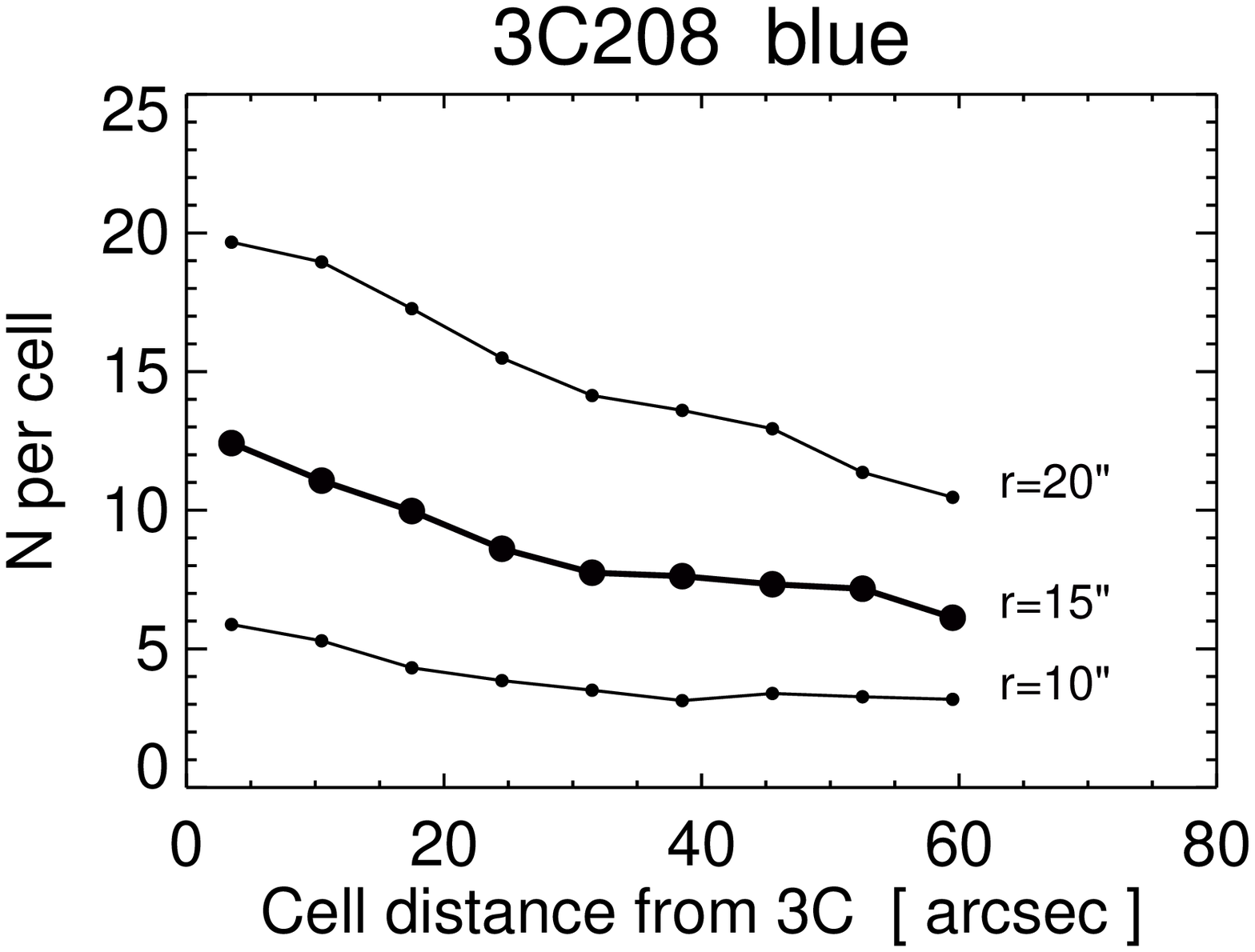}

                \caption{Surface density maps and radial density profiles of the remaining 3C fields, 
                  similar to 3C\,210 shown in Fig.~\ref{fig:sd_maps_3C210}.
                }
                \label{fig:sd_maps_1}
              \end{figure*}

\begin{figure*}
	        
                \hspace{-0mm}\includegraphics[width=0.245\textwidth, clip=true]{3C210_cell_density_all_20_26_ct_all_grid_15_1.map.ps}                 
                \includegraphics[width=0.245\textwidth, clip=true]{3C210_cell_density_red_20_26_ct_z_dep_1.7-2.3_grid_15_1.map.ps}       
                \includegraphics[width=0.245\textwidth, clip=true]{3C210_cell_density_green_20_26_ct_0.0_grid_15_1.map.ps}               
                \includegraphics[width=0.245\textwidth, clip=true]{3C210_cell_density_blue_20_26_ct_z_dep_1.7-2.3_grid_15_1.map.ps}      
                
                \hspace{-0mm}\includegraphics[width=0.245\textwidth, clip=true]{3C210_cell_radial_sd_all_20_26_ct_all_grid_xx_1.ps}                   
                \includegraphics[width=0.245\textwidth, clip=true]{3C210_cell_radial_sd_red_20_26_z_dep_grid_xx_1.ps}                    
                \includegraphics[width=0.245\textwidth, clip=true]{3C210_cell_radial_sd_green_20_26_ct_0.0_grid_xx_1.ps}                 
                \includegraphics[width=0.245\textwidth, clip=true]{3C210_cell_radial_sd_blue_20_26_z_dep_grid_xx_1.ps}

                \hspace{-0mm}\includegraphics[width=0.245\textwidth, clip=true]{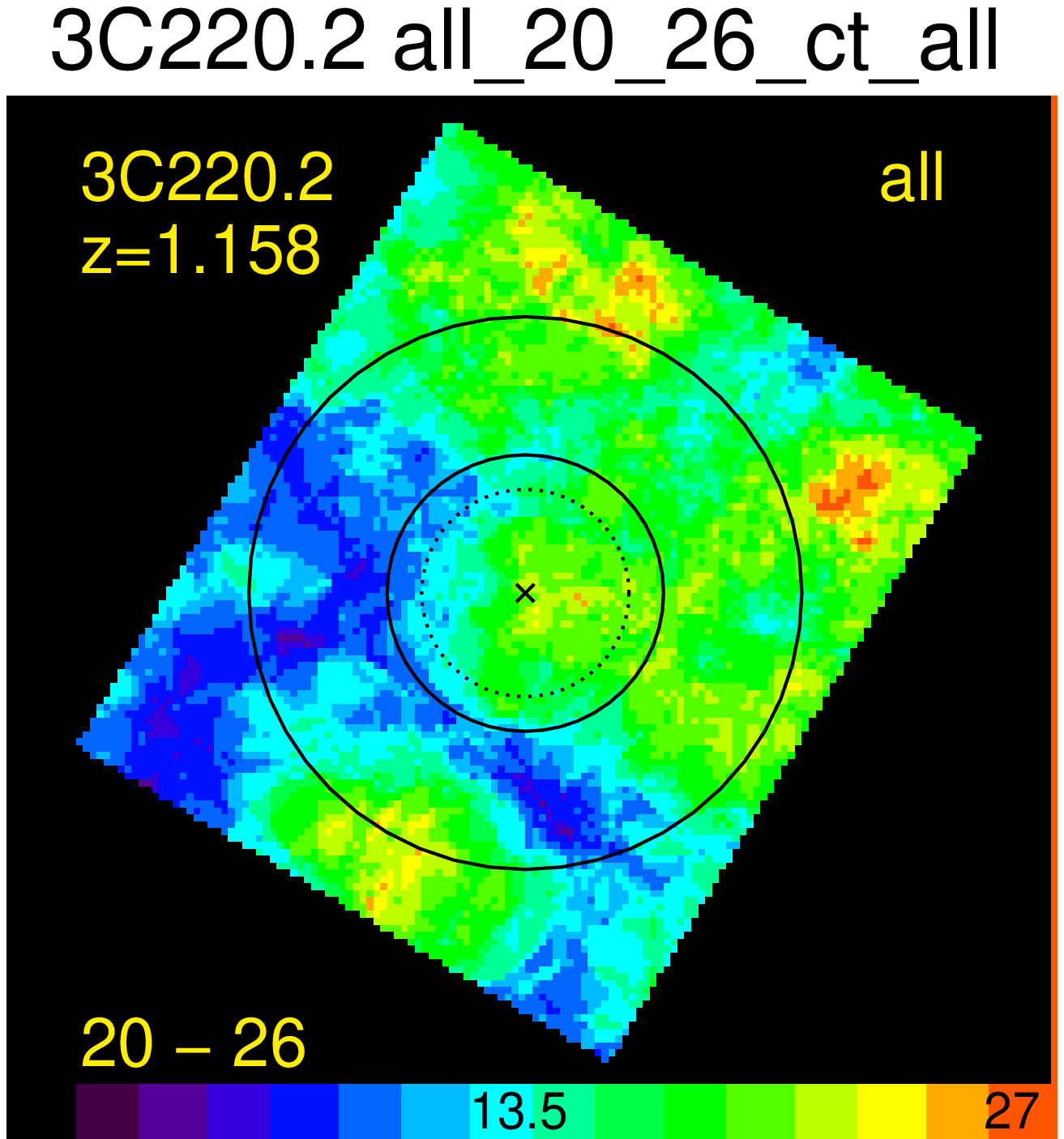}               
                \includegraphics[width=0.245\textwidth, clip=true]{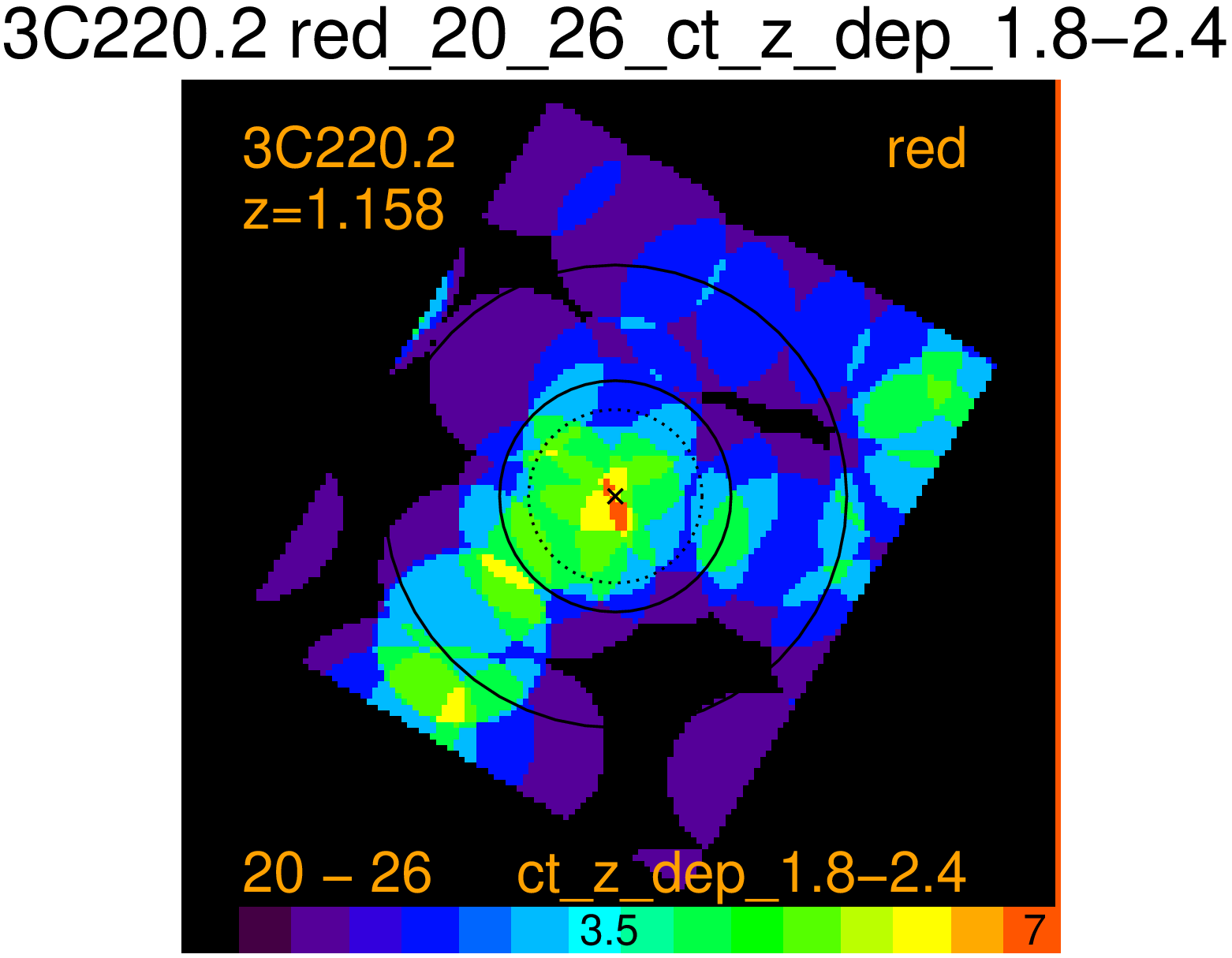}     
                \includegraphics[width=0.245\textwidth, clip=true]{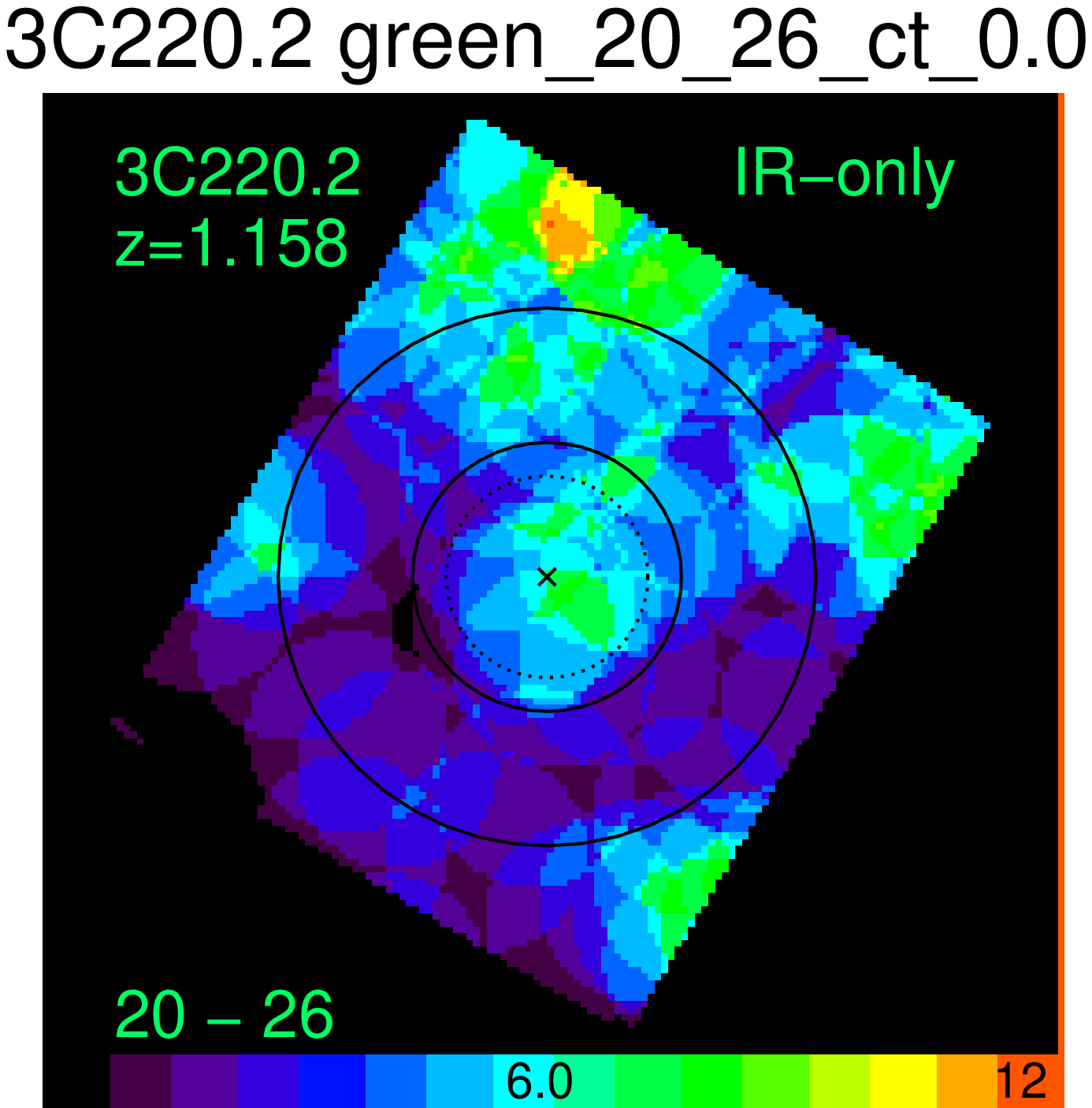}             
                \includegraphics[width=0.245\textwidth, clip=true]{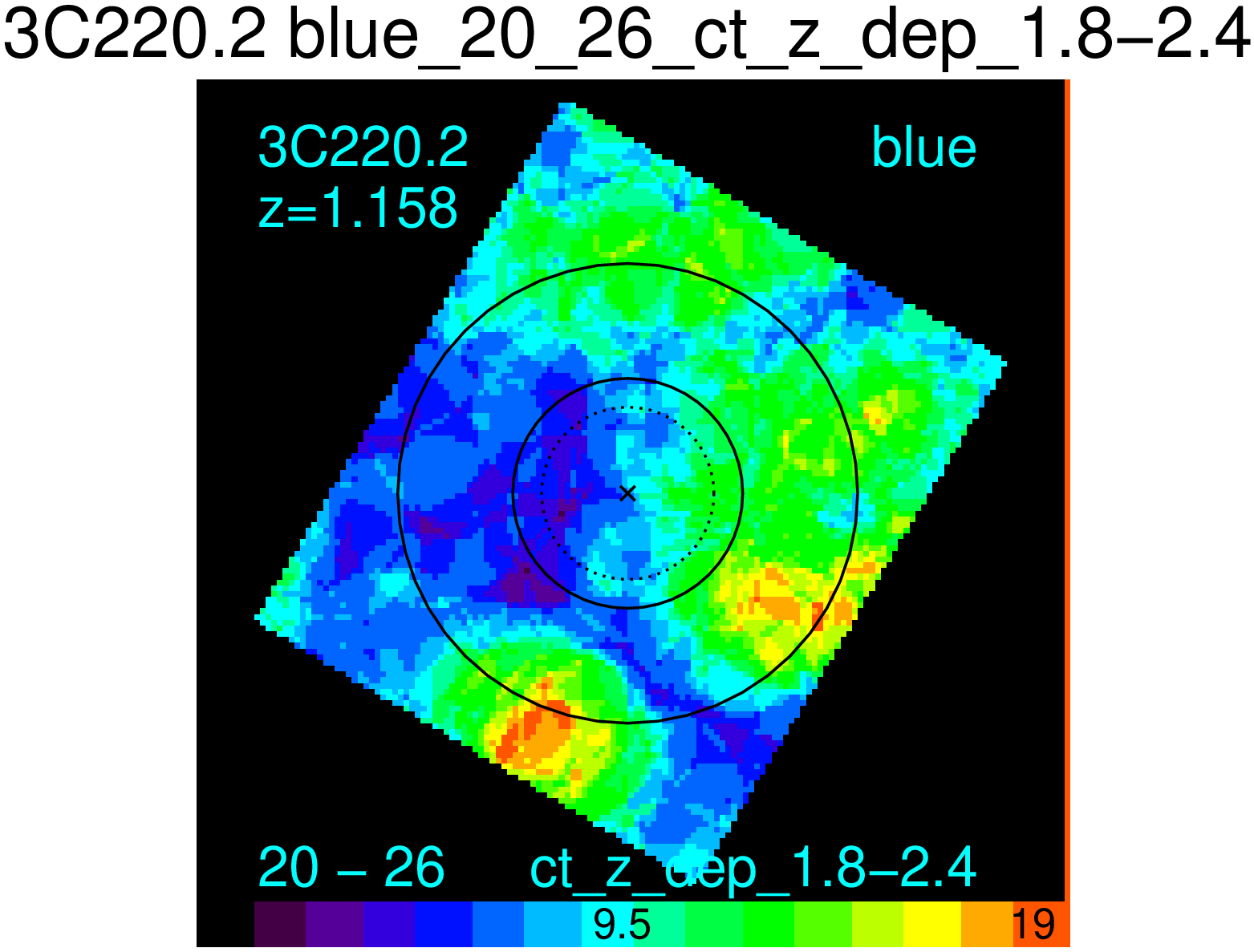}    
                
                \hspace{-0mm}\includegraphics[width=0.245\textwidth, clip=true]{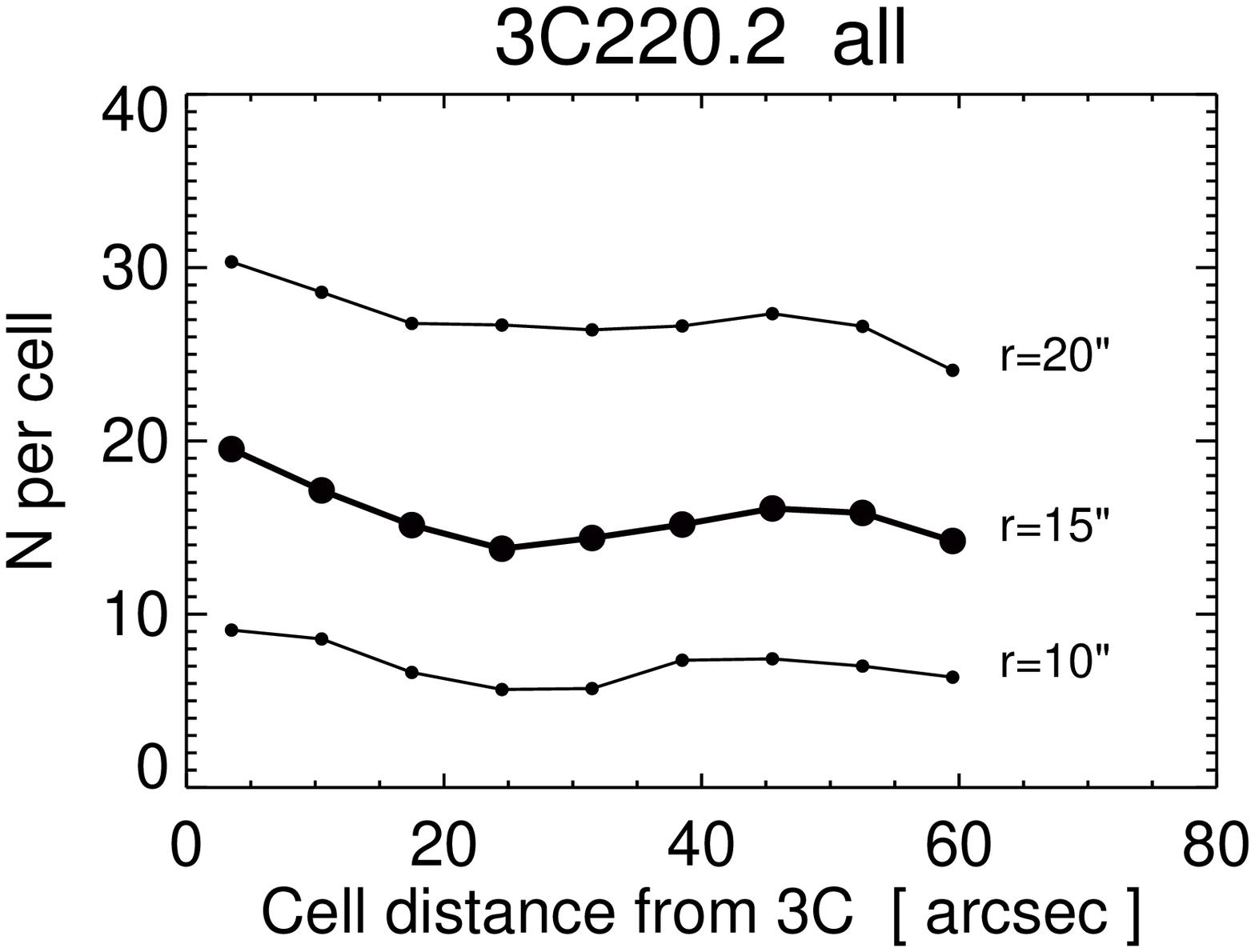}                 
                \includegraphics[width=0.245\textwidth, clip=true]{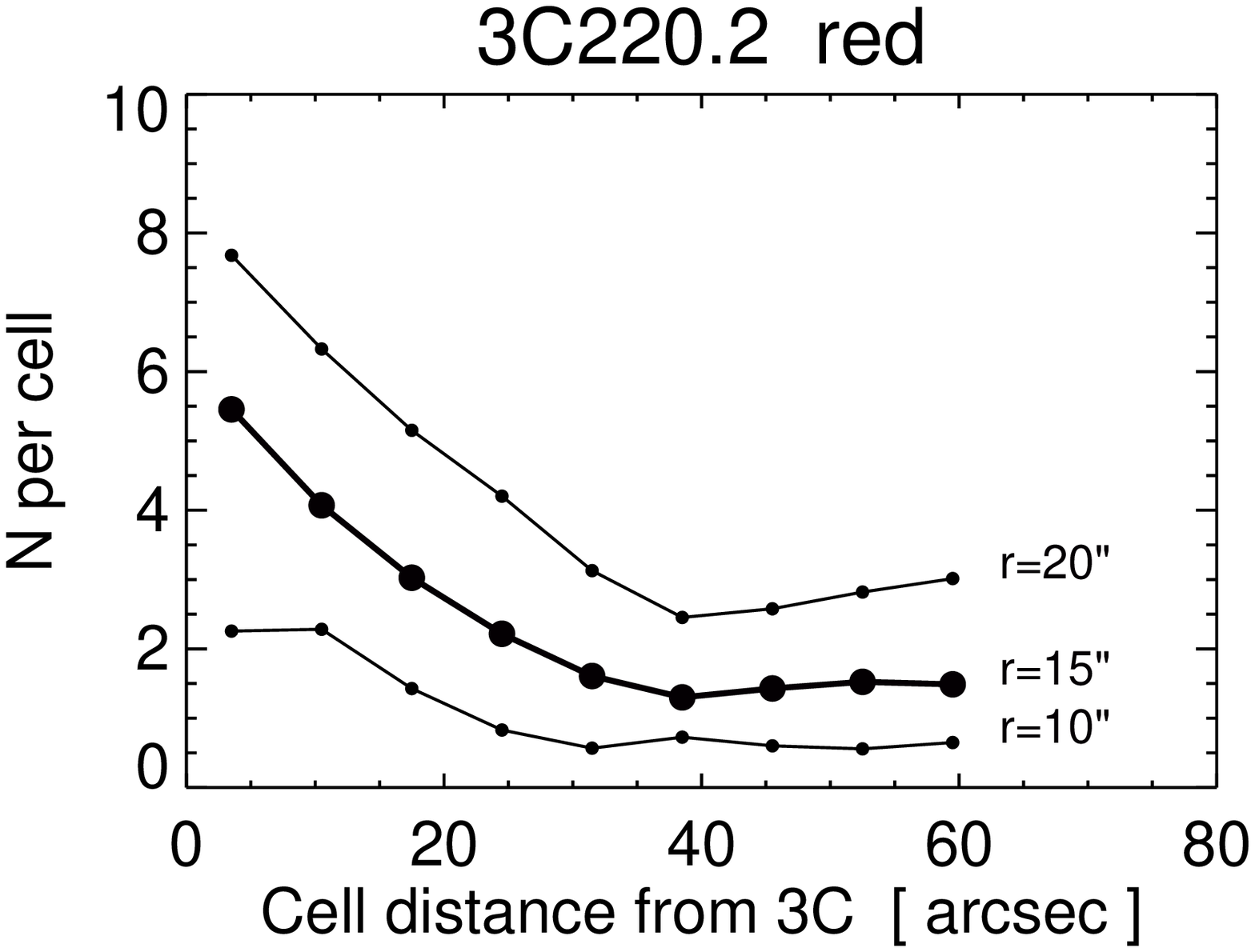}                  
                \includegraphics[width=0.245\textwidth, clip=true]{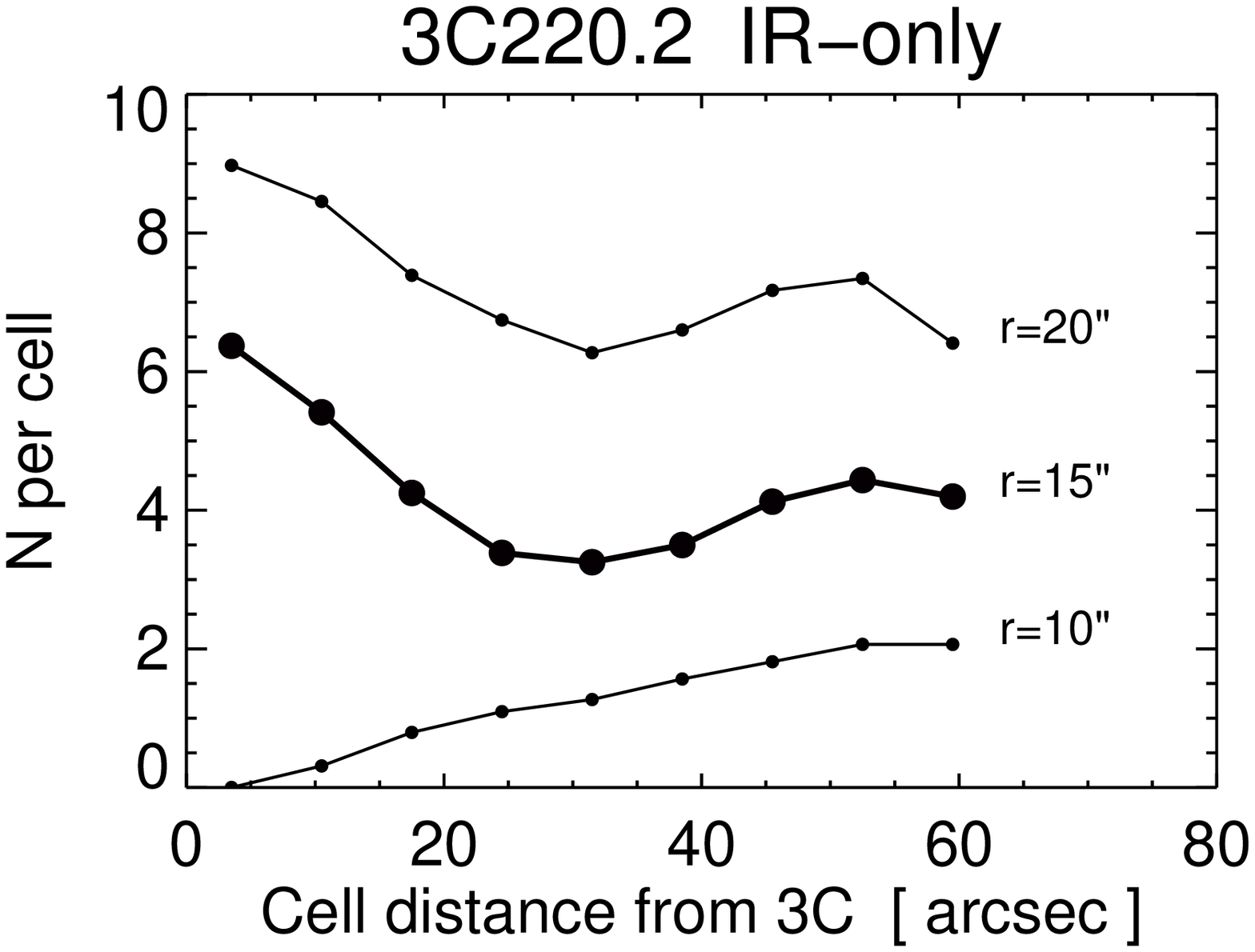}               
                \includegraphics[width=0.245\textwidth, clip=true]{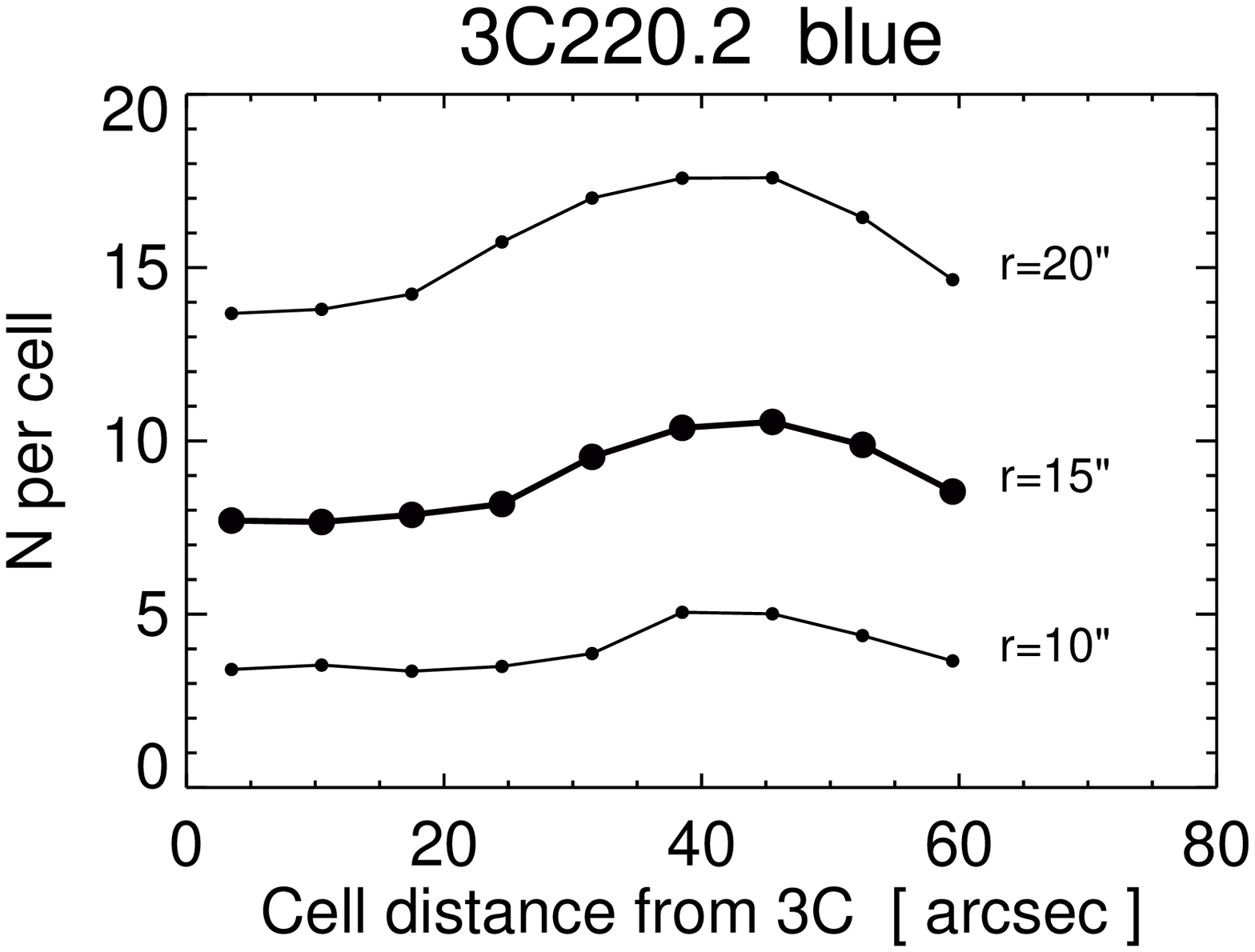}

                \hspace{-0mm}\includegraphics[width=0.245\textwidth, clip=true]{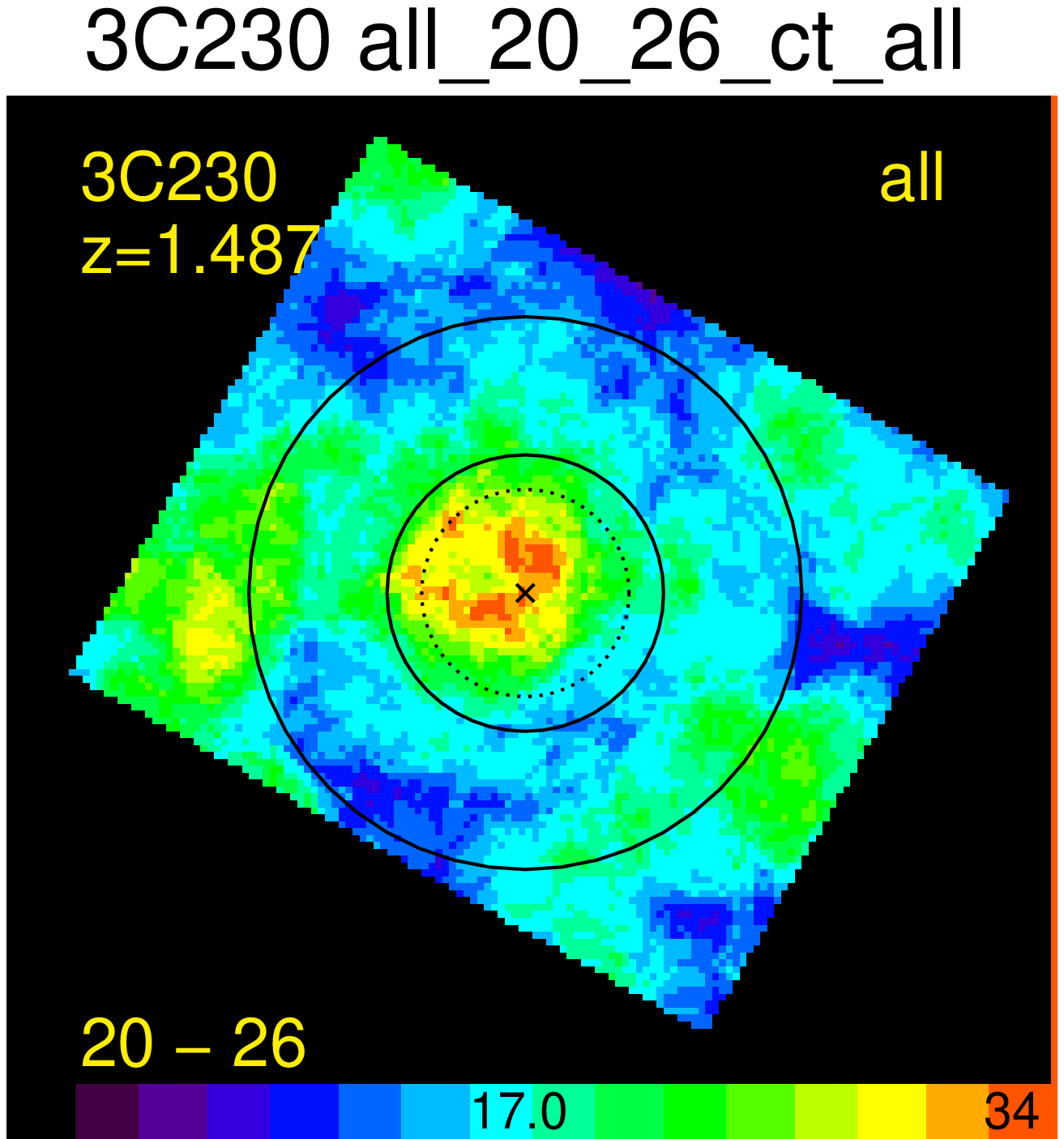}                 
                \includegraphics[width=0.245\textwidth, clip=true]{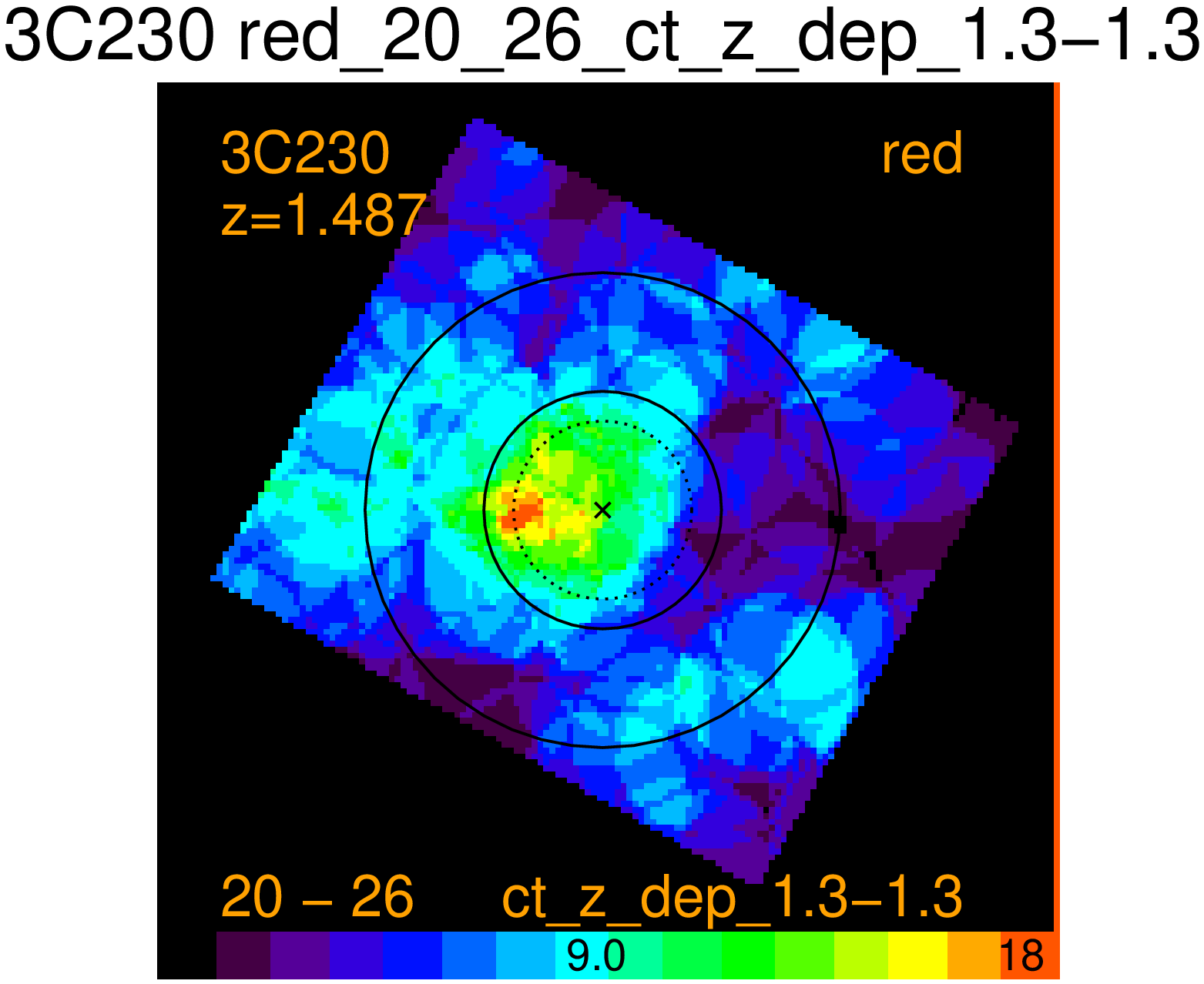}       
                \includegraphics[width=0.245\textwidth, clip=true]{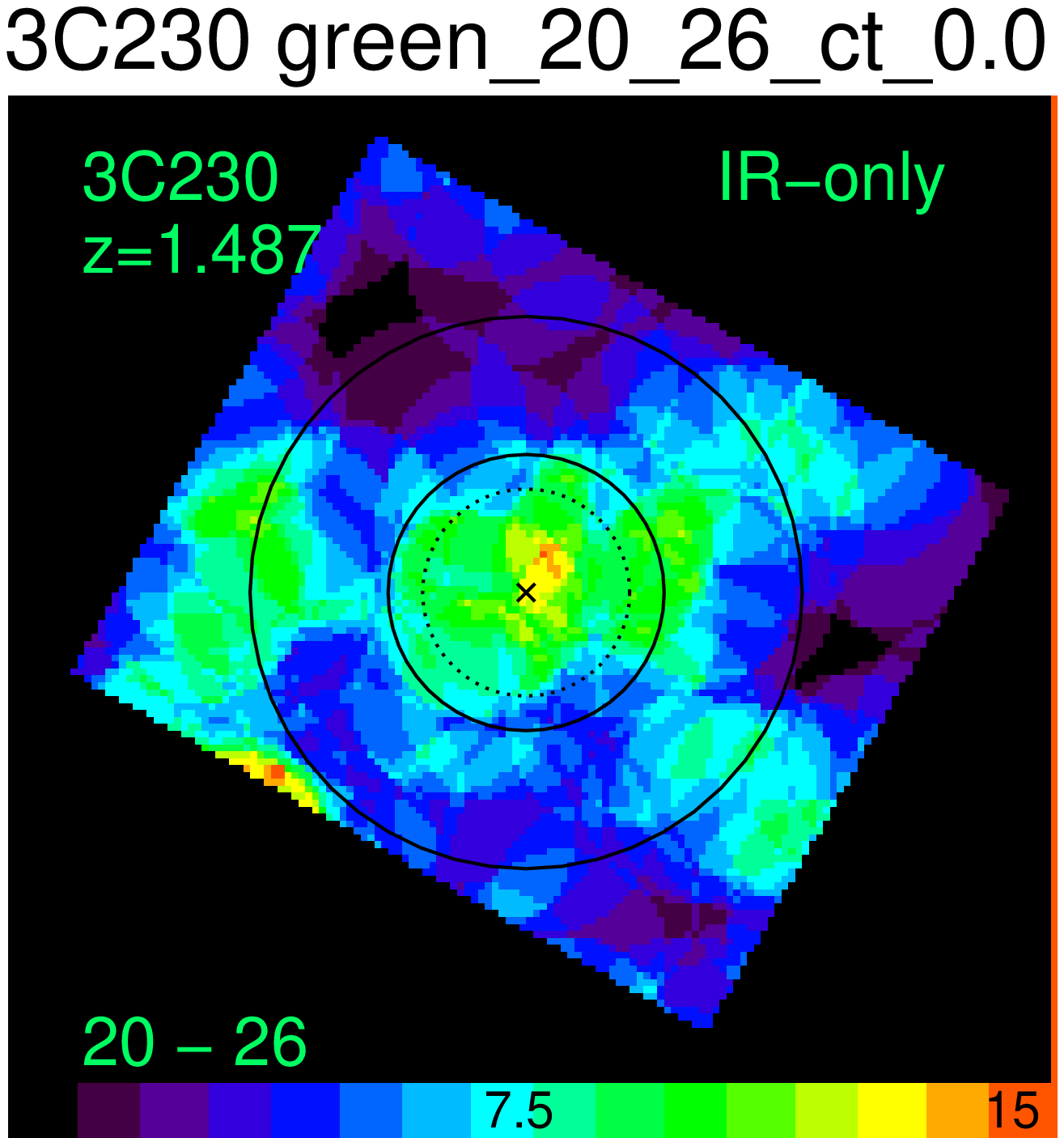}               
                \includegraphics[width=0.245\textwidth, clip=true]{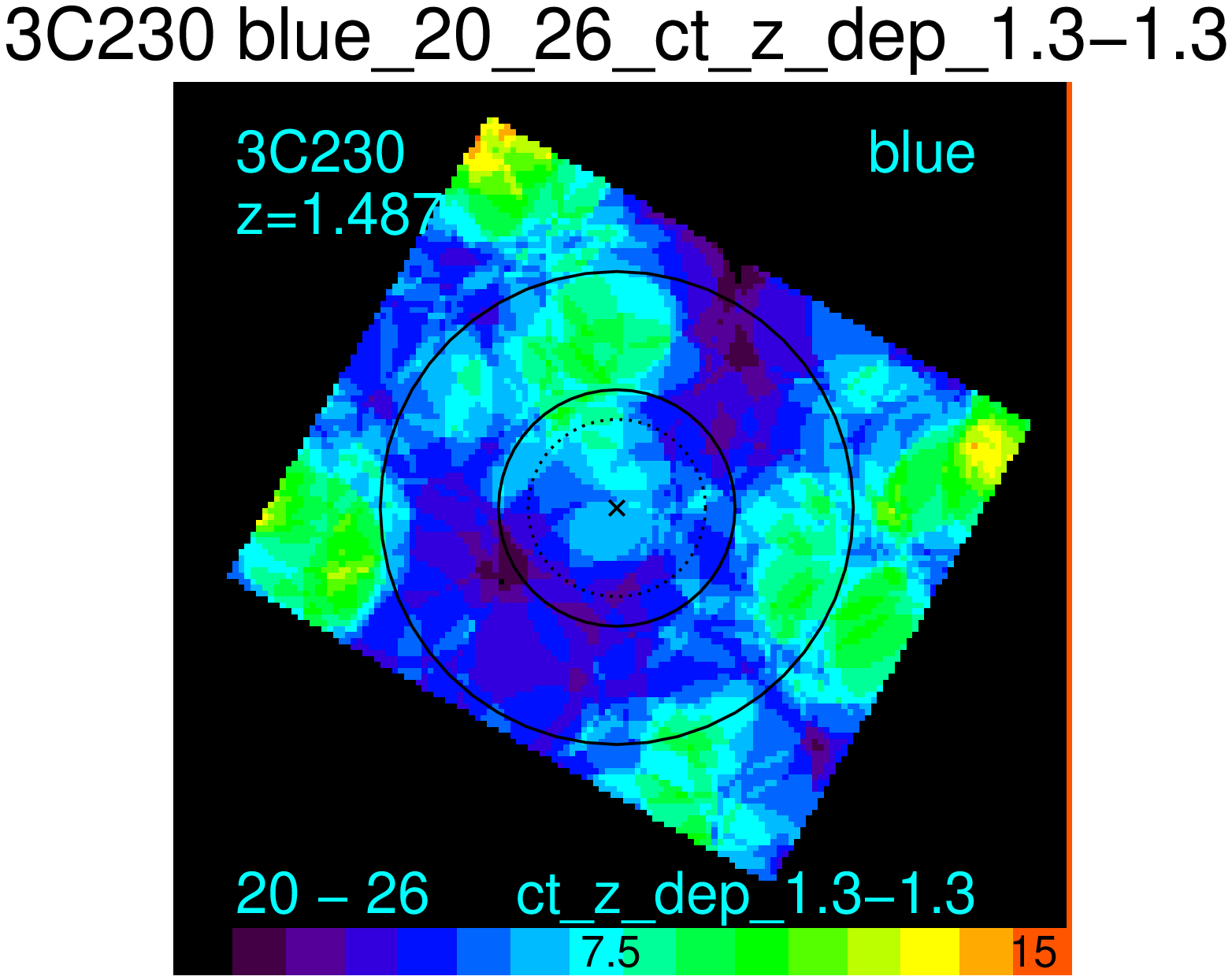}      
                
                \hspace{-0mm}\includegraphics[width=0.245\textwidth, clip=true]{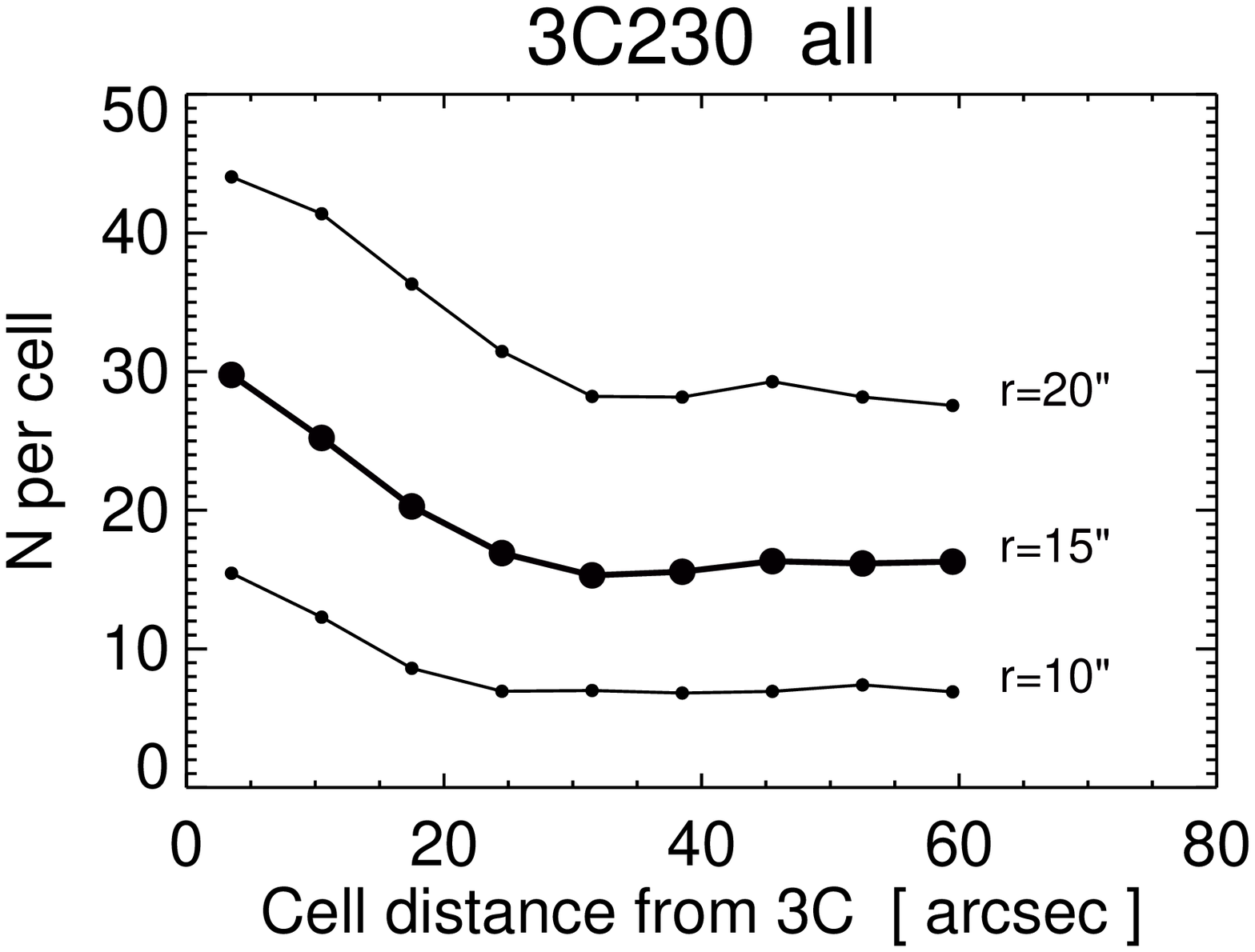}                   
                \includegraphics[width=0.245\textwidth, clip=true]{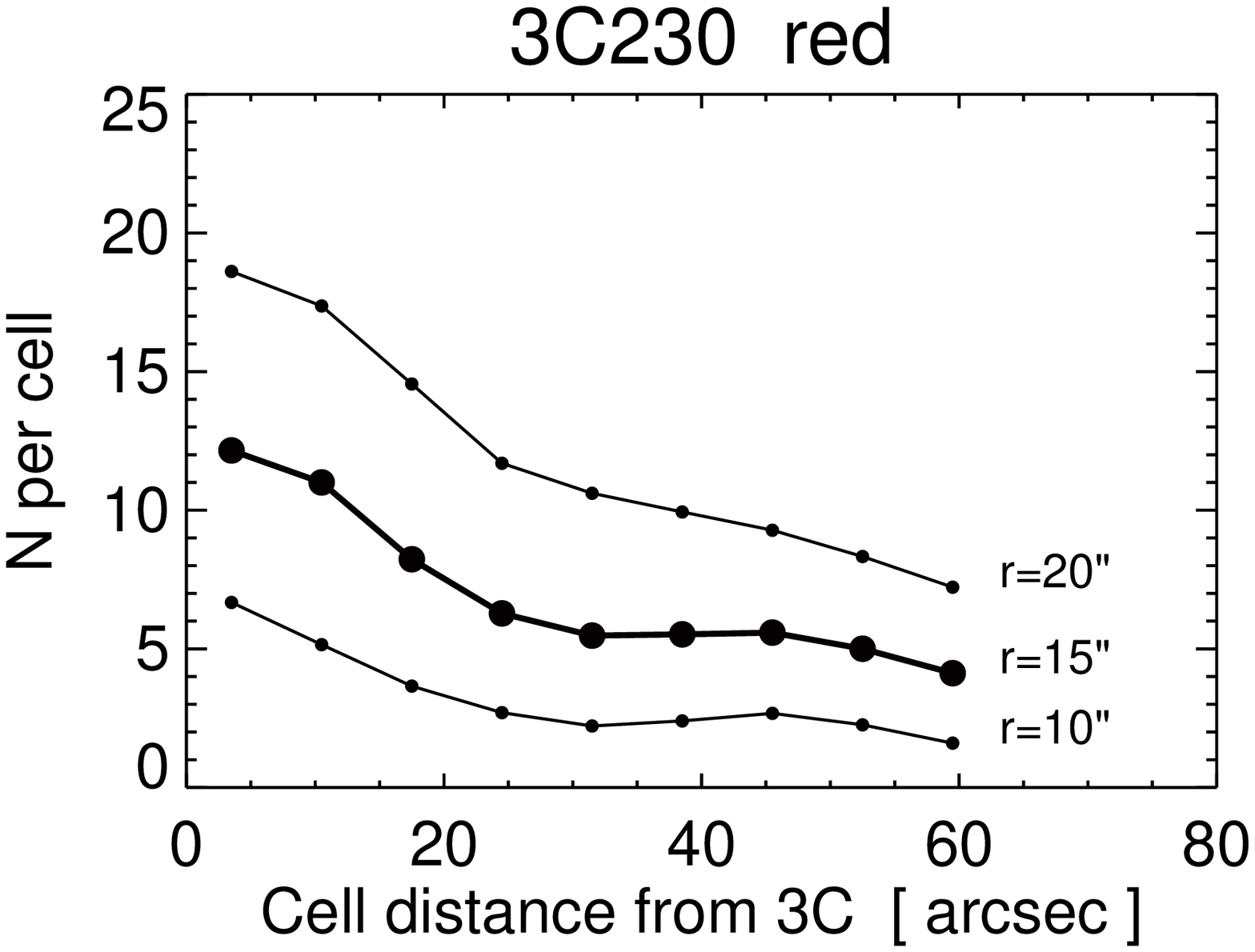}                    
                \includegraphics[width=0.245\textwidth, clip=true]{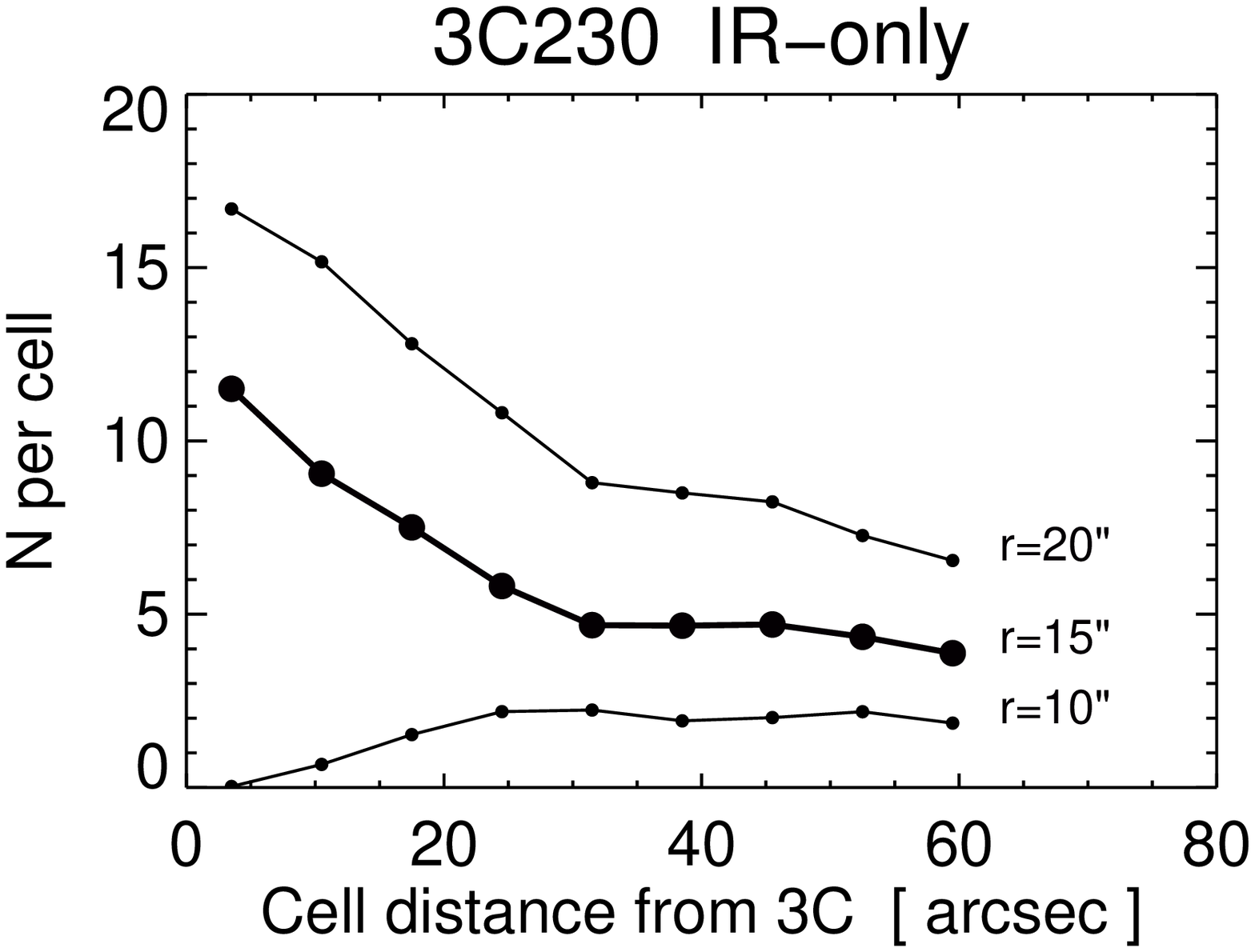}                 
                \includegraphics[width=0.245\textwidth, clip=true]{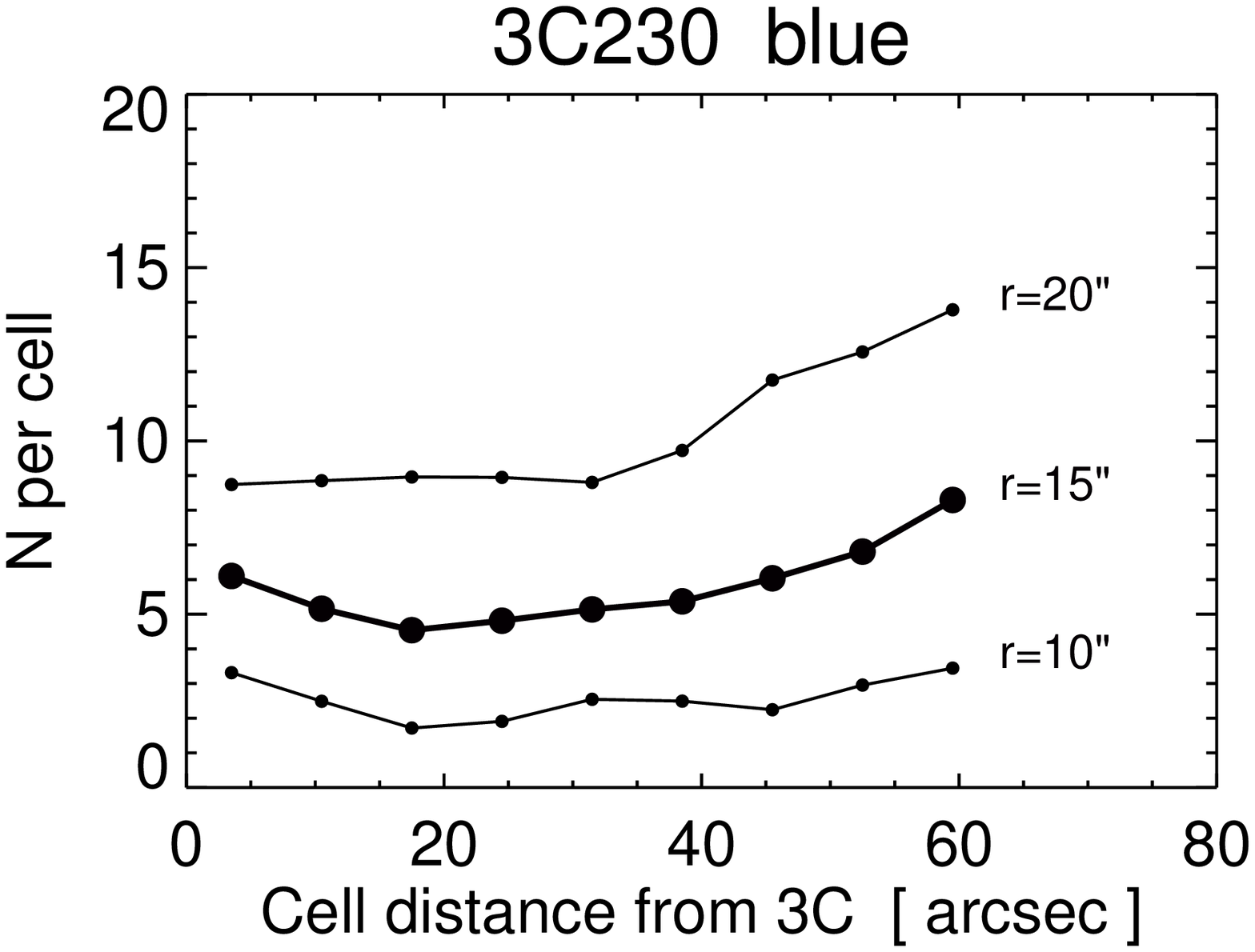}                   

                \caption{Surface density maps and radial density profiles of the 3C fields, continued.
                }
                \label{fig:sd_maps_2}
              \end{figure*}

\begin{figure*}

                \hspace{-0mm}\includegraphics[width=0.245\textwidth, clip=true]{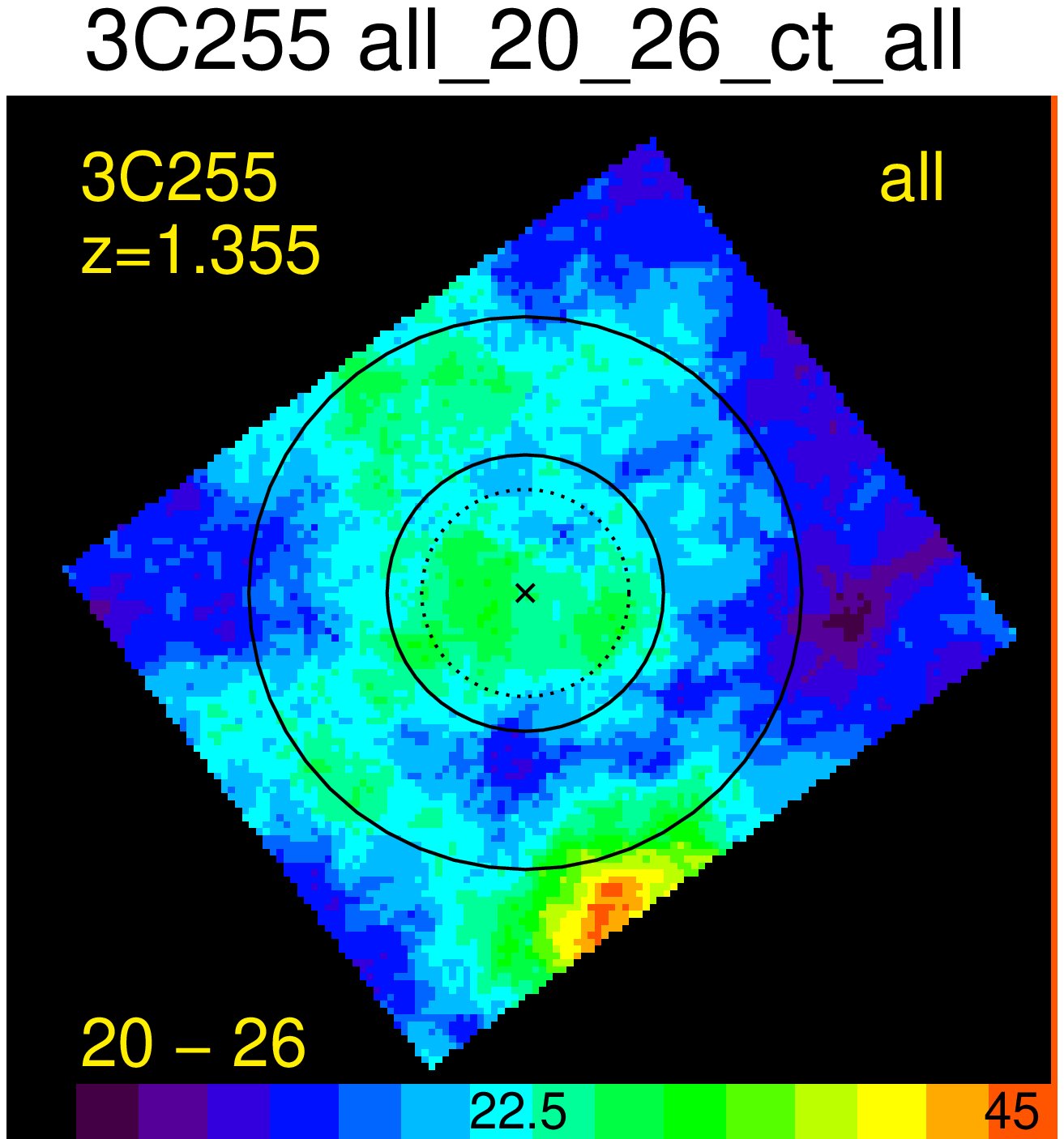}                 
                \includegraphics[width=0.245\textwidth, clip=true]{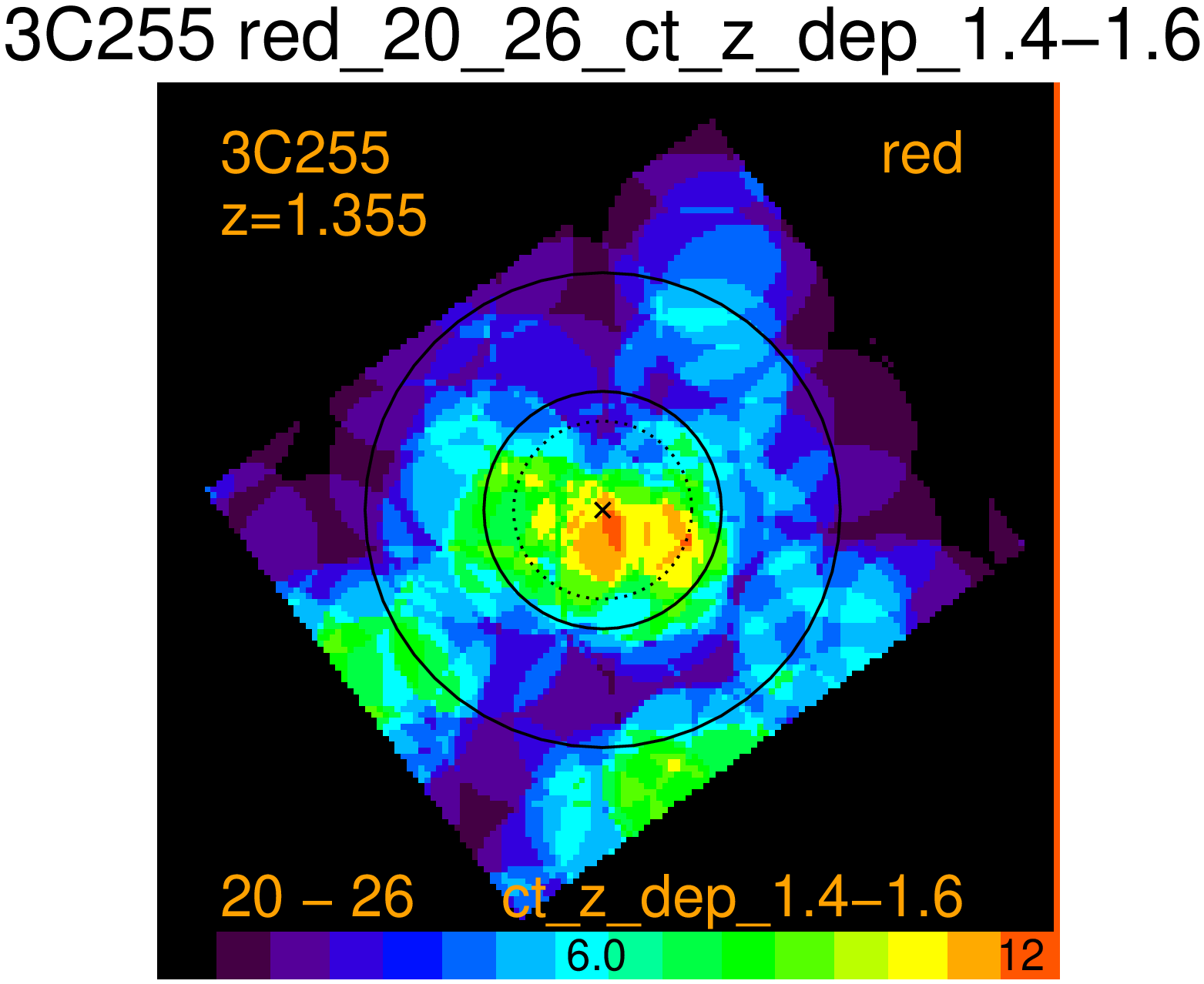}       
                \includegraphics[width=0.245\textwidth, clip=true]{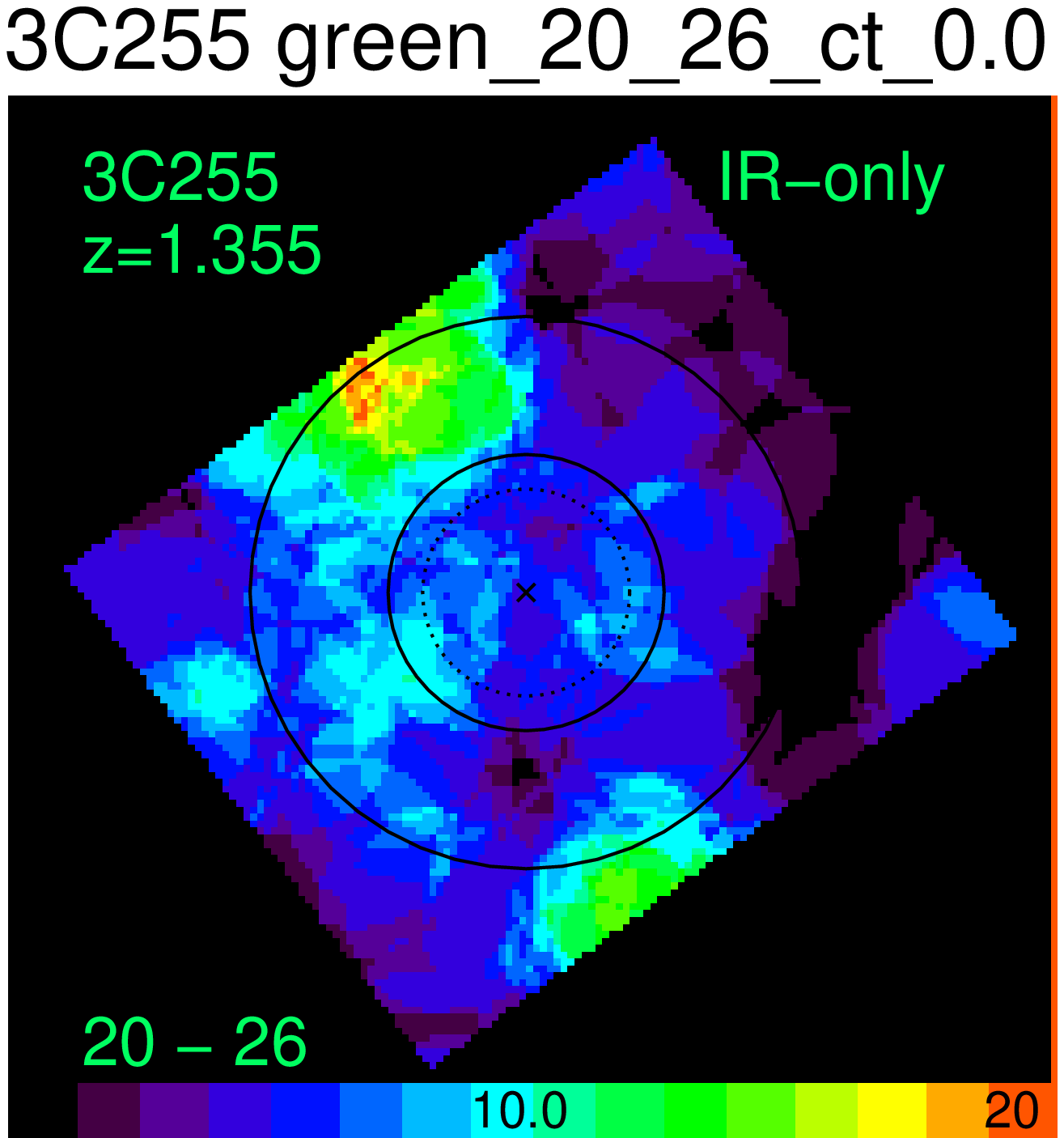}               
                \includegraphics[width=0.245\textwidth, clip=true]{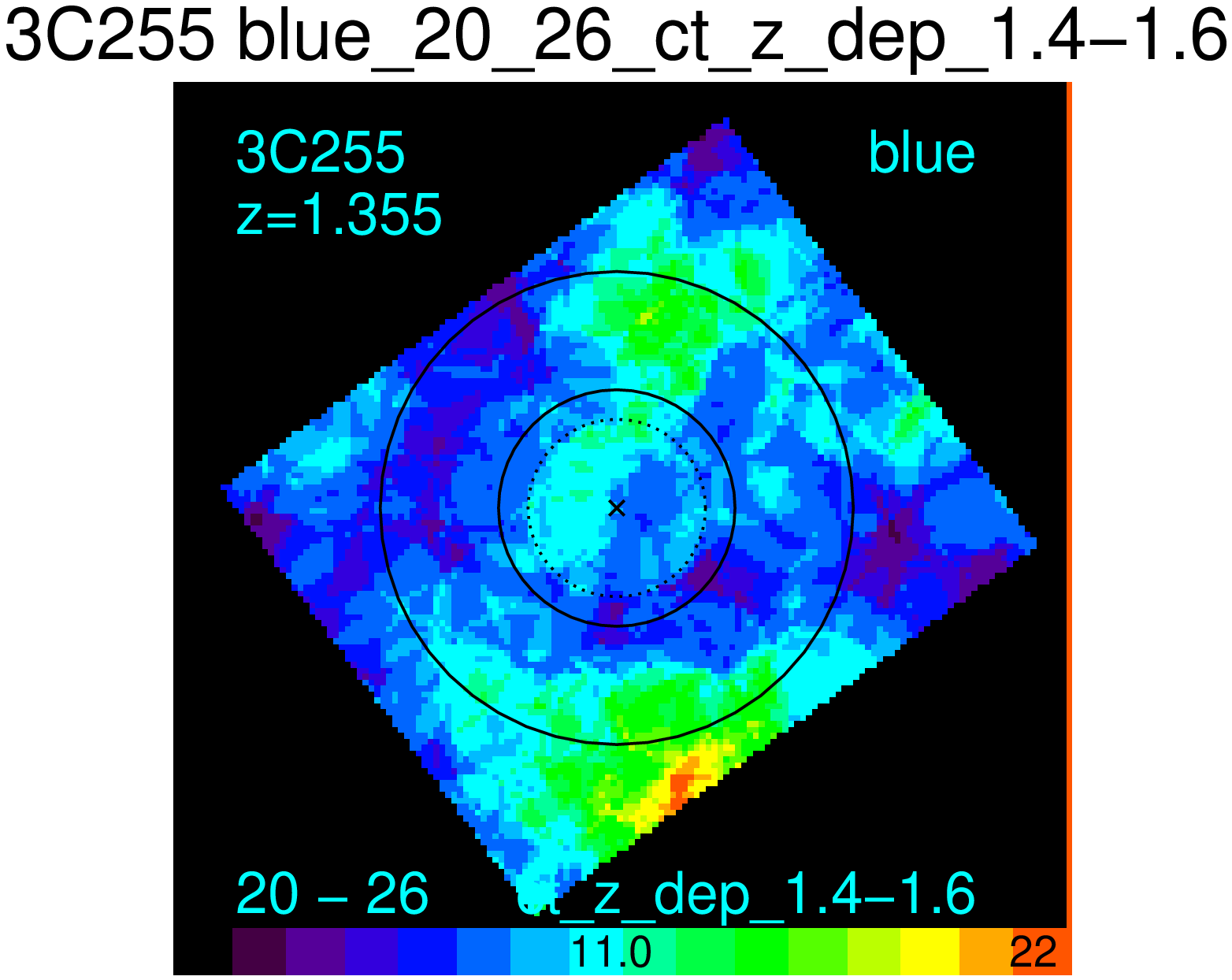}      
                
                \hspace{-0mm}\includegraphics[width=0.245\textwidth, clip=true]{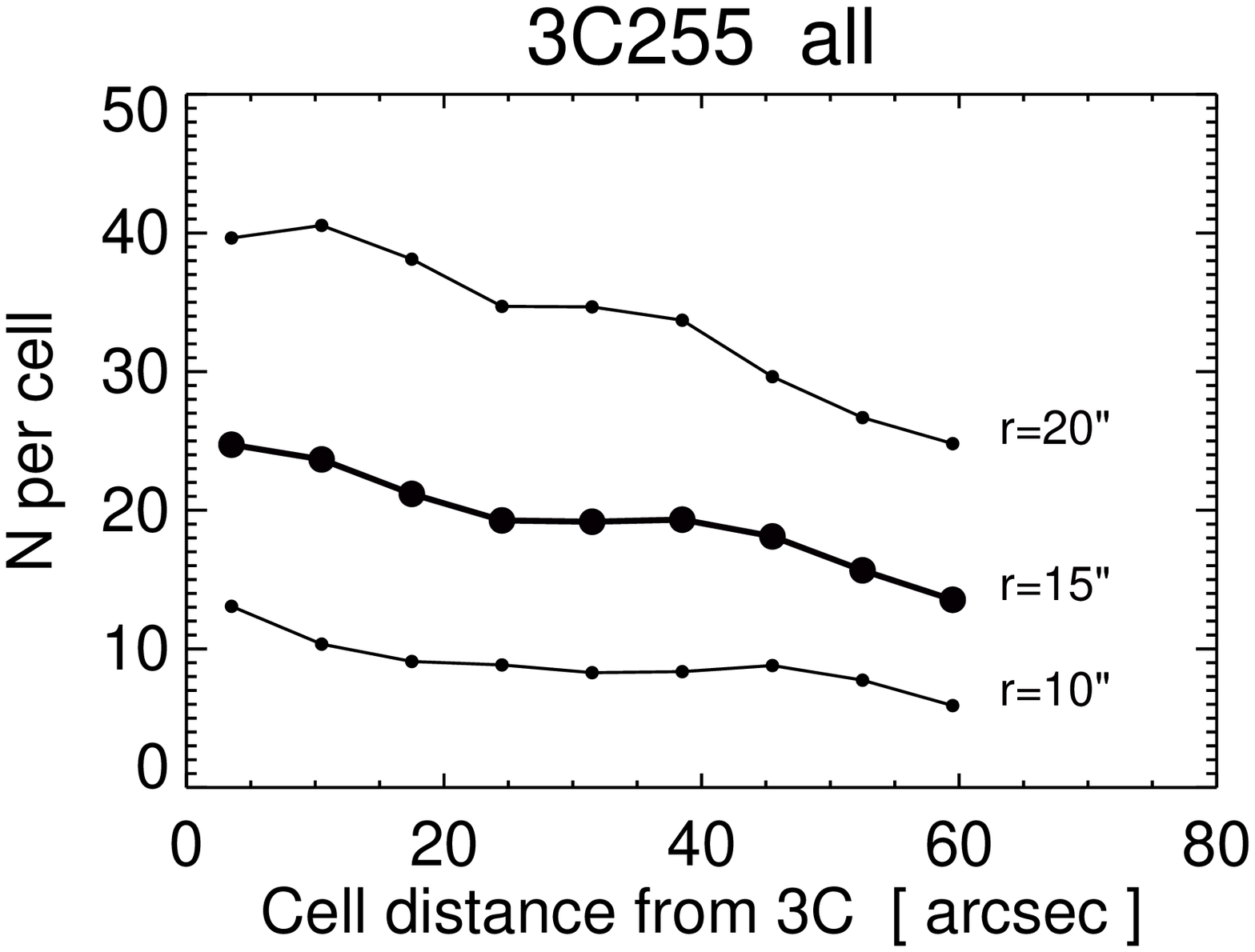}                   
                \includegraphics[width=0.245\textwidth, clip=true]{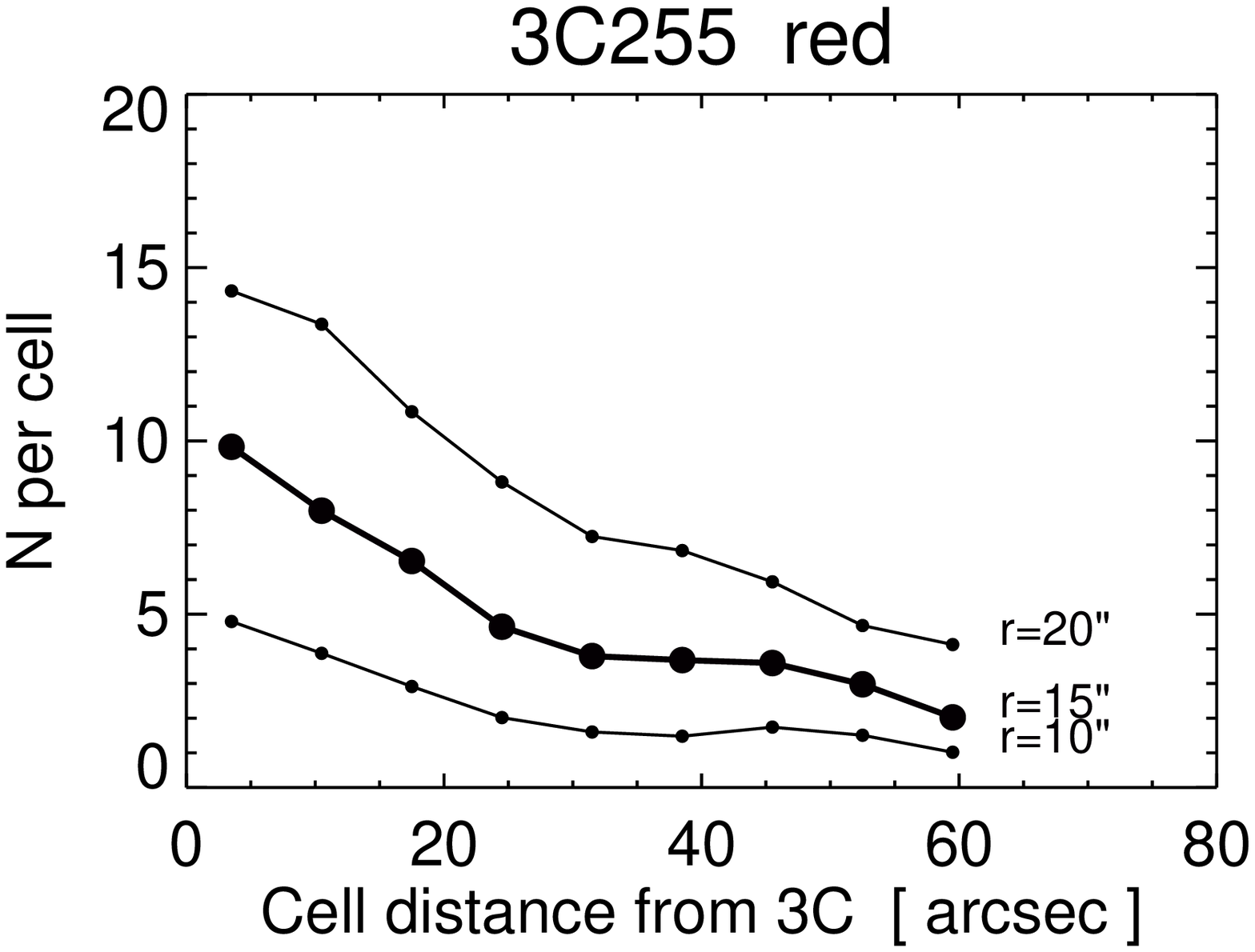}                    
                \includegraphics[width=0.245\textwidth, clip=true]{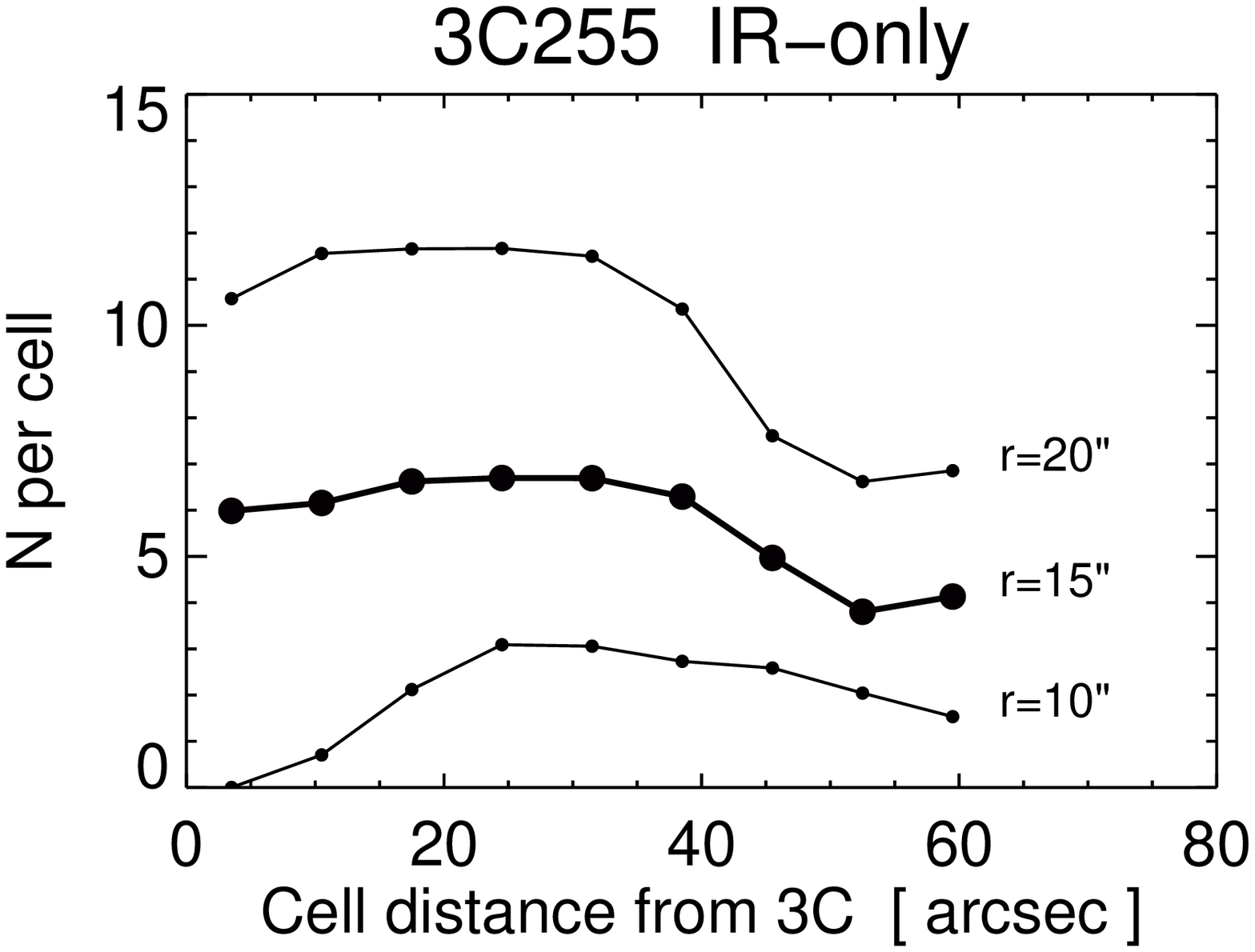}                 
                \includegraphics[width=0.245\textwidth, clip=true]{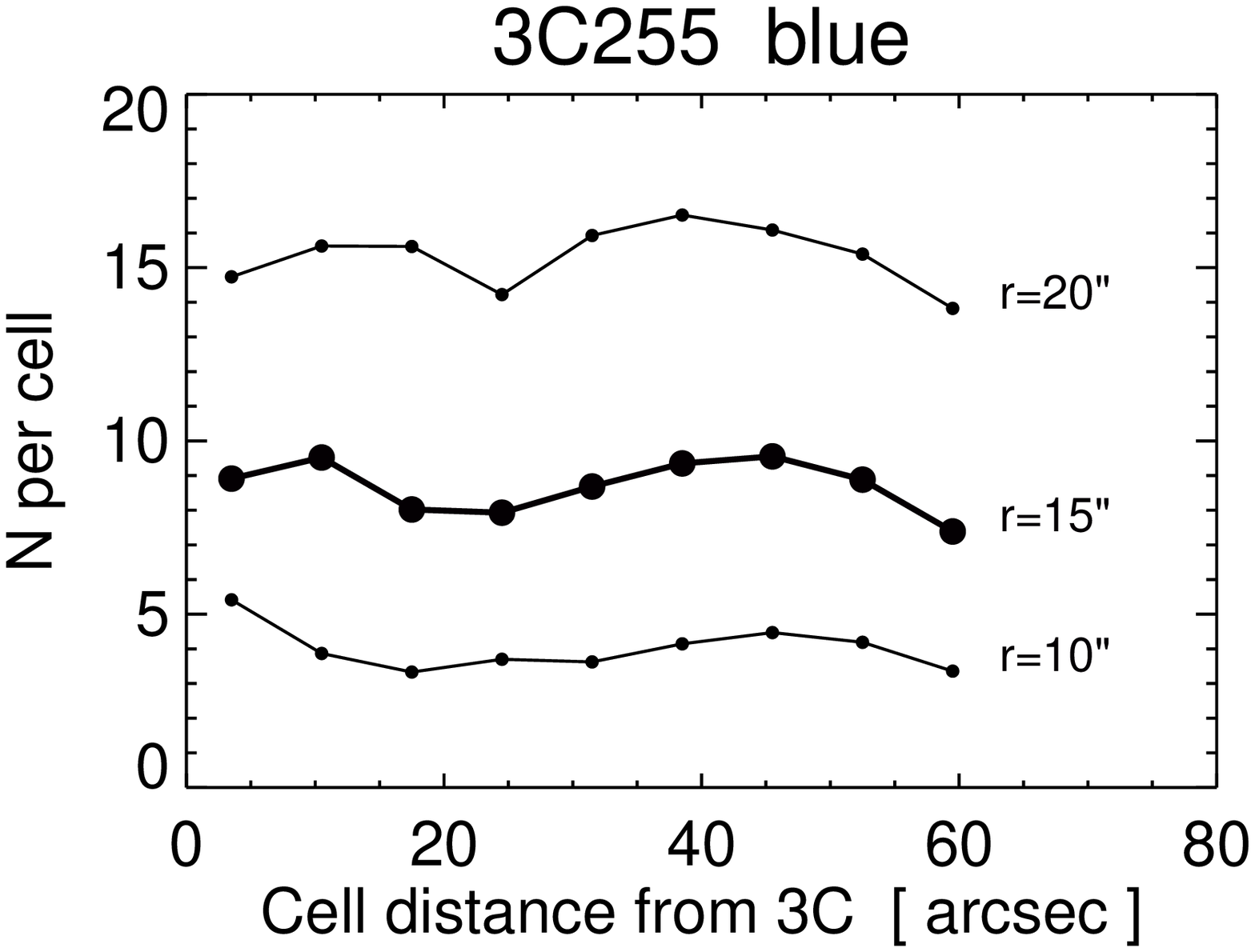}

                \hspace{-0mm}\includegraphics[width=0.245\textwidth, clip=true]{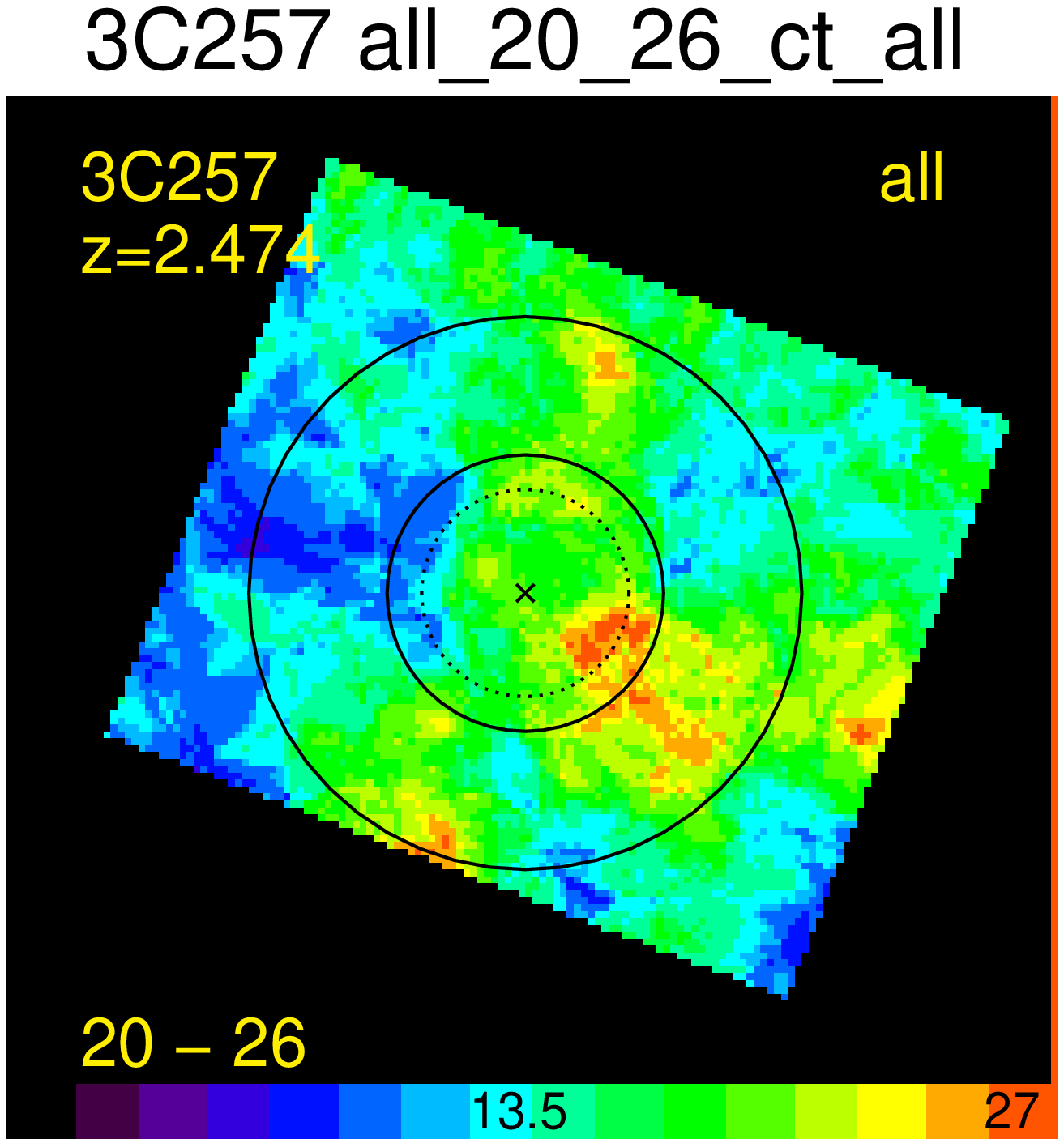}                 
                \includegraphics[width=0.245\textwidth, clip=true]{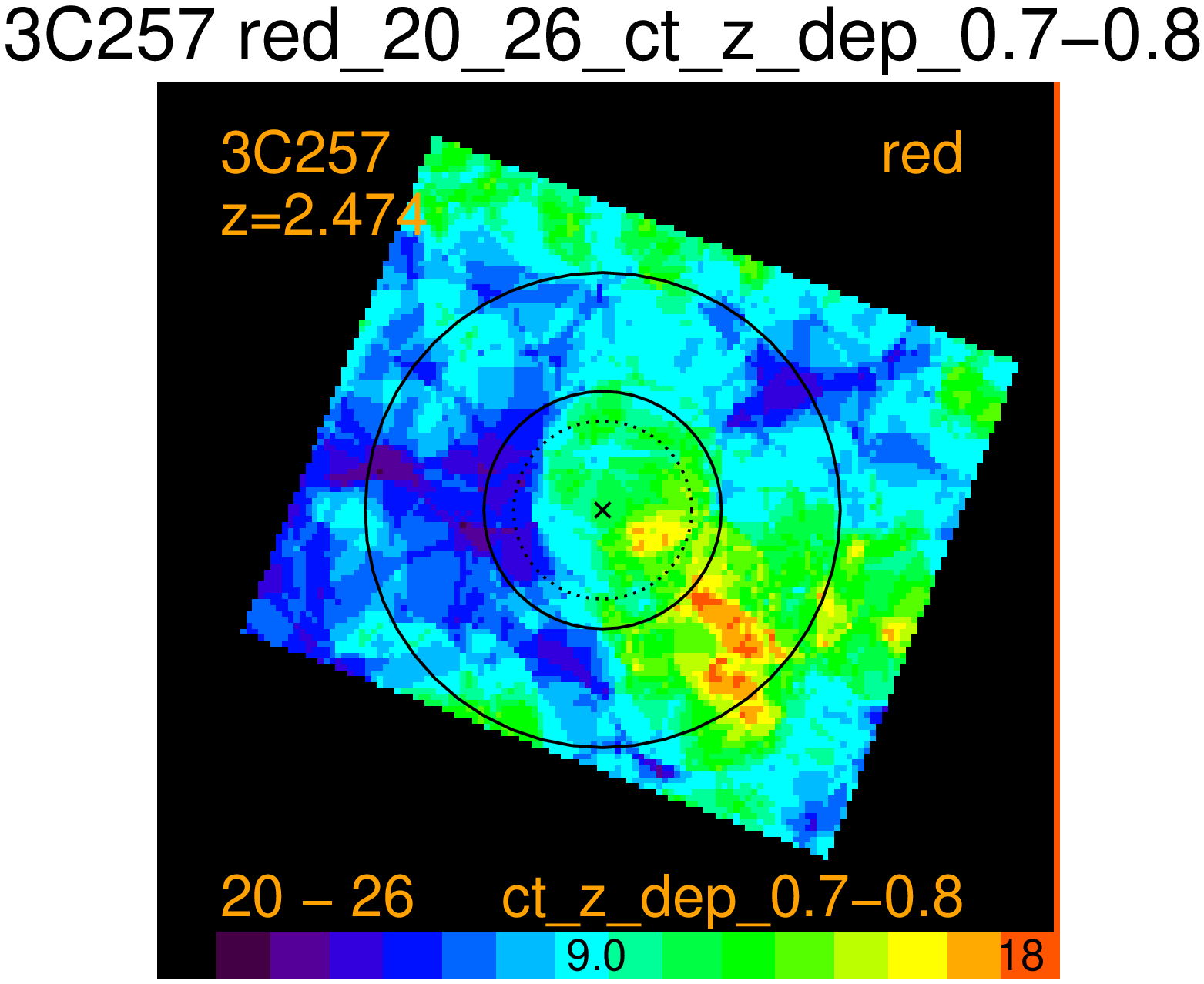}       
                \includegraphics[width=0.245\textwidth, clip=true]{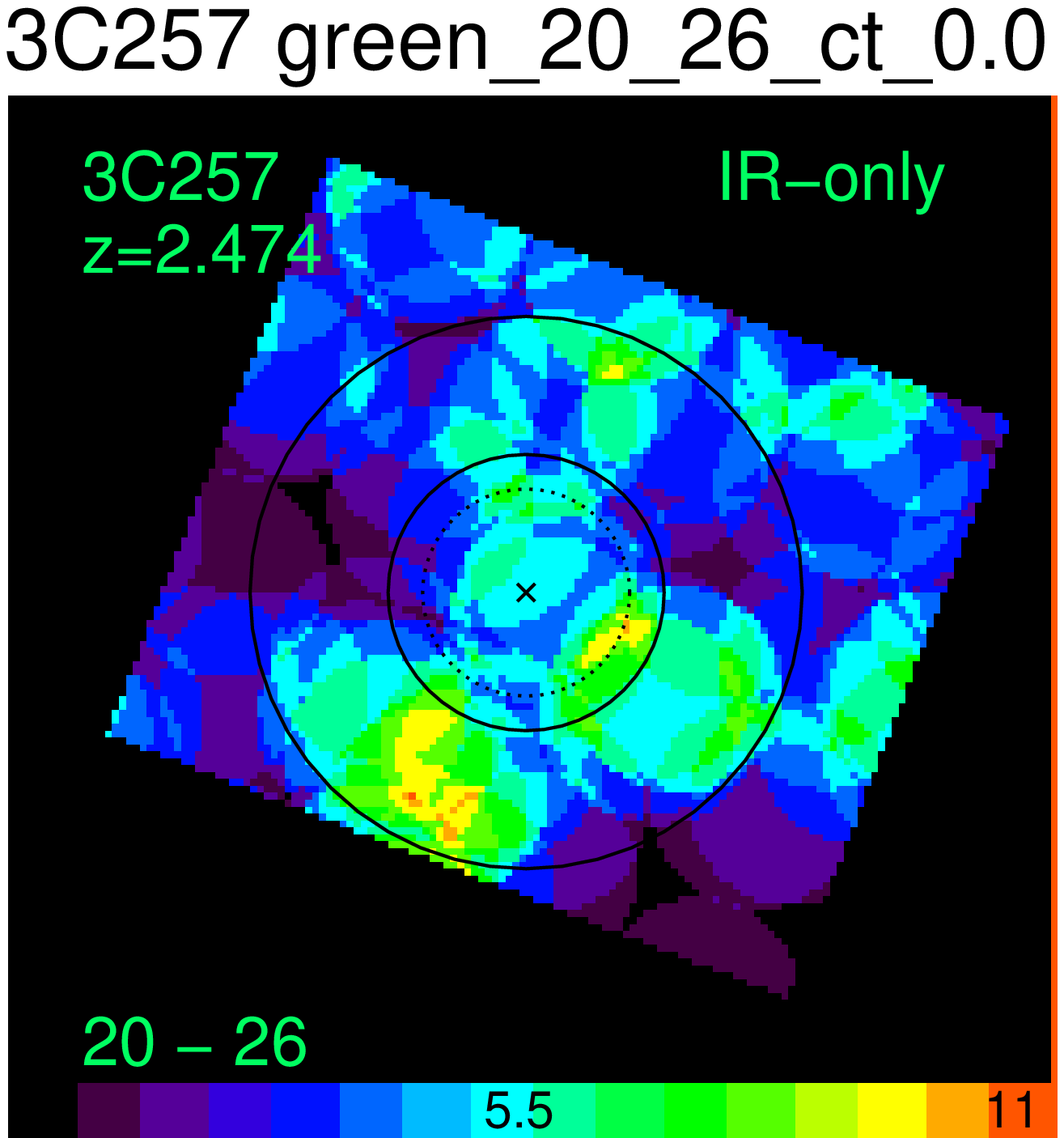}               
                \includegraphics[width=0.245\textwidth, clip=true]{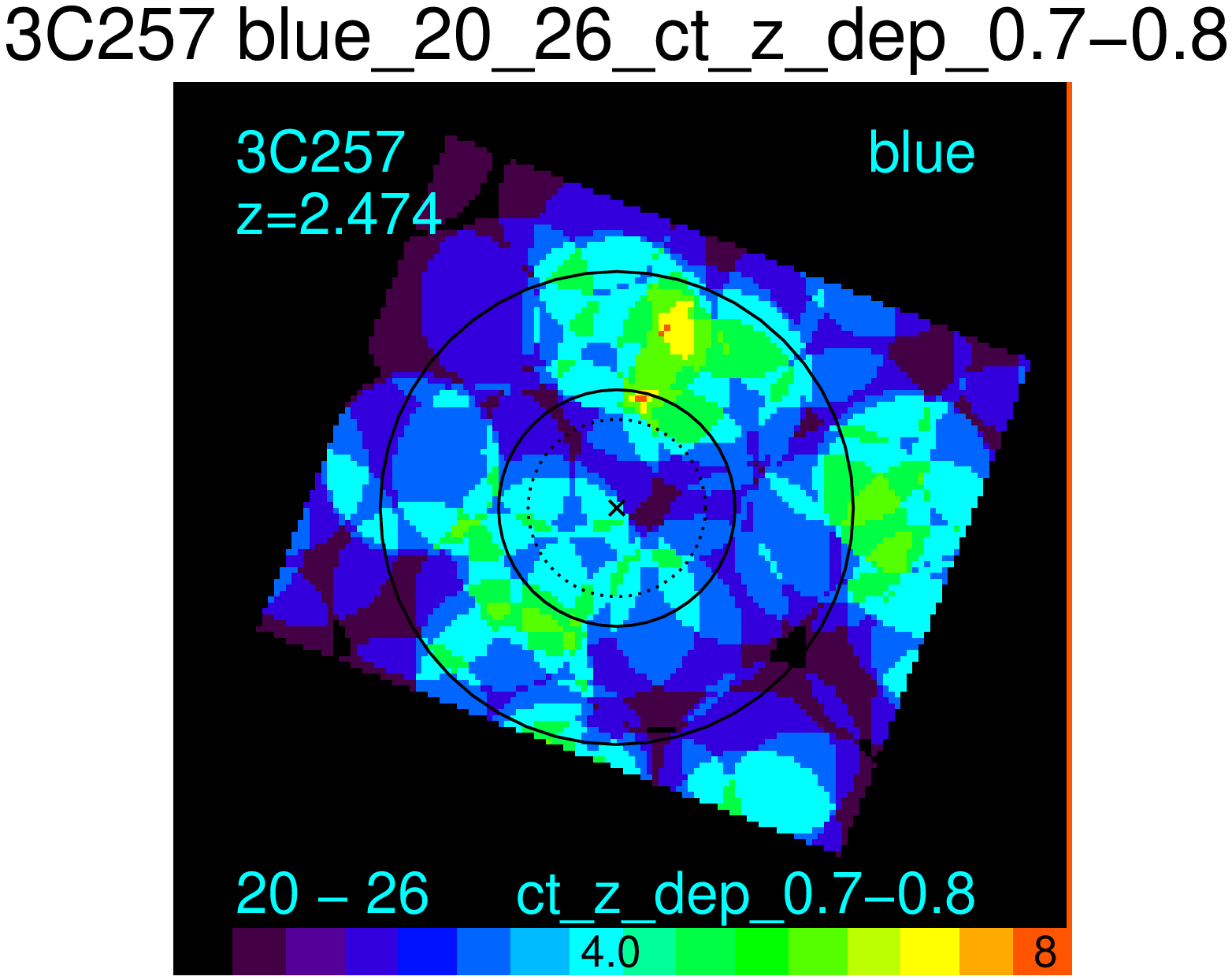}      
                
                \hspace{-0mm}\includegraphics[width=0.245\textwidth, clip=true]{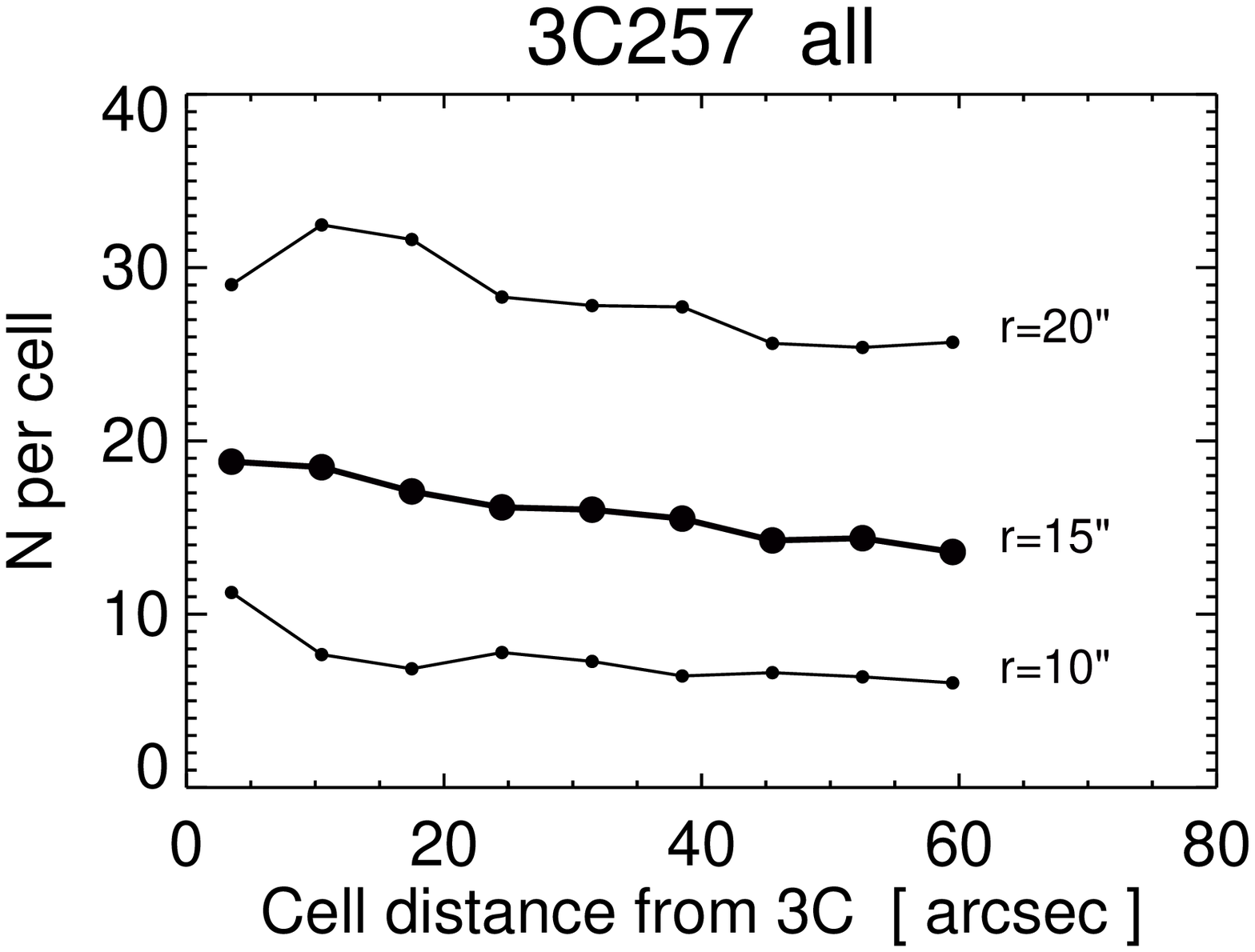}                   
                \includegraphics[width=0.245\textwidth, clip=true]{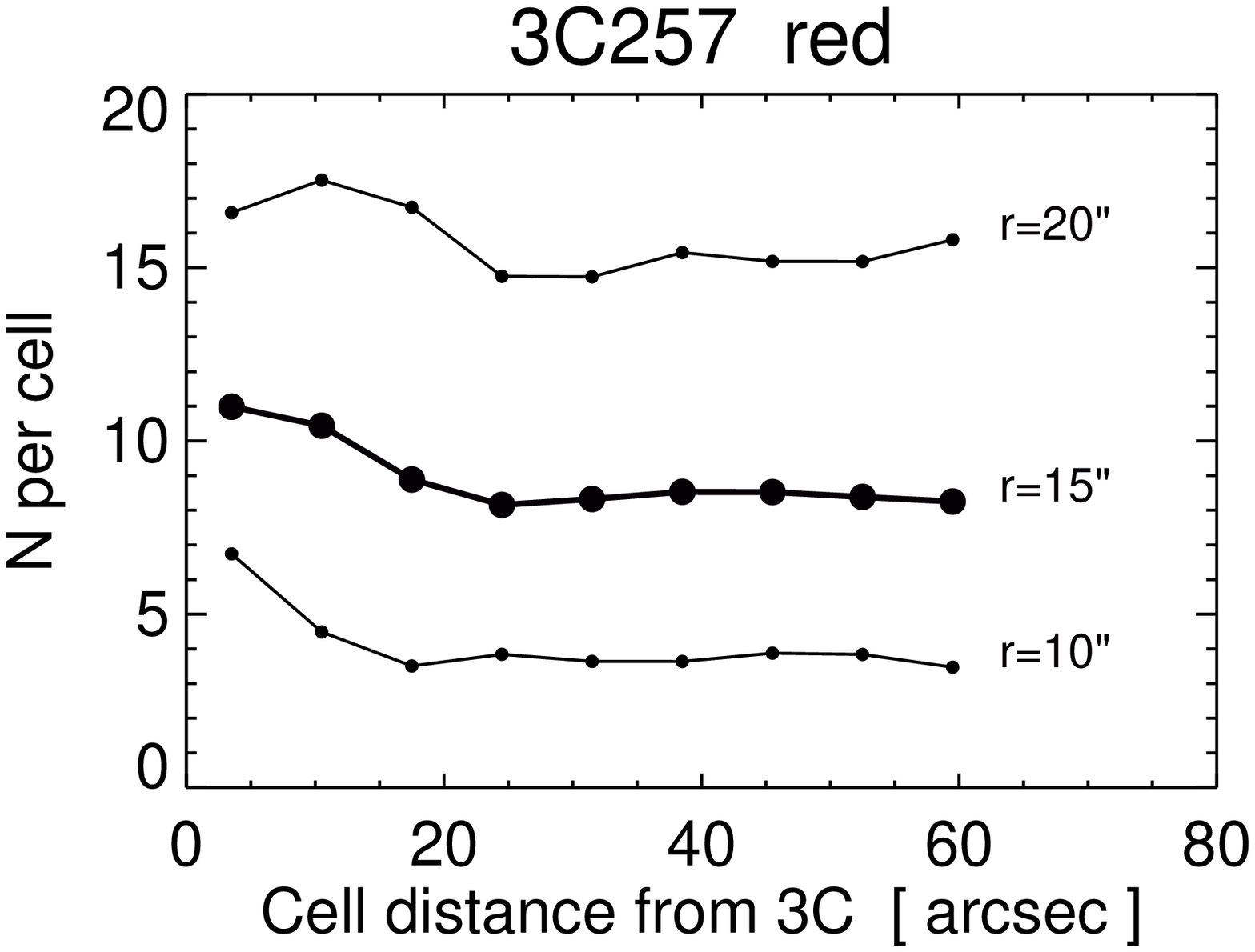}                    
                \includegraphics[width=0.245\textwidth, clip=true]{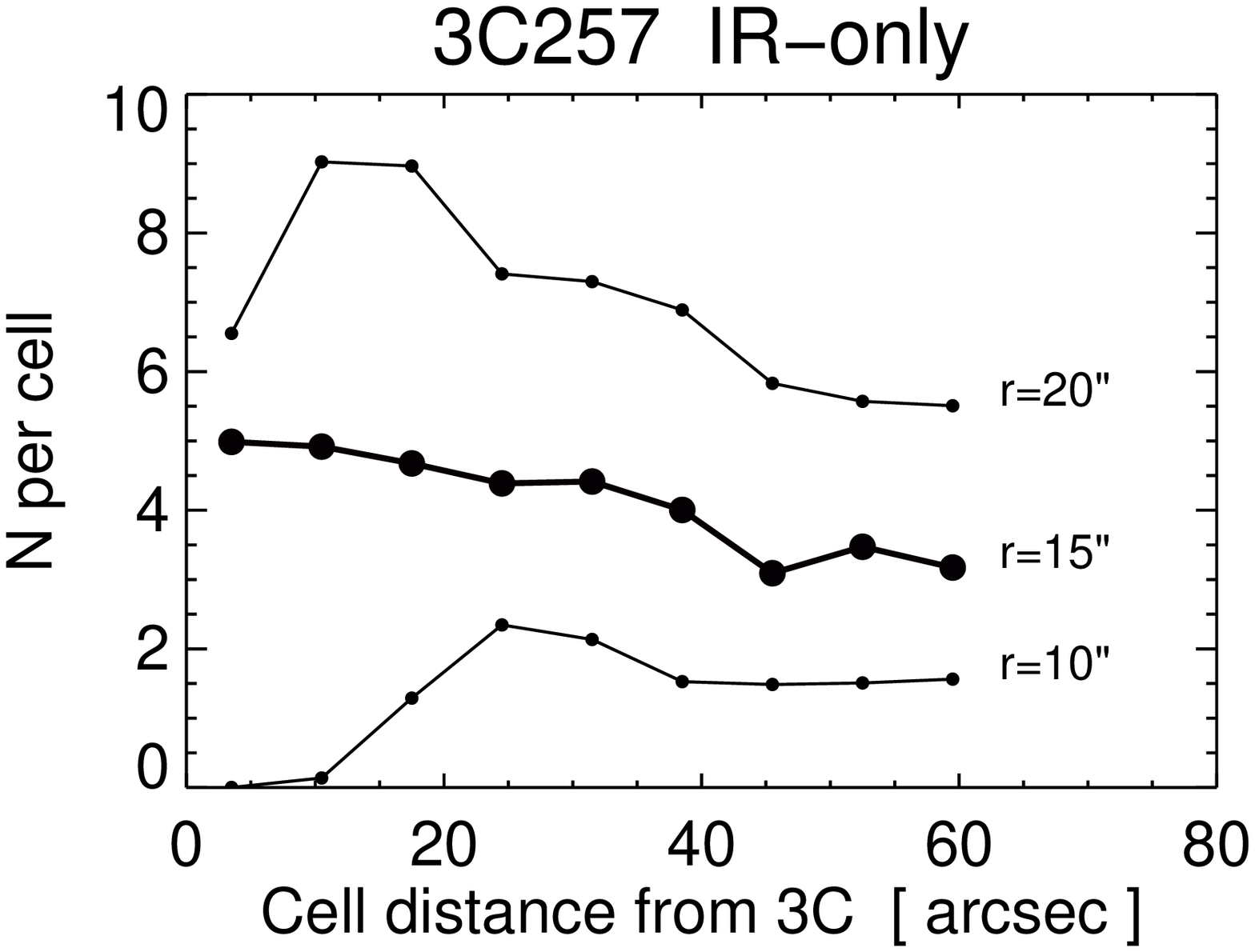}                 
                \includegraphics[width=0.245\textwidth, clip=true]{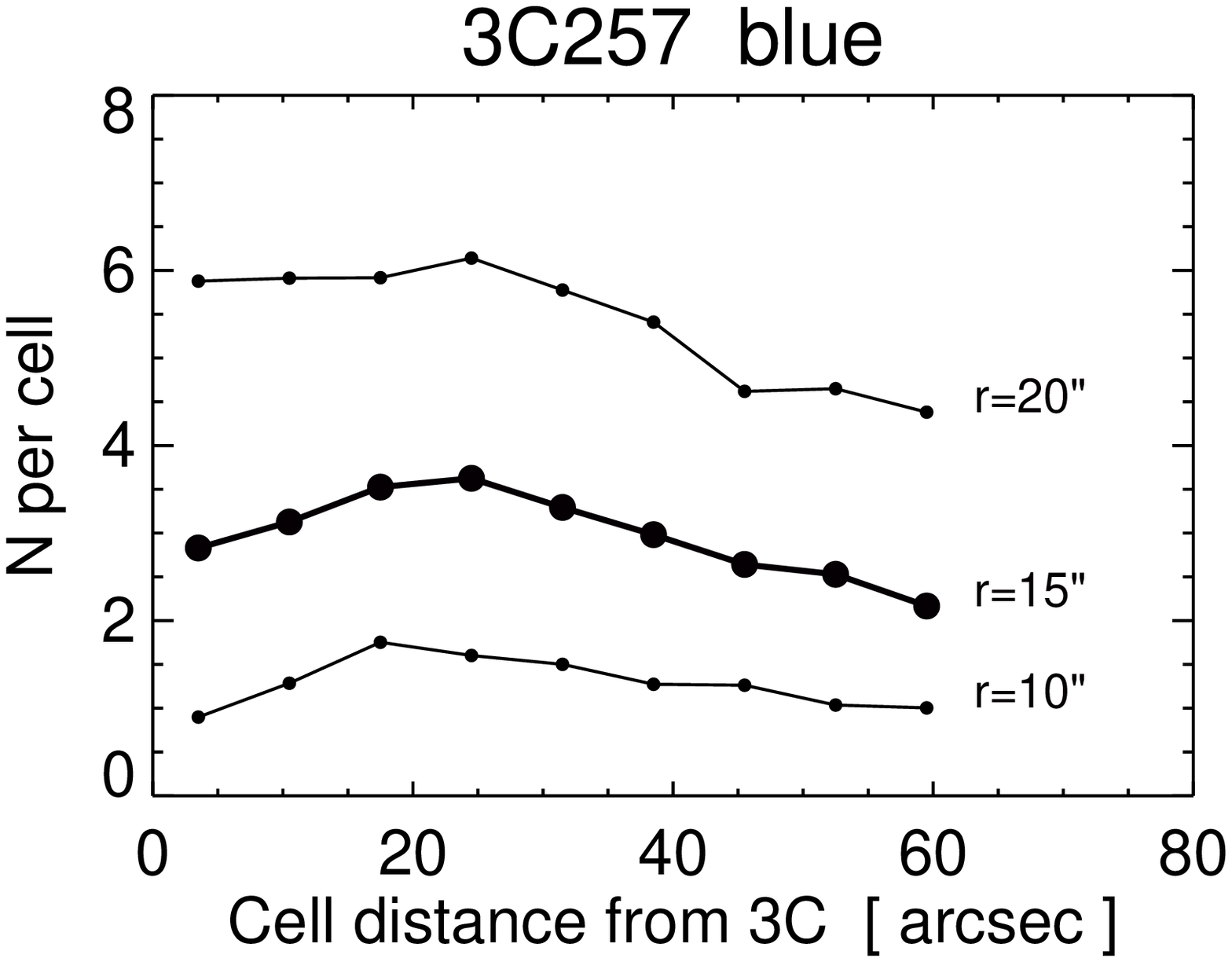}

                \hspace{-0mm}\includegraphics[width=0.245\textwidth, clip=true]{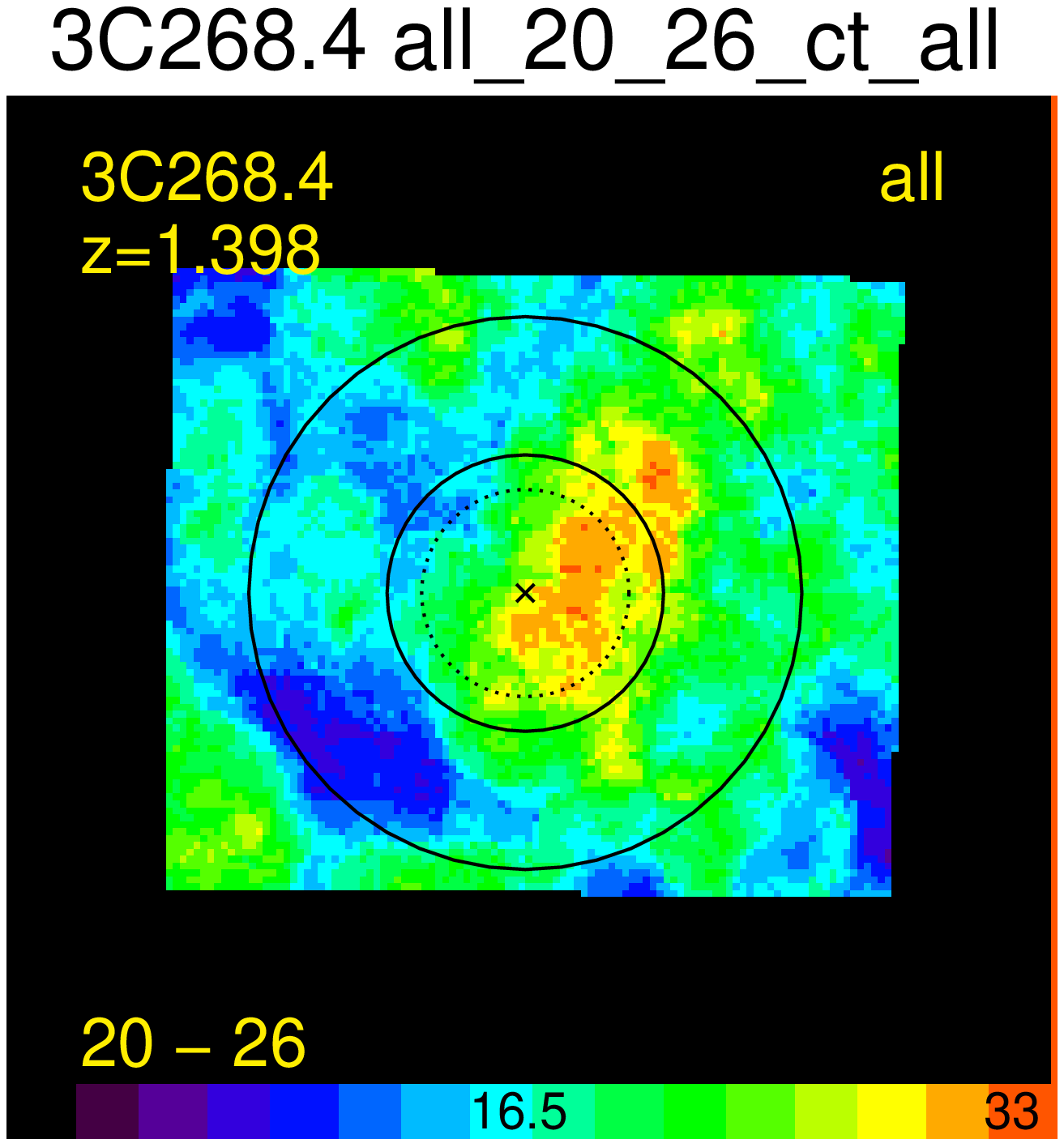}               
                \includegraphics[width=0.245\textwidth, clip=true]{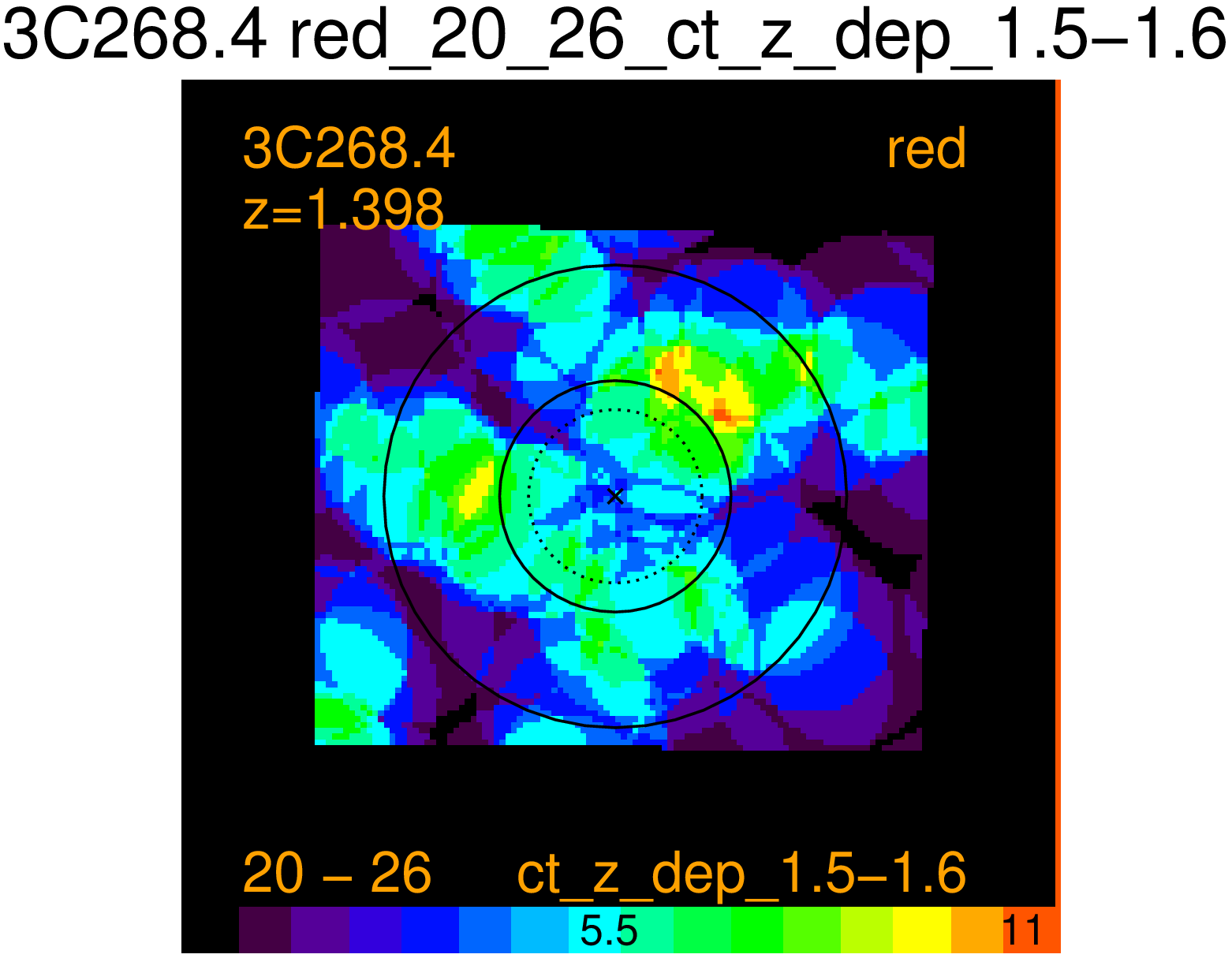}     
                \includegraphics[width=0.245\textwidth, clip=true]{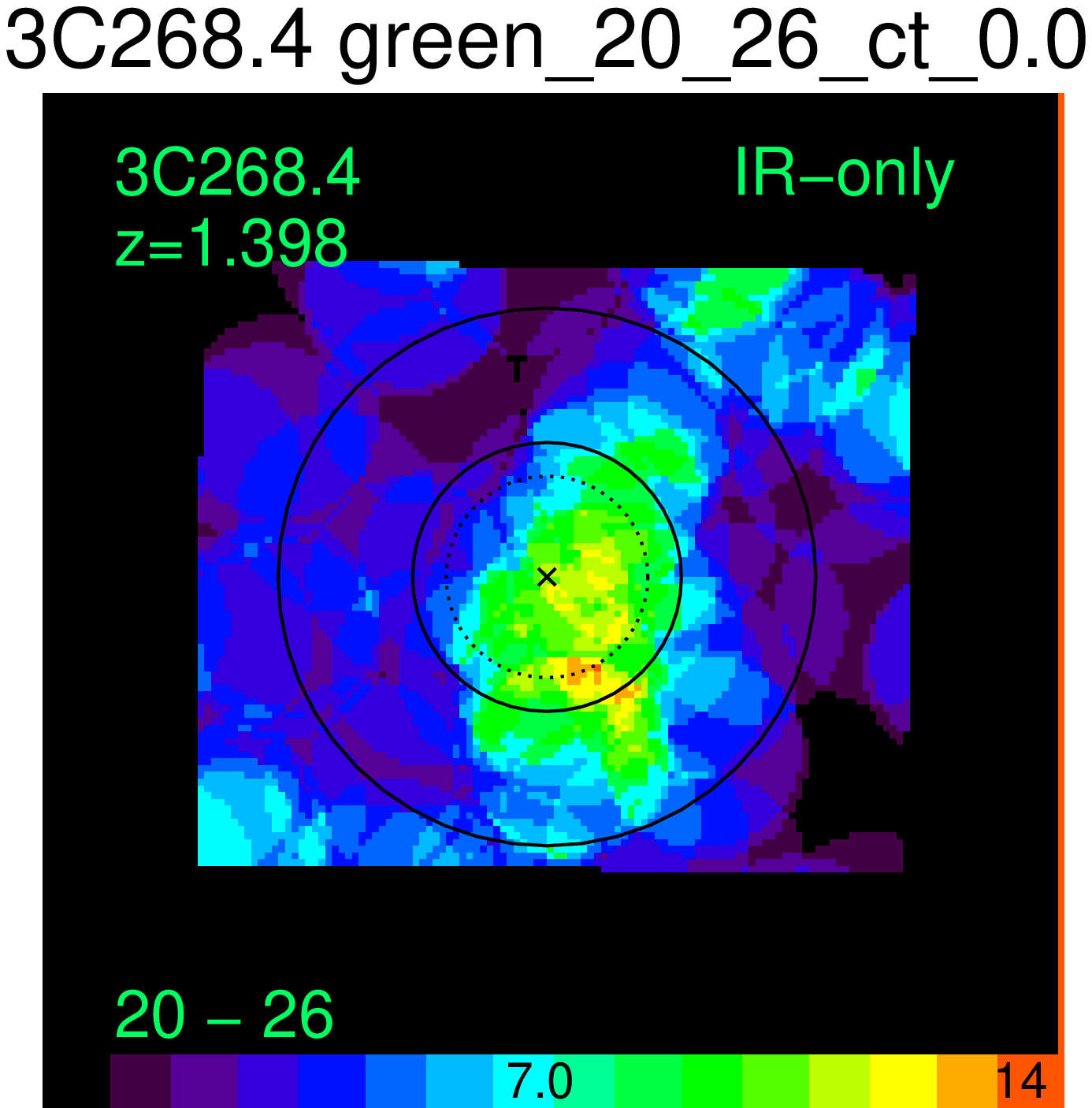}             
                \includegraphics[width=0.245\textwidth, clip=true]{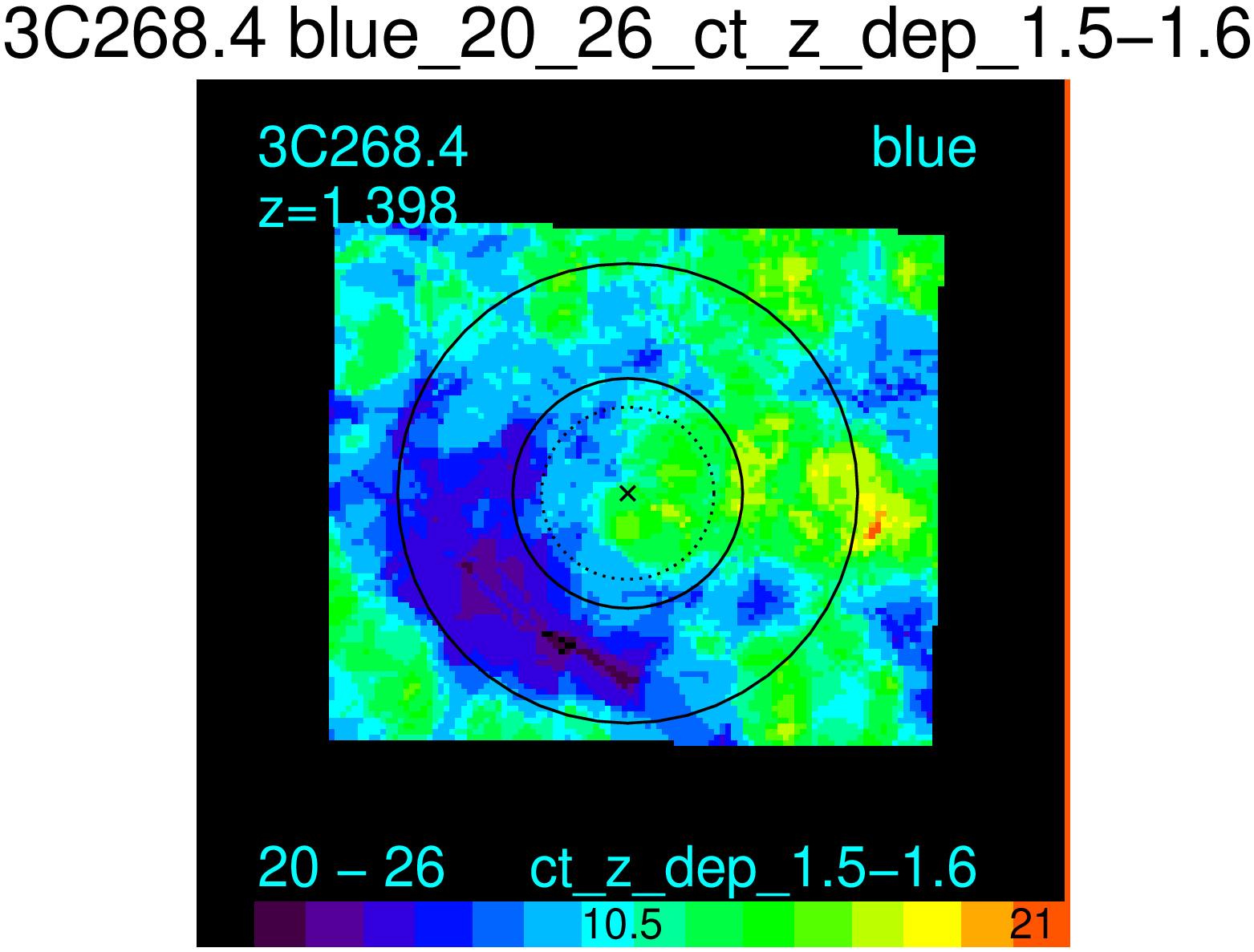}    
                
                \hspace{-0mm}\includegraphics[width=0.245\textwidth, clip=true]{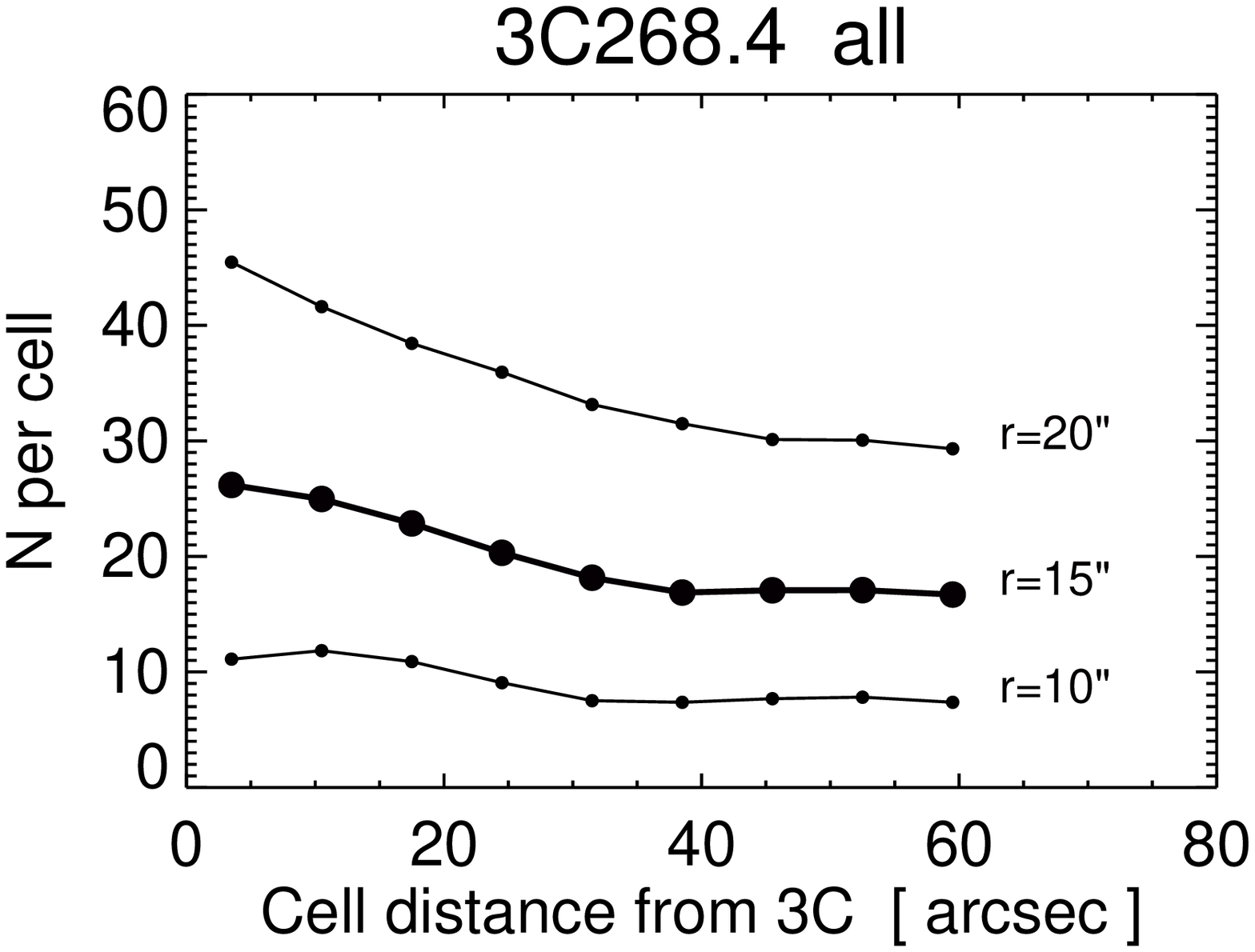}                 
                \includegraphics[width=0.245\textwidth, clip=true]{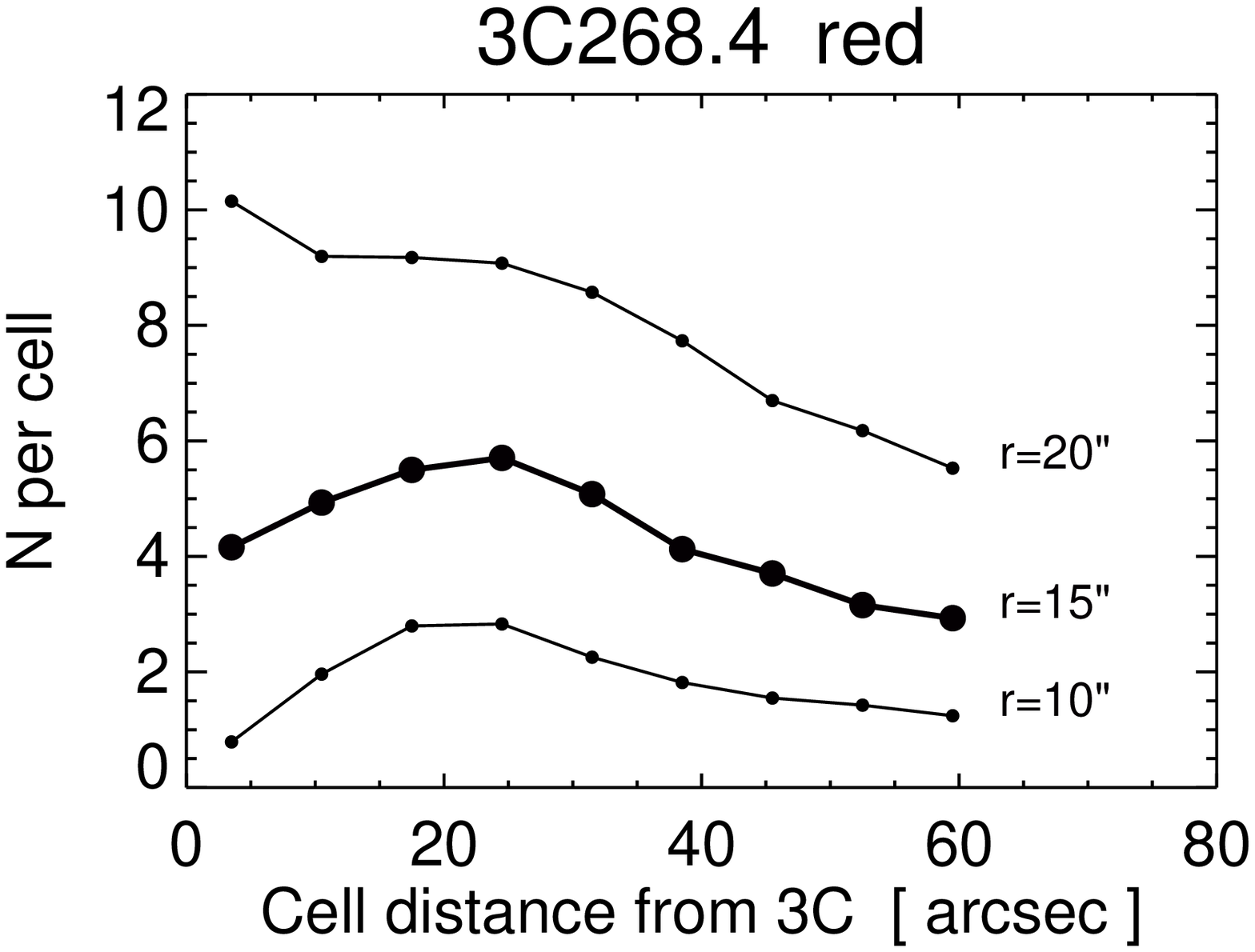}                  
                \includegraphics[width=0.245\textwidth, clip=true]{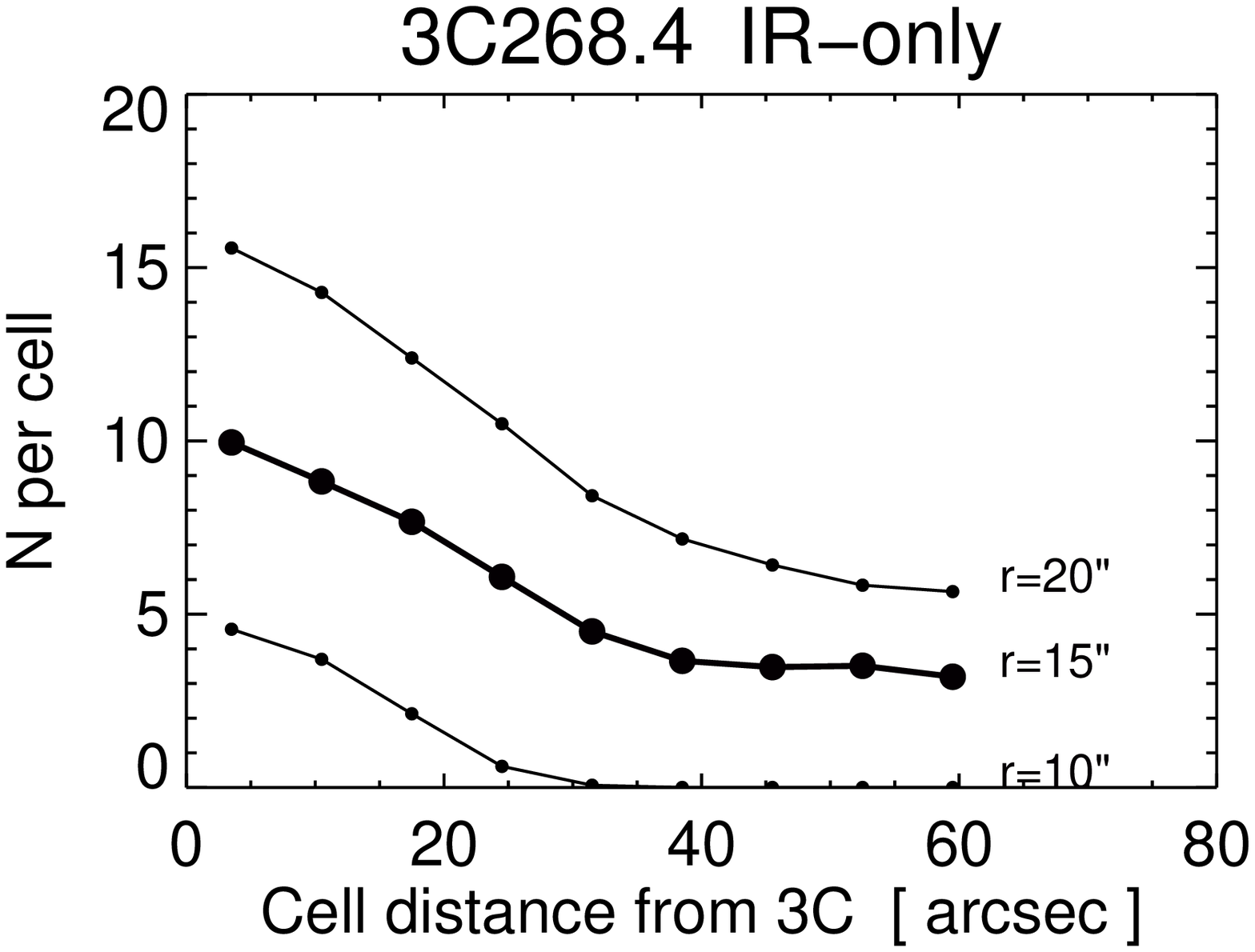}               
                \includegraphics[width=0.245\textwidth, clip=true]{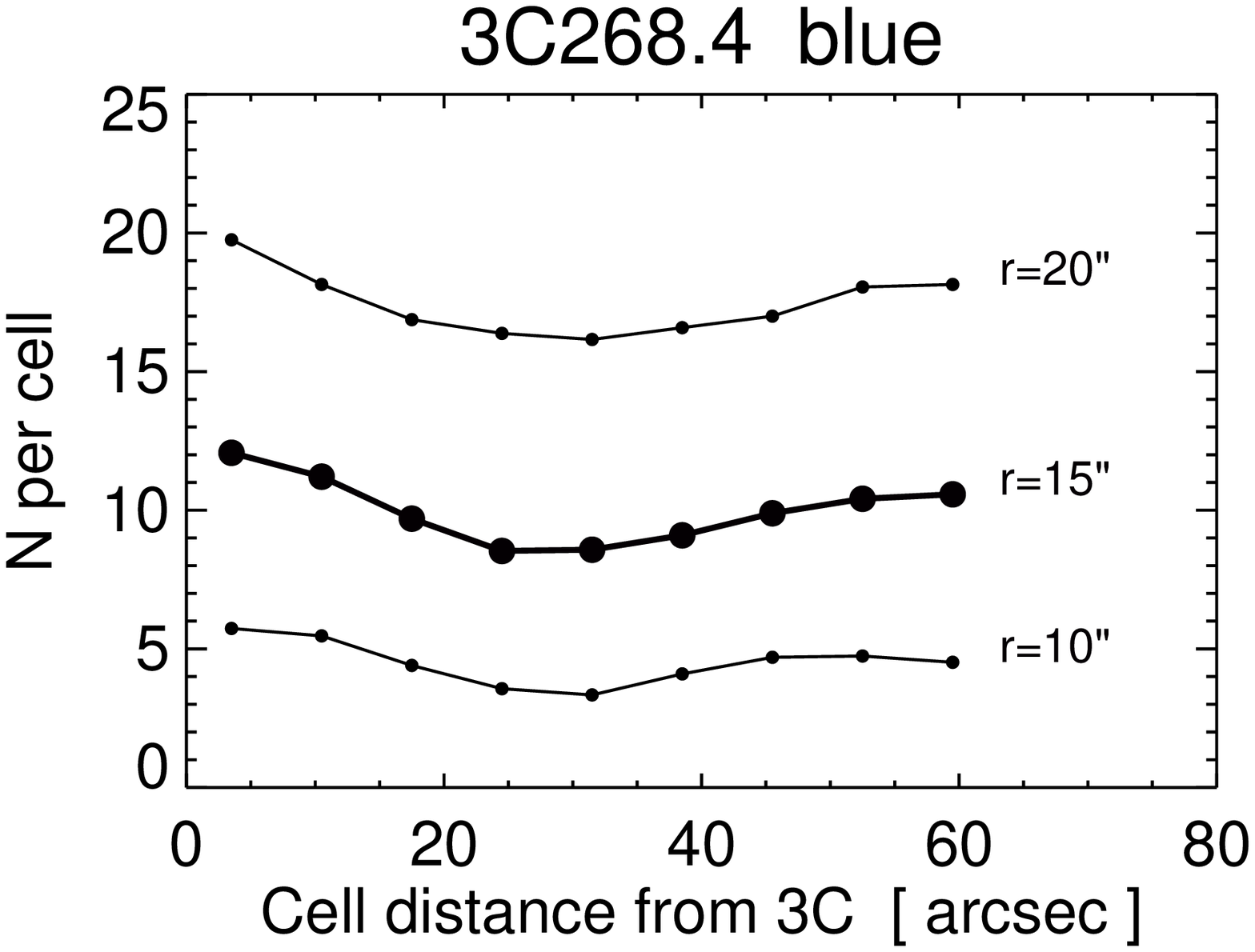}                 

                \caption{Surface density maps and radial density profiles of the 3C fields, continued.
                }
                \label{fig:sd_maps_3}
              \end{figure*}

\begin{figure*}

                \hspace{-0mm}\includegraphics[width=0.245\textwidth, clip=true]{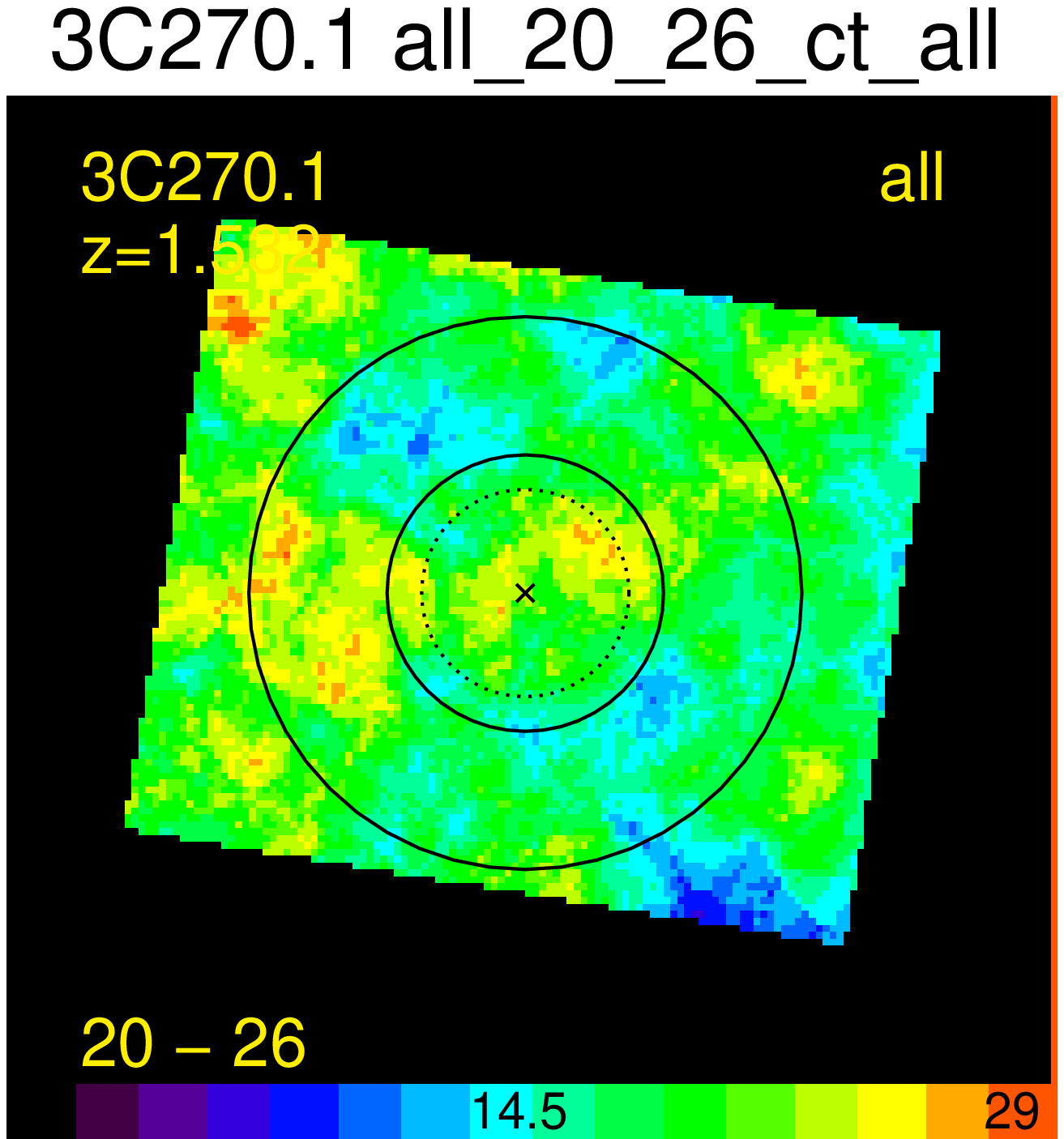}               
                \includegraphics[width=0.245\textwidth, clip=true]{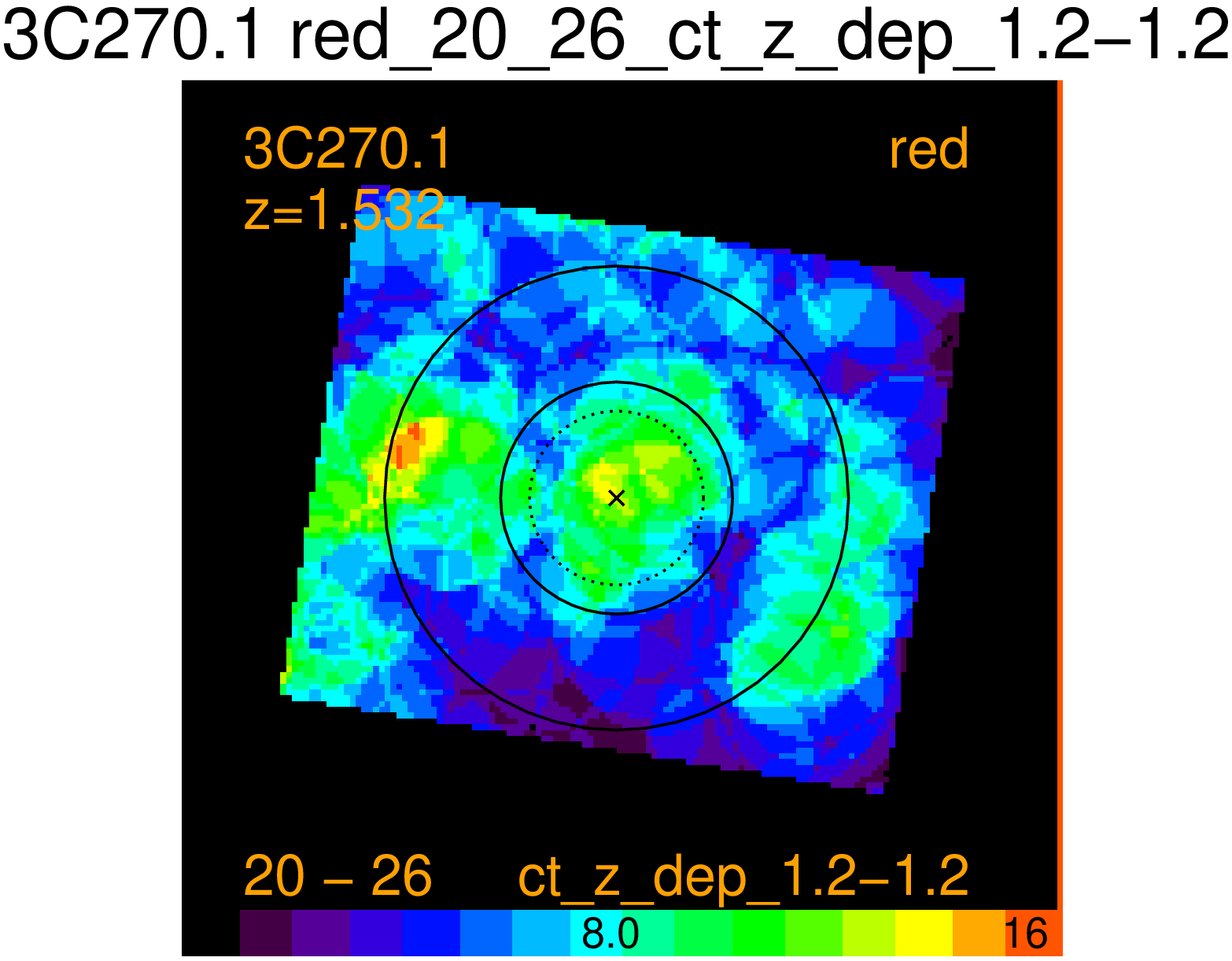}     
                \includegraphics[width=0.245\textwidth, clip=true]{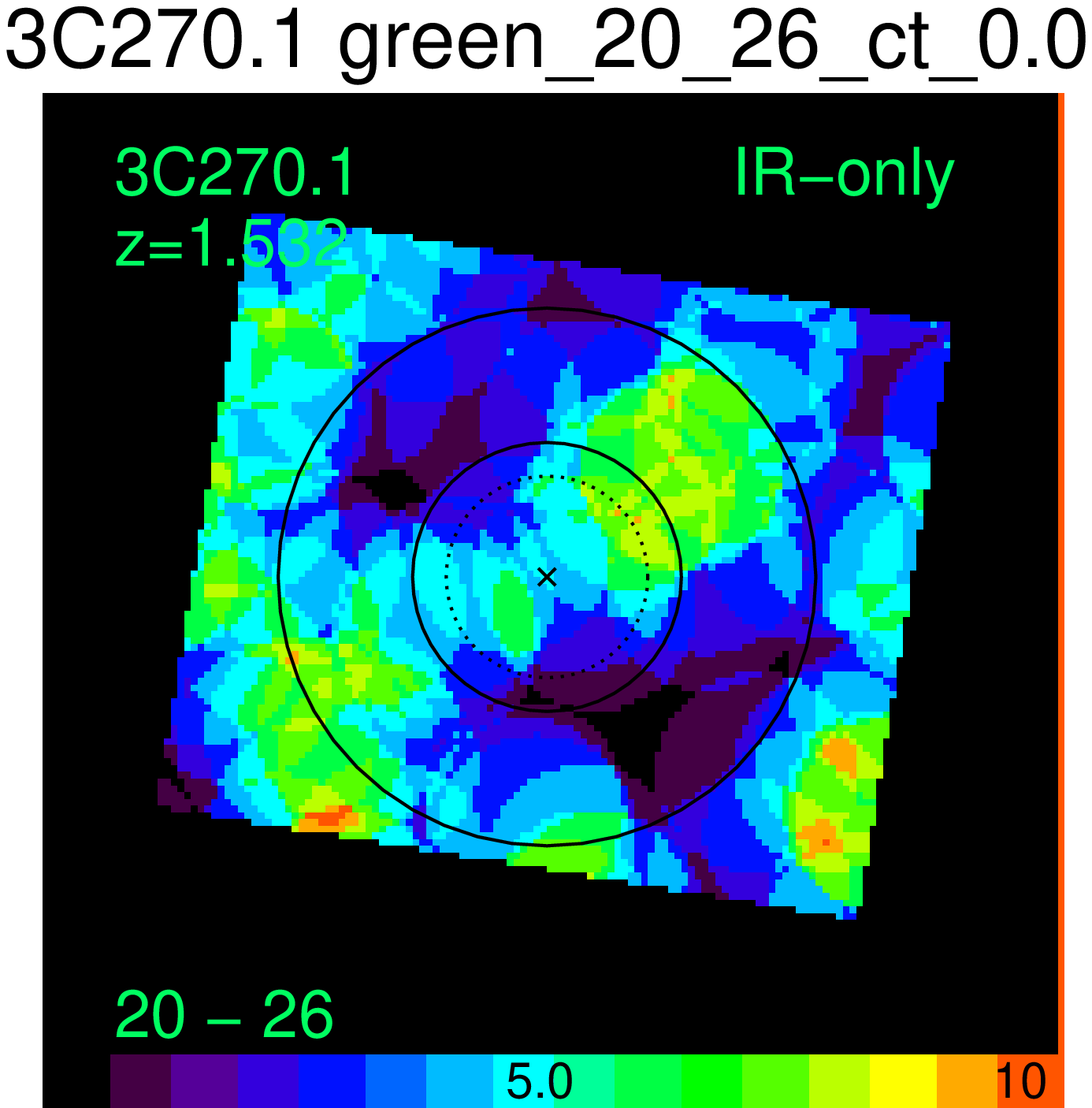}             
                \includegraphics[width=0.245\textwidth, clip=true]{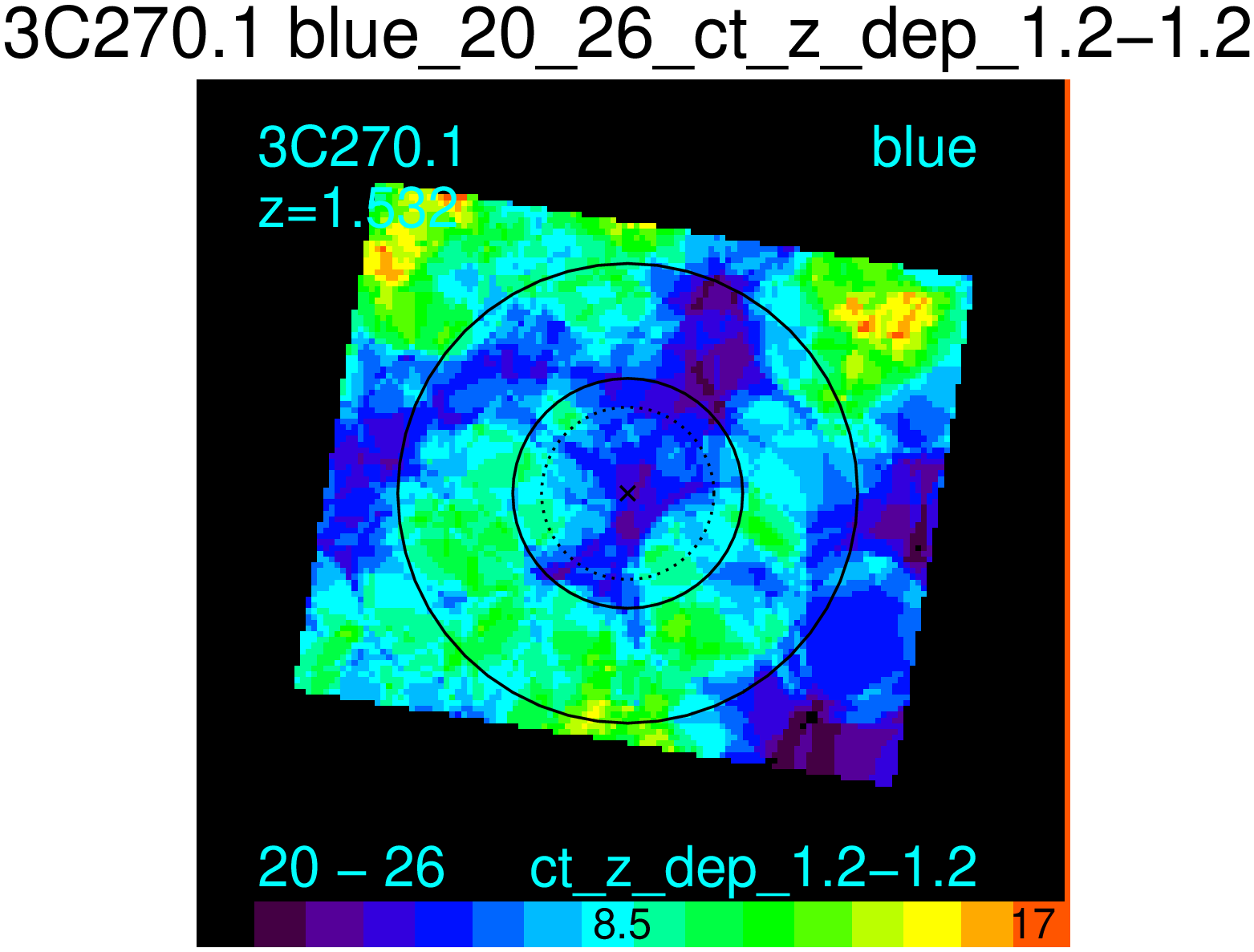}    
                
                \hspace{-0mm}\includegraphics[width=0.245\textwidth, clip=true]{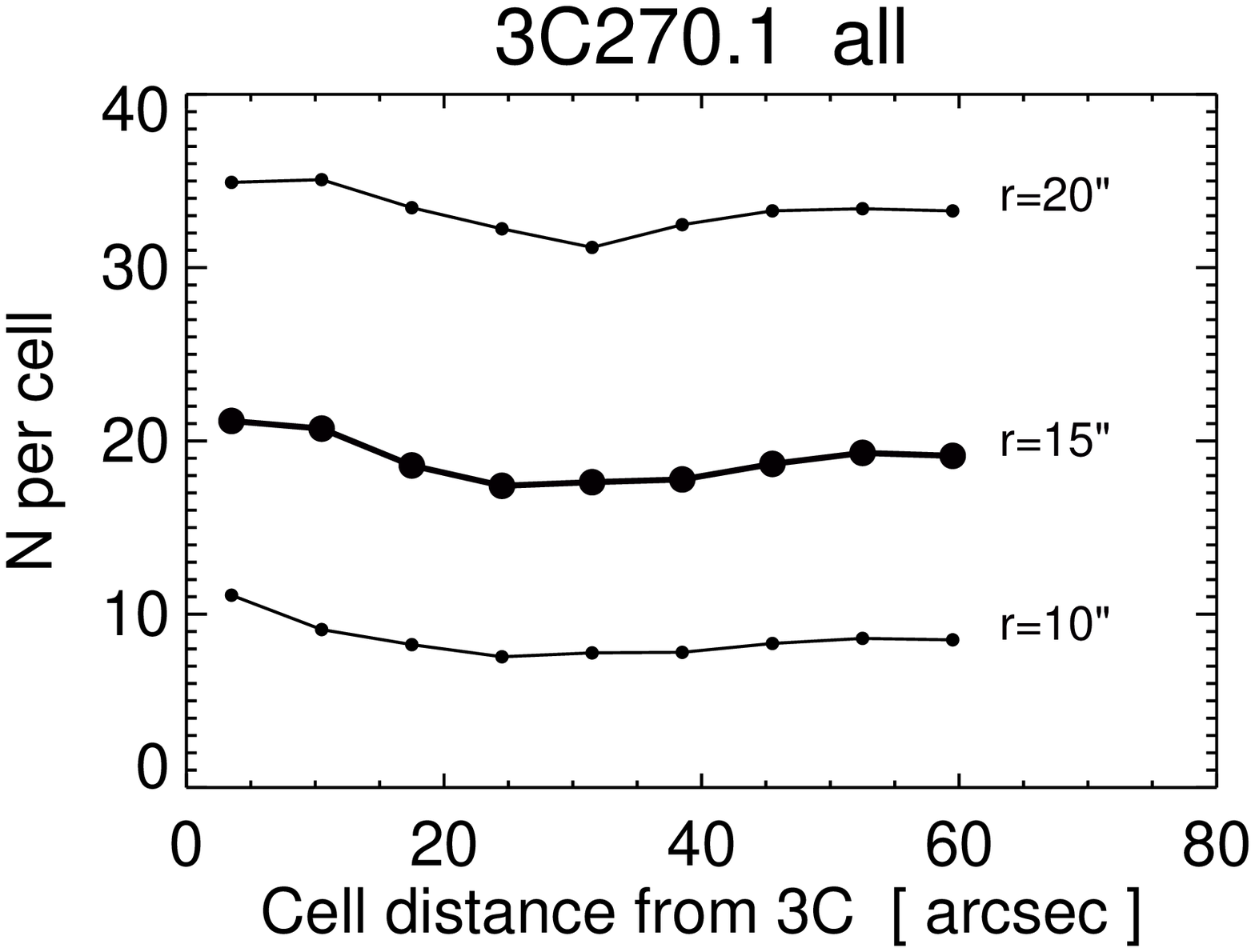}                 
                \includegraphics[width=0.245\textwidth, clip=true]{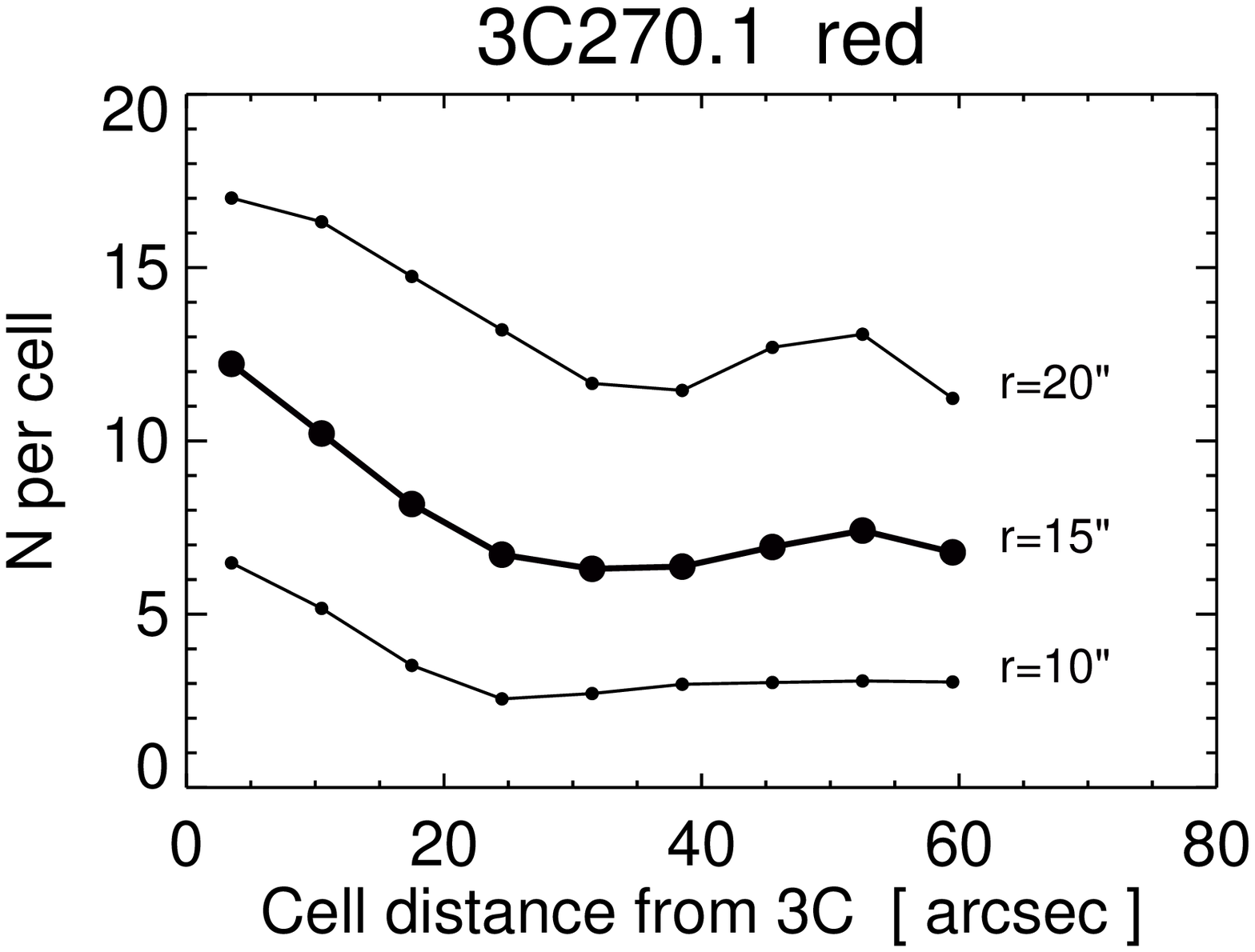}                  
                \includegraphics[width=0.245\textwidth, clip=true]{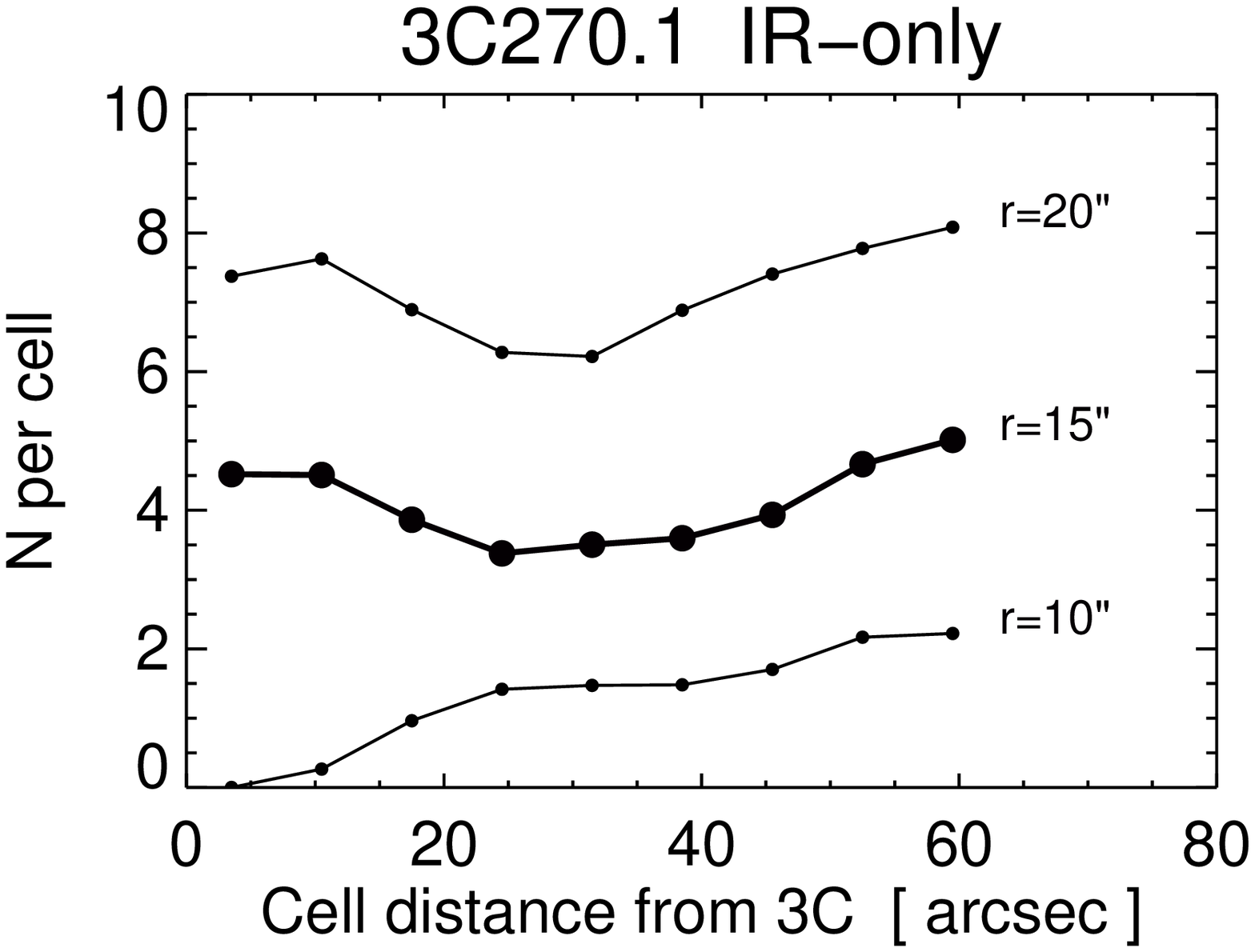}               
                \includegraphics[width=0.245\textwidth, clip=true]{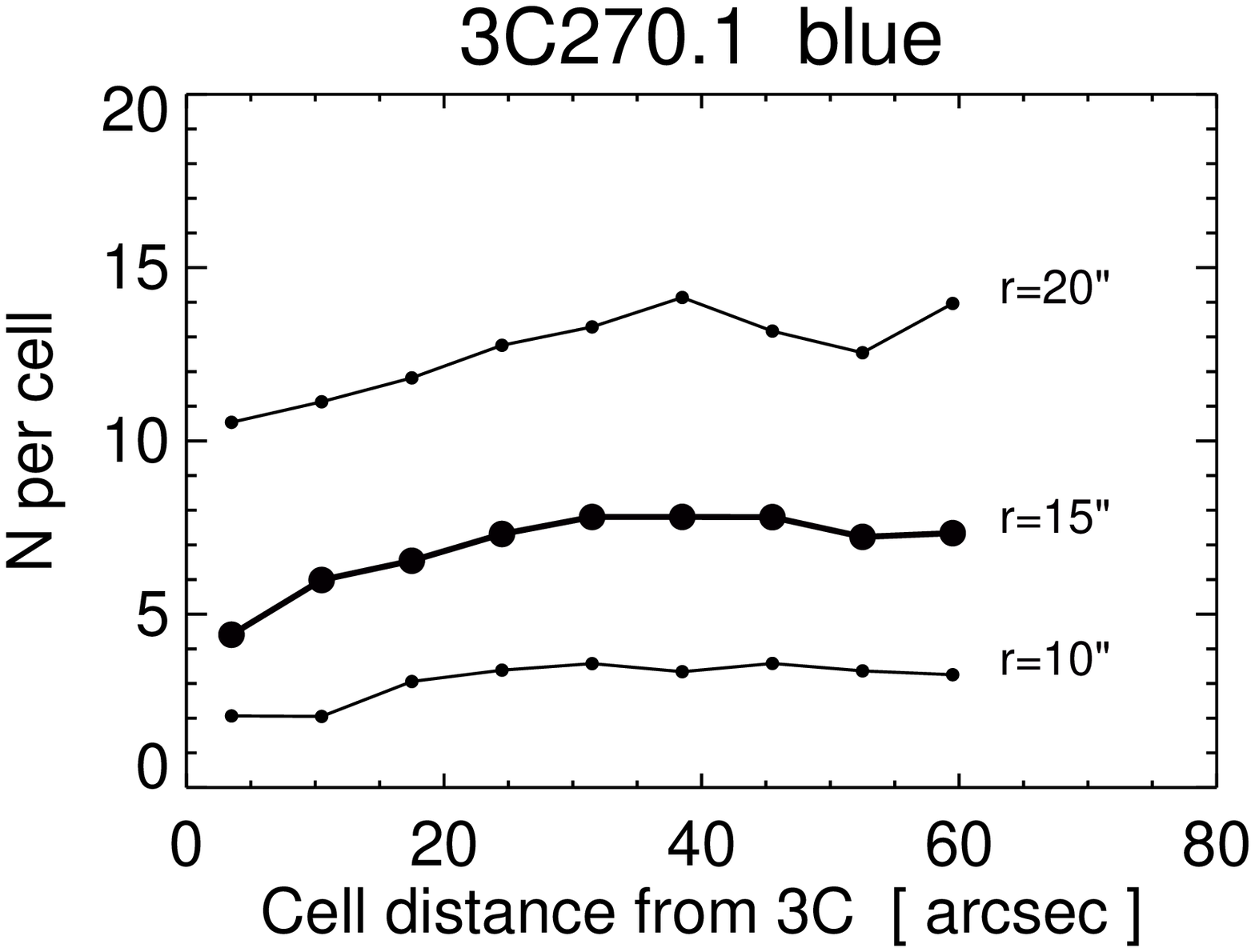}

                \hspace{-0mm}\includegraphics[width=0.245\textwidth, clip=true]{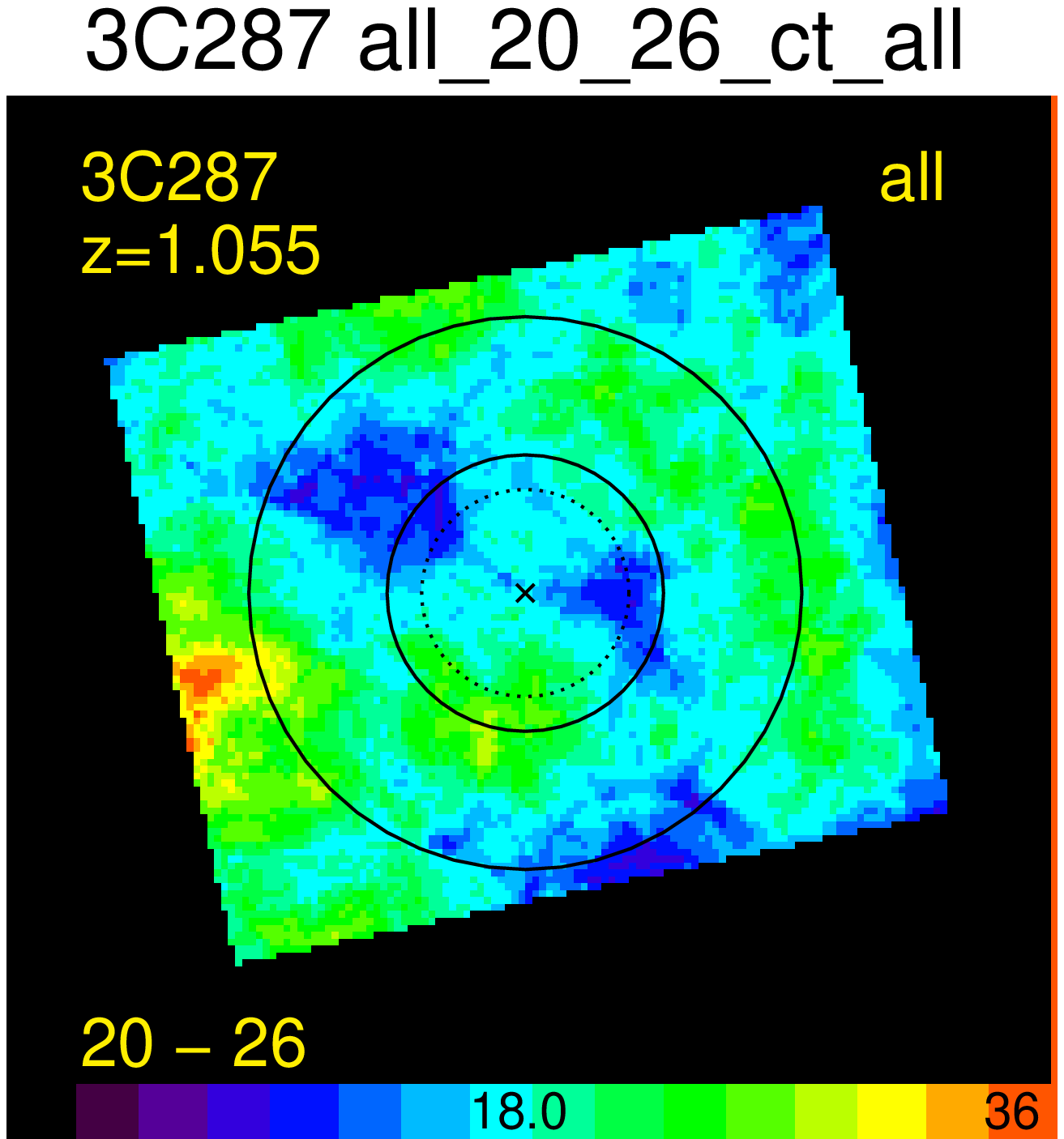}                 
                \includegraphics[width=0.245\textwidth, clip=true]{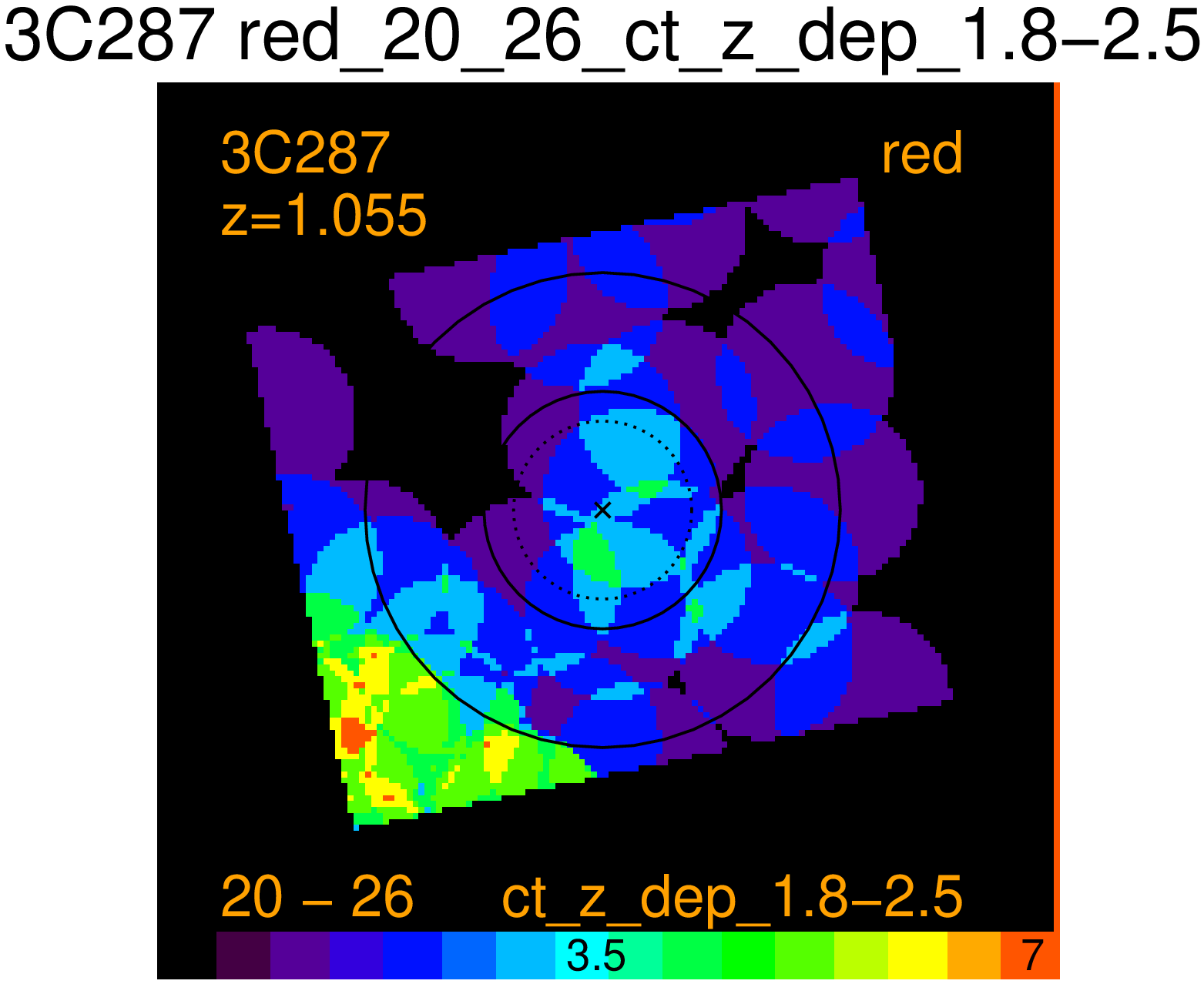}       
                \includegraphics[width=0.245\textwidth, clip=true]{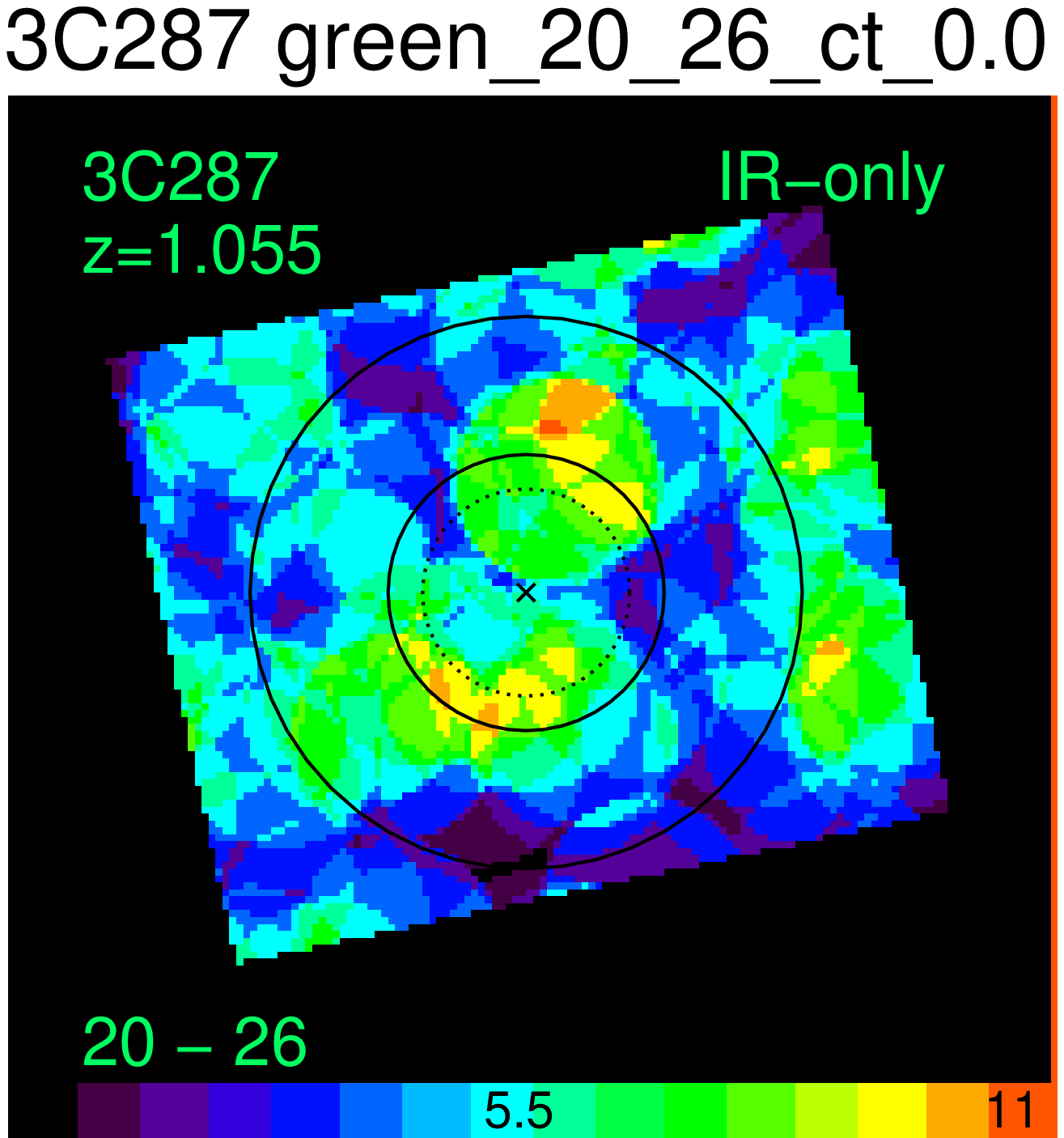}               
                \includegraphics[width=0.245\textwidth, clip=true]{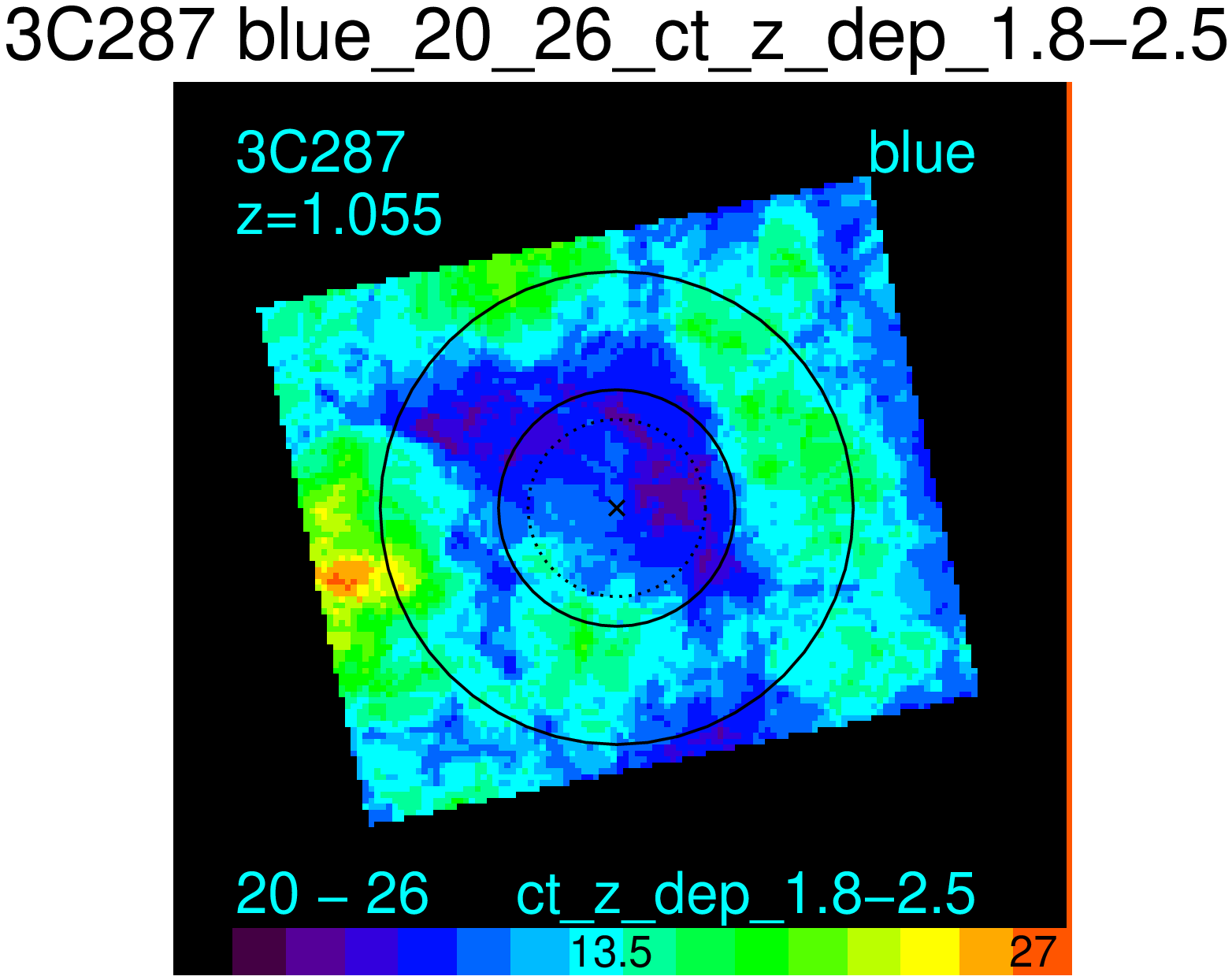}      
                
                \hspace{-0mm}\includegraphics[width=0.245\textwidth, clip=true]{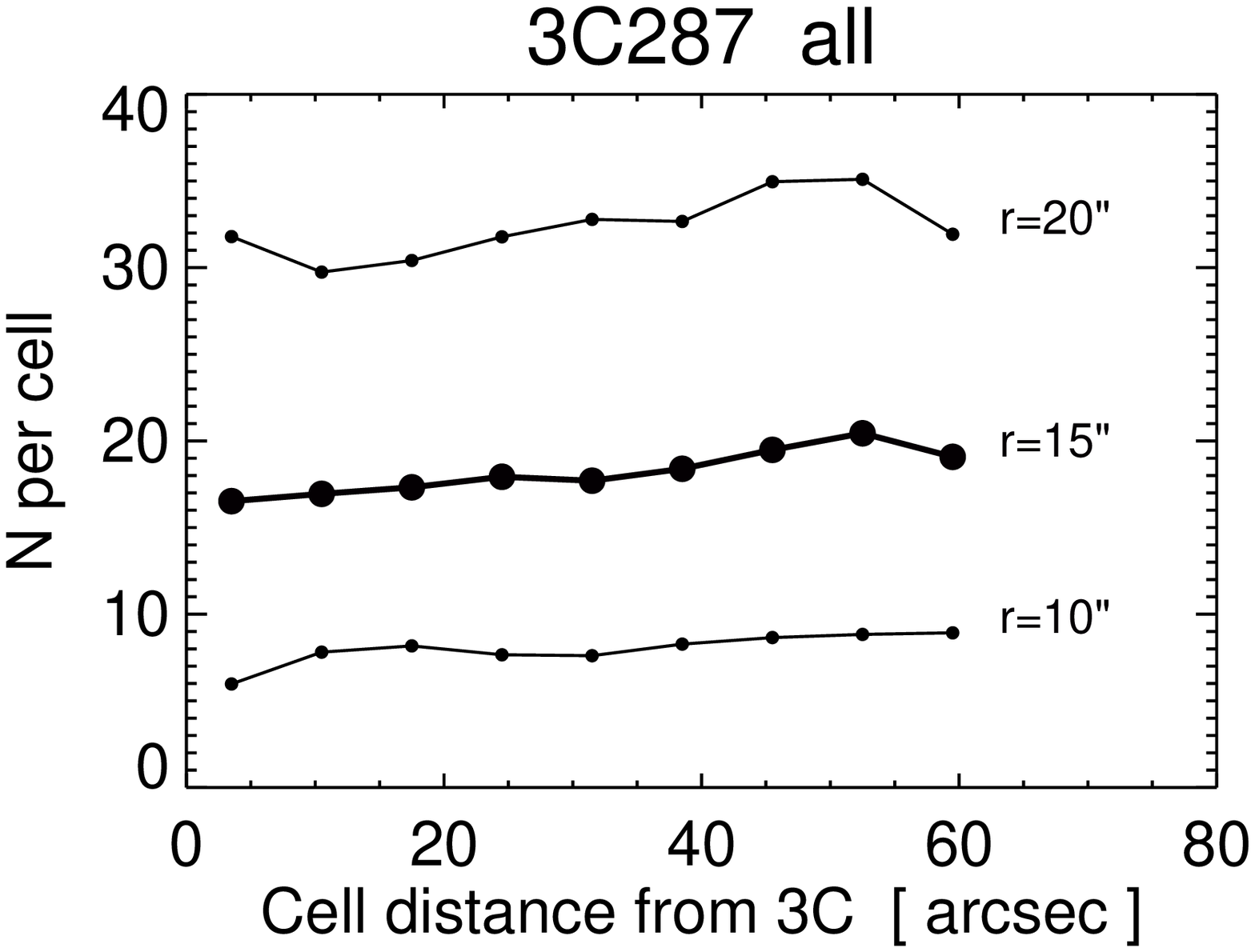}                   
                \includegraphics[width=0.245\textwidth, clip=true]{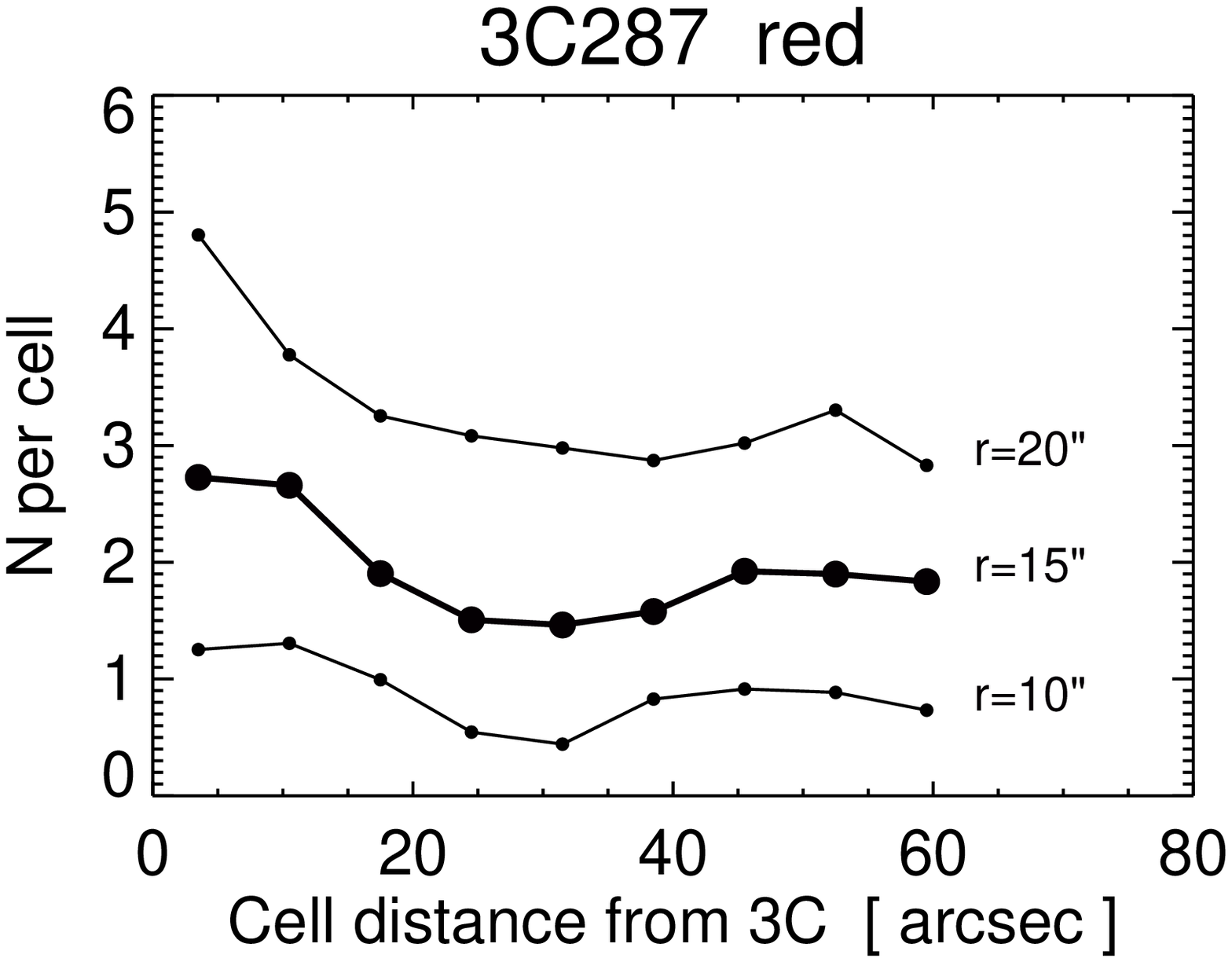}                    
                \includegraphics[width=0.245\textwidth, clip=true]{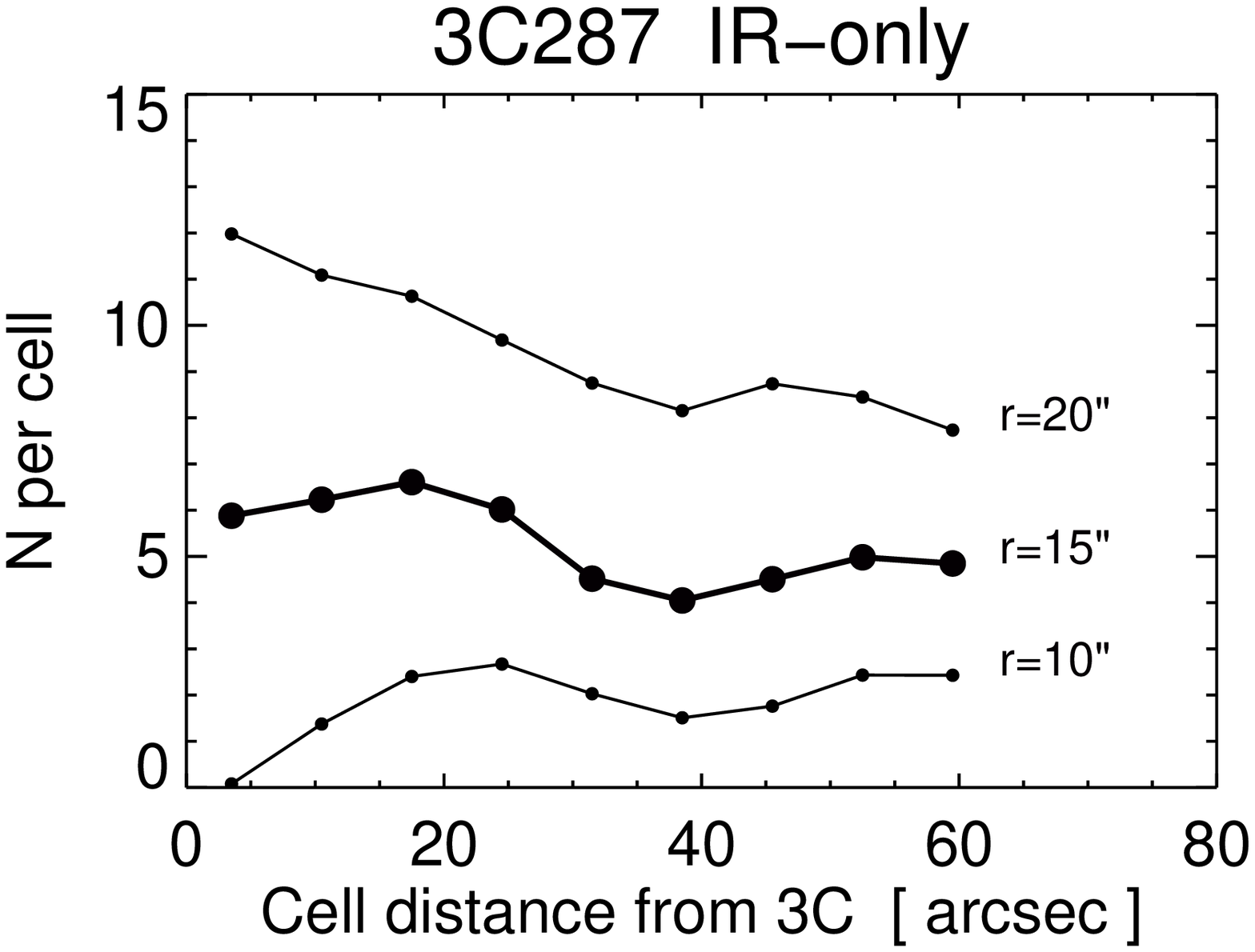}                 
                \includegraphics[width=0.245\textwidth, clip=true]{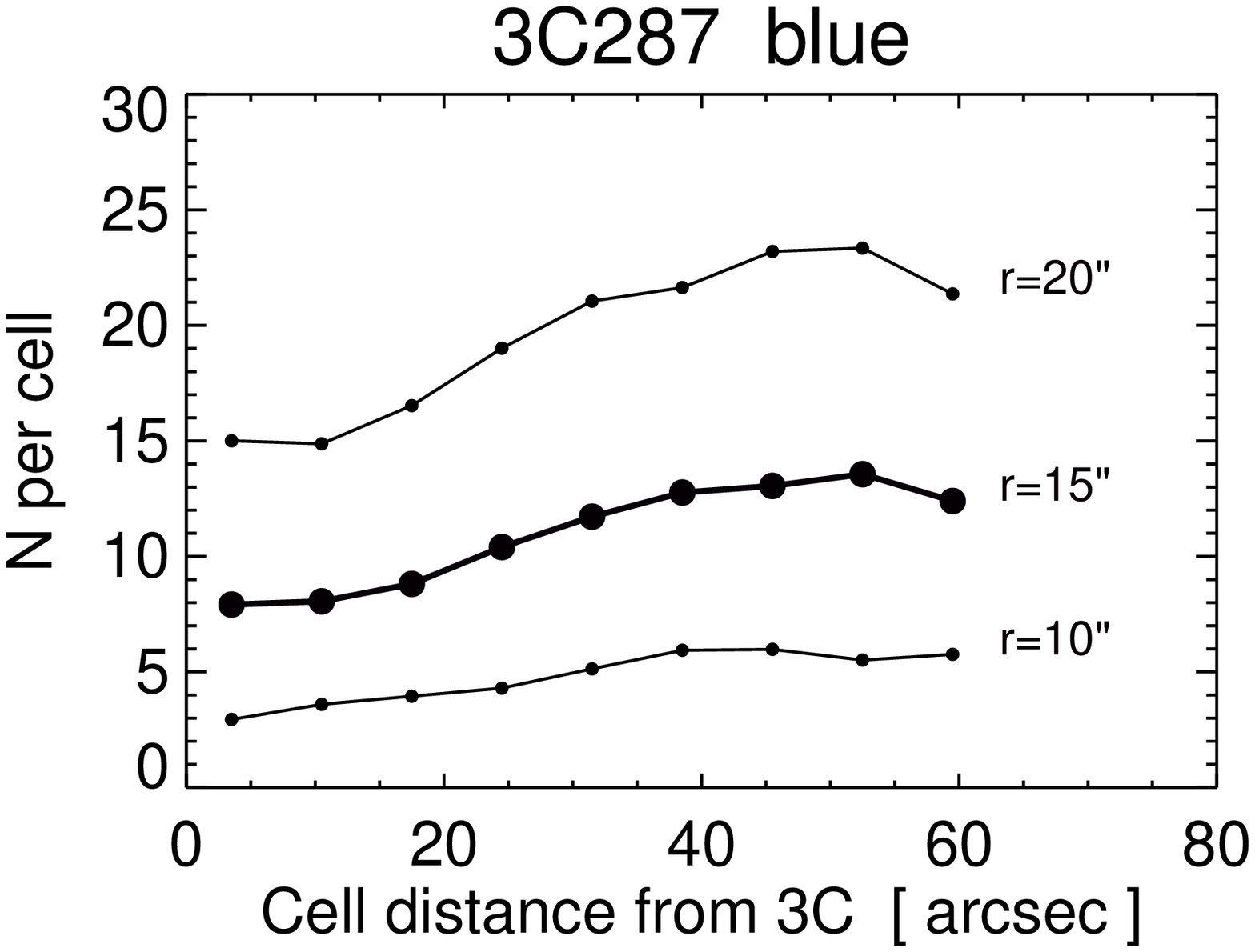}

                \hspace{-0mm}\includegraphics[width=0.245\textwidth, clip=true]{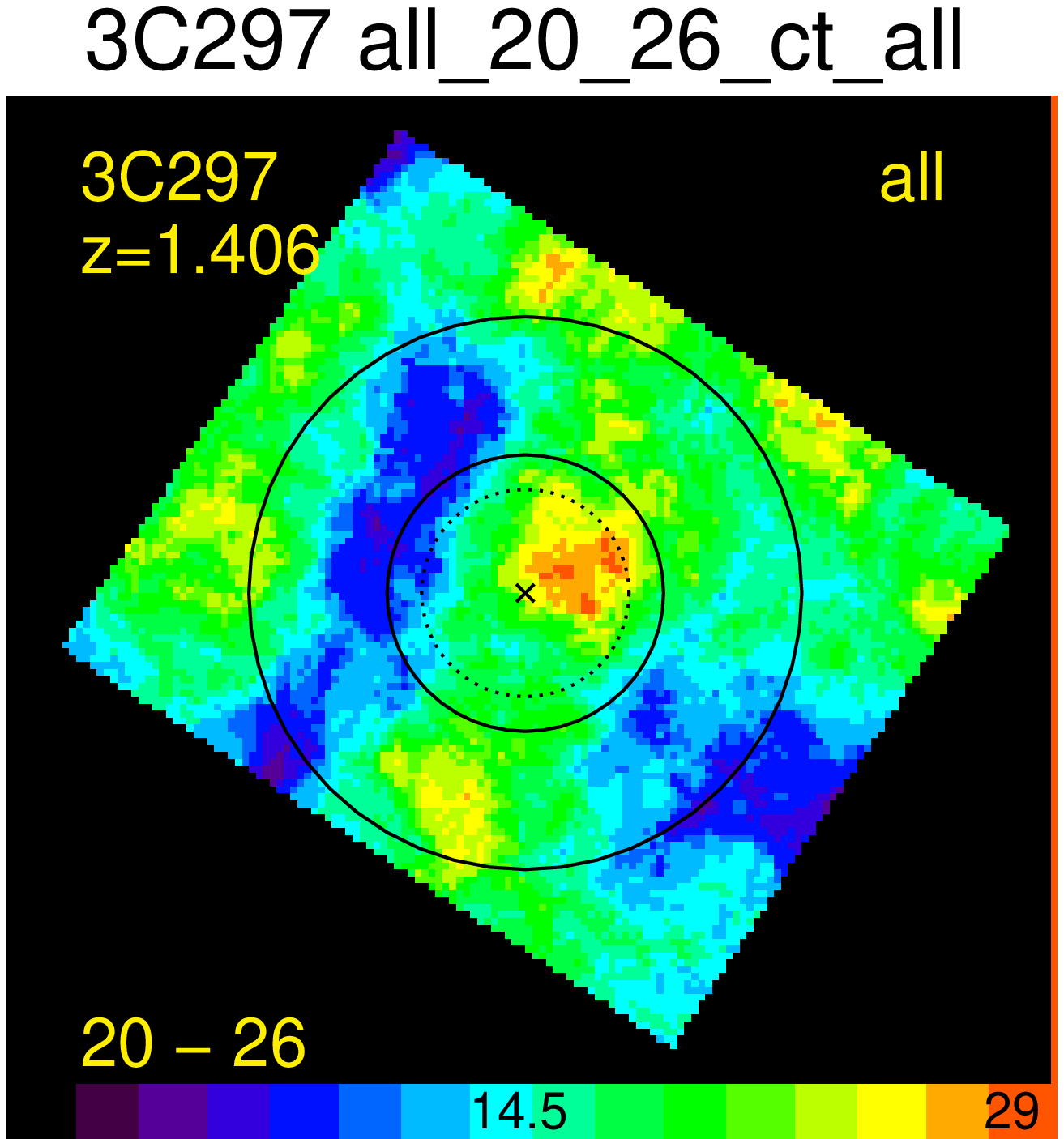}                 
                \includegraphics[width=0.245\textwidth, clip=true]{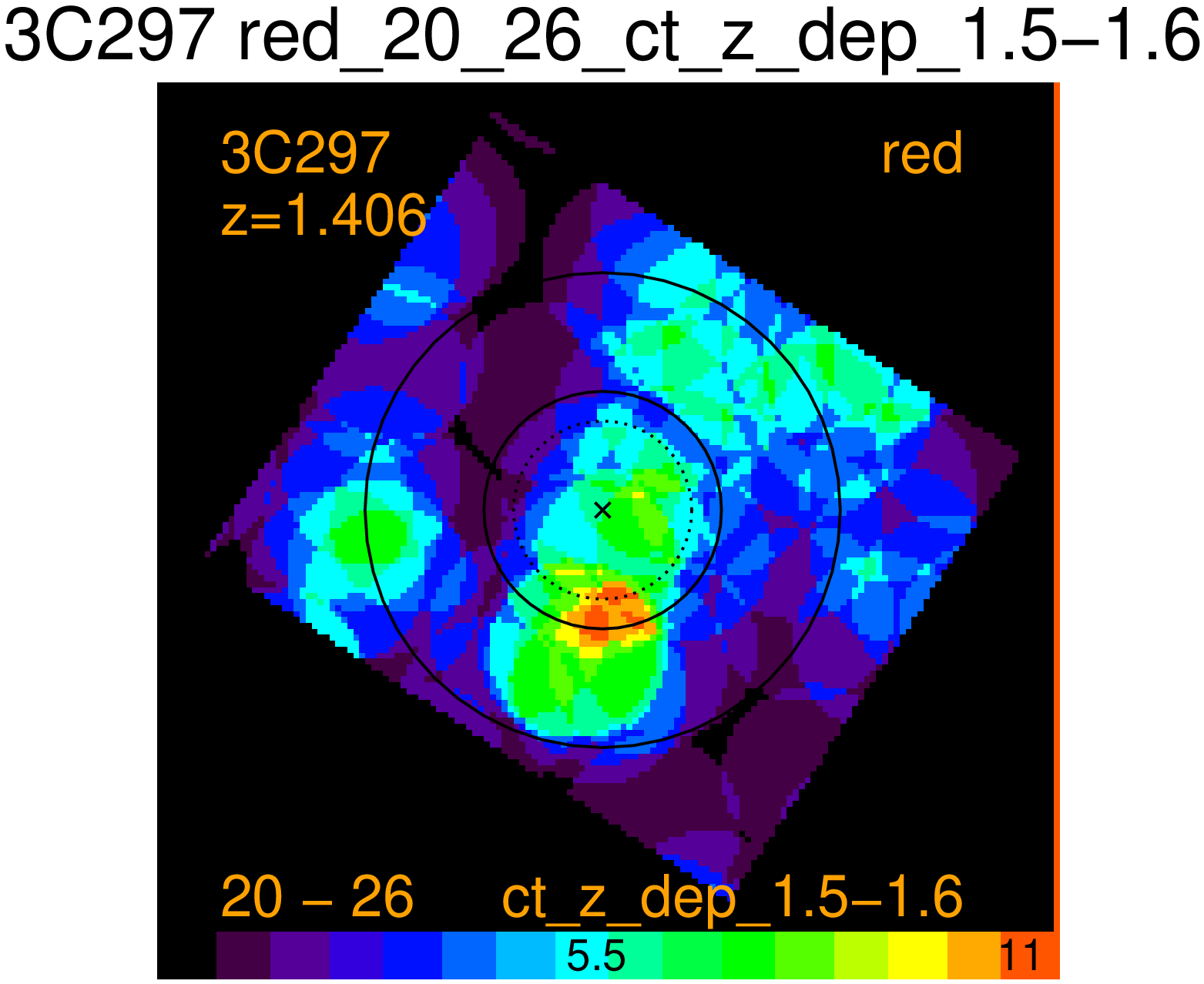}       
                \includegraphics[width=0.245\textwidth, clip=true]{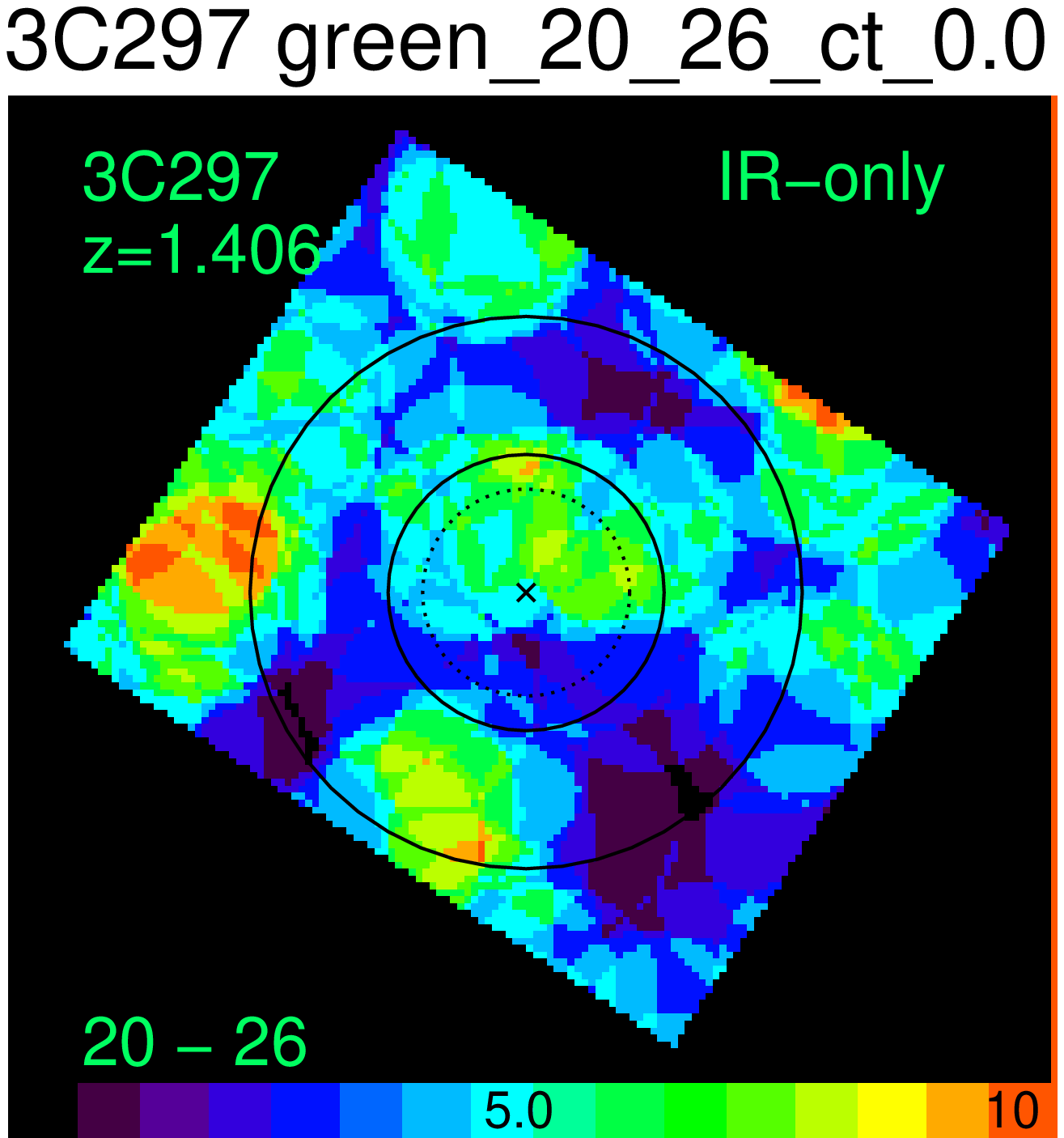}               
                \includegraphics[width=0.245\textwidth, clip=true]{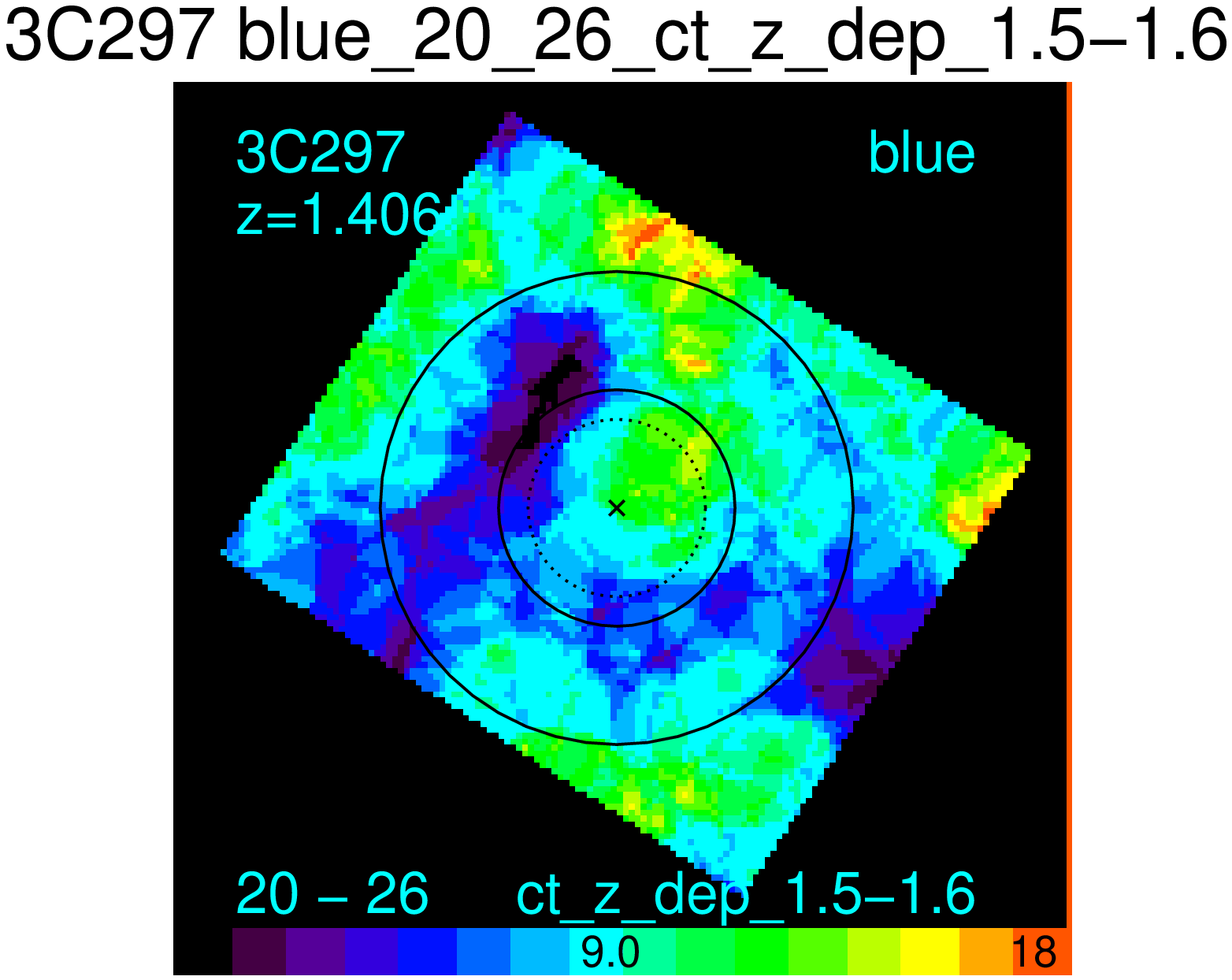}      
                
                \hspace{-0mm}\includegraphics[width=0.245\textwidth, clip=true]{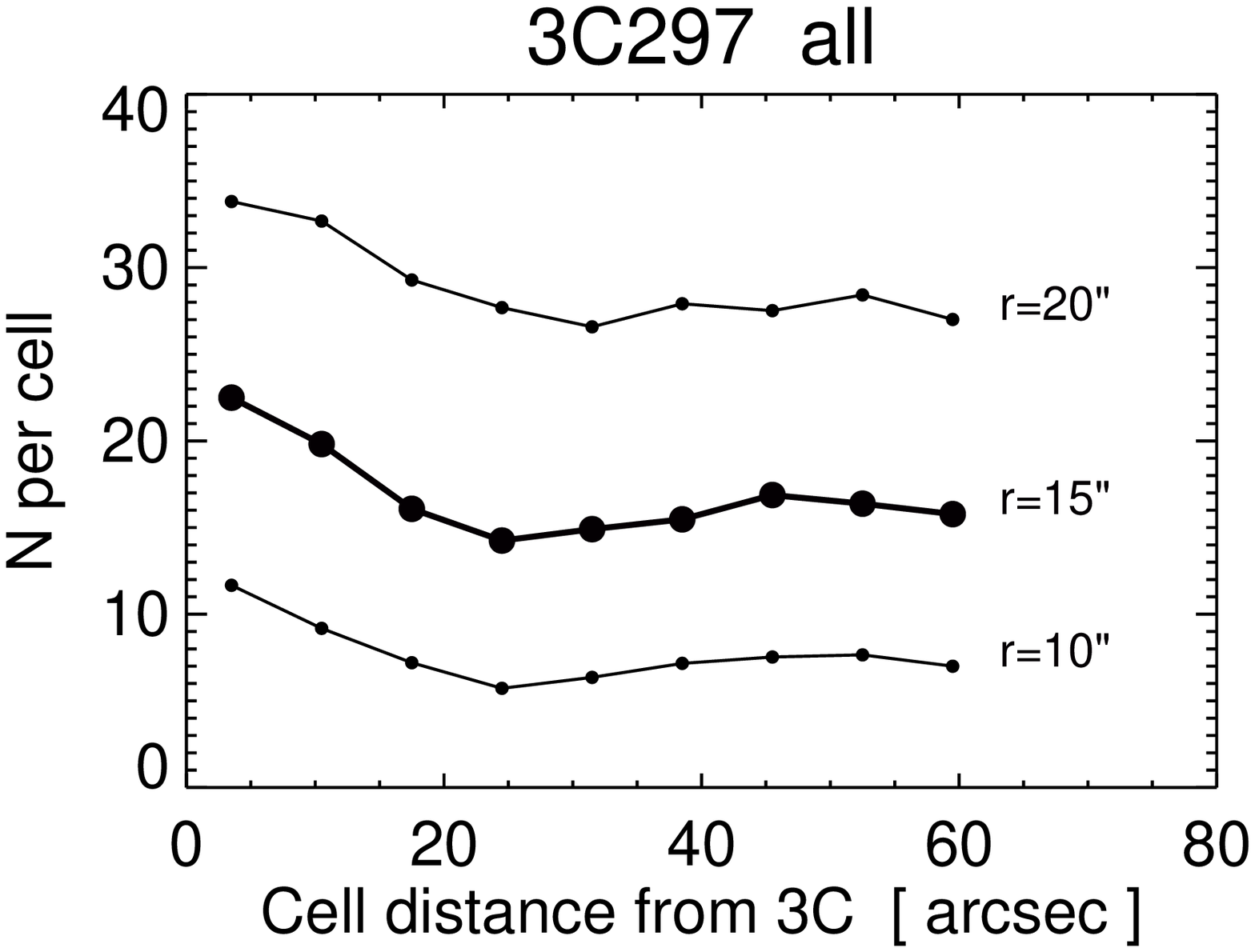}                   
                \includegraphics[width=0.245\textwidth, clip=true]{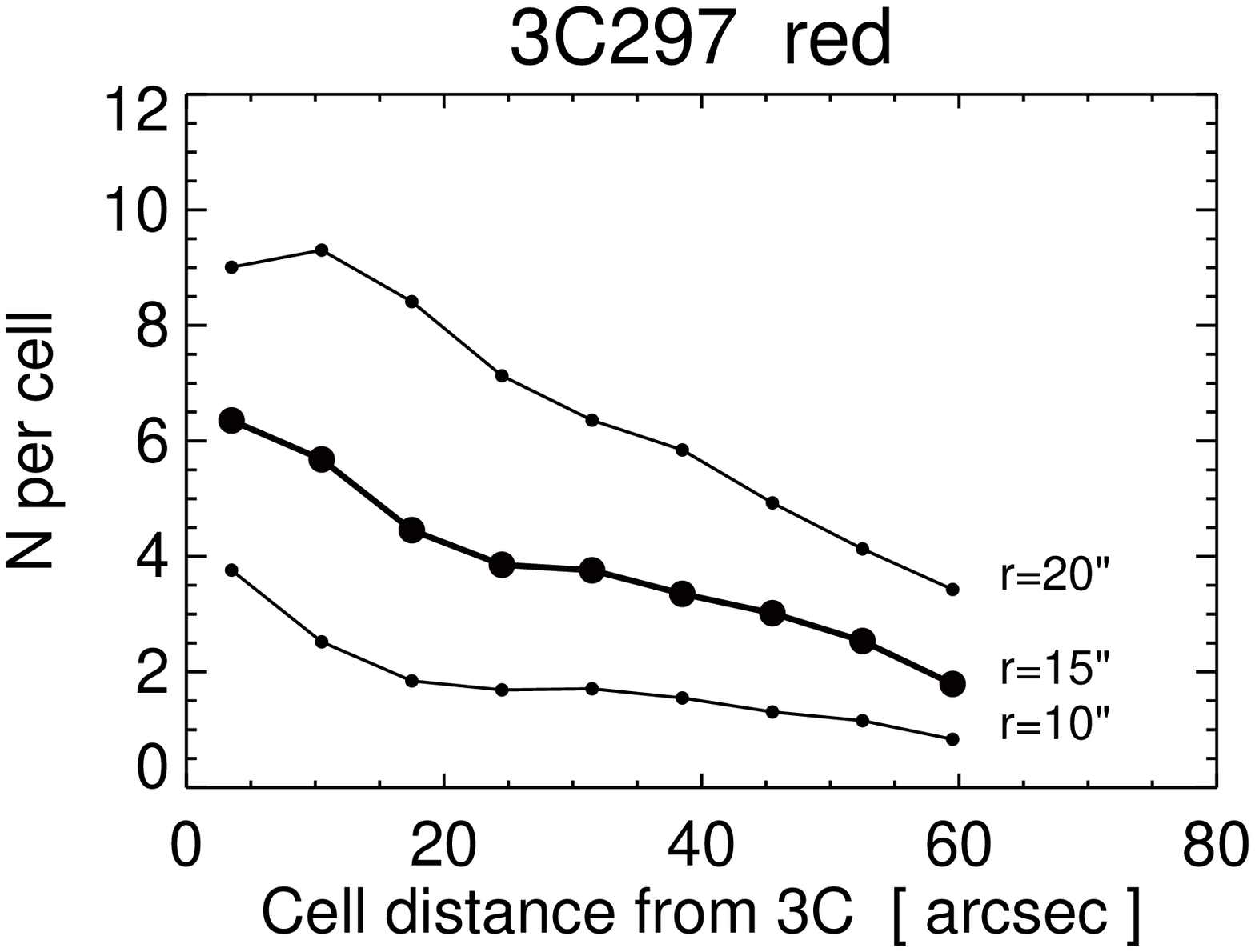}                    
                \includegraphics[width=0.245\textwidth, clip=true]{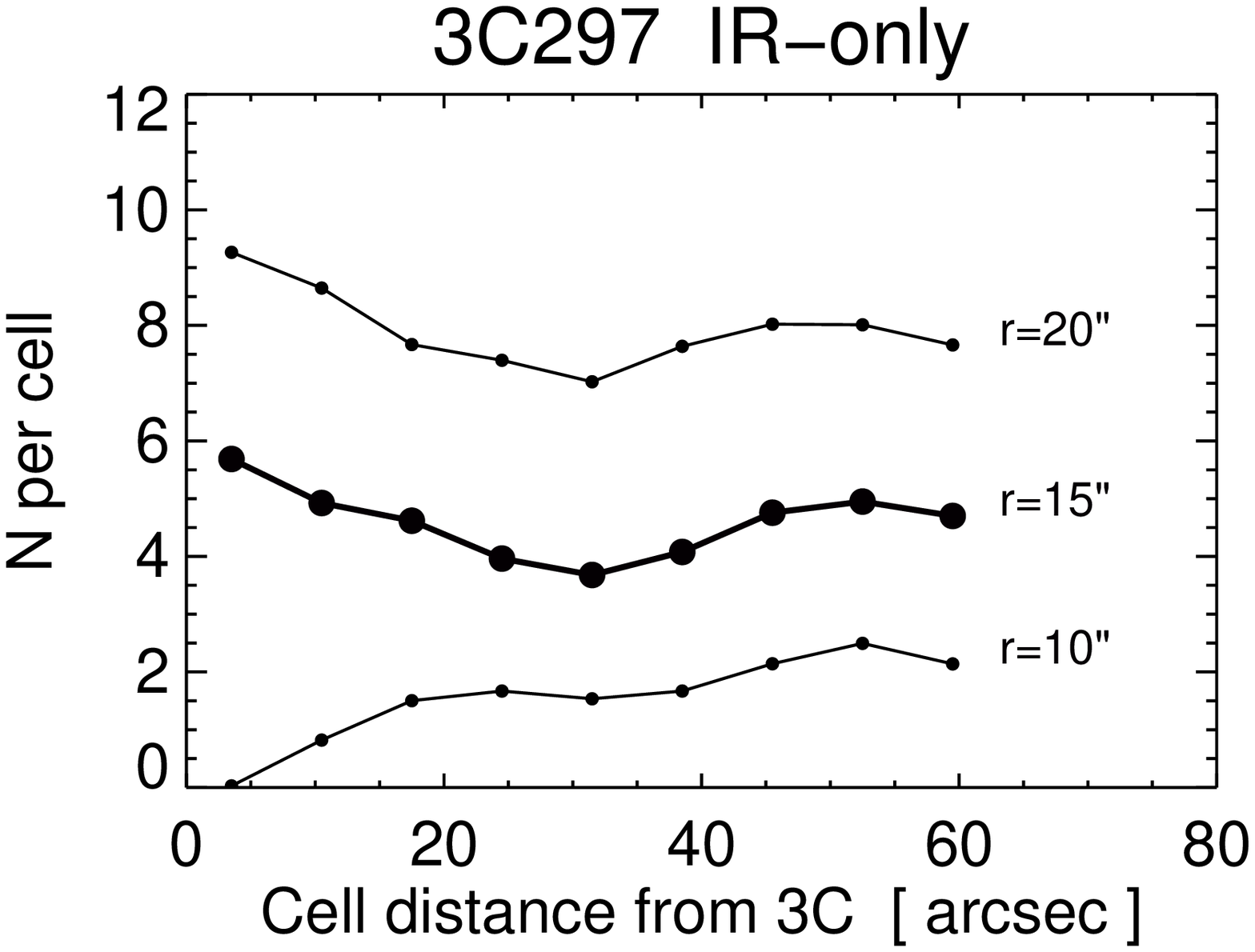}                 
                \includegraphics[width=0.245\textwidth, clip=true]{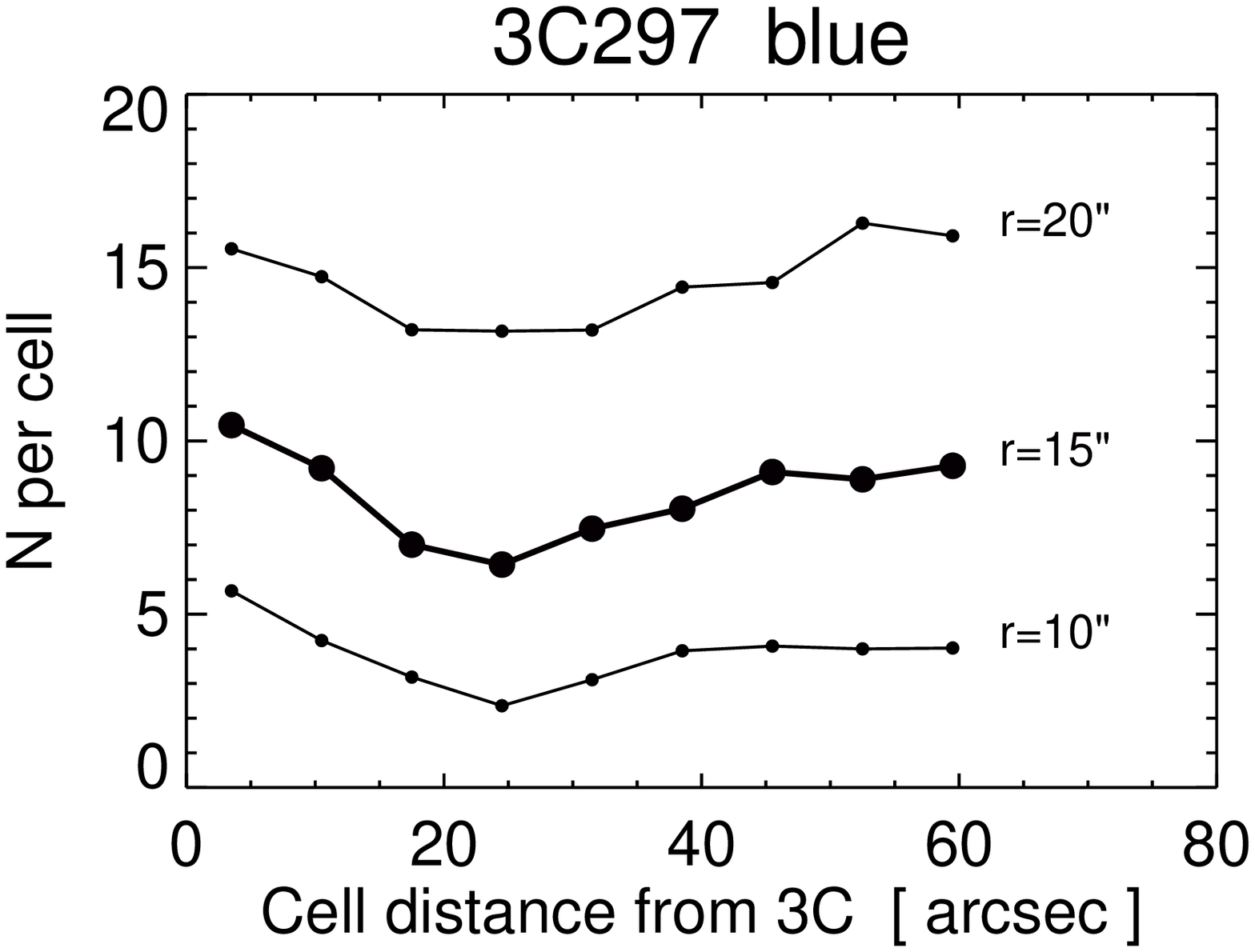}

                \caption{Surface density maps and radial density profiles of the 3C fields, continued.
                }
                \label{fig:sd_maps_4}
              \end{figure*}

\begin{figure*}

                \hspace{-0mm}\includegraphics[width=0.245\textwidth, clip=true]{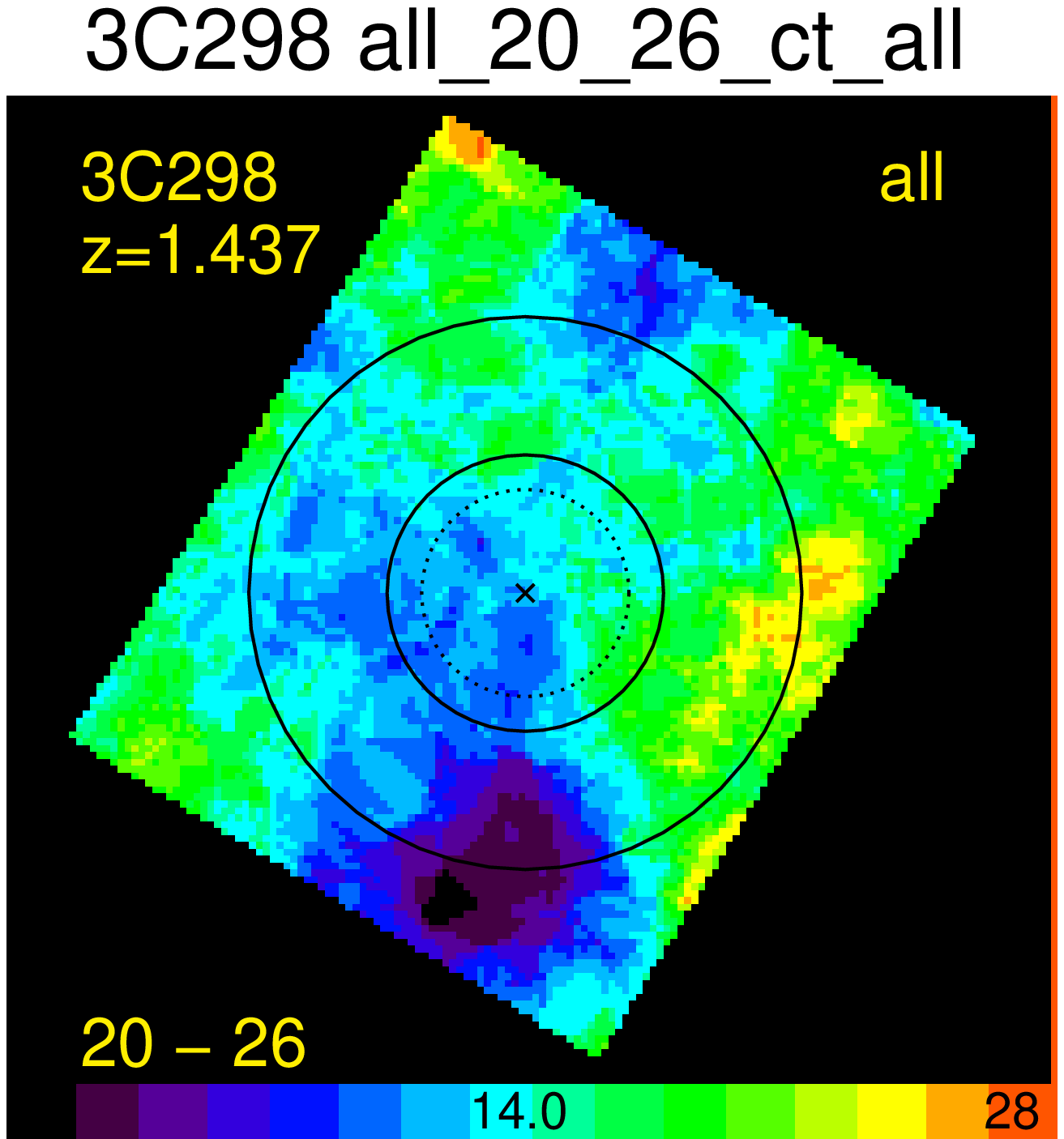}                 
                \includegraphics[width=0.245\textwidth, clip=true]{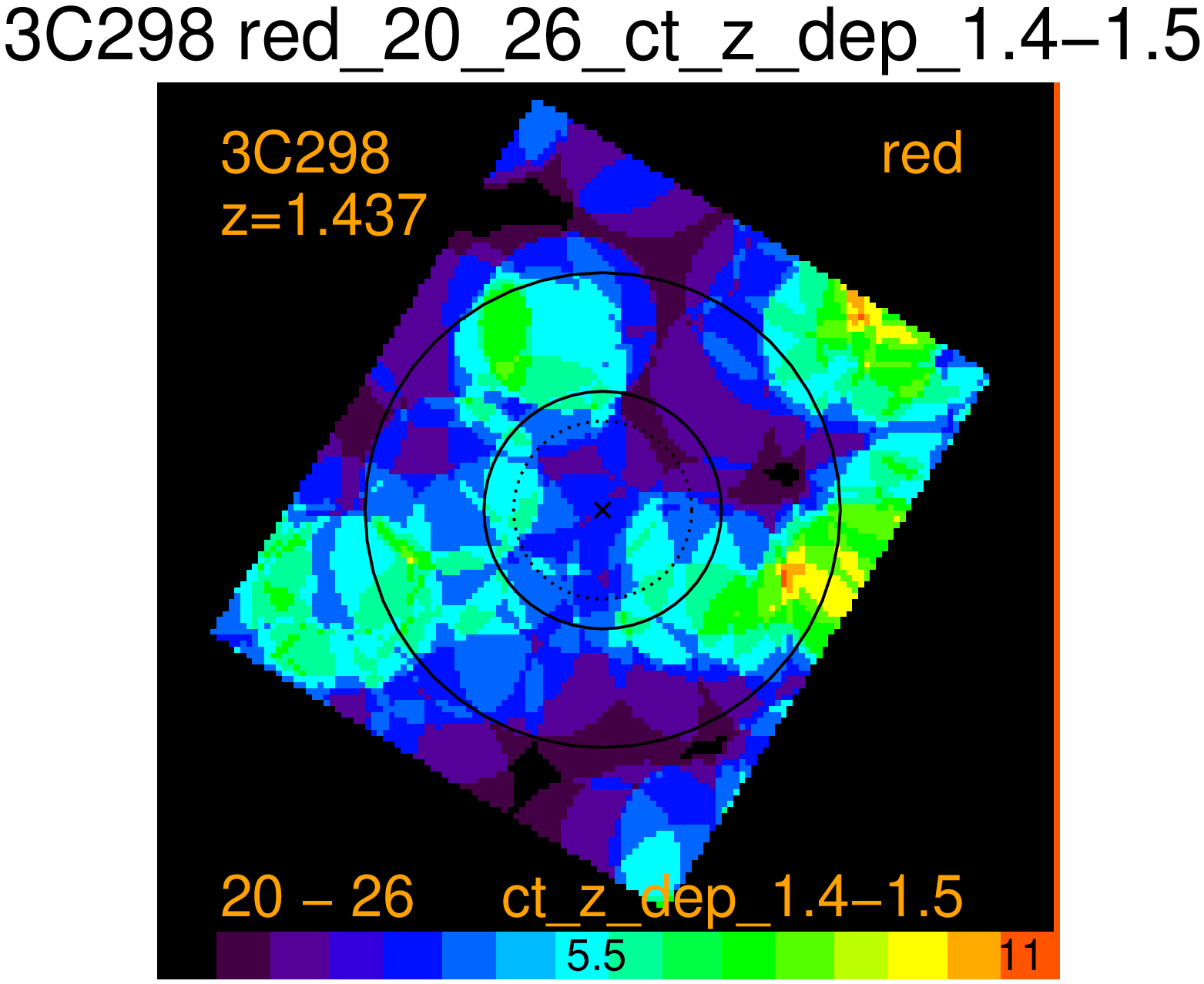}       
                \includegraphics[width=0.245\textwidth, clip=true]{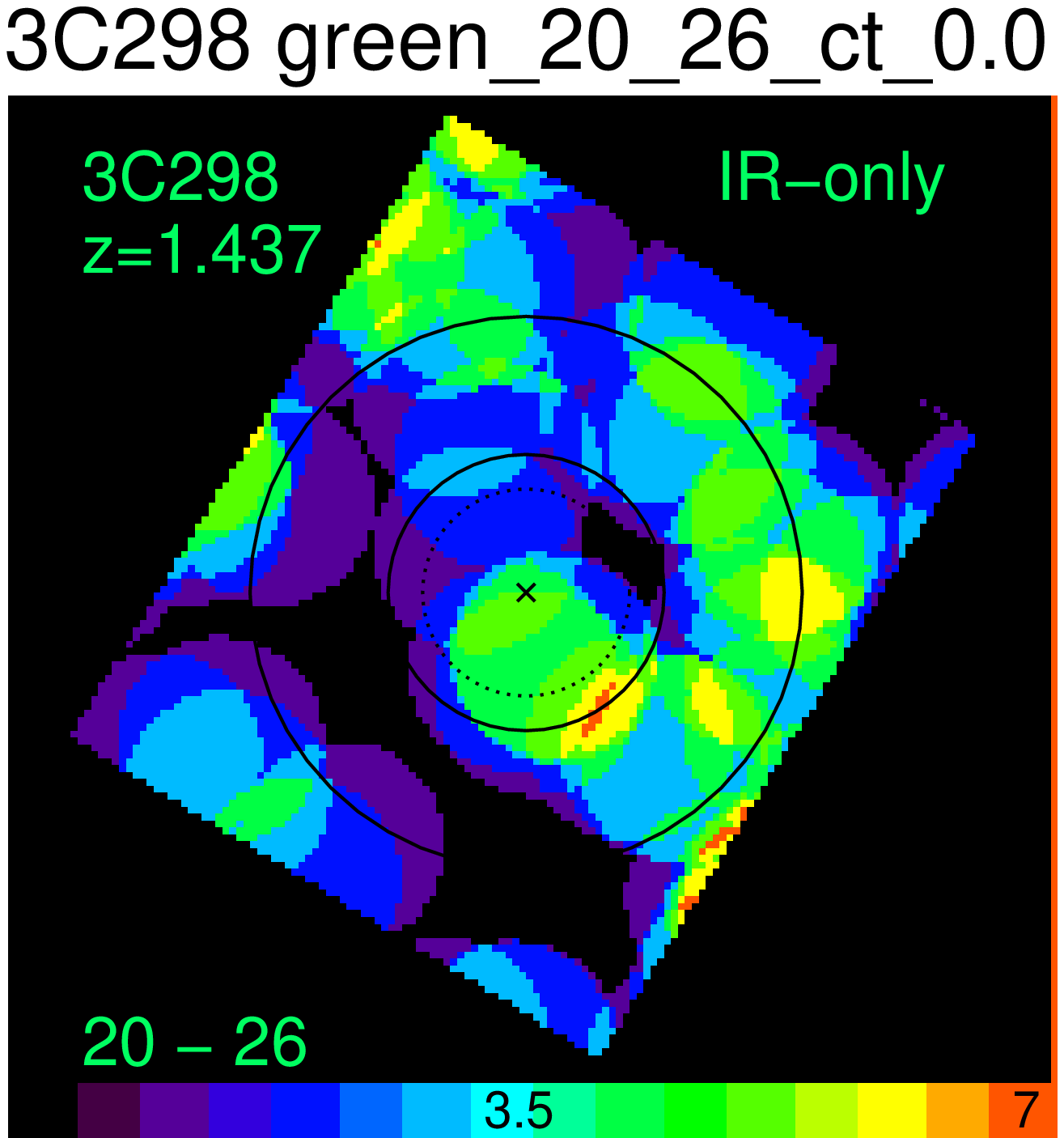}               
                \includegraphics[width=0.245\textwidth, clip=true]{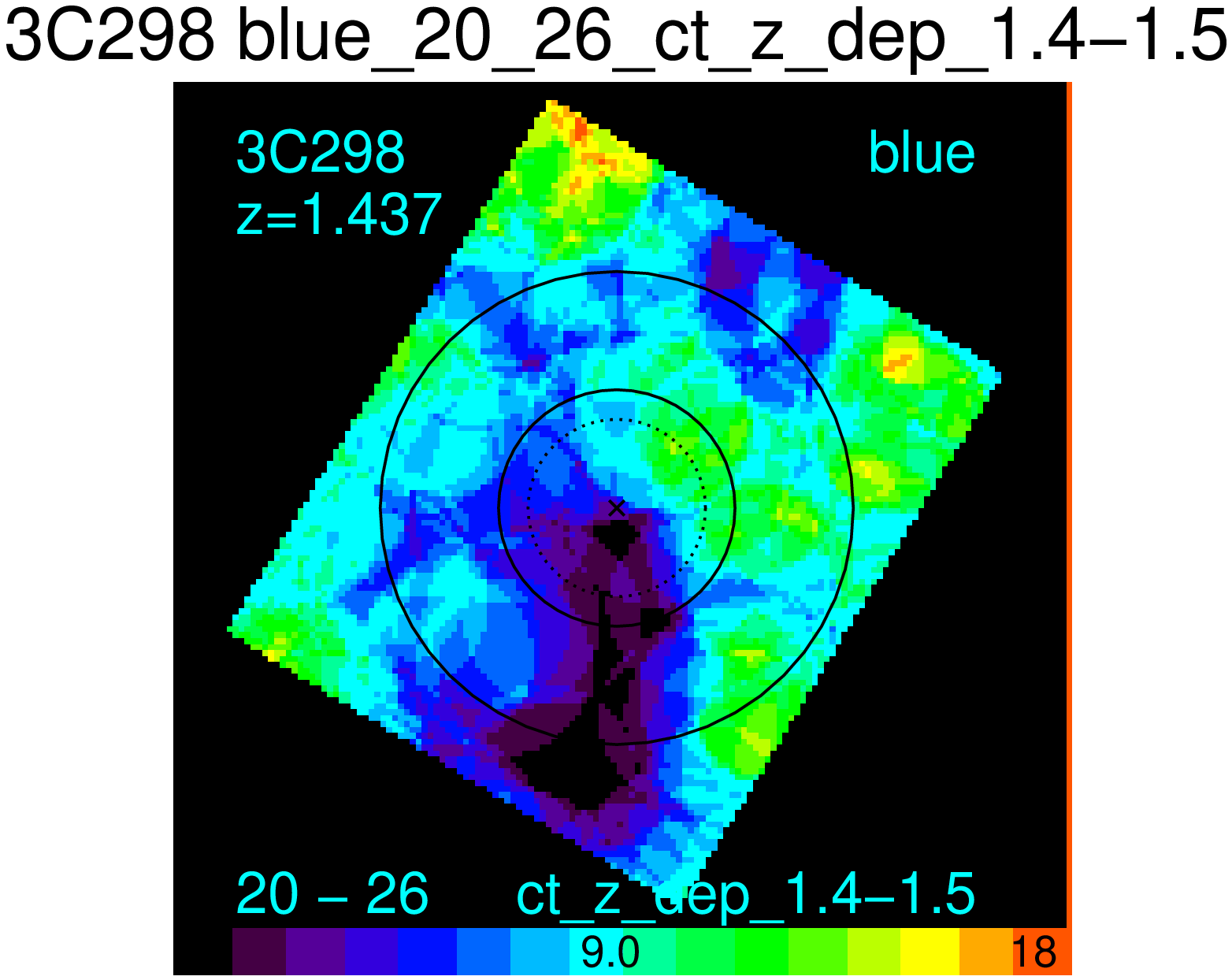}      
                
                \hspace{-0mm}\includegraphics[width=0.245\textwidth, clip=true]{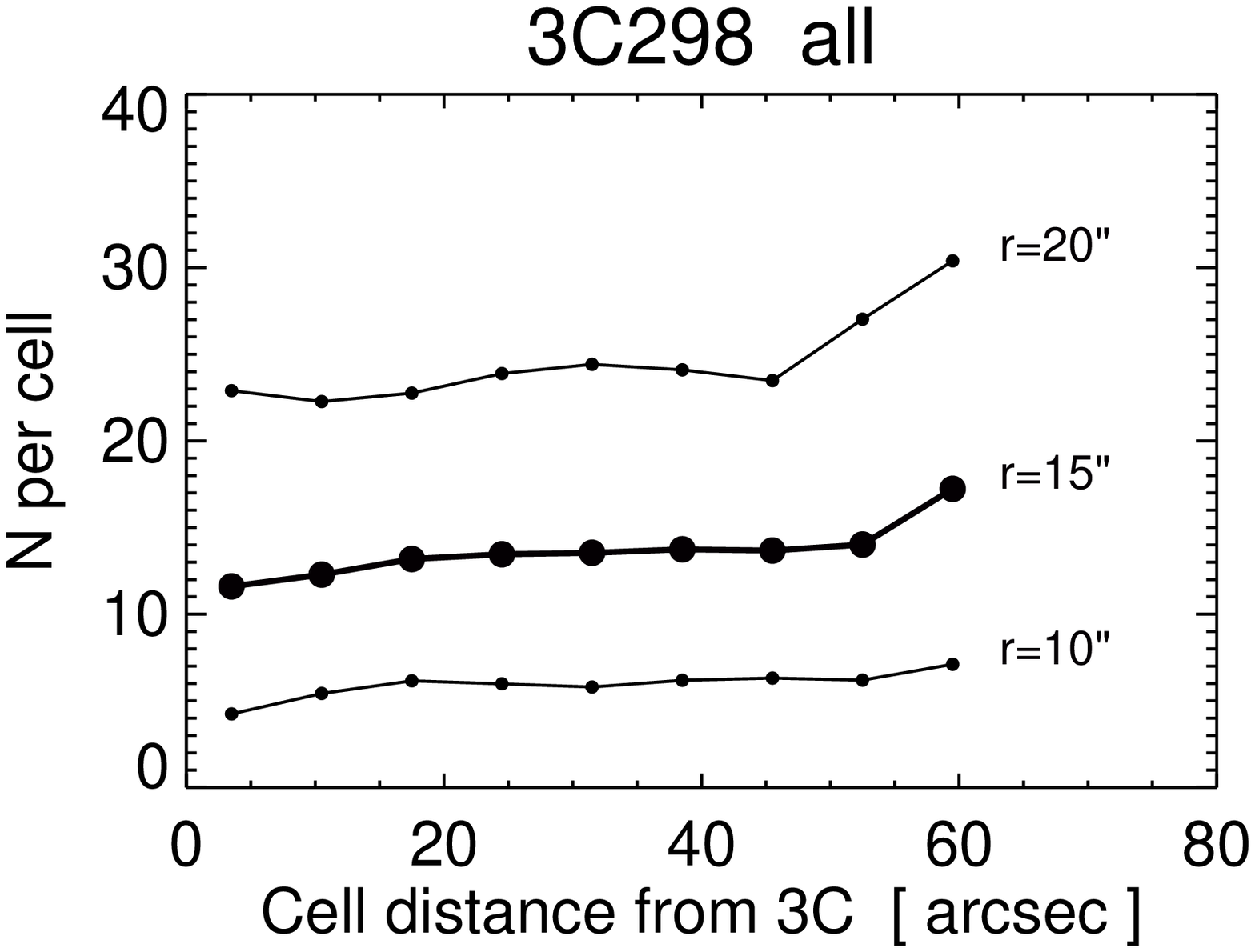}                   
                \includegraphics[width=0.245\textwidth, clip=true]{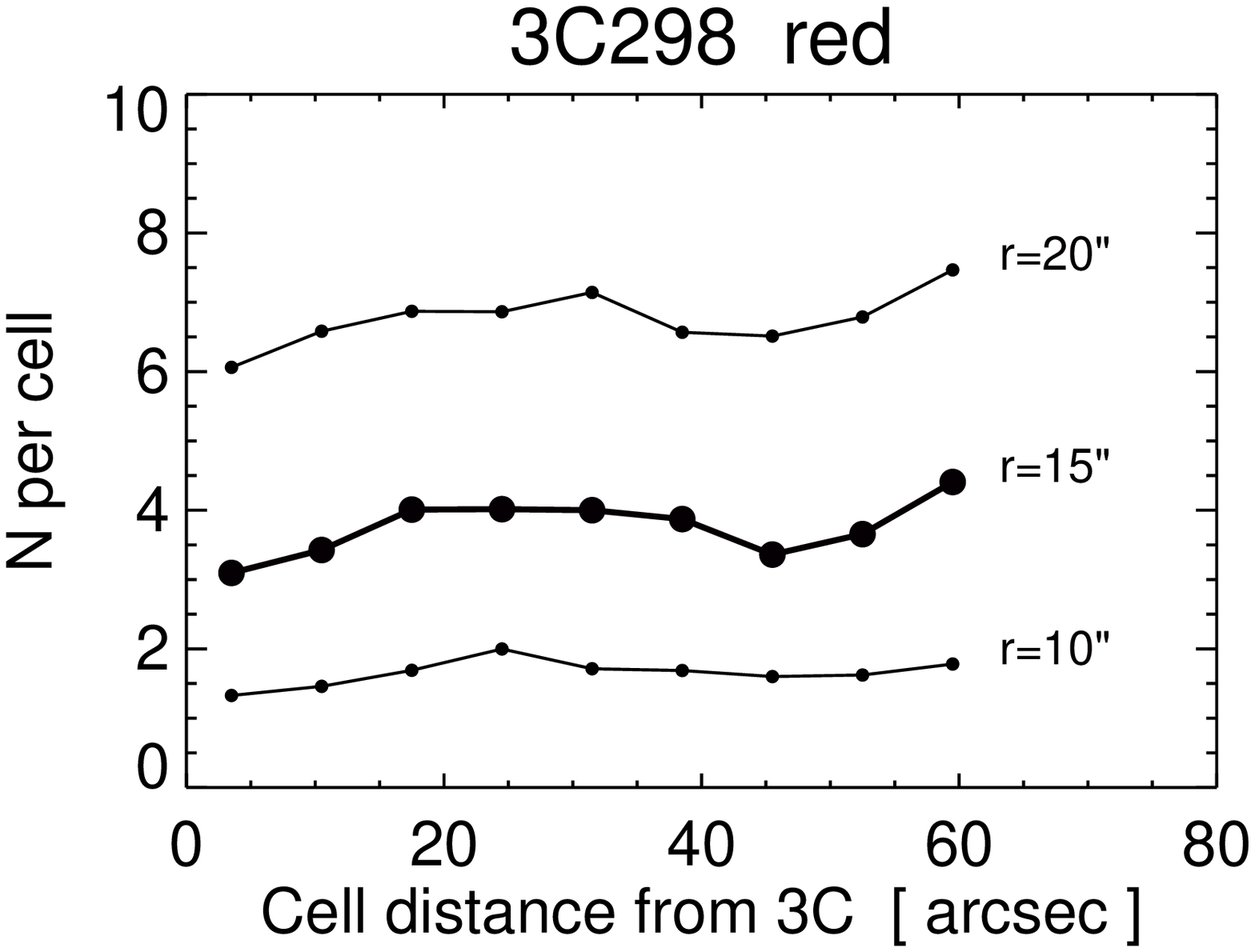}                    
                \includegraphics[width=0.245\textwidth, clip=true]{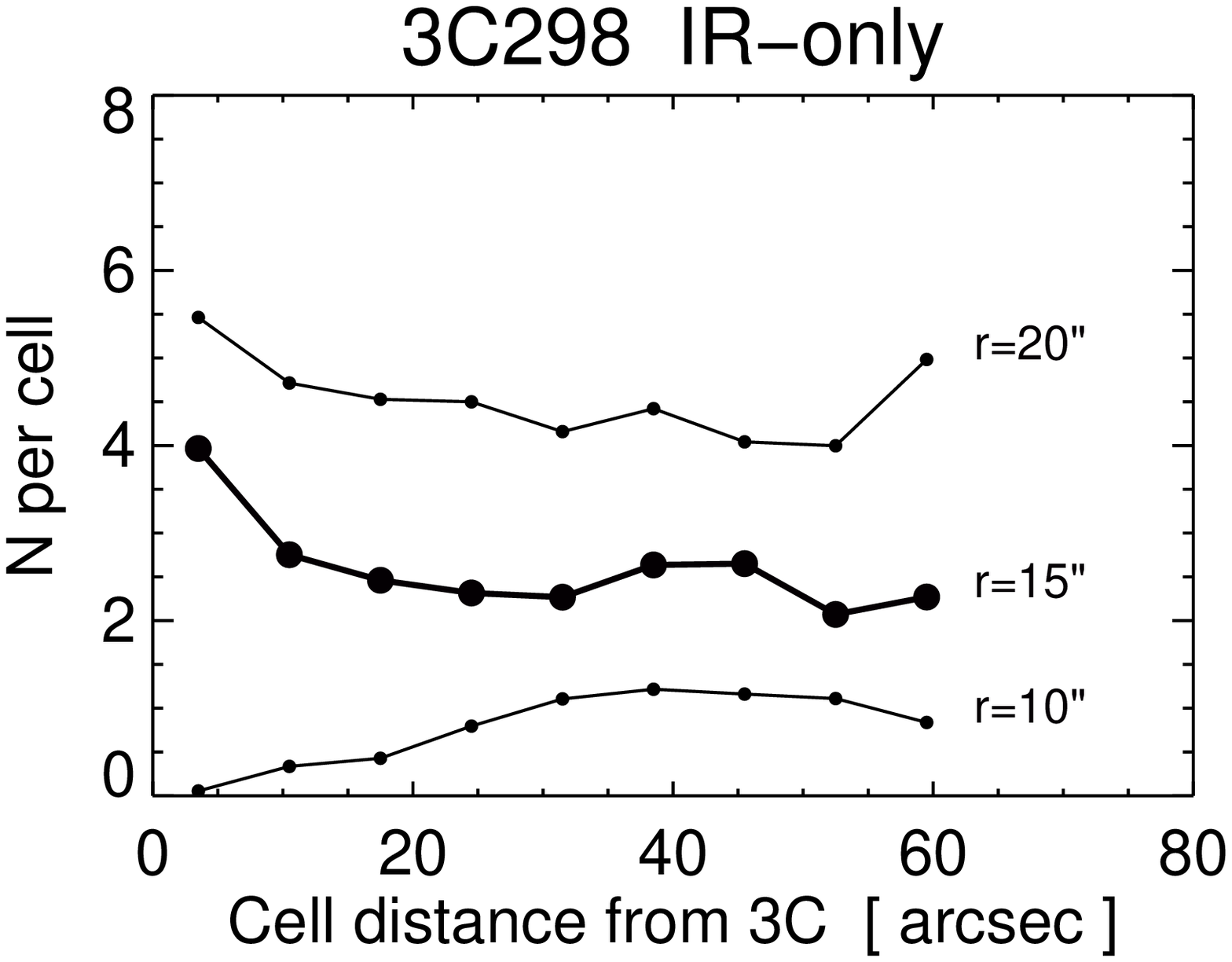}                 
                \includegraphics[width=0.245\textwidth, clip=true]{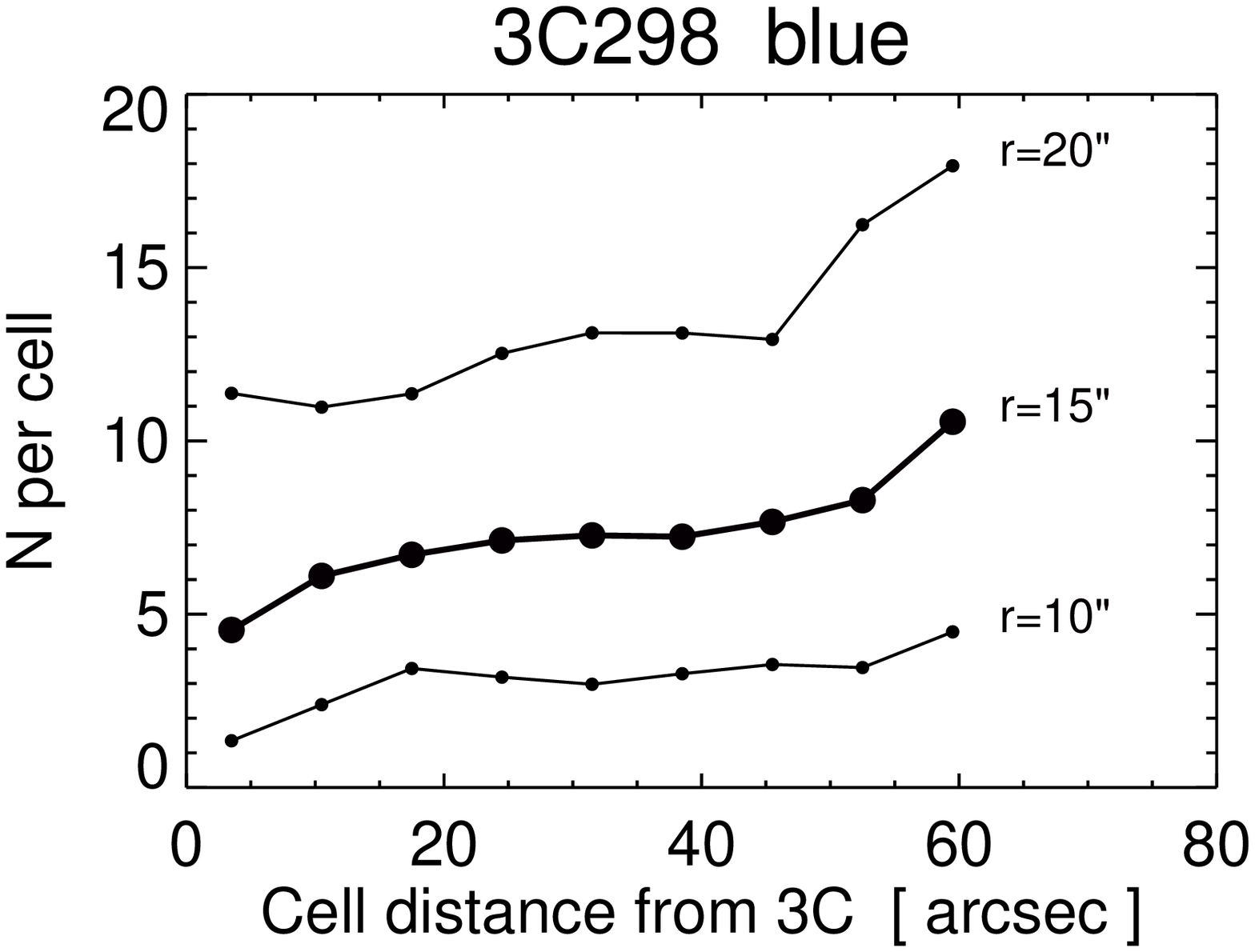}

                \hspace{-0mm}\includegraphics[width=0.245\textwidth, clip=true]{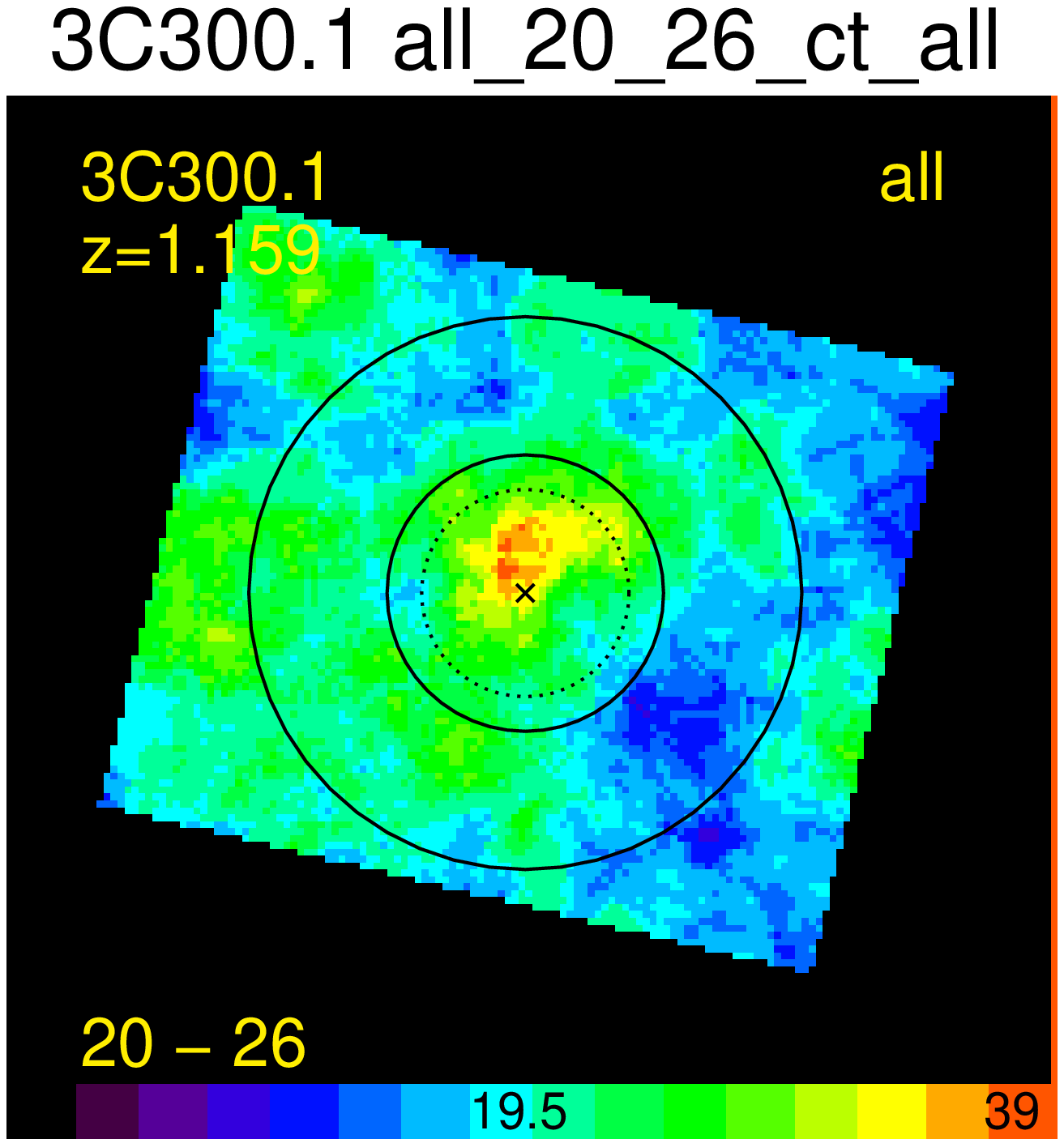}               
                \includegraphics[width=0.245\textwidth, clip=true]{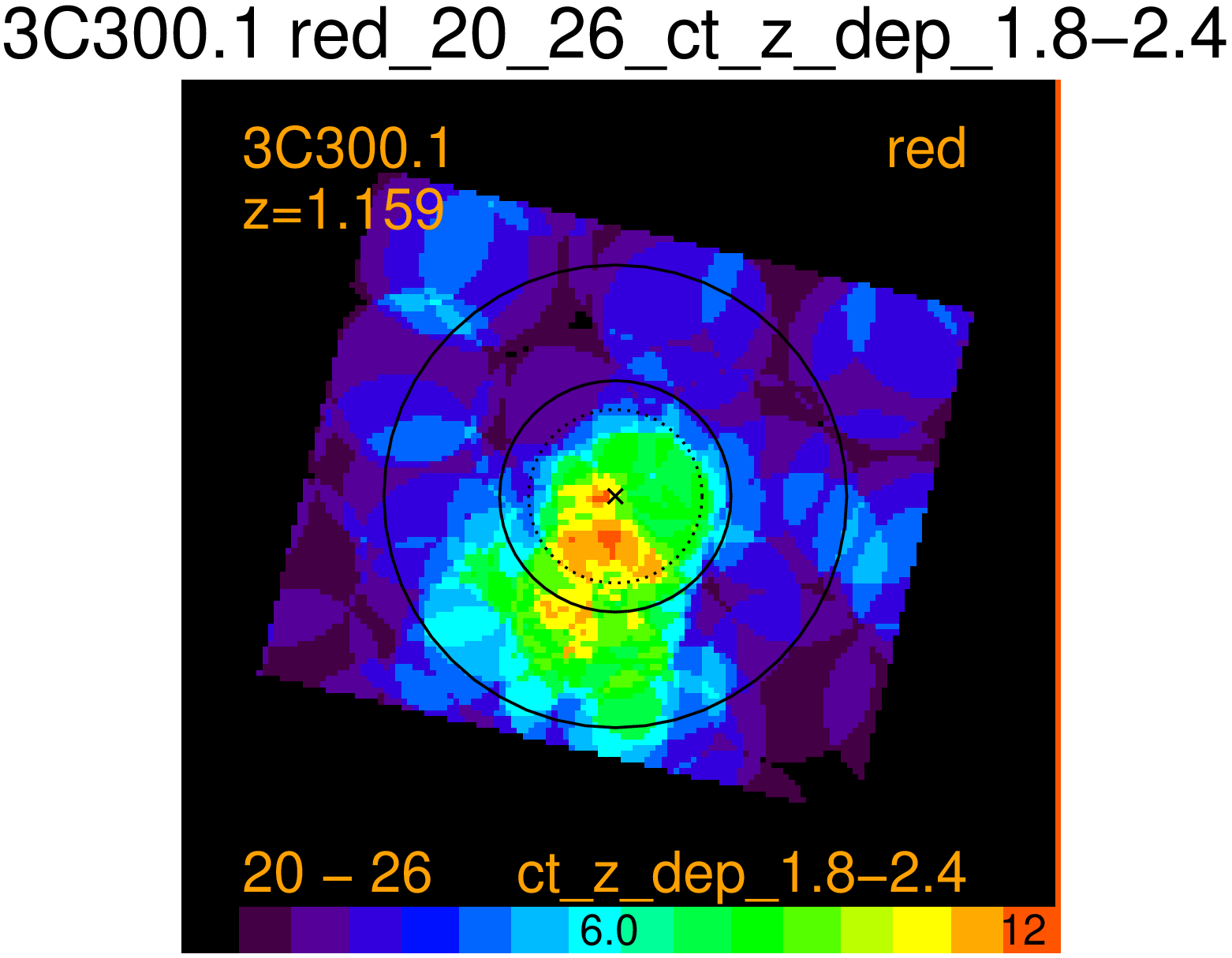}     
                \includegraphics[width=0.245\textwidth, clip=true]{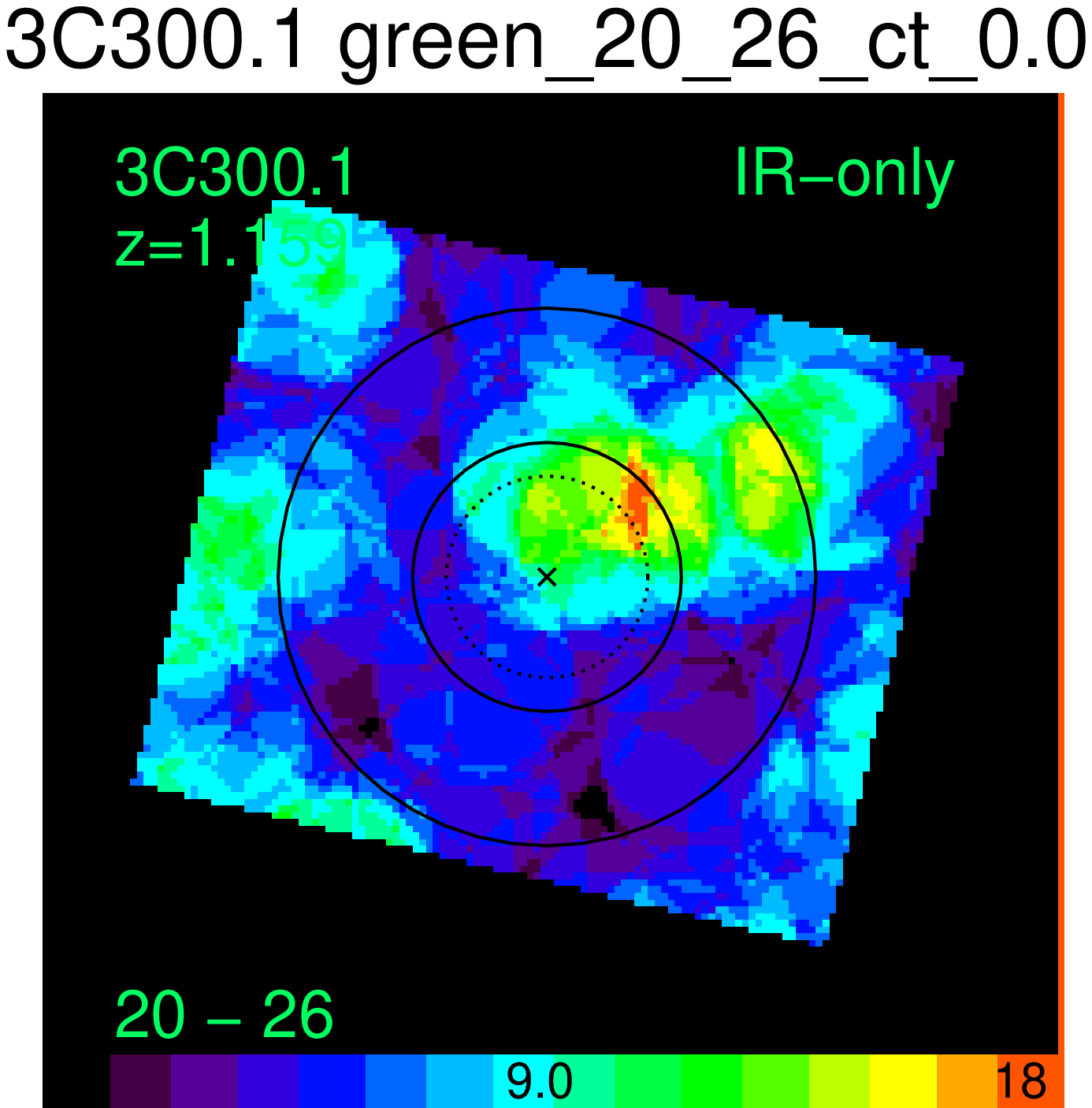}             
                \includegraphics[width=0.245\textwidth, clip=true]{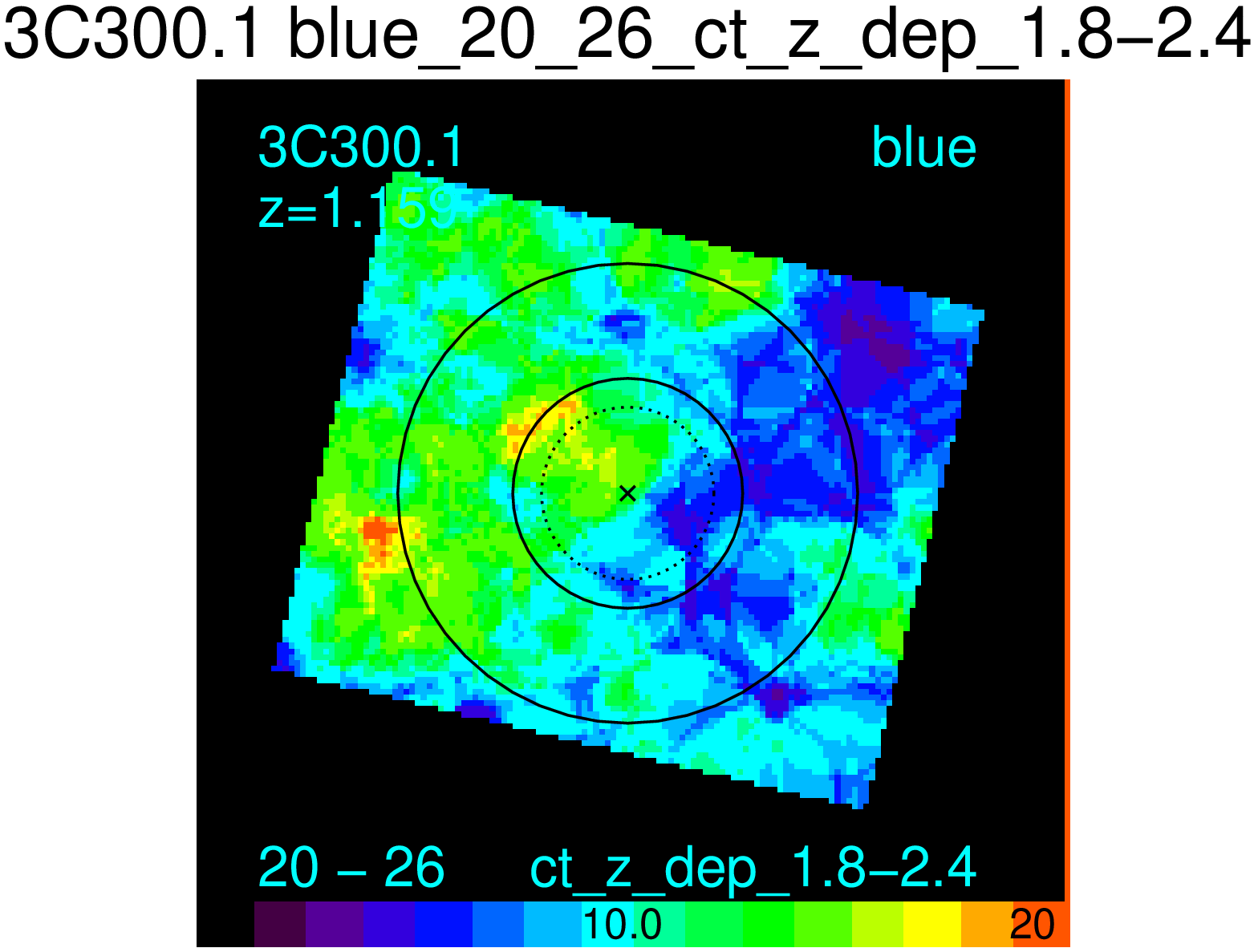}    
                
                \hspace{-0mm}\includegraphics[width=0.245\textwidth, clip=true]{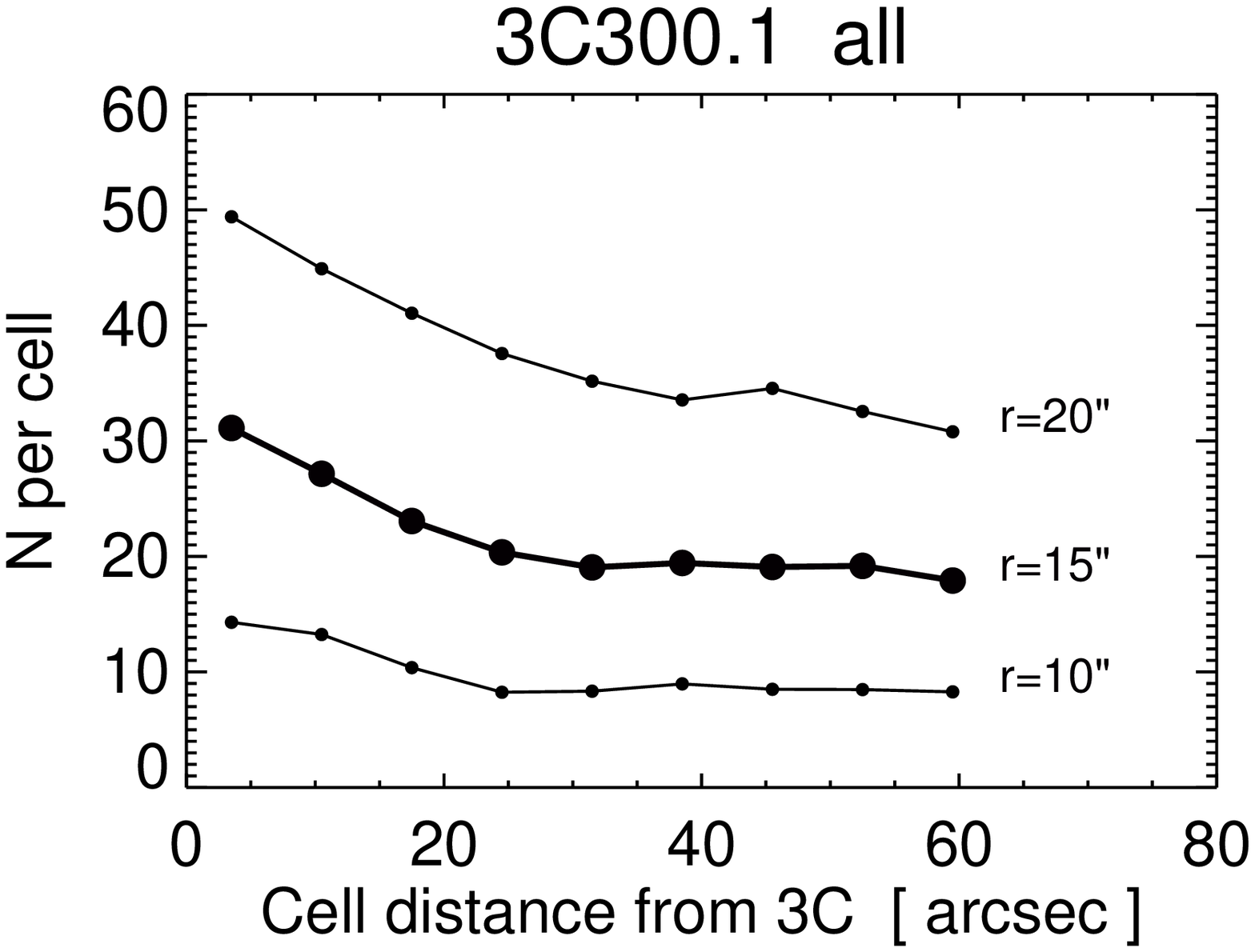}                 
                \includegraphics[width=0.245\textwidth, clip=true]{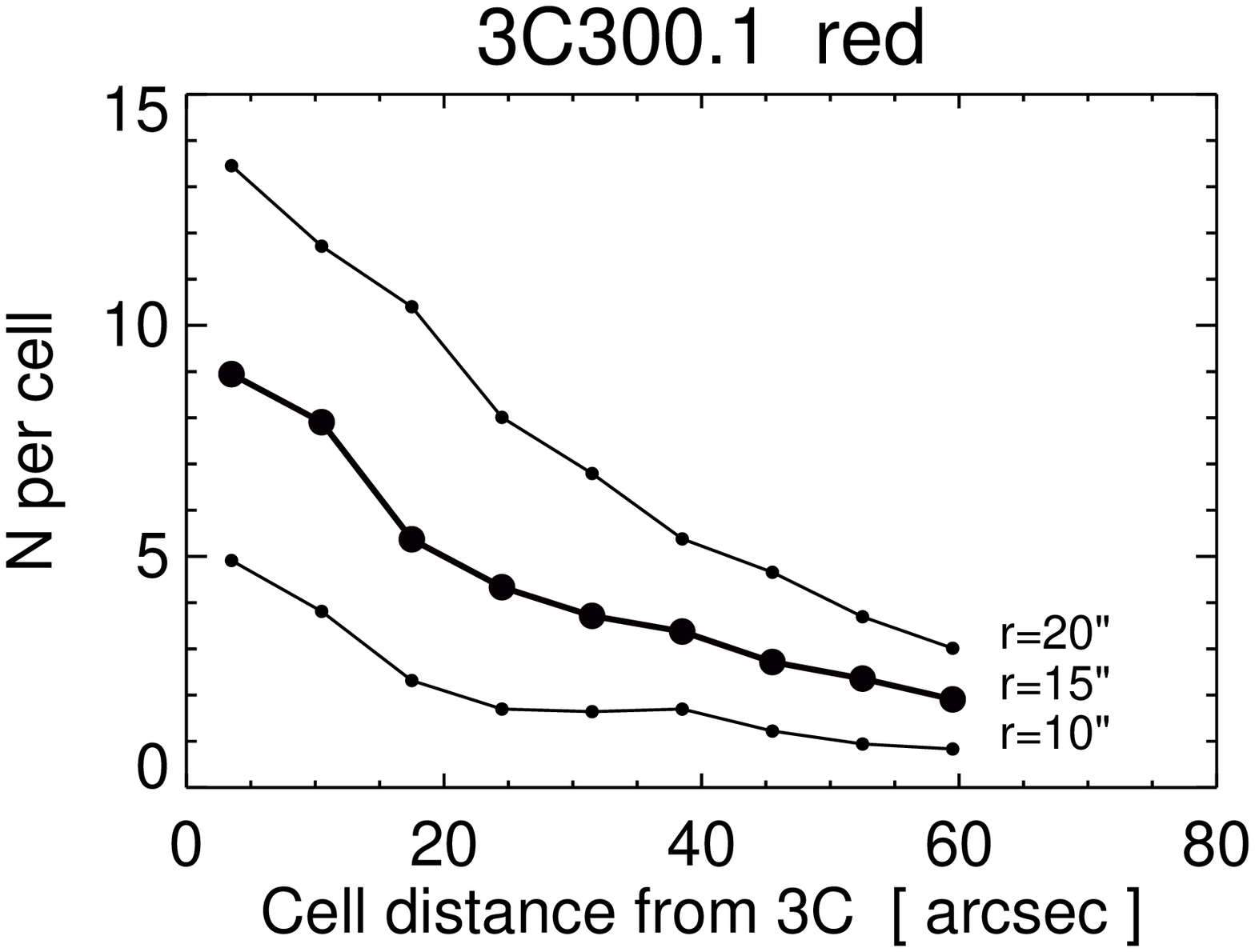}                  
                \includegraphics[width=0.245\textwidth, clip=true]{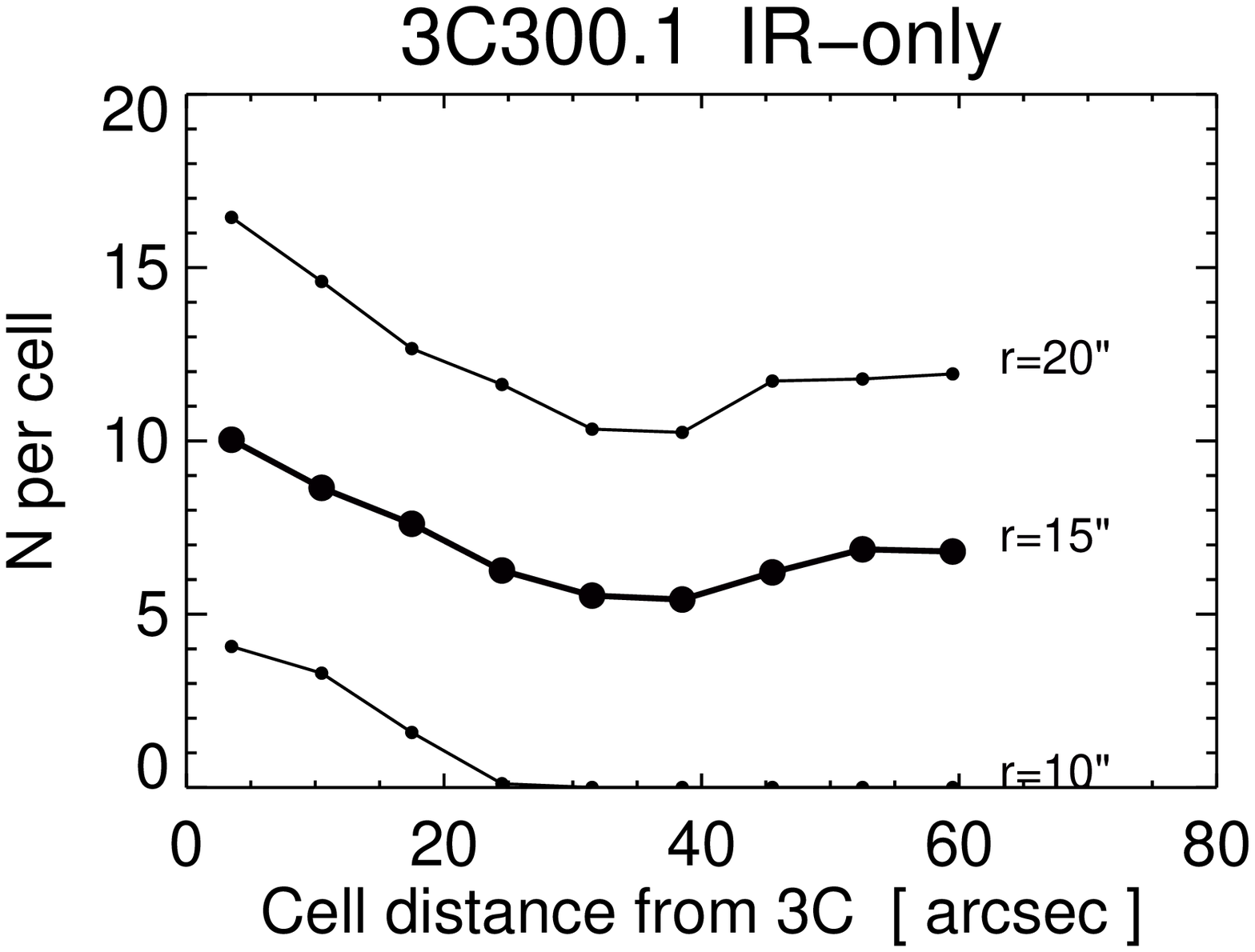}               
                \includegraphics[width=0.245\textwidth, clip=true]{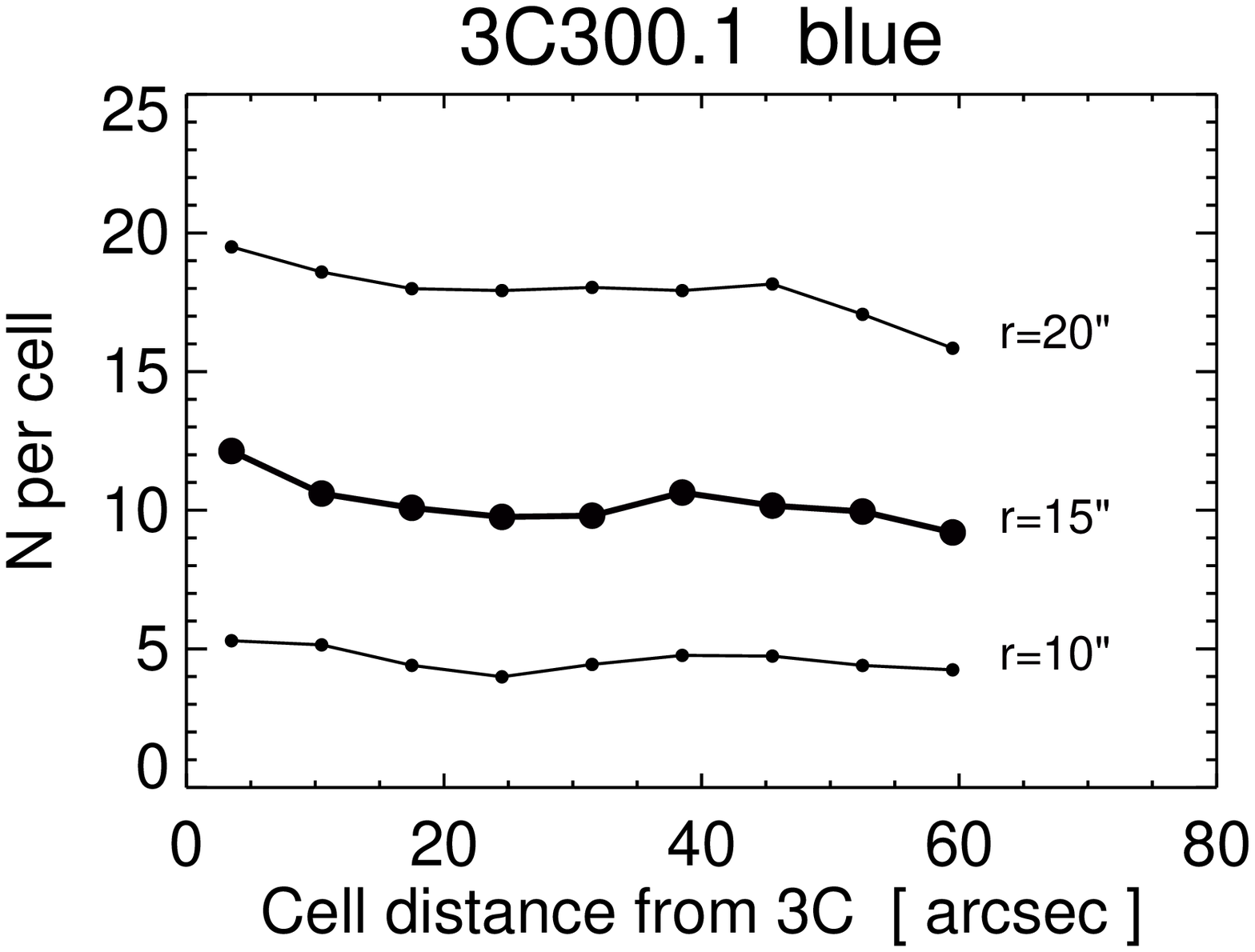}

                \hspace{-0mm}\includegraphics[width=0.245\textwidth, clip=true]{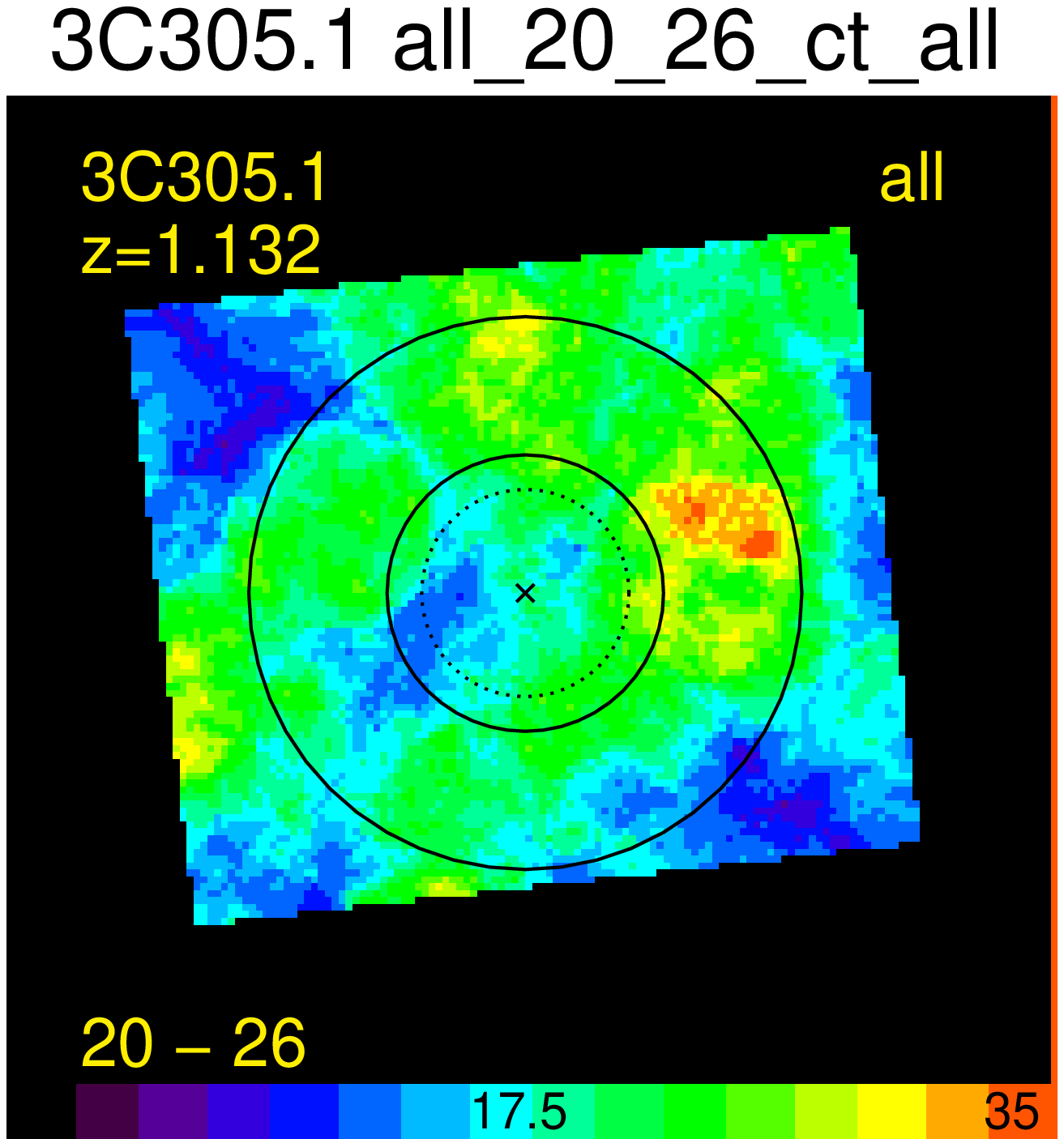}               
                \includegraphics[width=0.245\textwidth, clip=true]{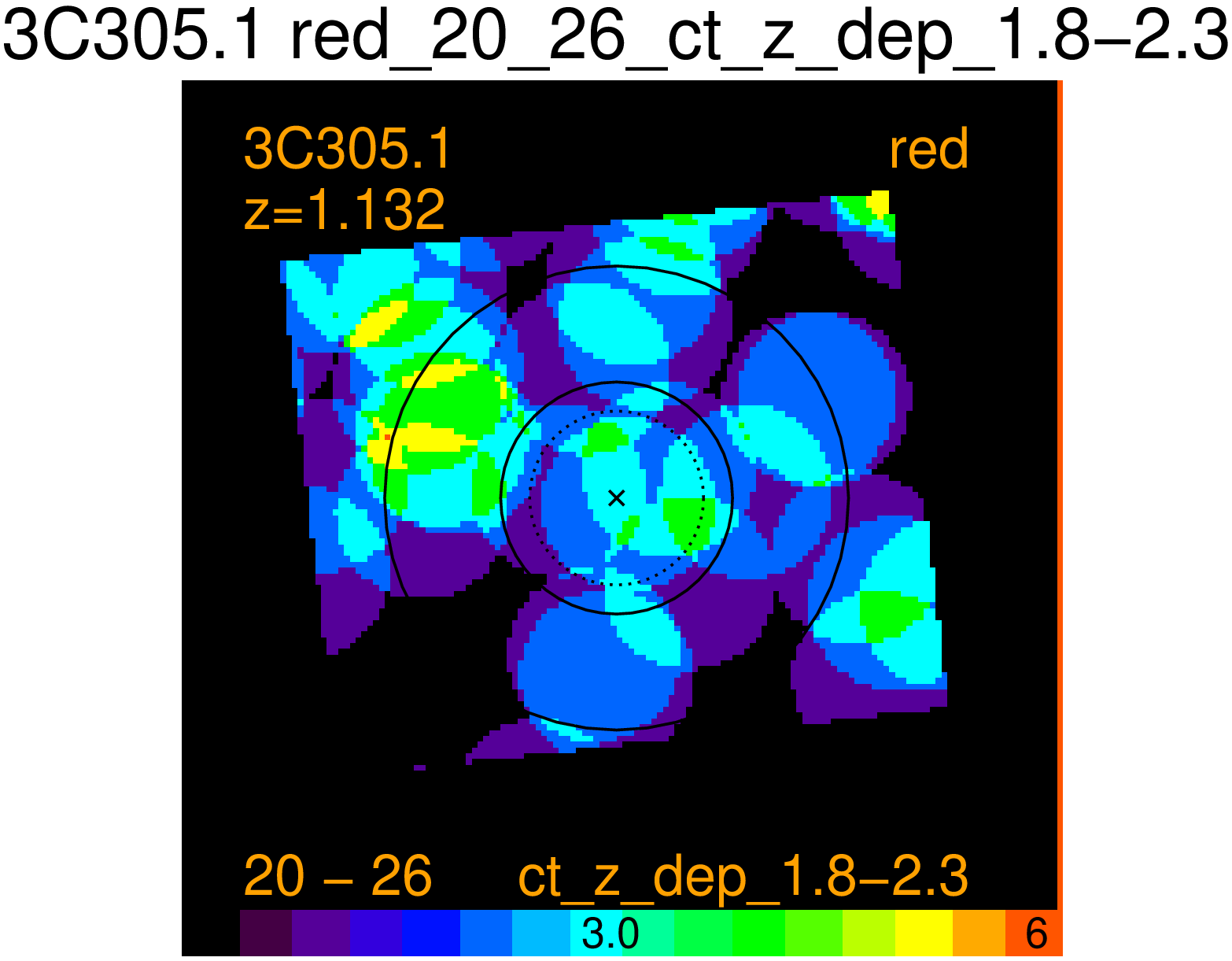}     
                \includegraphics[width=0.245\textwidth, clip=true]{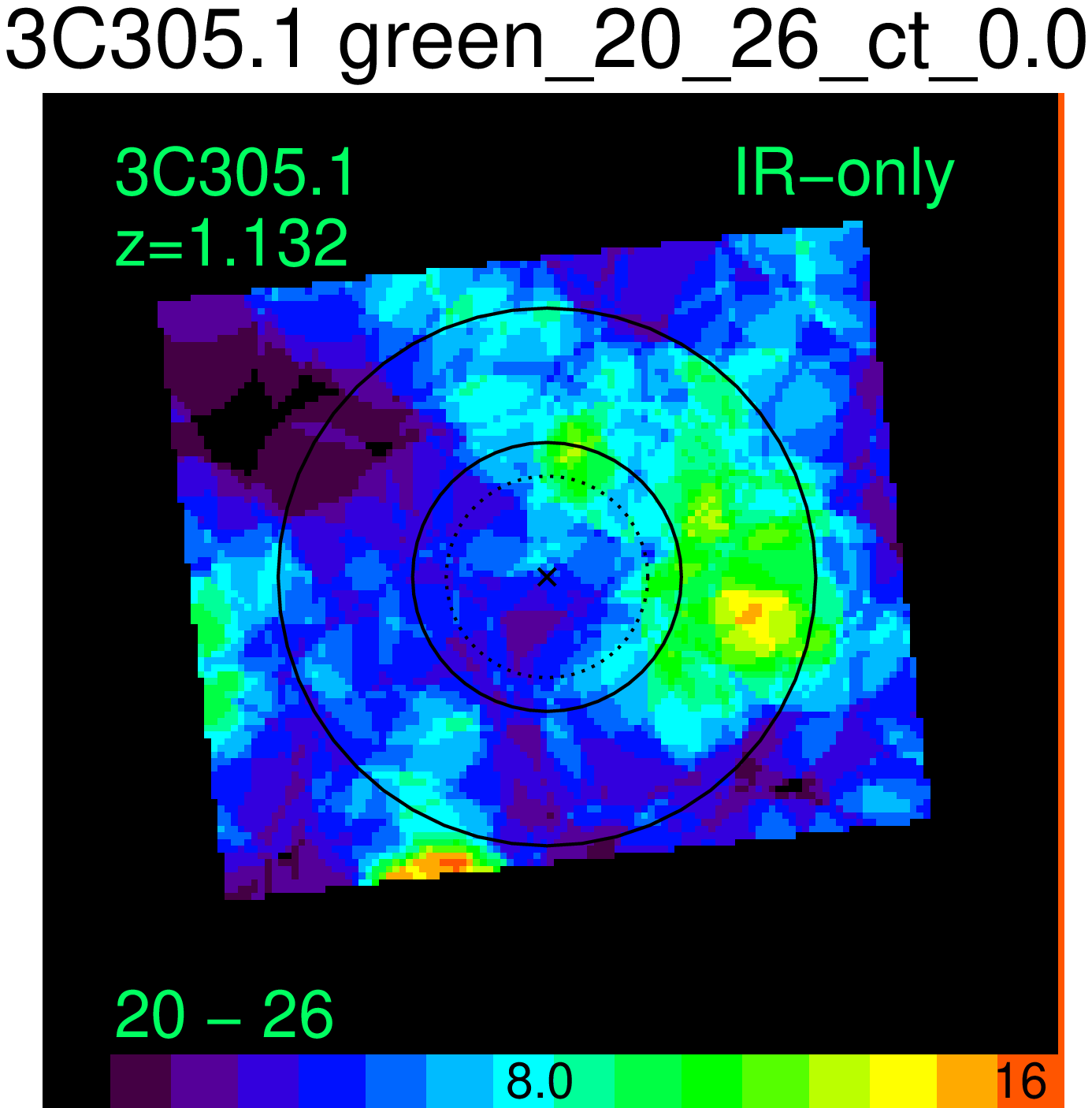}             
                \includegraphics[width=0.245\textwidth, clip=true]{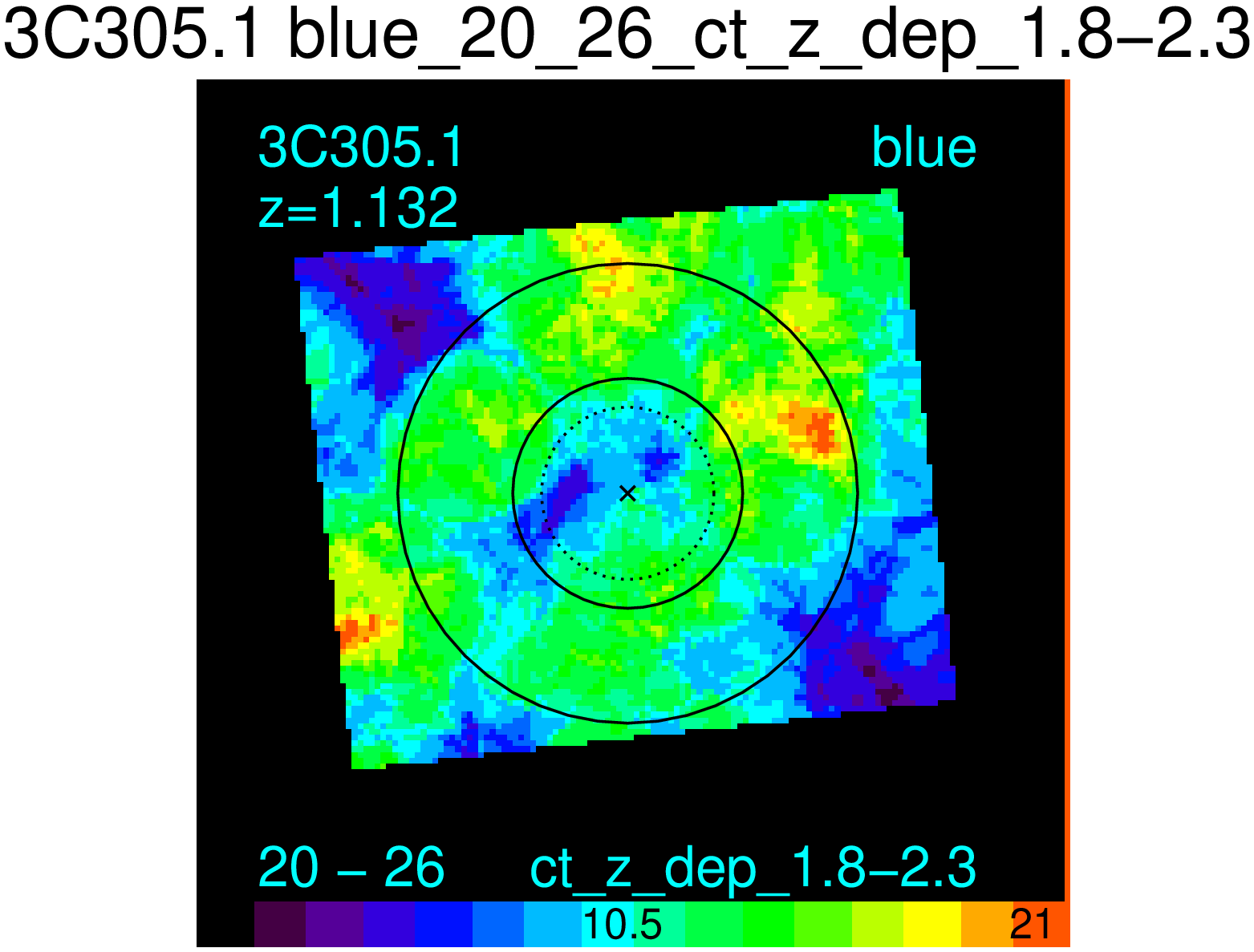}    
                
                \hspace{-0mm}\includegraphics[width=0.245\textwidth, clip=true]{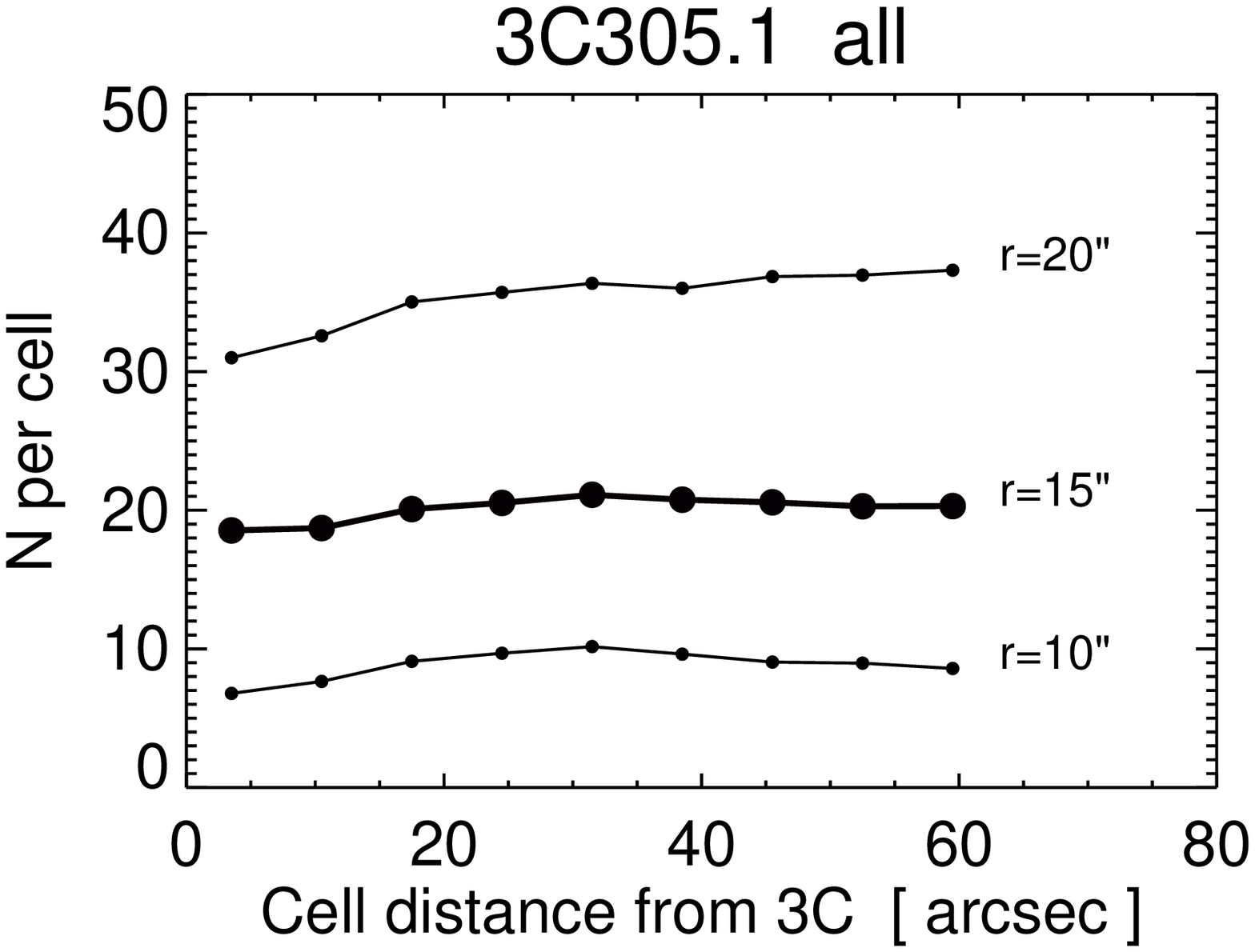}                 
                \includegraphics[width=0.245\textwidth, clip=true]{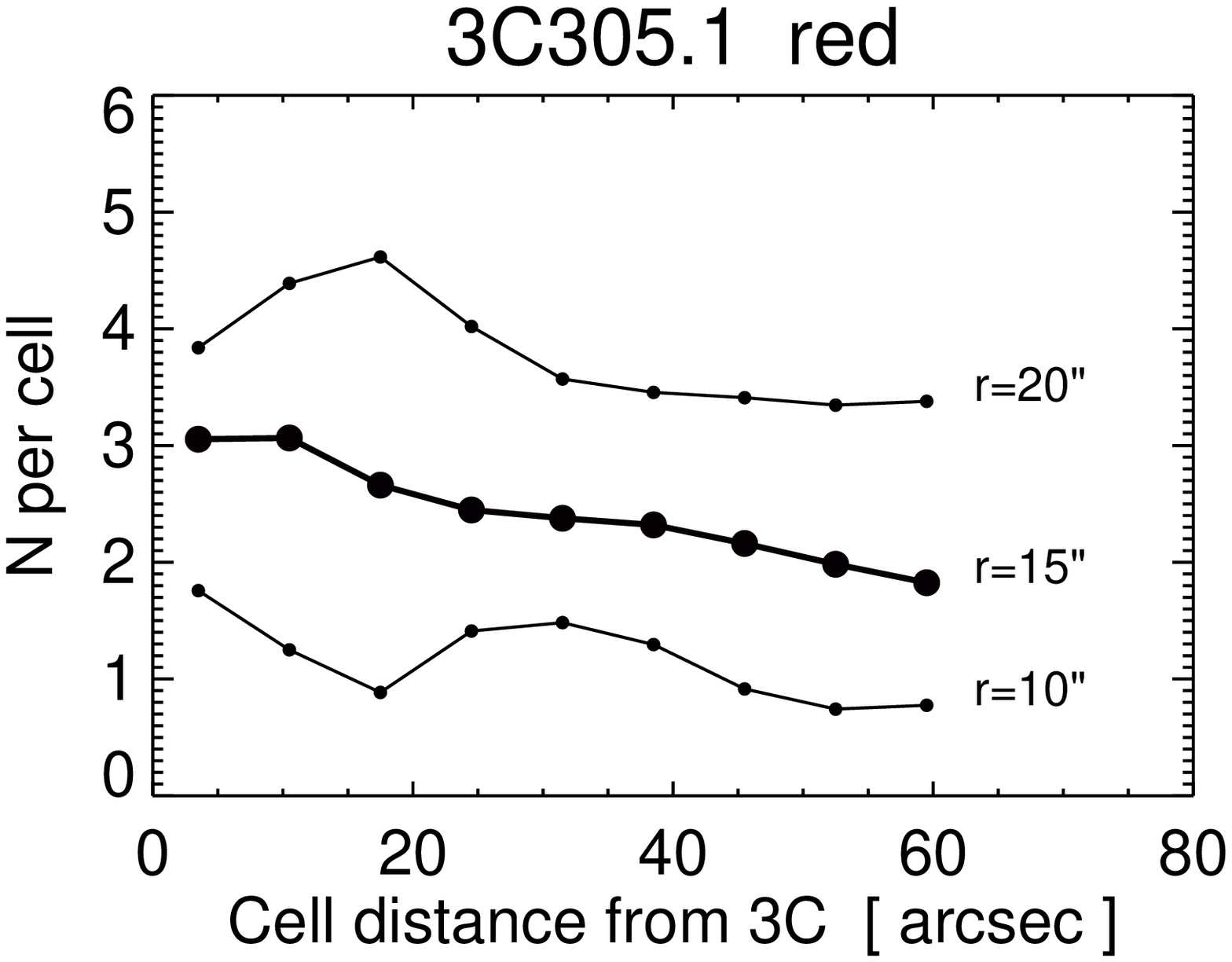}                  
                \includegraphics[width=0.245\textwidth, clip=true]{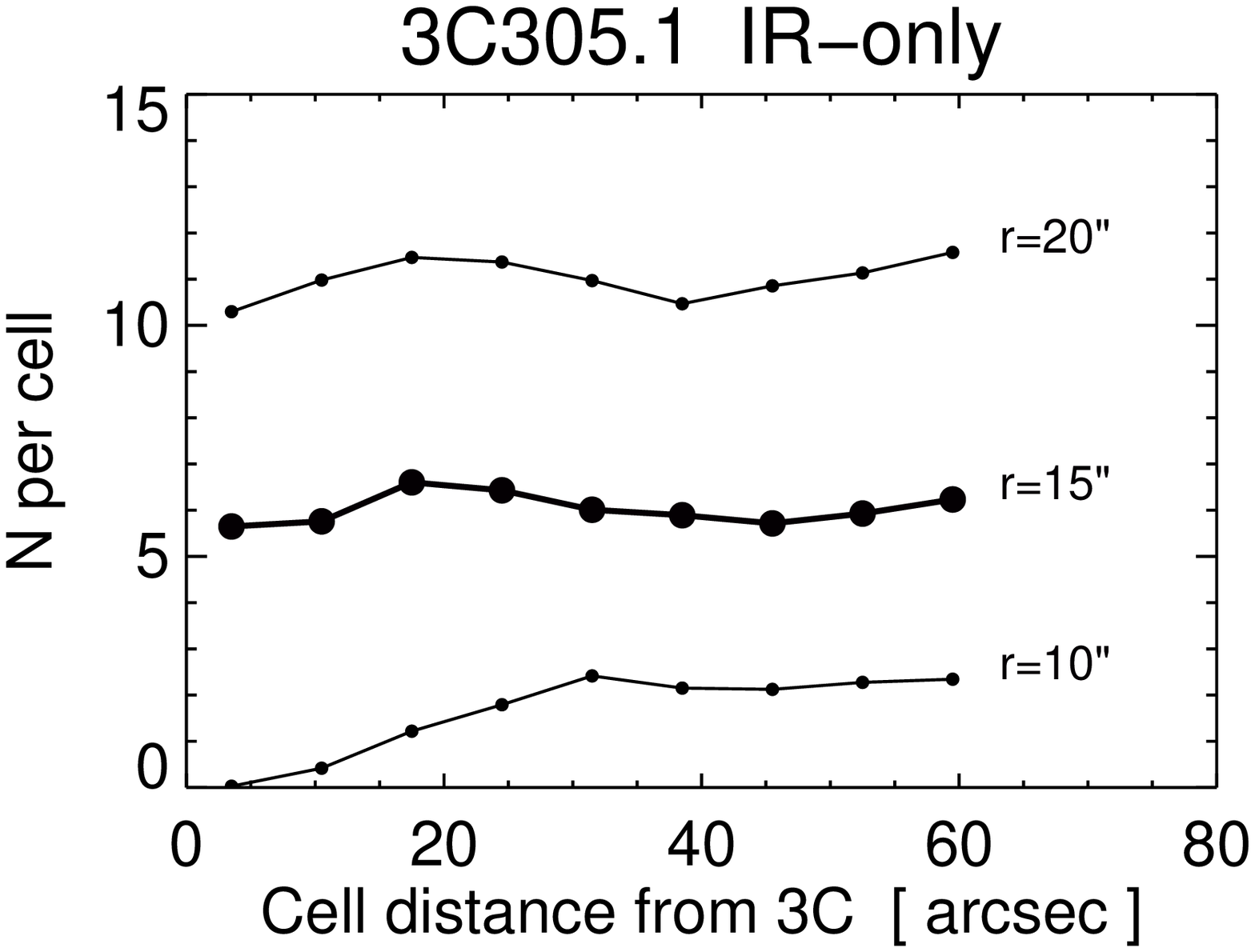}               
                \includegraphics[width=0.245\textwidth, clip=true]{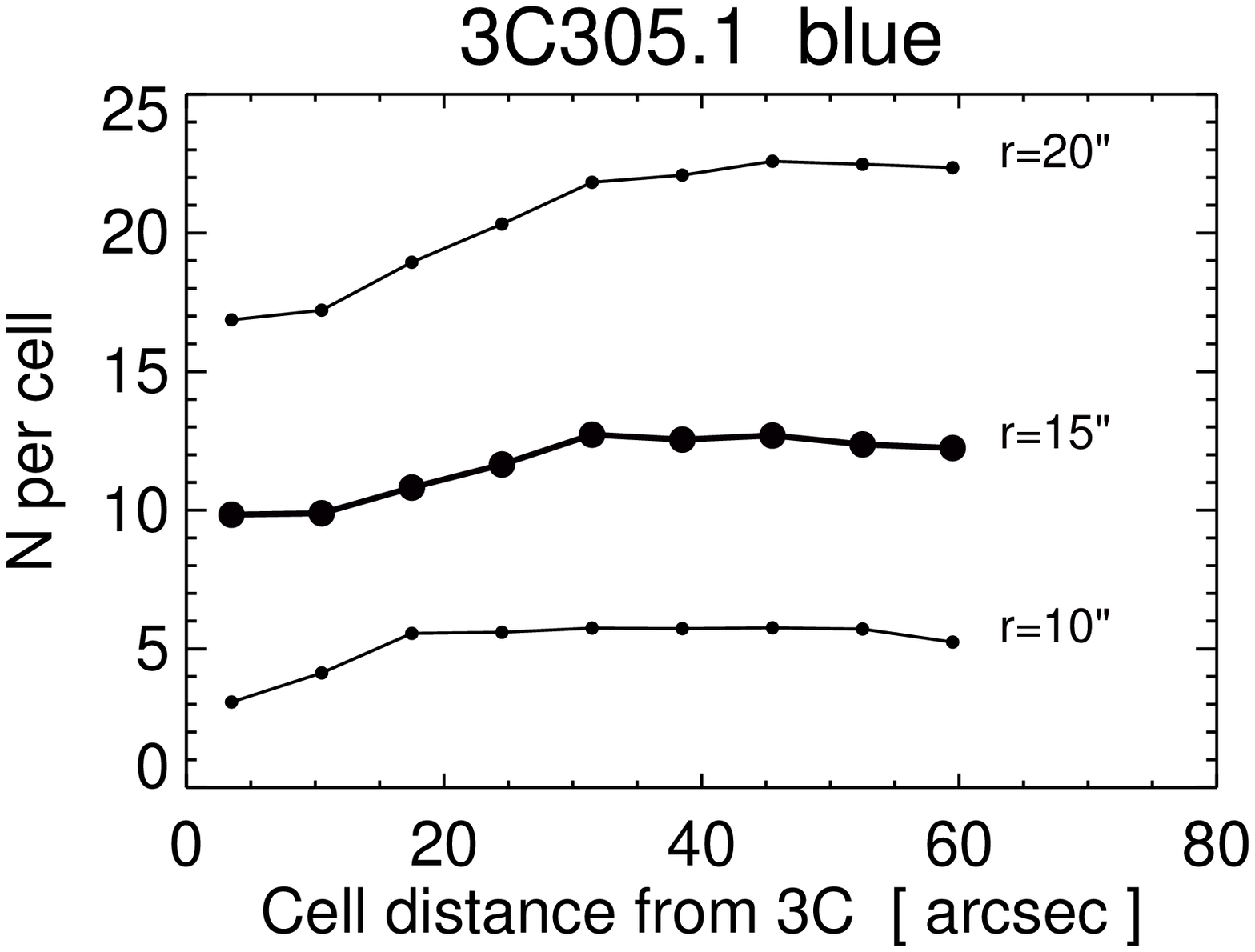}

                \caption{Surface density maps and radial density profiles of the 3C fields, continued.
                }
                \label{fig:sd_maps_5}
              \end{figure*}

\begin{figure*}

                \hspace{-0mm}\includegraphics[width=0.245\textwidth, clip=true]{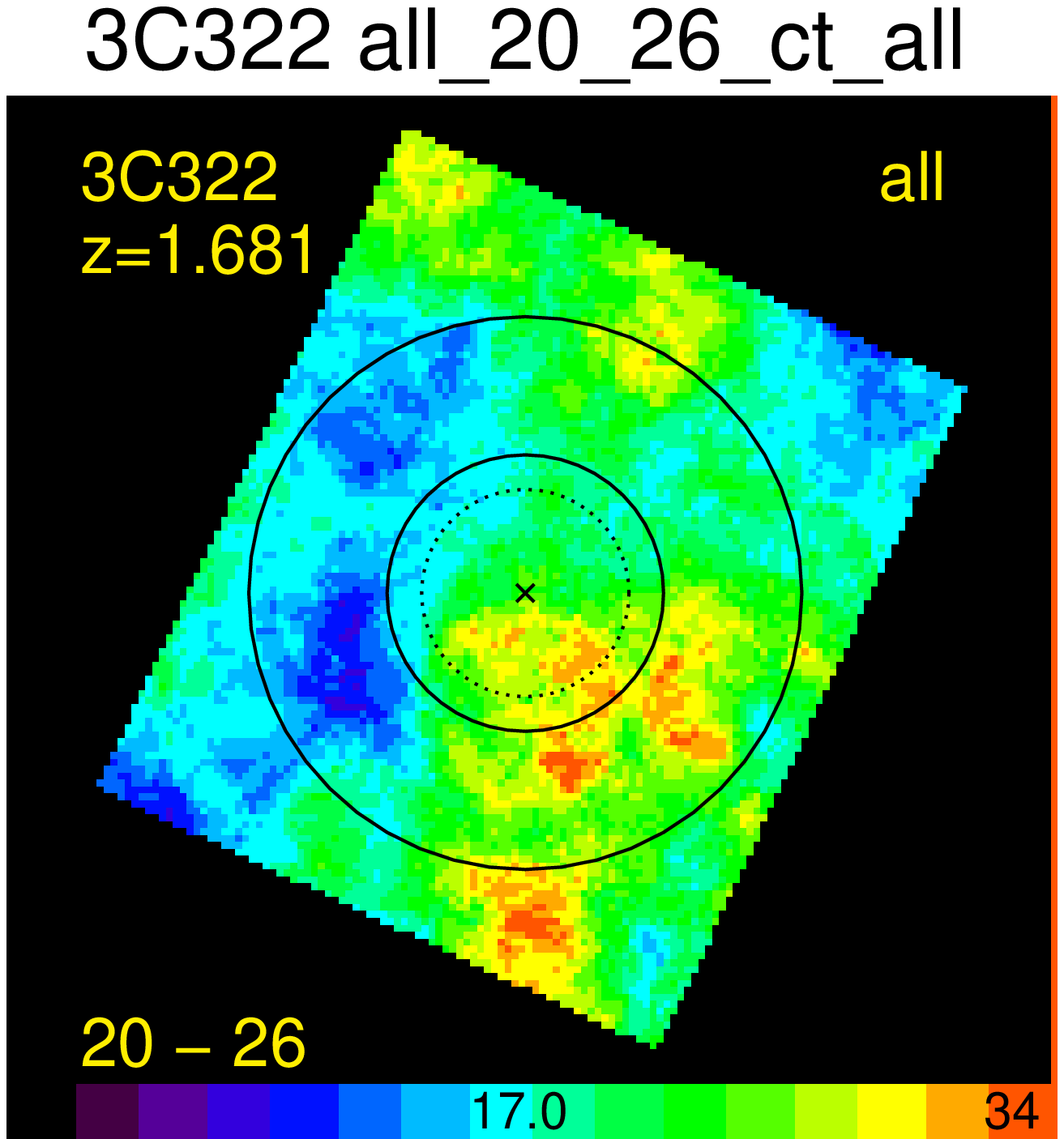}                 
                \includegraphics[width=0.245\textwidth, clip=true]{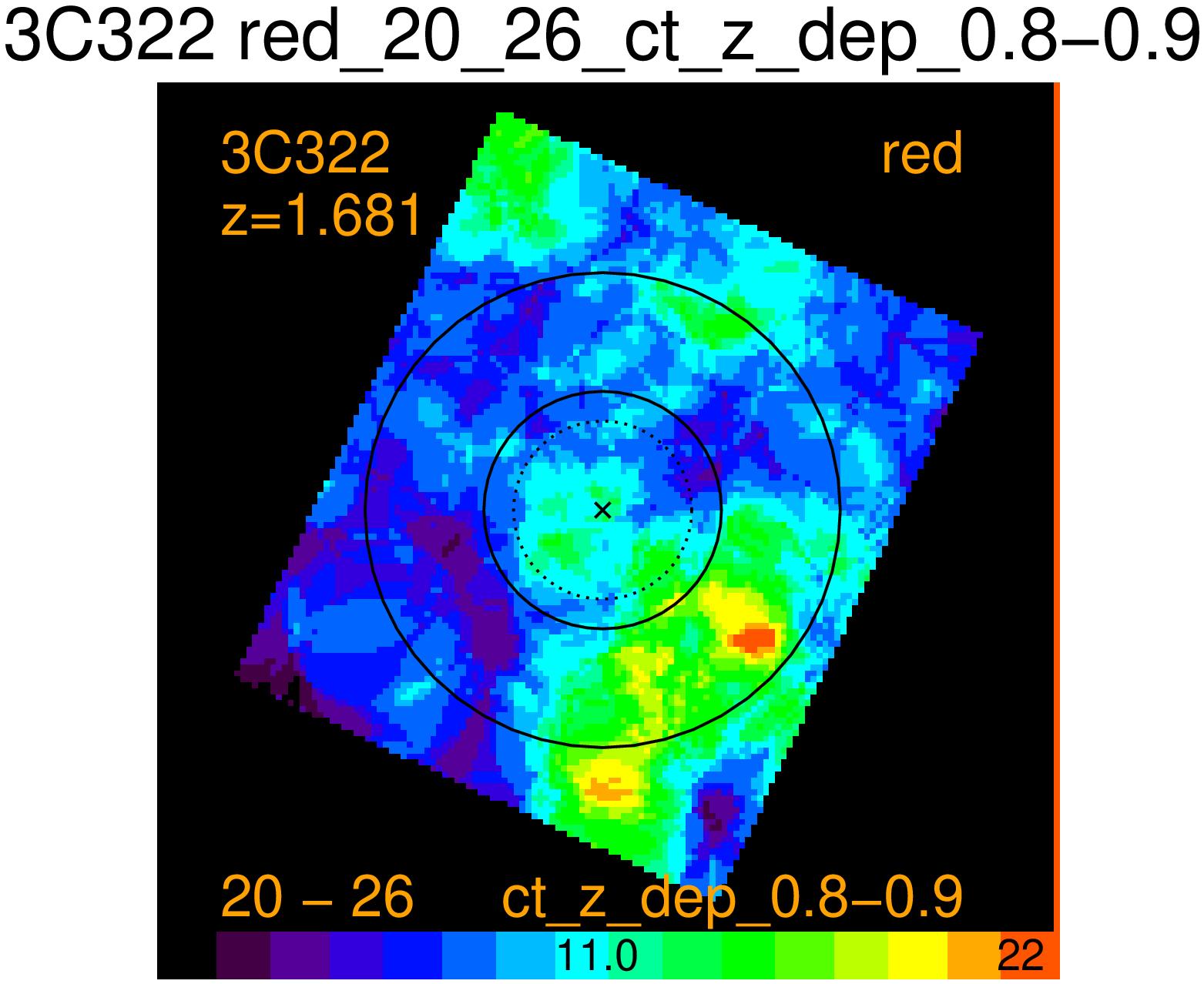}       
                \includegraphics[width=0.245\textwidth, clip=true]{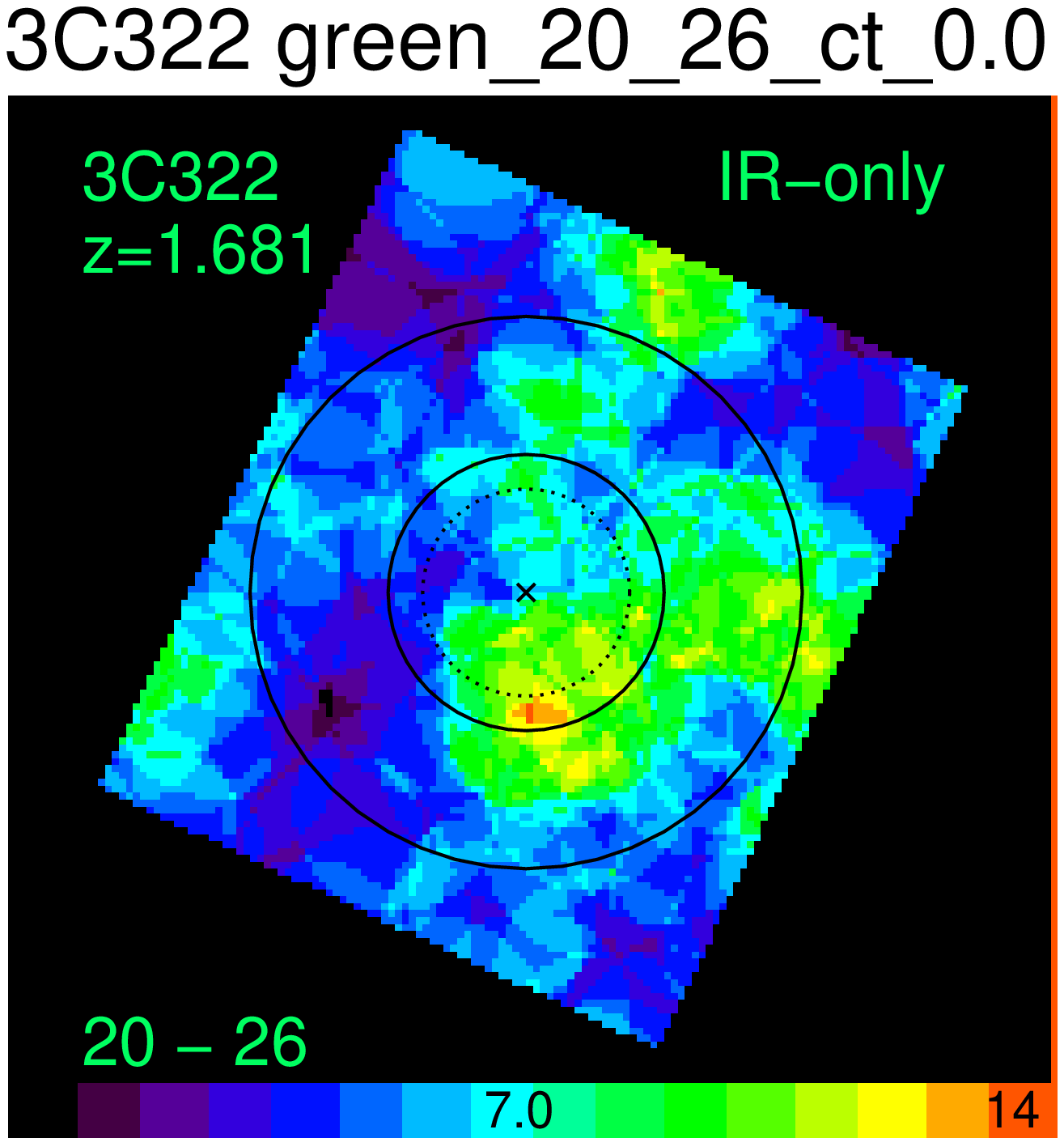}               
                \includegraphics[width=0.245\textwidth, clip=true]{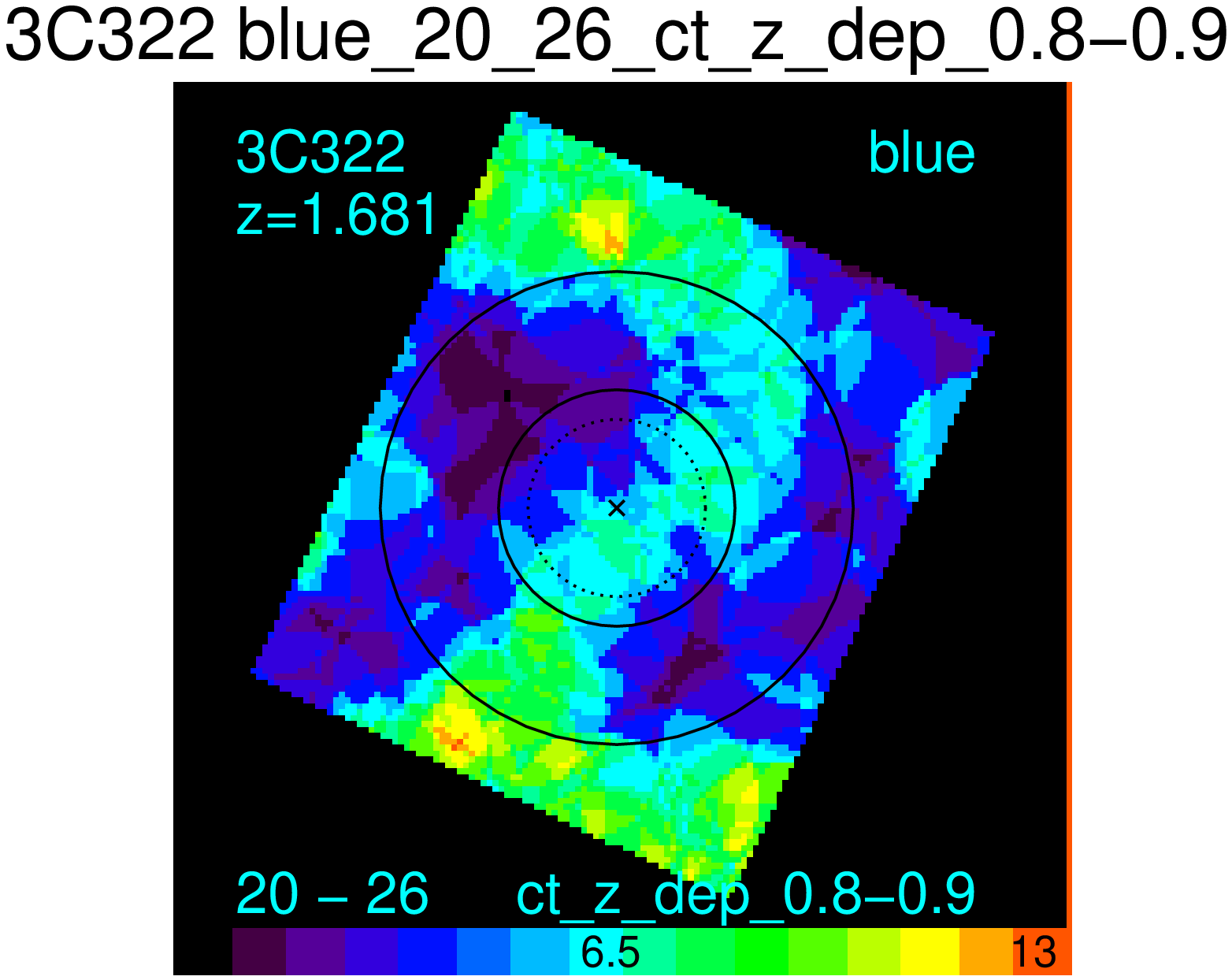}      
                
                \hspace{-0mm}\includegraphics[width=0.245\textwidth, clip=true]{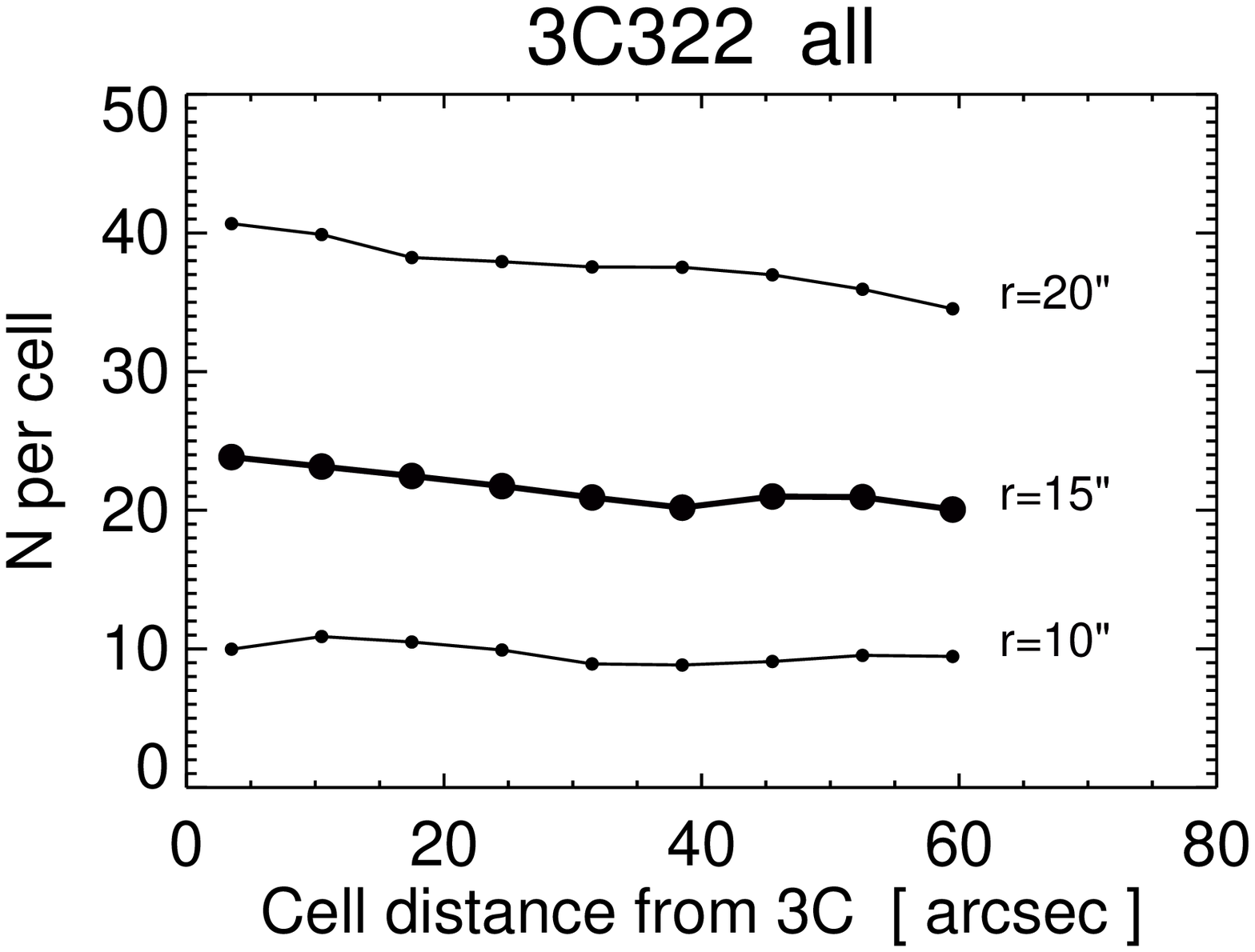}                   
                \includegraphics[width=0.245\textwidth, clip=true]{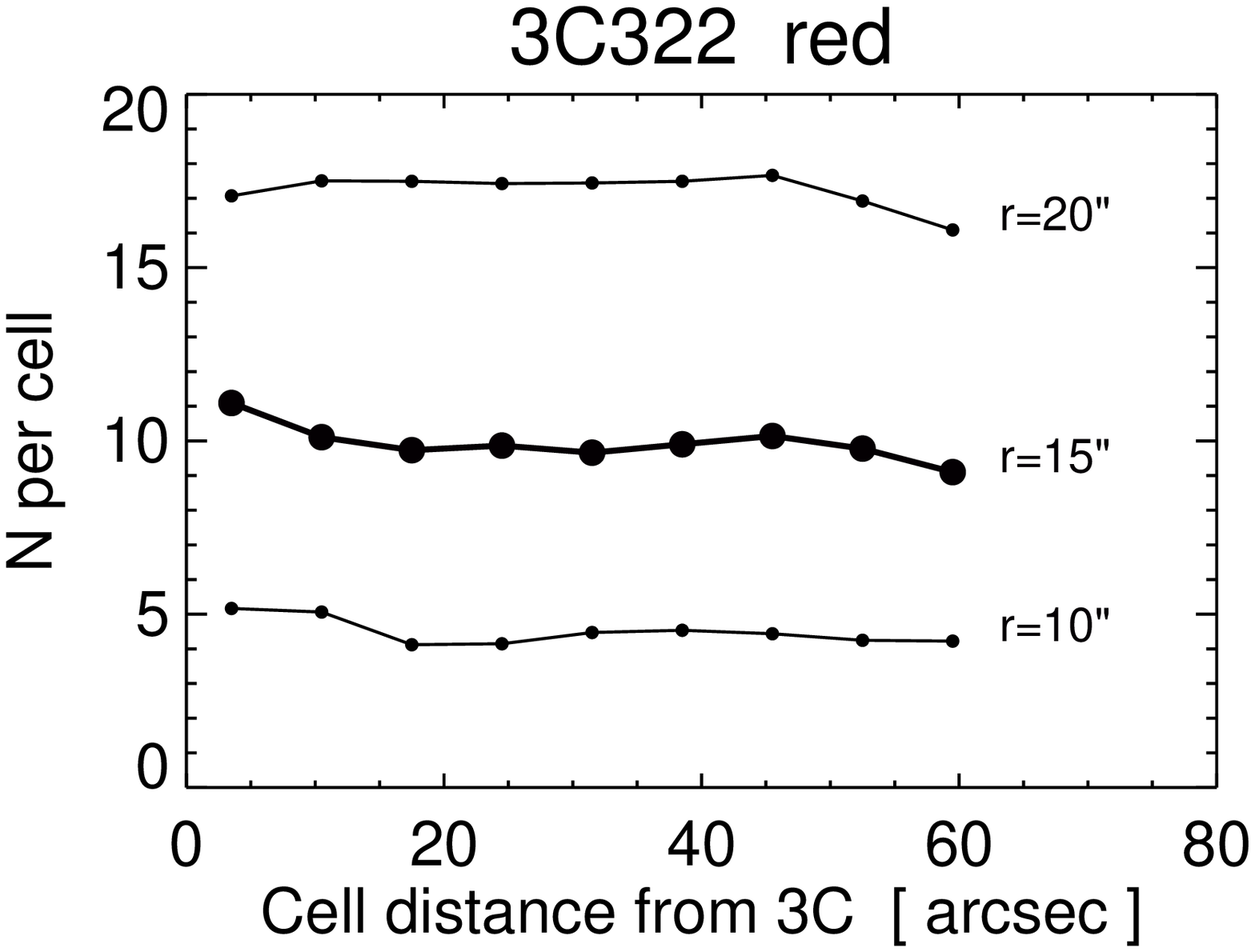}                    
                \includegraphics[width=0.245\textwidth, clip=true]{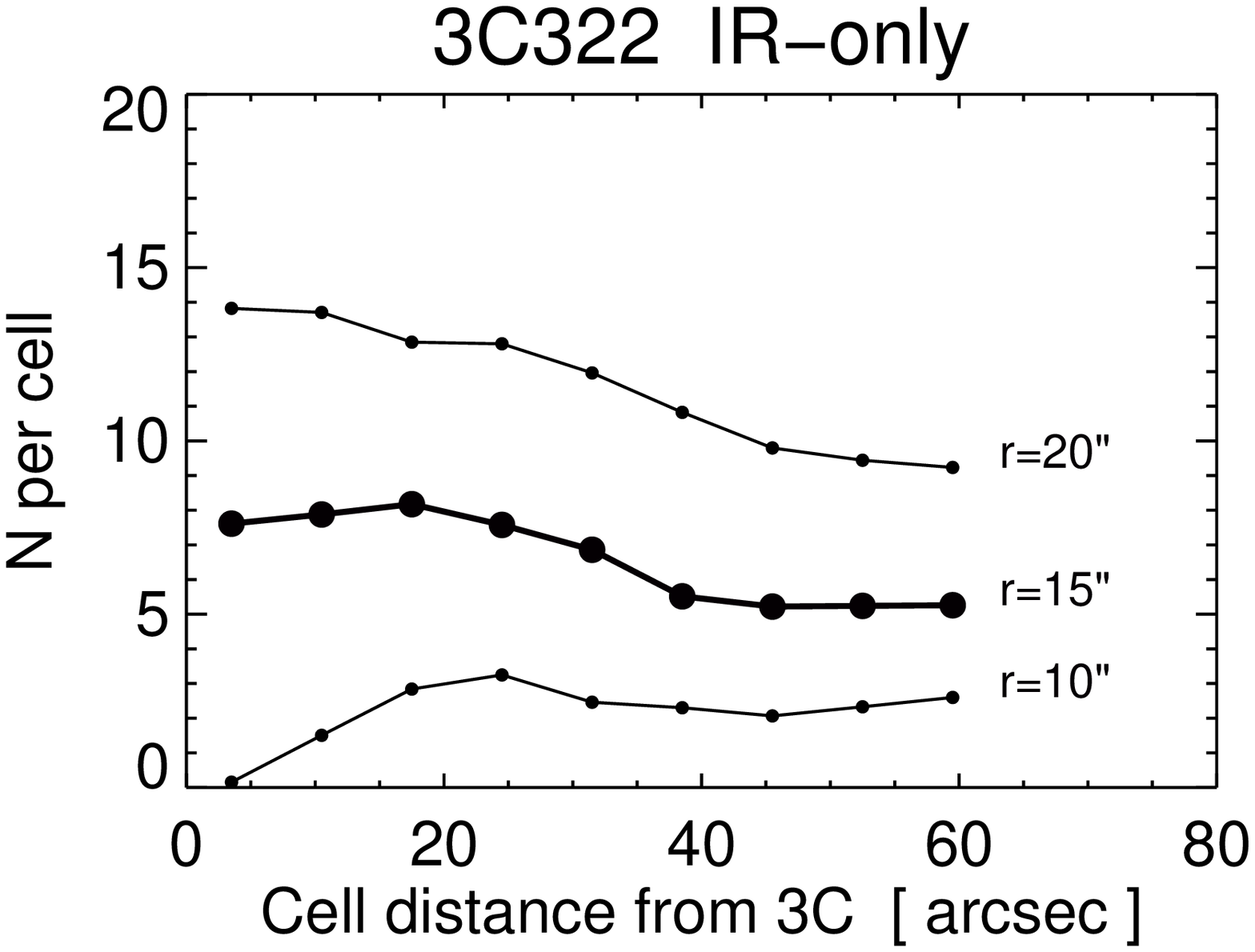}                 
                \includegraphics[width=0.245\textwidth, clip=true]{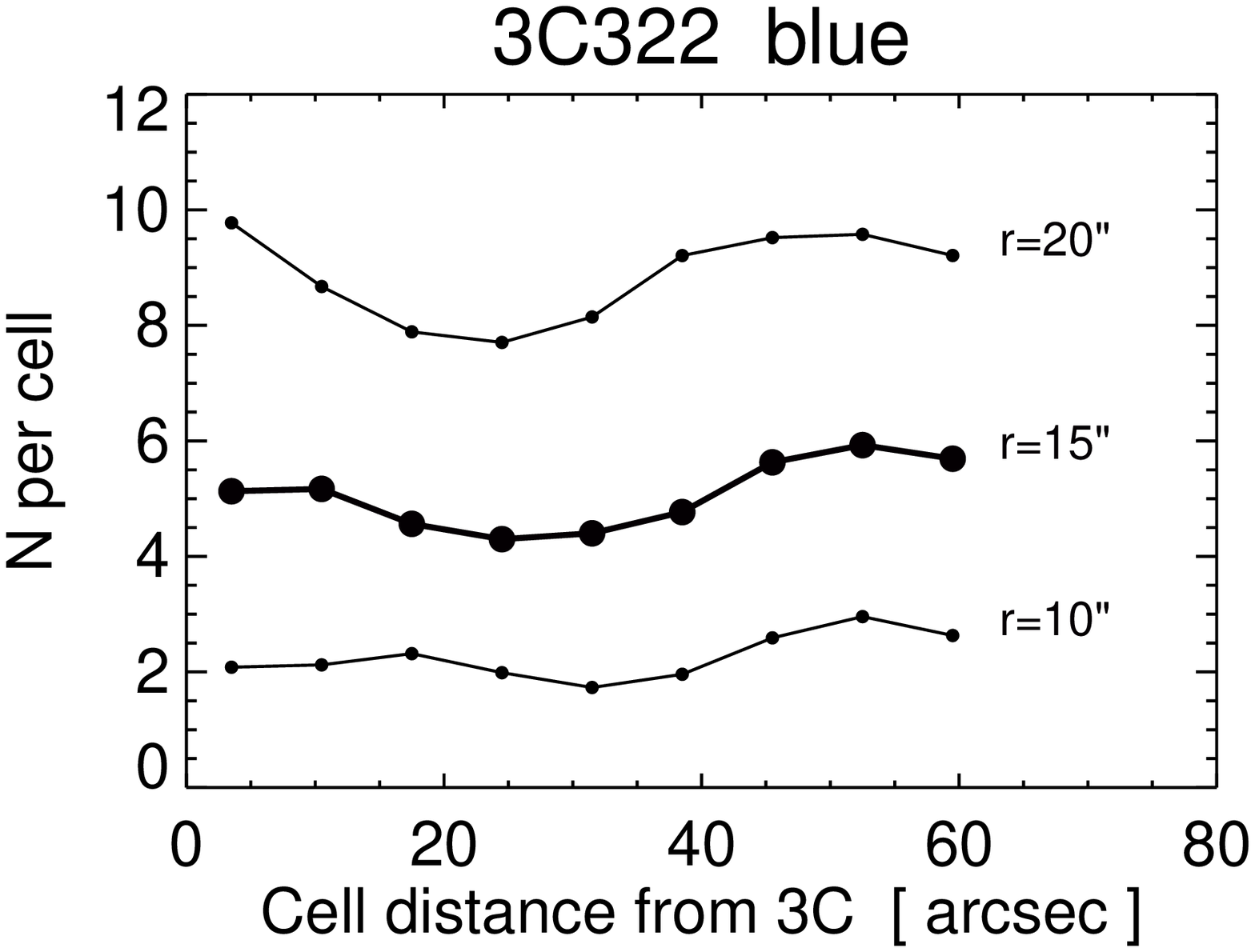}

                \hspace{-0mm}\includegraphics[width=0.245\textwidth, clip=true]{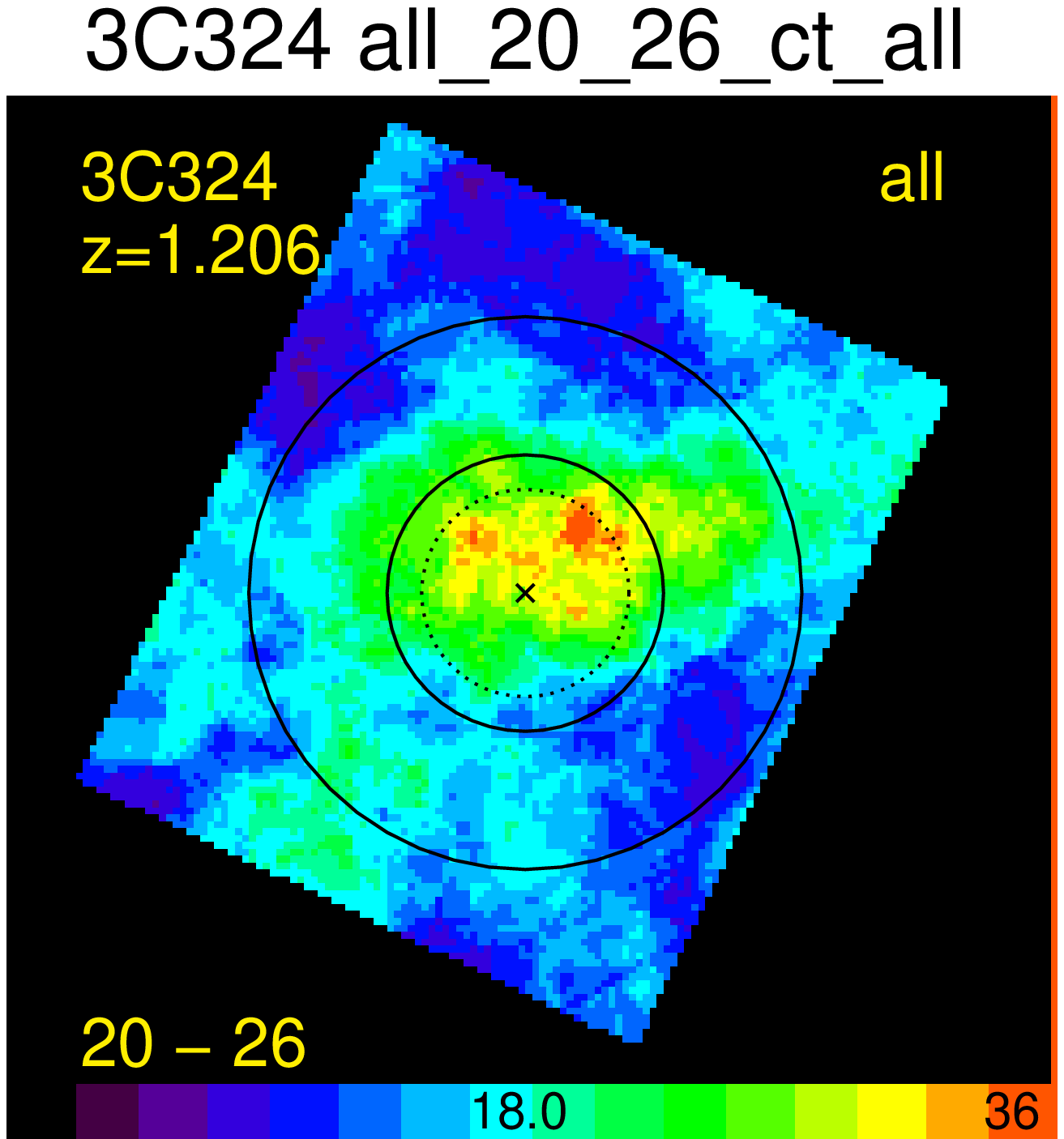}                 
                \includegraphics[width=0.245\textwidth, clip=true]{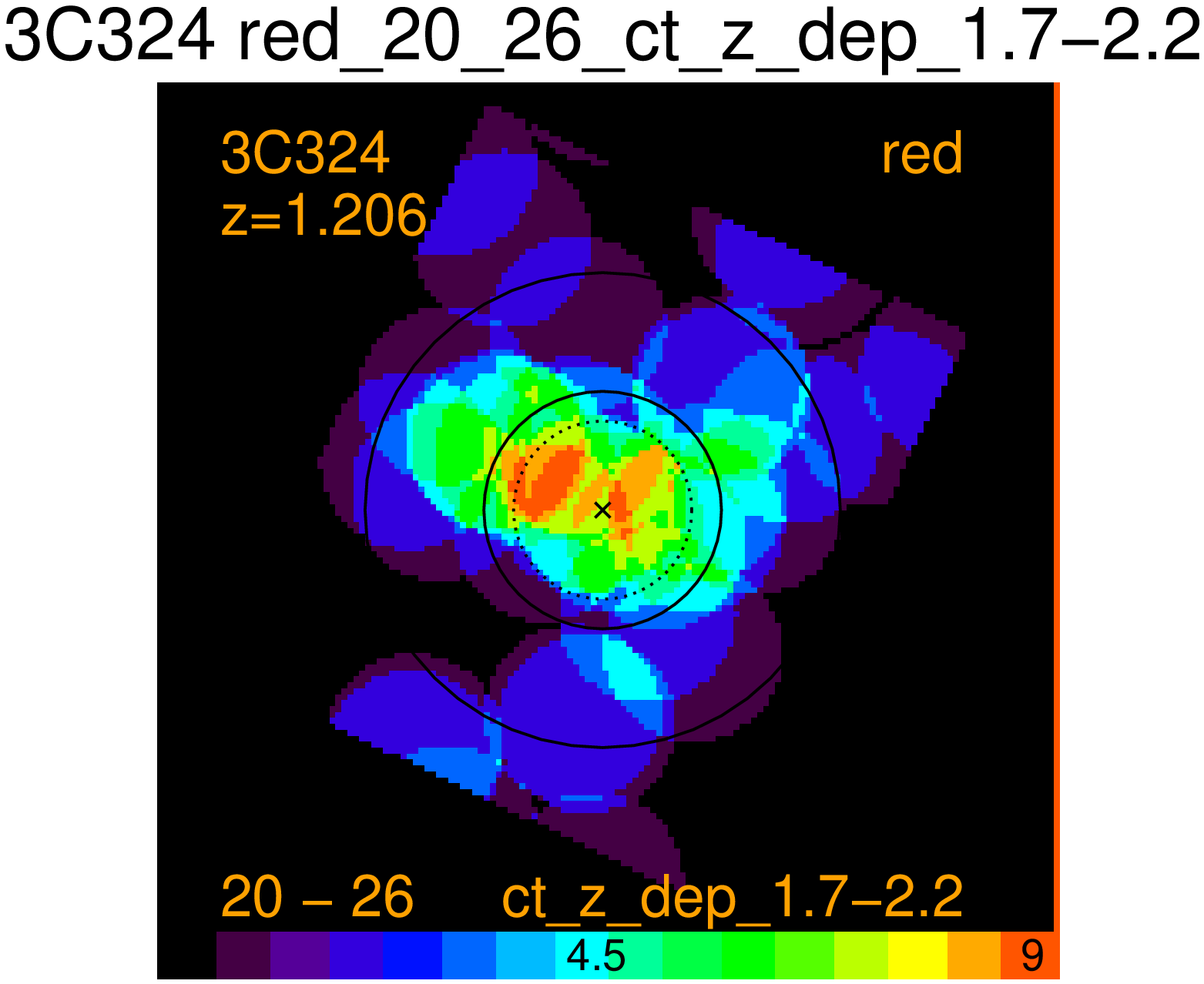}       
                \includegraphics[width=0.245\textwidth, clip=true]{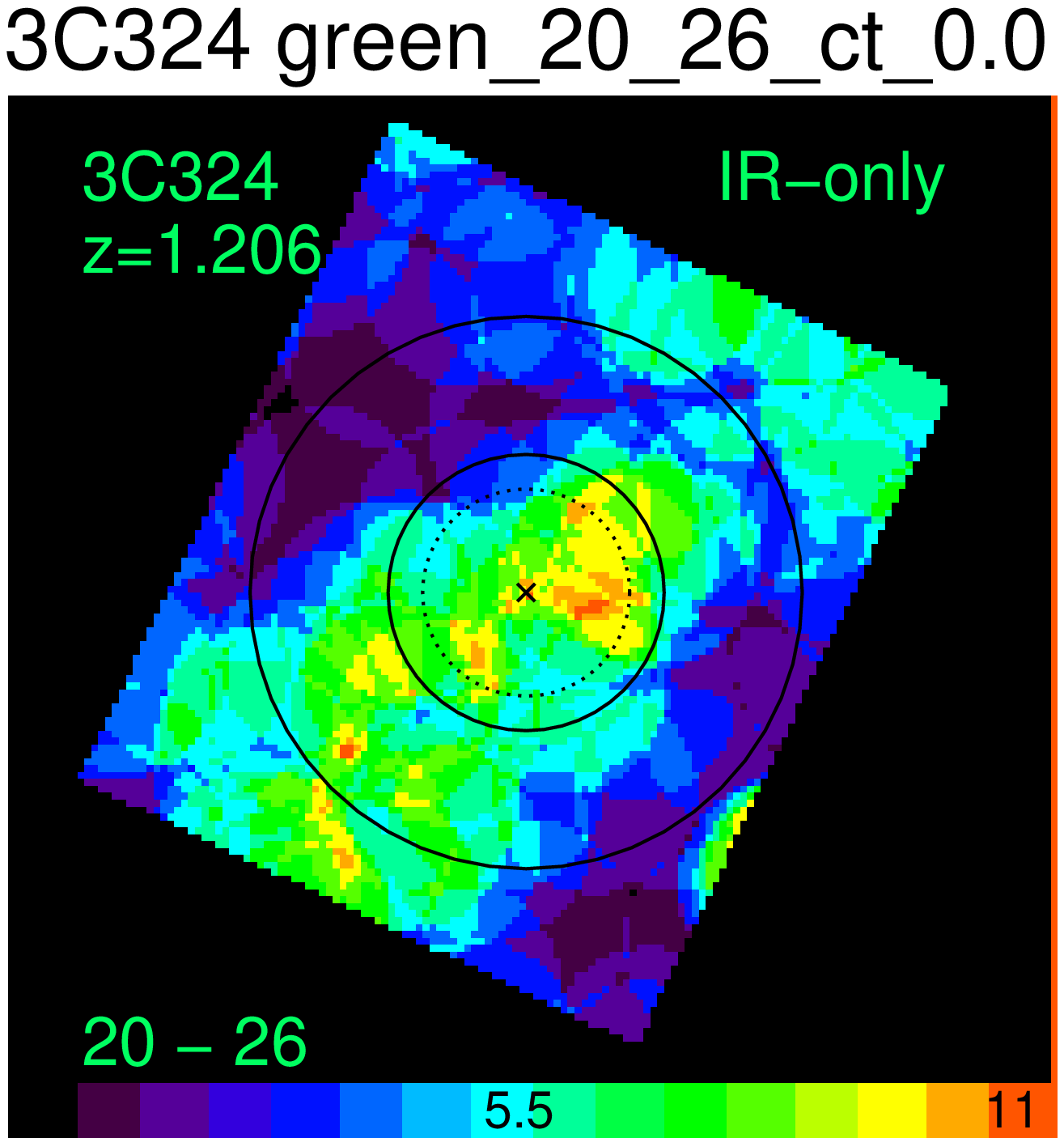}               
                \includegraphics[width=0.245\textwidth, clip=true]{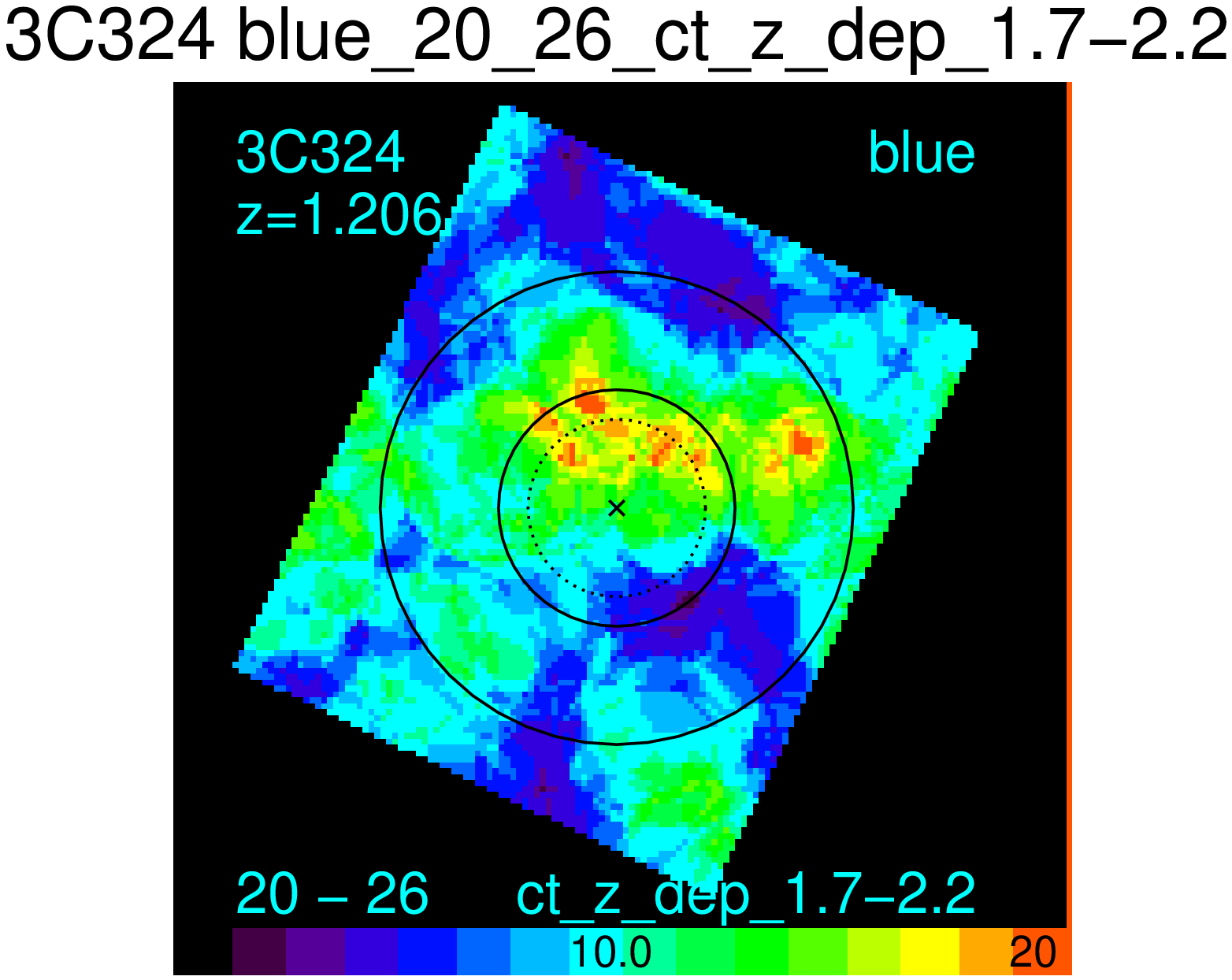}      
                
                \hspace{-0mm}\includegraphics[width=0.245\textwidth, clip=true]{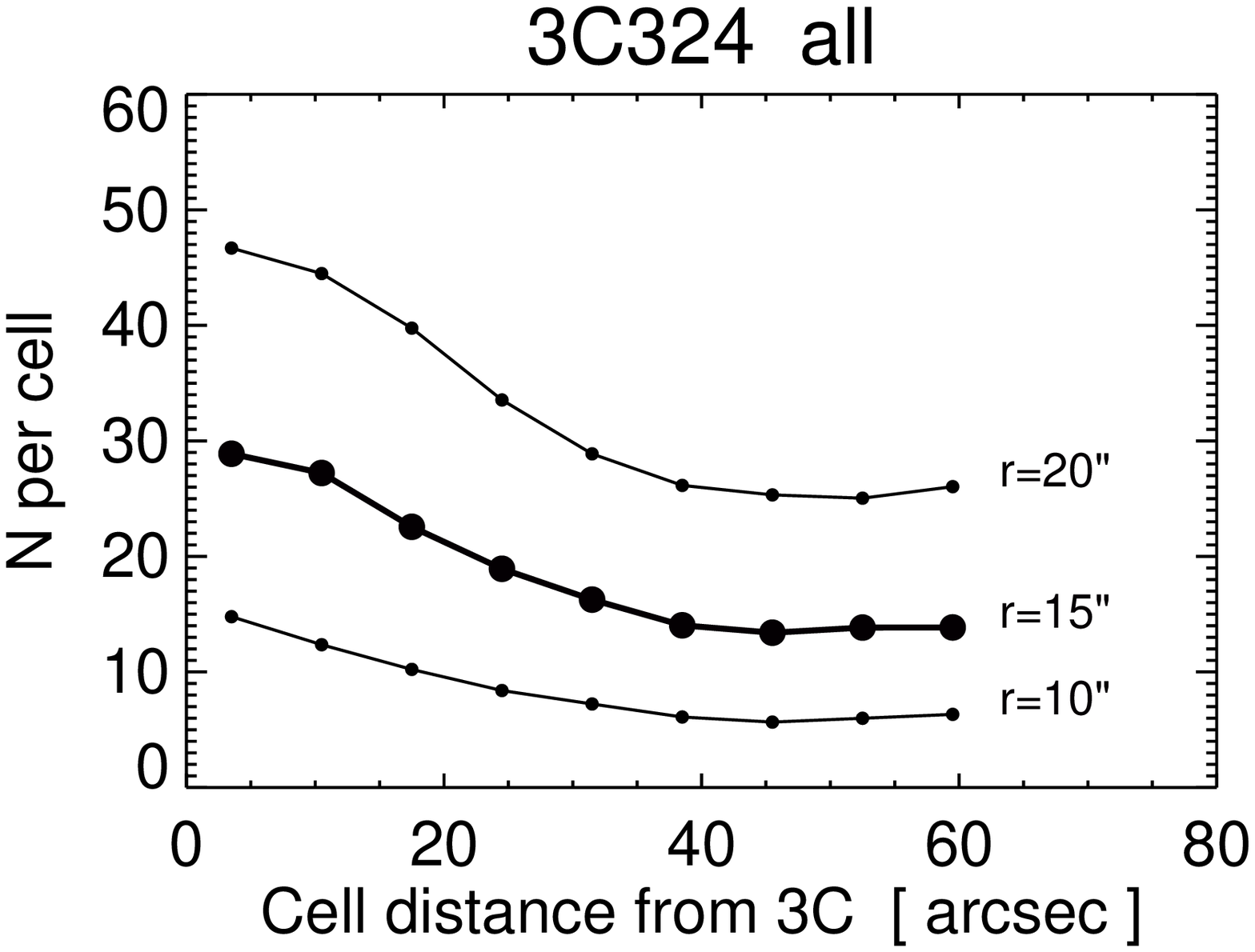}                   
                \includegraphics[width=0.245\textwidth, clip=true]{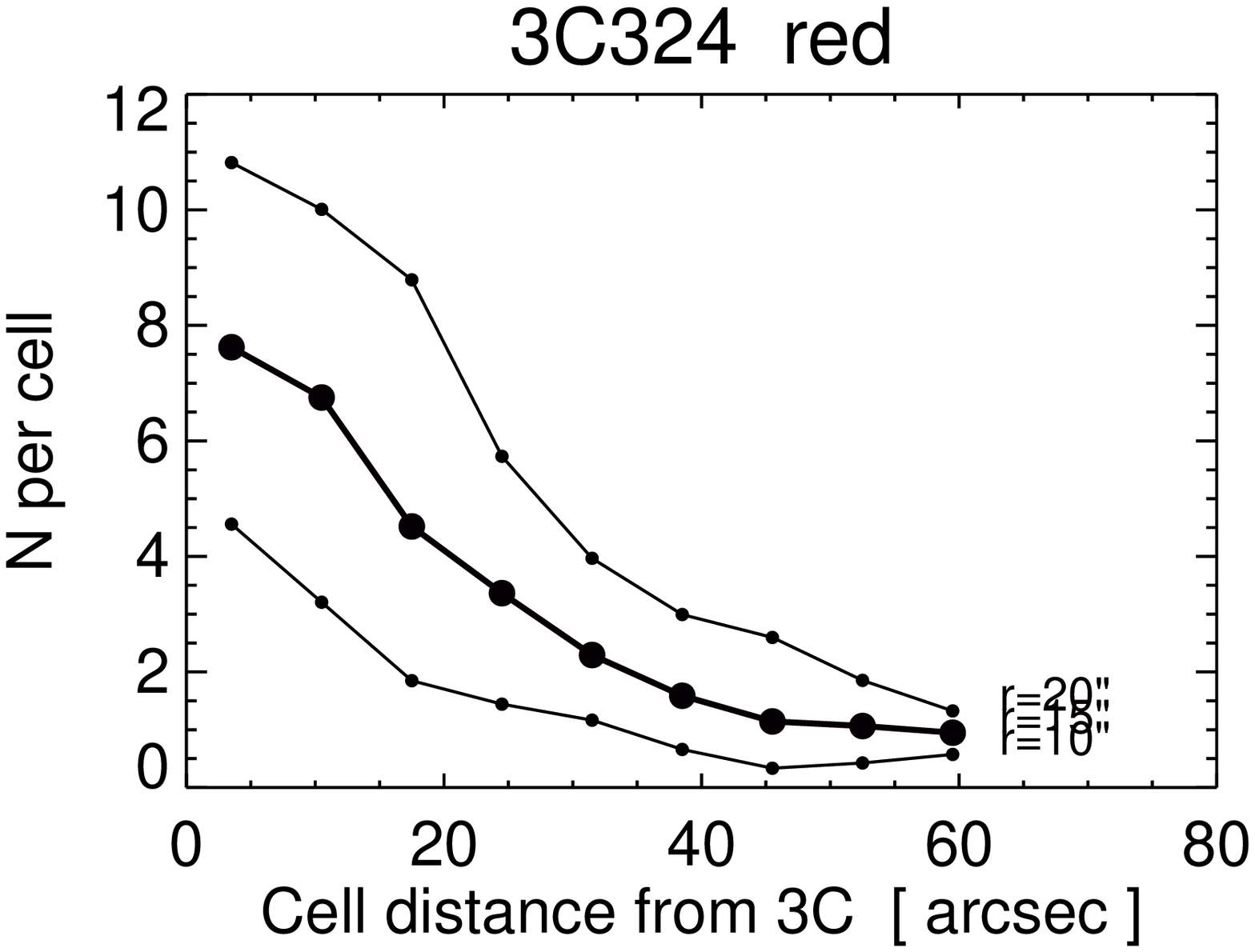}                    
                \includegraphics[width=0.245\textwidth, clip=true]{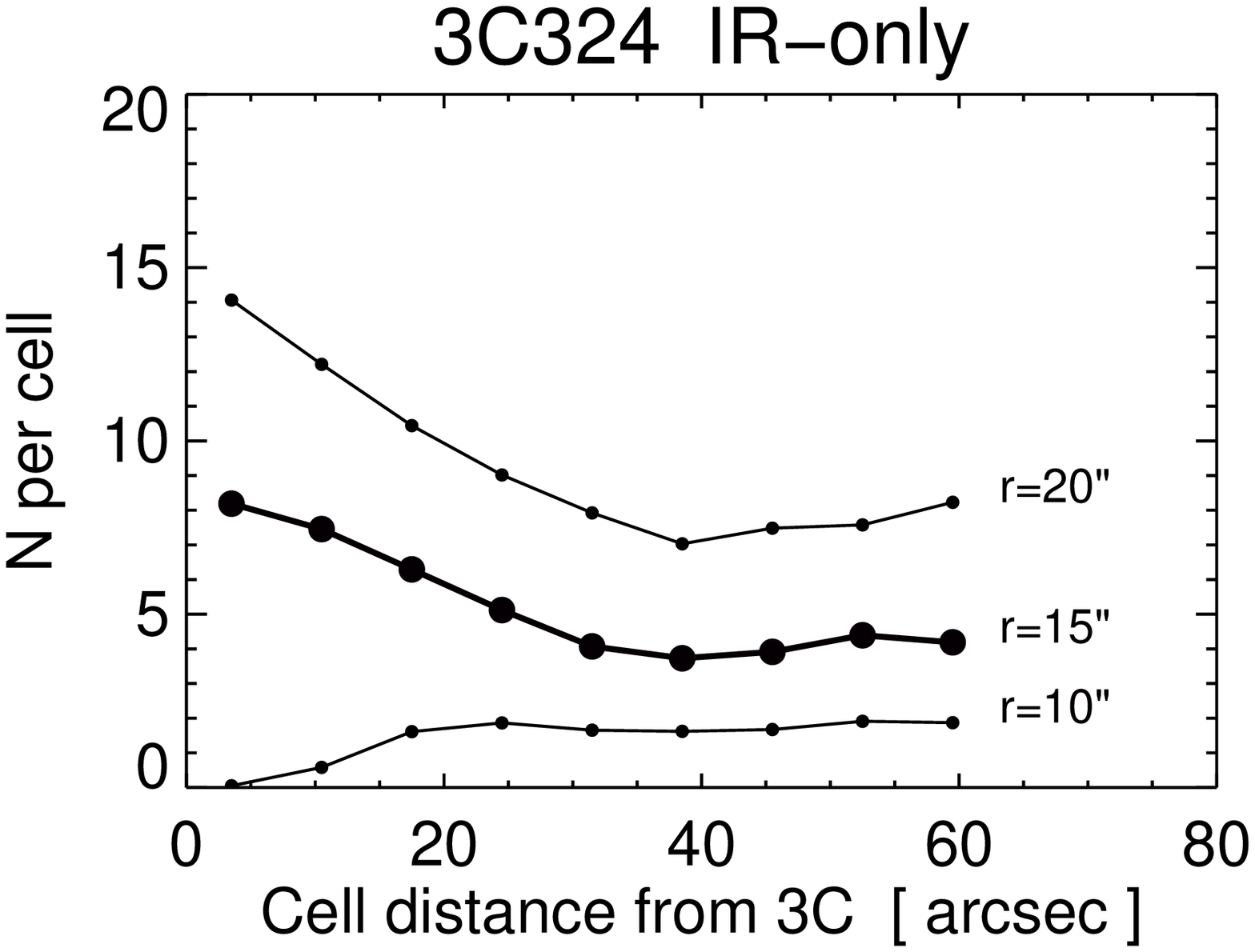}                 
                \includegraphics[width=0.245\textwidth, clip=true]{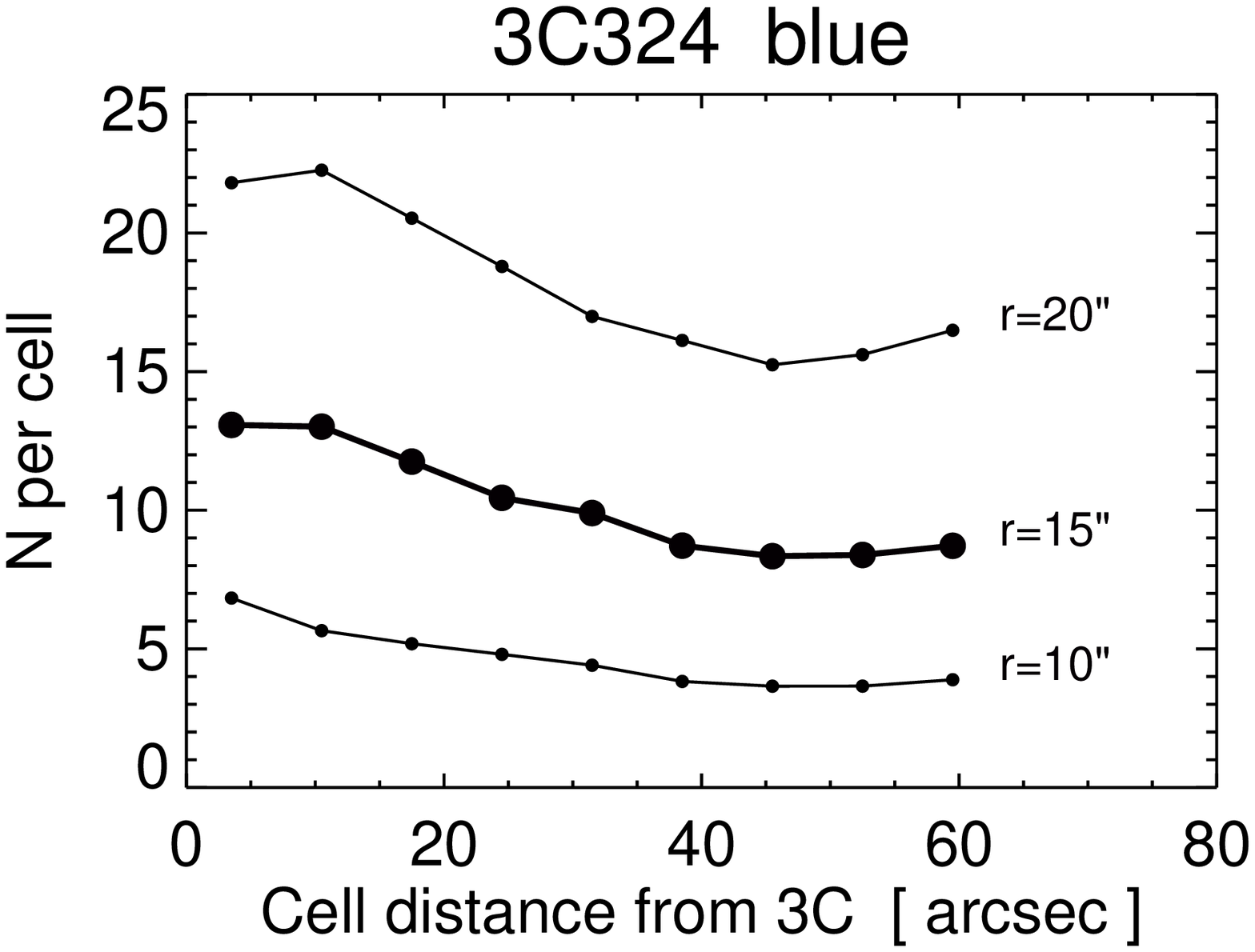}

                \hspace{-0mm}\includegraphics[width=0.245\textwidth, clip=true]{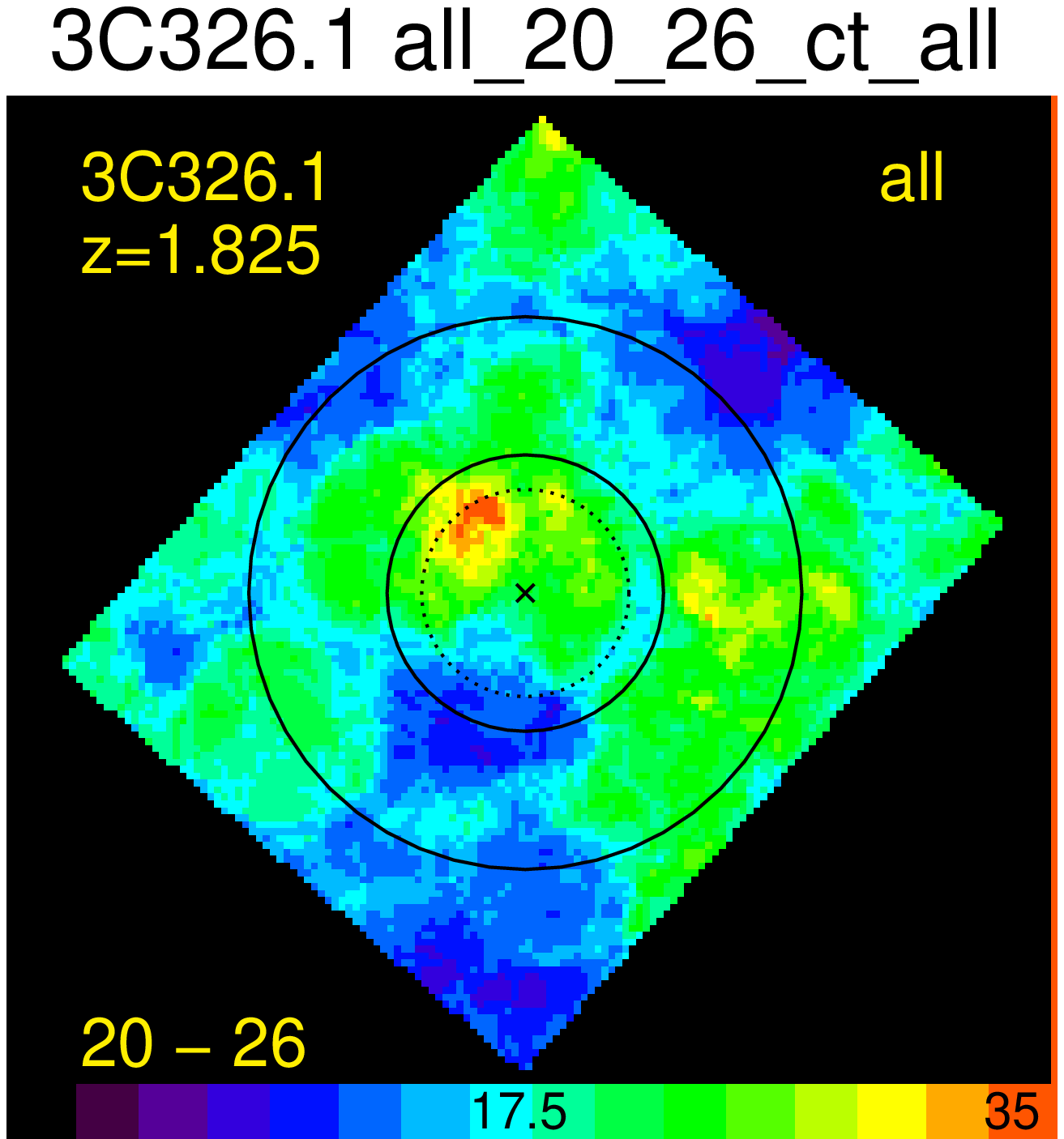}               
                \includegraphics[width=0.245\textwidth, clip=true]{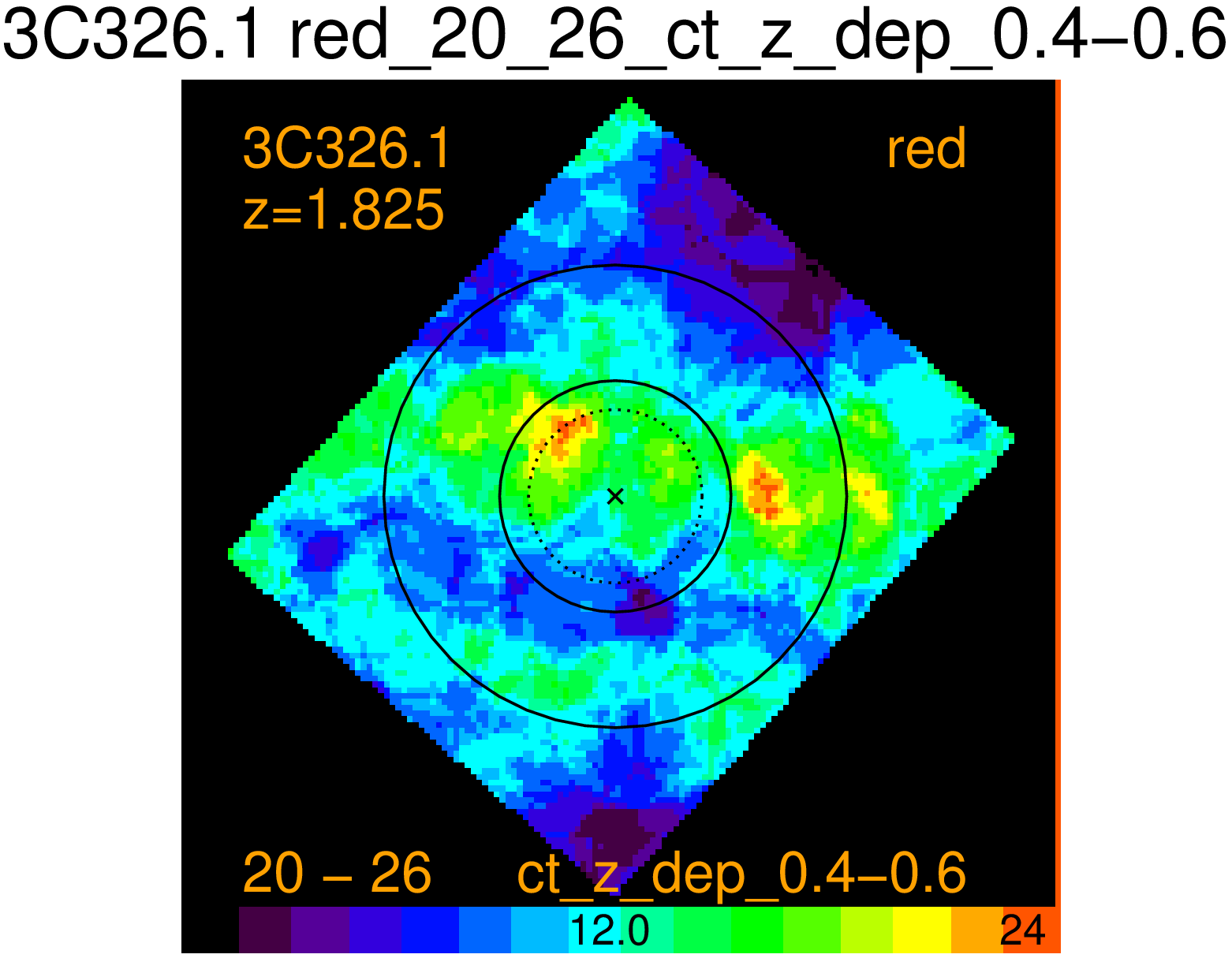}     
                \includegraphics[width=0.245\textwidth, clip=true]{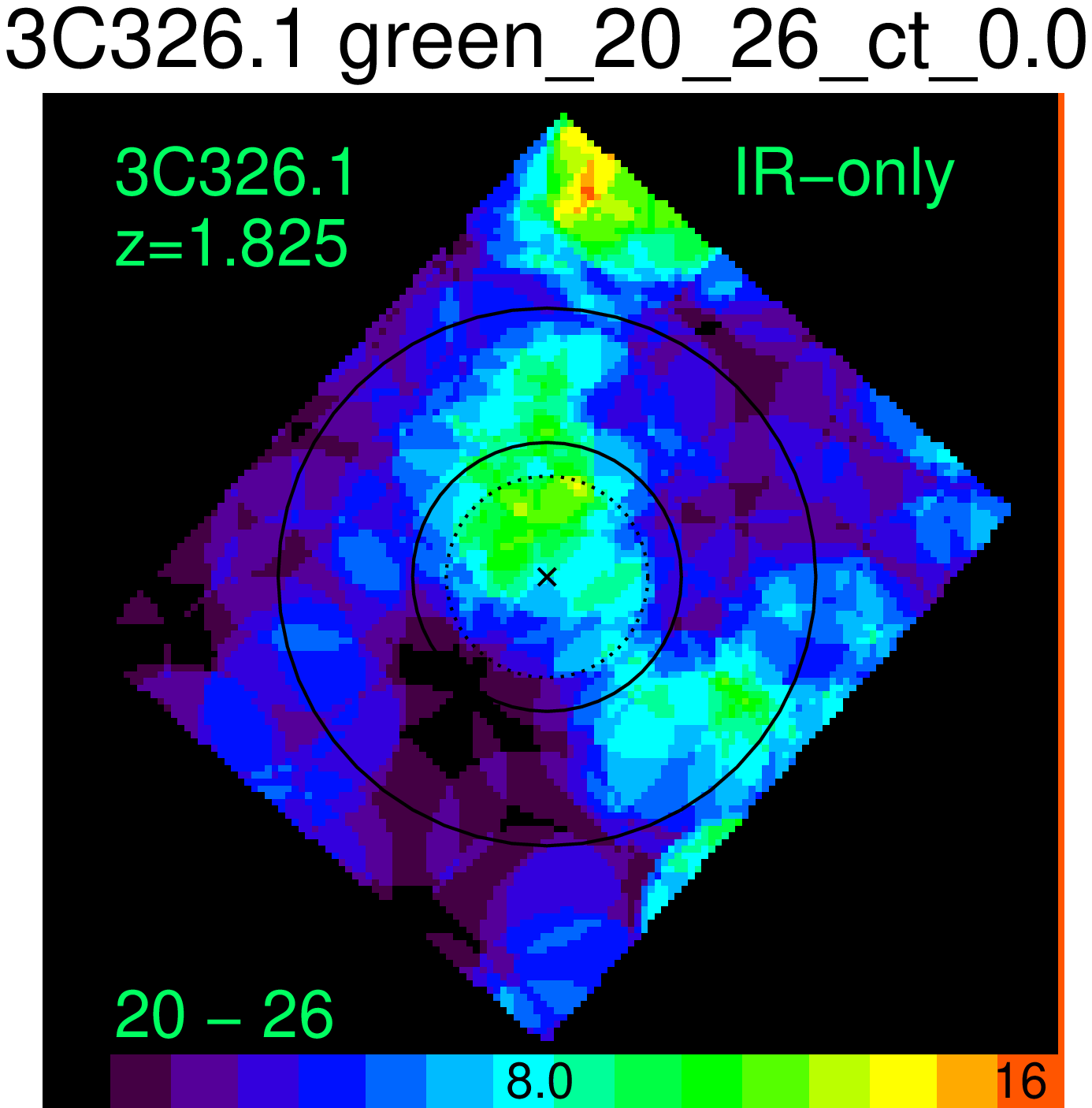}             
                \includegraphics[width=0.245\textwidth, clip=true]{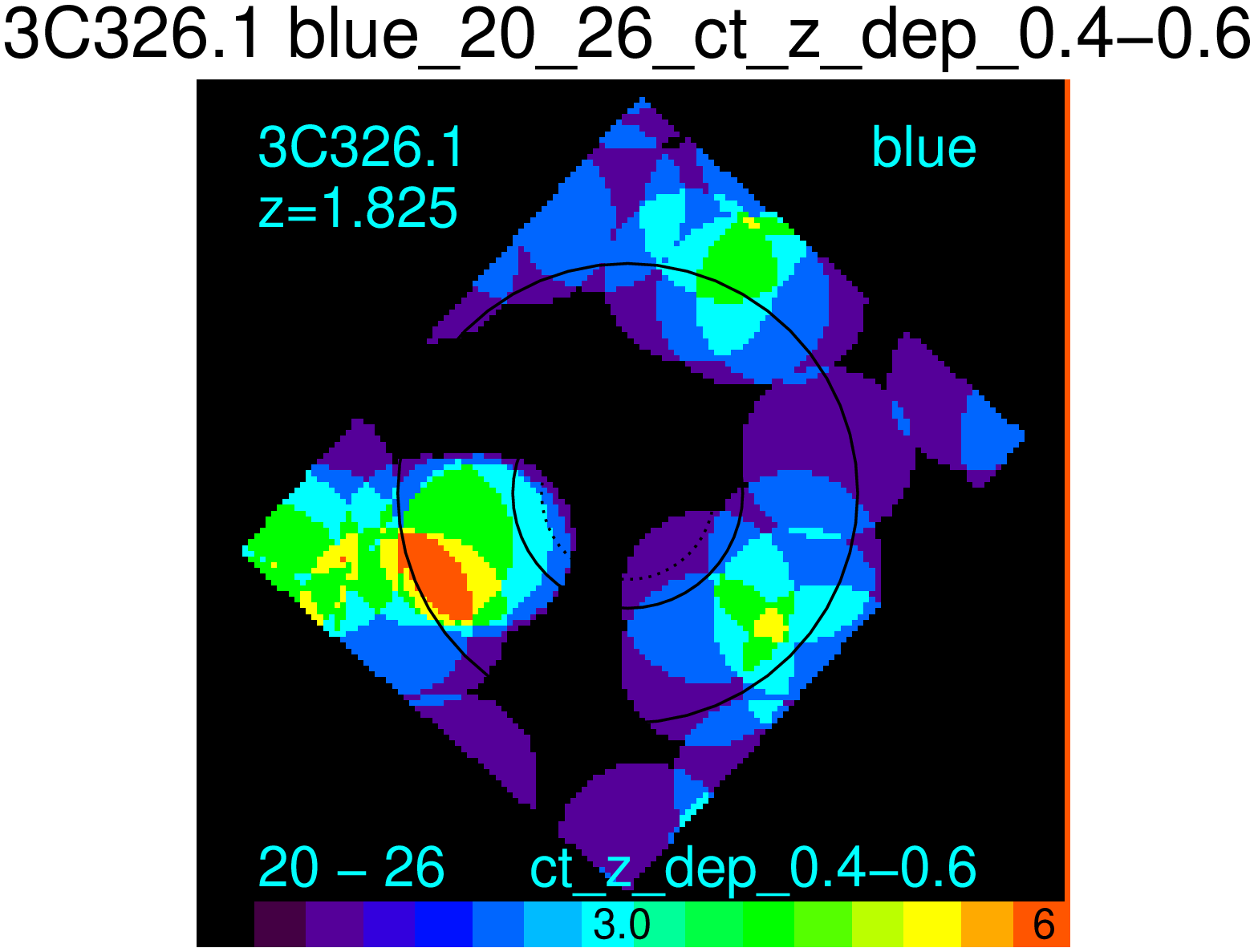}    
                
                \hspace{-0mm}\includegraphics[width=0.245\textwidth, clip=true]{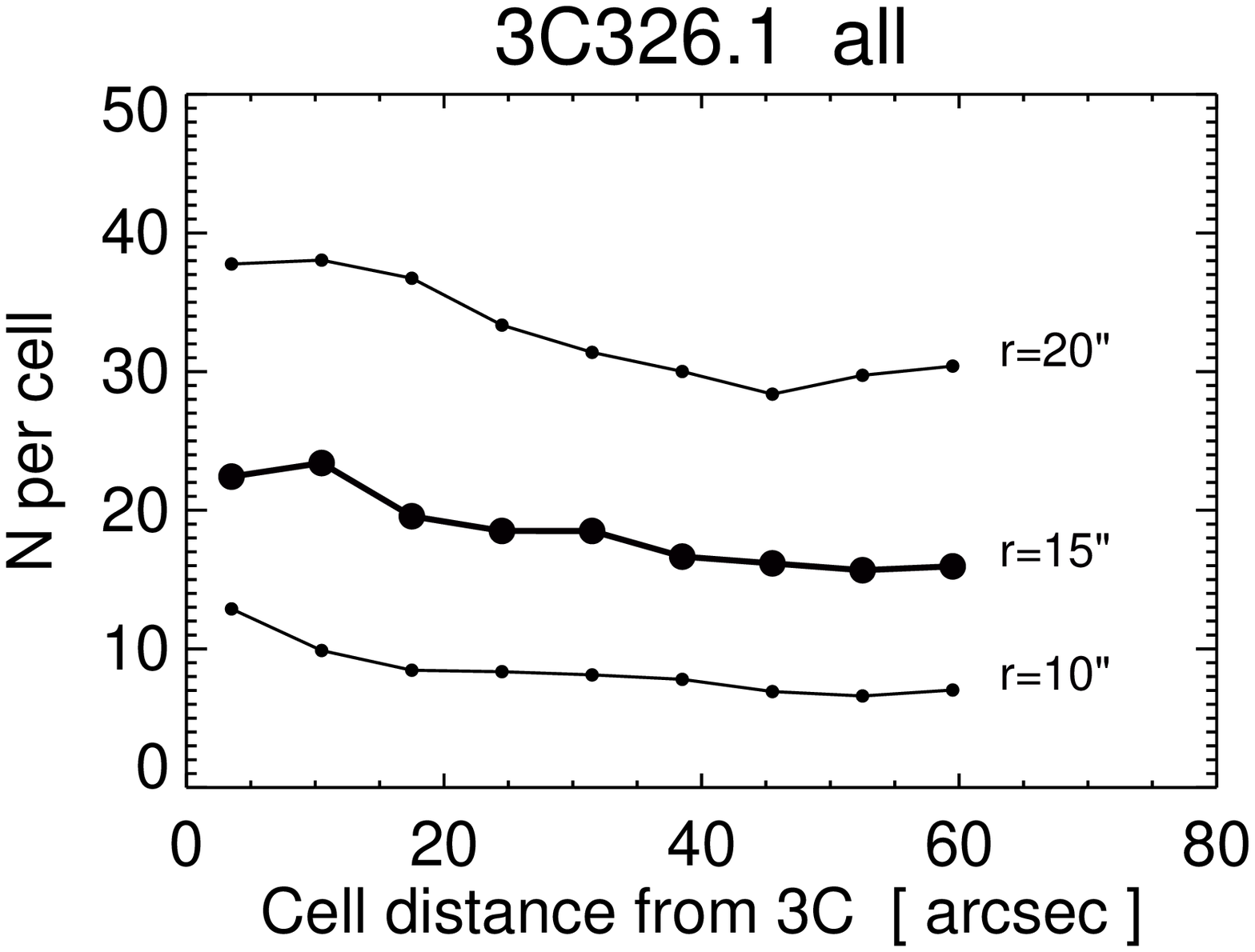}                 
                \includegraphics[width=0.245\textwidth, clip=true]{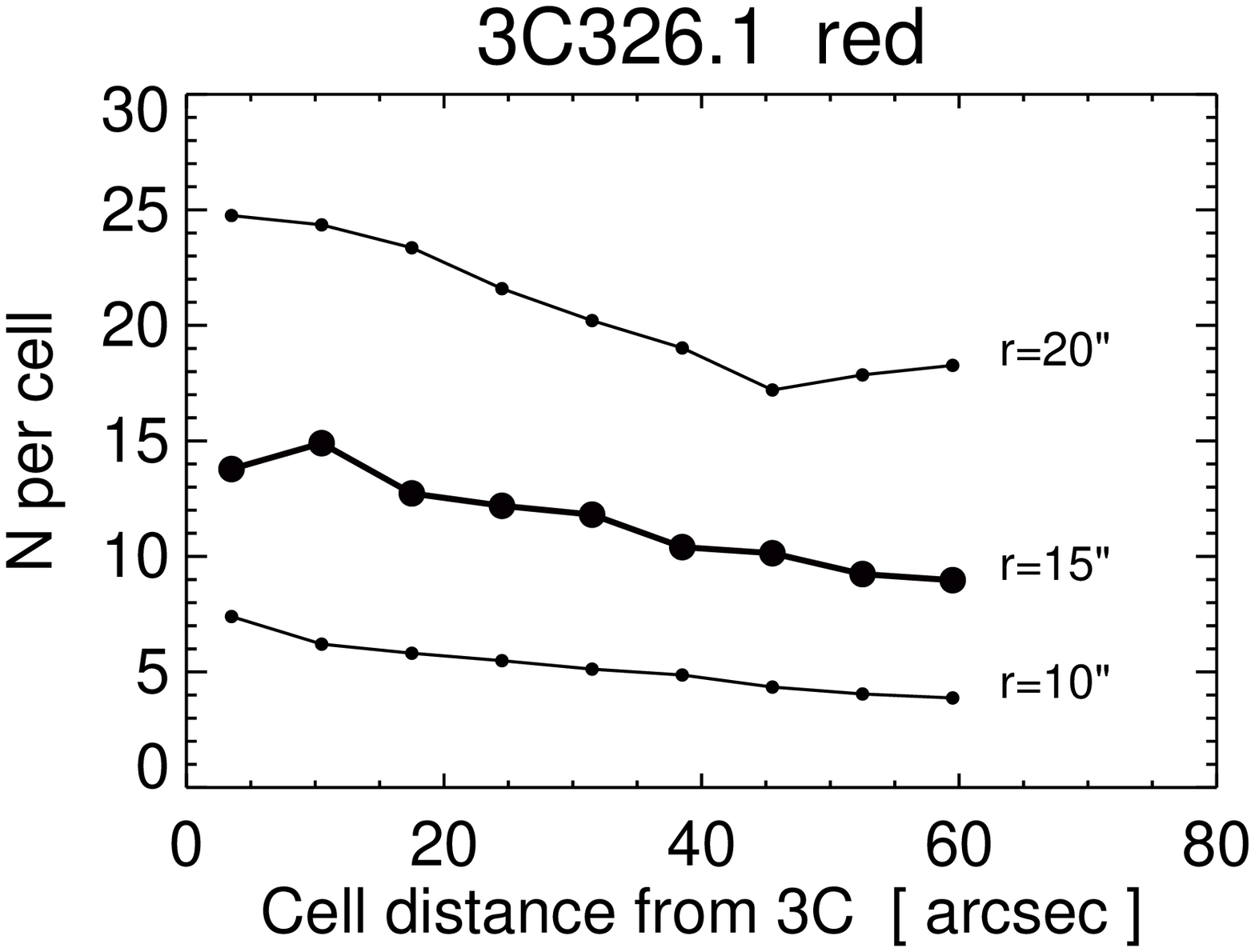}                  
                \includegraphics[width=0.245\textwidth, clip=true]{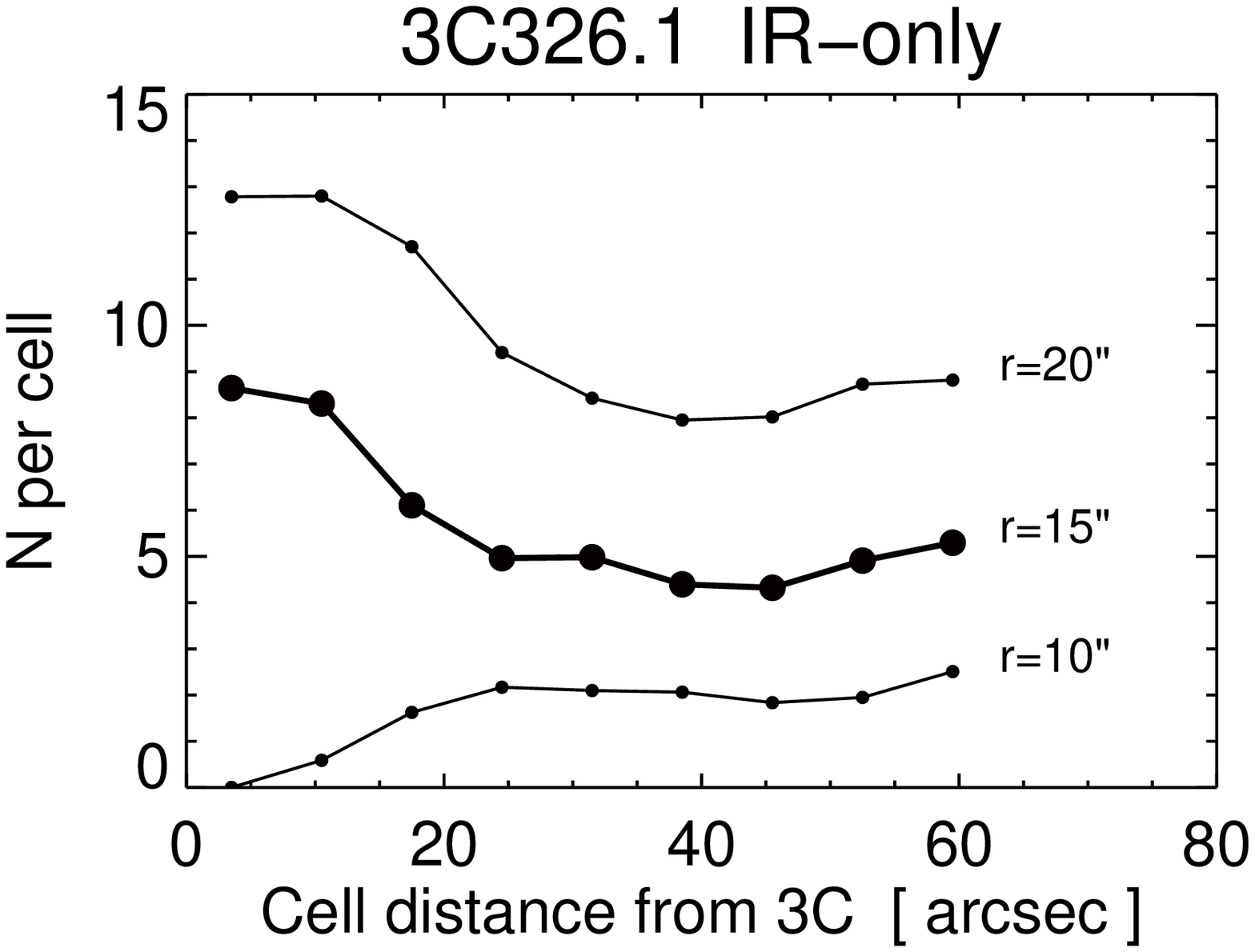}               
                \includegraphics[width=0.245\textwidth, clip=true]{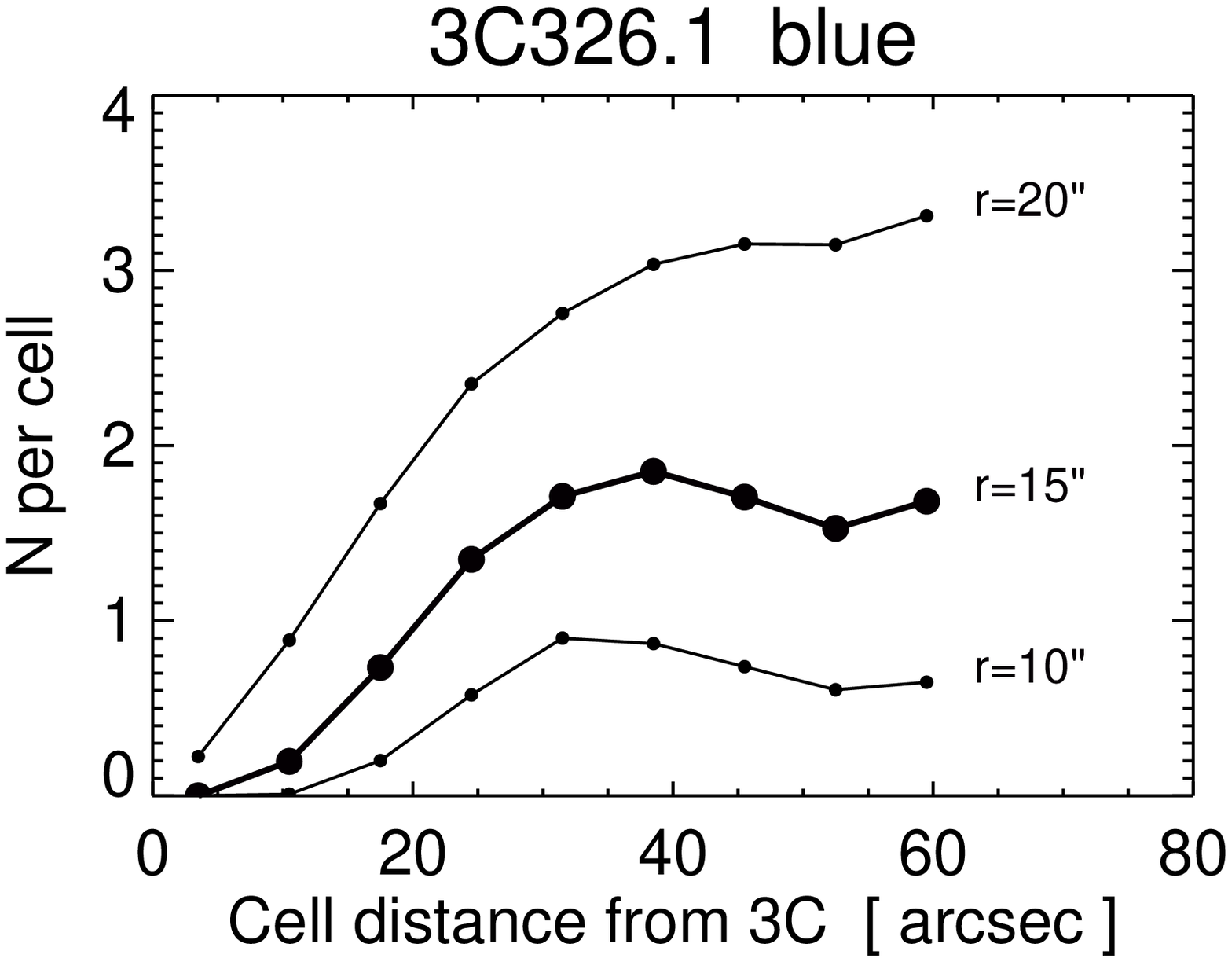}

                \caption{Surface density maps and radial density profiles of the 3C fields, continued.
                }
                \label{fig:sd_maps_6}
              \end{figure*}

\begin{figure*}

                \hspace{-0mm}\includegraphics[width=0.245\textwidth, clip=true]{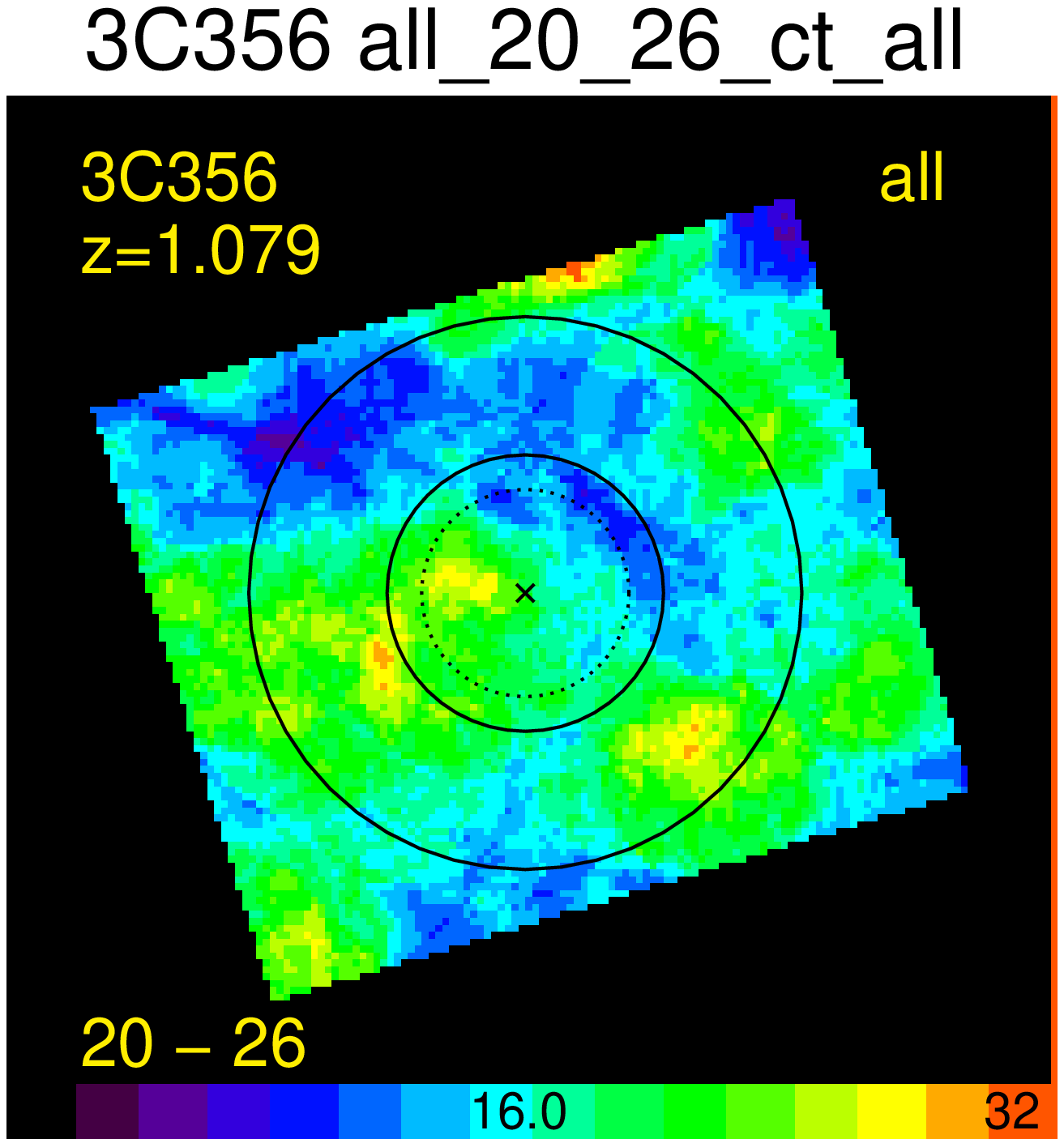}                 
                \includegraphics[width=0.245\textwidth, clip=true]{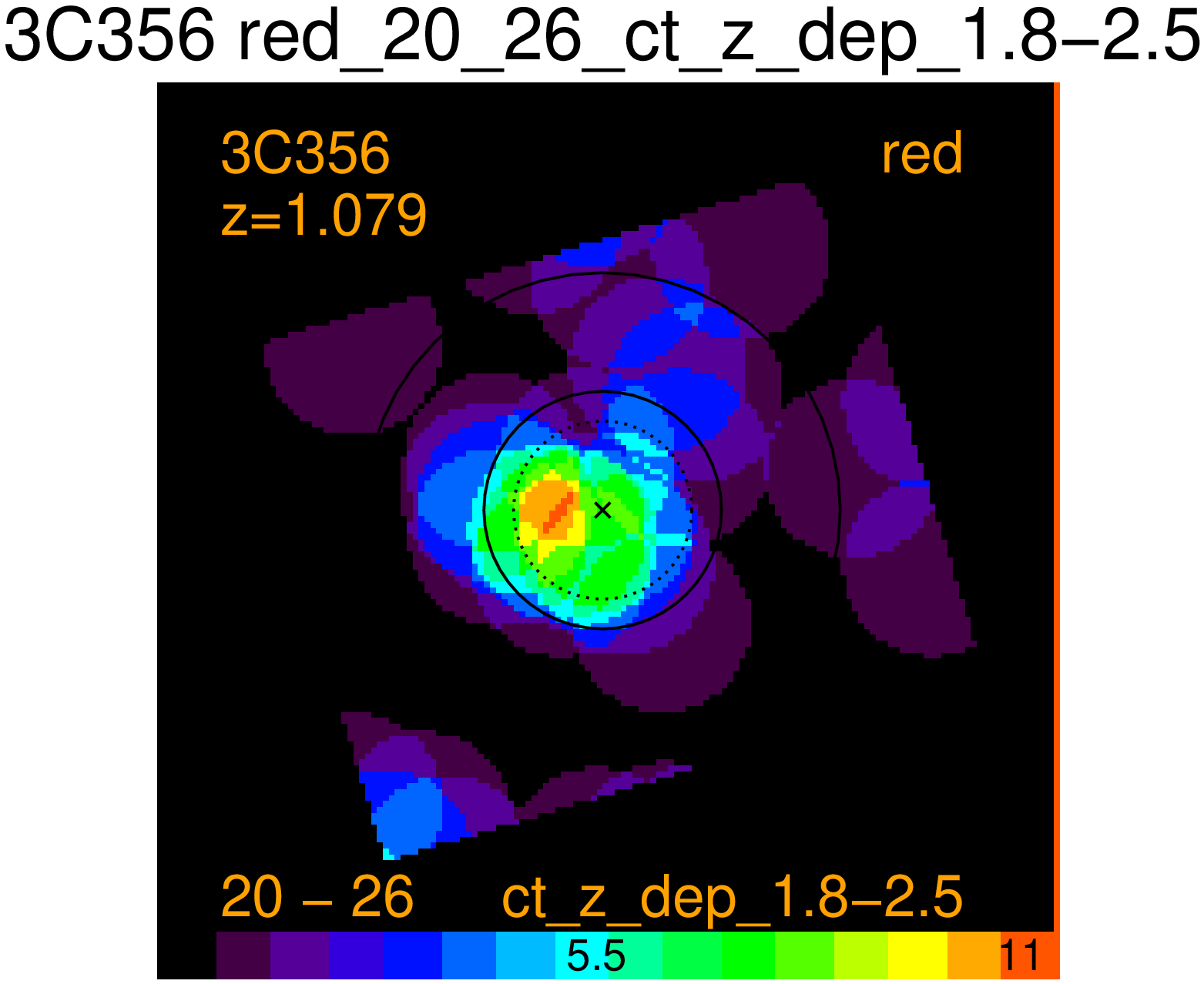}       
                \includegraphics[width=0.245\textwidth, clip=true]{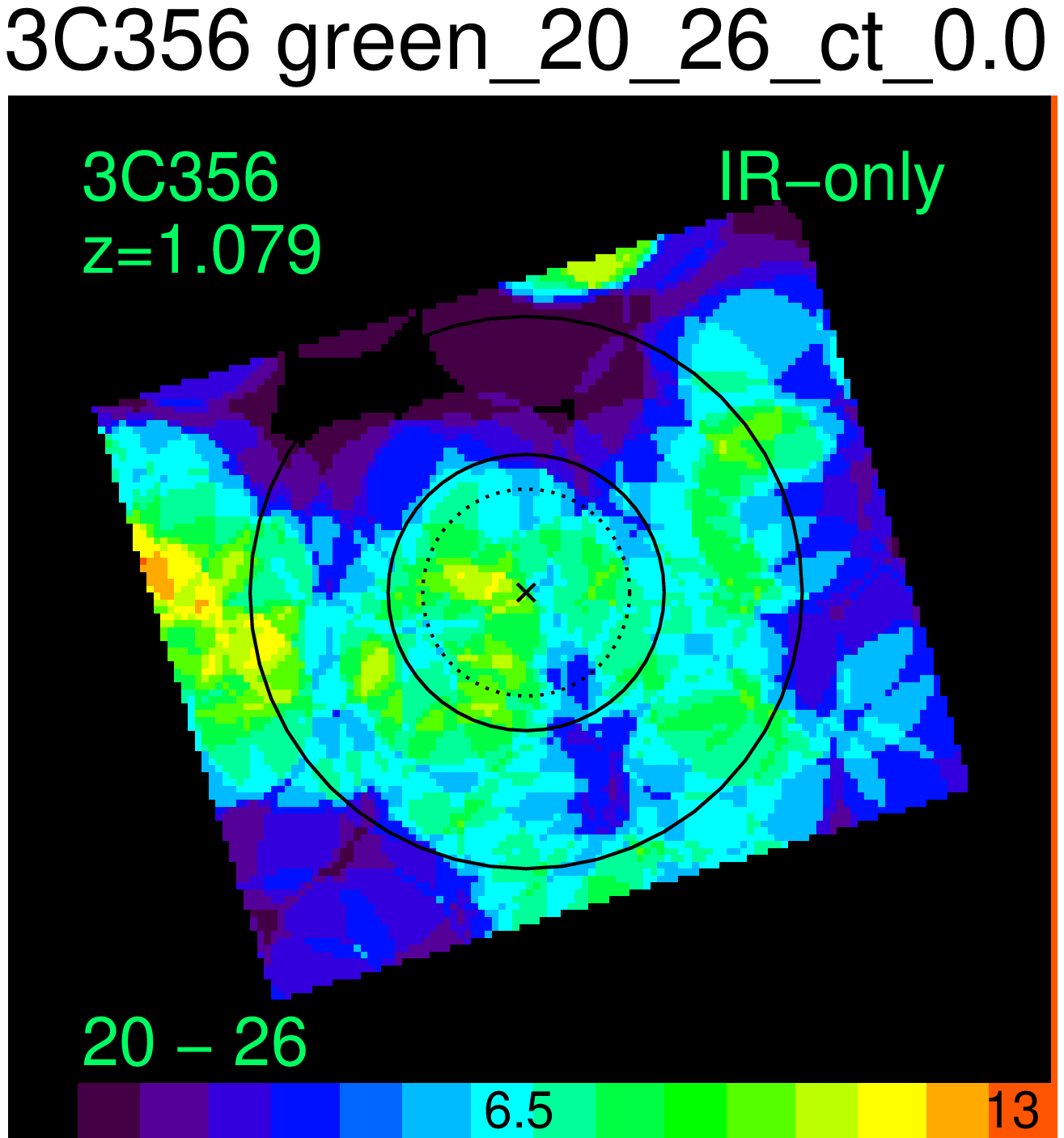}               
                \includegraphics[width=0.245\textwidth, clip=true]{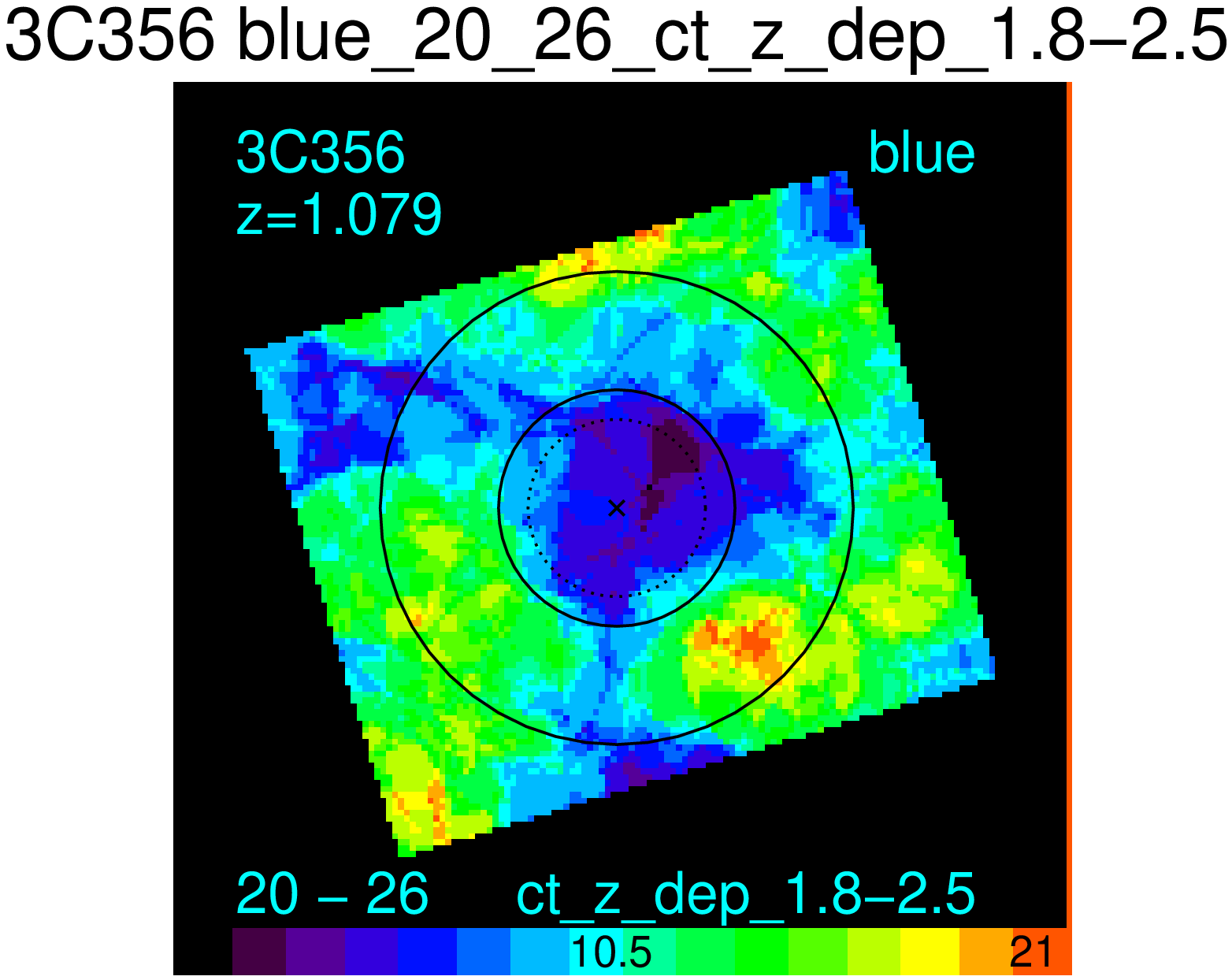}      
                
                \hspace{-0mm}\includegraphics[width=0.245\textwidth, clip=true]{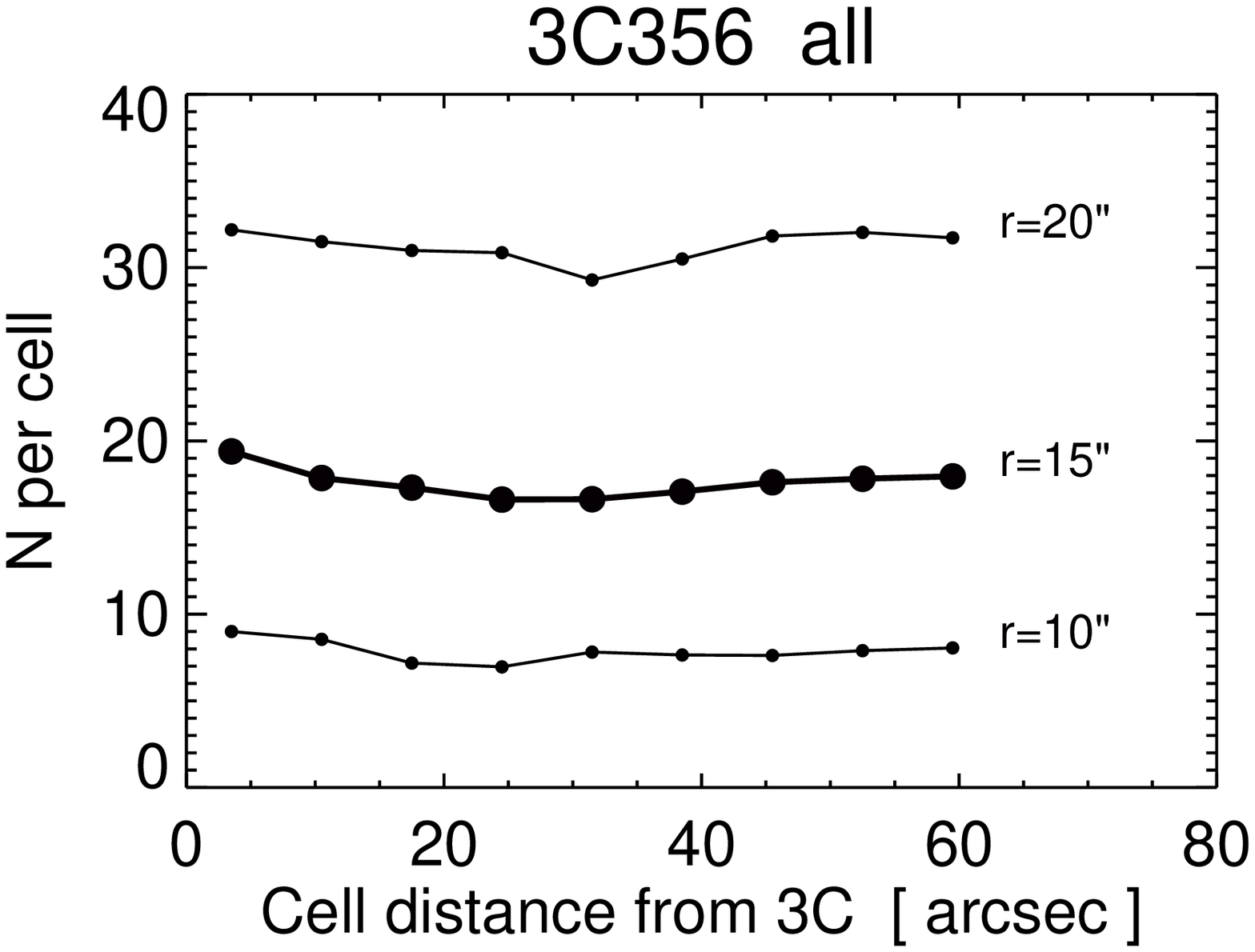}                   
                \includegraphics[width=0.245\textwidth, clip=true]{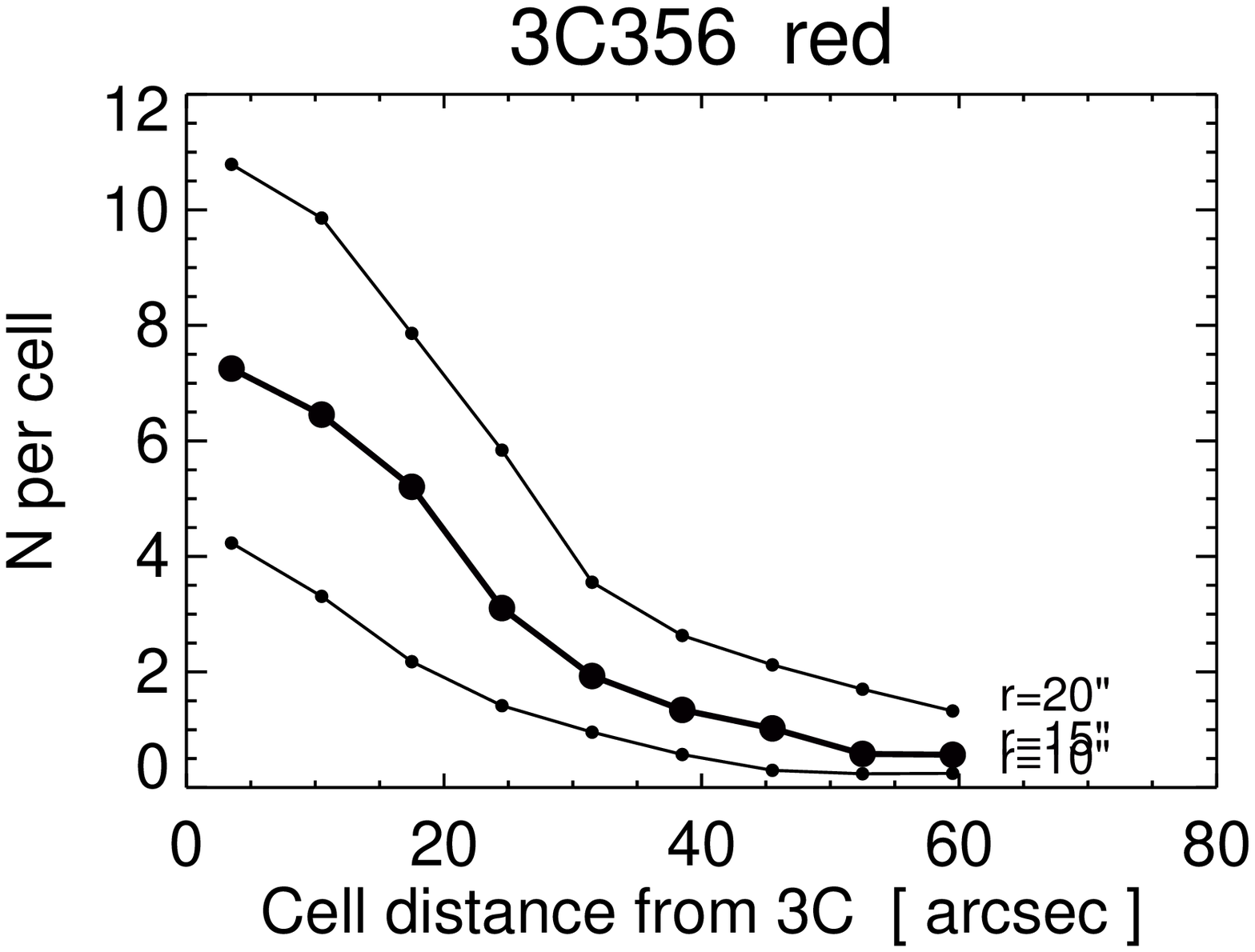}                    
                \includegraphics[width=0.245\textwidth, clip=true]{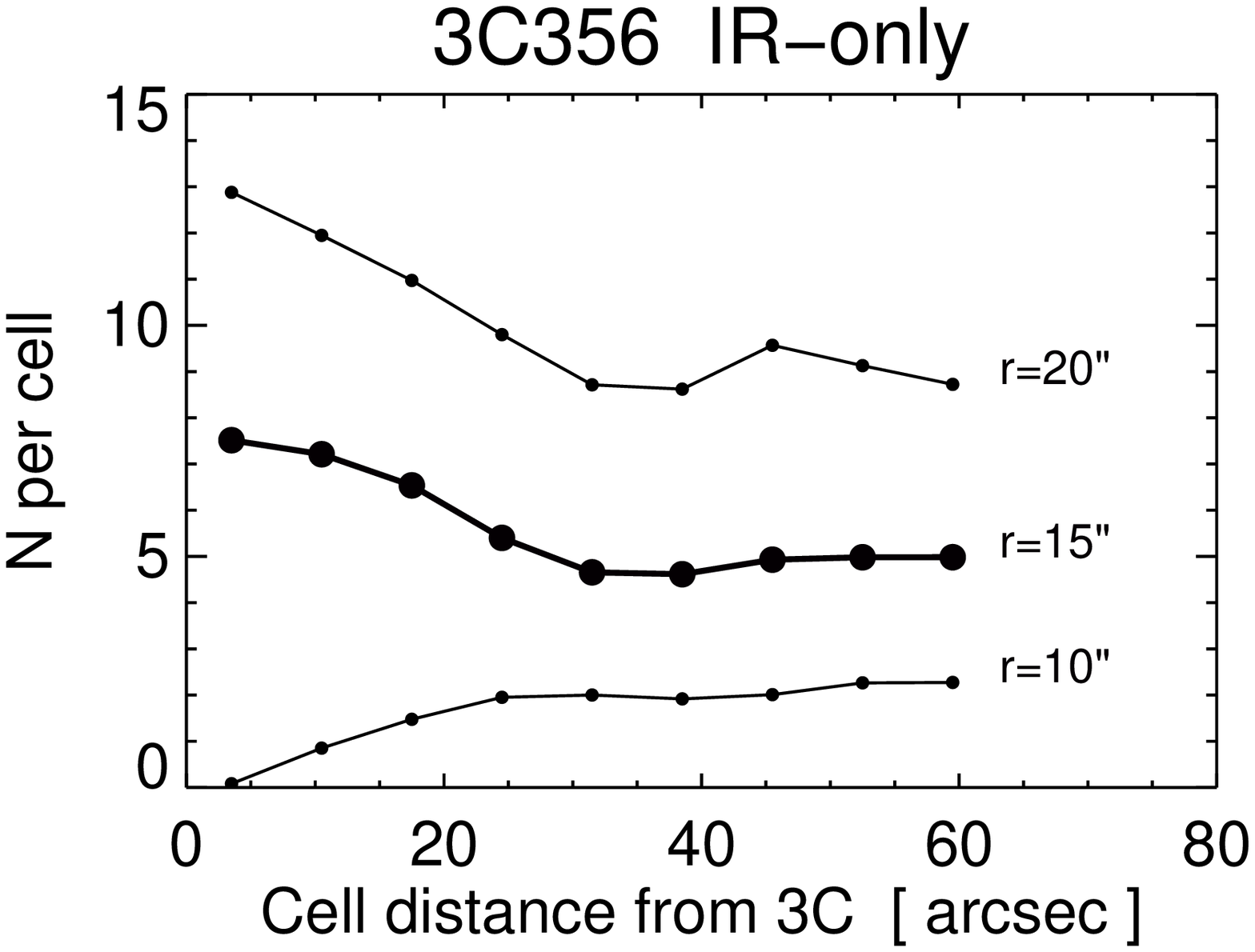}                 
                \includegraphics[width=0.245\textwidth, clip=true]{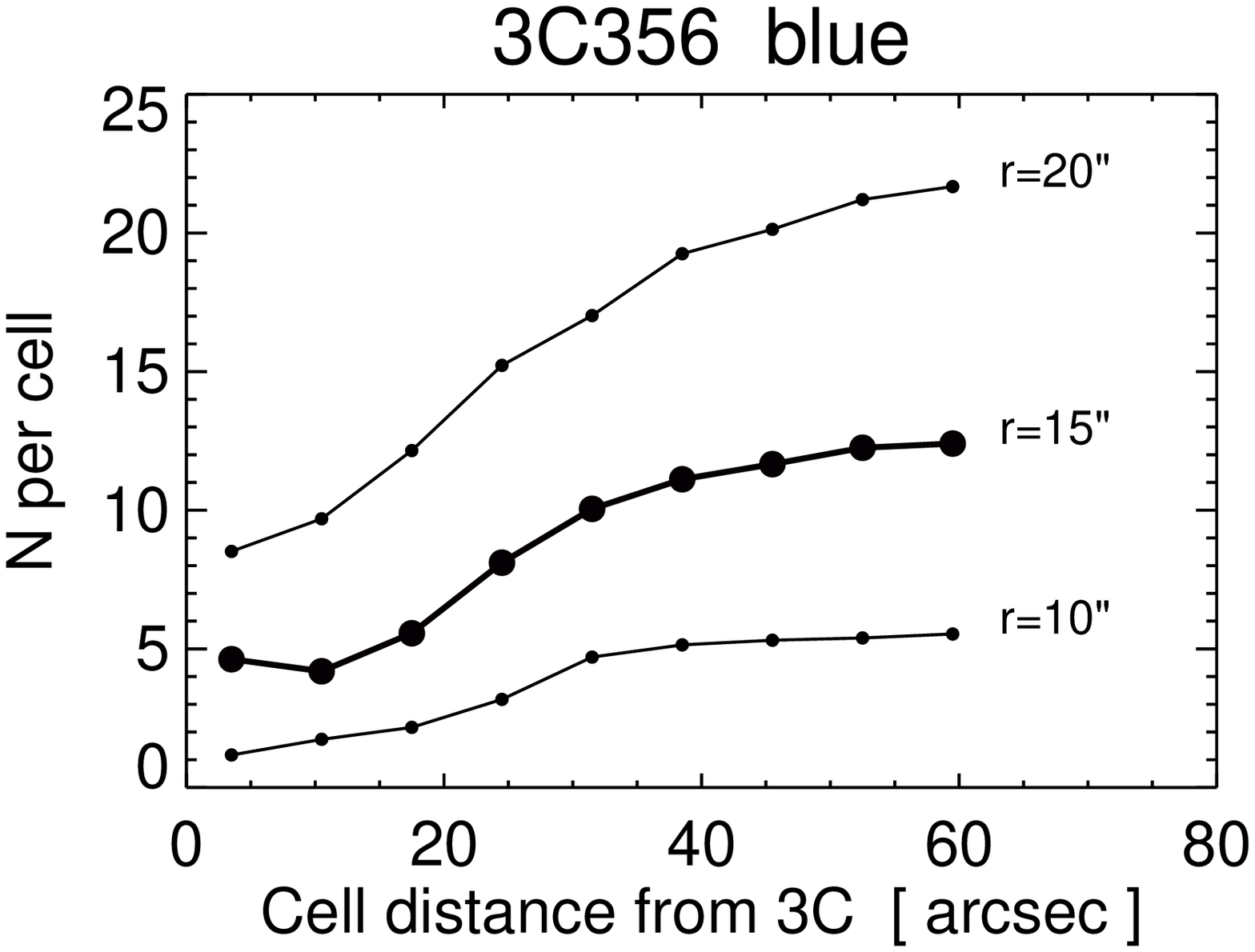}

                \hspace{-0mm}\includegraphics[width=0.245\textwidth, clip=true]{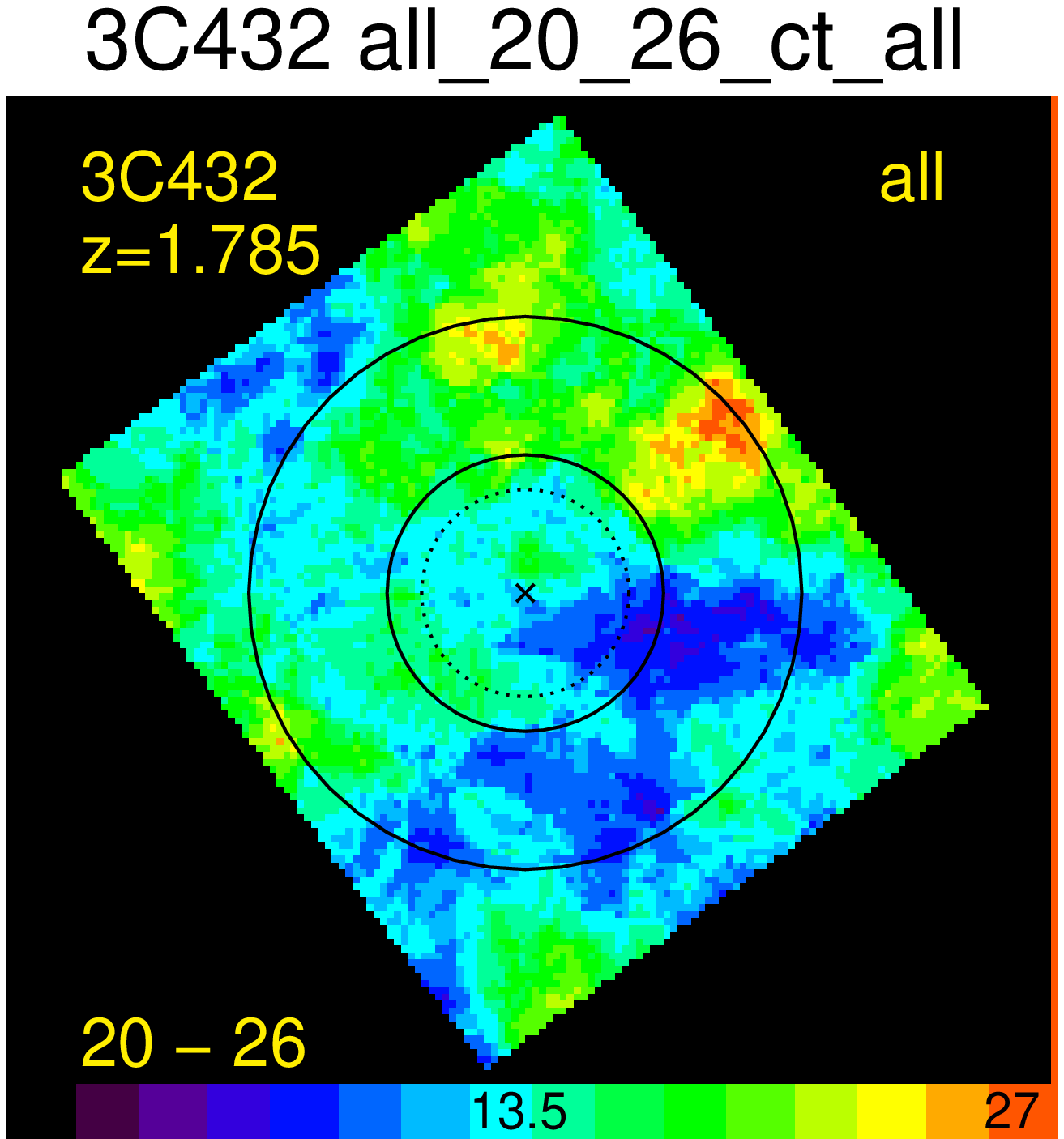}                 
                \includegraphics[width=0.245\textwidth, clip=true]{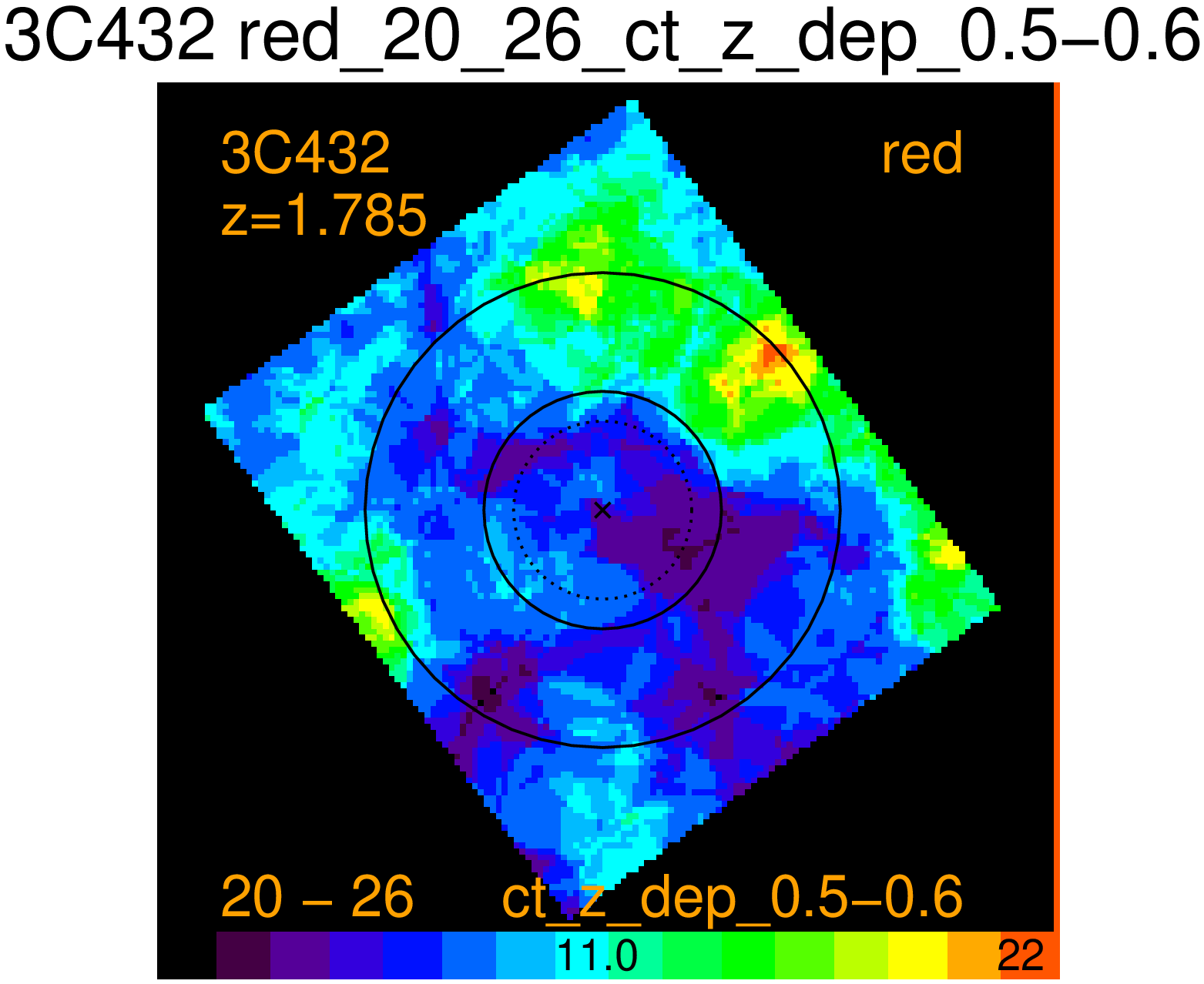}       
                \includegraphics[width=0.245\textwidth, clip=true]{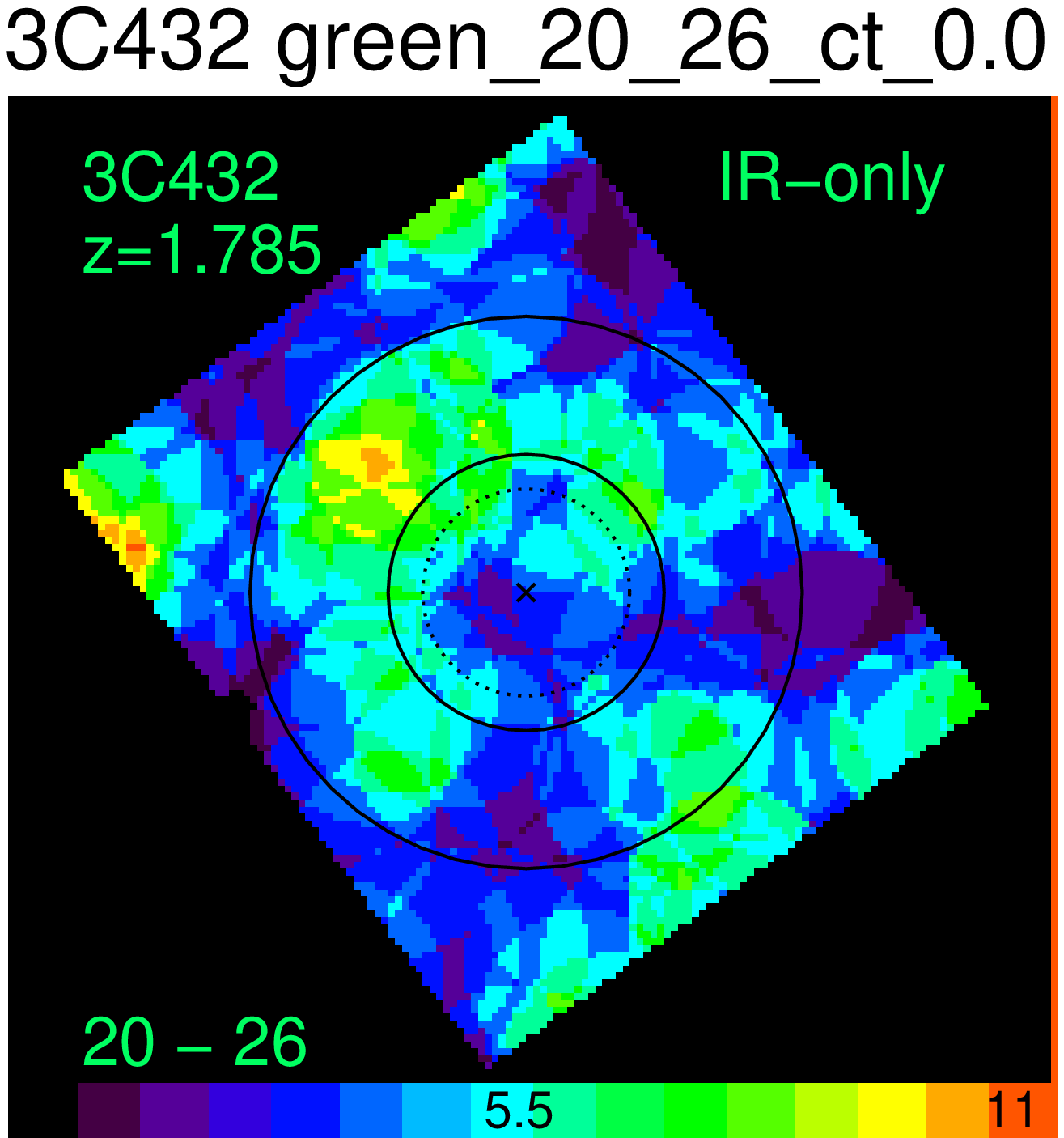}               
                \includegraphics[width=0.245\textwidth, clip=true]{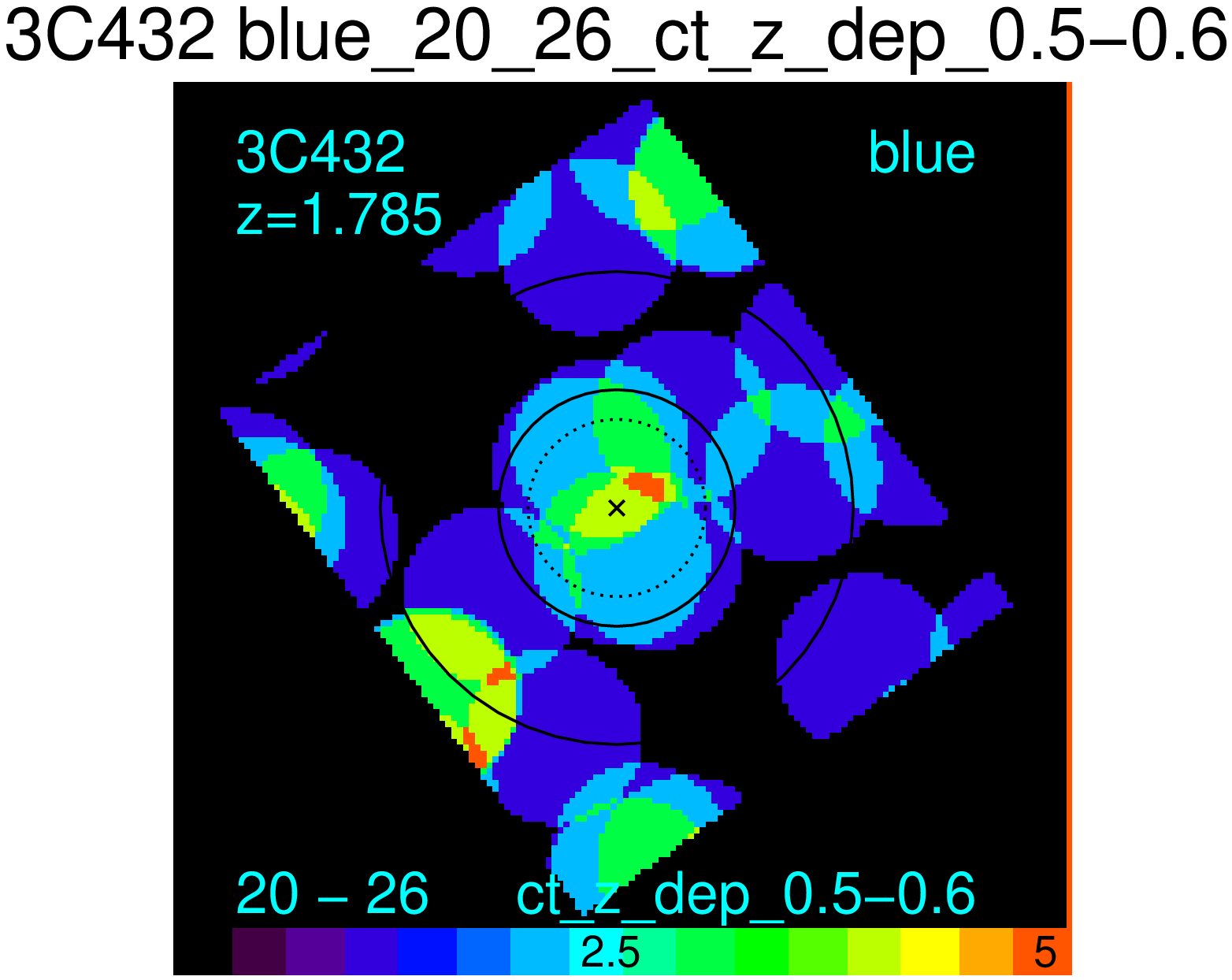}      
                
                \hspace{-0mm}\includegraphics[width=0.245\textwidth, clip=true]{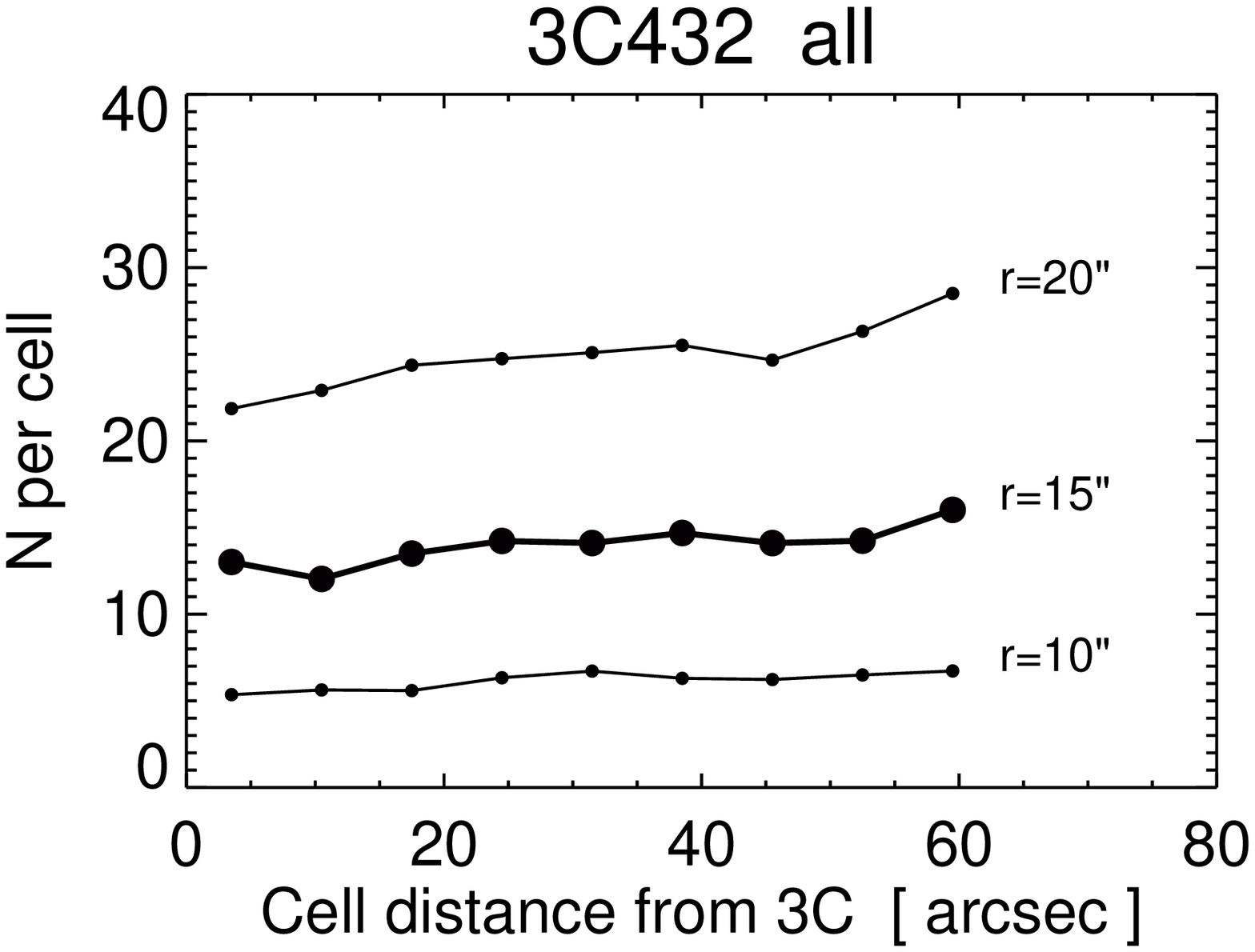}                   
                \includegraphics[width=0.245\textwidth, clip=true]{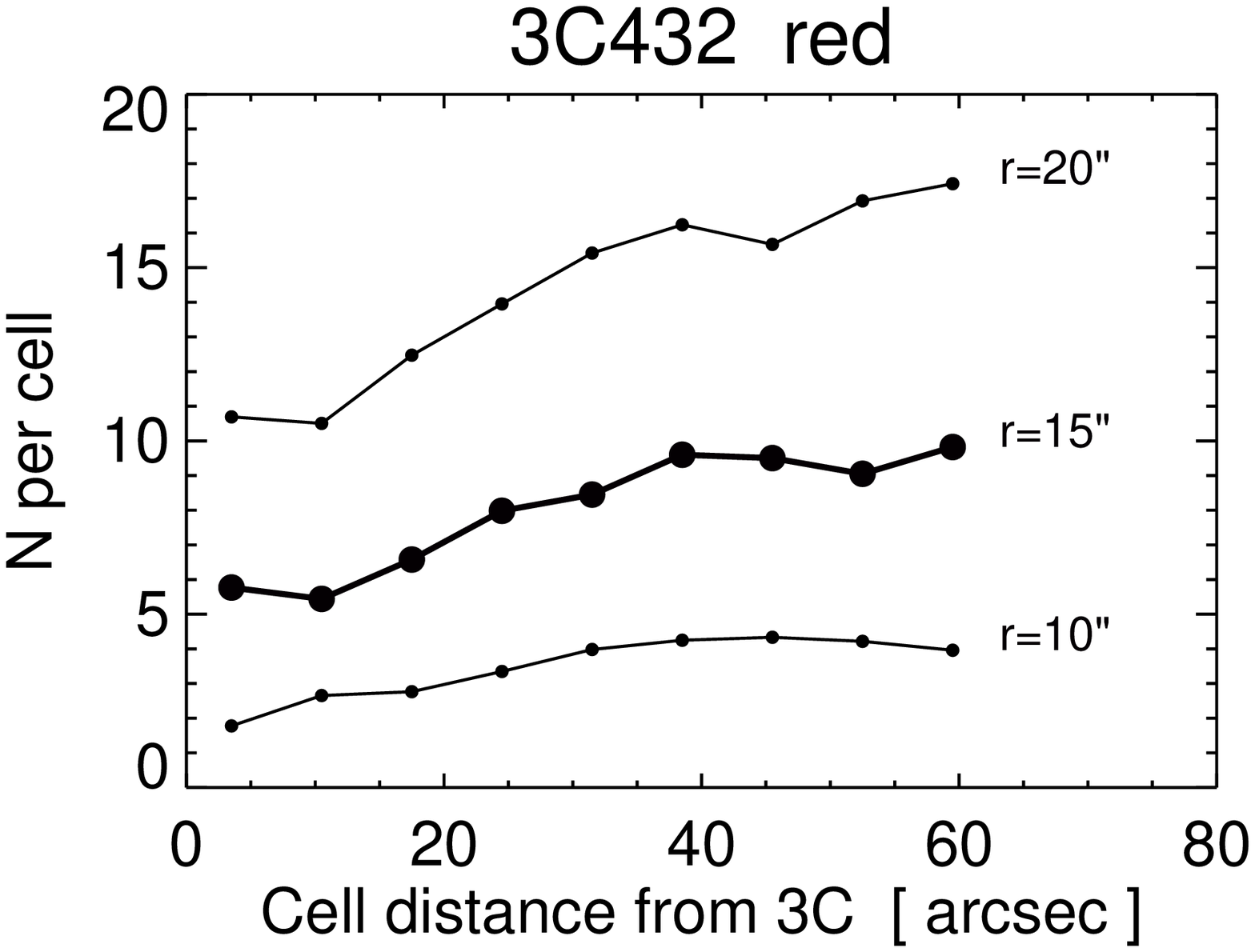}                    
                \includegraphics[width=0.245\textwidth, clip=true]{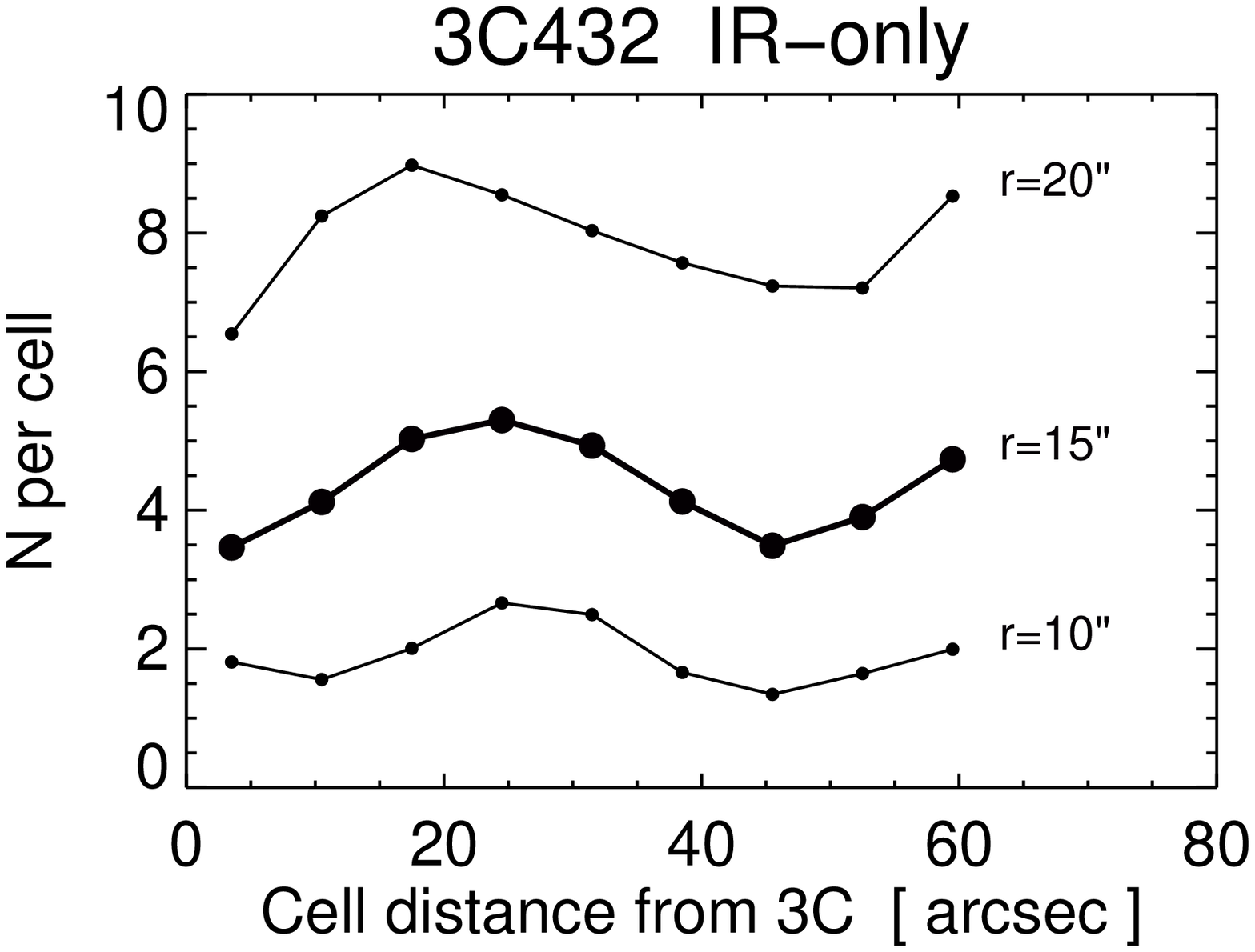}                 
                \includegraphics[width=0.245\textwidth, clip=true]{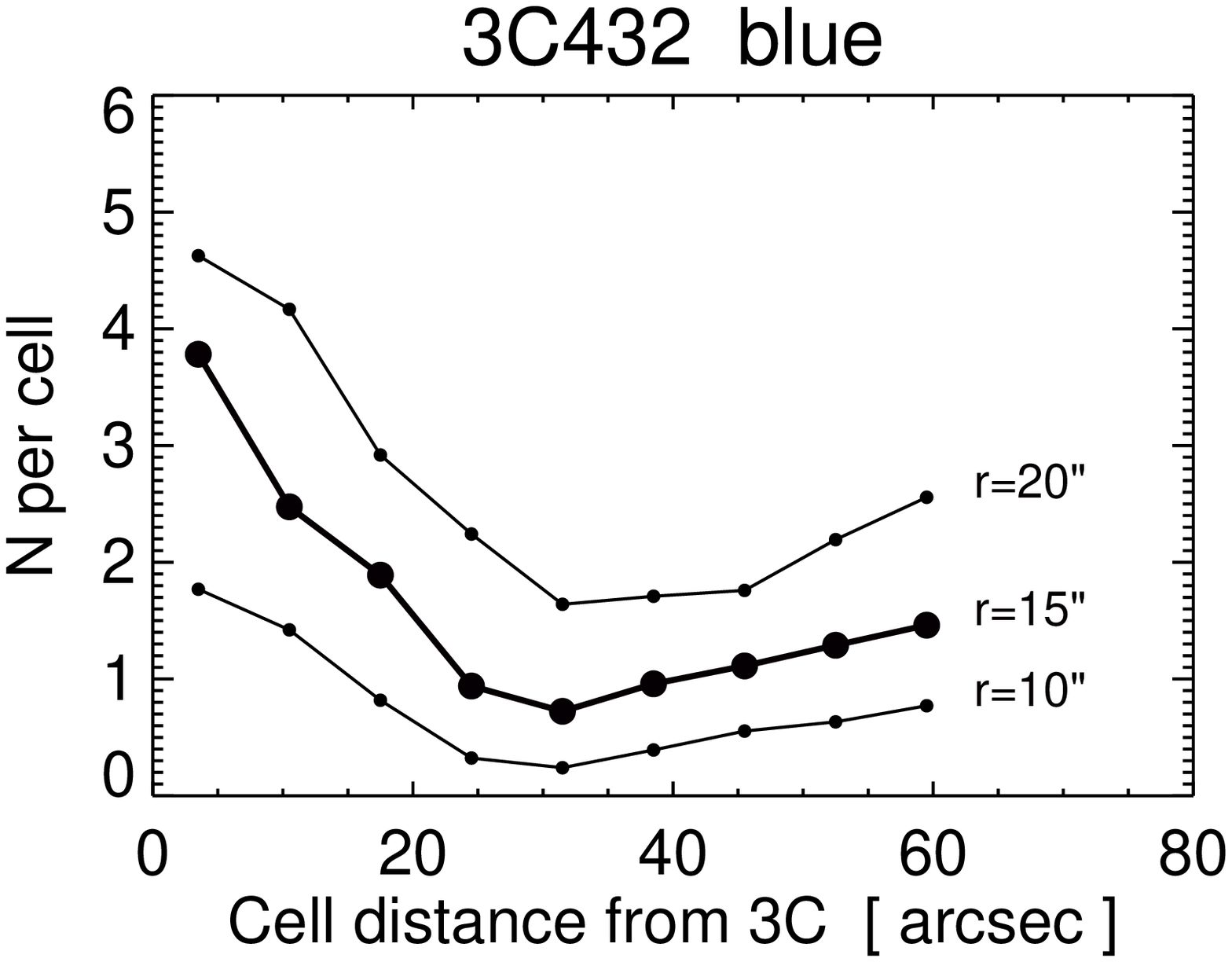}

                \hspace{-0mm}\includegraphics[width=0.245\textwidth, clip=true]{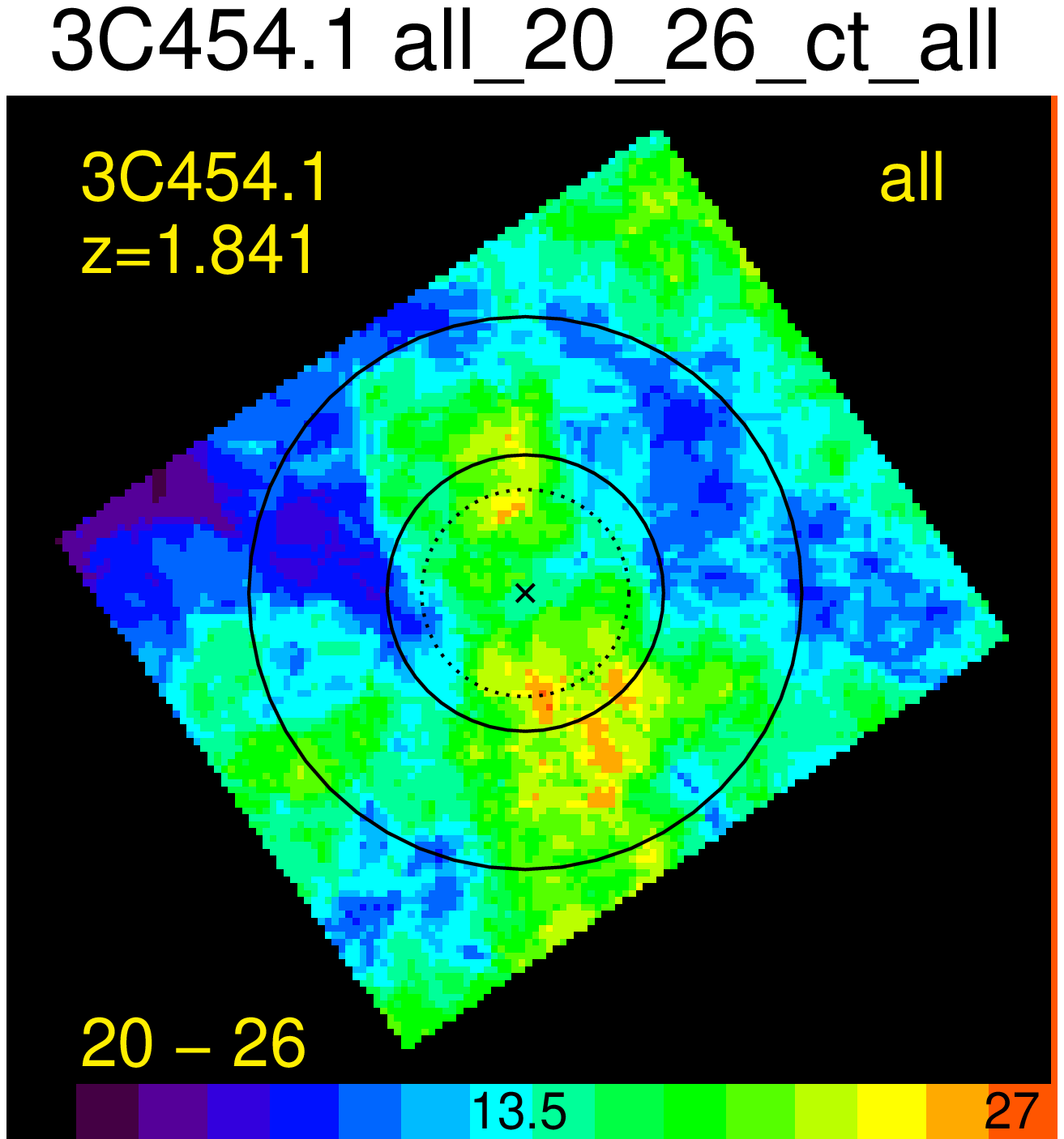}               
                \includegraphics[width=0.245\textwidth, clip=true]{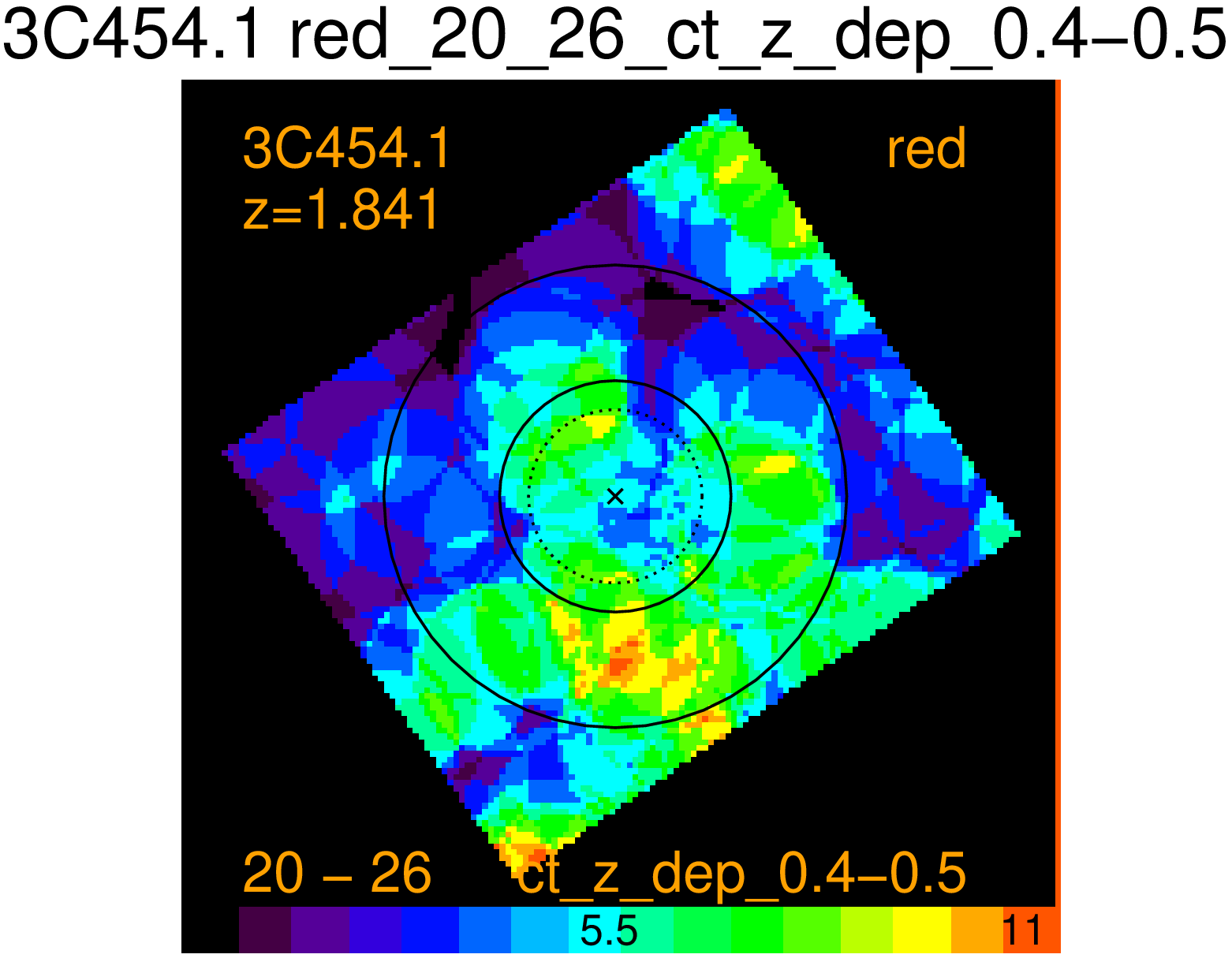}     
                \includegraphics[width=0.245\textwidth, clip=true]{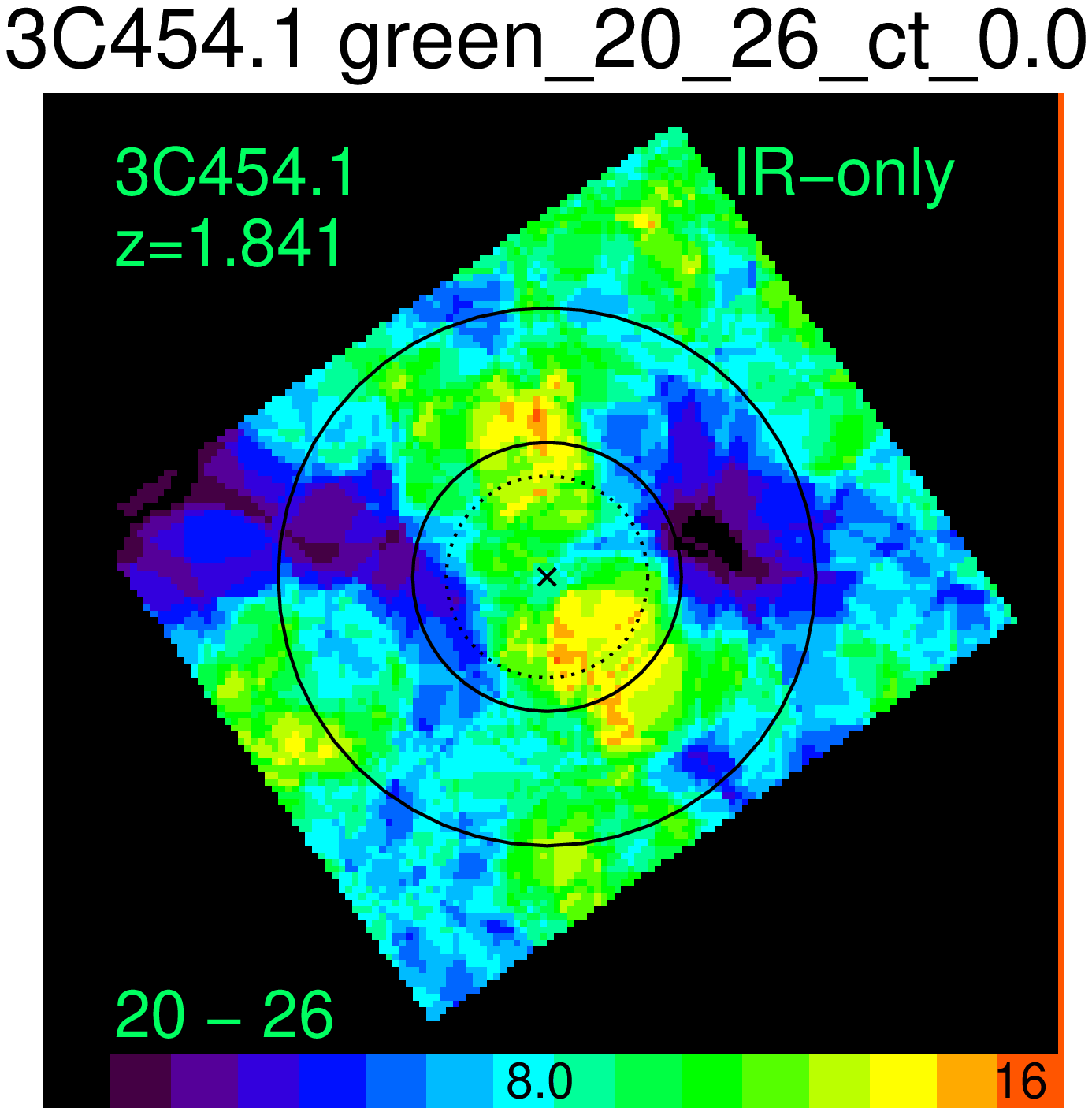}             
                \includegraphics[width=0.245\textwidth, clip=true]{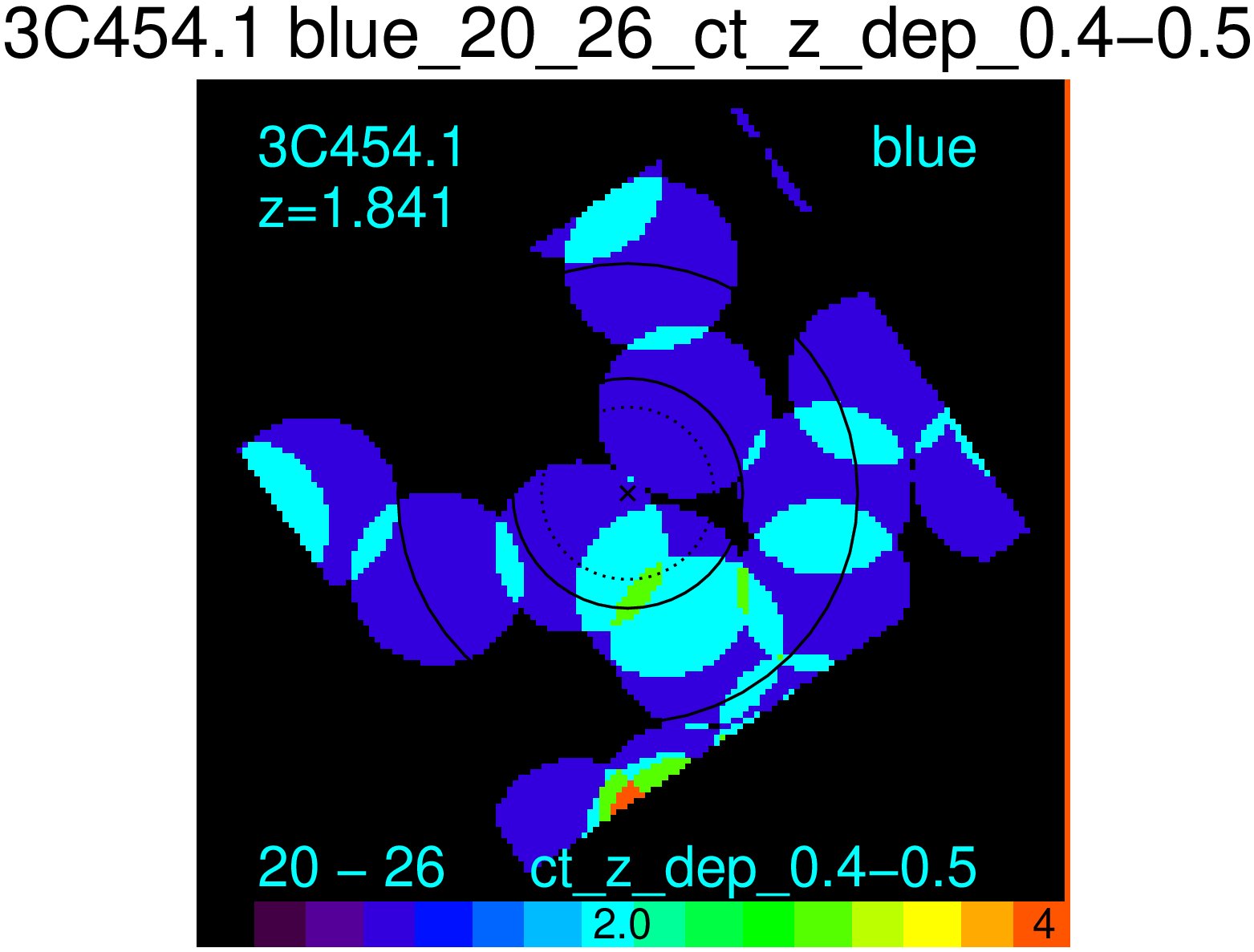}    
                
                \hspace{-0mm}\includegraphics[width=0.245\textwidth, clip=true]{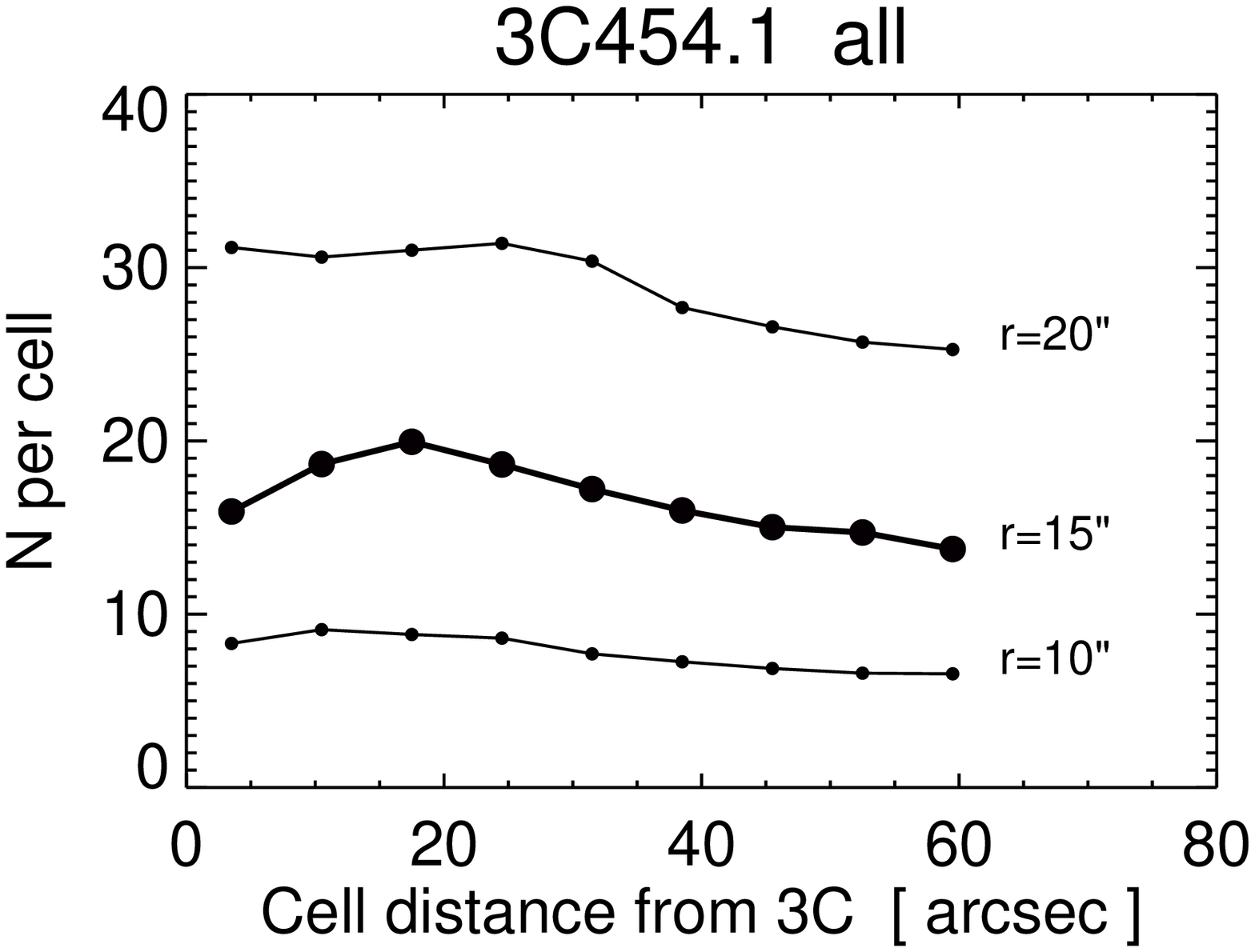}                 
                \includegraphics[width=0.245\textwidth, clip=true]{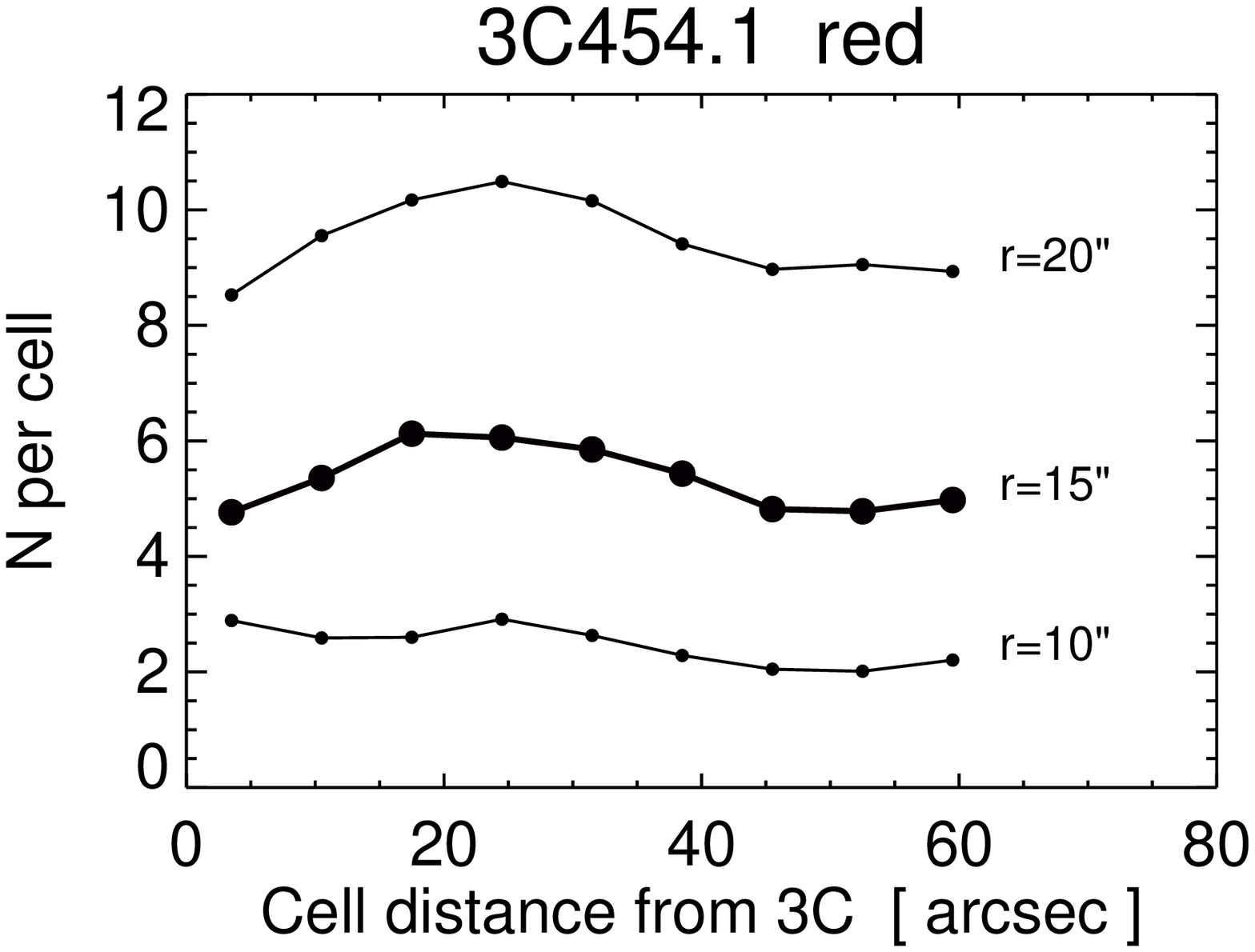}                  
                \includegraphics[width=0.245\textwidth, clip=true]{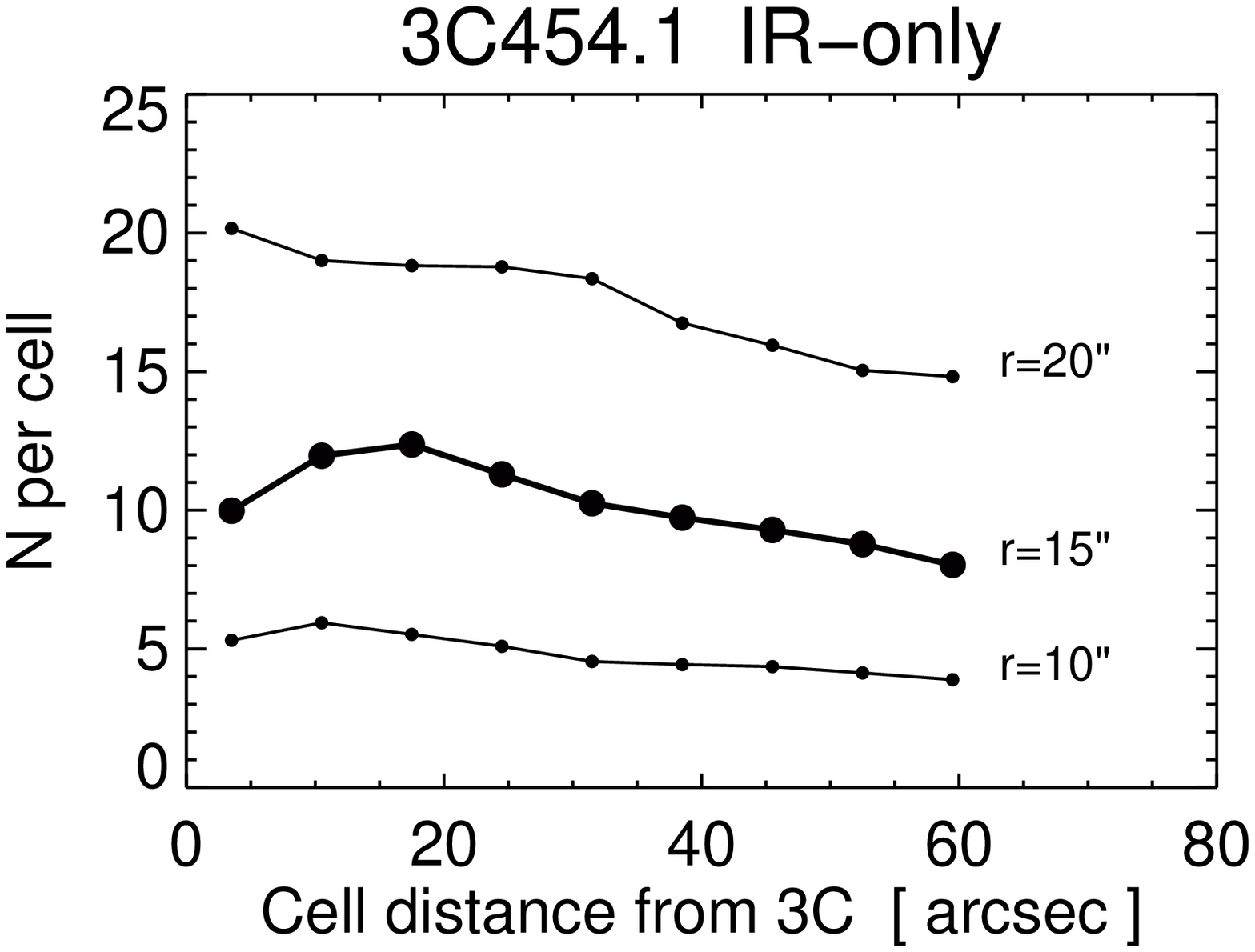}               
                \includegraphics[width=0.245\textwidth, clip=true]{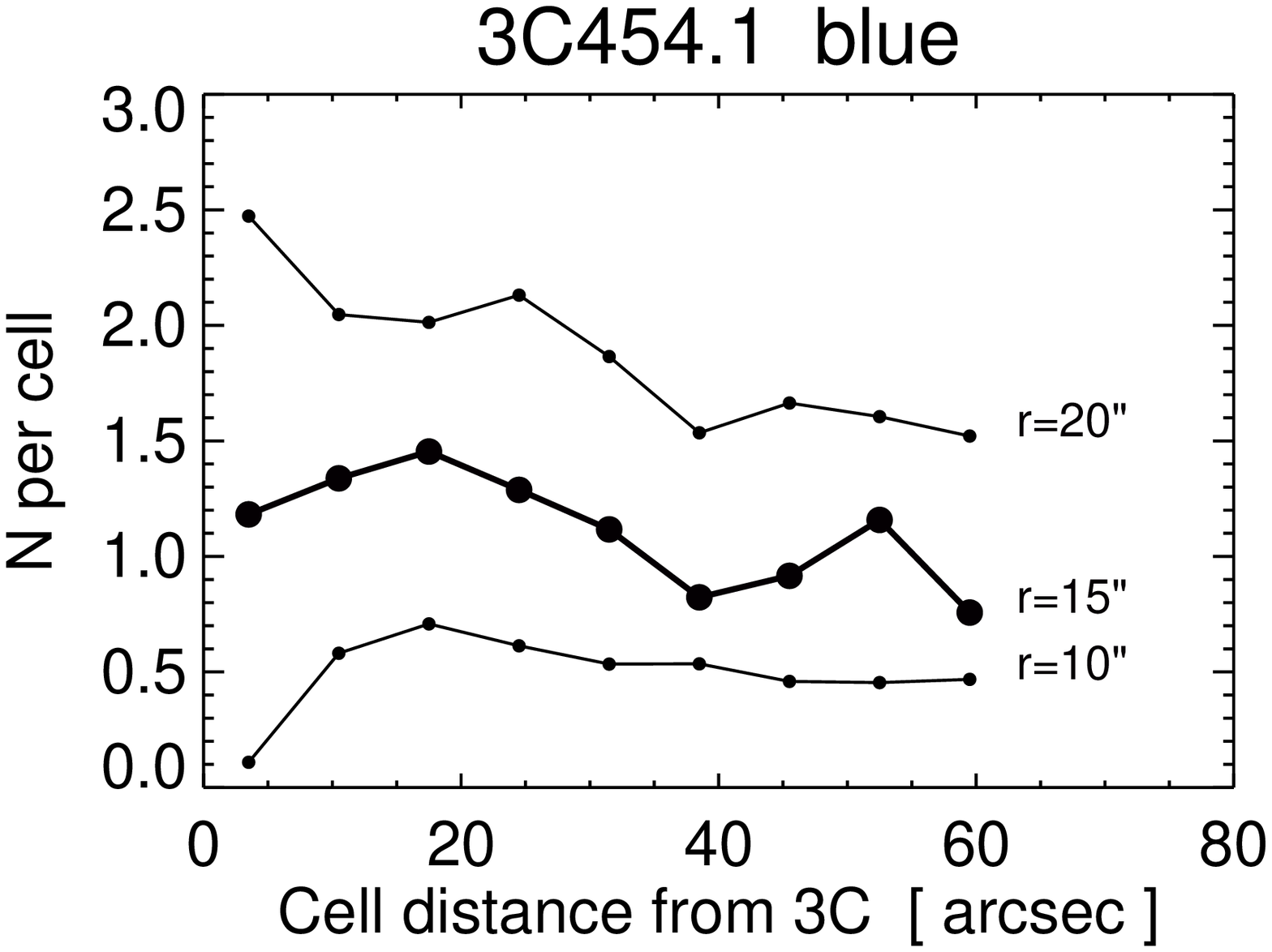}

                \caption{Surface density maps and radial density profiles of the 3C fields, continued.
                }
                \label{fig:sd_maps_7}
              \end{figure*}

              Figs.~\ref{fig:sd_maps_1} to~\ref{fig:sd_maps_7} show the surface density maps and radial profiles of the entire sample.
              Explanations are given in Sect.~\ref{sec:maps}.

              \section{Tables}\label{appendix_t}

              \begin{table*}
                \footnotesize
                \caption{Radial surface density profiles, as shown in Figs.~\ref{fig:sd_maps_1} to~\ref{fig:sd_maps_7}, using $20 < \rm{F140W} < 26$, cell radius $15\arcsec$,  and radial bins of 7$\arcsec$. 
                  Column 2 specifies the sample ($all$, $blue$, $IR$-$only$, $red$). Columns 3--11 list the average and standard deviation of the number $N(i)$ of galaxies per cell in the $i$-th radial bin. The $i$-th radial bin contains all cells whose central position has a distance $d$ from the 3C source of $(i-1) < d/7\arcsec < i $.
                }
                \label{tab_rsf}
                \setlength\tabcolsep{2.0pt}
                \begin{tabular}{llrrrrrrrrr}
                  (1)       & (2)  & (3)  & (4) & (5)  & (6)  & (7)  &  (8)  & (9)  & (10)  & (11) \\
                  Name       & color  & N1  & N2 & N3  & N4  & N5  &  N6  & N7  & N8  & N9 \\
                  \hline
                  &   &   &  &   &   &   &    &   &   &  \\

                  3C068.1  &  all    &  25.46$\pm$ 3.31  &  23.09$\pm$ 3.80  &  22.10$\pm$ 4.71  &  20.54$\pm$ 6.43  &  18.62$\pm$ 6.18  &  17.92$\pm$ 4.80  &  17.51$\pm$ 4.38  &  18.08$\pm$ 5.88  &  16.33$\pm$ 6.20  \\
                  3C068.1  &  red    &   4.14$\pm$ 2.04  &   3.90$\pm$ 2.64  &   3.13$\pm$ 2.88  &   3.16$\pm$ 2.90  &   3.31$\pm$ 2.46  &   3.16$\pm$ 1.73  &   3.34$\pm$ 1.70  &   3.63$\pm$ 2.01  &   3.02$\pm$ 2.04  \\
                  3C068.1  & IR-only &  11.51$\pm$ 2.10  &   9.05$\pm$ 2.04  &   7.91$\pm$ 2.05  &   6.12$\pm$ 2.62  &   4.59$\pm$ 3.18  &   4.57$\pm$ 2.42  &   4.63$\pm$ 1.72  &   5.23$\pm$ 2.08  &   4.97$\pm$ 2.26  \\
                  3C068.1  &  blue   &   9.80$\pm$ 1.43  &  10.14$\pm$ 2.45  &  11.07$\pm$ 3.26  &  11.25$\pm$ 3.44  &  10.72$\pm$ 2.87  &  10.19$\pm$ 2.50  &   9.54$\pm$ 2.80  &   9.22$\pm$ 3.82  &   8.35$\pm$ 3.52  \\
                  &   &   &  &   &   &   &    &   &   &  \\

                  3C186    &  all    &  15.55$\pm$ 1.16  &  16.38$\pm$ 2.13  &  15.64$\pm$ 2.90  &  14.90$\pm$ 4.43  &  15.02$\pm$ 5.01  &  15.57$\pm$ 4.44  &  16.35$\pm$ 4.28  &  16.86$\pm$ 4.11  &  16.97$\pm$ 3.79  \\
                  3C186    &  red    &   4.80$\pm$ 0.86  &   4.96$\pm$ 1.15  &   4.28$\pm$ 1.65  &   3.79$\pm$ 1.86  &   3.42$\pm$ 1.94  &   3.27$\pm$ 2.15  &   2.97$\pm$ 1.91  &   2.78$\pm$ 1.74  &   2.42$\pm$ 1.64  \\
                  3C186    & IR-only &   9.31$\pm$ 1.55  &   8.86$\pm$ 2.61  &   7.99$\pm$ 3.23  &   6.36$\pm$ 3.20  &   5.86$\pm$ 2.83  &   5.87$\pm$ 2.32  &   6.03$\pm$ 2.14  &   6.47$\pm$ 2.18  &   6.68$\pm$ 2.42  \\
                  3C186    &  blue   &   1.44$\pm$ 0.95  &   2.56$\pm$ 1.26  &   3.36$\pm$ 1.30  &   4.75$\pm$ 1.41  &   5.74$\pm$ 1.70  &   6.43$\pm$ 1.79  &   7.35$\pm$ 2.42  &   7.61$\pm$ 2.46  &   7.88$\pm$ 2.39  \\
                  &   &   &  &   &   &   &    &   &   &  \\
                  3C208.0  &  all    &  27.24$\pm$ 3.65  &  24.15$\pm$ 5.64  &  22.60$\pm$ 7.39  &  19.56$\pm$ 7.03  &  16.21$\pm$ 5.06  &  14.95$\pm$ 3.77  &  13.68$\pm$ 3.35  &  13.14$\pm$ 3.17  &  12.13$\pm$ 3.13  \\
                  3C208.0  &  red    &   7.72$\pm$ 1.47  &   5.95$\pm$ 1.25  &   5.78$\pm$ 1.46  &   4.34$\pm$ 1.62  &   2.64$\pm$ 1.72  &   1.93$\pm$ 1.63  &   1.39$\pm$ 1.26  &   1.62$\pm$ 1.31  &   1.60$\pm$ 1.64  \\
                  3C208.0  & IR-only &   7.09$\pm$ 2.20  &   7.14$\pm$ 3.26  &   6.85$\pm$ 4.44  &   6.61$\pm$ 4.61  &   5.82$\pm$ 3.32  &   5.40$\pm$ 2.41  &   4.96$\pm$ 1.97  &   4.35$\pm$ 2.05  &   4.41$\pm$ 1.99  \\
                  3C208.0  &  blue   &  12.42$\pm$ 1.47  &  11.07$\pm$ 2.77  &   9.97$\pm$ 2.69  &   8.61$\pm$ 2.60  &   7.75$\pm$ 2.61  &   7.62$\pm$ 3.09  &   7.33$\pm$ 3.34  &   7.17$\pm$ 3.23  &   6.12$\pm$ 2.78  \\
                  3C210    &  all    &  31.00$\pm$ 3.64  &  29.05$\pm$ 4.65  &  24.78$\pm$ 5.15  &  20.34$\pm$ 4.74  &  18.31$\pm$ 5.10  &  17.25$\pm$ 5.36  &  17.99$\pm$ 4.52  &  18.66$\pm$ 3.34  &  18.69$\pm$ 3.27  \\
                  3C210    &  red    &  14.48$\pm$ 0.89  &  12.21$\pm$ 1.89  &   7.49$\pm$ 2.63  &   3.84$\pm$ 1.62  &   2.70$\pm$ 1.37  &   2.20$\pm$ 1.59  &   2.14$\pm$ 1.77  &   2.84$\pm$ 1.68  &   3.01$\pm$ 1.58  \\
                  3C210    & IR-only &  10.15$\pm$ 2.60  &   9.47$\pm$ 3.06  &   8.60$\pm$ 3.27  &   6.59$\pm$ 3.48  &   5.35$\pm$ 2.65  &   4.93$\pm$ 1.96  &   5.46$\pm$ 2.14  &   5.68$\pm$ 2.31  &   5.90$\pm$ 2.43  \\
                  3C210    &  blue   &   6.37$\pm$ 1.35  &   7.36$\pm$ 1.88  &   8.70$\pm$ 1.83  &   9.91$\pm$ 2.49  &  10.25$\pm$ 3.37  &  10.12$\pm$ 3.67  &  10.39$\pm$ 2.98  &  10.14$\pm$ 2.66  &   9.79$\pm$ 2.82  \\
                  &   &   &  &   &   &   &    &   &   &  \\
                  3C220.2  &  all    &  19.53$\pm$ 1.80  &  17.15$\pm$ 2.59  &  15.14$\pm$ 3.06  &  13.78$\pm$ 3.80  &  14.40$\pm$ 4.14  &  15.18$\pm$ 4.18  &  16.10$\pm$ 4.41  &  15.85$\pm$ 4.07  &  14.23$\pm$ 4.73  \\
                  3C220.2  &  red    &   5.45$\pm$ 0.97  &   4.07$\pm$ 0.91  &   3.03$\pm$ 1.09  &   2.22$\pm$ 1.36  &   1.61$\pm$ 1.34  &   1.30$\pm$ 1.01  &   1.43$\pm$ 1.17  &   1.52$\pm$ 1.33  &   1.49$\pm$ 1.32  \\
                  3C220.2  & IR-only &   6.37$\pm$ 0.90  &   5.41$\pm$ 1.02  &   4.25$\pm$ 1.24  &   3.38$\pm$ 1.59  &   3.25$\pm$ 1.80  &   3.50$\pm$ 1.61  &   4.12$\pm$ 1.99  &   4.43$\pm$ 2.38  &   4.20$\pm$ 2.39  \\
                  3C220.2  &  blue   &   7.70$\pm$ 1.38  &   7.66$\pm$ 1.98  &   7.87$\pm$ 2.73  &   8.18$\pm$ 3.16  &   9.54$\pm$ 3.35  &  10.38$\pm$ 3.38  &  10.55$\pm$ 3.57  &   9.89$\pm$ 3.38  &   8.54$\pm$ 2.55  \\
                  &   &   &  &   &   &   &    &   &   &  \\
                  3C230    &  all    &  29.76$\pm$ 2.12  &  25.21$\pm$ 4.00  &  20.28$\pm$ 4.50  &  16.91$\pm$ 2.63  &  15.30$\pm$ 2.82  &  15.56$\pm$ 4.31  &  16.32$\pm$ 4.96  &  16.16$\pm$ 4.40  &  16.29$\pm$ 3.86  \\
                  3C230    &  red    &  12.16$\pm$ 1.66  &  11.00$\pm$ 2.93  &   8.23$\pm$ 3.59  &   6.28$\pm$ 2.49  &   5.48$\pm$ 1.84  &   5.52$\pm$ 2.08  &   5.59$\pm$ 1.85  &   5.00$\pm$ 1.88  &   4.12$\pm$ 1.62  \\
                  3C230    & IR-only &  11.50$\pm$ 1.45  &   9.05$\pm$ 1.37  &   7.51$\pm$ 1.74  &   5.81$\pm$ 1.90  &   4.68$\pm$ 1.92  &   4.67$\pm$ 2.46  &   4.70$\pm$ 3.02  &   4.35$\pm$ 2.32  &   3.87$\pm$ 1.52  \\
                  3C230    &  blue   &   6.10$\pm$ 0.58  &   5.16$\pm$ 1.19  &   4.54$\pm$ 1.86  &   4.81$\pm$ 2.08  &   5.14$\pm$ 2.20  &   5.37$\pm$ 1.73  &   6.04$\pm$ 2.22  &   6.81$\pm$ 2.06  &   8.30$\pm$ 2.58  \\
                  &   &   &  &   &   &   &    &   &   &  \\
                  3C255    &  all    &  24.73$\pm$ 2.24  &  23.67$\pm$ 3.18  &  21.18$\pm$ 2.79  &  19.26$\pm$ 3.30  &  19.17$\pm$ 3.40  &  19.32$\pm$ 5.34  &  18.11$\pm$ 7.60  &  15.66$\pm$ 6.47  &  13.54$\pm$ 3.74  \\
                  3C255    &  red    &   9.83$\pm$ 1.46  &   7.99$\pm$ 2.43  &   6.53$\pm$ 2.15  &   4.64$\pm$ 1.59  &   3.79$\pm$ 1.27  &   3.68$\pm$ 1.75  &   3.59$\pm$ 2.37  &   2.98$\pm$ 2.36  &   2.02$\pm$ 1.52  \\
                  3C255    & IR-only &   5.99$\pm$ 1.00  &   6.16$\pm$ 1.29  &   6.62$\pm$ 1.40  &   6.70$\pm$ 2.82  &   6.69$\pm$ 3.71  &   6.30$\pm$ 4.31  &   4.97$\pm$ 3.82  &   3.80$\pm$ 2.57  &   4.13$\pm$ 1.32  \\
                  3C255    &  blue   &   8.91$\pm$ 1.50  &   9.52$\pm$ 1.49  &   8.02$\pm$ 1.78  &   7.93$\pm$ 1.97  &   8.68$\pm$ 2.78  &   9.35$\pm$ 3.03  &   9.55$\pm$ 3.45  &   8.88$\pm$ 3.28  &   7.39$\pm$ 2.55  \\
                  &   &   &  &   &   &   &    &   &   &  \\
                  3C257    &  all    &  18.80$\pm$ 1.24  &  18.48$\pm$ 3.36  &  17.08$\pm$ 4.73  &  16.17$\pm$ 4.31  &  16.03$\pm$ 4.00  &  15.52$\pm$ 4.11  &  14.26$\pm$ 3.46  &  14.39$\pm$ 3.72  &  13.59$\pm$ 2.65  \\
                  3C257    &  red    &  10.99$\pm$ 1.29  &  10.44$\pm$ 2.59  &   8.89$\pm$ 3.35  &   8.15$\pm$ 3.28  &   8.32$\pm$ 2.94  &   8.53$\pm$ 2.88  &   8.52$\pm$ 2.53  &   8.38$\pm$ 2.30  &   8.25$\pm$ 2.10  \\
                  3C257    & IR-only &   4.99$\pm$ 0.57  &   4.92$\pm$ 1.31  &   4.68$\pm$ 1.82  &   4.39$\pm$ 1.94  &   4.41$\pm$ 2.21  &   4.00$\pm$ 2.09  &   3.09$\pm$ 1.41  &   3.47$\pm$ 1.53  &   3.17$\pm$ 1.55  \\
                  3C257    &  blue   &   2.83$\pm$ 1.21  &   3.13$\pm$ 0.98  &   3.53$\pm$ 1.17  &   3.63$\pm$ 1.24  &   3.29$\pm$ 1.29  &   2.98$\pm$ 1.36  &   2.64$\pm$ 1.29  &   2.53$\pm$ 1.33  &   2.17$\pm$ 1.29  \\
                  &   &   &  &   &   &   &    &   &   &  \\
                  3C268.4  &  all    &  26.19$\pm$ 2.88  &  24.97$\pm$ 4.28  &  22.86$\pm$ 4.81  &  20.31$\pm$ 5.26  &  18.16$\pm$ 5.52  &  16.87$\pm$ 4.05  &  17.07$\pm$ 4.11  &  17.08$\pm$ 4.35  &  16.70$\pm$ 3.09  \\
                  3C268.4  &  red    &   4.16$\pm$ 0.75  &   4.93$\pm$ 0.93  &   5.50$\pm$ 1.36  &   5.71$\pm$ 1.75  &   5.08$\pm$ 1.97  &   4.13$\pm$ 1.82  &   3.70$\pm$ 1.92  &   3.16$\pm$ 2.08  &   2.93$\pm$ 1.81  \\
                  3C268.4  & IR-only &   9.96$\pm$ 1.18  &   8.83$\pm$ 2.07  &   7.67$\pm$ 2.86  &   6.08$\pm$ 2.95  &   4.50$\pm$ 2.43  &   3.65$\pm$ 1.69  &   3.47$\pm$ 1.42  &   3.51$\pm$ 1.55  &   3.19$\pm$ 1.84  \\
                  3C268.4  &  blue   &  12.07$\pm$ 1.97  &  11.21$\pm$ 2.52  &   9.69$\pm$ 2.25  &   8.53$\pm$ 2.88  &   8.57$\pm$ 3.85  &   9.10$\pm$ 3.29  &   9.89$\pm$ 3.33  &  10.41$\pm$ 3.64  &  10.57$\pm$ 3.00  \\
                  &   &   &  &   &   &   &    &   &   &  \\
                  3C270.1  &  all    &  21.15$\pm$ 1.96  &  20.71$\pm$ 2.32  &  18.58$\pm$ 3.02  &  17.41$\pm$ 3.26  &  17.62$\pm$ 3.31  &  17.78$\pm$ 3.15  &  18.67$\pm$ 2.98  &  19.31$\pm$ 3.42  &  19.14$\pm$ 3.95  \\
                  3C270.1  &  red    &  12.22$\pm$ 1.10  &  10.21$\pm$ 1.52  &   8.18$\pm$ 1.62  &   6.72$\pm$ 1.91  &   6.31$\pm$ 1.96  &   6.37$\pm$ 2.33  &   6.94$\pm$ 2.78  &   7.41$\pm$ 2.35  &   6.78$\pm$ 1.94  \\
                  3C270.1  & IR-only &   4.52$\pm$ 0.70  &   4.51$\pm$ 1.46  &   3.86$\pm$ 2.14  &   3.38$\pm$ 2.20  &   3.50$\pm$ 2.22  &   3.60$\pm$ 2.02  &   3.93$\pm$ 1.65  &   4.66$\pm$ 1.59  &   5.01$\pm$ 1.75  \\
                  3C270.1  &  blue   &   4.41$\pm$ 0.90  &   5.99$\pm$ 1.32  &   6.54$\pm$ 1.73  &   7.31$\pm$ 2.01  &   7.80$\pm$ 2.08  &   7.81$\pm$ 2.15  &   7.80$\pm$ 2.13  &   7.23$\pm$ 2.72  &   7.34$\pm$ 3.49  \\
                  &   &   &  &   &   &   &    &   &   &  \\
                  3C287    &  all    &  16.52$\pm$ 1.85  &  16.94$\pm$ 2.77  &  17.32$\pm$ 4.47  &  17.93$\pm$ 3.99  &  17.70$\pm$ 3.23  &  18.39$\pm$ 4.00  &  19.48$\pm$ 4.38  &  20.44$\pm$ 4.39  &  19.08$\pm$ 4.46  \\
                  3C287    &  red    &   2.73$\pm$ 0.65  &   2.66$\pm$ 0.80  &   1.90$\pm$ 0.82  &   1.51$\pm$ 0.93  &   1.46$\pm$ 0.90  &   1.58$\pm$ 0.96  &   1.92$\pm$ 1.57  &   1.90$\pm$ 1.75  &   1.83$\pm$ 1.98  \\
                  3C287    & IR-only &   5.88$\pm$ 1.15  &   6.23$\pm$ 1.46  &   6.61$\pm$ 2.28  &   6.02$\pm$ 2.06  &   4.52$\pm$ 1.69  &   4.05$\pm$ 1.65  &   4.51$\pm$ 1.74  &   4.98$\pm$ 1.72  &   4.85$\pm$ 1.38  \\
                  3C287    &  blue   &   7.92$\pm$ 1.44  &   8.06$\pm$ 1.93  &   8.81$\pm$ 2.75  &  10.41$\pm$ 2.87  &  11.72$\pm$ 2.53  &  12.77$\pm$ 2.95  &  13.05$\pm$ 3.38  &  13.56$\pm$ 3.76  &  12.40$\pm$ 3.23  \\
                  &   &   &  &   &   &   &    &   &   &  \\
                  3C297    &  all    &  22.50$\pm$ 3.23  &  19.82$\pm$ 4.25  &  16.08$\pm$ 4.70  &  14.25$\pm$ 4.97  &  14.91$\pm$ 4.03  &  15.47$\pm$ 4.18  &  16.88$\pm$ 4.79  &  16.38$\pm$ 3.73  &  15.77$\pm$ 3.12  \\
                  3C297    &  red    &   6.35$\pm$ 1.02  &   5.68$\pm$ 1.67  &   4.45$\pm$ 2.73  &   3.86$\pm$ 2.45  &   3.76$\pm$ 1.92  &   3.35$\pm$ 1.82  &   3.02$\pm$ 1.76  &   2.54$\pm$ 1.53  &   1.79$\pm$ 1.07  \\
                  3C297    & IR-only &   5.69$\pm$ 1.34  &   4.93$\pm$ 1.80  &   4.62$\pm$ 1.61  &   3.96$\pm$ 1.44  &   3.68$\pm$ 1.89  &   4.08$\pm$ 2.22  &   4.76$\pm$ 2.14  &   4.95$\pm$ 2.10  &   4.70$\pm$ 1.49  \\
                  3C297    &  blue   &  10.46$\pm$ 1.58  &   9.21$\pm$ 2.47  &   7.01$\pm$ 3.21  &   6.43$\pm$ 3.03  &   7.47$\pm$ 2.34  &   8.04$\pm$ 2.41  &   9.10$\pm$ 3.29  &   8.89$\pm$ 2.96  &   9.28$\pm$ 2.35  \\
                \end{tabular}
              \end{table*}

              \addtocounter{table}{-1}

              \begin{table*}
                \footnotesize
                \caption{continued.
                }
                \label{tab_rsf_cont}
                \setlength\tabcolsep{2.0pt}
                \begin{tabular}{llrrrrrrrrr}
                  (1)       & (2)  & (3)  & (4) & (5)  & (6)  & (7)  &  (8)  & (9)  & (10)  & (11) \\
                  Name       & color  & N1  & N2 & N3  & N4  & N5  &  N6  & N7  & N8  & N9 \\
                  \hline
                  &   &   &  &   &   &   &    &   &   &  \\

                  3C298    &  all    &  11.60$\pm$ 1.77  &  12.28$\pm$ 2.25  &  13.18$\pm$ 2.86  &  13.46$\pm$ 3.34  &  13.54$\pm$ 4.55  &  13.74$\pm$ 5.00  &  13.67$\pm$ 5.51  &  14.02$\pm$ 4.92  &  17.23$\pm$ 3.63  \\
                  3C298    &  red    &   3.10$\pm$ 0.67  &   3.43$\pm$ 1.19  &   4.01$\pm$ 1.48  &   4.01$\pm$ 1.58  &   4.00$\pm$ 2.02  &   3.87$\pm$ 2.07  &   3.36$\pm$ 2.10  &   3.65$\pm$ 2.34  &   4.41$\pm$ 2.00  \\
                  3C298    & IR-only &   3.97$\pm$ 1.00  &   2.75$\pm$ 1.42  &   2.46$\pm$ 1.72  &   2.32$\pm$ 1.65  &   2.27$\pm$ 1.82  &   2.64$\pm$ 1.77  &   2.65$\pm$ 1.66  &   2.07$\pm$ 1.54  &   2.27$\pm$ 1.57  \\
                  3C298    &  blue   &   4.54$\pm$ 2.46  &   6.10$\pm$ 3.23  &   6.71$\pm$ 3.37  &   7.13$\pm$ 2.84  &   7.27$\pm$ 2.57  &   7.24$\pm$ 2.73  &   7.66$\pm$ 3.40  &   8.29$\pm$ 3.40  &  10.55$\pm$ 3.11  \\
                  &   &   &  &   &   &   &    &   &   &  \\
                  3C300.1  &  all    &  31.12$\pm$ 4.06  &  27.15$\pm$ 4.30  &  23.07$\pm$ 3.65  &  20.35$\pm$ 3.99  &  19.05$\pm$ 3.56  &  19.43$\pm$ 3.21  &  19.09$\pm$ 3.84  &  19.18$\pm$ 4.52  &  17.91$\pm$ 3.82  \\
                  3C300.1  &  red    &   8.95$\pm$ 1.39  &   7.91$\pm$ 2.17  &   5.37$\pm$ 2.75  &   4.33$\pm$ 2.68  &   3.71$\pm$ 2.16  &   3.37$\pm$ 1.53  &   2.71$\pm$ 1.21  &   2.36$\pm$ 1.06  &   1.91$\pm$ 1.09  \\
                  3C300.1  & IR-only &  10.03$\pm$ 2.03  &   8.65$\pm$ 3.64  &   7.61$\pm$ 4.26  &   6.26$\pm$ 3.24  &   5.54$\pm$ 2.82  &   5.42$\pm$ 2.86  &   6.21$\pm$ 2.29  &   6.87$\pm$ 2.09  &   6.81$\pm$ 1.82  \\
                  3C300.1  &  blue   &  12.14$\pm$ 2.94  &  10.60$\pm$ 3.37  &  10.09$\pm$ 3.26  &   9.76$\pm$ 3.09  &   9.80$\pm$ 2.62  &  10.64$\pm$ 3.04  &  10.17$\pm$ 3.29  &   9.95$\pm$ 3.30  &   9.20$\pm$ 2.24  \\
                  &   &   &  &   &   &   &    &   &   &  \\
                  3C305.1  &  all    &  18.54$\pm$ 1.17  &  18.71$\pm$ 1.57  &  20.08$\pm$ 2.81  &  20.53$\pm$ 3.59  &  21.11$\pm$ 3.63  &  20.76$\pm$ 4.37  &  20.57$\pm$ 3.89  &  20.28$\pm$ 3.77  &  20.30$\pm$ 4.00  \\
                  3C305.1  &  red    &   3.05$\pm$ 0.23  &   3.07$\pm$ 0.60  &   2.66$\pm$ 0.64  &   2.45$\pm$ 0.62  &   2.38$\pm$ 0.57  &   2.32$\pm$ 0.63  &   2.16$\pm$ 0.76  &   1.98$\pm$ 0.90  &   1.83$\pm$ 0.97  \\
                  3C305.1  & IR-only &   5.65$\pm$ 1.34  &   5.76$\pm$ 1.77  &   6.60$\pm$ 2.40  &   6.43$\pm$ 2.26  &   6.01$\pm$ 1.96  &   5.89$\pm$ 1.90  &   5.72$\pm$ 1.69  &   5.93$\pm$ 1.72  &   6.23$\pm$ 1.88  \\
                  3C305.1  &  blue   &   9.84$\pm$ 1.30  &   9.89$\pm$ 1.62  &  10.82$\pm$ 2.12  &  11.65$\pm$ 2.59  &  12.73$\pm$ 2.63  &  12.55$\pm$ 3.11  &  12.69$\pm$ 2.94  &  12.37$\pm$ 2.69  &  12.24$\pm$ 2.80  \\
                  &   &   &  &   &   &   &    &   &   &  \\
                  3C322    &  all    &  23.83$\pm$ 2.68  &  23.15$\pm$ 3.95  &  22.47$\pm$ 4.57  &  21.74$\pm$ 4.84  &  20.92$\pm$ 4.94  &  20.18$\pm$ 5.70  &  20.99$\pm$ 6.31  &  20.94$\pm$ 5.42  &  20.05$\pm$ 4.50  \\
                  3C322    &  red    &  11.09$\pm$ 1.02  &  10.11$\pm$ 1.59  &   9.73$\pm$ 2.18  &   9.86$\pm$ 3.04  &   9.66$\pm$ 3.50  &   9.90$\pm$ 4.24  &  10.14$\pm$ 4.57  &   9.78$\pm$ 3.71  &   9.10$\pm$ 3.08  \\
                  3C322    & IR-only &   7.61$\pm$ 1.67  &   7.88$\pm$ 1.97  &   8.17$\pm$ 2.46  &   7.58$\pm$ 2.26  &   6.86$\pm$ 2.12  &   5.51$\pm$ 1.90  &   5.22$\pm$ 1.97  &   5.24$\pm$ 2.20  &   5.26$\pm$ 2.25  \\
                  3C322    &  blue   &   5.13$\pm$ 1.39  &   5.17$\pm$ 1.42  &   4.56$\pm$ 1.49  &   4.30$\pm$ 1.65  &   4.40$\pm$ 2.15  &   4.77$\pm$ 2.10  &   5.63$\pm$ 2.61  &   5.93$\pm$ 2.49  &   5.69$\pm$ 2.55  \\
                  &   &   &  &   &   &   &    &   &   &  \\
                  3C324    &  all    &  28.89$\pm$ 2.36  &  27.22$\pm$ 4.48  &  22.57$\pm$ 5.51  &  18.93$\pm$ 4.77  &  16.26$\pm$ 4.14  &  14.04$\pm$ 3.73  &  13.40$\pm$ 3.35  &  13.84$\pm$ 3.39  &  13.85$\pm$ 2.96  \\
                  3C324    &  red    &   7.62$\pm$ 0.80  &   6.75$\pm$ 1.40  &   4.52$\pm$ 1.98  &   3.36$\pm$ 1.83  &   2.29$\pm$ 1.53  &   1.59$\pm$ 1.05  &   1.14$\pm$ 0.91  &   1.06$\pm$ 0.94  &   0.95$\pm$ 0.88  \\
                  3C324    & IR-only &   8.19$\pm$ 1.03  &   7.45$\pm$ 1.64  &   6.29$\pm$ 1.88  &   5.12$\pm$ 2.14  &   4.07$\pm$ 2.10  &   3.73$\pm$ 2.04  &   3.91$\pm$ 2.08  &   4.39$\pm$ 1.94  &   4.19$\pm$ 1.61  \\
                  3C324    &  blue   &  13.08$\pm$ 1.67  &  13.01$\pm$ 3.49  &  11.75$\pm$ 4.43  &  10.45$\pm$ 3.74  &   9.90$\pm$ 3.02  &   8.72$\pm$ 2.89  &   8.34$\pm$ 2.73  &   8.39$\pm$ 2.66  &   8.72$\pm$ 2.23  \\
                  &   &   &  &   &   &   &    &   &   &  \\
                  3C326.1  &  all    &  22.43$\pm$ 2.71  &  23.41$\pm$ 4.99  &  19.57$\pm$ 5.74  &  18.50$\pm$ 4.97  &  18.50$\pm$ 3.95  &  16.65$\pm$ 4.10  &  16.17$\pm$ 4.51  &  15.67$\pm$ 3.88  &  15.95$\pm$ 4.54  \\
                  3C326.1  &  red    &  13.78$\pm$ 1.73  &  14.90$\pm$ 3.32  &  12.73$\pm$ 4.22  &  12.19$\pm$ 3.97  &  11.81$\pm$ 3.40  &  10.40$\pm$ 3.19  &  10.14$\pm$ 3.84  &   9.24$\pm$ 3.51  &   8.97$\pm$ 3.59  \\
                  3C326.1  & IR-only &   8.64$\pm$ 1.35  &   8.31$\pm$ 2.40  &   6.11$\pm$ 2.70  &   4.96$\pm$ 2.56  &   4.98$\pm$ 2.23  &   4.40$\pm$ 1.91  &   4.32$\pm$ 1.89  &   4.91$\pm$ 2.75  &   5.30$\pm$ 3.56  \\
                  3C326.1  &  blue   &   0.00$\pm$ 0.00  &   0.20$\pm$ 0.40  &   0.73$\pm$ 1.00  &   1.35$\pm$ 1.36  &   1.71$\pm$ 1.55  &   1.85$\pm$ 1.67  &   1.71$\pm$ 1.34  &   1.53$\pm$ 1.19  &   1.68$\pm$ 1.33  \\
                  &   &   &  &   &   &   &    &   &   &  \\
                  3C356    &  all    &  19.39$\pm$ 2.38  &  17.86$\pm$ 3.75  &  17.31$\pm$ 4.54  &  16.61$\pm$ 4.03  &  16.64$\pm$ 4.49  &  17.07$\pm$ 4.38  &  17.61$\pm$ 4.75  &  17.82$\pm$ 4.59  &  17.95$\pm$ 4.35  \\
                  3C356    &  red    &   7.25$\pm$ 0.70  &   6.45$\pm$ 2.12  &   5.21$\pm$ 2.41  &   3.11$\pm$ 2.03  &   1.93$\pm$ 1.73  &   1.34$\pm$ 1.30  &   1.02$\pm$ 1.14  &   0.58$\pm$ 0.69  &   0.57$\pm$ 0.61  \\
                  3C356    & IR-only &   7.51$\pm$ 1.35  &   7.21$\pm$ 1.47  &   6.54$\pm$ 1.95  &   5.40$\pm$ 2.13  &   4.66$\pm$ 2.34  &   4.61$\pm$ 2.60  &   4.93$\pm$ 2.52  &   4.98$\pm$ 2.55  &   4.98$\pm$ 2.59  \\
                  3C356    &  blue   &   4.63$\pm$ 1.04  &   4.19$\pm$ 1.12  &   5.56$\pm$ 2.00  &   8.11$\pm$ 2.29  &  10.05$\pm$ 3.13  &  11.12$\pm$ 3.65  &  11.66$\pm$ 3.96  &  12.25$\pm$ 3.19  &  12.40$\pm$ 2.89  \\
                  &   &   &  &   &   &   &    &   &   &  \\
                  3C432    &  all    &  13.01$\pm$ 2.28  &  12.04$\pm$ 1.87  &  13.50$\pm$ 3.29  &  14.23$\pm$ 4.32  &  14.11$\pm$ 4.32  &  14.69$\pm$ 4.63  &  14.10$\pm$ 3.66  &  14.24$\pm$ 2.97  &  16.02$\pm$ 3.43  \\
                  3C432    &  red    &   5.77$\pm$ 1.50  &   5.44$\pm$ 1.52  &   6.58$\pm$ 2.31  &   7.98$\pm$ 3.31  &   8.45$\pm$ 3.87  &   9.60$\pm$ 4.38  &   9.50$\pm$ 3.53  &   9.05$\pm$ 2.32  &   9.82$\pm$ 2.66  \\
                  3C432    & IR-only &   3.46$\pm$ 1.01  &   4.12$\pm$ 1.15  &   5.03$\pm$ 1.44  &   5.30$\pm$ 1.58  &   4.93$\pm$ 1.88  &   4.13$\pm$ 1.58  &   3.49$\pm$ 1.35  &   3.90$\pm$ 1.66  &   4.74$\pm$ 2.19  \\
                  3C432    &  blue   &   3.78$\pm$ 0.75  &   2.47$\pm$ 0.66  &   1.89$\pm$ 0.57  &   0.94$\pm$ 0.66  &   0.72$\pm$ 0.87  &   0.96$\pm$ 1.15  &   1.11$\pm$ 1.07  &   1.29$\pm$ 1.10  &   1.46$\pm$ 1.24  \\
                  &   &   &  &   &   &   &    &   &   &  \\
                  3C454.1  &  all    &  15.93$\pm$ 1.62  &  18.66$\pm$ 2.69  &  19.95$\pm$ 2.92  &  18.64$\pm$ 2.86  &  17.22$\pm$ 3.39  &  15.99$\pm$ 3.87  &  15.02$\pm$ 3.81  &  14.72$\pm$ 3.69  &  13.76$\pm$ 3.65  \\
                  3C454.1  &  red    &   4.76$\pm$ 0.61  &   5.36$\pm$ 1.30  &   6.12$\pm$ 1.68  &   6.06$\pm$ 2.15  &   5.85$\pm$ 2.57  &   5.43$\pm$ 2.53  &   4.82$\pm$ 2.28  &   4.78$\pm$ 2.02  &   4.98$\pm$ 1.96  \\
                  3C454.1  & IR-only &   9.98$\pm$ 1.41  &  11.97$\pm$ 1.86  &  12.37$\pm$ 1.67  &  11.29$\pm$ 2.11  &  10.25$\pm$ 2.35  &   9.73$\pm$ 2.56  &   9.29$\pm$ 2.49  &   8.77$\pm$ 2.24  &   8.03$\pm$ 2.50  \\
                  3C454.1  &  blue   &   1.18$\pm$ 0.39  &   1.34$\pm$ 0.55  &   1.45$\pm$ 0.77  &   1.29$\pm$ 0.76  &   1.12$\pm$ 0.68  &   0.82$\pm$ 0.63  &   0.92$\pm$ 0.69  &   1.16$\pm$ 0.96  &   0.76$\pm$ 0.72  \\
                  \hline
                \end{tabular}
              \end{table*}


              \begin{table*}
                \footnotesize
                \caption{Average surface densities for center and periphery and resulting overdensities in units of $N$ galaxies per cell, 
                  using $20 < \rm{F140W} < 26$, cell radius 15$\arcsec$, and radial bins of 7$\arcsec$. 
                  Column 2 specifies the sample ($all$, $blue$, $IR$-$only$, $red$).
                  Columns 3--4 list the average surface densities for center and periphery.
                  Columns 5--7 list the resulting OD (SD cent $-$ SD peri), the standard deviation (1$\sigma$) of the periphery, and 
                  the significance of the OD in term of Nsigma~$=$~OD/$\sigma$.
                  Column 8--9 list the uncertainty of the periphery (error of the mean, EoM = $\sigma$/$\sqrt{N_{\rm ic}}$, 
                  $N_{\rm ic}$ = 13), and the signal-to-noise ratio (S/N) of the OD.
                  Column 10 lists if a significant OD or UD (negative OD) is present;
                  an OD or UD is significant, if abs(S/N) $> 3$ \,and abs(OD $> 1$).
                  Column 11 lists comments from individual inspection of the density maps in Appendix.
                  Columns 12--13 list for comparison, whether an OD or RS was found by K16, in the row $all$ and $red$, respectively,
                  and likewise for G17  in the row $red$ if an OD was found using the IRAC--PSO selection criterion (their Table~5, col.~11).
                }
                \label{tab_od}
                \begin{tabular}{ll|rrrrrrrrr|cc}
                  (1)       & (2)  & (3)  & (4) & (5)  & (6)  & (7)  &  (8)  & (9)  & (10)  & (11) & (12) & (13)\\
                  Name       &  color  & SD cent  & SD peri  & OD & Sig peri &  Nsigma & EoM & S/N &  OD/UD & comment                & K16  & G17 \\

                  \hline
                  &   &   &  &   &   &   &    &   &    &    \\
                  3C\,68.1  &    all  &    23.67   &   17.43   &    6.24   &    5.46   &    1.14   &    1.52   &    4.12   &   OD   &    &  OD  &  $-$   \\
                  3C\,68.1  &   blue  &    10.06   &    9.13   &    0.93   &    3.40   &    0.27   &    0.94   &    0.98   &   $-$  &    &  $-$ &  $-$   \\
                  3C\,68.1  &  IR-only&     9.65   &    4.93   &    4.72   &    2.01   &    2.35   &    0.56   &    8.48   &   OD   &    &  $-$ &  $-$   \\
                  3C\,68.1  &    red  &     3.96   &    3.37   &    0.60   &    1.92   &    0.31   &    0.53   &    1.12   &   $-$  &    &  RS  &  $-$   \\
                  &   &   &  &   &   &   &    &   &    &    \\

                  3C\,186  &    all  &    16.18   &   16.70   &   -0.51   &    4.11   &    0.12   &    1.14   &    0.45   &   $-$  &    &  OD  &  $-$    \\
                  3C\,186  &   blue  &     2.29   &    7.58   &   -5.29   &    2.43   &    2.17   &    0.68   &    7.84   &   UD   &    &  $-$ &  $-$    \\
                  3C\,186  &  IR-only&     8.97   &    6.36   &    2.61   &    2.24   &    1.16   &    0.62   &    4.20   &   OD   & moderate   &  $-$ &  $-$    \\
                  3C\,186  &    red  &     4.93   &    2.76   &    2.17   &    1.80   &    1.21   &    0.50   &    4.35   &   OD   & between 2 ODs    &  RS  &  OD     \\
                  &   &   &  &   &   &   &    &   &    &    \\

                  3C\,208  &    all  &    24.90   &   13.20   &   11.70   &    3.29   &    3.55   &    0.91   &   12.80   &   OD   & 20$\arcsec$ offset    &  OD  &  $-$  \\
                  3C\,208  &   blue  &    11.39   &    7.05   &    4.34   &    3.23   &    1.34   &    0.90   &    4.84   &   OD   &     &  $-$ &  $-$  \\
                  3C\,208  &  IR-only&     7.13   &    4.64   &    2.49   &    2.02   &    1.23   &    0.56   &    4.45   &   OD   & 20$\arcsec$ offset    &  $-$ &  $-$  \\
                  3C\,208  &    red  &     6.38   &    1.51   &    4.86   &    1.36   &    3.57   &    0.38   &   12.88   &   OD   &     &  $-$ &  OD   \\
                  &   &   &  &   &   &   &    &   &    &    \\

                  3C\,210  &    all  &    29.53   &   18.39   &   11.14   &    3.87   &    2.88   &    1.07   &   10.39   &   OD   & 10$\arcsec$ offset    &  OD  &  $-$   \\
                  3C\,210  &   blue  &     7.11   &   10.15   &   -3.04   &    2.84   &    1.07   &    0.79   &    3.85   &   UD   &     &  $-$ &  $-$   \\
                  3C\,210  &  IR-only&     9.64   &    5.64   &    4.00   &    2.28   &    1.75   &    0.63   &    6.33   &   OD   & 10$\arcsec$ offset    &  $-$ &  $-$   \\
                  3C\,210  &    red  &    12.78   &    2.60   &   10.18   &    1.73   &    5.87   &    0.48   &   21.16   &   OD   &     &  RS  &  OD    \\
                  &   &   &  &   &   &   &    &   &    &    \\

                  3C\,220.2  &    all  &    17.72   &   15.58   &    2.14   &    4.42   &    0.48   &    1.22   &    1.74   &   $-$  &    &  $-$ &  $-$      \\
                  3C\,220.2  &   blue  &     7.67   &    9.84   &   -2.17   &    3.38   &    0.64   &    0.94   &    2.31   &   $-$  &    &  $-$ &  $-$    \\
                  3C\,220.2  &  IR-only&     5.64   &    4.26   &    1.38   &    2.24   &    0.62   &    0.62   &    2.22   &   $-$  & marginal OD   &  $-$ &  $-$    \\
                  3C\,220.2  &    red  &     4.40   &    1.48   &    2.92   &    1.27   &    2.30   &    0.35   &    8.30   &   OD   & moderate   &  $-$ &  OD       \\
                  &   &   &  &   &   &   &    &   &    &    \\

                  3C\,230  &    all  &    26.29   &   16.26   &   10.03   &    4.59   &    2.19   &    1.27   &    7.89   &   OD   &     &  $-$ &  $-$   \\
                  3C\,230  &   blue  &     5.38   &    6.70   &   -1.32   &    2.37   &    0.56   &    0.66   &    2.00   &   $-$  &     &  $-$ &  $-$   \\
                  3C\,230  &  IR-only&     9.63   &    4.43   &    5.20   &    2.59   &    2.01   &    0.72   &    7.24   &   OD   &     &  $-$ &  $-$   \\
                  3C\,230  &    red  &    11.28   &    5.13   &    6.15   &    1.90   &    3.24   &    0.53   &   11.67   &   OD   &  15$\arcsec$ offset    &  RS  &  $-$   \\
                  &   &   &  &   &   &   &    &   &    &    \\

                  3C\,255  &    all  &    23.93   &   16.43   &    7.49   &    6.88   &    1.09   &    1.91   &    3.93   &   OD   &   &  $-$ &  $-$   \\
                  3C\,255  &   blue  &     9.38   &    8.94   &    0.44   &    3.34   &    0.13   &    0.93   &    0.48   &   $-$  &   &  $-$ &  $-$    \\
                  3C\,255  &  IR-only&     6.12   &    4.40   &    1.72   &    3.13   &    0.55   &    0.87   &    1.98   &   $-$  &   &  $-$ &  $-$    \\
                  3C\,255  &    red  &     8.43   &    3.10   &    5.33   &    2.31   &    2.31   &    0.64   &    8.33   &   OD   &   &  RS  &  $-$    \\
                  &   &   &  &   &   &   &    &   &    &    \\

                  3C\,257  &    all  &    18.56   &   14.18   &    4.38   &    3.43   &    1.28   &    0.95   &    4.60   &   OD   &  20$\arcsec$ offset   &  $-$ &  $-$    \\
                  3C\,257  &   blue  &     3.06   &    2.51   &    0.54   &    1.32   &    0.41   &    0.37   &    1.48   &   $-$  &    &  RS  &  $-$   \\
                  3C\,257  &  IR-only&     4.94   &    3.24   &    1.70   &    1.49   &    1.14   &    0.41   &    4.11   &   OD   &  marginal  &  $-$ &  $-$   \\
                  3C\,257  &    red  &    10.57   &    8.42   &    2.14   &    2.38   &    0.90   &    0.66   &    3.25   &   OD   &  extended    &  $-$ &  $-$   \\
                  &   &   &  &   &   &   &    &   &    &    \\

                  3C\,268.4  &    all  &    25.27   &   16.96   &    8.31   &    3.92   &    2.12   &    1.09   &    7.65   &   OD   & extended    &  OD&  $-$    \\
                  3C\,268.4  &   blue  &    11.42   &   10.26   &    1.16   &    3.35   &    0.35   &    0.93   &    1.25   &   $-$  &    &  $-$ &  $-$   \\
                  3C\,268.4  &  IR-only&     9.11   &    3.40   &    5.71   &    1.60   &    3.56   &    0.44   &   12.85   &   OD   & extended    &  $-$ &  $-$   \\
                  3C\,268.4  &    red  &     4.74   &    3.30   &    1.45   &    1.97   &    0.73   &    0.55   &    2.65   &   $-$  & 25$\arcsec$ offset?    &  $-$ &  $-$   \\
                  &   &   &  &   &   &   &    &   &    &    \\

                  3C\,270.1  &    all  &    20.82   &   19.02   &    1.81   &    3.45   &    0.52   &    0.96   &    1.89   &   $-$  &   &  $-$ &  $-$   \\
                  3C\,270.1  &   blue  &     5.61   &    7.48   &   -1.87   &    2.80   &    0.67   &    0.78   &    2.41   &   $-$  &   &  $-$ &  $-$   \\
                  3C\,270.1  &  IR-only&     4.51   &    4.49   &    0.02   &    1.73   &    0.01   &    0.48   &    0.04   &   $-$  &   &  $-$ &  $-$    \\
                  3C\,270.1  &    red  &    10.70   &    7.04   &    3.66   &    2.43   &    1.51   &    0.67   &    5.44   &   OD   &   &  $-$ &  $-$    \\
                  &   &   &  &   &   &   &    &   &    &    \\

                  3C\,287  &    all  &    16.83   &   19.71   &   -2.88   &    4.44   &    0.65   &    1.23   &    2.34   &   $-$  &     &  $-$ &  $-$   \\
                  3C\,287  &   blue  &     8.02   &   13.06   &   -5.04   &    3.50   &    1.44   &    0.97   &    5.20   &   UD   &     &  $-$ &  $-$   \\
                  3C\,287  &  IR-only&     6.13   &    4.75   &    1.38   &    1.67   &    0.83   &    0.46   &    2.99   &   $-$  &  between two faint ODs   &  $-$ &  $-$    \\
                  3C\,287  &    red  &     2.67   &    1.89   &    0.78   &    1.74   &    0.45   &    0.48   &    1.62   &   $-$  &     &  $-$ &  $-$    \\
                  &   &   &  &   &   &   &    &   &    &    \\

                  \hline
                \end{tabular}
              \end{table*}

              \addtocounter{table}{-1}

              \begin{table*}
                \footnotesize
                \caption{continued.}
                \label{tab_od_continued}
                \begin{tabular}{ll|rrrrrrrrr|cc}
                  (1)       & (2)  & (3)  & (4) & (5)  & (6)  & (7)  &  (8)  & (9)  & (10)  & (11) & (12) & (13)\\
                  Name       &  color  & SD cent  & SD peri  & OD & Sig peri &  Nsigma & EoM & S/N &  OD/UD & comment                & K16  & G17 \\
                  \hline
                  &   &   &  &   &   &   &    &       \\

                  3C\,297  &    all  &    20.46   &   16.51   &    3.96   &    4.20   &    0.94   &    1.16   &    3.40   &   OD   & 10$\arcsec$ offset     &  $-$ &  $-$    \\
                  3C\,297  &   blue  &     9.51   &    9.06   &    0.45   &    3.03   &    0.15   &    0.84   &    0.54   &   $-$  &     &  $-$ &  $-$    \\
                  3C\,297  &  IR-only&     5.11   &    4.82   &    0.29   &    2.03   &    0.14   &    0.56   &    0.52   &   $-$  &     &  $-$ &  $-$   \\
                  3C\,297  &    red  &     5.84   &    2.64   &    3.21   &    1.64   &    1.96   &    0.45   &    7.07   &   OD   & 20$\arcsec$ offset    &  $-$ &  $-$  \\
                  &   &   &  &   &   &   &    &   &    &    \\

                  3C\,298  &    all  &    12.12   &   14.41   &   -2.29   &    5.18   &    0.44   &    1.44   &    1.59   &   $-$  &     &  $-$  &  $-$  \\
                  3C\,298  &   blue  &     5.73   &    8.39   &   -2.66   &    3.50   &    0.76   &    0.97   &    2.73   &   $-$  &     &  $-$  &  $-$      \\
                  3C\,298  &  IR-only&     3.05   &    2.38   &    0.67   &    1.63   &    0.41   &    0.45   &    1.48   &   $-$  &     &  $-$  &  $-$       \\
                  3C\,298  &    red  &     3.35   &    3.65   &   -0.30   &    2.21   &    0.14   &    0.61   &    0.49   &   $-$  &     &  $-$  &  $-$  \\
                  &   &   &  &   &   &   &    &   &    &    \\

                  3C\,300.1  &    all  &    28.09   &   18.93   &    9.16   &    4.12   &    2.22   &    1.14   &    8.02   &   OD   &    &  OD   &  $-$     \\
                  3C\,300.1  &   blue  &    10.97   &    9.93   &    1.04   &    3.16   &    0.33   &    0.88   &    1.18   &   $-$  &    &  $-$  &  $-$     \\
                  3C\,300.1  &  IR-only&     8.97   &    6.55   &    2.43   &    2.17   &    1.12   &    0.60   &    4.02   &   OD   &  offset west   &  $-$  &  $-$   \\
                  3C\,300.1  &    red  &     8.15   &    2.45   &    5.70   &    1.17   &    4.86   &    0.32   &   17.54   &   OD   &  offset south  &  RS   &  OD    \\
                  &   &   &  &   &   &   &    &   &    &    \\

                  3C\,305.1  &    all  &    18.67   &   20.39   &   -1.72   &    3.88   &    0.44   &    1.08   &    1.60   &   $-$  &     &  $-$  &  $-$   \\
                  3C\,305.1  &   blue  &     9.87   &   12.45   &   -2.58   &    2.82   &    0.91   &    0.78   &    3.29   &   UD   &     &  $-$  &  $-$  \\
                  3C\,305.1  &  IR-only&     5.74   &    5.94   &   -0.20   &    1.77   &    0.11   &    0.49   &    0.41   &   $-$  &     &  $-$  &  $-$  \\
                  3C\,305.1  &    red  &     3.06   &    2.00   &    1.06   &    0.88   &    1.19   &    0.25   &    4.30   &   OD   &     marginal   &  $-$  &  $-$  \\
                  &   &   &  &   &   &   &    &   &    &    \\

                  3C\,322  &    all  &    23.32   &   20.66   &    2.66   &    5.45   &    0.49   &    1.51   &    1.76   &   $-$  &    &  OD   &  $-$     \\
                  3C\,322  &   blue  &     5.16   &    5.75   &   -0.59   &    2.55   &    0.23   &    0.71   &    0.84   &   $-$  &    &  $-$  &  $-$   \\
                  3C\,322  &  IR-only&     7.80   &    5.24   &    2.56   &    2.15   &    1.19   &    0.60   &    4.31   &   OD   &    at border of OD    &  $-$  &  $-$   \\
                  3C\,322  &    red  &    10.36   &    9.66   &    0.69   &    3.83   &    0.18   &    1.06   &    0.65   &   $-$  &    &  $-$  &  OD  \\
                  &   &   &  &   &   &   &    &   &    &    \\

                  3C\,324  &    all  &    27.65   &   13.65   &   14.00   &    3.30   &    4.24   &    0.92   &   15.29   &   OD   &    &  OD   &  $-$    \\
                  3C\,324  &   blue  &    13.03   &    8.43   &    4.60   &    2.62   &    1.76   &    0.73   &    6.33   &   OD   &    &  $-$  &  $-$    \\
                  3C\,324  &  IR-only&     7.64   &    4.14   &    3.50   &    1.96   &    1.78   &    0.54   &    6.43   &   OD   &    &  $-$  &  $-$   \\
                  3C\,324  &    red  &     6.97   &    1.08   &    5.90   &    0.92   &    6.44   &    0.25   &   23.22   &   OD   &    &  $-$  &  OD   \\
                  &   &   &  &   &   &   &    &   &    &    \\

                  3C\,326.1  &    all  &    23.17   &   15.95   &    7.23   &    4.30   &    1.68   &    1.19   &    6.06   &   OD   & between 2 ODs       &  OD   &  $-$   \\
                  3C\,326.1  &   blue  &     0.15   &    1.64   &   -1.49   &    1.29   &    1.15   &    0.36   &    4.16   &   UD   & between 2 ODs       &  $-$  &  $-$   \\
                  3C\,326.1  &  IR-only&     8.39   &    4.72   &    3.67   &    2.63   &    1.40   &    0.73   &    5.04   &   OD   & at border of OD     &  $-$  &  $-$   \\
                  3C\,326.1  &    red  &    14.64   &    9.59   &    5.04   &    3.71   &    1.36   &    1.03   &    4.91   &   OD   &                     &  $-$  &  OD   \\
                  &   &   &  &   &   &   &    &   &    &    \\

                  3C\,356  &    all  &    18.25   &   17.78   &    0.47   &    4.58   &    0.10   &    1.27   &    0.37   &   $-$  &    &  $-$  &  $-$     \\
                  3C\,356  &   blue  &     4.30   &   12.08   &   -7.77   &    3.43   &    2.26   &    0.95   &    8.16   &   UD   &    &  $-$  &  $-$    \\
                  3C\,356  &  IR-only&     7.29   &    4.96   &    2.33   &    2.55   &    0.91   &    0.71   &    3.29   &   OD   &    &  $-$  &  $-$   \\
                  3C\,356  &    red  &     6.66   &    0.74   &    5.92   &    0.89   &    6.63   &    0.25   &   23.91   &   OD   &    &  $-$  &  $-$   \\
                  &   &   &  &   &   &   &    &   &    &    \\

                  3C\,432  &    all  &    12.28   &   14.50   &   -2.22   &    3.46   &    0.64   &    0.96   &    2.31   &   $-$  &     &  $-$  &  $-$    \\
                  3C\,432  &   blue  &     2.80   &    1.24   &    1.56   &    1.12   &    1.39   &    0.31   &    5.03   &   OD   &  marginal?   &  $-$  &  $-$   \\
                  3C\,432  &  IR-only&     3.96   &    3.86   &    0.09   &    1.70   &    0.05   &    0.47   &    0.20   &   $-$  &     &  $-$  &  $-$  \\
                  3C\,432  &    red  &     5.52   &    9.39   &   -3.87   &    3.00   &    1.29   &    0.83   &    4.65   &   UD   &  at border of OD?    &  $-$  &  $-$  \\

                  &   &   &  &   &   &   &    &   &    &    \\

                  3C\,454.1   &    all  &    17.92   &   14.53   &    3.38   &    3.75   &    0.90   &    1.04   &    3.25   &   OD    & between two ODs   &  OD &  $-$   \\
                  3C\,454.1   &   blue  &     2.66   &    2.90   &   -0.24   &    1.80   &    0.13   &    0.50   &    0.48   &   $-$   &                   &  $-$  &  $-$  \\
                  3C\,454.1   &  IR-only&     4.93   &    4.61   &    0.31   &    2.13   &    0.15   &    0.59   &    0.53   &   $-$   & between two ODs   &  $-$  &  $-$  \\
                  3C\,454.1   &    red  &     5.20   &    4.85   &    0.34   &    2.09   &    0.16   &    0.58   &    0.59   &   $-$   & at border of OD   &  $-$  &  OD   \\
                  &   &   &  &   &   &   &    &   &    &       \\

                  \hline
                \end{tabular}
              \end{table*}


            \end{appendix}

\end{document}